\documentclass[12pt]{book}

\catcode`\@=11
\@addtoreset{equation}{section}

\usepackage{mathrsfs,amsbsy,amssymb,latexsym,amsfonts,amsmath,cite,bm}
\usepackage{graphicx,color}
\usepackage{mathtools}

\usepackage{docmute}

\global\arraycolsep=2pt 
\makeatletter
\newcommand*{\rmT}{{\mathpalette\@transpose{}}}
\newcommand*{\@transpose}[2]{\raisebox{\depth}{$\m@th#1\intercal$}}
\makeatother

\newcommand{\Exp}[1]{\operatorname{e}^{#1}}
\newcommand{\abs}[1]{\lvert {#1} \rvert}
\newcommand{\rmd}{{\mathrm{d}}}
\newcommand{\ii}{i}
\newcommand{\nn}{\nonumber}
\newcommand{\Lie}{\pounds}
\newcommand{\gLie}{\hat{\pounds}}

\newcommand{\sla}[1]{\setbox0=\hbox{$#1$} 
\dimen0=\wd0 \setbox1=\hbox{/} \dimen1=\wd1 
\ifdim\dimen0>\dimen1 \rlap{\hbox to \dimen0{\hfil/\hfil}} #1 
\else\rlap{\hbox to \dimen1{\hfil$#1$\hfil}} / \fi}

\newcommand{\WSa}{\alpha}
\newcommand{\WSb}{\beta}

\newcommand{\cA}{\mathcal A}
\newcommand{\cC}{\mathcal C}\newcommand{\cD}{\mathcal D}
\newcommand{\cE}{\mathcal E}\newcommand{\cF}{\mathcal F}
\newcommand{\cH}{\mathcal H}
\newcommand{\cI}{\mathcal I}
\newcommand{\cK}{\mathcal K}\newcommand{\cL}{\mathcal L}
\newcommand{\cM}{\mathcal M}
\newcommand{\cO}{\mathcal O}\newcommand{\cP}{\mathcal P}

\newcommand{\cS}{\mathcal S}\newcommand{\cT}{\mathcal T}

\newcommand{\cZ}{\mathcal Z}

\newcommand{\bmF}{{\bm{F}}}
\newcommand{\bmp}{{\bm{p}}}
\newcommand{\bmq}{{\bm{q}}}
\newcommand{\bmr}{{\bm r}}
\newcommand{\bms}{{\bm{s}}}

\newcommand{\bisC}{\hat{\bm{\cC}}}
\newcommand{\bisF}{\hat{\bm{\cF}}}

\newcommand{\OO}{\text{O}}

\newcommand{\SO}{\text{SO}}

\newcommand{\GL}{\mathrm{GL}}

\newcommand{\dd}{{\rm d}}

\newcommand{\diag}{{\rm{diag}} }

\newcommand{\ga}{\gamma}

\newcommand{\la}{\lambda}

\newcommand{\Tr}{{\rm Tr}}
\newcommand{\str}{{\rm STr}}

\newcommand{\alg}[1]{\mathfrak{#1}}

\newcommand{\nln}{\nonumber\\}

\newcommand{\Z}{\mathcal Z}

\newcommand*{\AdS}[1]{\ensuremath{\text{AdS}_{#1}}}

\newcommand{\bfr}{{\bm r}}
\newcommand{\bX}{{\bm X}}

\newcommand{\ket}[1]{\lvert {#1} \rangle}
\newcommand{\bra}[1]{\langle {#1} \rvert}
\newcommand{\bS}{{\mathring{\cS}}{}}
\newcommand{\bbS}{{\mathbb{S}}}
\newcommand{\bbX}{\mathbb{X}}

\newcommand{\Loa}{a}
\newcommand{\Lob}{b}
\newcommand{\Loc}{c}
\newcommand{\Lod}{d}
\newcommand{\Loe}{e}
\newcommand{\Lof}{f}
\newcommand{\Log}{g}
\newcommand{\Loh}{h}

\newcommand{\Lobra}{\bar{a}}
\newcommand{\Lobrb}{\bar{b}}
\newcommand{\Lobrc}{\bar{c}}
\newcommand{\Lobrd}{\bar{d}}

\newcommand{\GV}{V}
\newcommand{\brGV}{\bar{V}}

\newcommand{\brTheta}{\bar{\Theta}}
\newcommand{\brbrTheta}{\bar{\bar{\Theta}}}
\newcommand{\brtheta}{\bar{\theta}}
\newcommand{\brOmega}{\bar{\Omega}}
\newcommand{\ODDeta}{\eta}
\newcommand{\OG}{G}
\newcommand{\CG}{g}
\newcommand{\Einv}{\mathcal{E}}

\newcommand{\sfa}{{\mathsf{a}}}
\newcommand{\sfb}{{\mathsf{b}}}
\newcommand{\sfc}{{\mathsf{c}}}
\newcommand{\sfd}{{\mathsf{d}}}
\newcommand{\sfe}{{\mathsf{e}}}
\newcommand{\sff}{{\mathsf{f}}}
\newcommand{\sfi}{{\mathsf{i}}}
\newcommand{\sfj}{{\mathsf{j}}}
\newcommand{\sfk}{\mathsf{k}}

\newcommand{\sfD}{{\mathsf{D}}}
\newcommand{\sfP}{{\mathsf{P}}}
\newcommand{\sfJ}{{\mathsf{J}}}
\newcommand{\sfK}{{\mathsf{K}}}

\newcommand{\rmc}{\text{c}}
\newcommand{\rmC}{\text{C}}
\newcommand{\rmD}{\text{D}}
\newcommand{\rmS}{\text{S}}
\newcommand{\TT}{\text{T}}

\newcommand{\sfF}{\mathsf F}

\newcommand{\dlT}{T}
\newcommand{\Pg}{P}
\newcommand{\gga}{\gamma}

\newcommand\bre{\bar{e}}
\newcommand\breta{\bar{\eta}}
\newcommand\brgamma{\bar{\gamma}}
\newcommand\bromega{\bar{\omega}}
\newcommand\brPhi{\bar{\Phi}}

\newcommand{\SPa}{\alpha}
\newcommand{\SPb}{\beta}
\newcommand{\SPc}{\gamma}

\newcommand{\SPbra}{\bar{\SPa}}
\newcommand{\SPbrb}{\bar{\SPb}}
\newcommand{\SPbrc}{\bar{\SPc}}

\newcommand\brC{\bar{C}}
\newcommand\brP{\bar{P}}
\newcommand\brGamma{\bar{\Gamma}}

\newcommand{\no}{\nonumber}

\newcommand{\gP}{{\rm\bf P}}
\newcommand{\gJ}{{\rm\bf J}}
\newcommand{\gQ}{{\rm\bf Q}}

\newcommand{\bos}{\text{b}}
\newcommand{\fer}{\text{f}}
\newcommand{\Ad}{{\rm Ad}}
\newcommand{\sgn}{{\rm sgn}}
\newcommand{\CYBE}{{\rm CYBE}}
\newcommand{\YB}{{\rm YB}}
\newcommand{\Pf}{{\rm Pf}}
\newcommand{\Span}{{\rm span}}
\newcommand{\ST}{{\rm st}}

\newcommand{\bB}{{\bm B}}
\newcommand{\bbeta}{{\bm\beta}}
\newcommand{\bnabla}{{\mathring{\nabla}}{}}
\newcommand{\bGamma}{{\mathring{\Gamma}}{}}

\newcommand\Atop[2]{\genfrac{}{}{0pt}{}{#1}{#2}}

\makeatother

\begin{document}

%

\vspace*{3.0cm}

\begin{center}
{\Huge \bf
\scalebox{0.9}[1.1]{\hspace{-5mm}Integrable deformations of string sigma models }}\\
\vspace{3mm} 
{\Huge \bf and }\\
\vspace{5mm} 
{\Huge \bf
\scalebox{0.9}[1.1]{generalized supergravity}}

\vspace{20mm} 

{\Large \bf Jun-ichi Sakamoto%
\footnote{E-mail address: \texttt{quantumfield15@g-mail.com}}
}\\
\vspace*{1.0cm}

\vspace{5mm} 

\begin{figure}[h]
\begin{center}
 \includegraphics[clip,width=4.0cm]{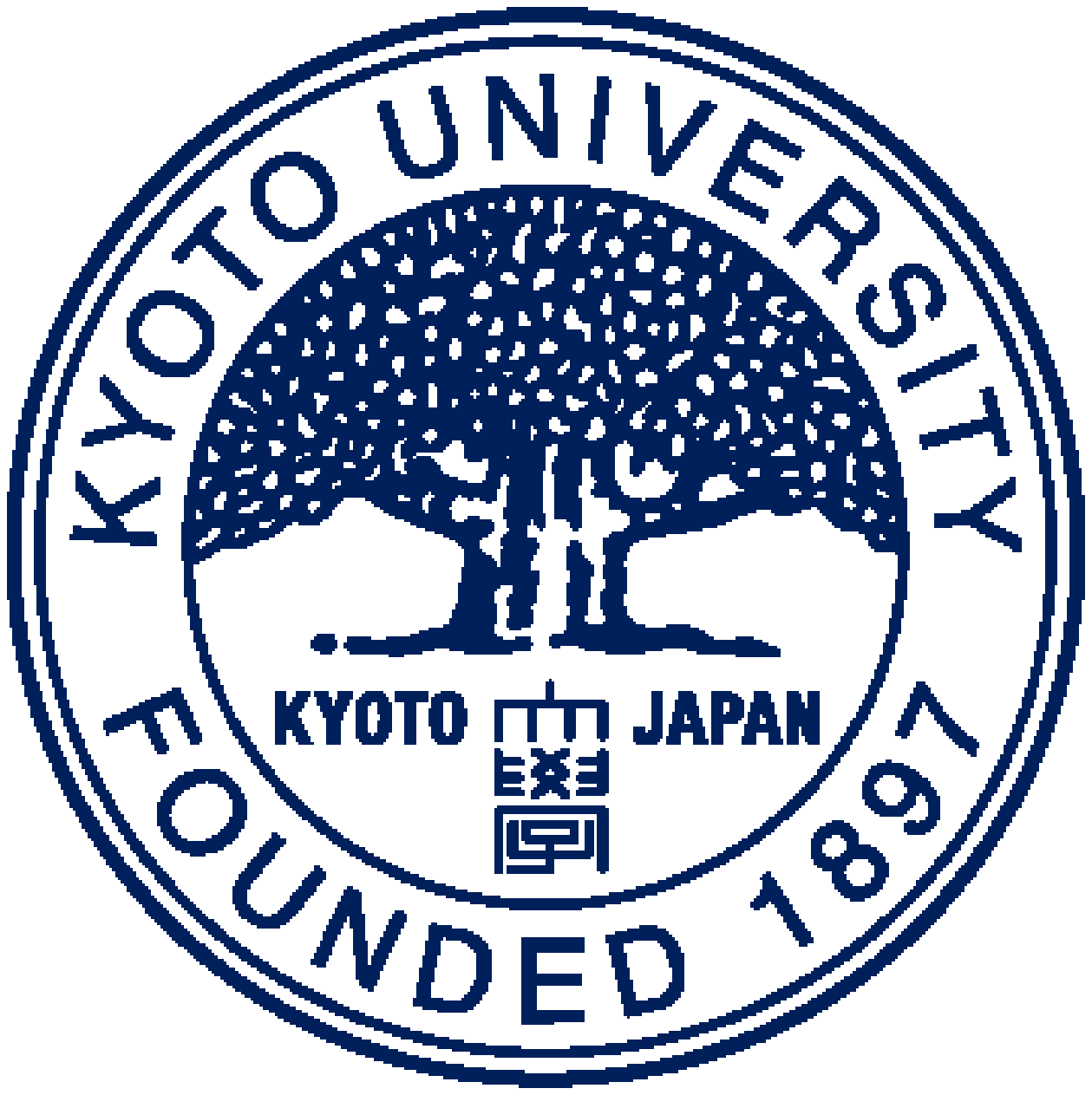}
\end{center}
\end{figure}
\end{center}

\vspace{-10mm} 

\begin{center}
{\large A Dissertation submitted to the Kyoto University\\
for the degree of Doctor of Philosophy}\\

\vspace{10mm} 

{\large January 2019}
\end{center}

\thispagestyle{empty}

\newpage
  
\thispagestyle{empty}
\newpage
                 
\pagenumbering{roman}
\pagestyle{plain}

\vspace{-5mm} 
\begin{center}
\section*{Abstract}
\end{center}
\vspace{0mm} 
{\normalsize

This thesis is mainly devoted to studying integrable deformations of the $\AdS{5} \times \rmS^5$ superstring
and generalized supergravity.\par
We start to give a brief review of the $\AdS{5} \times \rmS^5$ superstring formulated in the Green-Schwartz formalism,
and then introduce homogeneous Yang-Baxter (YB) deformations of the $\AdS{5} \times \rmS^5$ superstring based on $r$-matrices which are solutions to the homogeneous classical YB equation.
By performing a supercoset construction,
we derive the general formula for homogeneous YB deformed backgrounds associated with bosonic $r$-matrices.
The resulting deformed backgrounds are shown to be solutions of the standard type IIB supergravity or generalized supergravity.\par
Next, in chapter \ref{Ch:YB-duality}, we relate homogeneous YB deformations to string duality transformations.
We first review double field theory which is a $T$-duality covariant formulation for the massless sector of string theory.
After that, we explain that homogeneous YB deformation can be regarded as a kind of the $O(d,d)$ duality transformations.
Once homogeneous YB deformations are realized as duality transformations,
the corresponding $O(d,d)$ transformations are applied to almost all backgrounds.\par
Moreover, in chapter \ref{Ch:YB-T-fold}, we discuss spacetime structures of homogeneous YB deformed backgrounds
and clarify a T-fold structure of them by showing the associated $O(d,d; \mathbb{Z})$ $T$-duality monodromy.\par
In chapter \ref{Ch:Weyl-GSE},
we consider the Weyl invariance of string theories in generalized supergravity backgrounds.
We show that generalized supergravity can be reproduced from double field theory
with the dilaton depending on a linear dual coordinate.
From this result, we construct a possible counterterm to cancel out the Weyl anomaly of bosonic string theories on generalized supergravity backgrounds.
In particular, we show that the counterterm is definitely local.
In this sense, string theories can be consistently defined on generalized supergravity backgrounds. \par
Finally, in chapter \ref{Ch:concl}, we summarize the results presented in this thesis and conclude by
mentioning open problems for future directions.
}

\tableofcontents

　
　
\newpage
　\thispagestyle{empty}
\newpage

\pagenumbering{arabic}
\pagestyle{myheadings}


\chapter{Introduction}
\section{AdS/CFT correspondence}

The AdS/CFT correspondence \cite{Maldacena:1997re,Gubser:1998bc,Witten:1998qj} is a fascinating topic in string theory.
This duality is a conjecture which states an equivalence between quantum gravity on $d+1$-dimensional anti-de Sitter (AdS) and $d$-dimensional conformal field theory (CFT).
The correspondence is a concrete realization of the holographic principle originally proposed by t'Hooft \cite{tHooft:1973alw}.

\medskip

The most striking example is the $\AdS{5}/{\rm CFT}_4$ correspondence.
This is correspondence between the $SU(N)$ $\mathcal{N}=4$\,supersymmetric Yang-Mills theory (SYM) and
type IIB superstring theory on the $\AdS{5} \times \rmS^5$ background.
As a consistency check,
it is easy to see that both theories have the same global symmetry $SO(2,4)\times SO(6)$\,.
For the gauge theory side, $SO(2,4)$ and $SU(4)$ are realized as the four-dimensional conformal group and the $R$-symmetry of $\mathcal{N}=4$ supersymmetry, respectively.
On the other hand, the $\AdS{5} \times \rmS^5$ background has the global symmetry as an isometry on the spacetime.
Furthermore, if we consider the fermionic symmetries of both theories,
the global symmetry is extended to the supergroup $PSU(2,2|4)$\,.

\medskip

Typt IIB superstring theory in the $\AdS{5} \times \rmS^5$ background with the radius $R$ of $\AdS{5}$ and $\rmS^5$ is described by a non-linear sigma model with the dimensionless coupling constant which is the effective string tension
\begin{align}
T=\frac{R^2}{2\pi\alpha'}\no\,.
\end{align}
In the AdS/CFT correspondence, 
the effective tension $T$ and the string coupling $g_s$ are related to t'Hooft coupling $\lambda:=g^2_{\rm YM}N$ and the rank of the gauge group $N$ as
\begin{align}
g_s=\frac{\la}{4\pi\,N}\,,\qquad
T=\frac{\sqrt{\lambda}}{2\pi}\no\,.
\end{align}
In general parameter regions,
both theories are highly complicated, and the equivalence is extremely non-trivial.
For this issue, we shall take the t'Hooft limit
\begin{align}
N\to\infty\,,\qquad \lambda={\rm fixed}\no\,.
\end{align}
The limit suppresses non-planar interactions and corresponds to consider a free string on the $\AdS{5} \times \rmS^5$ background.

\medskip

Moreover, both theories in this limit display a remarkable property.
In certain aspects, integrable structures emerge
and indicate an infinite set of conserved charges.
The integrable structures of the gauge theory side were first discovered in a study of the spectral problem.
In a grand breaking work \cite{Minahan:2002ve},
Minahan and Zarembo showed that the dilatation operator of the $\mathcal{N}=4$ SYM acting on the gauge invariant single trace operators can be mapped to the Hamiltonian of an integrable spin chain model.
As a result, the scaling dimensions of the single trace operators are given by diagonalizing the integrable spin chain's Hamiltonian. 

\medskip

On the other hand, the $\AdS{5} \times \rmS^5$ superstring is described by a classical integrable non-linear sigma model.
The action is formulated in the Green-Schwarz formalism and
the target space has the supercoset\cite{Metsaev:1998it}
\begin{align}
\frac{PSU(2,2|4)}{SO(1,4)\times SO(5)}\no\,.
\end{align}
The $\mathbb{Z}_4$-grading structure of the supercoset allows for the Lax pair,
and therefore the string sigma model is classical integrable\cite{Bena:2003wd}
because the monodromy matrix constructed from the Lax pair can generate an infinite conserved charge.

\medskip

From the identification of the symmetry,
the global energy of a string state corresponds to the scaling dimension of a single trace operator.
These physical observables including non-BPS sectors can be computed even at finite t'Hooft coupling $\lambda$ by using integrability techniques, such as Thermodynamic Bethe Ansatz\cite{Gromov:2009bc,Bombardelli:2009ns,Arutyunov:2009ur}, Y-system\cite{Gromov:2009tv,Cavaglia:2010nm}, and the quantum spectral curve\cite{Gromov:2013pga,Gromov:2014caa}.
Therefore, we could directly verify the correspondence at the planar level.

\medskip

More recently, it has been proposed an integrability-based approach to compute higher-point functions in $\mathcal{N}=4$ SYM at a finite coupling.
The approach referred to as the hexagon formalism was originally proposed in \cite{Basso:2015zoa}, 
and it was shown that the technique could be also adapted to higher-point functions \cite{Eden:2016xvg,Fleury:2016ykk}.
In this way, integrability of the AdS/CFT correspondence plays an important role in the verification of the correspondence (for a comprehensive review, see \cite{Beisert:2010jr}).

\section{Integrable deformations of AdS/CFT correspondence}

\subsubsection*{Yang-Baxter deformation of the $\AdS{5} \times \rmS^5$ superstring}

As a next step,
generalizations to integrable systems with less supersymmetry or non-conformal symmetry have been considered by many authors.
In this thesis, we concentrate on deformations of the $\AdS{5} \times \rmS^5$ superstring preserving the classical integrability.
One of the approaches
\footnote{Another important class of integrable deformations is the $\lambda$-deformation \cite{Sfetsos:2013wia,Demulder:2015lva,Hollowood:2014qma,Hollowood:2014rla,Borsato:2016ose}.
As discussed in \cite{Vicedo:2015pna,Hoare:2015gda,Sfetsos:2015nya,Klimcik:2015gba}, this deformation is related to the $q$-deformation via the Poisson-Lie T-duality \cite{Klimcik:1995ux,Klimcik:1995jn,Klimcik:1995dy} ( see \cite{Hassler:2017yza,Jurco:2017gii,Hoare:2017ukq,Hoare:2018ebg,Severa:2018pag,Demulder:2018lmj,Sakatani:2019jgu} for recent studies on the Poisson-Lie T-duality).}
 is to use the Yang-Baxter (YB) deformation which is a systematic way describing integrable deformations of two-dimensional non-linear sigma models.
An important point is that a YB deformation can be specified by taking a skew-symmetric classical $r$-matrix
solving the classical Yang-Baxter equation (CYBE).

\medskip

The YB deformation was originally introduced as integrable deformations of the principal chiral model for a Lie group $G$ by Klimcik\cite{Klimcik:2002zj}
and its classical integrability was shown in \cite{Klimcik:2008eq}.
In the original case, it had been taken the Drinfeld-Jimbo type $r$-matrix \cite{Drinfeld:1985rx,Jimbo:1985zk} which is a solution of the modified CYBE.
The method had been generalized to symmetric coset sigma models \cite{Delduc:2013fga},
and furthermore to the $\AdS{5} \times \rmS^5$ string sigma model \cite{Delduc:2013qra,Delduc:2014kha} which is frequently called the $q$-deformed
$\AdS{5} \times \rmS^5$ superstring
\footnote{The $q$-deformed $\AdS{5} \times \rmS^5$ background is also called the $\eta$-deformed $\AdS{5} \times \rmS^5$ or the ABF background.
}
.

\medskip

The metric and NS-NS two-form of the $q$-deformed $\AdS{5} \times \rmS^5$ background have been obtained by performing a coset construction for the bosonic part\cite{Arutyunov:2013ega}.
The derivation of full $q$-deformed background was a challenging problem due to the technical difficulty of the supercoet construction.
In a pioneer work \cite{Arutyunov:2015qva}, Arutyunov, Borsato, and Frolov achieved doing the supercoset construction and the
full background was given.
A remarkable point is that the resulting background does not satisfy the equations of motion of type IIB supergravity
\footnote{In \cite{Hoare:2018ngg}, by using Drinfeld-Jimbo type $r$-matrices with different fermionic structures,
another $q$-deformed $\AdS{2} \times \rmS^2\times T^4$ and $\AdS{5} \times \rmS^5$ backgrounds were constructed
and shown to be solutions of the standard supergravity.}.
This fact led to the discovery of the generalized supergravity equations \cite{Arutyunov:2015mqj,Wulff:2016tju} explained in the next section.
Hereafter, we will refer to them as the generalized supergravity equations of motion (GSE).

\medskip

We can consider another type of the YB deformation based on the homogeneous CYBE.
We shall call it the homogeneous YB deformation and focus on it throughout the rest of this thesis.
The deformations of PCMs and symmetric coset sigma models had been developed in \cite{Matsumoto:2015jja}.
Moreover, it had been generalized to the $\AdS{5} \times \rmS^5$ superstring case in \cite{Kawaguchi:2014qwa}.
Some YB deformations led to various type IIB SUGRA backgrounds \cite{Matsumoto:2014nra,Matsumoto:2014gwa,Matsumoto:2015uja,Kyono:2016jqy} including the well-known examples such as Lunin-Maldacena-Frolov backgrounds\cite{Lunin:2005jy,Frolov:2005dj}, gravity duals of non-commutative gauge theories \cite{Hashimoto:1999ut,Maldacena:1999mh} and Schr\"odinger spacetime \cite{Herzog:2008wg,Maldacena:2008wh,Adams:2008wt}.

\medskip

The homogeneous YB deformations also give solutions of the GSE as with the $q$-deformation.
To obtain deformed backgrounds solving the usual supergravity equations,
we need to impose a condition, the unimodularity condition, on the associated $r$-matrix.
If we take a skew-symmetric $r$-matrix
\begin{align}
r&=\frac{1}{2}r^{ij}T_i\wedge T_j=r^{ij}T_i\otimes T_j\,,\no\\
 r^{ij}&=-r^{ji}={\rm\,const.}\,,\quad T_i\in \mathfrak{g}\,,\no
\end{align}
the unimodularity condition has a very simple form\cite{Borsato:2016ose}
\begin{align}
r^{ij}[T_i, T_j]=0\,.\no
\end{align}
Here, the Lie algebra $\mathfrak{g}$ is $\mathfrak{su}(2,2|4)$ when we consider YB deformations of the $\AdS{5} \times \rmS^5$ superstring.
In particular, if a given $r$-matrix doesn't satisfy the condition,
the associated background is a solution of generalized supergravity.

\subsubsection*{YB deformation as a duality transformation}

Remarkably, the homogeneous YB deformation can be regarded as a string duality transformation
\cite{Matsumoto:2014nra,Matsumoto:2014gwa,Matsumoto:2015uja,Osten:2016dvf,Orlando:2016qqu,Hoare:2016wsk,Borsato:2016pas,Borsato:2017qsx,Sakamoto:2017cpu,Sakamoto:2018krs,Araujo:2018rho,Borsato:2018spz,Hoare:2016wca,Sakamoto:2016ppx,Bakhmatov:2018apn,Lust:2018jsx,Araujo:2018rbc,Borsato:2018idb,Bakhmatov:2018bvp}.
Indeed, the above well-known backgrounds, such as Lunin-Maldacena-Frolov backgrounds\cite{Lunin:2005jy,Frolov:2005dj},
 can also be reproduced by the TsT-transformations
which is a combination of $T$-dualities and coordinate changes.
As observed in a series of works \cite{Matsumoto:2014nra,Matsumoto:2014gwa,Matsumoto:2015uja} and shown in \cite{Osten:2016dvf},
the homogeneous YB deformations describe TsT-transformations
when we take Abelian $r$-matrices,
\begin{align}
r=\frac{1}{2}r^{ij}T_i\wedge T_j\,,\qquad [T_i,T_j]=0\,.\no
\end{align}
The relationship to duality transformations is further extended to the case of non-Abelian $r$-matrices.
In \cite{Orlando:2016qqu}, it had been shown that a specific class of non-Abelian YB-deformed deformations could also be realized as a generalization of the TsT-transformation.
Furthermore, 
Hoare and Tseytlin\cite{Hoare:2016wsk} had conjectured that the homogeneous YB deformations are equivalent to non-Abelian $T$-dualities \cite{delaOssa:1992vci,Giveon:1993ai,Alvarez:1994np,Sfetsos:2010uq,Lozano:2011kb}
up to performing a gauge transformation of the $B$-field and appropriate field redefinitions.
Thereafter, the conjecture was shown in \cite{Borsato:2016pas,Borsato:2017qsx}.

\medskip

The above relations have been studied in terms of a manifestly $T$-duality covariant formulation for the massless sector of string theory, i.e. Double Field Theory (DFT).
The $T$-duality covariance is accomplished by introducing the dual coordinates.
In a series of works \cite{Sakamoto:2017cpu,Sakamoto:2018krs},
by reformulating the homogeneous YB deformations, it was clarified that the deformation is precisely equivalent to a $\beta$-transformation which is a kind of the 
$O(d,d)$ $T$-transformations.
Once the homogeneous YB deformations are realized as $\beta$-transformations,
we write down a very simple formula for general deformed backgrounds \cite{Sakamoto:2017cpu,Sakamoto:2018krs}.

\medskip

Such reformulation has some advantages.
If the original background is not described by a symmetric coset or supported by a non-trivial $H$-flux,
it is not straightforward to define a YB sigma model in general.
However, once the homogeneous YB deformations are realized as duality transformations,
we can apply ``YB deformations'' to almost all backgrounds such as Minkowski spacetime\cite{Borowiec:2015wua, Kyono:2015zeu,Matsumoto:2015ypa,Pachol:2015mfa,Fernandez-Melgarejo:2017oyu}, pp-wave and $\AdS{3} \times \rmS^3\times T^4$ with $H$-flux\cite{Sakamoto:2018krs,Araujo:2018rho,Borsato:2018spz}
(see \cite{Crichigno:2014ipa,Sakamoto:2016ppx,Negron:2018btz} for other backgrounds)
\footnote{As another application of YB deformations, the deformations of Jackiw-Teitelboim model \cite{Jackiw:1984je,Teitelboim:1983ux} were discussed in \cite{Kyono:2017jtc,Kyono:2017pxs, Okumura:2018xbh,Roychowdhury:2018clp,Lala:2018yib}.}.
In particular, the computation is straightforward.
In this way, the $\beta$-transformation is a new useful tool to generate solutions of (generalized) supergravity.

\medskip

Moreover, the DFT enables us to discuss the global structure of non-geometric spacetimes which are stringy geometries whose structure group contains $T$-duality transformations.
Such spacetime is called $T$-fold in some literature.
It was clarified that the homogeneous YB deformed background has the non-geometric structures\cite{Fernandez-Melgarejo:2017oyu}.

\section{Generalized supergravity}

The generalized supergravity equations was originally proposed to support a $q$-deformed $\AdS{5} \times \rmS^5$ background as a solution\cite{Arutyunov:2015mqj}.
The classical action has not been revealed yet, and only the equations of motion are presented.
The generalized supergravity includes an extra Killing vector field $I$ as well as the usual supergravity fields.
For YB deformed backgrounds,
the Killing vector field $I$ measures the violation of the unimodularity condition.
In particular, the Killing vector $I$ appearing in the GSE implies the existence of the trace of the $Q$-flux
which measures non-geometricity of a given background.
Therefore, generalized supergravity intrinsically has a $T$-fold structure.

\medskip

A remarkable feature of this theory is that the GSE can be reproduced from the requirement of the $\kappa$-symmetry in the Green-Swartz formalism.
It was originally well known that the on-shell constraints of type II supergravity ensure the $\kappa$-symmetry of the associated Green-Schwarz type string sigma model \cite{Howe:1983sra,Grisaru:1985fv}.
At the same time, it had been conjectured that the $\kappa$-symmetry requires the type II supergravity equations. 
However, after about 30 years, 
Tseytlin and Wulff \cite{Wulff:2016tju} had shown that the equations of motion of type II generalized supergravity could be reproduced by solving the $\kappa$-symmetry constraints on spacetime fields.
In this way, an old fundamental problem of string theory had been resolved.

\medskip

This result implies that, at the classical level,
string theory is consistently defined on generalized supergravity backgrounds.
However, the quantum consistency of string theories defined on such backgrounds is not apparent.

\subsubsection*{Weyl invariance for generalized supergravity background}

As explained in the previous section,
the homogeneous YB deformations could be reformulated in the DFT.
This implies that the DFT can reproduce both the usual and generalized supergravity from a single action.

\medskip

It is well known that the DFT can reproduce the standard supergravity equations by taking a solution of the section condition
which all spacetime fields don't depend on the dual coordinates.
For the GSE, this story is slightly modified.
As shown in \cite{Sakamoto:2017wor}, the associated dilaton has the linear dual coordinate dependence 
\begin{align}
\Phi_*=\Phi+I^i\tilde{Y}_i\,,\no
\label{intro:dilaton}
\end{align}
where $\Phi$ is the usual dilaton and $\tilde{Y}_i$ is the dual coordinate corresponding to the Killing direction $I$\,.
In particular, the linear dual coordinate dependence is consistent with the section condition.
Therefore, generalized supergravity is realized as a non-standard solution of the section condition in the DFT
.

\medskip

The modification of the dilaton teaches us how we generalize the well-known counterterm to cancel out the Weyl anomaly of the string sigma model. 
When we consider a string in the standard supergravity backgrounds,
the counterterm to cancel the Weyl anomaly was proposed by Fradkin and Tseytlin \cite{Fradkin:1984pq} and is given by 
\begin{align}
S_{\rm FT}=\int \rmd^2\sigma\,\sqrt{-\gamma}R^{(2)}\Phi\,,\no
\end{align}
where $R^{(2)}$ is the Ricci scalar for the world-sheet metric $\gamma^{\alpha\beta}$.
Since the counterterm has higher order degree in $\alpha'$ than the one of the classical action,
it should be regarded as a quantum correction.
From the discussion of the previous paragraph,
it is straightforward to generalize the Fradkin-Tseytlin term.
In \cite{Sakamoto:2017wor, Fernandez-Melgarejo:2018wpg},
Weyl invariance of bosonic string theory in generalized supergravity backgrounds was considered and
a possible counterterm to cancel the Weyl anomaly was proposed as
\footnote{Recently, Weyl invariance of type I superstring in generalized supergravity backgrounds was discussed in \cite{Muck:2019pwj}
and a possible local counterterm was constructed.}
\begin{align}
S_{\rm FT}^{(*)}=\int \rmd^2\sigma\,\sqrt{-\gamma}R^{(2)}\Phi_*\,.\no
\end{align}
It is also noted that the Killing vector $I$\,, which appears in the
equations of motion of generalized supergravity, does not appear in the classical string
sigma model action, but enters firstly as a quantum correction at stringy level.
In particular, the counterterm has a local expression as discussed in \cite{Fernandez-Melgarejo:2018wpg}.
In this sense, string theories may consistently be defined on the generalized supergravity backgrounds at the quantum level. 

\section{Outline of this thesis}

We shall present the outline of this thesis.

\medskip

In chapter \ref{Ch:YB-AdS5}, we start to give a small review the $\AdS{5} \times \rmS^5$ superstring
and then introduce the homogeneous Yang-Baxter deformations of the $\AdS{5} \times \rmS^5$ superstring.
The deformed action is shown to be classical integrable and invariant under the $\kappa$-symmetry transformation.
We present the general formula for YB deformed backgrounds associated with $r$-matrices which are only composed of bosonic generators.
The derivation of the formula is explained in the next chapter.
By using the formula, some examples of the homogeneous YB deformed $\AdS{5} \times \rmS^5$ backgrounds
associated with Abelian $r$-matrices and non-unimodular $r$-matrices are obtained.
The former leads to solutions of the standard supergravity,
and the latter gives solutions of the GSE.

\medskip

The main purpose of chapter \ref{Ch:YB-duality} is to relate the homogeneous YB deformations to a string duality transformation.
We first review DFT and explain the transformation rule of the spacetime fields under the $O(d,d)$ T-duality transformations.
Then, we directly derive the general formula for YB deformed backgrounds
from the deformed GS action.
After that, 
we show that a homogeneous YB deformation is equivalent to a $\beta$-transformation.
Once homogeneous YB deformations are regarded as duality transformations,
the deformations can be applied to arbitrary background with non-trivial isometry.
As an example, we consider $\beta$-deformations of the $\AdS{3} \times \rmS^3\times T^4$ background
supported by a $H$-flux.

\medskip

In chapter \ref{Ch:YB-T-fold}, we discuss the spacetime structures of the YB deformed backgrounds.
We clarify a T-fold structure of YB-deformed backgrounds by showing the associated $O(D,D;\mathbb{Z})$ T-duality monodromy. 
In particular, the Killing vector $I$ appearing in the GSE implies the existence of the trace of the non-geometric $Q$-flux.
Therefore, generalized supergravity is a theory which describes non-geoemtric backgrounds.

\medskip

In chapter \ref{Ch:Weyl-GSE},
we consider the Weyl invariance of string theories in generalized supergravity backgrounds.
We start to show that generalized supergravity can be reproduced from double field theory.
In this case, we see the dilaton has a linear dual coordinate dependence.
From this observation, we propose a possible counterterm to cancel out the Weyl anomaly of bosonic string theories in generalized supergravity backgrounds.
In particular, we can show that the counterterm is local.
In this sense, the string theories can be consistently defined on generalized supergravity backgrounds. 

\medskip

Finally, in chapter \ref{Ch:concl}, we summarize the results presented in this thesis and conclude by
mentioning open problems for future directions.


\chapter{Integrable deformations of the $\AdS{5}\times $S$^5$ superstring}
\label{Ch:YB-AdS5}

In this chapter, we discuss Yang-Baxter (YB) deformations of the $\AdS{5}\times $S$^5$ superstring.
In this thesis, we concentrate on the YB deformations based on the homogeneous CYBE.
This chapter is organized as follows.
Before considering the deformations,
we shall give a brief review of the $\AdS{5}\times $S$^5$ superstring.
In section \ref{sec:YB-sigma},
we then introduce the homogeneous YB deformations of the $\AdS{5}\times $S$^5$ superstring.
It is shown that the deformed system is classically integrable and has the $\kappa$-symmetry.
We present the simple formula for the deformed backgrounds which can be regarded as a kind of $O(10,10)$ $T$-duality transformations.
Finally, in section \ref{sec:ExampleYBAdS5}, we present various YB deformed $\AdS{5}\times $S$^5$ backgrounds by using the formula as presented in the previous section.
We show that when a given $r$-matrix satisfies the unimodularity condition,
the associated deformed background is a solution of the generalized supergravity equations.

\section{The $\AdS{5}\times $S$^5$ superstring}

In the section, we will briefly review some basic facts on the $\AdS{5}\times $S$^5$ superstring. 

\subsection{Metsaev--Tseytlin action}

The dynamics of the superstring on the $\AdS{5}\times $S$^5$ background is described by the supercoset
\begin{align}
 \frac{PSU(2,2|4)}{SO(1,4)\times SO(5)}\,.
\end{align}
The action of the $\AdS{5}\times $S$^5$ superstring in the Green-Shwartz formulation had been written down by Metsaev and Tseytlin \cite{Metsaev:1998it}.

\medskip

The action the $\AdS{5}\times $S$^5$ superstring is formulated in the Green-Shwartz formalism and has the form
\begin{align}
S&=-\frac{T}{2}\int \rmd^2\sigma\,\Pg_-^{\WSa\WSb} \, \str \bigl[\,A_{\alpha} \, d_-(A_{\beta})\,\bigr]\,,
\label{AdS5S5-action}
\end{align}
where $T \equiv R^2/2\pi\alpha'$ ($R$\,: the radius of $\AdS{5}$ and $\rmS^5$) is the effective tension of a string.
$\Pg_{\pm}^{\WSa\WSb}$ are a linear combination of the metric on the world-sheet $\gamma^{\alpha\beta}$
and the anti-symmetric tensor $\epsilon^{\alpha\beta}$\,,
\begin{align}
 \Pg_{\pm}^{\WSa\WSb}\equiv\,\frac{\gga^{\WSa\WSb}\pm \varepsilon^{\WSa\WSb}}{2}\,.
\end{align}
We take the conformal gauge $\gamma^{\alpha\beta}={\rm diag}(-1,1)$ and normalize $\varepsilon^{\alpha\beta}$ as $\varepsilon^{\tau\sigma}=\frac{1}{\sqrt{-\gamma}}$\,.
$A$ is the left-invariant $1$-form for an element $g$ of $SU(2,2|4)$ defined by
\begin{align}
 A=g^{-1}\,\rmd g \,,\qquad g\in SU(2,2|4)\,,
\end{align}
and satisfies the Maurer--Cartan equation
\begin{align}
 \rmd A + A\wedge A = 0\,. 
\label{eq:MC}
\end{align}
Here, we introduce projection operators $P^{(i)}\,(i=0,1,2,3)$ on each $\mathbb{Z}_4$-graded subspaces of $\alg{g}\equiv\alg{su}(2,2|4)$\,.
Then, the projection operators $d_\pm$ are defined as a linear combination of $P^{(i)}$
\begin{align}
 d_{\pm} \equiv \mp P^{(1)}+2\,P^{(2)}\pm P^{(3)} \,,
\label{eq:dpm}
\end{align}
and satisfies a relation
\begin{align}
 \str \bigl[ X\,d_\pm(Y) \bigr] = \str \bigl[ d_\mp (X)\, Y \bigr] \,.
\label{eq:dpm-transpose}
\end{align}

\medskip

If we expand the left-invariant $1$-form $A$ as
\begin{align}
 A = A^{(0)} +A^{(1)} +A^{(2)} + A^{(3)}\,,\qquad A^{(i)}=P^{(i)}(A) \,,
\end{align}
the action (\ref{AdS5S5-action}) can be rewritten as
\begin{align}
S=\frac{T}{2}\int \str\bigl(A^{(2)}\wedge *_{\gga} A^{(2)} - A^{(1)}\wedge A^{(3)} \bigr)\,,
\end{align}
where is the summation of the kinematic and the Wess-Zumino (WZ) term.
The ratio of the coefficient of the kinematic term to the WZ term is determined by the requirement of the $\kappa$-symmetry of the action (\ref{AdS5S5-action}).

\subsection{Classical integrability}

The equations of motion of (\ref{AdS5S5-action}) and the flatness condition (\ref{eq:MC}) are packed into the flatness condition (\ref{flat-super}) of the Lax pair with a parameter $u$\,.
Then, we can obtain the infinite conserved charges from the monodromy matrix constructed by the Lax pair.
In this sense, the $\AdS{5}\times \text{S}^5$ superstring is classical integrable.
In this subsection, we show the classical integrability of the $\AdS{5}\times \text{S}^5$ superstring by constructing the Lax pair.

\subsubsection*{Equations of motion}

For this purpose, let us first present the equations of motion of the action (\ref{AdS5S5-action}).

\medskip

The equations of motion of the $\AdS{5}\times \text{S}^5$ superstring are given by
\begin{align}
\cE=\cD_\alpha d_-(A^\alpha_{(-)})+\cD_\alpha d_+(A_{(+)}^\alpha)
+[A_{(+)\alpha}\,,d_-(A_{(-)}^\alpha)]+[A_{(-)\alpha}\,,d_+(A_{(+)}^\alpha)]=0\,.
\label{uEOM}
\end{align}
where $\cD_{\alpha}$ is the covariant derivative associated with $\gamma^{\alpha\beta}$ and worldsheet vectors with $(\pm)$ are projected with the projection operator $\Pg_{\pm\WSa}{}^{\WSb}$, like
\begin{eqnarray}
A_{(\pm)}^\alpha=P^{\alpha\beta}_\pm A_\beta\,.
\end{eqnarray}
The flatness condition (\ref{eq:MC}) of $A$ is rewritten as
\begin{align}
\Z&=\frac{1}{2}\sqrt{-\gamma}\,\varepsilon^{\alpha\beta}(\partial_\alpha A_\beta-\partial_\beta A_\alpha+[A_\alpha\,,A_\beta])\no \\
&=\cD_\alpha A_{(+)}^\alpha-\cD_\alpha A_{(-)}^\alpha+[A_{(-)\alpha}\,,A_{(+)}^\alpha]=0\,.
\label{uflat}
\end{align}
For later convenience, we will decompose the equations of motion (\ref{uEOM}) and the flatness condition (\ref{uflat}) into each $\mathbb{Z}_4$-graded components.
The bosonic parts are
\begin{align}
\mathsf{B}_1&:=\Z^{(0)}
=\cD_\alpha A_{(+)}^{\alpha(0)}-\cD_\alpha A_{(-)}^{\alpha(0)}
+[A_{(-)\alpha}^{(0)}\,,A_{(+)}^{\alpha(0)}]+[A_{(-)\alpha}^{(2)}\,,A_{(+)}^{\alpha(2)}]\no\\
&\qquad\qquad\qquad+[A_{(-)\alpha}^{(1)}\,,A_{(+)}^{\alpha(3)}]+[A_{(-)\alpha}^{(3)}\,,A_{(+)}^{\alpha(1)}]=0\,,\\
\mathsf{B}_2&:=\frac{1}{4}(\cE^{(2)}+2\,\Z^{(2)})
=\cD_\alpha A_{(+)}^{\alpha(2)}+[A_{(-)\alpha}^{(0)}\,,A_{(+)}^{\alpha(2)}]+[A_{(-)\alpha}^{(3)}\,,A_{(+)}^{\alpha(3)}]=0\,,\\ 
\mathsf{B}_3&:=\frac{1}{4}(\cE^{(2)}-2\,\Z^{(2)})
=\cD_\alpha A_{(-)}^{\alpha(2)}-[A_{(-)\alpha}^{(2)}\,,A_{(+)}^{\alpha(0)}]-[A_{(-)\alpha}^{(1)}\,,A_{(+)}^{\alpha(1)}]=0\,,
\end{align}
and the fermionic parts are
\begin{align}
\mathsf{F}_1&:=\frac{1}{4}(3\Z^{(1)}-\cE^{(1)})\no\\
&=\cD_\alpha A_{(+)}^{\alpha(1)}-\cD_\alpha A_{(-)}^{\alpha(1)}+[A_{(-)\alpha}^{(0)}\,,A_{(+)}^{\alpha(1)}]+[A_{(-)\alpha}^{(1)}\,,A_+^{\alpha(0)}]+[A_{-\alpha}^{(2)}\,,A_+^{\alpha(3)}]=0\,,\\ 
\mathsf{F}_2&:=\frac{1}{4}(3\Z^{(3)}+\cE^{(3)})\no\\
&=\cD_\alpha A_{(+)}^{\alpha(3)}-\cD_\alpha A_{(-)}^{\alpha(3)}+[A_{(-)\alpha}^{(0)}\,,A_{(+)}^{\alpha(3)}]+[A_{(-)\alpha}^{(1)}\,,A_+^{\alpha(2)}]+[A_{(-)\alpha}^{(3)}\,,A_{(+)}^{\alpha(0)}]=0\,,\\
\mathsf{F}_3&:=\frac{1}{4}(\cE^{(1)}+\Z^{(1)})
=[A_{(-)\alpha}^{(3)}\,,A_{(+)}^{\alpha(2)}]=0\,,\\ 
\mathsf{F}_4&:=\frac{1}{4}(-\cE^{(3)}+\Z^{(3)})
=[A_{(-)\alpha}^{(2)}\,,A_{(+)}^{\alpha(1)}]=0 \,.
\end{align}

\subsubsection*{A construction of the Lax pair}

Now, we shall present the Lax pair of the $\AdS{5}\times \text{S}^5$ superstring.
This is given by\cite{Bena:2003wd}
\begin{eqnarray}
\cL_\alpha\equiv M_{(-)\alpha}+L_{(+)\alpha}\,,
\label{uLax}
\end{eqnarray}
where $M^\alpha_{(-)}$ and $L^\alpha_{(+)}$ are
\begin{align}
M^\alpha_{(-)}&=A^{\alpha(0)}_{(-)}+u A_{(-)}^{\alpha(1)}+u^{2}A_{(-)}^{\alpha(2)}+u^{-1} A_{(-)}^{\alpha(3)}\,, \\
L^\alpha_{(+)}&=A^{\alpha(0)}_{(+)}+u A_{(+)}^{\alpha(1)}+u^{-2}A_{(+)}^{\alpha(2)}+u^{-1} A_{(+)}^{\alpha(3)}\,.
\end{align}
Here $u$ is the spectral parameter.
It is easy to show that the flatness condition of the Lax pair (\ref{uLax}) 
\begin{eqnarray}
\frac{1}{2}\epsilon^{\alpha\beta}
(\partial_\alpha \cL_\beta-\partial_\beta \cL_\alpha+[\cL_\alpha\,,\cL_\beta])
=0\,,
\label{flat-super}
\end{eqnarray}
is equivalent to the equations of motion (\ref{uEOM}) and the flatness condition (\ref{uflat}).
Indeed, the left-hand side of (\ref{flat-super}) can be rewritten as
\begin{align}
{\rm LHS\,~of}\,~(\ref{flat-super})=&u^0\,\mathsf{B}_1+u^{-2}\,\mathsf{B}_2-u^2\,\mathsf{B}_3\no \\
&+u\,\mathsf{F}_1+u^{-1} \mathsf{F}_2+u^{-3}\,\mathsf{F}_3+u^3\,\mathsf{F}_4 \,.
\end{align}
Therefore, the Lax pair (\ref{uLax}) is on-shell flat current.

\medskip

By using the Lax pair (\ref{uLax}), we can define the monodromy matrix
\begin{align}
T(u)=\mathsf{P}\exp\left(\int_{C} \rmd\sigma\,\cL_\sigma(u)\right)\,,
\end{align}
where the symbol $\mathsf{P}$ expresses the equal-time path ordering in terms of $\sigma$ and
$C$ is a closed path on the worldsheet.
The flatness condition allows us to deform the contour freely and
implies that $T(u)$ do not depend on $\tau$.
Therefore, expanding the monodormy matrix in $u$\,,
we obtain the infinite conserved charges.

\subsection{$\kappa$-symmetry}\label{sec:u-kappa}

The action (\ref{AdS5S5-action}) of the $\AdS{5}\times \text{S}^5$ superstring has the $\kappa$-symmetry \cite{Metsaev:1998it}.
In the subsection, we will demonstrate it.

\medskip

The $\kappa$-symmetry is realized as a combination of the variation of a group element $g$ and the world-sheet metric $\gamma^{\alpha\beta}$ like
\begin{align}
g^{-1}\delta_{\kappa}g&=
P^{\alpha\beta}_-\{\gQ^1\kappa_{1\alpha},A_{\beta}^{(2)}\}
+P^{\alpha\beta}_+\{\gQ^2\kappa_{2\alpha},A_{\beta}^{(2)}\}
\,, \label{ukappa1} \\
\delta_\kappa(\sqrt{-\ga} \ga^{\alpha\beta})&=
\frac{1}{4}\sqrt{-\ga}\,{\rm Str}\biggl[\Upsilon\Bigl([\gQ^1\kappa^\alpha_{1(+)},A_{+(+)}^{(1)\beta}]
+[\gQ^2\kappa^\alpha_{2(-)},A_{-(-)}^{(3)\beta}]\Bigr)+(\alpha\leftrightarrow \beta)\biggr]\,,
\label{ukappa2}
\end{align}
where  $\kappa_{I\alpha}\,(I=1,2)$ are local fermionic parameters and we have defined $\Upsilon\:={\rm diag}\,(1_4,\,-1_4)$\,.
In the subsection, we show the action (\ref{AdS5S5-action}) is invariant under the $\kappa$-symmetry transformation.

\medskip

We first decompose the variation of the action into two parts
\begin{align}
\delta_{\kappa}S=\delta_gS+\delta_{\gamma}S\,.
\end{align}
The variation $\delta_g S$ in a group element $g$ is
\begin{align}
\delta_g S&=\frac{T}{2}\int \rmd^2\sigma\,\sqrt{-\ga}\,\str(g^{-1}\delta\,g\,\cE)\,,
\end{align}
where $\cE$ is the equations of motion (\ref{uEOM}).
If we take $g^{-1}\delta\,g=\epsilon=\epsilon^{(1)}+\epsilon^{(3)}$ as
\begin{align}
\epsilon^{(1)}=P^{\alpha\beta}_-\{\gQ^1\kappa_{1\alpha},A_{\beta}^{(2)}\}\,,\qquad
\epsilon^{(3)}=P^{\alpha\beta}_+\{\gQ^2\kappa_{2\alpha},A_{\beta}^{(2)}\}\,,
\label{kappaan}
\end{align}
$\delta_g S$ can be rewritten as
\begin{align}
\delta_g S&=\frac{T}{2}\int \rmd^2\sigma\,
\sqrt{-\ga}\,\str\left[\epsilon^{(1)}\left(\cE^{(3)}-\cZ^{(3)}\right)+\epsilon^{(3)}\left(\cE^{(1)}+\cZ^{(1)}\right)\right]\no \\
&=-2T\int \rmd^2\sigma\,\sqrt{-\ga}\,\str\left(\epsilon^{(1)}[A_{-\alpha}^{(2)},A_+^{\alpha(1)}]+\epsilon^{(3)}[A_{+\alpha}^{(2)},A_-^{\alpha(3)}]\right)\,,
\label{eq:deltag-S-ukappa}
\end{align}
where we have used 
\begin{eqnarray}
\cE^{(1)}+\cZ^{(1)}=-4[A_{+\alpha}^{(2)}\,,A_-^{\alpha(3)}]\,,\qquad 
\cE^{(3)}-\cZ^{(3)}=-4[A_{-\alpha}^{(2)}\,,A_+^{\alpha(1)}]\,.
\end{eqnarray}
By using the expression\,(\ref{kappaan}), we can rewrite the each terms in (\ref{eq:deltag-S-ukappa}) as
\begin{align}
\begin{split}
\str\left(\epsilon^{(1)}[A_{-\alpha}^{(2)},A_+^{\alpha(1)}]\right)
&=\str\left(A_{-\alpha}^{(2)}A_{-\beta}^{(2)}[A_{+\alpha}^{(1)}\,,\gQ^1\kappa_{+1}^\beta]\right)\,,\\
\str\left(\epsilon^{(3)}[A_{+\alpha}^{(2)},A_-^{\alpha(3)}]\right)
&=\str\left(A_{+\alpha}^{(2)}A_{+\beta}^{(2)}[A_{-\alpha}^{(3)}\,,\gQ^2\kappa_{-2}^\beta]\right)\,.
\end{split}
\label{eq:kappa-para-str}
\end{align}
Moreover, an arbitrary grad-$2$ traceless element $A^{(2)}$ of $\mathfrak{su}(2,2|4)$ holds a relation
\begin{eqnarray}
A^{(2)}_{\alpha\pm}A^{(2)}_{\beta\pm}
=\frac{1}{8}\Upsilon\,\str(A^{(2)}_{\alpha\pm}A^{(2)}_{\beta\pm})+c_{\alpha\beta} Z\,,
\label{symformula}
\end{eqnarray}
where $Z$ is a central charge of $\mathfrak{su}(2,2|4)$ and $c_{\alpha\beta}$ is a symmetric function in $\alpha$ and $\beta$\,.
By using these expressions (\ref{eq:kappa-para-str}), (\ref{symformula}),
$\delta_g S$ becomes
\begin{align}
\delta_g S&=\frac{T}{4}\int \rmd^2\sigma\,
\sqrt{-\ga}\,\biggl[\str\left(A^{(2)}_{\alpha-}A^{(2)}_{\beta-}\right)
\str\left([\gQ^1\kappa_{+1}^{\beta}\,,A_+^{\alpha(1)}]\right)\nonumber \\
&\qquad\qquad\qquad+\str\left(A^{(2)}_{\alpha+}A^{(2)}_{\beta+}\right)
\str\left([\gQ^2\kappa_{-2}^{\beta}\,,A_-^{\alpha(3)}]\right)\biggr]\,.
\label{ukappag}
\end{align}

\medskip

It is straightforward to evaluate the variation $\delta_{\gamma}S$ in the world-sheet metric $\ga^{\alpha\beta}$\,.
From (\ref{ukappa2}), $\delta_{\gamma}S$ is
\begin{eqnarray}
\delta_\ga S&=&-\frac{T}{4}\int \rmd^2\sigma\,
\sqrt{-\ga}\,\str(A_\alpha^{(2)} A_\beta^{(2)})
\str\left[\Upsilon\left([\gQ^1\kappa_{+1}^{\beta}\,,A_+^{(1)\alpha}]
+[\gQ^2\kappa_{-2}^{\beta}\,,A_-^{(3)\alpha}]\right)\right]\no \\
&=&-\frac{T}{4}\int \rmd^2\sigma\,
\sqrt{-\ga}\,\biggl[\str\left(A^{(2)}_{\alpha-}A^{(2)}_{\beta-}\right)
\str\left([\gQ^1\kappa_{+1}^{\beta}\,,A_+^{\alpha(1)}]\right)\no \\
&&\qquad\qquad\qquad\qquad
+\str\left(A^{(2)}_{\alpha+}A^{(2)}_{\beta+}\right)\str\left([\gQ^2\kappa_{-2}^{\beta}\,,A_-^{\alpha(3)}]\right)\biggr]\,.
\label{ukappaga}
\end{eqnarray}
Here, we used a relation
\begin{eqnarray}
A_{\alpha\pm}B^\alpha=A_{\alpha\pm}B_\mp^\alpha\,,
\label{eq:AB-relation}
\end{eqnarray}
where $A_\alpha\,,B_\alpha$ are arbitrary vectors.
This variation manifestly cancels out $\delta_g S$,
\begin{eqnarray}
\delta_\kappa S=\delta_gS+\delta_{\gamma}S=0\,.
\end{eqnarray}
In this way, the action (\ref{AdS5S5-action}) is invariant under the $\kappa$-symmetry transformation (\ref{ukappa1}), (\ref{ukappa2}).

\subsection{The $\AdS{5} \times \rmS^5$ background from the GS action}
\label{subsec:AdS5S5-from-GS}

Next, let us explain how we read off the $\AdS{5} \times \rmS^5$ background from the action (\ref{AdS5S5-action}).

\subsubsection{The canonical form of the GS action}

To read off the target space from the action of the $\AdS{5} \times \rmS^5$ superstring,
we need to introduce the canonical form of the GS action.

\medskip

The canonical form of the type II superstring Lagrangian at second order in $\theta$\cite{Cvetic:1999zs} is
\begin{align}
\begin{split}
 \cL &= - \Pg_{-}^{\WSa\WSb}\, (\CG_{mn}+B_{mn})\, \partial_{\WSa}X^m\,\partial_{\WSb}X^n
\\ 
 &\quad - \ii\, \bigl(\Pg_{+}^{\WSa\WSb} \,\partial_{\WSa} X^m\, \brTheta_{1}\, \Gamma_m\, D_{+\WSb}\Theta_{1} 
  + \Pg_{-}^{\WSa\WSb} \,\partial_{\WSa} X^m\, \brTheta_{2}\, \Gamma_m\, D_{-\WSb}\Theta_{2} \bigr)
\\
 &\quad + \frac{\ii}{8}\, \Pg_{+}^{\WSa\WSb}\, \brTheta_{1}\, \Gamma_m\, \bisF \, \Gamma_n\, \Theta_{2} \, \partial_{\WSa} X^m\, \partial_{\WSb} X^n \,,
\end{split}
\label{eq:GS-action-canonical-YBsec}
\end{align}
where $\Gamma_a$\,($\Gamma_m=e_m{}^a\Gamma_a$) are the $32\times 32$ gamma matrices.
The differential operators $D_{\pm \WSa}$ are defined by
\begin{align}
 D_{\pm \WSa}& \equiv \partial_{\WSa} + \frac{1}{4}\, \partial_{\WSa} X^m\, \omega_{\pm m}{}^{\Loa\Lob}\, \Gamma_{\Loa\Lob}\,,\\
\omega_{\pm m\Loa\Lob}&\equiv \omega_{m\Loa\Lob} \pm \frac{1}{2}\, e_m{}^{\Loc}\,H_{\Loc\Loa\Lob} \,.
\end{align}
where $\omega^{\Loa\Lob}=\omega_m{}^{\Loa\Lob}\,\rmd X^m$ is a spin connection on the target space.
$\Theta_I$ and $\brTheta_I$ are the $32$-component Majorana spinors and its conjugate, respectively (see Appendix \ref{app:psu-algebra} for the details).

\medskip

The metric $g_{mn}$ and the $B$-field $B_{mn}$ of the target space are read off from the first line in (\ref{eq:GS-action-rewriting}).
Furthermore, we can read off the dilaton $\Phi$ and the R-R field strengths $\hat{F}_{a_1\dots a_n}$
from the R-R bispinor $\bisF$ which is defined by
\begin{align}
\bisF=\sum_{p}\frac{1}{p!}e^{\Phi}\,\hat{F}_{a_1\dots a_p}\Gamma^{a_1\dots a_p}
\,.
\end{align}
If the Lagrangian (\ref{eq:GS-action-rewriting}) describes type IIB superstring,
the summation is over $p=1,3,5,7,9$.
Each R--R field strengths satisfy
\begin{align}
 \hat{F}_p = (-1)^{\frac{p(p-1)}{2}}\, * \hat{F}_{10-p} \,,
\end{align}
where the hodge star $*$ is defined in Appendix \ref{app-section:conventions-form}.
The detail of our convention for R--R fields is explained in the section \ref{sec:DFT}.
In this way, 
by comparing the action (\ref{AdS5S5-action}) expanded in terms of $\theta$ with the canonical form (\ref{eq:GS-action-rewriting}),
we can obtain the explicit expression of the $\AdS{5} \times \rmS^5$ superstring.

\subsubsection{Group parametrization}

To derive the $\AdS{5} \times \rmS^5$ background from the GS action,
we introduce a coordinate system through a parametrization of a group element $g$\,.

\medskip

We first decompose the group element into the bosonic and the fermionic parts\,,
\begin{align}
 g =g_{\bos}\cdot g_{\fer}\in SU(2,2|4)\,.
\label{bf-decomposition}
\end{align}
We will parametrize the bosonic part $g_{\bos}$ as
\begin{align}
\begin{split}
 &g_{\bos} = g_{\AdS5}\cdot g_{\rmS^5}\,,\\
 &g_{\AdS5} \equiv \exp(x^\mu\,P_\mu)\cdot \exp(\ln z \,D)\,,
\\
 &g_{\rmS^5} \equiv \exp(\phi_1\, h_1+\phi_2\,h_2+\phi_3\,h_3)\cdot \exp(\xi\,\gJ_{56})\cdot \exp(r\,\gP_5)\,.
\end{split}
\label{eq:group-parameterization}
\end{align}
Here, $P_{\mu}$ ($\mu,\nu=0,\dotsc,3$) and $D$ are the translation and dilatation generators in the conformal algebra $\alg{so}(2,4)$\,. $h_{i}\, (i=1,2,3)$ are Cartan generators of the $\alg{so}(6)$ algebra and we defined them as
\begin{align}
 h_1 \equiv \gJ_{57}\,,\qquad 
 h_2 \equiv \gJ_{68}\,,\qquad 
 h_3 \equiv \gP_9\,. 
\end{align}
On the other hand, we parameterize the fermionic part $g_{\fer}$ as
\begin{align}
 g_{\fer} = \exp(\gQ^I\, \theta_I) \,, \qquad 
 \gQ^I\,\theta_I=(\gQ^I)^{\check{\SPa}\hat{\SPa}}\,\theta_{I\check{\SPa}\hat{\SPa}} \,,
\label{eq:group-parameterization-fermi}
\end{align}
where the supercharges $(\gQ^I)^{\check{\SPa}\hat{\SPa}}$ $(I=1,2)$ are labeled by two indices $(\check{\SPa}\,,\hat{\SPa}=1,\dotsc, 4)$ and $\theta_{I\check{\SPa}\hat{\SPa}}\,(I=1,2)$ are $16$-components Majorana--Weyl fermions.
 A matrix representation of the above generators of $\mathfrak{su}(2,2|4)$ is presented in Appendix \ref{app:psu-algebra}.

\subsubsection{Expansion of the left-invariant current}

Next, we will expand the left-invariant current $A$ to second order in the spacetime fermion $\theta$ like
\begin{align}
A&=A_{(0)}+A_{(1)}+A_{(2)}+\cO(\theta^3)\,.
\end{align}
Now, since we chose the parametrization (\ref{bf-decomposition}), (\ref{eq:group-parameterization-fermi}) of $g$,
the left-invariant current $A$ can be expanded as
\begin{align}
 A &= g_{\fer}^{-1}\,A_{(0)}\,g_{\fer} + \gQ^I\,\rmd\theta_I
\no\\
 &= A_{(0)} + [A_{(0)},\,\gQ^I\,\theta_I] + \frac{1}{2}\,\bigl[[A_{(0)},\,\gQ^I\,\theta_I],\,\gQ^J\,\theta_J\bigr] + \gQ^I\,\rmd\theta_I + \cO(\theta^3)\,,
\label{eq:A-AdS5xS5}
\end{align}
where $A_{(p)}$ are defined as $\cO(\theta^p)$ of the left-invariant current $A$
and $A_{(0)}$ is given by
\begin{align}
 A_{(0)}&\equiv g_{\bos}^{-1}\,\rmd g_{\bos} = \Bigl(e_m{}^{\Loa}\,\gP_{\Loa} - \frac{1}{2}\, \omega_m{}^{\Loa\Lob}\,\gJ_{\Loa\Lob}\Bigr)\,\rmd X^m \,.
\end{align}
The vielbein $e^{\Loa}=e_{m}{}^{\Loa}\,\rmd X^m$ has the form
\begin{align}
 e^{\Loa} &= \biggl(\frac{\rmd x^0}{z},\frac{\rmd x^1}{z} ,\frac{\rmd x^2}{z},\frac{\rmd x^3}{z}, \frac{\rmd z}{z},
\rmd r, \sin r\,\rmd \xi, \sin r\,\cos\xi\,\rmd\phi_1,\sin r\,\sin\xi\,\rmd\phi_2,\cos r\,\rmd\phi_3\biggr)\,,
\end{align}
and $\omega^{\Loa\Lob}=\omega_m{}^{\Loa\Lob}\,\rmd X^m$ are the associated spin connection.

\medskip

Moreover, by using the commutation relations of $\mathfrak{su}(2,2|4)$ (see Appendix \ref{app:conventions} for our conventions),
the each commutators in (\ref{eq:A-AdS5xS5}) can be evaluated as
\begin{align}
 &[A_{(0)},\,\gQ^I\,\theta_I]= \gQ^I\,\Bigl(\frac{1}{4}\,\delta^{IJ}\,\omega^{\Loa\Lob} \,\gamma_{\Loa\Lob} 
  + \frac{\ii}{2}\,\epsilon^{IJ}\, e^{\Loa}\, \hat{\gamma}_{\Loa} \Bigr)\,\theta_J \label{A0Q}\,,
\\
 &\bigl[[A_{(0)},\,\gQ^I\,\theta_I],\,\gQ^J\,\theta_J\bigr] 
 = \ii\,\brtheta_I\,\hat{\gamma}^{\Loa}\,\Bigl(\frac{1}{4}\,\delta^{IJ}\,\omega^{\Loc\Lod}\, \gamma_{\Loc\Lod} 
  + \frac{\ii}{2}\,\epsilon^{IJ}\, e^{\Lob}\, \hat{\gamma}_{\Lob} \Bigr)\,\theta_J\,\gP_{\Loa} \no
\\
 &\qquad\qquad\qquad\qquad\qquad + \frac{1}{4}\,\epsilon^{IK}\, \brtheta_I\, \gamma^{\Loc\Lod}\,\Bigl(\frac{1}{4}\,\delta^{KJ}\,\omega^{\Loa\Lob}\, \gamma_{\Loa\Lob} 
  + \frac{\ii}{2}\,\epsilon^{KJ}\, e^{\Loa}\, \hat{\gamma}_{\Loa} \Bigr)\,\theta_J\,\rmd X^m\, R_{\Loc\Lod}{}^{\Loe\Lof}\,\gJ_{\Loe\Lof} 
\no\\
 &\qquad\qquad\qquad\qquad\qquad + \text{(irrelevant terms proportional to the central charge $Z$)}\,. \label{A0QQ}
\end{align}
Here, $R_{\Loa\Lob\Loc\Lod}$ is the Riemann tensor in the tangent space of the $\AdS{5} \times \rmS^5$ background.
For the derivation of (\ref{A0QQ}),
we have used $\delta^{IJ}\,\brtheta_I\,\hat{\gamma}^{\Loa}\,\rmd \theta_J = 0$ and
$\epsilon^{IJ}\,\brtheta_I\,\gamma^{\Loa\Lob}\,\rmd \theta_J = 0$\,.

\medskip

From the above calculations, the left-invariant current $A$ up to the second order in $\theta$ becomes
\begin{align}
 A &= \Bigl(e^{\Loa}+\frac{\ii}{2}\,\brtheta_I\,\hat{\gamma}^{\Loa}\,D^{IJ}\theta_J\Bigr)\,\gP_{\Loa} 
-\frac{1}{2}\,\Bigl(\omega^{\Loa\Lob}-\frac{1}{4}\,\epsilon^{IK}\,\brtheta_I\,\gamma^{\Loc\Lod}\,R_{\Loc\Lod}{}^{\Loa\Lob}\,D^{KJ}\theta_J\Bigr)\,\gJ_{\Loa\Lob}\no\\
 &\quad+\gQ^I\,D^{IJ}\theta_J + \cO(\theta^3)\,,
\end{align}
where we defined the differential operator
\begin{align}
 D^{IJ} &\equiv \delta^{IJ}\,\Bigl(\rmd + \frac{1}{4}\,\omega^{\Loa\Lob}\,\gamma_{\Loa\Lob}\Bigr)+\frac{\ii}{2}\,\epsilon^{IJ}\,e^{\Loa}\,\hat{\gamma}_{\Loa}\,.
\end{align}
In particular, $A_{(1)}$\,, $A_{(2)}$ are given by
\begin{align}
 A_{(1)} &= \gQ^I\,D^{IJ}\theta_J\,, \\
A_{(2)}&=\frac{\ii}{2}\,\brtheta_I\,\hat{\gamma}^{\Loa}\,D^{IJ}\theta_J\,\gP_{\Loa} 
+\frac{1}{8}\,\epsilon^{IK}\,\brtheta_I\,\gamma^{\Loc\Lod}\,R_{\Loc\Lod}{}^{\Loa\Lob}\,D^{KJ}\theta_J\,\gJ_{\Loa\Lob}\,.
\end{align}

\subsubsection{Evaluation of the bi-linear current part}

Using the expansion \eqref{eq:A-AdS5xS5}, we can straightforwardly obtain
\begin{align}
 \frac{1}{2}\,\str\bigl[\,A_{\WSa}\,d_-(A_{\WSb})\,\bigr]
 &= \eta_{\Loa\Lob}\,e_{\WSa}{}^{\Loa}\,e_{\WSb}{}^{\Lob} 
  + \ii\, \bigl[\,e_{\WSb}{}^{\Loa}\, (\brtheta_1\,\hat{\gamma}_{\Loa}\, \partial_{\WSa} \theta_1) 
  + e_{\WSa}{}^{\Loa}\, (\brtheta_2\,\hat{\gamma}_{\Loa}\, \partial_{\WSb} \theta_2) \,\bigr]
\no\\
 &\quad + \frac{\ii}{4}\,\Bigl[e_{\WSb}{}^{\Lob}\, e_{\WSa}{}^{\Loa}\,\omega_{\Loa}{}^{\Loc\Lod} \,(\brtheta_1\,\hat{\gamma}_{\Lob}\,\gamma_{\Loc\Lod} \,\theta_1)
 + e_{\WSa}{}^{\Loa}\, e_{\WSb}{}^{\Lob}\, \omega_{\Lob}{}^{\Loc\Lod} \, (\brtheta_2\,\hat{\gamma}_{\Loa}\,\gamma_{\Loc\Lod}\,\theta_2) \Bigr]
\no\\
 &\quad - e_{\WSb}{}^{\Loa} \,e_{\WSa}{}^{\Lob}\, \brtheta_1\,\hat{\gamma}_{\Loa}\, \hat{\gamma}_{\Lob}\,\theta_2+\cO(\theta^3)\,,
\end{align}
where $e_{\WSa}{}^{\Loa}\equiv e_m{}^{\Loa}\,\partial_{\WSa}X^m$\,. 
Further using \eqref{eq:lift-32-AdS5-1}, \eqref{eq:lift-32-AdS5-2}, and \eqref{eq:lift-32-AdS5-3}, we obtain
\begin{align}
 \frac{1}{2}\,\str\bigl[\,A_{\WSa}\,d_-(A_{\WSb})\,\bigr]
 &= \CG_{mn}\,\partial_{\WSa} X^m\,\partial_{\WSb} X^n 
  + \ii\, \bigl[\,e_{\WSb}{}^{\Loa}\, \brTheta_1\,\Gamma_{\Loa}\, \partial_{\WSa} \Theta_1 
  + e_{\WSa}{}^{\Loa}\, \brTheta_2\,\Gamma_{\Loa}\, \partial_{\WSb} \Theta_2 \,\bigr]
\no\\
 &\quad + \frac{\ii}{4}\,\Bigl[e_{\WSb}{}^{\Lob}\, e_{\WSa}{}^{\Loa}\,\omega_{\Loa}{}^{\Loc\Lod} \, \brTheta_1\,\Gamma_{\Lob}\,\Gamma_{\Loc\Lod} \,\Theta_1 
 + e_{\WSa}{}^{\Loa}\, e_{\WSb}{}^{\Lob}\, \omega_{\Lob}{}^{\Loc\Lod} \, \brTheta_2\,\Gamma_{\Loa}\,\Gamma_{\Loc\Lod}\,\Theta_2 \Bigr]
\no\\
 &\quad - \frac{\ii}{8}\,e_{\WSb}{}^{\Loa} \,e_{\WSa}{}^{\Lob}\, \brTheta_I\,\Gamma_{\Loa}\,\bisF_5\,\Gamma_{\Lob}\,\Theta_J
+\cO(\Theta^3)\,,
\label{STrAA}
\end{align}
where $\bisF_5$ is a bispinor
\begin{align}
 \bisF_5 \equiv 4\,(\Gamma^{01234}+\Gamma^{56789})\,. 
\end{align}
This describes the R--R $5$-form field strength in the tangent space of the $\AdS{5} \times \rmS^5$ background.

\medskip

The expression (\ref{STrAA}) implies that the action \eqref{AdS5S5-action} takes the canonical form of the GS action.
Therefore, the target space of the action \eqref{AdS5S5-action} is
the familiar $\AdS{5} \times \rmS^5$ background with the RR $5$-form,
\begin{align}
 \rmd s^2 &=\rmd s_{\AdS{5}}^2+\rmd s_{\rmS^5}^2\,, \\
\Exp{\Phi}\hat{F}_5 &= 4\,\bigl(\omega_{\AdS5}+\omega_{\rmS^5}\bigr) \,.
\label{eq:uF5}
\end{align}
Under  the parametrization (\ref{eq:group-parameterization}) of $g_{b}$, the metric of $\AdS5$ and $\rmS^5$ are
\begin{align}
\rmd s_{\AdS{5}}^2&=\frac{-(\rmd x^0)^2+(\rmd x^1)^2+(\rmd x^2)^2+(\rmd x^3)^2 + \rmd z^2}{z^2}\,,\\
\rmd s_{\rmS^5}^2&=\rmd r^2 + \sin^2 r\, \rmd\xi^2 + \cos^2\xi\,\sin^2 r\, \rmd\phi_1^2 + \sin^2r\,\sin^2\xi\, \rmd\phi_2^2 + \cos^2r\, \rmd\phi_3^2\,,
\end{align}
and the volume forms $\omega_{\AdS5}$\,$\omega_{\rmS^5}$ of $\AdS5$ and $\rmS^5$ are given by
\begin{align}
 &\omega_{\AdS5} \equiv - \frac{\rmd x^0\wedge\rmd x^1\wedge \rmd x^2\wedge\rmd x^3\wedge\rmd z}{z^5}\,, 
\\
 &\omega_{\rmS^5} \equiv \sin^3r \cos r \sin\xi \cos\xi\,\rmd r\wedge \rmd\xi\wedge\rmd \phi_1\wedge \rmd\phi_2\wedge\rmd \phi_3\qquad (\omega_{\AdS{5}} = *_{10}\omega_{\rmS^5}) \,. 
\end{align}
In the following discussion, we set the dilaton as $\Phi=0$\,.

\subsection{Killing vectors}

For later convenience, let us calculate the Killing vectors $\hat{T}_i\equiv\hat{T}_i^m\,\partial_m$ associated with the bosonic symmetries $T_i$ of the $\AdS{5}$ background. 
From the general formula \eqref{eq:Killing-formula} explained in Appendix \ref{app:homogeneous-space}, the Killing vectors can be expressed as
\begin{align}
 \hat{T}_{i}=\hat{T}_{i}{}^{m}\,\partial_{m} 
 =\bigl[\Ad_{g_{\bos}^{-1}}\bigr]_{i}{}^{\Loa}\, e_{\Loa}{}^{m}\, \partial_{m}
 = \str\bigl(g_{\bos}^{-1}\,T_{i}\,g_{\bos}\,\gP_{\Loa}\bigr)\,e^{\Loa m}\, \partial_{m}\,,
\label{eq:Killing-Formula}
\end{align}
where we introduced a notation $g\,T_i\,g^{-1}\equiv [\Ad_{g}]_{i}{}^{j}\,T_j$\,. 
By using our parameterization \eqref{eq:group-parameterization}, the Killing vectors on the $\AdS{5}$ background are given by
\begin{align}
\begin{split}
 \hat{P}_\mu &\equiv \str\bigl(g_{\bos}^{-1}\,P_{\mu}\,g_{\bos}\, \gP_{\Loa} \bigr)\,e^{\Loa m}\,\partial_m = \partial_\mu \,,
\\
 \hat{K}_\mu &\equiv \str\bigl(g_{\bos}^{-1}\,K_{\mu}\,g_{\bos}\, \gP_{\Loa} \bigr)\,e^{\Loa m}\,\partial_m = \bigl(x^\nu\,x_\nu +z^2\bigr)\,\partial_\mu - 2\,x_\mu\,(x^\nu\,\partial_\nu+z\,\partial_z)\,,
\\
 \hat{M}_{\mu\nu} &\equiv \str\bigl(g_{\bos}^{-1}\,M_{\mu\nu}\,g_{\bos}\, \gP_{\Loa} \bigr)\,e^{\Loa m}\,\partial_m = x_\mu\,\partial_\nu -x_\nu\,\partial_\mu \,,
\\
 \hat{D} &\equiv \str\bigl(g_{\bos}^{-1}\,D\,g_{\bos}\, \gP_{\Loa} \bigr)\,e^{\Loa m}\,\partial_m = x^\mu\,\partial_\mu + z\,\partial_z \,.
\end{split}
\end{align}
The Lie brackets of these vector fields satisfy the same commutation relations \eqref{eq:so(2-4)-algebra} as the conformal algebra $\alg{so}(2,4)$ (with negative sign, $[\hat{T}_i,\,\hat{T}_j]=-f_{ij}{}^k\,\hat{T}_k$):
\begin{align}
\begin{split}
 &[\hat{P}_\mu,\, \hat{K}_\nu]= -2\,\bigl(\eta_{\mu\nu}\, \hat{D} - \hat{M}_{\mu\nu}\bigr)\,,\quad 
 [\hat{D},\, \hat{P}_{\mu}]= -\hat{P}_\mu\,,\quad [\hat{D},\,\hat{K}_\mu]= \hat{K}_\mu\,,
\\
 &[\hat{M}_{\mu\nu},\, \hat{P}_\rho] = -\eta_{\mu\rho}\, \hat{P}_\nu +\eta_{\nu\rho}\, \hat{P}_\mu \,,\quad 
 [\hat{M}_{\mu\nu},\, \hat{K}_\rho] = -\eta_{\mu\rho}\,\hat{K}_\nu + \eta_{\nu\rho}\,\hat{K}_\mu\,, 
\\
 &[\hat{M}_{\mu\nu},\,\hat{M}_{\rho\sigma}] = -\eta_{\mu\rho}\,\hat{M}_{\nu\sigma}+\eta_{\mu\sigma}\,\hat{M}_{\nu\rho} + \eta_{\nu\rho}\,\hat{M}_{\mu\sigma}-\eta_{\nu\sigma}\,\hat{M}_{\mu\rho}\,.
\end{split}
\end{align}

\section{YB deformation of the $\AdS{5} \times \rmS^5$ superstring}
\label{sec:YB-sigma}

In this section, we shall introduce the homogeneous YB deformations of the $\AdS{5} \times \rmS^5$ superstring.

\subsection{The action of YB deformed $\AdS{5} \times \rmS^5$ superstring}

The action of YB deformed $\AdS{5} \times \rmS^5$ superstring is given by\cite{Kawaguchi:2014qwa}
\footnote{In \cite{Benitez:2018xnh}, in the pure spinor formalism, the action of homogeneous YB deformed $\AdS{5} \times \rmS^5$ superstring was constructed.}
\begin{align}
 S_{\YB}=-\frac{\dlT}{2}\int \rmd^2\sigma\,\Pg_{-}^{\WSa\WSb}\, \str\bigl[A_{\WSa}\, d_-\circ\cO_-^{-1}(A_{\WSb})\bigr]\,, 
\label{eq:YBsM}
\end{align}
where $\eta\in\mathbb{R}$ is a deformation parameter and the linear operator $\cO_{\pm}$ are defined by
\begin{align}
 \cO_{\pm}=1\pm\eta\, R_g\circ d_\pm\,.
\end{align}
When $\eta=0$\,, the deformed action (\ref{eq:YBsM}) reduces to the undeformed $\AdS{5}\times \rmS^5$ superstring sigma model action \eqref{AdS5S5-action}.
A key ingredient of the YB deformations is the $R$-operator.
The $R$-operator is a skew-symmetric linear operator $R:\alg{g} \to \alg{g}$
and solves the homogeneous CYBE
\begin{align}
\begin{split}
 \CYBE(X,Y) &\equiv [R(X),\,R(Y)] - R([R(X),\,Y]+[X,\,R(Y)])\\
&=0\,,\qquad X,\,Y \in\alg{g}\,.
\label{eq:CYBE}
\end{split}
\end{align}
Also, the dressed $R$-operator $R_g$ is defined by
\begin{align}
 R_g(X):= g^{-1}\, R(g\,X\,g^{-1})\,g ={\rm Ad}_g^{-1}\circ R \circ {\rm Ad}_g(X)\,,\qquad g\in SU(2,2|4)\,.
\end{align}
The operator $R_g$ is also a solution of the homogeneous CYBE \eqref{eq:CYBE}
\begin{align}
 \CYBE_g(X,Y) \equiv [R_g(X),\,R_g(Y)] - R_g([R_g(X),\,Y]+[X,\,R_g(Y)])=0\,,
\label{eq:CYBE-g}
\end{align}
if the linear operator $R$ satisfies the homogeneous CYBE.
It is easily seen from a relation $\CYBE_g(X,Y)={\rm Ad}_g^{-1}\CYBE({\rm Ad}_g(X),{\rm Ad}_g(Y))$\,.

\medskip

It is useful to rewrite the $R$-operator in terms of the tensorial notation.
Then, the $R$-operator can be expressed by using a skew-symmetric classical $r$-matrix $r\in \mathfrak{g}\otimes \mathfrak{g}$\,.
Introducing a $r$-matrix, 
\begin{align}
 r=\frac{1}{2}\,r^{ij}\,T_{i}\wedge T_{j}\,, \qquad r^{ij}=-r^{ji}\,, \qquad T_{i}\in \alg{g}\,,
\end{align}
the action of the $R$-operator can be defined as 
\begin{align}
 R(X) = r^{ij}\, T_{i}\,\str(T_{j}\,X) \,, \qquad X\in\alg{g} \,.
\label{eq:R-operator}
\end{align}
Therefore, a YB deformation is specified by taking a classical $r$-matrix.
In terms of the $r$-matrix, the homogeneous CYBE \eqref{eq:CYBE} can be rewritten by
\begin{align}
 f_{l_1l_2}{}^i\,r^{jl_1}\,r^{kl_2} + f_{l_1l_2}{}^j\,r^{kl_1}\,r^{il_2} + f_{l_1l_2}{}^k\,r^{il_1}\,r^{jl_2} =0\,,
\label{eq:CYBE-r}
\end{align}
where $f_{ij}{}^k$ are the structure constants $[T_i,\,T_j]=f_{ij}{}^k\,T_k$ of $\mathfrak{g}$\,.

\subsection{Classical integrability}

The deformed action (\ref{eq:YBsM}) also admits a Lax pair.
Therefore, the YB deformation describes integrable deformations of the $\AdS{5} \times \rmS^5$ superstring.
To show this, we will present the Lax pair of the deformed system.

\subsubsection*{Equations of motion and the flatness condition}

To demonstrate the classical integrability of (\ref{eq:YBsM}),
we give the equations of motion of the deformed action (\ref{eq:YBsM}).

\medskip

In this subsection, it is useful to introduce deformed currents and the projected ones
\begin{align}
J_{\alpha}&=\cO_-^{-1}A_{\alpha}\,,\qquad
\tilde{J}_{\alpha}=\cO_+^{-1}A_{\alpha}\,,\\
J_{(\pm)}^{\alpha}&=P_{\pm}^{\alpha\beta}J_{\beta}\,,\qquad
\tilde{J}_{(\pm)}^{\alpha}=P_{\pm}^{\alpha\beta}\tilde{J}_{\beta}\,.
\end{align}
Then, the equations of motion of the deformed action (\ref{eq:YBsM}) are
\begin{eqnarray}
\tilde{\cE}=\cD_\alpha d_-(J^\alpha_{(-)})+\cD_\alpha d_+(\tilde{J}_{(+)}^\alpha)
+[\tilde{J}_{(+)\alpha}\,,d_-(J_{(-)}^\alpha)]+[J_{(-)\alpha}\,,d_+(\tilde{J}_{(+)}^\alpha)]=0\,.
\label{mEOM}
\end{eqnarray}
The flatness condition for the left-invariant current is 
\begin{eqnarray}
\tilde{\cZ}&=&\frac{1}{2}\epsilon^{\alpha\beta}(\cD_\alpha A_\beta-\cD_\beta A_\alpha+[A_\alpha\,,A_\beta])\nonumber \\
&=&\cD_\alpha \tilde{J}_{(+)}^\alpha-\cD_\alpha J_{(-)}^\alpha+[J_{(-)\alpha}\,,\tilde{J}_{(+)}^\alpha]
+\eta R_g(\cE)=0\,.
\label{mflat}
\end{eqnarray}
As with the undeformed case,
we decompose the equations of motion (\ref{mEOM}) and the flatness condition (\ref{mflat}) into the $\mathbb{Z}_4$ graded components.
The bosonic parts are
\begin{align}
\tilde{\mathsf{B}}_1&:=\Z^{(0)}=\cD_\alpha \tilde{J}_+^{\alpha(0)}-\cD_\alpha J_-^{\alpha(0)}
+[J_{-\alpha}^{(0)}\,,\tilde{J}_+^{\alpha(0)}]\no \\
&\qquad+[J_{-\alpha}^{(2)}\,,\tilde{J}_+^{\alpha(2)}]
+[J_{-\alpha}^{(1)}\,,\tilde{J}_+^{\alpha(3)}]+[J_{-\alpha}^{(3)}\,,\tilde{J}_+^{\alpha(1)}]=0\,,\qquad \\ 
\tilde{\mathsf{B}}_2&:=\frac{1}{4}(\cE^{(2)}+2\,\Z^{(2)})
=\cD_\alpha \tilde{J}_+^{\alpha(2)}+[J_{-\alpha}^{(0)}\,,\tilde{J}_+^{\alpha(2)}]
+[J_{-\alpha}^{(3)}\,,\tilde{J}_+^{\alpha(3)}]=0\,,\\
\tilde{\mathsf{B}}_3&:=\frac{1}{4}(\cE^{(2)}-2\,\Z^{(2)})
=\cD_\alpha J_-^{\alpha(2)}-[J_{-\alpha}^{(2)}\,,\tilde{J}_+^{\alpha(0)}]-[J_{-\alpha}^{(1)}\,,\tilde{J}_+^{\alpha(1)}]=0\,,
\label{bmEOM}
\end{align}
and the fermionic parts are given by
\begin{align}
\tilde{\mathsf{F}}_1&:=\frac{1}{4}(3\tilde{\Z}^{(1)}-\tilde{\cE}^{(1)})\no\\
&=\cD_\alpha \tilde{J}_{(+)}^{\alpha(1)}-\cD_\alpha J_{(-)}^{\alpha(1)}+[J_{(-)\alpha}^{(0)}\,,\tilde{J}_{(+)}^{\alpha(1)}]+[J_{(-)\alpha}^{(1)}\,,\tilde{J}_{(+)}^{\alpha(0)}]+[J_{(-)\alpha}^{(2)}\,,\tilde{J}_{(+)}^{\alpha(3)}]=0\,,\\
\tilde{\mathsf{F}}_2&:=\frac{1}{4}(3\Z^{(3)}+\cE^{(3)})\no\\
&=\cD_\alpha \tilde{J}_{(+)}^{\alpha(3)}-\cD_\alpha J_{(-)}^{\alpha(3)}+[J_{(-)\alpha}^{(0)}\,,\tilde{J}_{(+)}^{\alpha(3)}]+[J_{(-)\alpha}^{(3)}\,,\tilde{J}_{(+)}^{\alpha(0)}]+[J_{(-)\alpha}^{(1)}\,,\tilde{J}_{(+)}^{\alpha(2)}]=0\,,\\
\tilde{\mathsf{F}}_3&:=\frac{1}{4}(\cE^{(1)}+\Z^{(1)})
=[J_{(-)\alpha}^{(3)}\,,\tilde{J}_{(+)}^{\alpha(2)}]=0\,,\\ 
\tilde{\mathsf{F}}_4&:=\frac{1}{4}(-\cE^{(3)}+\Z^{(3)})
=[J_{(-)\alpha}^{(2)}\,,\tilde{J}_{(+)}^{\alpha(1)}]=0\,.
\label{fmEOM}
\end{align}

\subsubsection*{A construction of the Lax pair}

Now, let us present the Lax pair of the homogeneous YB deformed AdS$_5\times$S$^5$ superstring.
The Lax pair is given by\cite{Kawaguchi:2014qwa}
\begin{align}
\tilde{\cL}_\alpha=\tilde{L}_{(+)\alpha}+\tilde{M}_{(-)\alpha}\,,
\label{eta-Lax}
\end{align}
where $\tilde{L}^\alpha_{(+)}$ and $\tilde{M}^\alpha_{(-)}$ are
\begin{align}
\tilde{L}^\alpha_{(+)}=&\tilde{J}_{(+)}^{\alpha(0)}+u\tilde{J}_{(+)}^{\alpha(1)}
+u^{-2}\tilde{J}_{(+)}^{\alpha(2)}+u^{-1}\tilde{J}_{(+)}^{\alpha(3)}\no \\
\tilde{M}^\alpha_{(-)}=&J_{(-)}^{\alpha(0)}+\la J_{(-)}^{\alpha(1)}+u^2J_{(-)}^{\alpha(2)}+u^{-1}J_{(-)}^{\alpha(3)}\,.
\end{align}
The expression has a similar form as in the undeformed one (\ref{uLax}).

\medskip

To confirm it, we evaluate the flatness condition of the Lax pair
\begin{align}
\epsilon^{\alpha\beta}\left(\partial_{\alpha} \tilde{\cL}_{\beta}-\partial_{\beta} \tilde{\cL}_{\alpha}+[\tilde{\cL}_{\alpha}\,,\tilde{\cL}_{\beta}]\right)=0\,.
\label{YB-flat}
\end{align}
The left-hand side of the equation can be rewritten as
\begin{align}
{\rm LHS\,~of}\,~(\ref{YB-flat})=&u^0\,\tilde{\mathsf{B}}_1+u^{-2}\,\tilde{\mathsf{B}}_2-u^2\,\tilde{\mathsf{B}}_3\no \\
&+u\,\tilde{\mathsf{F}}_1+u^{-1} \tilde{\mathsf{F}}_2+u^{-3}\,\tilde{\mathsf{F}}_3+u^3\,\tilde{\mathsf{F}}_4 \,.
\end{align}
Therefore, the flatness condition (\ref{YB-flat}) is equivalent to the equations of motion (\ref{mEOM}) and the flatness condition (\ref{mflat}) at the on-shell.
As we discussed in the previous section,
the deformed system (\ref{eq:YBsM}) is also classical integrable.

\subsection{The $\kappa$-symmetry of the YB deformed action}
\label{sec:kappa-YB}

The deformed action (\ref{eq:YBsM}) is also invariant under the $\kappa$-symmetry transformation\cite{Kawaguchi:2014qwa}.
We shall show the $\kappa$-invariance of the deformed action in the subsection.

\medskip

The $\kappa$-symmetry transformation is given by\cite{Kawaguchi:2014qwa}
\begin{align}
\mathcal{O}_-^{-1}g^{-1}\delta_{\kappa}g&=
P^{\alpha\beta}_-\{\gQ^1\kappa_{1\alpha},J_{-\beta}^{(2)}\}
+P^{\alpha\beta}_+\{\gQ^2\kappa_{2\alpha},J_{+\beta}^{(2)}\}
\,, \label{kappa1} \\
\delta_\kappa(\sqrt{-\ga} \ga^{\alpha\beta})&=
\frac{1}{4}\sqrt{-\ga}\,{\rm Str}\biggl[\Upsilon\Bigl([\gQ^1\kappa^\alpha_{1(+)},J_{+(+)}^{(1)\beta}]
+[\gQ^2\kappa^\alpha_{2(-)},J_{-(-)}^{(3)\beta}]\Bigr)+(\alpha\leftrightarrow \beta)\biggr]\,.
\label{kappa2}
\end{align}
It is easy to see that when we take $\eta=0$\,, this reduces to the undeformed one (\ref{ukappa1}), (\ref{ukappa2}).
As with the undeformed case,
we decompose the variation of the deformed action (\ref{eq:YBsM}) under the $\kappa$-symmetry transformation to 
\begin{align}
\delta_\kappa S_{\YB}\equiv \delta_gS_{\YB}+\delta_\ga S_{\YB}\,.
\end{align}
where $\delta_g S_{\YB}$ and $\delta_\ga S_{\YB}$ are the variations with respect to the group element and the world-sheet metric, respectively.

\medskip

Let us first consider $\delta_gS_{\YB}$\,.
By using (\ref{kappa1}), this is given by
\begin{align}
\delta_g S_{\YB}&=\frac{T}{2}\,\int \rmd^2\sigma\,
\sqrt{-\gamma}\,\str\left[\epsilon^{(1)}P_3\circ(1+\eta R_g)(\tilde{\cE})
+\epsilon^{(3)}P_1\circ(1-\eta R_g)(\tilde{\cE})\right]\no\\
&=-2\sqrt{T}\,\int \rmd^2\sigma\,\sqrt{-\gamma}\,\str\left(\epsilon^{(1)}[J_{-\alpha}^{(2)}\,,\tilde{J}_+^{\alpha(1)}]+\epsilon^{(3)}[\tilde{J}_{+\alpha}^{(2)}\,,J_-^{\alpha(3)}]\right)\,,
\label{eq:kappa-g-vari}
\end{align}
where $\epsilon^{(1)}$ and $\epsilon^{(3)}$ are
\begin{align}
\epsilon^{(1)}=(1+\eta R_g)P^{\alpha\beta}_-\{\gQ^1\kappa_{1\alpha},J_{-\beta}^{(2)}\}\,,\qquad
\epsilon^{(3)}=(1-\eta R_g)P^{\alpha\beta}_+\{\gQ^2\kappa_{2\alpha},J_{+\beta}^{(2)}\}\,.
\label{mkappaan}
\end{align}
In the second equation, we have used 
\begin{align}
P_1\circ(1-\eta R_g)(\tilde{\cE})&=-4[\tilde{J}_{+\alpha}^{(2)}\,,J_-^{\alpha(3)}]-\tilde{\cZ}^{(1)}\,,\\
P_3\circ(1+\eta R_g)(\tilde{\cE})&=-4[J_{-\alpha}^{(2)}\,,\tilde{J}_+^{\alpha(1)}]+\tilde{\cZ}^{(3)}\,.
\end{align}
and ignored the flatness condition $\tilde{\cZ}$\,.
The each terms in (\ref{eq:kappa-g-vari}) can be rewritten as
\begin{align}
\begin{split}
\str\left(\epsilon^{(1)}[J_{-\alpha}^{(2)}\,,\tilde{J}_+^{\alpha(1)}]\right)
=\str\left(J_{-\alpha}^{(2)}J_{-\beta}^{(2)}[\tilde{J}_{+\alpha}^{(1)}\,,\gQ^1\kappa_+^{\beta}]\right)\,, \\
\str\left(\epsilon^{(3)}[\tilde{J}_{+\alpha}^{(2)}\,,J_-^{\alpha(3)}]\right)
=\str\left(\tilde{J}_{+\alpha}^{(2)}\tilde{J}_{+\beta}^{(2)}[J_{-\alpha}^{(3)}\,,\gQ^2\kappa^{\beta}_-]\right)\,.
\end{split}
\end{align}
By using the formula (\ref{symformula}), the variation $\delta_g\,S_{\YB}$ becomes
\begin{align}
\delta_g S_{\YB}&=\frac{T}{4}\,\int \rmd^2\sigma\,
\sqrt{-\gamma}\,\biggl[\str\left(J^{(2)}_{\alpha-}J^{(2)}_{\beta-}\right)\str\left([\gQ^1\kappa_+^{\beta(1)}\,,\tilde{J}_+^{\alpha}]\right)\no \\
&\qquad\qquad\qquad\qquad+\str\left(\tilde{J}^{(2)}_{\alpha+}\tilde{J}^{(2)}_{\beta+}\right)
\str\left([\gQ^2\kappa_-^{\beta}\,,J_-^{\alpha(3)}]\right)\biggr]\,.
\label{gmkappa}
\end{align}

\medskip

Next, let us consider the variation $\delta_{\gamma}\,S_{\YB}$.
By using the relation (\ref{eq:AB-relation}), $\delta_{\gamma}\,S_{\YB}$ is given by
\begin{align}
\delta_\ga S_{\YB}&=-\frac{T}{4}\,\int \rmd^2\sigma\,\sqrt{-\gamma}\,
\str(J_\alpha^{(2)} J_\beta^{(2)})
\str\left[\Upsilon\left([\gQ^1\kappa_+^{\beta}\,,\tilde{J}_+^{(1)\alpha}]
+[\gQ^2\kappa_-^{\beta}\,,J_-^{(3)\alpha}]\right)\right]\no \\
&=-\frac{T}{4}\,\int \rmd^2\sigma\,\sqrt{-\gamma}\,
\biggl[\str\left(J^{(2)}_{\alpha-}J^{(2)}_{\beta-}\right)
\str\left([\gQ^1\kappa_+^{\beta}\,,\tilde{J}_+^{\alpha(1)}]\right)\no \\
&\qquad\qquad\qquad\qquad+\str\left(\tilde{J}^{(2)}_{\alpha+}\tilde{J}^{(2)}_{\beta+}\right)
\str\left([\gQ^2\kappa_-^{\beta}\,,J_-^{\alpha(3)}]\right)\biggr]\,.
\label{gamkappa}
\end{align}
This obviously cancel out the variation $\delta_g\,S_{\YB}$\,,
\begin{align}
\delta_\kappa S_{\YB}=(\delta_g+\delta_\ga)S_{\YB}=0\,.
\end{align}
As a result, the deformed action (\ref{eq:YBsM}) has the $\kappa$-symmetry.

\subsection{General formula for YB deformed backgrounds}

As explained in the subsection \ref{subsec:AdS5S5-from-GS},
to read off the YB deformed backgrounds from the deformed action (\ref{eq:YBsM}),
we need to expand the action up to second order in the spacetime fermion $\theta_I$,
\begin{align}
 S_{\YB}=S_{(0)}+S_{(2)}+\cO(\theta^4)\,,
\label{eq:YBsM-expansion}
\end{align}
and compare the expanded action with the canonical form of the type IIB GS action.
In a pioneering work \cite{Arutyunov:2015qva},
the explicit expression of a $q$-deformed $\AdS{5}\times \rmS^5$ background had been given by using the method.
Thereafter, it was generalized to the case of the homogeneous YB deformations in \cite{Kyono:2016jqy}.
In this subsection, we only present a formula for the YB deformed AdS$_5\times$S$^5$ backgrounds
and its derivation will be explained in the section \ref{sec:YB-beta-deform}.

\medskip

For the simplicity, we assume that the classical $r$-matrix is only composed of bosonic generators of $\mathfrak{su}(2,2|4)$
\footnote{Rewriting of the YB sigma model action to the standard GS form based on the $\kappa$-symmetry was done in \cite{Borsato:2016ose} to full order in fermionic variables.
Moreover, the deformed background associated with a general $r$-matrix was determined.} \,.
Then, the YB deformed AdS$_5\times$S$^5$ backgrounds are given by the following formula:
\begin{align}
\begin{split}
 &\CG_{mn}'+B_{mn}'=\bigl[(G^{-1}-\beta)\bigr]^{-1}_{mn}\,,
\qquad  \Exp{-2\Phi'} = \frac{\sqrt{-\CG}}{\sqrt{-\CG'}} \Exp{-2\Phi}\,,
\\
 &\hat{F}'=\Exp{-B_2'\wedge}\Exp{-\beta\vee}\hat{F}_5\,,
\label{pre-beta-formula-noB}
\end{split}
\end{align}
where $G_{mn}$ is the metric of the original AdS$_5\times$S$^5$ background and $\beta$ is given by
\begin{align}
\beta^{mn}(x) =2\,\eta\,r^{ij}\,\hat{T}^{m}_i(x)\,\hat{T}_j^{n}(x) \,.
\label{eq:YB-beta}
\end{align}
Here, $\hat{T}^{m}_i(x)$ are Killing vector fields associated with generators $T_i$ appearing in the $r$-matrix.
$\hat{F}_5$ is the undeformed R--R $5$-form field strength (\ref{eq:uF5}) and
$\hat{F}'$ is a polyform which is a formal summation of the all deformed R--R fields strengths
\begin{align}
\hat{F}':=\sum_{p=1,3,5,7,9}\hat{F}_p\,,\qquad 
\hat{F}_p:=\frac{1}{p!}\hat{F}_{m_1\cdots m_p}\dd x^{m_1}\wedge \cdots \wedge \rmd x^{m_p}\,.
\end{align}
The operator $\beta\vee$ which acts on an arbitrary $p$-form $A_p$ is defined as
\begin{align}
\beta\vee A_p=\frac{1}{2}\beta^{mn}\iota_m\iota_n A_p\,,
\end{align}
where $\iota_m$ is the inner production along the  $x^m$-direction.

\medskip

Here, we should comment on relations between the YB deformations and a duality transformation.
An important observation of \cite{Sakamoto:2017cpu} is that the transformation (\ref{pre-beta-formula-noB}) of the AdS$_5\times$S$^5$ background is nothing but the $\beta$-transformation with $\beta$-field (\ref{eq:YB-beta}) as explained in \ref{sec:YB-beta-deform}.
The $\beta$-transformation is a kind of the $O(d,d)$ $T$-duality transformations.
Therefore, the YB-deformations can be regarded as a string duality transformation.

\subsection{The unimodularity condition}

In general, YB deformations give solutions of non only the usual supergravity but also generalized supergravity.
Therefore, to obtain supergravity solutions from YB deformations,
we need to impose further constraints on the classical $r$-matrices.
Borsato and Wulff gave the condition for the classical $r$-matrices \cite{Borsato:2016ose} which is called the unimodularity condition,
\begin{align}
r^{ij}[T_i,T_j]=0\,,\qquad T_i\in\mathfrak{su}(2,2|4)\,.
\label{unimodular1}
\end{align}
If a given $r$-matrix satisfies this condition, we call it the unimodular $r$-matrix.

\medskip

Let us briefly explain the origin of the name of the condition (\ref{unimodular1}).
For simplicity, we will consider a bosonic subalgebra $\mathfrak{so}(2,4)\oplus \mathfrak{so}(6)$ of $\mathfrak{su}(2,2|4)$
\footnote{Recently, in \cite{vanTongeren:2019dlq}, homogeneous YB deformations associated with unimoular $r$-matrices including fermionic generators were considered, and the associated deformed backgrounds explicitly were constructed.}.
To explain it, we will note on an important theorem \cite{Stolin1, Stolin2}.
The theorem states that a constant solution of the homogeneous CYBE (\ref{eq:CYBE-r}) for a Lie algebra $\mathfrak{g}$ corresponds to
a subalgebra $\mathfrak{f}\subset \mathfrak{g}$ in one-to-one correspondence.
Here, a constant solution means that $r^{ij}$ is the constant skew-symmetric matrix.
Furthermore, restricting the range of the indices of $r^{ij}$ to $i, j=1,\dots , {\rm dim}\,\mathfrak{f}$\,,
the matrices $r^{ij}$ are always invertible.
This implies that $\mathfrak{f}$ is always even dimensional.
Then, we can show that $\mathfrak{f}$ is a unimodular Lie algbera by using the homogeneous CYBE,
In this way, we call (\ref{unimodular1}) the unimodularity condition.

\medskip

To see the unimodularity of $\mathfrak{f}$,
we introduce a bi-linear map $\omega: \mathfrak{g}\times \mathfrak{g}\to \mathbb{R}$ and the components are defined by
\begin{align}
\omega(T_i, T_j):=(r^{-1})_{ij}\,.
\end{align}
The bi-linear map $\omega$ satisfies the $2$-cocycle condition
\begin{align}
\begin{split}
&\omega(x,y)=-\omega(y,x)\,,\\
&\omega([x,y],z)+\omega([x,y],z)+\omega([x,y],z)=0\,,
\end{split}
\end{align}
where $x,y,z\in \mathfrak{f}$\,.
The second condition can be shown by using the homogeneous CYBE.
Then, $\mathfrak{f}$ is called a quasi-Frobenius Lie algebra.
The $2$-cocycle condition can be rewritten as
\begin{align}
(r^{-1})_{i[j}f_{kl]}{}^{i}=0\,.
\label{2-cocycle-r}
\end{align}
By taking a contraction $r^{kl}$ with (\ref{2-cocycle-r}), we obtain
\begin{align}
0=3(r^{-1})_{i[j}f_{kl]}{}^{i}r^{kl}
=(r^{-1})_{ij}f_{kl}{}^{i}r^{kl}+2\,f_{ij}{}^{i}\,.
\end{align}
If the $r$-matrix satisfies the unimodularity condition, the equation becomes  
\begin{align}
f_{ij}{}^{i}=-\frac{1}{2}(r^{-1})_{ij}f_{kl}{}^{i}r^{kl}=0\,,
\label{eq:uni-r-structure}
\end{align}
This shows that $\mathfrak{f}$ is also a unimodular Lie algbera.

\subsection{A classification of $r$-matrices}

Now, it is useful to explain the classification of the homogeneous YB deformations. 
An $r$-matrix
\begin{align}
r =\frac{1}{2}\, r^{ij}\,T_i\wedge T_j\,,
\end{align}
is called \emph{Abelian} if it consists of a set of generators which commute with each other $[T_i,\,T_j]=0$, 
and otherwise called \emph{non-Abelian}. 
Most of homogeneous YB deformations studied in the literature are based on Abelian $r$-matrices. 
The Abelian $r$-matrices are obviously unimodular.
Moreover, when $\mathfrak{g}$ is a compact Lie algebra (for example $\mathfrak{su}(N)$\,, $\mathfrak{so}(N)$),
all quasi-Frobenius Lie subalgebra $\mathfrak{f}$ is Abelian \cite{Abelian-compact}.
Therefore, non-Abelian unimodular $r$-matrices only exist for a non-compact Lie algebra $\mathfrak{g}$.

\subsubsection*{Non-Abelian unimodular $r$-matrices}

We shall discuss the classification of non-Abelian unimodular $r$-matrices. 
Although the classification is very complicated in general, 
it is mainly classified by the unimodularity condition \eqref{unimodular1}. 
If the rank of an $r$-matrix is defined as rank $r^{ij} := {\rm dim}\,\mathfrak{f}$\,, 
non-Abelian unimodular $r$-matrices with lower rank are classified well. 
Obviously, the rank-2 unimodular $r$-matrix is Abelian.

\medskip

\paragraph{\bf rank-$4$}
The rank-$4$ unimodular $r$-matrix for the bosonic isometry of $\AdS5$ has been classified in \cite{Borsato:2016ose}. 
If we take a rank-$4$ $r$-matrix
\begin{align}
r=T_1\wedge T_2+T_3\wedge T_4\,,\qquad T_{1,\dots ,4}\in \mathfrak{so}(2,4)\,,
\end{align}
such $r$-matrix can be classified like
\begin{enumerate}
\renewcommand{\labelenumi}{(\roman{enumi})}
\setlength{\itemsep}{-5mm}
\item $\mathfrak{h}_3\oplus \mathbb{R}\qquad [T_1, T_2]=T_3$\\
\item $\mathfrak{r}_{3,-1}\oplus \mathbb{R}\qquad [T_1,T_2]=T_2\,,\quad [T_1,T_3]=-T_3$\\
\item $\mathfrak{r}_{3,0}\oplus \mathbb{R}\qquad [T_1,T_2]=-T_3\,,\quad [T_1,T_3]=-T_2$\\
\item $\mathfrak{n}_4\qquad [T_1,T_2]=-T_4\,,\quad [T_4,T_2]=-T_3$
\end{enumerate}
One can see the details of the discussion on the classification of the rank-4 unimodular $r$-matrices in \cite{Borsato:2016ose}.

\medskip

The three at the top of the list is called an Almost abelian $r$-matrix \cite{vanTongeren:2016eeb}, 
which covers most of the rank-4 examples studied in \cite{Borsato:2016ose}. 
As argued in \cite{Borsato:2016ose,vanTongeren:2016eeb}, this class of YB deformations can be realized as a sequence of 
non-commuting TsT-transformations (see \cite{Borsato:2016ose} for the explicit form in the rank-4 examples), 
which consists of the usual TsT-transformations and diffeomorphisms that make the Killing vectors as coordinate basis. 
On the other hand, the final class cannot be generated by performing non-commuting TsT-transformations.

\medskip

\paragraph{\bf rank-$6$}
There is not a compete classification of the rank $6$ unimodular $r$-matrices.
At this moment, some examples have only been given \cite{Borsato:2016ose}.

\medskip

\paragraph{\bf rank-$8$}
It was shown that the rank-$8$ non-Abelian unimodular $r$-matrices do not exit as discussed in \cite{Borsato:2016ose}.

\subsubsection*{Non-unimodular $r$-matrices}

Finally, we will briefly comment on non-unimodular $r$-matrices.
Most simple example is the rank-$2$ $r$-matrix
\begin{align}
r=\frac{1}{2}T_1\wedge T_2\,,\qquad [T_1, T_2]=T_2\,.
\end{align}
It is easy to see that the $r$-matrix does not satisfy the unimodularity condition (\ref{unimodular1}).
The $r$-matrix is called a Jordanian $r$-matrix.
A generalization of the rank-$2$ Jordanian $r$-matrix was given in \cite{Tolstoy2004}.

\medskip

The deformed YB deformed backgrounds associated with non-unimodular $r$-matrices are solutions of the generalized supergravity equations \cite{Borsato:2016ose}.
In next subsection, we shall review the generalized supergravity.

\subsection{Generalized supergravity equations}
\label{sec:GSE-YBsec}

The generalized supergravity had originally been discovered in \cite{Arutyunov:2015mqj}.
A remarkable property of the effective theory is that the requirement of the $\kappa$-symmetry derive the generalized supergravity equations \cite{Wulff:2016tju}.

\medskip

Our conventions for the type II GSE \cite{Arutyunov:2015mqj,Wulff:2016tju,Sakatani:2016fvh,Baguet:2016prz,Sakamoto:2017wor} are as follows:
\begin{align}
 &R_{mn}- \frac{1}{4}\,H_{mpq}\,H_n{}^{pq} + 2\,\sfD_m \partial_n \Phi + \sfD_m U_n +\sfD_n U_m = T_{mn} \,,
\nn\\
 &R + 4\,\sfD^m \partial_m \Phi - 4\,\abs{\partial \Phi}^2 - \frac{1}{2}\,\abs{H_3}^2 
  - 4\,\bigl(I^m I_m+U^m U_m + 2\,U^m\,\partial_m \Phi - \sfD_m U^m\bigr) =0 \,,
\nn\\
 &-\frac{1}{2}\,\sfD^k H_{kmn} + \partial_k\Phi\,H^k{}_{mn} + U^k\,H_{kmn} + \sfD_m I_n - \sfD_n I_m = \cK_{mn} \,,
\\
 &\rmd *\hat{F}_n -H_3\wedge * \hat{F}_{n+2} -\iota_I B_2 \wedge * \hat{F}_n -\iota_I * \hat{F}_{n-2} =0 \,,
\nn
\label{eq:GSEsecAdS}
\end{align}
where we have defined $\abs{\alpha_p}^2\equiv \frac{1}{p!}\alpha_{m_1\cdots m_p}\,\alpha^{m_1\cdots m_p}$\,.
$\sfD_m$ is the usual covariant derivative associated with $\CG_{mn}$\,,
and $T_{mn}\,, \cK_{mn}$ are defined by
\begin{align}
\begin{split}
 T_{mn} &\equiv \frac{1}{4}\Exp{2\Phi} \sum_p \biggl[ \frac{1}{(p-1)!}\, 
 \hat{F}_{(m}{}^{k_1\cdots k_{p-1}} \hat{F}_{n) k_1\cdots k_{p-1}} - \frac{1}{2}\, 
 \CG_{mn}\,\abs{\hat{F}_p}^2 \biggr] \,,
\\
 \cK_{mn}&\equiv \frac{1}{4}\Exp{2\Phi} \sum_p \frac{1}{(p-2)!}\, \hat{F}_{k_1\cdots k_{p-2}}\, 
 \hat{F}_{mn}{}^{k_1\cdots k_{p-2}}  \,. 
\end{split}
\end{align}
The relation between the R--R field strengths and potential is expressed as (see section \ref{sec:R-R(m)DFT} for details)
\begin{align}
 \hat{F}_n&\equiv \rmd \hat{C}_{n-1} + H_3\wedge \hat{C}_{n-3} - \iota_I B_2 \wedge \hat{C}_{p-1} -\iota_I \hat{C}_{n+1}\,.
\end{align}
The Killing vector $I=I^m\,\partial_m$ is defined to satisfy
\begin{align}
 \Lie_I \CG_{mn} = 0\,, \qquad 
 \Lie_I B_2 + \rmd \bigl(U -\iota_I B_2\bigr) = 0\,,\qquad 
 \Lie_I \Phi =0 \,, \qquad 
 I^m\,U_m = 0\,. 
\end{align}
When $I=0$, the GSE reduce to the usual supergravity equations of motion. 
We usually choose a particular gauge $U_m=I^nB_{nm}$\,(see \cite{Arutyunov:2015mqj,Sakamoto:2017wor} for the details).
Therefore, in the generalized supergravity, the deformation is characterized only by the Killing vector $I^m$\,. 
It is also noted that due to the presence of a Killing vector, solutions of the GSE are effectively nine dimensional.
Here we have ignored spacetime fermions, but one can see the full explicit expression of type IIB generalized supergravity equations in \cite{Wulff:2016tju}.

\medskip

The Killing vector $I$ in the GSE doesn't appear in the classical action of the string sigma model.
As discussed in \cite{Sakamoto:2017wor,Fernandez-Melgarejo:2018wpg},
$I$ appears in the counterterm which is introduced to cancel the Weyl anomaly of the string sigma model defined on generalized supergravity backgrounds (see section \ref{sec:Weyl}).

\medskip

Finally, we present an experimental formula of the Killing vector $I^m$ for YB deformed backgrounds.
As discovered in \cite{Araujo:2017jkb}, the formula has a very simple form
\begin{align}
 I^m = \sfD_n \bmr^{nm} \,,
\label{div-formula}
\end{align}
where $\sfD_n$ is the covariant derivatives associated with the original metric.
Here $\bmr^{mn}$ is given by (\ref{eq:YB-beta}).
If the $r$-matrix gives a non-zero $I$\,,
this implies the violation of the unimodular condition (see also subsection \ref{sec:Killing-I-divformula}). 
To see this, we shall consider non-unimodular $r$-matrices satisfying
\begin{align}
 \cI =\eta\,r^{ij}\,[T_i,\,T_j] = \eta\,r^{ij}\,f_{ij}{}^k\,T_k \neq 0 \,. 
\label{eq:cI-formula}
\end{align}
By using the concrete expression (\ref{eq:YB-beta}) of $\bmr^{mn}$,
the divergent formula (\ref{div-formula}) can be rewritten as
\begin{align}
\begin{split}
 &I^m =- \eta\,r^{ij}\,[\hat{T}_i,\,\hat{T}_j]^m = \eta\,r^{ij}\,f_{ij}{}^k\,\hat{T}_k^m
\equiv\hat{\cI} 
\,,
\end{split}
\label{eq:experimental}
\end{align}
where the Killing vectors $\hat{T}_i$ satisfy
\begin{align}
\,[\hat{T}_i,\,\hat{T}_j]^m = \Lie_{\hat{T}_i}\hat{T}_j^m= - f_{ij}{}^k\,\hat{T}_k^m\,.
\end{align}
In the next section, we can see that the formula works well for various non-unimodular $r$-matrices.

\section{Examples of YB deformed  $\AdS{5} \times \rmS^5$ backgrounds}
\label{sec:ExampleYBAdS5}

\subsection{Abelian $r$-matrices}

We will consider here YB deformations of the $\AdS{5} \times \rmS^5$ background associated with Abelian $r$-matrices.

\subsubsection{The Maldacena-Russo background}

To demonstrate how to use the formula \eqref{pre-beta-formula-noB},
let us consider a YB-deformed $\AdS{5}\times S^5$ background associated 
with a classical $r$-matrix \cite{Matsumoto:2014gwa},
\begin{align}
 r=\frac{1}{2}\,P_1\wedge P_2\,. \label{MR-r}
\end{align}
This $r$-matrix is Abelian and satisfies the homogeneous CYBE \eqref{eq:CYBE}. 
The associated YB deformed background is derived in \cite{Matsumoto:2014gwa,Kyono:2016jqy}. 

\medskip

The classical $r$-matrix \eqref{MR-r} leads to the associated $\beta$-field,
\begin{align}
 \beta =\eta\,\hat{P}_1\wedge \hat{P}_2=\eta\,\partial_1\wedge \partial_2\,.
\end{align}
Then, the $\AdS{5}$ part of a $10\times 10$ matrix $(G^{-1}-\beta)$ is
\begin{align}
 \bigl(G^{-1}-\beta\bigr)^{mn}=
 \begin{pmatrix}
 z^2&0&0&0&0\\
 0&-z^2&0&0&0\\
 0&0&z^2&-\eta&0\\
 0&0&\eta&z^2&0\\
 0&0&0&0&z^2
 \end{pmatrix} \,,
\label{G-beta}
\end{align}
where we have ordered the coordinates as $(z\,,x^0\,,x^1\,,x^2\,,x^3)$\,.
By using the inverse of the matrix \eqref{G-beta} and the formula \eqref{pre-beta-formula-noB},
we obtain the NS-NS fields of the YB-deformed background,
\begin{align}
\begin{split}
 \rmd s^2&=\frac{\rmd z^2-(\rmd x^0)^2+(\rmd x^3)^2}{z^2}+\frac{z^2\,[(\rmd x^1)^2+(\rmd x^2)^2]}{z^4+\eta^2} +\rmd s_{\rm S^5}^2\,,
\\
 B_2&=\frac{\eta}{z^4+\eta^2}\,\rmd x^1\wedge \rmd x^2\,,\qquad
 \Phi=\frac{1}{2}\,\ln\biggl[\frac{z^4}{z^4+\eta^2}\biggr]\,.
\label{MR-NS}
\end{split}
\end{align}

\medskip

The next task is to derive the R-R fields of the deformed background.
From the undeformed R-R $5$-form field strength \eqref{eq:uF5} of the $\AdS5\times\rmS^5$ background,
``the R-R fields'' $F:=\Exp{-\beta\vee}\hat{F}_5$ are given by
\begin{align}
\begin{split}
 F&=\Exp{-\beta\vee}\hat{F}_5
 =4\,\bigl(\omega_{\AdS5}+\omega_{\rmS^5}\bigr)-4\,\beta\vee\omega_{\AdS5}
\\
 &=4\,\bigl(\omega_{\AdS5}+\omega_{\rmS^5}\bigr)-4\,\eta\,\frac{\rmd z\wedge \rmd x^0\wedge \rmd x^3}{z^5}\,.
\end{split}
\end{align}
This is nothing but a linear combination of the deformed R--R field strengths with different rank. 
Hence we can readily read off the following expressions: 
\begin{align}
 F_3=-4\,\eta\,\frac{\rmd z\wedge \rmd x^0\wedge \rmd x^3}{z^5}\,,\qquad
 F_5=4\,\bigl(\omega_{\AdS5}+\omega_{\rmS^5}\bigr)\,.
\end{align}
Furthermore, the deformed R--R fields $\hat{F}'$ can be computed as
\begin{align}
 \hat{F}'&=\Exp{-B_2\wedge}F
\no\\
 &=-4\,\eta\,\frac{\rmd z \wedge \rmd x^0\wedge \rmd x^3}{z^5}
 +4\,\biggl(\frac{z^4}{z^4+\eta^2}\,\omega_{\AdS5}+\omega_{\rmS^5}\biggr)
 -4\,B_2\wedge\omega_{\rmS^5}\,.
\end{align}
Namely, we obtain
\begin{align}
\begin{split}
 \hat{F}_1'&=0\,,\qquad \hat{F}_3'=-4\,\eta\,\frac{\rmd z \wedge \rmd x^0\wedge \rmd x^3}{z^5}\,,
\\
 \hat{F}_5'&=4\,\biggl(\frac{z^4}{z^4+\eta^2}\,\omega_{\AdS5}+\omega_{\rmS^5}\biggr)\,,
\\
 \hat{F}_7'&=-4\,B_2\wedge\omega_{\rmS^5}\,.
\label{MR-RR}
\end{split}
\end{align}
The full deformed background, given by \eqref{MR-NS} and \eqref{MR-RR}, is a solution of 
the standard type IIB supergravity.
This background is nothing but 
a gravity dual of non-commutative gauge theory \cite{Hashimoto:1999ut,Maldacena:1999mh}. 

\medskip 

Thus, nowadays, we do not have to perform supercoset construction to obtain 
the full expression of YB-deformed background. 
Just by using a simple formula \eqref{pre-beta-formula-noB}, given a classical $r$-matrix, 
the full background can easily be derived.

\subsubsection{Lunin-Maldacena-Frolov background}

Next, we will consider $r$-matrix 
\begin{align}
r=\frac{1}{2}(\mu_3\,h_1\wedge h_2+\mu_1\,h_2\wedge h_3+\mu_2\,h_3\wedge h_1)\,.
\label{abelian}
\end{align}
The $r$-matrix is composed of the Cartan generators $h_1$\,, $h_2$\,, $h_3$ in $\mathfrak{su}(4)$\,.
Here, $\mu_i$~($i=1,2,3$) are deformation parameters.
The metric and $B$--filed were computed in \cite{Matsumoto:2014nra} and
the full background was reproduced in \cite{Kyono:2016jqy} by performing the supercoset construction.

\medskip

The associated $\beta$-field is
\begin{align}
\beta=2\eta\,(\mu_3\,\partial_{\phi_1}\wedge \partial_{\phi_2}+\mu_1\,\partial_{\phi_2}\wedge \partial_{\phi_3}+\mu_2\,\partial_{\phi_3}\wedge \partial_{\phi_1})\,.
\end{align}
By using the formula (\ref{pre-beta-formula-noB}),
we obtain the deformed background
\begin{align}
\begin{split}
\rmd s^2 &= \rmd s^2_{\AdS5}+\sum_{i=1}^{3} \left( {d\rho_i}^2+G(\hat{\gamma}_i) {\rho_i}^2 {d\phi_i}^2 \right)
+G(\hat{\gamma}_i){\rho_1}^2 {\rho_2}^2 {\rho_3}^2 \left( \sum_{i=1}^{3} \hat{\gamma}_i d{\phi_i} \right)^2\,,\\
B_2&=G(\hat{\gamma}_i)\,(\hat{\gamma}_3\,{\rho_1}^2{\rho_2}^2 d\phi_1\wedge d\phi_2
+\hat{\gamma}_1\,{\rho_2}^2{\rho_3}^2 d\phi_2\wedge d\phi_3
+\hat{\gamma}_2\,{\rho_3}^2{\rho_1}^2 d\phi_3\wedge d\phi_1)\,,\\
\Phi&=\frac{1}{2}\log\,G(\hat{\gamma}_i)\,,\\
\hat{F}_3&=-4\,\sin^3\alpha\,\cos\,\alpha\,\sin\theta\,\cos\theta\left(\sum_{i=1}^3\hat{\gamma}_i\,d\phi_i\right)\wedge d \alpha \wedge d \theta\,,\\
\hat{F}_5&=4\,\left(\omega_{{\rm AdS}_5}+G(\hat{\gamma}_i)\,\omega_{{\rm S}^5}\right)\,,
\label{3p-Lunin-Maldacena}
\end{split}
\end{align}
where we defined new coordinates $\rho_i~(i=1,2,3)$ as
\begin{align}
\rho_1= \sin r\cos  \zeta\,,\qquad \rho_2= \sin r\sin\zeta\,,\qquad\rho_3 =\cos r\,.
\end{align}
The deformation parameters $\hat{\gamma}_i$ are defined by 
\begin{align}
\hat{\gamma}_i = 8\,\eta\mu_i\,,
\end{align}
and the scalar function $G(\hat{\gamma}_i)$ is 
\begin{align}
G^{-1}(\hat{\gamma}_i)& \equiv1+\sin^2 r(\hat{\gamma}_1^2\cos^2 r\sin^2\zeta+\hat{\gamma}_2^2\cos^2 r\cos^2\zeta
+\hat{\gamma}_3^2\sin^2 r\sin^2\zeta\cos^2\zeta)\,.
\end{align}
The background (\ref{3p-Lunin-Maldacena}) is originally given in \cite{Frolov:2005dj}.
The supersymmetry is completely broken by the deformation.

\medskip 

In particular, when all deformation parameters $\hat{\gamma}_i$ are set equal,
\begin{align}
\hat{\gamma}_1=\hat{\gamma}_2=\hat{\gamma}_3 \equiv \hat{\gamma}\,, 
\end{align}
the above background become
\begin{align}
\begin{split}
ds^2 &= \sum_{i=1}^{3} \left( {d\rho_i}^2+G {\rho_i}^2 {d\phi_i}^2 \right)
+G \hat{\gamma}^2{\rho_1}^2 {\rho_2}^2 {\rho_3}^2 \left( \sum_{i=1}^{3}d{\phi_i}\right)^2\,,\\
B_2&=G\,\hat{\gamma}({\rho_1}^2{\rho_2}^2 d\phi_1\wedge d\phi_2
+{\rho_2}^2{\rho_3}^2 d\phi_2\wedge d\phi_3+{\rho_3}^2{\rho_1}^2 d\phi_3\wedge d\phi_1)\,,\\
\Phi&=\frac{1}{2}\log\,G\,,\\
\hat{F}_3&=-4\hat{\gamma}\,\sin^3\alpha\,\cos\,\alpha\,\sin\theta\,\cos\theta\left(\sum_{i=1}^3\,d\phi_i\right)\wedge d \alpha \wedge d \theta\,,\\
\hat{F}_5&=4\,\left(\omega_{{\rm AdS}_5}+G\,\omega_{{\rm S}^5}\right)\,,
\label{Lunin-Maldacena}
\end{split}
\end{align}
where the scalar function $G$ is defined by 
\begin{align}
G^{-1} \equiv 1+\frac{\hat{\gamma}^2}{4}(\sin^22r+\sin^4r\sin^22\zeta)\,.
\end{align}
In the special case, the background has $8$ supercharges.
The gauge dual of the background (\ref{Lunin-Maldacena}) is known as the $\beta$-deformed $\mathcal{N}$=4 SYM \cite{Lunin:2005jy} which is one of exactly marginal deformations of $\mathcal{N}$=4 SYM \cite{Leigh:1995ep} preserving $\mathcal{N}=1$ supersymmetry.

\subsubsection{The Schr\"odinger spacetime}

Finally, we consider the following Abelian $r$-matrix\cite{Matsumoto:2015uja}:
\begin{align}
r=\frac{1}{2}\,P_-\wedge(h_1+h_2+h_3)\,,\label{r-Sch}
\end{align}
where $P_-=(P_0-P_3)/\sqrt{2}$ is a light-cone transformation generator in $\mathfrak{so}(2,4)$\,.
By using the following coordinate system (\ref{S1overCP2}), (\ref{S1overCP21}),
the $\beta$-filed is expressed as
\begin{align}
\beta=-\eta\,\partial_-\wedge \partial_{\chi}\,.
\end{align}

\medskip

By using the formula (\ref{pre-beta-formula-noB}), the resulting deformed background is given by
\begin{align}
\begin{split}
\rmd s^2&=\frac{-2dx^+dx^-+(dx^1)^2+(dx^2)^2+dz^2}{z^2}-\eta^2\frac{(dx^+)^2}{z^4}
+ds_{\text{S}^5}^2\,, \\
B_2&=\frac{\eta}{z^2}\,dx^+\wedge (d\chi+\omega)\,,\qquad
\Phi=0\,,\qquad
\hat{F}_5=4\,\left(\omega_{{\rm AdS}_5}+\omega_{{\rm S}^5}\right)\,.
\end{split}
\end{align}
Here the metric of S$^5$ is described as $S^1$-fibration over $\mathbb{C}\text{P}^2$
and its explicit form is
\begin{align} 
\rmd s^2_{\rm S^5}&=(d\chi+\omega)^2 +ds^2_{\rm \mathbb{C}P^2}\,, \nln 
\rmd s^2_{\rm \mathbb{C}P^2}&= d\mu^2+\sin^2\mu\,
\bigl(\Sigma_1^2+\Sigma_2^2+\cos^2\mu\,\Sigma_3^2\bigr)\,,
\label{S1overCP2}
\end{align}
where $\chi$ is the fiber coordinate and $\omega$ is a one-form potential of the K\'ahler form on $\mathbb{C}$P$^2$\,.
$\Sigma_i ~(i=1,2,3)$ and $\omega$ are defined by
\begin{align}
\begin{split}
\Sigma_1&\equiv \tfrac{1}{2}(\cos\psi\, d\theta +\sin\psi\sin\theta\, d\phi)\,, \\
\Sigma_2&\equiv \tfrac{1}{2}(\sin\psi\, d\theta -\cos\psi\sin\theta\, d\phi)\,, \\
\Sigma_3&\equiv \tfrac{1}{2}(d\psi +\cos\theta\, d\phi)\,, 
\qquad 
\omega \equiv \sin^2\mu\, \Sigma_3\,.
\label{S1overCP21}
\end{split}
\end{align}
It is noted that the ${\rm S}^5$ part of the metric, the R-R $5$-form field strength and the dilaton remain the undeformed one.

\medskip

The deformed background is called the Schr\"odinger spacetime
and a gravity dual of dipole CFTs\cite{Herzog:2008wg,Maldacena:2008wh,Adams:2008wt}.
The spectral problem is recently studied in \cite{Guica:2017mtd} by using the integrability methods.

%
%
%
%
%
%

\subsection{Non-unimodular classical $r$-matrices}
\label{subsec:Non-unimodular-sol}

In this section, we consider YB deformations associated with non-unimodular $r$-matrix.
As explained in the previous sections,
the deformed backgrounds are solutions of the generalized supergravity equations.
Furthermore, we show that some of them reduce to the original $\AdS{5} \times S^5$ background
after performing a generalized TsT transformation.

\subsubsection{\mbox{\boldmath $1.~r=P_1\wedge D$}}

As a first example, let us consider the following non-Abelian classical $r$-matrix: 
\begin{equation}
  \label{eq:space-r-matrix}
r=\frac{1}{2}P_1\wedge D\,.
\end{equation}
This is a solution of the homogeneous CYBE which was already used to study 
a Yang--Baxter deformation of four-dimensional Minkowski spacetime~\cite{Matsumoto:2015ypa}. 

\medskip 

The corresponding $\beta$-field is
\begin{align}
\beta=\eta\,\partial_1 \wedge (t\,\partial_t+z\partial_z)\,,
\end{align}
where we have rewritten the four-dimensional Cartesian coordinates as:
\begin{align}
  \label{eq:cartesian-coord-x0x2x3}
  x^0 =t\sinh\phi\,,  \qquad x^2 =t\cosh\phi\cos\theta\,,  \qquad x^3 =t\cosh\phi\sin\theta\,.
\end{align}
Then, the deformed background is found to be\footnote{The metric and NS-NS two-form 
were computed in~\cite{vanTongeren:2015uha}.} 
\begin{equation}
  \begin{aligned}
    \dd{s}^2 &= \frac{z^2[\dd{t}^2+(\dd{x^1})^2+\dd{z}^2]+\eta^2(\dd{t}-t
      z^{-1}\dd{z})^2}{z^4+\eta^2(z^2+t^2)}
    +\frac{t^2(-\dd{\phi}^2+\cosh^2\phi \dd{\theta}^2)}{z^2}
    +\dd{s_{{\rm S}^5}^2}\,, \\
    B_2 &=-\eta\,\frac{t \dd{t} \wedge \dd{x^1} + z \dd{z}\wedge \dd{x^1}}{z^4+\eta^2(t^2+z^2)}\,, 
\qquad  \Phi = \frac{1}{2}\log
    \left[\frac{z^4}{z^4+\eta^2(t^2+z^2)}\right]\,,
\\
    \hat{F}_3 &= -\frac{4\eta\, t^2\cosh\phi}{z^4}\left[\dd{t} \wedge
      \dd{\theta} \wedge \dd{\phi}
      -\frac{t}{z}\dd{\theta}\wedge \dd{\phi} \wedge \dd{z}  \right]\,, \\
    \hat{F}_5 &=
    4\left[\frac{z^4}{z^4+\eta^2(t^2+z^2)}\omega_{\AdS5}+\omega_{{\rm
          S}^5}\right]\,,\qquad I=-\eta\,\partial_1\,.
  \end{aligned}
\label{space}
\end{equation}
Note here that the $\phi$ direction has the time-like signature. 
These fields \emph{do not} satisfy the equations of motion of type IIB
supergravity, but solve the equations of the generalized type IIB supergravity.
The GSE has the the Killing vector $I=\eta\,\partial_1$\,, and the vector satisfies the divergence formula,
\begin{align}
 I^1 = \eta = -\sfD_m\beta^{1m} \,.
\end{align}

\medskip

Let us now perform T-dualities for the deformed background
(\ref{space}). Following~\cite{Arutyunov:2015mqj}, the extra fields are traded for
a linear term in the dual dilaton. 
T-dualising along the $x^1$ and $\phi_3$ directions, we find:
\begin{equation}
  \label{T-4.3}
  \begin{aligned}
   \dd{s}^2 ={}& z^2\,(\dd{x^1})^2+\frac{1}{z^2}\Bigl[ (\dd{t} + \eta t
    \dd{x^1})^2+(\dd{z}+\eta z \dd{x^1})^2
    - t^2\dd{\phi}^2+t^2\cosh^2\phi \dd{\theta}^2\Bigr] \\
    &+ \dd{r}^2 +\sin^2 r \dd{\xi}^2+\cos^2\xi \sin^2 r\,
    \dd{\phi_1}^2
    + \sin^2r\sin^2\xi \dd{\phi_2}^2 + \frac{ \dd{\phi_3}^2}{\cos^2r}\,,\\
    \hat{\mathcal{F}}_5 ={}&
    \frac{4t^2\cosh\phi }{z^4\cos r} (\dd{t}+\eta t \dd{x^1})\wedge
    (\dd{z}+\eta z \dd{x^1})
    \wedge \dd{\theta}\wedge \dd{\phi}\wedge \dd{\phi_3}  \\
    & +
    2z\sin^3 r\sin2\xi \dd{x^1}\wedge \dd{r}\wedge \dd{\xi} \wedge
    \dd{\phi_1}\wedge \dd{\phi_2}
    \,, \\
    \Phi ={}& \eta x^1+\log\left[\frac{z}{\cos
        r}\right]\,,
  \end{aligned}
\end{equation}
where $\hat{\mathcal{F}}_5=e^{\Phi}\hat{F}$\,.
Remarkably, this is a solution 
of the \emph{usual} type IIB supergravity equations rather than the generalized ones. 
Note that the dilaton has a linear dependence on \(x^1\). 
This result is similar to the fact that 
the Hoare--Tseytlin solution~\cite{Hoare:2015wia} is ``T-dual'' to the $\eta$-deformed background. 
Indeed, we have followed the same strategy as in~\cite{Hoare:2015wia}. 

\medskip

The ``T-dualized'' background in \eqref{T-4.3} is a solution to the standard type
IIB equations and has a remarkable property: it is \emph{locally equivalent} to 
undeformed \(\AdS{5} \times S^5\). Let us first perform the following change of coordinates:
\begin{align}
  t = \tilde{t}(1-\eta\,\tilde{x}^1)\,, \qquad 
                                  z = \tilde{z}(1-\eta\, \tilde{x}^1)\,, \qquad 
                                       x^1 = -\frac{1}{\eta}\log(1-\eta\,\tilde{x}^1)\,.
\end{align}
Note that the new coordinate system does {\it not} cover all
of spacetime: the new coordinate $\tilde{x}^1$ has to be restricted to the region $\tilde{x}^1<\eta^{-1}\,$.
The signature of $\eta$ is fixed when we have chosen the deformation. 
This change of coordinates achieves the following points:
\begin{itemize}
\item it diagonalizes the metric;
\item it absorbs the \(x^1\)-dependence of the dilaton into the
  \(\tilde z\) variable, such that \(\partial_1\) is now a symmetry of the
  full background.
\end{itemize}
Explicitly, we find
\begin{equation}
  \begin{aligned}
    \dd{s}^2 ={}&\tilde{z}^2\,(\dd{\tilde{x}^1})^2+\frac{1}{\tilde{z}^2}\Bigl[
    \dd{\rho}^2+\dd{\tilde{z}}^2
    -\rho^2\dd{\phi}^2+\rho^2\cosh^2\phi \dd{\theta}^2\Bigr] \\
    & +\dd{r}^2 +\sin^2 r \dd{\xi}^2+\cos^2\xi \sin^2 r\,
    \dd{\phi_1}^2
    + \sin^2r\sin^2\xi \dd{\phi_2}^2 + \frac{ \dd{\phi_3}^2}{\cos^2r}\,,\\
    \hat{\mathcal{F}}_5 ={}&
    \frac{4\rho^2\cosh\phi }{\tilde{z}^4\cos r} \dd{\rho}\wedge
    \dd{\tilde{z}}
    \wedge \dd{\theta}\wedge \dd{\phi}\wedge \dd{\phi_3} \\
    & +
    2\tilde{z}\sin^3 r\sin2\xi \dd{\tilde{x}^1}\wedge \dd{r}\wedge
    \dd{\xi} \wedge \dd{\phi_1}\wedge \dd{\phi_2} \\
    \Phi ={}& \log\left[\frac{\tilde{z}}{\cos r}\right]\,.
  \end{aligned}
\end{equation}
Now we can perform again the two standard T-dualities along \(\tilde{x}^1\) and
\(\phi_3\) to find, as mentioned above, the \emph{undeformed} \(\AdS{5} \times \rmS^5\) background.\footnote{The
  usual Poincar\'e coordinates are found using the same change of
  coordinates as in \eqref{eq:cartesian-coord-x0x2x3}.}

\medskip 

Here, let us summarize what we have done.
We have started with a Yang--Baxter deformation of \(\AdS{5}\) described by the non-Abelian \(r\)-matrix \eqref{eq:space-r-matrix}. Using the formula (\ref{pre-beta-formula-noB}),
we have found the corresponding deformed background \eqref{space} which is a solution to the
generalized equations described in section~\ref{sec:GSE-YBsec}.
Then we have ``T-dualized'' this background using the rules of~\cite{Arutyunov:2015mqj}
to find a new background \eqref{T-4.3} which solves the \emph{standard}
supergravity equations, but whose dilaton depends linearly on one of
the T-dual variable.
Finally, we have observed that after a change of variables,
this last background is locally equivalent to the T-dual of the undeformed \(\AdS{5} \times S^5\). 
This result implies that the Yang--Baxter deformation with the classical $r$-matrix in \eqref{eq:space-r-matrix} 
can be interpreted as an integrable twist, just like in the case of Abelian classical $r$-matrices 
(see for example~\cite{Frolov:2005dj,Alday:2005ww,Vicedo:2015pna,Kameyama:2015ufa}).

\subsubsection{\mbox{\boldmath $2.~r=P_0\wedge D$}}

Our next example is the classical $r$-matrix
\begin{equation}
r=\frac{1}{2}P_0\wedge D\,,
\end{equation}
where \(P_0\) is the generator of time translations. This is a solution of the homogeneous CYBE 
which was originally utilized to study 
a Yang--Baxter deformation of 4D Minkowski spacetime~\cite{Matsumoto:2015ypa}. 

\medskip 

Then the $\beta$-field is
\begin{align}
\beta=\eta\,\partial_0 \wedge (\rho\,\partial_{\rho}+z\,\partial_z)\,.
\end{align}
and the divergence of the bi-vector is given by 
\begin{align}
 I^0 = -\eta = \sfD_m\beta^{0m} \,.
\end{align}
Here, we introduced the polar coordinates
\begin{align}
  x^1 &=\rho\sin\theta \cos\phi\,,& x^2 &=\rho\sin\theta\sin\phi\,, & x^3 &=\rho\cos\theta\,.
\end{align}
Therefore, 
the associated background is found to be\footnote{This background was studied in~\cite{vanTongeren:2015uha}, 
but only the metric and NS-NS two-form were computed therein.}
\begin{equation}
  \label{time}
  \begin{aligned}
    \dd{s}^2 ={}&\frac{z^2[-(\dd{x^0})^2+\dd{z}^2+\dd{\rho}^2]-\eta^2(\dd{\rho}-\rho
      z^{-1}\dd{z})^2}{z^4-\eta^2(z^2+\rho^2)}
    +\frac{\rho^2(\dd{\theta}^2+\sin^2\theta \dd{\phi}^2)}{z^2}
    +\dd{s_{{\rm S}^5}^2}\,, \\
    B_2 =-{}&\eta\,\frac{\dd{x^0}\wedge (\rho \dd{\rho}+z\,\dd{z})}{z^4-\eta^2(z^2+\rho^2)}\,, 
   \qquad \Phi ={} \frac{1}{2}\log
    \left[\frac{z^4}{z^4-\eta^2(z^2+\rho^2)}\right]
\\
    \hat{F}_3 ={}&\frac{4\eta\,\rho^2\sin\theta}{z^5}(z\,\dd{\rho}-\rho\,\rmd z)\wedge d
      \theta\wedge \dd{\phi}\,, \\
    \hat{F}_5 ={}&4 \left[\frac{z^4}{z^4-\eta^2(z^2+\rho^2)}\,\omega_{\AdS5} +
      \omega_{{\rm S}^5}\right]\,,\qquad
  \qquad I=-\eta\,\partial_0\,.
  \end{aligned}
\end{equation}
The above background becomes a solution of the GSE.

\medskip

Our next task is to perform ``T-dualities'' for the background
(\ref{time}) in the $x^0$ and $\phi_3$ directions. 
We find a solution of the standard type IIB supergravity
equations\footnote{Having performed a time-like T-duality we
  necessarily find a purely imaginary five-form flux.} with a dilaton that depends linearly on \(x^0\): 
\begin{equation}
  \label{T-4.2}
  \begin{aligned}
    \dd{s}^2 ={}& -z^2\,(\dd{x^0})^2+\frac{1}{z^2}\Bigl[ (\dd{\rho}-\eta
    \rho \dd{x^0})^2+(\dd{z}-\eta z \dd{x^0})^2
    + \rho^2\dd{\theta}^2 + \rho^2\sin^2\theta \dd{\phi}^2\Bigr]  \\
    & + \dd{r}^2+\sin^2 r \dd{\xi}^2+\cos^2\xi \sin^2 r \dd{\phi_1}^2
    + \sin^2r\sin^2\xi \dd{\phi_2}^2+\frac{ \dd{\phi_3}^2}{\cos^2r}\,,\\
    \hat{\mathcal{F}}_5 ={}& -\frac{4i\rho^2\sin\theta }{z^4\cos r}
    (\dd{\rho}-\eta\rho \dd{x^0})\wedge(\dd{z}-\eta z \dd{x^0})
    \wedge \dd{\theta}\wedge \dd{\phi}\wedge \dd{\phi_3} \\
    &+ 2i z\sin^3 r\sin2\xi\,
    \dd{x^0}\wedge \dd{r}\wedge \dd{\xi} \wedge \dd{\phi_1}\wedge \dd{\phi_2}\,, \\
    \Phi ={}& -\eta\, x^0+\log\left[\frac{z}{\cos
        r}\right]\,. 
  \end{aligned}
\end{equation}

\medskip

Finally, let us show that the ``T-dualized'' background (\ref{T-4.2}) is again equivalent to 
the undeformed \(\AdS{5} \times S^5\). 
First of all, we perform the following coordinate transformations: 
\begin{align}
\rho = \tilde{\rho} (1+\eta\,\tilde{x}^0)\,, \qquad 
z = \tilde{z} (1+\eta\,\tilde{x}^0)\,, \qquad
x^0 = \frac{1}{\eta}\log(1+\eta \tilde{x}^0)\,.
\end{align}
Note here that the new coordinate $\tilde{x}^0$ is restricted to the region $\tilde{x}^0<-\eta^{-1}\,$. 
Just like in the previous case,
the metric is diagonal and the dilaton does not depend anymore on
\(\tilde{x}^0\), such that \(\partial_0\) is an isometry:
\begin{equation}
  \begin{aligned}
    \dd{s}^2 ={}&
    -\tilde{z}^2\,(\dd{\tilde{x}^0})^2+\frac{1}{\tilde{z}^2}\Bigl[
    \dd{\tilde{\rho}}^2+\dd{\tilde{z}}^2
    + \tilde{\rho}^2\dd{\theta}^2 + \tilde{\rho}^2\sin^2\theta \dd{\phi}^2\Bigr] \\
    & + \dd{r}^2+\sin^2 r \dd{\xi}^2+\cos^2\xi \sin^2 r \dd{\phi_1}^2
    + \sin^2r\sin^2\xi \dd{\phi_2}^2+\frac{ \dd{\phi_3}^2}{\cos^2r}\,,\\
   \hat{\mathcal{F}}_5 ={}& -\frac{4i\tilde{\rho}^2\sin\theta
    }{\tilde{z}^4\cos r}\, \dd{\tilde{\rho}}\wedge \dd{\tilde{z}}
    \wedge \dd{\theta}\wedge \dd{\phi}\wedge \dd{\phi_3} \\
    &+ 2i \tilde{z}\sin^3 r\sin2\xi\,
    \dd{\tilde{x}^0}\wedge \dd{r}\wedge \dd{\xi} \wedge \dd{\phi_1}\wedge \dd{\phi_2}\,, \\
    \Phi ={}& \log\left[\frac{\tilde{z}}{\cos r}\right]\,.
  \end{aligned}
\end{equation}
Again, by performing two T-dualities along the $\tilde{x}^0$ and $\phi_3$
directions, we go back to 
the undeformed \(\AdS{5} \times S^5\) background.

\subsubsection{\mbox{\boldmath $3.~r=(P_0-P_3)\wedge (D+M_{03})$}}

Let us consider now the classical \(r\)-matrix
\begin{equation}
  r = \frac{1}{2\sqrt{2}} (P_0-P_3)\wedge (D+M_{03})\,,
  \label{r-4.1}
\end{equation}
where \(M_{03}\) is the generator of the Lorentz rotation in the plane
\((x^0, x^3)\).
Then the $\beta$-field is
\begin{align}
\beta=\eta\,\partial_- \wedge (\rho\,\rmd \rho+z\,\rmd z)\,.
\end{align}
where the Cartesian coordinates of the four-dimensional Minkowski
spacetime $x^{\mu}$ are
\begin{align}
  x^\pm =\frac{1}{\sqrt{2}} (x^0\pm x^3)\,, \qquad 
 x^1 =\rho \cos\theta\,, \qquad x^2 &=\rho \sin\theta\,. 
\end{align}
The divergence of the $\beta$-field is given by 
\begin{align}
 I^- = -2\eta = \sfD_m\beta^{-m} \,.
\end{align}

\medskip 

Performing the supercoset construction~\cite{Kyono:2016jqy}, we obtain the corresponding
background:
\begin{equation}
  \label{KY-sol}
  \begin{aligned}
    \dd{s^2} &= \frac{-2\dd{x^+}\dd{x^-} + \dd{\rho^2} + \rho^2\dd{\theta^2} + \dd{z^2}}{z^2}
    -\eta^2\left[\frac{\rho^2}{z^6}+\frac{1}{z^4}\right](\dd{x^+})^2 + \dd{s_{S^5}^2}\,, \\
    B_2 &= -\eta\,\frac{\dd{x^+}\wedge (\rho \dd{\rho}+z\,\dd{z})}{z^4}\,, \\
    \hat{F}_3 &= 4\eta \left[\frac{\rho^2}{z^5} \dd{x^+} \wedge \dd{\theta} \wedge
      \dd{z}
      +\frac{\rho}{z^4} \dd{ x^+}\wedge \dd{\rho} \wedge \dd{\theta} \right]\,, \\
    \hat{F}_5 &= 4 (\omega_{\AdS5}+\omega_{{\rm S}^5}) \,, \\
    \Phi &= \Phi_0 \text{ (constant),} \qquad
 I=-2\eta\, \partial_-\,,
  \end{aligned}
\end{equation}
This background is a solution of the generalized supergravity equations characterized by the Killing vectors $I=-2\eta\, \partial_-$\,.

\medskip

Let us perform four ``T-dualities'' along the $x^+\,,x^-\,,\phi_1$ and $\phi_2$ directions\footnote{
To perform the T-dualities in the two light-like directions one can equivalently pass to 
Cartesian coordinates \((x^0, x^3)\), T-dualize in these and finally introduce light-like combinations 
for the T-dual variables.}. 
The resulting background is given by 
\begin{equation}
  \label{T-4.1}
  \begin{aligned}
    \dd{s}^2 ={}& -2z^2\dd{x^+}\dd{x^-}
    +\frac{(\dd{\rho}+\eta \rho\dd{x^-})^2+\rho^2\dd{\theta}^2+(\dd{z}+\eta z\dd{x^-})^2}{z^2} \\
     &+ \dd{r}^2 + \sin^2 r\, \dd{\xi}^2+\frac{ \dd{\phi_1}^2}{\cos^2\xi
      \sin^2 r}
    +\frac{ \dd{\phi_2}^2}{\sin^2r\sin^2\xi}+\cos^2r \dd{\phi_3}^2\,,   \\
    \hat{\mathcal{F}}_5 ={}& \frac{4 i \rho}{z^3\sin\xi \cos\xi\sin^2 r}
    (\dd{\rho}+\eta \rho\dd{x^-})\wedge \dd{\theta} \wedge (\dd{z}+\eta z\dd{x^-})
    \wedge \dd{\phi_1}\wedge \dd{\phi_2}  \\
    & + 4i z^2\sin r\cos r \dd{x^+}\wedge \dd{x^-}\wedge \dd{r} \wedge \dd{\xi}\wedge \dd{\phi_3}\,, \\
    \Phi ={}& 2\eta x^-+\log \left[\frac{z^2}{\sin^2r
        \sin\xi\cos\xi}\right]\,,
  \end{aligned}
\end{equation}
where the other components are zero.

\medskip

The ``T-dualized'' background in \eqref{T-4.1} is a solution to the standard type
IIB equations and is again  \emph{locally} equivalent to 
undeformed \(\AdS{5} \times S^5\). Let us first change the coordinates as follows:
\begin{align}
  x^- = -\frac{1}{2 \eta} \log( 1-2\eta\,\tilde{x}^-)\, , \qquad \rho &= \tilde \rho \sqrt{1-2\eta\,\tilde{x}^-}  \, , 
\qquad   z = \tilde z \sqrt{1-2\eta\,\tilde{x}^-} \, .
\end{align}
Explicitly, we find
\begin{equation}
  \label{eq:light-like-IIB-solution}
  \begin{aligned}
    \dd{s}^2 ={}& -2\tilde{z}^2\dd{x^+} \dd{\tilde{x}^-} +\frac{\dd{\tilde{\rho}}^2
      +\tilde{\rho}^2\dd{\theta}^2+\dd{\tilde{z}}^2}{\tilde{z}^2} \\
    &+ \dd{r}^2 + \sin^2 r \dd{\xi}^2+\frac{ \dd{\phi_1}^2}{\cos^2\xi \sin^2 r}
    +\frac{ \dd{\phi_2}^2}{\sin^2r\sin^2\xi}+\cos^2r \dd{\phi_3}^2\,,   \\
    \hat{\mathcal{F}}_5 ={}& \frac{4 i \tilde{\rho}}{\tilde{z}^3\sin\xi
      \cos\xi\sin^2 r}\,
    \dd{\tilde{\rho}}\wedge \dd{\theta}\wedge \dd{\tilde{z}}\wedge \dd{\phi_1}\wedge \dd{\phi_2} \\
    &+ 4i \tilde{z}^2\sin r\cos r\,
    \dd{x^+}\wedge \dd{\tilde{x}^-}\wedge \dd{r} \wedge \dd{\xi}\wedge \dd{\phi_3}\,,  \\
    \Phi ={}& \log \left[\frac{\tilde{z}^2}{\sin^2 r
        \sin\xi\cos\xi}\right]\,.
  \end{aligned}
\end{equation}
Now, rewriting the light-like coordinates in terms of the Cartesian
coordinates as
\begin{align}
  x^+ &\equiv \frac{1}{\sqrt{2}}(\tilde{x}^0 + \tilde{x}^3)\,, &
\tilde{x}^- &\equiv \frac{1}{\sqrt{2}}(\tilde{x}^0 - \tilde{x}^3) \,,
\end{align}
and performing four T-dualities along $\tilde{x}^0$, $\tilde{x}^3$, $\phi_1$ and $\phi_2$, 
we reproduce the \emph{undeformed} \(\AdS{5} \times S^5\) background.

\paragraph{Mixing of Abelian and non-Abelian classical $r$-matrices.}

This example admits a generalization, obtained by mixing Abelian and non-Abelian classical $r$-matrices:
\begin{equation}
  r = \frac{1}{2\sqrt{2}}(P_0-P_3)\wedge \left[a_1 (D+M_{03})+a_2 M_{12}\right]\,. 
\label{4.7-r}
\end{equation}
When $a_2=0 $, the classical $r$-matrix reduces to the one
described above; when $a_1=0$, the $r$-matrix becomes Abelian 
and the associated background is the Hubeny--Rangamani--Ross solution
of~\cite{Hubeny:2005qu}, as shown in~\cite{Kyono:2016jqy}. 

\medskip 

In~\cite{Kyono:2016jqy} it was shown that with a supercoset construction, one
finds the following ten-dimensional background
\begin{equation}
  \begin{aligned}
    \dd{s}^2 &=
    \frac{-2\dd{x^+}\dd{x^-}+\dd{\rho}^2+\rho^2\dd{\theta}^2+\dd{z}^2}{z^2}
    -\eta^2\left[(a^2_1+a^2_2)\frac{\rho^2}{z^6}+\frac{a_1^2}{z^4}\right](\dd{x^+})^2+\dd{s}_{\rm S^5}^2\,, \\
    B_2 &= -\frac{\eta}{z^4}\dd{x^+}\wedge \left[ a_1 ( \rho \dd{\rho} + z \dd{z} )
      - a_2 \rho^2 \dd{\theta} \right]\,, \\
    \hat{F}_3 &= \frac{4\eta\rho}{z^5}\dd{x^+}\wedge \left[ a_1 ( z \dd{\rho} - \rho \dd{z} ) \wedge \dd{\theta}
      + a_2 \dd{\rho}\wedge \dd{z} \right]\,, \\
    \hat{F}_5 &= 4 (\omega_{\AdS5} + \omega_{{\rm S}^5})\,, \\
    \Phi &= \Phi_0 \text{ (constant)}\,,\qquad
   I=-2\eta\,a_1\partial_-\,.
  \end{aligned}
\end{equation}
This background is still a solution of the generalized equations with the Killing vectors $I=-2\eta\,a_1\partial_-$\,.
The background can be reproduced by using the formula (\ref{pre-beta-formula-noB}) with
\begin{align}
\beta=\eta\,\partial_- \wedge[a_1\,(\rho\,\rmd \rho+z\,\rmd z)-a_2\,\partial_{\theta}]\,.
\end{align}
The Killing vector $I$ also satisfies
\begin{align}
 I^- = -2\eta\,a_1 = \sfD_m\beta^{-m} \,.
\end{align}
In the special case $a_1=0$, the above background reduces to 
a solution of standard type IIB supergravity. 

\medskip

Let us next take four ``T-dualities'' along the $x^+\,,x^-\,,\phi_1$ and $\phi_2$ directions.
Then we can obtain a solution of the usual type IIB supergravity as 
\begin{equation}
  \label{T-4.7}
  \begin{aligned}
    \dd{s}^2 ={}&-2z^2\dd{x^+}\dd{x^-} +\frac{(\dd{\rho}+\eta a_1
      \rho\dd{x^-})^2+\rho^2(\dd{\theta}-\eta a_2\dd{x^-})^2
      +(\dd{z}+\eta a_1 z\dd{x^-})^2}{z^2} \\
    & + \dd{r}^2 + \sin^2 r \dd{\xi}^2+\frac{
      \dd{\phi_1}^2}{\cos^2\xi \sin^2 r}
    +\frac{ \dd{\phi_2}^2}{\sin^2r\sin^2\xi}+\cos^2r \dd{\phi_3}^2\,,  \\
    \hat{\mathcal{F}}_5 ={}& \frac{4 i \rho}{z^3\sin\xi \cos\xi\sin^2 r}
    (\dd{\rho}+\eta a_1 \rho\dd{x^-})\wedge (\dd{\theta}-\eta
    a_2\dd{x^-})
    \wedge (\dd{z}+\eta a_1 z\dd{x^-})\wedge \dd{\phi_1}\wedge \dd{\phi_2}  \\
    & + 4i z^2\sin r\cos r \dd{x^+}\wedge \dd{x^-}\wedge \dd{r} \wedge \dd{\xi}\wedge \dd{\phi_3}\,,  \\
    \Phi ={}& 2\eta a_1x^-+\log\left[\frac{z^2}{\sin^2r
        \sin\xi\cos\xi}\right]\,, 
  \end{aligned}
\end{equation}
where the other components are zero.
It is easy to see that this is just a twist of the previous solution
(in \eqref{eq:light-like-IIB-solution}) and in fact there is a
change of variables 
\begin{align}
  \rho &= \tilde{\rho}\,{\rm e}^{-\eta a_1\, x^-}\,, &
                                                      z &= \tilde{z}\,{\rm e}^{-\eta a_1\, x^-}\,, &
\theta &= \tilde{\theta}-\eta\, a_2 x^-\,, &
x^- &= -\frac{1}{2\eta a_1}\log(1-2\eta a_1\, \tilde{x}^-)\,,
\end{align}
that maps this background to the same local form:
\begin{equation}
  \begin{aligned}
    \dd{s}^2 ={}& -2\tilde{z}^2\dd{x^+}d\tilde{x}^-+\frac{\dd{\tilde{\rho}}^2
      +\tilde{\rho}^2\dd{\tilde{\theta}}^2+\dd{\tilde{z}}^2}{\tilde{z}^2} \\
    & + \dd{r}^2 + \sin^2 r \dd{\xi}^2+\frac{
      \dd{\phi_1}^2}{\cos^2\xi \sin^2 r}
    +\frac{ \dd{\phi_2}^2}{\sin^2r\sin^2\xi}+\cos^2r \dd{\phi_3}^2\,,   \\
    \hat{\mathcal{F}}_5 ={}& \frac{4 i \tilde{\rho}}{\tilde{z}^3\sin\xi
      \cos\xi\sin^2 r}\,
    \dd{\tilde{\rho}}\wedge \dd{\tilde{\theta}}\wedge \dd{\tilde{z}}\wedge \dd{\phi_1}\wedge \dd{\phi_2}  \\
    & + 4i \tilde{z}^2\sin r\cos r\,
    \dd{x^+}\wedge \dd{\tilde{x}^-}\wedge \dd{r} \wedge \dd{\xi}\wedge \dd{\phi_3}\,, \\
    \Phi ={}&\log\left[\frac{\tilde{z}^2}{\sin^2r
        \sin\xi\cos\xi}\right]\,,
  \end{aligned}
\end{equation}
which is a T-dual of the undeformed $\AdS{5}\times {\rm S}^5$ background.

\subsubsection{\mbox{\boldmath $4.~r=(P_0-P_3)\wedge D$}}

Let us now consider the non-Abelian classical $r$-matrix given by 
\begin{equation}
  r=\frac{1}{2\sqrt{2}} ( P_0 - P_3 ) \wedge D\,,
\end{equation}
which is another solution of the homogeneous CYBE. 
The associated $\beta$-field is
\begin{align}
\beta=\eta\,\partial_- \wedge (\rho\,\rmd \rho+z\,\rmd z+x^+\,\partial_{+})\,.
\end{align}
Here the following new coordinates have been introduced:
\begin{eqnarray}
x^0=\frac{x^++x^-}{\sqrt{2}}\,,\qquad
x^3=\frac{x^+-x^-}{\sqrt{2}}\,,\qquad
x^1=\rho\cos\theta\,,\qquad
x^2=\rho\sin\theta
\,.\label{lccoord}
\end{eqnarray}

\medskip 

Using the formula (\ref{pre-beta-formula-noB}), 
the associated background is found to be\footnote{The metric and NS-NS two-form were 
computed in~\cite{vanTongeren:2015uha}.} 
\begin{equation}
  \label{4.3}
  \begin{aligned}
    \dd{s}^2 ={}&\frac{1}{z^4-\eta^2(x^+)^2}\biggl[z^2(-2\dd{x^+}\dd{x^-}+\dd{z}^2)
    +2\eta^2z^{-2}x^+\rho \dd{x^+}\dd{\rho}-\eta^2 z^{-2}\rho^2(\dd{x^+})^2 \\
    &-\eta^2(\dd{x^+}-x^+z^{-1}\dd{z})^2\biggr]+\frac{\dd{\rho}^2+\rho^2\dd{\theta}^2}{z^2}
    +\dd{s_{{\rm S}^5}^2}\,, \\
    B_2 ={}&- \eta\,\frac{\dd{x^+} \wedge (z\dd{z}+\rho \dd{\rho}-x^+\dd{x^-})}{z^4-\eta^2(x^+)^2}\,, \\
    \hat{F}_3 ={}& 4\eta\,\frac{\rho}{z^4}\left[\frac{\rho}{z}\dd{x^+}\wedge \dd{\theta}\wedge \dd{z}
      + \dd{x^+}\wedge \dd{\rho}\wedge \dd{\theta}
      -\frac{ x^+}{z}\dd{\rho}\wedge \dd{\theta} \wedge \dd{z}\right]\,, \\
    \hat{F}_5 ={}& 4\left[\frac{z^4}{z^4-\eta^2(x^+)^2}\,\omega_{\AdS5} + \omega_{{\rm S}^5}\right]\,, \\
    \Phi ={}& \frac{1}{2}\log \left[\frac{z^4}{z^4-\eta^2(x^+)^2}\right]\,,\qquad
    I=-\eta\,\partial_-\,.
  \end{aligned}
\end{equation}
This background satisfies the GSE with the Killing vector $I=-\eta\,\partial_-$\,.
In particular, the divergence formula (\ref{div-formula}) works well.

\medskip

As of now, we have not found an appropriate T-dual frame
in which this background is a solution to the standard type IIB equations with a linear dilaton.
However, the deformed background can be reproduced by the generalized diffeomorphism (\ref{eq:gdiff-PmD})
which is a gauge symmetry of double field theory.
We will explain this point in section \ref{sec:YB-gdiff}.

\subsubsection{\mbox{\boldmath $5.~$}A scaling limit of the Drinfeld--Jimbo $r$-matrix 
\label{scaling:sec}}

Our last example is the  classical $r$-matrix
\begin{equation}
  r = \frac{1}{2}[P_0\wedge D+P^{i}\wedge (M_{0i}+M_{1i})]
\,. \label{HvT-r}
\end{equation}
It was originally studied in~\cite{Hoare:2016hwh} in relation to a scaling limit of the classical $r$-matrix 
of Drinfeld-Jimbo type. 

\medskip 

By using the polar coordinates $(\rho,\theta)\,,$
\begin{eqnarray}
x^2=\rho\cos\theta\,,\qquad
x^3=\rho\sin\theta
\,,
\end{eqnarray}
the $\beta$-field is expressed as
\begin{align}
\beta=\eta\,(-\rho\,\partial_1\wedge \partial_{\rho}+z\,\partial_0\wedge \partial_z)\,.
\end{align}
The divergence of the $\beta$-field generates the Killing vector
\begin{align}
 I^0 = -4\,\eta = \sfD_m\beta^{0m} \,,\qquad  I^1 = -2\,\eta = \sfD_m\beta^{1m} \,.
\end{align}
Therefore, by using the formula (\ref{pre-beta-formula-noB}), 
the full background is determined to be\footnote{The metric and NS-NS two-form 
were computed in~\cite{Hoare:2016hwh} without the total derivative term in $B_2$.}
\begin{equation}
  \label{HvT}
  \begin{aligned}
    \dd{s}^2 &= \frac{-(\dd{x^0})^2+\dd{z}^2}{z^2-\eta^2}
    +\frac{z^2\left[(\dd{x^1})^2+\dd{\rho}^2\right]}{z^4+\eta^2\rho^2}
    +\frac{\rho^2\dd{\theta}^2}{z^2}+ \dd{s}^2_{{\rm S}^5}\,,  \\
    B_2 &= -\frac{\eta }{z(z^2-\eta^2)} \dd{x^0}\wedge \dd{z}
    -\frac{\eta\, \rho}{z^4+\eta^2 \rho^2} \dd{x^1}\wedge \dd{\rho}\,, \\
    \hat{F}_{1} &= \frac{4\eta^2\rho^2}{z^4}\dd{\theta}\,, \\
    \hat{F}_{3} &= \frac{4\eta \rho^2}{z^3(z^2-\eta^2)} \dd{x^0}\wedge
    \dd{\theta} \wedge \dd{z}
    +\frac{4\eta \rho}{z^4+\eta^2\rho^2} \dd{x^1}\wedge \dd{\rho} \wedge \dd{\theta}\,,  \\
    \hat{F}_{5} &=
    4\left[\frac{z^6}{(z^2-\eta^2)(z^4+\eta^2\rho^2)}\,\omega_{\AdS5}
      +\omega_{{\rm S}^5}\right] \,, \\
    \Phi &=
    \frac{1}{2}\log\left[\frac{z^6}{(z^2-\eta^2)(z^4+\eta^2\rho^2)}\right]\,,\qquad
    I=-4\,\eta\,\partial_0-2\,\eta\, \partial_1\,.
  \end{aligned}
\end{equation}
Again, this background becomes a solution of the generalized equations with the Killing vector $I= -4\,\eta\,\partial_0-2\,\eta\, \partial_1$\,.

\medskip

Performing ``T-dualities'' for all of the $U(1)$ directions, 
we can obtain a solution of the standard type IIB supergravity:  
\begin{equation}
  \begin{aligned}
    \dd{s}^2 ={}& -z^2\,(\dd{x^0})^2+z^2\,(\dd{x^1})^2
    +\frac{(\dd{\rho}-\eta\rho \dd{x^1})^2+(\dd{z}-\eta z
      \dd{x^0})^2}{z^2}
    +\frac{z^2\dd{\theta}^2}{\rho^2} \\
    & + \dd{r}^2+\sin^2 r \dd{\xi}^2+ \frac{\dd{\phi_1}^2}{ \sin^2
      r\cos^2\xi} + \frac{\dd{\phi_2}^2}{\sin^2r\sin^2\xi}
    +\frac{ \dd{\phi_3}^2}{\cos^2r}\,,\\
    \hat{\mathcal{F}}_5 ={}&
    \frac{-4i}{z^2\sin^2r\cos r\sin\xi\cos\xi}\, \left(
      \dd{\rho}-\eta\, \rho \dd{x^1}\right) \wedge
    \left(\dd{z}-\eta\, z \dd{x^0} \right) \wedge
    \dd{\phi_1}\wedge \dd{\phi_2}\wedge \dd{\phi_3} \\
    & + 4i
    \frac{z^3}{\rho}\sin r \dd{x^0}\wedge \dd{x^1}\wedge \dd{\theta}
     \wedge \dd{r} \wedge \dd{\xi}\,, \\
    \Phi ={}& -4\eta\, x^0 +2\eta\, x^1+\log\left[\frac{z^3}{\rho
        \sin^2 r \cos r \sin\xi \cos\xi}\right]\,,
  \end{aligned}
\label{T-HvT}
\end{equation}
where the other components are zero. 

\medskip 

Just like in the first examples that we have discussed, there is a
simple change of coordinates
\begin{align}
  \rho =\tilde{\rho}\,{\rm e}^{\eta\, x^1}\,, \qquad z =\tilde{z}\,{\rm e}^{\eta\, x^0}\,, 
\end{align}
that diagonalizes the metric:
\begin{equation}
  \begin{aligned}
    \dd{s}^2 ={}& {\rm e}^{2\eta
      x^0}\tilde{z}^2\left[-(\dd{x^0})^2+(\dd{x^1})^2\right]
    +\frac{ {\rm e}^{-2\eta(x^0-x^1)}\dd{\tilde{\rho}}^2+\dd{\tilde{z}}^2}{\tilde{z}^2}
    +{\rm e}^{2\eta(x^0-x^1)}\frac{\tilde{z}^2\dd{\theta}^2}{\tilde{\rho}^2} \\
    & + \dd{r}^2+\sin^2 r \dd{\xi}^2+ \frac{\dd{\phi_1}^2}{ \sin^2
      r\cos^2\xi} + \frac{\dd{\phi_2}^2}{\sin^2r\sin^2\xi}
    +\frac{ \dd{\phi_3}^2}{\cos^2r}\,,\\
    \hat{\mathcal{F}}_5 ={}&
    \frac{-4i {\rm e}^{-\eta (x^0-x^1)}}{\tilde{z}^2\sin^2r\cos
      r\sin\xi\cos\xi}\,
    \dd{\tilde{\rho}} \wedge \dd{\tilde{z}} \wedge \dd{\phi_1}
    \wedge \dd{\phi_2}\wedge \dd{\phi_3} \\
    & + 4i
    \frac{ {\rm e}^{\eta\,(3x^0-x^1)}\tilde{z}^3}{\tilde{\rho}}\sin r\,
    \dd{x^0}\wedge \dd{x^1}\wedge \dd{\theta} \wedge \dd{r} \wedge \dd{\xi}\,,  \\
    \Phi ={}& -\eta\,( x^0 -
    x^1)+\log\left[\frac{\tilde{z}^3}{\tilde{\rho} \sin^2 r \cos r
        \sin\xi \cos\xi}\right]\,.
  \end{aligned}
\end{equation}
In this case, however, the linear dependence of the dilaton on the T-dual variables remains.

\medskip 

It may be helpful to recall the result for the case of 3D Schr\"odinger spacetime~\cite{Kawaguchi:2011wt,Kawaguchi:2012ug,Kawaguchi:2013lba}. 
An affine symmetry algebra is given by a twisted Yangian, which is called an exotic symmetry 
in~\cite{Kawaguchi:2011wt,Kawaguchi:2012ug,Kawaguchi:2013lba}, and can be mapped to the standard Yangian by undoing the integrable twist. 
Then, however, the target spacetime is not mapped to the undeformed AdS$_3$\,. 
The resulting geometry is described by a dipole-like coordinate system and
 hence it is very close to, but not identical to AdS$_3$\,.

\subsubsection*{The case without the total derivative of $B_2$}

It would be interesting to study also the case without the total derivative term of $B_2$ in (\ref{HvT})\,. 
Then the ``T-dualized'' background is different from the one of  (\ref{T-HvT}) 
and the resulting background is given by
\begin{eqnarray}
\dd{s}^2 &=& -(z^2-\eta^2)\,(\dd{x^0})^2+z^2\,(\dd{x^1})^2
+\frac{(\dd{\rho}-\eta\rho \dd{x^1})^2}{z^2}+\frac{\dd{z}^2}{z^2-\eta^2}
+\frac{z^2\dd{\theta}^2}{\rho^2}\no \\
&& + \dd{r}^2+\sin^2 r \dd{\xi}^2+ \frac{\dd{\phi_1}^2}{ \sin^2 r\cos^2\xi} 
+ \frac{\dd{\phi_2}^2}{\sin^2r\sin^2\xi}
+\frac{ \dd{\phi_3}^2}{\cos^2r}\,,\no\\ 
\mathcal{F}_5 &=& 
\frac{-4i}{z^2\sin^2r\cos r\sin\xi\cos\xi}\,
\left( \dd{\rho}-\eta\, \rho \dd{x^1}\right) \wedge 
\left(-\frac{z^2\dd{z}}{z^2-\eta^2}-\eta\, z \dd{x^0} \right) \wedge 
\dd{\phi_1}\wedge \dd{\phi_2}\wedge \dd{\phi_3}\no \\
&& + 4i  
\frac{z^3}{\rho}\sin r\, \left(\dd{x^0}+\frac{\eta\dd{z}}{z(z^2-\eta^2)}\right)
\wedge \dd{x^1}\wedge \dd{\theta} \wedge \dd{r} \wedge \dd{\xi}\,, \no \\
\Phi &=& 4\eta\, x^0 +2\eta\, x^1
+\log\left[\frac{(z^2-\eta^2)^2}{\rho z \sin^2 r \cos r \sin\xi \cos\xi}\right]\,. 
\label{T-HvTno}
\end{eqnarray}
This background is a solution of the usual type IIB supergravity, 
{and agrees with the one obtained 
in~\cite{Hoare:2016hwh} after fixing some typos. 

\medskip

Now it is natural to ask whether (\ref{T-HvT}) and (\ref{T-HvTno}) are equivalent or not,
 and if so, whether this equivalence holds locally or globally.  
The local equivalence can be shown explicitly by using the coordinate 
transformations
\begin{eqnarray}
x^0&\to&-x^0-\frac{1}{2\eta}\log\left[\frac{z^2-\eta^2}{z^2}\right]\qquad z>\eta\,,\no \\
x^0&\to&-x^0-\frac{1}{2\eta}\log\left[\frac{\eta^2-z^2}{z^2}\right]\qquad z<\eta\,.
\end{eqnarray}
For the case of the global equivalence, some subtleties arise. 
{The background (\ref{T-HvT}) is regular while the one (\ref{T-HvTno}) has a coordinate singularity 
at $z=\eta$ and so are the coordinate transformations. }
Moreover, two types of time directions have to be introduced. 
Due to these observations, a more involved analysis is necessary in order to argue the global equivalence.


\chapter{YB deformations as duality transformations}
\label{Ch:YB-duality}

This chapter is devoted to study of connections between the YB deformations and string duality transformations. 
To be more precise, we show that,
the homogeneous YB deformed $\AdS5\times \rmS^5$ background can be regarded as a $\beta$-transformed $\AdS5\times \rmS^5$ background. 
During the proof, we perform a suitable identification of the deformed vielbein and make a redefinition of the fermionic variable. 
These procedures can be clearly explained by using the double-vielbein formalism of DFT \cite{Siegel:1993xq,Jeon:2011cn,Jeon:2011vx,Jeon:2011sq,Jeon:2012kd,Jeon:2012hp}. 
Therefore, we start to review the double-vielbein formalism of DFT and find a simple $\beta$-transformation rule for the Ramond--Ramond (R--R) fields in section \ref{sec:DFT}. 
We also find the action of the double sigma model for type II superstring that reproduces the conventional GS superstring action. 
In section \ref{sec:YB-beta-deform}, we show the equivalence of homogeneous YB deformations and local $\beta$-transformations. 
Furthermore, in section \ref{sec:YB-gdiff},
we demonstrate that the YB deformations can also be realized as generalized diffeomorphisms which are gauge symmetries of the DFT.
Once the YB deformation is shown to be a kind of string duality transformations,
we expect that the equivalence between YB deformations and $\beta$-deformations will hold in more general backgrounds beyond the $\AdS5\times \rmS^5$ background. 
As a non-trivial example, in section \ref{sec:AdS3-YB},
we study local $\beta$-deformations of the $\AdS3\times \rmS^3\times \TT^4$ background that contains a non-vanishing $H$-flux.

\section{Local $\beta$-deformations in DFT}
\label{sec:DFT}

In this section, we review the basics of the type II DFT and find a simple transformation rule for bosonic fields under local $\beta$-deformations. 
We also find a manifestly $\OO(10,10)$-invariant superstring action that reproduces the conventional GS type II superstring action. 

\subsection{DFT fields and their parameterizations}

Bosonic fields in DFT are the generalized metric $\cH_{MN}$, the $T$-duality-invariant dilaton $d$, and the R--R potential $\bisC$, which is an $\OO(1,D-1)\times \OO(D-1,1)$ bispinor. 
In this subsection, we review their definitions and basic properties by turning off fermions (such as gravitino). 
Here, we employ the double-vielbein formalism developed in \cite{Siegel:1993xq,Jeon:2011cn,Jeon:2011vx,Jeon:2011sq,Jeon:2012kd,Jeon:2012hp} (see also \cite{Hohm:2010xe}), which is quite suitable for discussing YB deformations.

\subsubsection{NS--NS fields:}

The generalized metric $\cH_{MN}$ $(M,N=0,\dotsc,2D-1)$ is defined as
\begin{align}
\begin{split}
 \cH &\equiv (\cH_{MN}) \equiv E\,\mathsf{S}\, E^\rmT \,,\qquad E = (E_M{}^N) \in \OO(D,D)\,,
\\
 \mathsf{S}&\equiv (\mathsf{S}_{MN}) \equiv \diag(\underbrace{-1,\,+1,\cdots,+1}_{D},\,\underbrace{-1,\,+1,\dotsc,+1}_{D}) \,,
\end{split}
\end{align}
where the $\OO(D,D)$ property of the generalized vielbein $E$ is defined as
\begin{align}
 &E \, \ODDeta\, E^\rmT = \ODDeta = E^\rmT\,\ODDeta\, E \,,\\ 
 &\ODDeta\equiv (\ODDeta_{MN})\equiv \begin{pmatrix} 0 & \delta_m^n \\ \delta^m_n & 0 \end{pmatrix} \qquad 
 (m,n=0,\dotsc,D-1). 
\end{align}
The familiar properties of the generalized metric
\begin{align}
 \cH^\rmT = \cH \,,\qquad \cH^\rmT\,\ODDeta\,\cH = \ODDeta\,,
\label{eq:H-properties}
\end{align}
follow from the above definitions. 
When an $\OO(D,D)$ matrix $h$ satisfies 
\begin{align}
h^\rmT\,\mathsf{S}\,h = \mathsf{S}\,,
\end{align}
$h$ is an element of $\OO(1,D-1)\times \OO(D-1,1)$\,.
Then, it is easy to see that both $E$ and $E\,h$ give the same generalized metric $\cH$\,. 
Thus, the generalized metric $\cH$ can be regarded as a representative of a coset
\begin{align}
 \frac{\OO(D,D)}{\OO(1,D-1)\times \OO(D-1,1)} \,, 
\end{align}
where $\OO(1,D-1)\times \OO(D-1,1)$ is known as the double local Lorentz group \cite{Jeon:2011sq}. 

\medskip

In the following discussion, we raise or lower the $\OO(D,D)$ indices by using the $\OO(D,D)$-invariant metric $\ODDeta$ like $\cH_M{}^N \equiv \cH_{MP}\,\ODDeta^{PN}$\,.
Then, \eqref{eq:H-properties} indicates that the matrix $\cH_M{}^N$ has eigenvalues $\pm1$\,. 
Therefore, we introduce the double (inverse) vielbeins $\GV_{\Loa}{}^{M}$ and $\brGV_{\Lobra}{}^M$ ($\Loa,\,\Lobra=0,\dotsc,D-1$) as the eigenvectors
\begin{align}
 \cH^{M}{}_{N}\,\GV_{\Loa}{}^N =+\GV_{\Loa}{}^M\,,\qquad 
 \cH^{M}{}_{N}\,\brGV_{\Lobra}{}^N=-\brGV_{\Lobra}{}^M\,. 
\end{align}
Since the eigenvalues are different, they are orthogonal to each other
\begin{align}
 \cH_{MN}\,\GV_{\Loa}{}^M\,\brGV_{\Lobrb}{}^{N}=0 \,,\qquad 
 \ODDeta_{MN}\,\GV_{\Loa}{}^M\,\brGV_{\Lobrb}{}^{N}=0\,. 
\label{eq:H-eta-orthogonal}
\end{align}
Following \cite{Jeon:2011cn,Jeon:2011vx,Jeon:2011sq,Jeon:2012kd,Jeon:2012hp}, we normalize the double vielbeins as
\begin{align}
\begin{split}
 \eta_{\Loa\Lob}&=\ODDeta_{MN}\,\GV_{\Loa}{}^M \,\GV_{\Lob}{}^{N} = \cH_{MN}\,\GV_{\Loa}{}^M \,\GV_{\Lob}{}^{N} 
 = \diag(-1,\,+1,\dotsc,+1) \,,
\\
 \breta_{\Lobra\Lobrb}&=\ODDeta_{MN}\,\brGV_{\Lobra}{}^M \,\brGV_{\Lobrb}{}^{N} =-\cH_{MN}\,\brGV_{\Lobra}{}^M \,\brGV_{\Lobrb}{}^{N} 
 = \diag(+1,\,-1,\dotsc,-1) \,. 
\end{split}
\end{align}
By introducing $2D\times 2D$ matrices,
\begin{align}
 (\GV_A{}^M) \equiv \begin{pmatrix} \GV_{\Loa}{}^{M}\\ \brGV_{\Lobra}{}^{M} \end{pmatrix}\,,\qquad 
 (\eta_{AB}) \equiv \begin{pmatrix} \eta_{\Loa\Lob}& 0\\ 0& \breta_{\Lobra\Lobrb} \end{pmatrix} \,,\qquad 
 (\cH_{AB}) \equiv \begin{pmatrix} \eta_{\Loa\Lob}& 0\\ 0& -\breta_{\Lobra\Lobrb} \end{pmatrix} \,,
\end{align}
where $\{A\}\equiv\{\Loa,\,\Lobra\}$, the above orthonormal conditions are summarized as
\begin{align}
 \eta_{AB} = \GV_A{}^M\, \ODDeta_{MN}\, (\GV^\rmT)^N{}_B \,, \qquad 
 \cH_{AB} = \GV_A{}^M\, \cH_{MN}\, (\GV^\rmT)^N{}_B \,. 
\label{eq:DV-ortho-normal}
\end{align}
The matrix $\GV_A{}^M$ is always invertible and the inverse matrix is given by
\begin{align}
 (\GV^{-1})_M{}^A = \ODDeta_{MN}\,(\GV^\rmT)^N{}_B\,\eta^{BA} \,,
\end{align}
which indeed satisfies
\begin{align}
 \GV_A{}^M\,(\GV^{-1})_M{}^B = \ODDeta_{MN}\, \GV_A{}^M\,\GV_C{}^N \,\eta^{CB} = \delta_A^B \,. 
\end{align}
As long as we raise or lower the indices $M,\,N$ with $\ODDeta_{MN}$ and $A,\,B$ with $\eta_{AB}$ (namely, $\Loa,\,\Lob$ and $\Lobra,\,\Lobrb$ with $\eta_{\Loa\Lob}$ and $\eta_{\Lobra\Lobrb}$, respectively), there is no difference between $\GV_A{}^M$ and $(\GV^{-\rmT})_A{}^M\equiv \eta_{AB}\,\ODDeta^{MN}\,(\GV^{-\rmT})^B{}_N$\,. 
Thus, in the following, we may not show the inverse or the transpose explicitly. 

\medskip

When $D\times D$ matrices, $\GV^m{}_{\Loa}$ and $\brGV^m{}_{\Lobra}$, are invertible, we can parameterize the double vielbeins as
\begin{align}
 (\GV^M{}_{\Loa}) =\frac{1}{\sqrt{2}} \begin{pmatrix} (e^{-\rmT})^m{}_{\Loa} \\ E_{mn}\,(e^{-\rmT})^n{}_{\Loa} \end{pmatrix} \,, \qquad 
 (\brGV^M{}_{\Lobra}) =\frac{1}{\sqrt{2}} \begin{pmatrix} (\bre^{-\rmT})^m{}_{\Lobra} \\ \bar{E}_{mn}\,(\bre^{-\rmT})^n{}_{\Lobra} \end{pmatrix} \,,
\end{align}
where we introduced matrix notations, $e\equiv (e_m{}^{\Loa})$ and $\bre\equiv (\bre_m{}^{\Lobra})$. 
From \eqref{eq:DV-ortho-normal}, we find
\begin{align}
 \bar{E}_{mn} = - E^\rmT_{mn}\,,\qquad 
 \CG_{mn} \equiv E_{(mn)} = (e \,\eta \, e^\rmT)_{mn} = - (\bre\,\breta \, \bre^\rmT)_{mn}\,.
\end{align}
By denoting $B_{mn}\equiv E_{[mn]}$, the parameterization of the double vielbeins becomes
\begin{align}
 &(\GV^M{}_{\Loa}) =\frac{1}{\sqrt{2}} \begin{pmatrix} (e^{-\rmT})^m{}_{\Loa} \\ (\CG+B)_{mn}\,(e^{-\rmT})^n{}_{\Loa} \end{pmatrix} \,, \qquad 
 (\brGV^M{}_{\Lobra}) =\frac{1}{\sqrt{2}} \begin{pmatrix} (\bre^{-\rmT})^m{}_{\Lobra} \\ -(\CG-B)_{mn}\,(\bre^{-\rmT})^n{}_{\Lobra} \end{pmatrix} \,.
\label{eq:dV-geometric}
\end{align}
Then the generalized metric is expressed as
\begin{align}
\cH &=\begin{pmatrix} (\CG-B\,\CG^{-1}\,B)_{mn} & (B\,\CG^{-1})_m{}^n \\ -(\CG^{-1}\,B)^m{}_n & \CG^{mn} \end{pmatrix} 
= \begin{pmatrix} \delta_m^p & B_{mp} \\ 0 & \delta^m_p \end{pmatrix} 
 \begin{pmatrix} \CG_{pq} & 0 \\ 0 & \CG^{pq} \end{pmatrix} 
 \begin{pmatrix} \delta^q_n & 0 \\ -B_{qn} & \delta^q_n \end{pmatrix} 
\label{eq:H-geometric}
\end{align}
In this way, the generalized metric $\cH_{MN}$ is parametrized by the metric $g_{mn}$ and the Kalb-Ramond 2-form $B_{mn}$\,.

\medskip

In addition, when $\GV_m{}^{\Loa}$ and $\brGV_m{}^{\Lobra}$ are invertible,
we can introduce the dual parametrization of these fields.
The parametrization is given by
\begin{align}
\begin{split}
 \GV_M{}^{\Loa} &=\frac{1}{\sqrt{2}} \begin{pmatrix} \tilde{e}_m{}^{\Loa} \\ (\OG^{-1}-\beta)^{mn}\, \tilde{e}_n{}^{\Loa} \end{pmatrix} \,, \qquad 
 \brGV_N{}^{\Lobra} =\frac{1}{\sqrt{2}} \begin{pmatrix} \tilde{\bre}_m{}^{\Lobra} \\ -(\OG^{-1}+\beta)^{mn}\, \tilde{\bre}_n{}^{\Lobra} \end{pmatrix} \,,
\\
 \OG_{mn}&\equiv (\tilde{e}\,\eta\, \tilde{e}^\rmT)_{mn} = - (\tilde{\bre} \,\breta \, \tilde{\bre}^\rmT)_{mn}\,,\qquad \beta^{mn}=-\beta^{nm} \,.
\end{split}
\end{align}
Here, we introduced the dual metric $G_{mn}$ and the $\beta$-field $\beta^{mn}$\,.
In terms of these fields, the generalized metric can be parametrized as
\begin{align}
 \cH = \begin{pmatrix} \OG_{mn} & (\OG\, \beta)_m{}^{n} \\
 -(\beta\,\OG)^m{}_{n} & (\OG^{-1}-\beta\,\OG\,\beta)^{mn} 
 \end{pmatrix} 
 = \begin{pmatrix} \delta_m^p & 0 \\
 - \beta^{mp} & \delta^m_p 
 \end{pmatrix} \begin{pmatrix} \OG_{pq} & 0 \\ 0 & \OG^{pq} 
 \end{pmatrix} 
 \begin{pmatrix} \delta^p_n & \beta^{pn} \\
 0 & \delta_p^n 
 \end{pmatrix} \,. 
\label{eq:H-non-geometric}
\end{align}
The dual parametrization is also referred as a no-geometric parametrization and useful when we discuss the non-geometric structure of a given background (see chapter \ref{Ch:YB-T-fold}).

\medskip

When both parameterizations are possible, comparing \eqref{eq:H-geometric} and \eqref{eq:H-non-geometric}, we obtain
\begin{align}
\begin{split}
 E^{mn} &\equiv (E^{-1})^{mn} = \OG^{mn} - \beta^{mn} \qquad \bigl(E_{mn} \equiv \CG_{mn} + B_{mn}\bigr)\,,
\\
 \CG_{mn}&= E_{mp}\,E_{nq}\,\OG^{pq}\,,\qquad B_{mn}= E_{mp}\,E_{nq}\,\beta^{pq}\,. 
\end{split}
\label{eq:relation-open-closed}
\end{align}
In the following, we raise or lower the indices of $\{e_m{}^{\Loa},\,\bre_m{}^{\Lobra},\,\tilde{e}_m{}^{\Loa},\,\tilde{\bre}_m{}^{\Lobra}\}$ as
\begin{align}
\begin{alignedat}{2}
 e^m{}_{\Loa}&=\CG^{mn}\,e_n{}^{\Lob}\,\eta_{\Lob\Loa}\,,&\qquad 
 \bre^m{}_{\Lobra}&=\CG^{mn}\,\bre_n{}^{\Lobrb}\,\breta_{\Lobrb\Lobra}\,,
\\
 \tilde{e}^m{}_{\Loa}&=\OG^{mn}\,\tilde{e}_n{}^{\Lob}\,\eta_{\Lob\Loa}\,,&\qquad 
 \tilde{\bre}^m{}_{\Lobra}&=\OG^{mn}\,\tilde{\bre}_n{}^{\Lobrb}\,\breta_{\Lobrb\Lobra}\,, 
\end{alignedat}
\end{align}
and then we obtain relations like $(e^{-\rmT})^m{}_{\Loa}=e^m{}_{\Loa}$\,. 
We can then omit the inverse or the transpose without any confusions as long as the indices are shown explicitly. 
By using the two metrics, $\CG_{mn}$ and $\OG_{mn}$, we also introduce two parameterizations of the dilaton $d$,
\begin{align}
 \sqrt{\abs{\OG}}\,\Exp{-2\tilde{\phi}} = \Exp{-2d} = \sqrt{\abs{\CG}}\,\Exp{-2\Phi} \,. 
\label{eq:DFT-dilaton}
\end{align}

\subsubsection{Ramond--Ramond fields}

In order to study the ten-dimensional type II supergravity, let us consider the case $D=10$\,. 
Associated with the double local Lorentz group $\OO(1,9)\times\OO(9,1)$, we introduce two sets of gamma matrices, $(\Gamma^{\Loa})^{\SPa}{}_{\SPb}$ and $(\brGamma^{\Lobra})^{\SPbra}{}_{\SPbrb}$, satisfying
\footnote{The reader should not confuse indices of the gamma matrices $(\Gamma^a)^{\alpha}{}_{\beta}$ with world-sheet indices.}
\begin{align}
\begin{split}
 &\{\Gamma^{\Loa},\,\Gamma^{\Lob}\} =2\,\eta^{\Loa\Lob}\,,\qquad 
  \{\brGamma^{\Lobra},\,\brGamma^{\Lobrb}\} = 2\,\breta^{\Lobra\Lobrb} \,,
\\
 &(\Gamma^{\Loa})^\dagger = - \Gamma^{0}\,\Gamma^{\Loa}\,(\Gamma^{0})^{-1} 
 = \mp \Gamma^{\Loa}\,, \qquad \Loa=\biggl\{\begin{array}{l} 0 \\[-1mm] 1,\dotsc,9 \end{array} ,
\\
 &(\brGamma^{\Lobra})^\dagger = + \brGamma^{0}\,\brGamma^{\Lobra}\,(\brGamma^{0})^{-1} 
 = \pm \brGamma^{\Lobra}\,, \qquad \Lobra=\biggl\{\begin{array}{l} 0 \\[-1mm] 1,\dotsc,9 \end{array} .
\end{split}
\end{align}
We also introduce the chirality operators
\begin{align}
\begin{split}
 &\Gamma^{11}\equiv \Gamma^{012\cdots 9}\,,\qquad 
 \brGamma^{11}\equiv \brGamma^{012\cdots 9}\,,\qquad (\Gamma^{11})^\dagger=\Gamma^{11}\,,\qquad
 (\brGamma^{11})^\dagger=\brGamma^{11}\,,
\\
 &\{\Gamma^{\Loa},\,\Gamma^{11}\} =0\,,\qquad \{\brGamma^{\Lobra},\,\brGamma^{11}\} = 0\,,\qquad
 (\Gamma^{11})^2 =1\,,\qquad (\brGamma^{11})^2=1\,,
\end{split}
\end{align}
and the charge conjugation matrices $C_{\SPa\SPb}$ and $\brC_{\SPbra\SPbrb}$ satisfying%
\footnote{In order to follow the convention of \cite{Arutyunov:2015qva}, we employ the charge conjugation matrices $C_-$ and $\bar{C}_-$ of \cite{Jeon:2012kd} rather than $C_+$ and $\bar{C}_+$. They are related as $C_-=C_+\,\Gamma^{11}$ and $\bar{C}_- = \bar{C}_+\,\brGamma^{11}$.}
\begin{align}
\begin{alignedat}{2}
 &C\,\Gamma^{\Loa}\,C^{-1} = -(\Gamma^{\Loa})^\rmT \,, &\qquad
  C &= - C^\rmT = - C^{-1}\,, \qquad C^*=C \,,
\\
 &\brC\,\brGamma^{\Lobra}\,\brC^{-1} = -(\brGamma^{\Lobra})^\rmT \,,&\qquad 
 \brC &= -\brC^\rmT = - \brC^{-1}\,, \qquad \brC^*=\brC \,.
\end{alignedat}
\label{eq:CC-properties}
\end{align}
We can show $C\,\Gamma^{11}\,C^{-1}=-\Gamma^{11}$ and $\brC\,\brGamma^{11}\,\brC^{-1}=-\brGamma^{11}$ by using
\begin{align}
 C\,\Gamma^{\Loa_1\cdots\Loa_n}\,C^{-1} = (-1)^{\frac{n(n+1)}{2}} \,(\Gamma^{\Loa_1\cdots\Loa_n})^\rmT \,, \qquad
 \brC\,\brGamma^{\Lobra_1\cdots\Lobra_n}\,\brC^{-1} = (-1)^{\frac{n(n+1)}{2}} \,(\brGamma^{\Lobra_1\cdots\Lobra_n})^\rmT \,. 
\end{align}
We raise or lower the spinor indices by using the charge conjugation matrices like
\begin{align}
\begin{alignedat}{2}
 (\Gamma^{\Loa})_{\SPa\SPb} &\equiv (\Gamma^{\Loa})^{\SPc}{}_{\SPb}\, C_{\SPc\SPa}\,,&\qquad 
 (\Gamma^{\Loa})^{\SPa\SPb} &\equiv C^{\SPb\SPc}\,(\Gamma^{\Loa})^{\SPa}{}_{\SPc} \,,
\\
 (\brGamma^{\Lobra})_{\SPbra\SPbrb} &\equiv (\brGamma^{\Lobra})^{\SPbrc}{}_{\SPbrb}\,\brC_{\SPbrc\SPbra}\,,&\qquad 
 (\brGamma^{\Lobra})^{\SPbra\SPbrb} &\equiv \brC^{\SPbrb\SPbrc}\,(\brGamma^{\Lobra})^{\SPbra}{}_{\SPbrc} \,,
\end{alignedat}
\end{align}
and then from \eqref{eq:CC-properties} we have
\begin{align}
 (\Gamma^{\Loa})_{\SPa\SPb} = (\Gamma^{\Loa})_{\SPb\SPa}\,,\qquad 
 (\brGamma^{\Lobra})^{\SPbra\SPbrb} = (\brGamma^{\Lobra})^{\SPbrb\SPbra} \,.
\end{align}

We define the R--R potential as a bispinor $\bisC^{\SPa}{}_{\SPbrb}$ with a definite chirality
\begin{align}
 \Gamma^{11}\,\bisC\,\brGamma^{11} = \pm\, \bisC \,,
\end{align}
where the sign is for type IIA/IIB supergravity. 
The R--R field strength is defined as
\begin{align}
\begin{split}
 &\bisF{}^{\SPa}{}_{\SPbrb} \equiv \cD_+ \bisC{}^{\SPa}{}_{\SPbrb} \equiv \frac{1}{\sqrt{2}}\,\bigl(\Gamma^M\,\cD_M \bisC + \Gamma^{11}\,\cD_M \bisC\,\brGamma^M\bigr){}^{\SPa}{}_{\SPbrb}\,,
\\
 &\Gamma^M\equiv \GV^M{}_{\Loa}\,\Gamma^{\Loa}\,,\qquad 
 \brGamma^M\equiv \brGV^M{}_{\Lobra}\,\brGamma^{\Lobra}\,,
\end{split}
\label{eq:F-C-relation}
\end{align}
where $\cD_+$ is a nilpotent operator introduced in \cite{Jeon:2012kd}, and the covariant derivative $\cD_M$ for a bispinor $\bm{\cT}^{\SPa}{}_{\SPbrb}$ and the spin connections are defined as \cite{Jeon:2011cn,Jeon:2011vx,Jeon:2012kd}
\begin{align}
\begin{split}
 &\cD_M \bm{\cT}^{\SPa}{}_{\SPbrb} \equiv \partial_M \bm{\cT}^{\SPa}{}_{\SPbrb} + \Phi_{M}{}^{\SPa}{}_{\SPc}\,\bm{\cT}^{\SPc}{}_{\SPbrb} - \bm{\cT}^{\SPa}{}_{\SPbrc}\,\brPhi_{M}{}^{\SPbrc}{}_{\SPbrb}\,,
\\
 &\Phi_{M}{}^{\SPa}{}_{\SPb} \equiv \frac{1}{4}\,\Phi_{M\Loc\Lod}\,(\Gamma^{\Loc\Lod})^{\SPa}{}_{\SPb}\,,\qquad 
 \brPhi_{M}{}^{\SPbra}{}_{\SPbrb}\equiv \frac{1}{4}\,\brPhi_{M\Lobrc\Lobrd}\,(\Gamma^{\Lobrc\Lobrd}){}^{\SPbra}{}_{\SPbrb}
\\
 &\Phi_{M\Loc\Lod} \equiv \GV^N{}_{\Loc}\,\nabla_M \GV_{N\Lod}
 = \GV^N{}_{\Loc}\,\bigl(\partial_M \GV_{N\Lod} - \Gamma_{M}{}^P{}_N\,\GV_{P\Lod} \bigr) \,,
\\
 &\brPhi_{M\Lobrc\Lobrd} \equiv \brGV^N{}_{\Lobrc}\,\nabla_M \brGV_{N\Lobrd}
 = \brGV^N{}_{\Lobrc}\,\bigl(\partial_M \brGV_{N\Lobrd} - \Gamma_{M}{}^P{}_N\,\brGV_{P\Lobrd} \bigr) \,,
\end{split}
\end{align}
where $\nabla_M$ is the (semi-)covariant derivative in DFT \cite{Jeon:2010rw,Jeon:2011cn,Hohm:2011si} (see also \cite{Sakatani:2016fvh} which employs the same convention as this thesis). 
Since $\cD_+$ flips the chirality, $\bisF$ has the opposite chirality to $\bisC$ \cite{Jeon:2012kd}
\begin{align}
 \Gamma^{11}\,\bisF\,\brGamma^{11} = \mp\, \bisF \,. 
\end{align}
As it has been shown in \cite{Jeon:2012kd}, $\bisF$ transforms covariantly under the $\OO(1,9)\times\OO(9,1)$ double Lorentz transformations, and transforms as a scalar under generalized diffeomorphisms. 
Further, from the nilpotency of $\cD_+$, $\bisF$ is invariant under gauge transformations of R--R potential
\begin{align}
 \delta \bisC = \cD_+ \bm{\lambda} \,,\qquad \Gamma^{11}\,\bm{\lambda}\,\brGamma^{11} = \mp \bm{\lambda} \,,
\end{align}
and the Bianchi identity is given by
\begin{align}
 \cD_+ \bisF = 0 \,.
\end{align}
As in the case of the democratic formulation \cite{Fukuma:1999jt,Bergshoeff:2001pv}, the self-duality relation
\begin{align}
 \bisF = -\Gamma^{11}\,\bisF \ \bigl(= \pm\, \bisF \,\brGamma^{11}\bigr)\,,
\label{eq:F-self-dual}
\end{align}
for type IIA/IIB supergravity is imposed by hand at the level of the equations of motion.

\subsection{Section condition and gauge symmetry}

In this subsection, we briefly explain the section condition and gauge symmetry of the DFT.

\subsubsection{Section condition}

In DFT, we consider a gravitational theory on a \emph{doubled space} with coordinates 
\begin{align}
 (x^M)=(x^m,\,\tilde{x}_m) \qquad (M=1,\dotsc,2D;\, m=1,\dotsc,D)\,,
\end{align}
where $x^m$ are the standard ``physical'' $D$-dimensional coordinates and $\tilde{x}_m$ are the dual coordinates. 
For the consistency of DFT, we require that arbitrary fields or gauge parameters $A(x)$ and $B(x)$ satisfy the so-called section condition \cite{Siegel:1993th,Hull:2009mi,Hull:2009zb},
\begin{align}
 &\ODDeta^{MN}\,\partial_M A(x)\,\partial_N B(x) = 0\,,\label{eq:strong-const}\\
 &\ODDeta^{MN}\,\partial_M \partial_N A(x) = 0 \,. \label{eq:weak-const}
\end{align}
In general, under this condition, supergravity fields can depend on at most $D$ coordinates out of the $2D$ coordinates $x^M$. 
We frequently choose the ``canonical solution'' where all fields and gauge parameters are independent of the dual coordinates; $\tilde{\partial}^m \equiv \frac{\partial}{\partial\tilde{x}_m}=0$\,. 
In this case, DFT reduces to the conventional supergravity. 
Instead, if all fields depend on $(D-1)$ coordinates $x^i$ and only the dilaton $d(x)$ has an additional linear dependence on a dual coordinates $\tilde{z}$,
DFT reduces to the generalized supergravity\cite{Sakatani:2016fvh,Sakamoto:2017wor}.

\subsubsection{Gauge symmetry of DFT}

A generalized diffeomorphism in the doubled space is generated by the \emph{generalized Lie derivative} \cite{Siegel:1993th,Hull:2009zb}
\begin{align}
 \gLie_V W^M \equiv V^N\,\partial_N W^M - \bigl(\partial_N V^M -\partial^M V_N \bigr)\,W^N\,. 
\end{align}
Concretely, the generalized Lie derivative acts on $\cH_{MN}(x)$ and $d(x)$ as
\begin{align}
\begin{split}
 \gLie_V \cH_{MN} &= V^K\,\partial_K \cH_{MN} 
 + \big(\partial_M V^K -\partial^K V_M\big)\, \cH_{KN} 
 + \big(\partial_N V^K -\partial^K V_N\big)\, \cH_{MK} \,, 
\\
 \gLie_V \Exp{-2d} &= \partial_M \bigl(\Exp{-2d}V^M\bigr) \,. 
\end{split}
\end{align}
This transformation is a gauge symmetry of DFT 
as long as the diffeomorphism parameter $V^M$ satisfies the weak constraint
\begin{align}
\partial_N \partial^N V^M=0\,,
\end{align}
and the strong constraint
\begin{align}
\partial_N V^M\,\partial^N A=0\,,
\end{align}
where $A$ represents the parameter $V^M$ or the supergravity fields. 
A finite generalized diffeomorphism is realized by $\Exp{\gLie_V}$ \cite{Hohm:2012gk}. 

\medskip

The generalized diffeomorphism satisfies 
the gauge algebra $[\gLie_{V_1},\,\gLie_{V_2}]=\gLie_{[V_1,\,V_2]_\rmC}$ which is governed 
by the \emph{C-bracket}, 
\begin{align}
 [V_1,\,V_2]_\rmC \equiv \frac{1}{2}\,\bigl(\gLie_{V_1}V_2-\gLie_{V_2}V_1\bigr)\,.
\end{align}
This symmetry is interpreted as diffeomorphisms in the doubled spacetime, $x^M\to x^M + V^M(x)$\,. 
Indeed, under the canonical section $\tilde{\partial}^m=0$\,, this symmetry consists of the conventional diffeomorphisms and $B$-field gauge transformations. 
If we parameterize the diffeomorphism parameter as $(V^M)=(v^m,\,\tilde{v}_m)$, the vector $v^m$ corresponds to the $D$-dimensional diffeomorphism parameter while the 1-form $\tilde{v}_m$ corresponds to the gauge parameter of the $B$-field gauge transformation, $B_2\to B_2 + \rmd \tilde{v}_1$\,. 
In particular, for the usual vectors $V_a^M=(v_a^m,\,0)$ $(a=1,2)$ satisfying 
$\frac{\partial}{\partial \tilde{x}_m}V_a^N=0$, 
the C-bracket gives rise to the usual Lie bracket,
\begin{align}
 [V_1,\,V_2]_\rmC=[v_1, v_2]\,.
\end{align}
Under the canonical section, this is the whole gauge symmetry, but if we choose a different section, the generalized diffeomorphism may generate other local $\OO(D,D)$ transformations, such as $\beta$-transformations. 
For more details, the reader may consult a concise review \cite{Hohm:2013bwa}.

\subsection{Diagonal gauge fixing}
\label{sec:diagonal-gauge}

In this subsection, we review the diagonal gauge fixing introduced in \cite{Jeon:2011cn,Jeon:2012kd}. 

\subsubsection{NS--NS fields}

In order to constrain the redundantly introduced two vielbeins $e_{m}{}^{\Loa}$ and $\bre_{m}{}^{\Lobra}$\,, we implement the diagonal gauge fixing 
\begin{align}
 e_{m}{}^{\Loa} = \bre_{m}{}^{\Lobra} \,, 
\label{eq:diagonal-gauge}
\end{align}
which is important to reproduce the conventional supergravity. 
Before the diagonal gauge fixing, the double vielbeins transform as
\begin{align}
 \GV_{\Loa}{}^M \to h^M{}_N\,\,\GV_{\Loa}{}^N\,,\qquad
 \brGV_{\Lobra}{}^M \to h^M{}_N\,\,\brGV_{\Lobra}{}^N \,,
\end{align}
under a global $\OO(10,10)$ rotation or a finite generalized diffeomorphism. 
We parameterize the $\OO(10,10)$ matrix $h^M{}_N$ as
\begin{align}
\begin{split}
 &h_M{}^N = \begin{pmatrix} \bmp_{m}{}^{n} & \bmq_{mn} \\ \bmr^{mn} & \bms^{m}{}_{n} \end{pmatrix} \,,\qquad 
 h^M{}_N = \begin{pmatrix} \bms^{m}{}_{n} & \bmr^{mn} \\ \bmq_{mn} & \bmp_{m}{}^{n} \end{pmatrix} 
\\
 &\bigl(\bmp\,\bms^\rmT + \bmq\,\bmr^\rmT =\bm{1}\,,\quad 
 \bmr\,\bms^\rmT + \bms\,\bmr^\rmT =0 \,,\quad 
 \bmp\,\bmq^\rmT + \bmq\,\bmp^\rmT =0 \bigr)\,, 
\end{split}
\end{align}
and then obtain the following transformation rule:
\begin{align}
\begin{alignedat}{2}
 e_m{}^{\Loa} &\to \bigl[\bigl(\bms^\rmT + E^\rmT\,\bmr^\rmT\bigr)^{-1}\bigr]_m{}^n\, e_n{}^{\Loa}\,,&\qquad 
 \bre_m{}^{\Lobra} &\to \bigl[\bigl(\bms^\rmT - E\,\bmr^\rmT\bigr)^{-1}\bigr]_m{}^n\, \bre_n{}^{\Lobra}\,,
\\
 \tilde{e}_m{}^{\Loa} &\to \bigl(\bmp + \bmq\, E^{-1}\bigr)_m{}^n\,\tilde{e}_n{}^{\Loa}\,,&\qquad 
 \tilde{\bre}_m{}^{\Lobra} &\to \bigl(\bmp - \bmq\, E^{-\rmT}\bigr)_m{}^n\,\tilde{\bre}_n{}^{\Lobra}\,, 
\\
 E_{mn} &\to [(\bmq+\bmp\,E)\,(\bms+\bmr\,E)^{-1}]_{mn} \,,&\qquad 
 E^{mn} &\to [(\bmr + \bms\,E^{-1})\,(\bmp + \bmq\,E^{-1})^{-1}]^{mn} \,. 
\label{eq:double-vielbein-transf}
\end{alignedat}
\end{align}
At the same time, the dilaton transforms as
\begin{align}
 \Exp{-2d} \to \abs{\det(\bmp_{m}{}^{n})}\,\Exp{-2d} \,, 
\end{align}
and the bispinors of R--R fields, $\bisC$ and $\bisF$, are invariant. 

\medskip

As we can see from \eqref{eq:double-vielbein-transf}, under a (geometric) subgroup (where $\bmr^{mn}=0$),
\begin{align}
 h_M{}^N = \begin{pmatrix} \bmp_{m}{}^{n} & \bmq_{mn} \\ 0 & (\bmp^{-\rmT})^{m}{}_{n} \end{pmatrix} \,,\qquad 
 h^M{}_N = \begin{pmatrix} (\bmp^{-\rmT})^{m}{}_{n} & 0 \\ \bmq_{mn} & \bmp_{m}{}^{n} \end{pmatrix} 
 \qquad 
 \bigl(\bmp\,\bmq^\rmT = - \bmq\,\bmp^\rmT \bigr)\,,
\end{align}
$e_m{}^{\Loa}$ and $\bre_m{}^{\Lobra}$ transform in the same manner. 
However, if we perform a general $\OO(10,10)$ transformation with $\bmr^{mn}\neq 0$, even if we choose the diagonal gauge in the original duality frame $(e_m{}^{\Loa}=\bre_m{}^{\Lobra})$, after the transformation, $e_m{}^{\Loa}\to e'_m{}^{\Loa}$ and $\bre_m{}^{\Lobra}\to \bre'_m{}^{\Lobra}$, we obtain
\begin{align}
 \bre'_m{}^{\Lobra} = (\Lambda^{-1})^{\Lobra}{}_{\Lob}\,e'_m{}^{\Lob} \,,\qquad 
 \Lambda^{\Loa}{}_{\Lobrb} \equiv \bigl[e^{\rmT}\,(\bms + \bmr\,E)^{-1}\,(\bms - \bmr\,E^\rmT)\,e^{-\rmT} \bigr]^{\Loa}{}_{\Lobrb} \in \OO(9,1) \,.
\label{eq:Lambda-def}
\end{align}
In order to maintain the diagonal gauge \eqref{eq:diagonal-gauge}, we shall simultaneously perform an $\OO(9,1)$ local Lorentz transformation for barred tensors that compensates the deviation of $\bre_m{}^{\Lobra}$ from $e_m{}^{\Loa}$. 
Namely, we modify the $\OO(10,10)$ transformation as \cite{Jeon:2012kd}
\begin{align}
 \GV_M{}^{\Loa} \to h_M{}^N\, \GV_{N}{}^{\Loa}\,,\qquad
 \brGV_M{}^{\Lobra} \to h_M{}^N\, \Lambda^{\Lobra}{}_{\Lobrb}\,\brGV_{N}{}^{\Lobrb} \,. 
\label{eq:modified-O(10-10)}
\end{align}
After the diagonal gauge fixing, since there is no more distinction between $\{\Loa,\,\SPa\}$ and $\{\Lobra,\,\SPbra\}$, we may simply replace $\{\Lobra,\,\SPbra\}$ by $\{\Loa,\,\SPa\}$. 
In this replacement, we should be careful about the signature
\begin{align}
 \breta_{\Loa\Lob} = -\eta_{\Loa\Lob}\,,\qquad \brC_{\SPa\SPb} = C_{\SPa\SPb} \,.
\end{align}
In addition, we relate the two sets of gamma matrices as
\begin{align}
 \brGamma^{\Loa}=\Gamma^{11}\,\Gamma^{\Loa}\qquad 
 \bigl(\,\{\brGamma^{\Loa},\,\brGamma^{\Lob}\} = - \{\Gamma^{\Loa},\,\Gamma^{\Lob}\} = 2\,\breta^{\Loa\Lob}\, \bigr)\,,\qquad 
 \brGamma^{11}=-\Gamma^{11}\,.
\end{align}

\subsubsection{R--R fields}

According to the diagonal gauge fixing, there is no distinction between the two spinor indices $\SPa$ and $\SPbra$, and we can convert the bispinors into polyforms:
\begin{align}
 \bisC^{\SPa}{}_{\SPb} = \sum_{n} \frac{1}{n!}\,\hat{\cC}_{\Loa_1\cdots \Loa_n}\,(\Gamma^{\Loa_1\cdots \Loa_n})^{\SPa}{}_{\SPb} \,,\qquad 
 \bisF^{\SPa}{}_{\SPb} = \sum_{n} \frac{1}{n!}\,\hat{\cF}_{\Loa_1\cdots \Loa_n}\,(\Gamma^{\Loa_1\cdots \Loa_n})^{\SPa}{}_{\SPb} \,. 
\end{align}
From the identity,
\begin{align}
 \Gamma^{11}\,\Gamma^{\Loa_1\cdots \Loa_p} = \frac{(-1)^{\frac{p(p+1)}{2}}}{(10-p)!}\,\epsilon^{\Loa_1\cdots \Loa_p \Lob_1\cdots \Lob_{10-p}}\,\Gamma_{\Lob_1\cdots \Lob_{10-p}} \,,
\end{align}
where $\epsilon_{0\cdots 9}=-\epsilon^{0\cdots 9}=1$\,, the self-duality relation \eqref{eq:F-self-dual} can be expressed as
\begin{align}
 \hat{\cF}_p = (-1)^{\frac{p(p-1)}{2}}\, * \hat{\cF}_{10-p} \,.
\end{align}
Here, we have defined
\begin{align}
\begin{alignedat}{2}
 \hat{\cF} &\equiv \sum_p \hat{\cF}_p\,,&\qquad 
 \hat{\cF}_p &\equiv \frac{1}{p!}\,\hat{\cF}_{m_1\cdots m_p}\,\rmd x^{m_1}\wedge \cdots \wedge\rmd x^{m_p}\,,
\\
 \hat{\cC} &\equiv \sum_p \hat{\cC}_p\,,&\qquad 
 \hat{\cC}_p &\equiv \frac{1}{p!}\,\hat{\cC}_{m_1\cdots m_p}\,\rmd x^{m_1}\wedge \cdots \wedge\rmd x^{m_p}\,,
\end{alignedat}
\end{align}
where the R--R fields with the curved indices are defined as
\begin{align}
 \hat{\cF}_{m_1\cdots m_p} \equiv e_{m_1}{}^{\Loa_1}\cdots e_{m_p}{}^{\Loa_p}\,\hat{\cF}_{\Loa_1\cdots \Loa_p}\,,\qquad
 \hat{\cC}_{m_1\cdots m_p} \equiv e_{m_1}{}^{\Loa_1}\cdots e_{m_p}{}^{\Loa_p}\,\hat{\cC}_{\Loa_1\cdots \Loa_p}\,.
\label{eq:B-curved-flat}
\end{align}
In addition, if we define the components of the spin connections as
\begin{align}
\begin{split}
 \Phi_{\Loa\Loc\Lod} &\equiv \GV^M{}_{\Loa}\,\Phi_{M\Loc\Lod} \,,\qquad 
 \brPhi_{\Lobra\Lobrc\Lobrd} \equiv \brGV^M{}_{\Lobra}\,\brPhi_{M\Lobrc\Lobrd} \,,
\\
 \Phi_{\Lobra\Loc\Lod} &\equiv \brGV^M{}_{\Lobra}\,\Phi_{M\Loc\Lod} \,,\qquad 
 \brPhi_{\Loa\Lobrc\Lobrd} \equiv \GV^M{}_{\Loa}\,\brPhi_{M\Lobrc\Lobrd} \,,
\end{split}
\end{align}
and compute their explicit forms under the canonical section $\tilde{\partial}^m=0$ as
\begin{align}
\begin{split}
 &\!\begin{alignedat}{2}
 \Phi_{\Lobra\Lob\Loc} &= \frac{1}{\sqrt{2}}\,\Bigl(\omega_{\Loa\Lob\Loc} + \frac{1}{2}\, H_{\Loa\Lob\Loc}\Bigr) \,,&\qquad 
 \brPhi_{\Loa\Lobrb\Lobrc} &= \frac{1}{\sqrt{2}}\,\Bigl(-\omega_{\Loa\Lob\Loc} + \frac{1}{2}\, H_{\Loa\Lob\Loc}\Bigr)\,,
\\
 \Phi_{[\Loa\Lob\Loc]} &= \frac{1}{\sqrt{2}}\,\Bigl(\omega_{[\Loa\Lob\Loc]} + \frac{1}{6}\,H_{\Loa\Lob\Loc}\Bigr)\,,&\qquad 
 \brPhi_{[\Lobra\Lobrb\Lobrc]} &= \frac{1}{\sqrt{2}}\,\Bigl(-\omega_{[\Loa\Lob\Loc]} + \frac{1}{6}\,H_{\Loa\Lob\Loc}\Bigr)\,,
\end{alignedat}
\\
 &\eta^{\Loa\Lob}\,\Phi_{\Loa\Lob\Loc} = \frac{1}{\sqrt{2}}\,\bigl(\eta^{\Loa\Lob}\,\omega_{\Loa\Lob\Loc} -2\,e_{\Loc}{}^m\,\partial_m\Phi\bigr) 
 = \breta^{\Lobra\Lobrb}\,\brPhi_{\Lobra\Lobrb\Lobrc} \,,
\\
 &\omega_{\Loa\Lob\Loc}\equiv e_{\Loa}{}^{m}\,\omega_{m\Lob\Loc} \,,\qquad 
  \omega_m{}^{\Loa\Lob} \equiv 2\,e^{n[\Loa}\,\partial_{[m} e_{n]}{}^{\Lob]} - e^{\Loa p}\,e^{\Lob q}\,\partial_{[p} e_{q]}{}^{\Loc}\,e_{m\Loc} \,,
\\
 &H_{\Loa\Lob\Loc}\equiv e_{\Loa}{}^{m}\,e_{\Lob}{}^{n}\,e_{\Loc}{}^{p}\,H_{mnp} \,, \qquad 
  H_{mnp}\equiv 3\,\partial_{[m}B_{np]}\,,
\end{split}
\label{eq:spin-connections}
\end{align}
we can show that the relation \eqref{eq:F-C-relation} between $\bisF$ and $\bisC$ can be expressed as \cite{Jeon:2012kd}\footnote{Here, we have used the following identities for type IIA/IIB theory:
\begin{align*}
\begin{split}
 &\frac{1}{2}\, \bigl(\Gamma^{m}\,\partial_{m}\bisC \mp \partial_m\bisC\,\Gamma^{m}\bigr) = \sum_n \frac{1}{n!}\,(\rmd \cC)_{\Loa_1\cdots\Loa_n}\,\Gamma^{\Loa_1\cdots\Loa_n} \,,
\\
 &\frac{1}{2}\, \partial_m\Phi\,\bigl(\Gamma^{m}\,\bisC \mp \bisC\,\Gamma^{m}\bigr) = \sum_n \frac{1}{n!}\,(\rmd\Phi \wedge \cC)_{\Loa_1\cdots\Loa_n}\,\Gamma^{\Loa_1\cdots\Loa_n} \,,
\\
 &\frac{1}{8}\, \omega_{m\Loa\Lob}\,\bigl[\Gamma^{\Loa}\,(\Gamma^{\Lob\Loc}\,\bisC-\bisC\,\Gamma^{\Lob\Loc})\mp (\Gamma^{\Lob\Loc}\,\bisC-\bisC\,\Gamma^{\Lob\Loc})\,\Gamma^{\Loa}\bigr] = -\sum_n \frac{\omega_{[\Loa_1}{}^{\Lob}{}_{\Loa_2}\,\cC_{|\Lob|\Loa_3\cdots\Loa_n]}}{2!\,(n-2)!} \,\Gamma^{\Loa_1\cdots\Loa_n} \,,
\\
 &\frac{1}{16}\,H_{\Loa\Lob\Loc}\,\Bigl[\frac{1}{3}\, \bigl(\Gamma^{\Loa\Lob\Loc}\,\bisC \mp \bisC\,\Gamma^{\Loa\Lob\Loc}\bigr) 
 + \bigl(\Gamma^{\Loa}\,\bisC\,\Gamma^{\Lob\Loc} \mp \Gamma^{\Lob\Loc}\,\bisC\,\Gamma^{\Loa}\bigr) \Bigr]
 = \sum_n \frac{1}{n!}\,(H_3\wedge \cC)_{\Loa_1\cdots\Loa_n}\,\Gamma^{\Loa_1\cdots\Loa_n} \,. 
\end{split}
\end{align*}
}
\begin{align}
 \hat{\cF} = \rmd \hat{\cC} - \rmd \Phi\wedge \hat{\cC} + H_3\wedge \hat{\cC} \,. 
\label{eq:cF-cC-relation}
\end{align}
It is noted here that as explained in \cite{Sakamoto:2017wor},
the relation (\ref{eq:cF-cC-relation}) is modified in the case of generalized supergravity backgrounds.
We will come back to this point in section \ref{sec:R-R(m)DFT}.

\medskip

Originally, the R--R fields were invariant under global $\OO(10,10)$ transformations or generalized diffeomorphisms, but after the diagonal gauge fixing, according to the modified transformation rule \eqref{eq:modified-O(10-10)}, they transform as
\begin{align}
 \bisC \to \bisC \, \Omega^{-1} \,,\qquad 
 \bisF \to \bisF \, \Omega^{-1} \,,
\label{eq:R-R-beta-transf}
\end{align}
where $\Omega$ is the spinor representation of the local Lorentz transformation \eqref{eq:Lambda-def},
\begin{align}
 \Omega^{-1}\,\brGamma^{\Loa}\,\Omega = \Lambda^{\Loa}{}_{\Lob}\, \brGamma^{\Lob} \qquad 
 \bigl(\Lambda^{\Loa}{}_{\Lob} = \bigl[e^{\rmT}\,(\bms + \bmr\,E)^{-1}\,(\bms - \bmr\,E^\rmT)\,e^{-\rmT} \bigr]^{\Loa}{}_{\Lob} \bigr)\,.
\label{eq:Omega-def}
\end{align}

\medskip

For later convenience, we here introduce several definitions of R--R fields that can be summarized as follows:
\begin{align}
\begin{split}
 (\check{C},\,\check{F};\,\check{\bm{C}},\,\check{\bm{F}}) \quad \xleftarrow[\quad \beta\text{-untwist}\quad]{\Exp{\beta\vee}}\quad 
 (A,\,F;&\,\sla{A},\,\sla{F}) \quad \xrightarrow[\quad B\text{-untwist}\quad]{\Exp{-B_2\wedge}}\quad 
 (\hat{C},\,\hat{F};\,\hat{\bm{C}},\,\hat{\bm{F}})
\\
 \text{\scriptsize$\Exp{\Phi}$ }~\!\!\!\Big\downarrow\text{\scriptsize$\Phi$-untwist}\hspace{39mm}
 \text{\scriptsize$\Exp{\Phi}$ }~&\!\!\!\Big\downarrow\text{\scriptsize$\Phi$-untwist}\hspace{39mm}
 \text{\scriptsize$\Exp{\Phi}$ }~\!\!\!\Big\downarrow\text{\scriptsize$\Phi$-untwist}
\\
 (\check{\cC},\,\check{\cF};\,\check{\bm{\cC}},\,\check{\bm{\cF}}) \quad \xleftarrow[\quad \beta\text{-untwist}\quad]{\Exp{\beta\vee}}\quad \hspace{6mm}
 (\cA,&\,\,\cF) \quad\hspace{6mm} \xrightarrow[\quad B\text{-untwist}\quad]{\Exp{-B_2\wedge}}\quad 
 (\hat{\cC},\,\hat{\cF};\,\bisC,\,\bisF)
\label{eq:R-R-diagram}
\end{split}
\end{align}
The quantities at the lower right, polyforms $(\hat{\cC},\,\hat{\cF})$ and bispinors $(\bisC,\,\bisF)$, are already defined, which we call $(B,\,\Phi)$-untwisted fields. 
There, the curved indices and flat indices are interchanged by using the usual vielbein $e_m{}^{\Loa}$ like \eqref{eq:B-curved-flat}. 

\medskip

The quantities at the upper right, which we call the $B$-untwisted fields, are defined as
\begin{align}
\begin{split}
 \hat{\bm{C}} \equiv \Exp{-\Phi}\bisC\,,\qquad 
 \hat{\bm{F}} \equiv \Exp{-\Phi}\bisF\,,\\
 \hat{C}\equiv \Exp{-\Phi}\hat{\cC}\,,\qquad 
 \hat{F}\equiv \Exp{-\Phi}\hat{\cF}\,. 
\end{split}
\end{align}
The curved and flat indices are again related as
\begin{align}
 \hat{C}_{m_1\cdots m_n}\equiv e_{m_1}{}^{\Loa_1}\cdots e_{m_n}{}^{\Loa_n}\, \hat{C}_{\Loa_1\cdots \Loa_n}\,,\qquad
 \hat{F}_{m_1\cdots m_n}\equiv e_{m_1}{}^{\Loa_1}\cdots e_{m_n}{}^{\Loa_n}\, \hat{F}_{\Loa_1\cdots \Loa_n}\,. 
\end{align}
The $B$-untwisted fields are rather familiar R--R fields satisfying
\begin{align}
 \hat{F} = \rmd \hat{C} + H_3\wedge \hat{C} \,,
\end{align}
which can be shown from \eqref{eq:cF-cC-relation}. 
We also define a polyform $A$ and its field strength $F$ as
\begin{align}
 A= \Exp{-\Phi} \Exp{B_2\wedge} \hat{\cC} = \Exp{B_2\wedge} \hat{C} \,,\qquad 
 F= \Exp{-\Phi} \Exp{B_2\wedge} \hat{\cF} = \Exp{B_2\wedge} \hat{F} \,. 
\end{align}
These are utilized in \cite{Fukuma:1999jt,Hassan:1999mm,Hohm:2011dv} to define R--R fields as $\OO(D,D)$ spinors (see also \cite{Sakamoto:2017wor})
\begin{align}
 \sla{A} \equiv \sum_n \frac{1}{n!}\, A_{m_1\cdots m_n}\,\gamma^{m_1\cdots m_n}\vert0\rangle\,,\qquad 
 \sla{F} \equiv \sum_n \frac{1}{n!}\, F_{m_1\cdots m_n}\,\gamma^{m_1\cdots m_n}\vert0\rangle\,. 
\end{align}
By using the dual fields $(\tilde{e}_m{}^{\Loa},\,\beta^{mn},\,\tilde{\phi})$\,, we can also introduce the dual R--R fields,
\begin{center}
\begin{tabular}{ll}
 \underline{$\bullet$ $\beta$-untwisted fields:} & polyforms $(\check{C},\,\check{F})$ and bispinors $(\check{\bm{C}},\,\check{\bm{F}})$\,,
\\[2mm]
 \underline{$\bullet$ $(\beta,\,\tilde{\phi})$-untwisted fields:} \qquad\quad& polyforms $(\check{\cC},\,\check{\cF})$ and bispinors $(\check{\bm{\cC}},\,\check{\bm{\cF}})$\,.
\end{tabular}
\end{center}
By introducing an operator $\beta\vee F\equiv \frac{1}{2}\,\beta^{mn}\,\iota_m\,\iota_n F$, we define these polyforms as
\begin{align}
 \check{C} \equiv \Exp{\beta\vee} A \,,\qquad 
 \check{F} \equiv \Exp{\beta\vee} F \,. \qquad
 \check{\cC} \equiv \Exp{\tilde{\phi}} \Exp{\beta\vee} A \,,\qquad 
 \check{\cF} \equiv \Exp{\tilde{\phi}} \Exp{\beta\vee} F \,,
\end{align}
and their flat components as
\begin{align}
\begin{split}
 \check{C}_{\Loa_1\cdots \Loa_p} &\equiv \tilde{e}_{\Loa_1}{}^{m_1}\cdots \tilde{e}_{\Loa_p}{}^{m_p}\,\check{C}_{m_1\cdots m_p}\,,\qquad 
 \check{F}_{\Loa_1\cdots \Loa_p} \equiv \tilde{e}_{\Loa_1}{}^{m_1}\cdots \tilde{e}_{\Loa_p}{}^{m_p}\,\check{F}_{m_1\cdots m_p}\,,
\\
 \check{\cC}_{\Loa_1\cdots \Loa_p} &\equiv \tilde{e}_{\Loa_1}{}^{m_1}\cdots \tilde{e}_{\Loa_p}{}^{m_p}\,\check{\cC}_{m_1\cdots m_p}\,,\qquad 
 \check{\cF}_{\Loa_1\cdots \Loa_p} \equiv \tilde{e}_{\Loa_1}{}^{m_1}\cdots \tilde{e}_{\Loa_p}{}^{m_p}\,\check{\cF}_{m_1\cdots m_p}\,,
\end{split}
\label{eq:dual-RR-flat-components}
\end{align}
by using the dual vielbein $\tilde{e}_m{}^{\Loa}$\,. 
Their corresponding bispinors are defined as
\begin{align}
\begin{split}
 \check{\bm{C}} &\equiv \sum_n \frac{1}{n!}\,\check{C}_{\Loa_1\cdots \Loa_n}\,\Gamma^{\Loa_1\cdots \Loa_n} \,,\qquad
 \check{\bm{F}} \equiv \sum_n \frac{1}{n!}\,\check{F}_{\Loa_1\cdots \Loa_n}\,\Gamma^{\Loa_1\cdots \Loa_n} \,,
\\
 \check{\bm{\cC}} &\equiv \sum_n \frac{1}{n!}\,\check{\cC}_{\Loa_1\cdots \Loa_n}\,\Gamma^{\Loa_1\cdots \Loa_n} \,,\qquad
 \check{\bm{\cF}} \equiv \sum_n \frac{1}{n!}\,\check{\cF}_{\Loa_1\cdots \Loa_n}\,\Gamma^{\Loa_1\cdots \Loa_n} \,.
\end{split}
\end{align}

\subsubsection{Single $T$-duality}

As a simple application of the formula \eqref{eq:R-R-beta-transf}, let us explain how the R--R fields transform under a single $T$-duality along the $x^z$-direction,
\begin{align}
 (h^M{}_N) = \begin{pmatrix} \bm{{1_{10}}} - \bm{e}_z & \bm{e}_z \\ \bm{e}_z & \bm{{1_{10}}} - \bm{e}_z \end{pmatrix}\,, \qquad 
 \bm{e}_z \equiv \diag (0,\dotsc,0,\, \overset{z\text{-th}}{1},\,0,\dotsc,0)\,. 
\end{align}
In this case, the vielbein and the dilaton transform as
\begin{align}
 e' = \bigl[\bm{{1_{10}}} - (\bm{{1_{10}}}-E^\rmT)\, \bm{e}_z \bigr]^{-1} \, e
 = \bigl[\bm{{1_{10}}} + \CG_{zz}^{-1}\, (\bm{{1_{10}}}-E^\rmT)\, \bm{e}_z \bigr] \, e \,, \qquad 
 \Exp{\Phi'} = \frac{1}{\sqrt{\CG_{zz}}} \Exp{\Phi} \,,
\label{eq:abelian-NS-NS}
\end{align}
and the Lorentz transformation matrix is
\begin{align}
 \Lambda \equiv (\Lambda^{\Loa}{}_{\Lob}) = e^{\rmT}\,\bigl[\bm{{1_{10}}} - \bm{e}_z \,(\bm{{1_{10}}} - E)\bigr]^{-1}\,\bigl[\bm{{1_{10}}} - \bm{e}_z \,(\bm{{1_{10}}}+ E^\rmT)\bigr]\,e^{-\rmT} \,.
\end{align}
This can be simplified as
\begin{align}
 \Lambda^{\Loa}{}_{\Lob} = \delta^{\Loa}_{\Lob} - 2\,\frac{e_z{}^{\Loa} \,e_{z\Lob}}{\CG_{zz}} \,,
\end{align}
and we can easily see that the R--R field transforms under the $T$-duality as \cite{Hassan:1999mm}
\begin{align}
 \bisC' = \bisC\, \Omega_z^{-1} \,,\qquad 
 \Omega_z \equiv \frac{e_{z \Loa}}{\sqrt{\CG_{zz}}}\,\brGamma^{\Loa}\,\brGamma^{11} = \frac{1}{\sqrt{\CG_{zz}}}\,\Gamma_z = \Omega_z^{-1} \quad (\Gamma_m\equiv e_m{}^{\Loa}\,\Gamma_{\Loa})\,,
\end{align}
where we have supposed $\CG_{zz}\geq 0$. 

\medskip

From the identity \eqref{eq:Gamma-n-1}, we obtain
\begin{align}
 \bisC' = \bisC\, \Omega_z^{-1}
 = \frac{1}{\sqrt{\CG_{zz}}}\sum_n \frac{1}{n!}\biggl( n\,\hat{\cC}_{[\Loa_1\cdots\Loa_{n-1}}\,e_{\Loa_n]z} + \hat{\cC}_{\Loa_1\cdots\Loa_n\Lob}\,e_z{}^{\Lob} \biggr)\,\Gamma^{\Loa_1\cdots\Loa_n}\,.
\label{eq:C-transformation-single}
\end{align}
By using the $B$-untwisted R--R potentials, $\hat{\bm{C}} = \Exp{-\Phi}\bisC$ and $\hat{C}=\Exp{-\Phi}\hat{\cC}$, \eqref{eq:C-transformation-single} is expressed as
\begin{align}
 \hat{\bm{C}}' = \sum_n \frac{1}{n!}\biggl( n\,\hat{C}_{[\Loa_1\cdots\Loa_{n-1}}\,e_{\Loa_n]z} + \hat{C}_{\Loa_1\cdots\Loa_n\Lob}\,e_z{}^{\Lob} \biggr)\,\Gamma^{\Loa_1\cdots\Loa_n}\,,
\end{align}
where we have used \eqref{eq:abelian-NS-NS}. 
For the curved components, using the transformation rule of the vielbein \eqref{eq:abelian-NS-NS}, we obtain
\begin{align}
 \hat{C}'_{m_1\cdots m_n} 
 &= e'_{m_1}{}^{\Loa_1}\cdots e'_{m_n}{}^{\Loa_n}\, \bigl(n\,\hat{C}_{[\Loa_1\cdots\Loa_{n-1}}\,e_{\Loa_n]z} + \hat{C}_{\Loa_1\cdots\Loa_n\Lob}\,e_z{}^{\Lob} \bigr)
\nn\\
 &=n\,\hat{C}_{[m_1\cdots m_{n-1}}\,\CG_{m_n]z} + \hat{C}_{m_1\cdots m_{n}z}
\nn\\
 &\quad +n\,\CG_{zz}^{-1}\,
 \bigl[\hat{C}_{[m_1\cdots m_{n-1}}\,\CG_{zz}-(n-1)\,\hat{C}_{[m_1\cdots m_{n-2}|z|}\,\CG_{m_{n-1}|z|}\bigr]\,\bigl(\delta^z_{m_n]}-E^{\rmT}_{m_n]z}\bigr)
\nn\\
 &= \hat{C}_{m_1\cdots m_{n}z} +n\, \biggl[\hat{C}_{[m_1\cdots m_{n-1}} -(n-1)\,\frac{\hat{C}_{[m_1\cdots m_{n-2}|z|}\,\CG_{m_{n-1}|z|}}{\CG_{zz}}\biggr]\,\bigl(\delta^z_{m_n]}+B_{m_n]z}\bigr)\,.
\end{align}
This reproduces the famous transformation rule,
\begin{align}
\begin{split}
 &\hat{C}'_{i_1\cdots i_n} = \hat{C}_{i_1\cdots i_{n}z} +n\,\hat{C}_{[i_1\cdots i_{n-1}}\,B_{i_n]z} + n\,(n-1)\,\frac{\hat{C}_{[i_1\cdots i_{n-2}|z|}\,B_{i_{n-1}|z|}\,\CG_{i_n]z}}{\CG_{zz}} \,,
\\
 &\hat{C}'_{i_1\cdots i_{n-1}z} 
 = \hat{C}_{i_1\cdots i_{n-1}} - (n-1)\,\frac{\hat{C}_{[i_1\cdots i_{n-2}}\,\CG_{i_{n-1}]z}}{\CG_{zz}} \,,
\end{split}
\end{align}
where we have decomposed the coordinates as $\{x^m\}=\{x^i,\,x^z\}$. 

\medskip

It is also noted that, under the single $T$-duality after taking the diagonal gauge, an arbitrary $\OO(1,9)$ spinor $\Psi_1^{\SPa}$ and an $\OO(9,1)$ spinor $\Psi_2^{\SPbra}$ transform as
\begin{align}
 \Psi_1 \ \to \ \Psi'_1=\Psi_1 \,,\qquad \Psi_2 \ \to \ \Psi'_2=\Omega \,\Psi_2 = \frac{e_{z \Loa}}{\sqrt{\CG_{zz}}}\, \brGamma^{\Loa}\,\brGamma^{11}\,\Psi_2 \,. 
\label{eq:Psi-Tdual}
\end{align}
When we consider a single $T$-duality connecting type IIA and type IIB superstring, these transformations are applied to the spacetime fermions $\Theta_1$ and $\Theta_2$ introduced later. 

\subsection{$\beta$-transformation of R--R fields}

In this subsection, we consider local $\beta$-transformations
\begin{align}
 h_M{}^N = \begin{pmatrix} \bm{{1_{10}}} & \bm{{0_{10}}} \\ \bmr^{mn}(x) & \bm{{1_{10}}} \end{pmatrix} \,,\qquad 
 h^M{}_N = \begin{pmatrix} \bm{{1_{10}}} & \bmr^{mn}(x) \\ \bm{{0_{10}}} & \bm{{1_{10}}} \end{pmatrix} \qquad \bigl(\bmr^{mn}=-\bmr^{nm}\bigr) \,. 
\end{align}
From the general transformation rule \eqref{eq:R-R-beta-transf}, the R--R fields should transform as $\bisC \to \bisC'=\bisC\, \Omega^{-1}$ and $\bisF \to \bisF' = \bisF\, \Omega^{-1}$\,. 
We here find an explicit form of $\Omega$ associated with $\beta$-transformations [the final result is obtained in \eqref{eq:Omega-general}]. 

\subsubsection{Gauge fixing for dual fields}

Let us first specify the dual vielbein $\tilde{e}_m{}^{\Loa}$ explicitly. 
As we can see from \eqref{eq:double-vielbein-transf}, under $\beta$-transformations, we have the following transformation rules:\footnote{The transformation rule of $e_m{}^{\Loa}$ given \eqref{eq:beta-rule-NS} make sense only when $(E^{-\rmT})^{mn}$ is not singular. When $(E^{-\rmT})^{mn}$ is singular, we should express it as $e_m{}^{\Loa} \to e'_m{}^{\Loa}=\bigl[\bigl(\bm{1} - E^\rmT\, \bmr\bigr)^{-1}\bigr]_m{}^n\, e_n{}^{\Loa}$\,. When both $E_{mn}$ and $E^{mn}$ are singular, we should choose another parameterization of the double vielbein, although we do not consider such cases in this thesis.}
\begin{align}
\begin{split}
 e_m{}^{\Loa} &\ \to\ e'_m{}^{\Loa}=\bigl[\bigl(E^{-\rmT} - \bmr\bigr)^{-1}\,E^{-\rmT}\bigr]_m{}^n\, e_n{}^{\Loa}\,, 
\\
 \tilde{e}_m{}^{\Loa} &\ \to\ \tilde{e}'_m{}^{\Loa}=\tilde{e}_m{}^{\Loa} \,, \qquad
 E^{mn}\ \to\ E'^{mn} = E^{mn} + \bmr^{mn} \,. 
\end{split}
\label{eq:beta-rule-NS}
\end{align}
Then, we can consistently relate $e_m{}^{\Loa}$ and $\tilde{e}_m{}^{\Loa}$ as
\begin{align}
 \tilde{e}_{m}{}^{\Loa} = E_{mn}\,e^n{}_{\Lob}\,\eta^{\Lob\Loa} \,. 
\label{eq:tilde-e-gauge}
\end{align}
This is equivalent to a direct identification of two parameterizations,
\begin{align}
 \frac{1}{\sqrt{2}} \begin{pmatrix} e^m{}_{\Lob}\,\eta^{\Lob\Loa} \\ (\CG+B)_{mn}\,e^n{}_{\Lob}\,\eta^{\Lob\Loa} \end{pmatrix} = \GV^{M\Loa}= \frac{1}{\sqrt{2}} \begin{pmatrix} (\OG^{-1}-\beta)^{mn}\, \tilde{e}_n{}^{\Loa} \\ \tilde{e}_m{}^{\Loa} \end{pmatrix} \,, 
\end{align}
and consistent with the relation \eqref{eq:relation-open-closed}. 
If we introduce the flat components of $E^{mn}$ as
\begin{align}
 \Einv^{\Loa\Lob} \equiv \tilde{e}_m{}^{\Loa}\, \tilde{e}_n{}^{\Lob}\,E^{mn} \equiv \eta^{\Loa\Lob} - \beta^{\Loa\Lob} \,,
\label{eq:E-cEinv-cE}
\end{align}
we obtain
\begin{align}
 E_{mn} = \tilde{e}_m{}^{\Loa}\,\tilde{e}_n{}^{\Lob}\,(\Einv^{-1})_{\Loa\Lob} = e_m{}^{\Loa}\,e_n{}^{\Lob}\, (\Einv^{\rmT})_{\Loa\Lob}\,. 
\end{align}
Namely, we have simple expressions,
\begin{align}
\begin{alignedat}{2}
 \CG_{mn} &= e_{m}{}^{\Loa}\,e_{n}{}^{\Lob}\,\eta_{\Loa\Lob} \,,&\qquad 
 B_{mn} &= e_{m}{}^{\Loa}\,e_{n}{}^{\Lob}\,\beta_{\Loa\Lob}\,,
\\
 \OG_{mn} &= \tilde{e}_{m}{}^{\Loa}\,\tilde{e}_{n}{}^{\Lob}\,\eta_{\Loa\Lob} \,,&\qquad 
 \beta^{mn} &= \tilde{e}^m{}_{\Loa}\,\tilde{e}^n{}_{\Lob}\,\beta^{\Loa\Lob}\,. 
\end{alignedat}
\label{eq:g-G-B-beta}
\end{align}
In terms of $\Einv^{\Loa\Lob}$, the relation \eqref{eq:tilde-e-gauge} can also be expressed as
\begin{align}
 e_m{}^{\Loa} = \tilde{e}_{m}{}^{\Lob}\,(\Einv^{-\rmT})_{\Lob}{}^{\Loa}\,. 
\label{e-etilde-Einv}
\end{align}
From \eqref{e-etilde-Einv}, the relation \eqref{eq:DFT-dilaton} between the two dilatons, $\Phi$ and $\tilde{\phi}$, can be expressed as
\begin{align}
 \Exp{\Phi} = (\det\Einv_{\Loa}{}^{\Lob})^{-\frac{1}{2}} \Exp{\tilde{\phi}}\,. 
\label{eq:two-dilatons}
\end{align}

\subsubsection{Relation between untwisted R--R fields}

From \eqref{eq:R-R-diagram},
the relation between $(B,\,\Phi)$-untwisted R--R polyforms $(\hat{\cC},\,\hat{\cF})$ and the $(\beta,\,\tilde{\phi})$-untwisted R--R polyforms $(\check{\cC},\,\check{\cF})$ can be expressed as
\begin{align}
\begin{alignedat}{2}
 \check{\cF} &= \Exp{\tilde{\phi}-\Phi}\Exp{\beta\vee} \Exp{B_2\wedge} \hat{\cF} \,, \qquad& 
 \hat{\cF} &= \Exp{\Phi-\tilde{\phi}}\Exp{-B_2\wedge} \Exp{-\beta\vee} \check{\cF} \,, \\
 \check{\cC} &= \Exp{\tilde{\phi}-\Phi}\Exp{\beta\vee} \Exp{B_2\wedge} \hat{\cC} \,, \qquad& 
 \hat{\cC} &= \Exp{\Phi-\tilde{\phi}}\Exp{-B_2\wedge} \Exp{-\beta\vee} \check{\cC} \,. 
\end{alignedat}
\label{eq:R-R-relation1}
\end{align}
As we show in Appendix \ref{app:equivRR} by a brute force calculation, if rephrased in terms of bispinors, these relations have quite simple forms
\begin{align}
\begin{alignedat}{2}
 &\bisF = \check{\bm{\cF}}\,\Omega_0^{-1}\,,\qquad 
 \bisC = \check{\bm{\cC}}\,\Omega_0^{-1}\,,\qquad &
 &\check{\bm{\cF}} = \bisF \,\Omega_0\,,\qquad 
 \check{\bm{\cC}} = \bisC \,\Omega_0\,, 
\label{eq:R-R-relation2}
\\
 &\Omega_0^{-1} =(\det \Einv_{\Loc}{}^{\Lod})^{-\frac{1}{2}} \text{\AE}\bigl(-\tfrac{1}{2}\,\beta^{\Loa\Lob}\,\Gamma_{\Loa\Lob}\bigr)\,,\qquad&
 &\Omega_0 =(\det \Einv_{\Loc}{}^{\Lod})^{-\frac{1}{2}} \text{\AE}\bigl(\tfrac{1}{2}\,\beta^{\Loa\Lob}\,\Gamma_{\Loa\Lob}\bigr)\,, 
\end{alignedat}
\end{align}
where $\text{\AE}$ is an exponential-like function with the gamma matrices totally antisymmetrized \cite{Hassan:1999mm}
\begin{align}
 \text{\AE}\bigl(\tfrac{1}{2}\,\beta^{\Loa\Lob}\,\Gamma_{\Loa\Lob}\bigr) \equiv \sum^5_{p=0}\frac{1}{2^{p}\,p!}\, \beta_{\Loa_1\Loa_2}\cdots\beta_{\Loa_{2p-1}\Loa_{2p}}\,\Gamma^{\Loa_1\cdots \Loa_{2p}}\,. 
\end{align}
In fact, this $\Omega_0$ is a spinor representation of a local Lorentz transformation,\footnote{Note that $\brGamma^{\Loa}=\Gamma^{11}\,\Gamma^{\Loa}$ also satisfies the same relation, $\Omega_0^{-1} \,\brGamma^{\Loa}\,\Omega_0 = \bigl(\Einv^{-1}\, \Einv^\rmT\bigr)^{\Loa}{}_{\Lob}\, \brGamma^{\Lob}$\,.}
\begin{align}
 \Omega_0^{-1} \,\Gamma^{\Loa}\,\Omega_0 = \bigl(\Einv^{-1}\, \Einv^\rmT\bigr)^{\Loa}{}_{\Lob}\, \Gamma^{\Lob} \,, 
\end{align}
as we can show by employing the formula provided below \cite{Hassan:1999mm} (see Appendix \ref{app:omega} for a proof). 
In this sense, the $(B,\,\Phi)$-untwisted fields and the $(\beta,\,\tilde{\phi})$-untwisted fields are related by a local Lorentz transformation. 

\medskip

\noindent\textbf{\underline{Formula:}} For an arbitrary antisymmetric matrix $a_{\Loa\Lob}$, the spinor representation of a local Lorentz transformation
\begin{align}
 \Lambda^{\Loa}{}_{\Lob} \equiv \bigl[(\eta+a)^{-1}\,(\eta-a)\bigr]^{\Loa}{}_{\Lob} = \bigl[(\eta-a)\,(\eta+a)^{-1}\bigr]^{\Loa}{}_{\Lob}\ \in \OO(1,D-1)\,,
\end{align}
is given by
\begin{align}
\begin{split}
 \Omega_{(a)} &=\bigl[\det (\delta_{\Loc}^{\Lod}\pm a_{\Loc}{}^{\Lod})\bigr]^{-\frac{1}{2}} \text{\AE}\bigl(-\tfrac{1}{2}\,a_{\Loa\Lob}\,\Gamma^{\Loa\Lob}\bigr)\,,
\\
 \Omega_{(a)}^{-1} &=\bigl[\det (\delta_{\Loc}^{\Lod}\pm a_{\Loc}{}^{\Lod})\bigr]^{-\frac{1}{2}} \text{\AE}\bigl(\tfrac{1}{2}\,a_{\Loa\Lob}\,\Gamma^{\Loa\Lob}\bigr) \,, \qquad 
 \Omega_{(a)}^{-1} \,\Gamma^{\Loa}\,\Omega_{(a)} = \Lambda^{\Loa}{}_{\Lob}\, \Gamma^{\Lob} \,. 
\end{split}
\label{eq:Hassan-formula}
\end{align}

\subsubsection{General formula for $\Omega$}

Now, let us find the explicit form of $\Omega$ for $\beta$-transformations [recall \eqref{eq:Omega-def}], satisfying
\begin{align}
 \Omega^{-1}\,\brGamma^{\Loa}\,\Omega = \Lambda^{\Loa}{}_{\Lob}\, \brGamma^{\Lob} \qquad 
 \bigl[\Lambda=e^{\rmT}\,(1+ \bmr\,E)^{-1}\,(1- \bmr\,E^\rmT)\,e^{-\rmT}\bigr]\,.
\end{align}
A key observation is that by using \eqref{eq:tilde-e-gauge}, \eqref{eq:E-cEinv-cE}, and \eqref{e-etilde-Einv}, $\Lambda^{\Loa}{}_{\Lob}$ can be decomposed into a product of two Lorentz transformations,
\begin{align}
\Lambda 
 = \Lambda'\, \Lambda^{-1}_{0} \,,\qquad
\Lambda' \equiv \eta^{-1}\,\Einv'^{-1}\, \Einv'^\rmT\,\eta \,,\qquad 
  \Lambda_{0}\equiv \eta^{-1}\,\Einv^{-1}\, \Einv^\rmT\,\eta\,,
\label{eq:Lambda-decomp}
\end{align}
where $\Einv'$ is defined by
\begin{align}
 \Einv'^{\Loa\Lob} \equiv \bigl[\tilde{e}^\rmT\,(E^{-1}+\bmr)\,\tilde{e}\bigr]^{\Loa\Lob} \equiv \eta^{\Loa\Lob} - \beta'^{\Loa\Lob} \qquad 
 \bigl(\beta'^{\Loa\Lob}\equiv \beta^{\Loa\Lob}- \bmr^{mn}\,\tilde{e}_m{}^{\Loa}\,\tilde{e}_n{}^{\Lob} \bigr)\,.
\end{align}
Then, we can check the following relations associated with $\Einv'^{\Loa\Lob}$:
\begin{align}
\begin{alignedat}{2}
 \Einv'^{\Loa\Lob}&= \tilde{e}_m{}^{\Loa}\,\tilde{e}_n{}^{\Lob}\, E'^{mn}
  = e'^{\Loa m}\,e'^{\Lob n}\, E'^\rmT_{mn} \,,&\qquad 
 E'^{mn}&=E^{mn}+\bmr^{mn} \equiv \OG'^{mn} - \beta'^{mn}\,,
\\
 e'_m{}^{\Loa} &\equiv E'^\rmT_{mn}\,\tilde{e}^{n\Loa} = (\Einv'^{-\rmT})_{\Lob}{}^{\Loa}\,\tilde{e}_{m}{}^{\Lob}\,,&\qquad 
 E'_{mn} &= \bigl[(E^{-1}+\bmr)^{-1}\bigr]_{mn} \equiv \CG'_{mn} + B'_{mn} \,,
\end{alignedat}
\end{align}
where $\CG'_{mn}$ and $B'_{mn}$ are the $\beta$-transformed metric and $B$-field, respectively. 
From the invariance of $d$, $\tilde{e}_m{}^{\Loa}$, and $\tilde{\phi}$ under $\beta$-transformations, the dilaton $\Phi$ in the $\beta$-transformed background becomes
\begin{align}
 \Exp{\Phi'} = (\det\Einv'_{\Loa}{}^{\Lob})^{-\frac{1}{2}} \Exp{\tilde{\phi}} 
 = \frac{(\det\Einv'_{\Loa}{}^{\Lob})^{-\frac{1}{2}}}{(\det\Einv_{\Loc}{}^{\Lod})^{-\frac{1}{2}}} \Exp{\Phi} \,. 
\label{eq:Phi-beta}
\end{align}

\medskip

Corresponding to the decomposition \eqref{eq:Lambda-decomp}, we can also decompose $\Omega$ as
\begin{align}
\begin{split}
 \Omega &= \Omega' \,\Omega_{0}^{-1} = \bigl[\det (\Einv'\,\Einv)_{\Loe}{}^{\Lof}\bigr]^{-\frac{1}{2}} \, \text{\AE}\bigl(\tfrac{1}{2}\,\beta'^{\Loa\Lob}\,\Gamma_{\Loa\Lob}\bigr)\,\text{\AE}\bigl(-\tfrac{1}{2}\,\beta^{\Loc\Lod}\,\Gamma_{\Loc\Lod}\bigr)\,, 
\\
 \Omega^{-1} &= \Omega_0 \,\Omega'^{-1} = \bigl[\det (\Einv'\,\Einv)_{\Loe}{}^{\Lof}\bigr]^{-\frac{1}{2}} \,\text{\AE}\bigl(\tfrac{1}{2}\,\beta^{\Loa\Lob}\,\Gamma_{\Loa\Lob}\bigr)\, \text{\AE}\bigl(-\tfrac{1}{2}\,\beta'^{\Loc\Lod}\,\Gamma_{\Loc\Lod}\bigr)\,. 
\end{split}
\label{eq:Omega-general}
\end{align}
where we have defined
\begin{align}
 \Omega' \equiv (\det \Einv'_{\Loc}{}^{\Lod})^{-\frac{1}{2}} \text{\AE}\bigl(\tfrac{1}{2}\,\beta'^{\Loa\Lob}\,\Gamma_{\Loa\Lob}\bigr)\,,\qquad 
 \Omega'^{-1} = (\det \Einv'_{\Loc}{}^{\Lod})^{-\frac{1}{2}} \text{\AE}\bigl(-\tfrac{1}{2}\,\beta'^{\Loa\Lob}\,\Gamma_{\Loa\Lob}\bigr)\,.
\end{align}
This gives the desired local Lorentz transformation,
\begin{align}
 \Omega^{-1} \,\brGamma^{\Loa}\,\Omega = \Omega_{0}\,\Omega'^{-1} \,\brGamma^{\Loa}\,\Omega'\,\Omega_{0}^{-1} 
 = (\Lambda'\,\Lambda^{-1}_{0})^{\Loa}{}_{\Lob}\, \brGamma^{\Lob} = \Lambda^{\Loa}{}_{\Lob}\, \brGamma^{\Lob} \,. 
\end{align}
The $\beta$-transformed R--R field is then expressed as
\begin{align}
 \bisF' = \bisF\, \Omega^{-1} \,. 
\label{eq:beta-F-transf}
\end{align}
In terms of the differential form, we can express the same transformation rule as
\begin{align}
 \hat{\cF}' = \Exp{\Phi'-\Phi}\Exp{-B'_2\wedge} \Exp{\bmr\vee} \Exp{B_2\wedge} \hat{\cF} \qquad
 \bigl(\hat{\cF}'_{m_1\cdots m_n} \equiv e'_{m_1}{}^{\Loa_1}\cdots e'_{m_n}{}^{\Loa_n}\,\hat{\cF}'_{\Loa_1\cdots \Loa_n}\bigr)\,. 
\end{align}
In terms of the $B$-untwisted field $\hat{F}$, the $\beta$-untwisted field $\check{F}$, and the $(\beta,\,\tilde{\phi})$-untwisted field $\check{\cF}$, we can express the above formula as
\begin{align}
 \hat{F}' = \Exp{-B'_2\wedge} \Exp{\bmr\vee} \Exp{B_2\wedge} \hat{F} \,,\qquad 
 \check{F}' = \check{F}\,,\qquad \check{\cF}' = \check{\cF} \,. 
\end{align}
Namely, the $\beta$- or $(\beta,\,\tilde{\phi})$-untwisted field is invariant under $\beta$-transformations, which has been shown in \cite{Sakamoto:2017cpu} (see also \cite{Sakamoto:2017wor}) by treating the R--R fields, $A$ and $F$, as $\OO(D,D)$ spinors. 

\medskip

Specifically, if the $B$-field and the dilaton $\Phi$ are absent before the $\beta$-transformation, we have $\beta^{\Loa\Lob}=0$\,, $\Einv_{\Loa}{}^{\Lob}=\delta_{\Loa}^{\Lob}$\,, and $\beta'^{\Loa\Lob}= - \bmr^{mn}\,\tilde{e}_m{}^{\Loa}\,\tilde{e}_n{}^{\Lob}$. 
Then, \eqref{eq:Omega-general} becomes
\begin{align}
 \Omega = (\det\Einv'_{\Loc}{}^{\Lod})^{-\frac{1}{2}}\,\text{\AE}\bigl(-\tfrac{1}{2}\,\bmr^{\Loa\Lob}\,\Gamma_{\Loa\Lob}\bigr) \qquad 
 (\bmr^{\Loa\Lob}\equiv \bmr^{mn}\,\tilde{e}_m{}^{\Loa}\,\tilde{e}_n{}^{\Lob}) \,. 
\end{align}
In section \ref{sec:YB-beta-deform}, we see that this $\Omega$ plays an important role in YB deformations of $\AdS{5}\times\rmS^5$ superstring [see Eq.~\eqref{eq:omega} where $2\,\eta\,\lambda^{\Loa\Lob}$ plays the same role as $\bmr^{\Loa\Lob}$ here].

\subsection{$T$-duality-invariant Green--Schwarz action}

In section \ref{sec:YB-beta-deform}, we show that YB deformations are equivalent to $\beta$-deformations of the target space. 
In order to show the equivalence, it will be useful to manifest the covariance of the GS superstring theory under $\beta$-transformations. 
In this section, we provide a manifestly $\OO(10,10)$ $T$-duality-covariant formulation of the GS type II superstring theory. 

\medskip

A manifestly $T$-duality covariant formulations of string theory, the so-called double sigma model (DSM), has been developed in \cite{Duff:1989tf,Tseytlin:1990nb,Tseytlin:1990va,Hull:2004in,Hull:2006va,Copland:2011wx,Lee:2013hma} for the bosonic string. 
More recently, the DSM for the GS type II superstring theory was formulated in \cite{Park:2016sbw} (see also \cite{Hull:2006va,Blair:2013noa,Bandos:2015cha,Driezen:2016tnz,Bandos:2016jez} for other approaches to supersymmetric DSMs). 
The action by Park, in our convention, is given by
\begin{align}
 S &= \frac{1}{4\pi\alpha'}\int \Bigl[\,\frac{1}{2}\, \cH_{MN}\,\Pi^M\wedge *_{\gga} \Pi^N - D X^M\wedge \bigl(\cA_{M} + \Sigma_{M} \bigr)\Bigr]
\nn\\
 &= -\frac{1}{4\pi\alpha'}\int \sqrt{-\gga}\,\rmd^2\sigma\,
 \Bigl[\,\frac{1}{2}\,\gga^{\WSa\WSb}\,\cH_{MN}\,\Pi_{\WSa}^M\,\Pi_{\WSb}^N
  + \varepsilon^{\WSa\WSb}\, D_{\WSa}X^M\,\bigl(\cA_{\WSb M} + \Sigma_{\WSb M} \bigr)\Bigr] \,,
\label{eq:GS-DSM-Park}
\end{align}
where $\gga_{\WSa\WSb}$ is the intrinsic metric on the string worldsheet and
\begin{align}
\begin{split}
 &\Pi^M \equiv DX^M + \Sigma^M \,,\qquad DX^M \equiv \rmd X^M-\cA^M \,,\qquad 
 \varepsilon^{01}\equiv \frac{1}{\sqrt{-\gga}}\,,
\\
 &(X^M)\equiv \begin{pmatrix} X^m \\ \tilde{X}_m \end{pmatrix} \,,\qquad 
 \Sigma^M \equiv \begin{pmatrix} \Sigma^m \\ \tilde{\Sigma}_m \end{pmatrix} \equiv \frac{\ii}{\sqrt{2}}\,\bigl(\brTheta_1\,\Gamma^M\,\rmd \Theta_1 + \brbrTheta_2\,\brGamma^M\,\rmd \Theta_2\bigr) \,,
\end{split}
\end{align}
and a worldsheet 1-form $\cA^M(\sigma)$ is defined to satisfy,
\begin{align}
 \cA^M \, \partial_M T = 0\,,\qquad \cA^M\,\cA_M = 0\,,
\label{eq:cA-condition}
\end{align}
for arbitrary supergravity fields or gauge parameters $T(x)$\,. 
Here, the Dirac conjugates for the spacetime fermions $\Theta_1^{\SPa}$ and $\Theta_2^{\SPbra}$ are defined respectively as
\begin{align}
 \brTheta_1 \equiv \Theta_1^\dagger\,\Gamma^0\,,\qquad 
 \brbrTheta_2 \equiv -\Theta_2^\dagger\,\brGamma^0\,,
\end{align}
which indeed transform as
\begin{align}
 \brTheta_1 \ \to \ \brTheta_1 \, \Exp{-\frac{1}{4}\,\omega_{\Loa\Lob}\,\Gamma^{\Loa\Lob}}\,,\qquad
 \brbrTheta_2 \ \to \ \brbrTheta_2 \, \Exp{-\frac{1}{4}\,\bromega_{\Lobra\Lobrb}\,\brGamma^{\Lobra\Lobrb}} \,,
\end{align}
under a double Lorentz transformation $\Theta_1\to \Exp{\frac{1}{4}\,\omega_{\Loa\Lob}\,\Gamma^{\Loa\Lob}} \Theta_1$ and $\Theta_2\to \Exp{-\frac{1}{4}\,\bromega_{\Lobra\Lobrb}\,\brGamma^{\Lobra\Lobrb}} \Theta_2$\,. 
The Majorana--Weyl conditions are defined as\footnote{The non-standard factor $-\brGamma^{11}$ is introduced in the Majorana condition for $\Theta_2$ such that the condition becomes the standard Majorana condition after the diagonal gauge fixing; $\Theta_2 = C\, (\Gamma^0)^\rmT\,\Theta_2^*$\,.}
\begin{align}
\begin{alignedat}{2}
 &\Theta_1 = C\,(\Gamma^0)^\rmT\,\Theta_1^*\,,\qquad& 
 &\Theta_2 = -\brGamma^{11}\,\brC\,(\brGamma^0)^\rmT\,\Theta_2^* \,, 
\\
 &\Gamma^{11}\,\Theta_1 = \Theta_1\,,\qquad& 
 &\brGamma^{11}\,\Theta_2 = \pm \Theta_2\qquad (\text{IIA/IIB})\,,
\end{alignedat}
\end{align}
and then we obtain
\begin{align}
 \brTheta_1 = \Theta_1^\dagger\,\Gamma^0 = \Theta_1^\rmT\,C \,,\qquad 
 \brbrTheta_2 = -\Theta_2^\dagger\,\brGamma^0 = -\Theta_2^\rmT\,\brC\, \brGamma^{11} \,. 
\end{align}
In \cite{Park:2016sbw}, the target space was assumed to be flat and have no the R--R fluxes,
but here we generalize the action to any curved backgrounds including the R--R fluxes.

\medskip

In order to consider the superstring action in the presence of fluxes, such as the $H$-flux and the R--R fluxes, we introduce generalized tensors,
\begin{align}
\begin{split}
 \cK_{MN}^{(1)}&\equiv -\frac{\ii}{\sqrt{2}}\,\GV_{(M}{}^{\Loa}\,\brGV_{N)}{}^{\Lobrb}\, \brTheta_1\,\Gamma_{\Loa}\,\Gamma^{\Loc\Lod}\,\Theta_1 \, \Phi_{\Lobrb\Loc\Lod} \,,
\\
 \cK_{MN}^{(2)}&\equiv -\frac{\ii}{\sqrt{2}}\,\brGV_{(M}{}^{\Lobra}\,\GV_{N)}{}^{\Lob}\, \brbrTheta_2\,\brGamma_{\Lobra}\,\brGamma^{\Lobrc\Lobrd}\,\Theta_2 \, \brPhi_{\Lob\Lobrc\Lobrd} \,,
\\
 \cK^{\text{\tiny(RR)}}_{MN} &\equiv \frac{\ii}{4}\,\GV_{(M}{}^{\Loa}\,\brGV_{N)}{}^{\Lobrb}\,\brTheta_{1}\, \Gamma_{\Loa}\, \bisF \, \brGamma_{\Lobrb} \Theta_{2} \,. 
\end{split}
\end{align}
Then, we add the following term to the DSM action \eqref{eq:GS-DSM-Park}:
\begin{align}
 \Delta S \equiv \frac{1}{8\pi\alpha'}\int \cK_{MN}\,\Pi^M\wedge *_{\gga} \Pi^N\,, \qquad 
 \cK_{MN}\equiv \cK_{MN}^{(1)}+\cK_{MN}^{(2)}+\cK_{MN}^{\text{\tiny(RR)}}\,. 
\end{align}
By choosing the diagonal gauge, the explicit form of $\cK_{MN}$ becomes
\begin{align}
 \cK_{MN} &= \begin{pmatrix}
  -(\CG\,\kappa^{\text{s}}\,\CG+B\,\kappa^{\text{s}}\,B+B\,\kappa^{\text{a}}\,\CG+\CG\,\kappa^{\text{a}}\,B)_{mn} & (B\,\kappa^{\text{s}} +\CG\,\kappa^{\text{a}})_m{}^n \\ - (\kappa^{\text{s}}\,B+\kappa^{\text{a}}\,\CG)^m{}_n & (\kappa^{\text{s}})^{mn}
 \end{pmatrix}
 \,,
\\
 \kappa_{mn} &\equiv -\frac{\ii}{4}\,\Bigl(\sqrt{2}\,\brTheta_1\,\Gamma_{m}\,\Gamma^{\Loa\Lob}\,\Theta_1\,\Phi_{n\Loa\Lob} 
 + \sqrt{2}\,\brbrTheta_2\,\brGamma_{n}\,\brGamma^{\Lobra\Lobrb}\,\Theta_2 \, \brPhi_{m\Lobra\Lobrb}
 - \frac{1}{2}\,\brTheta_{1}\, \Gamma_m\, \bisF \, \brGamma_n \Theta_{2}\Bigr)\,,
\end{align}
where we defined $\kappa^{\text{s}}_{mn}\equiv \kappa_{(mn)}$ and $\kappa^{\text{a}}_{mn}\equiv \kappa_{[mn]}$ and their indices are raised or lowered with the metric $\CG_{mn}$\,. 
Note that $\cK_{MN}$ is an $\OO(10,10)$ matrix up to quadratic order in $\Theta_I$ ($I=1,2$). 

\medskip

The modification of the DSM action, $S\to S+\Delta S$, is equivalent to the replacement of the generalized metric
\begin{align}
 \cH_{MN} \ \to\ \cM_{MN}\equiv \cH_{MN} + \cK_{MN} \,. 
\label{eq:replacement}
\end{align}
The explicit form of $\cM_{MN}$ is given by
\begin{align}
 (\cM_{MN}) &= \begin{pmatrix}
  \delta_m^p & B_{mp} \\ 0 & \delta^m_p 
 \end{pmatrix}
 \begin{pmatrix}
  \CG_{pq}-\kappa^{\text{s}}_{pq} & (\kappa^{\text{a}})_{p}{}^q \\ -(\kappa^{\text{a}})^p{}_q & \CG^{pq}+(\kappa^{\text{s}})^{pq}
 \end{pmatrix}
 \begin{pmatrix}
  \delta^q_n & 0 \\ -B_{qn} & \delta_q^n
 \end{pmatrix} 
\nn\\
 &= \begin{pmatrix}
  \delta_m^p & \hat{B}_{mp} \\ 0 & \delta^m_p 
 \end{pmatrix}
 \begin{pmatrix}
  \hat{\CG}_{pq} & 0 \\ 0 & (\hat{\CG}^{-1})^{pq} 
 \end{pmatrix}
 \begin{pmatrix}
  \delta^q_n & 0 \\ -\hat{B}_{qn} & \delta_q^n
 \end{pmatrix} + \cO(\Theta^4)\,,
\end{align}
where we defined
\begin{align}
 \hat{\CG}_{mn}\equiv \CG_{mn}-\kappa^{\text{s}}_{mn}\,,\qquad 
 \hat{B}_{mn}\equiv B_{mn}+\kappa^{\text{a}}_{mn} \,. 
\end{align}
Then, we consider an action
\begin{align}
 S = \frac{1}{4\pi\alpha'}\int \Bigl[\,\frac{1}{2}\, \cM_{MN}\,\Pi^M\wedge *_{\gga} \Pi^N - D X^M\wedge \bigl(\cA_{M} + \Sigma_{M} \bigr)\Bigr]\,.
\label{eq:GS-DSM-ours}
\end{align}

\medskip

Before we choose the diagonal gauge, spinors $\Theta_I$, R--R fields $\cF^{\SPa}{}_{\SPbrb}$, and the spin connections $\Phi_{\Lobra\Lob\Loc}$, $\brPhi_{\Loa\Lobrb\Lobrc}$ are invariant under global $\OO(10,10)$ transformations or (finite) generalized diffeomorphisms, while $\cH_{MN}$, $\cK_{MN}$, $\Pi^M$, and $\Sigma^M$ transform covariantly and the action is invariant.%
\footnote{More precisely, as discussed in \cite{Lee:2013hma,Park:2016sbw}, $\cA^M$ does not transform covariantly because $\rmd X^M$ does not transform covariantly, and $D X^M\wedge \cA_{M}$ is not invariant under generalized diffeomorphisms. However, the variation is only the total-derivative term and the action is invariant under generalized diffeomorphisms.}
The global double Lorentz symmetry is manifest but the local one is not manifest because $\Sigma^M$ contains non-covariant quantity $\rmd\Theta_I$ (or $\cK_{MN}$ contains the spin connection). 
The local symmetry becomes manifest only after eliminating the auxiliary fields. 
On the other hand, if we choose the diagonal gauge fixing, although the global $\OO(10,10)$ transformations are manifest, the covariance under generalized diffeomorphisms are lost, because the barred indices are rotated under the compensating local Lorentz transformation. 
Indeed, the transformation rule of fermionic fields after the diagonal gauge fixing is
\begin{align}
 \Theta_1 \ \to \ \Theta_1 \,,\qquad 
 \Theta_2 \ \to \ \Omega\, \Theta_2 \,, 
\label{eq:Theta-transf}
\end{align}
where $\Omega$ is the one given in \eqref{eq:Omega-def}, and in general, it is non-constant. 
Accordingly, $\rmd\Theta_2$ (and thus $\Sigma^M$ also) does not transform covariantly. 

\medskip

It is interesting to note that all information on the curved background is contained in the generalized metric $\cM_{MN}$\,. 
The usual generalized metric $\cH_{MN}$ contains only the $P$-$P$ or $\brP$-$\brP$ components [see \eqref{eq:H-eta-orthogonal}] while other quantities such as the R--R fluxes are contained in the $P$-$\brP$ or $\brP$-$P$ components $\cK_{MN}$\,. 

\medskip

In the following, we show that the action \eqref{eq:GS-DSM-ours} reproduces the conventional GS superstring action \cite{Cvetic:1999zs} up to quadratic order in fermions $\Theta_I$. 

\subsubsection{Classical equivalence to the type II GS action}

In order to reproduce the conventional action, we choose the canonical section $\tilde{\partial}^m =0$. 
Then, the condition \eqref{eq:cA-condition} for $\cA^M$ indicates that $\cA^M$ takes the form $(\cA^M)=(0,\,A_m)$ and $DX^M$ becomes
\begin{align}
 DX^M = \begin{pmatrix} \rmd X^m \\ \rmd \tilde{X}_m -A_m \end{pmatrix} \equiv \begin{pmatrix} \rmd X^m \\ P_m \end{pmatrix} \,,
\end{align}
where, for simplicity, we defined $P_m$ and treated it as a fundamental variable rather than $A_m$\,. 
The action then becomes
\begin{align}
 S &= \frac{1}{4\pi\alpha'}\int \Bigl[\,\frac{1}{2}\, \cM_{MN}\,\Pi^M\wedge *_{\gga} \Pi^N - P_{m}\wedge \bigl(\rmd X^m+\Sigma^{m}\bigr)\Bigr]
\nn\\
 &\quad - \frac{1}{4\pi\alpha'}\int \bigl(\rmd X^m\wedge \tilde{\Sigma}_m + \rmd X^m\wedge \rmd \tilde{X}_m \bigr) \,.
\end{align}
We can expand the first line as
\begin{align}
 &\frac{1}{2}\, \cM_{MN}\,\Pi^M\wedge *_{\gga} \Pi^N - P_{m}\wedge \bigl(\rmd X^m+\Sigma^{m}\bigr)
\nn\\
 &=\frac{1}{2}\, \hat{\CG}_{mn}\,\bigl(\rmd X^m+\Sigma^{m}\bigr)\wedge *_{\gga} \bigl(\rmd X^n+\Sigma^n\bigr)
\nn\\
 &\quad +\frac{1}{2}\, \hat{\CG}^{mn}\,\bigl[P_m+\tilde{\Sigma}_{m} -\hat{B}_{mp}\,(\rmd X^p+\Sigma^{p})\bigr]\wedge *_{\gga} \bigl[P_n+\tilde{\Sigma}_{n} -\hat{B}_{nq}\,(\rmd X^q+\Sigma^{q})\bigr]
\nn\\
 &\quad - P_{m}\wedge \bigl(\rmd X^m+\Sigma^{m}\bigr)
\nn\\
 &= \hat{\CG}_{mn}\,\bigl(\rmd X^m+\Sigma^{m}\bigr)\wedge *_{\gga} \bigl(\rmd X^n+\Sigma^n\bigr)
  + \bigl[\tilde{\Sigma}_n + \hat{B}_{mn}\, (\rmd X^m+\Sigma^{m})\bigr]\wedge \bigl(\rmd X^n+\Sigma^n\bigr) 
\nn\\
 &\quad +\frac{1}{2}\, \hat{\CG}^{mn}\,\bigl[P_m+\tilde{\Sigma}_{m} -\hat{B}_{mp}\,(\rmd X^p+\Sigma^{p})-\hat{\CG}_{mp}\,*_{\gga} (\rmd X^p+\Sigma^{p})\bigr]
\nn\\
 &\quad\qquad\qquad \wedge *_{\gga} \bigl[P_n+\tilde{\Sigma}_{n} -\hat{B}_{nq}\,(\rmd X^q+\Sigma^{q}) -\hat{\CG}_{nq}\,*_{\gga} (\rmd X^q+\Sigma^{q})\bigr]\,,
\end{align}
and eliminating the auxiliary fields $P_m$, we obtain
\begin{align}
 S &= \frac{1}{4\pi\alpha'}\int 
  \bigl[\,\hat{\CG}_{mn}\,\bigl(\rmd X^m+\Sigma^{m}\bigr)\wedge *_{\gga} \bigl(\rmd X^n+\Sigma^n\bigr)
  + \hat{B}_{mn}\,\bigl(\rmd X^m+\Sigma^{m}\bigr)\wedge \bigl(\rmd X^n+\Sigma^n\bigr) 
\nn\\
 &\qquad\qquad\quad - 2\, \rmd X^m\wedge \tilde{\Sigma}_{m}- \Sigma^{m} \wedge \tilde{\Sigma}_{m} - \rmd X^m\wedge \rmd \tilde{X}_m \bigr] \,.
\end{align}
By using the explicit expression for $\Sigma^M$,
\begin{align}
 \Sigma^M = \begin{pmatrix} \Sigma^m \\ 
 \tilde{\Sigma}_m 
\end{pmatrix}
 \equiv \begin{pmatrix} \Sigma^m \\ 
 \hat{\Sigma}_m + B_{mn}\, \Sigma^n
\end{pmatrix}\,,\quad
 \begin{pmatrix} \Sigma^m \\ \hat{\Sigma}_m \end{pmatrix}
 \equiv \frac{\ii}{2} \begin{pmatrix} \brTheta_1\,\Gamma^m\,\rmd \Theta_1 + \brbrTheta_2\,\brGamma^m\,\rmd \Theta_2 \\
 \brTheta_1\,\Gamma_m\,\rmd \Theta_1 - \brbrTheta_2\,\brGamma_m\,\rmd \Theta_2 \end{pmatrix} \,,
\end{align}
and neglecting quartic terms in $\Theta$ and the topological term, the action becomes
\begin{align}
\begin{split}
 S &= \frac{1}{2\pi\alpha'}\int \rmd^2\sigma \sqrt{-\gga}\,\cL \,,
\\
 \cL&=-\frac{1}{2}\,\bigl({\gga}^{\WSa\WSb}- \varepsilon^{\WSa\WSb}\bigr)\,(\hat{\CG}_{mn}+ \hat{B}_{mn})\, \partial_{\WSa}X^m \, \partial_{\WSb}X^n 
 - \CG_{mn}\, \partial_{\WSa}X^m \, \bigl(\gga^{\WSa\WSb}\,\Sigma_{\WSb}^n + \varepsilon^{\WSa\WSb}\, \hat{\Sigma}_{\WSb}^n \bigr) \,. 
\end{split}
\end{align}

\medskip

In order to compare the obtained action with the conventional GS superstring action, let us further expand the Lagrangian as
\begin{align}
\begin{split}
 \cL &=- \Pg_{-}^{\WSa\WSb}\,(\CG_{mn}+ B_{mn})\, \partial_{\WSa}X^m \, \partial_{\WSb}X^n 
\\
 &\quad - \ii\,\Pg_{+}^{\WSa\WSb}\, \partial_{\WSa}X^m \, \brTheta_1\,\Gamma_m\,\Bigl(\partial_{\WSb} \Theta_1 +\frac{1}{4}\, \partial_{\WSb}X^n\,\omega_{+n\Loa\Lob} \, \Gamma^{\Loa\Lob} \,\Theta_1\Bigr)
\\
 &\quad - \ii\,\Pg_{-}^{\WSa\WSb}\, \partial_{\WSa}X^m \, 
  \brTheta_2\,\Gamma_m\,\Bigl(\partial_{\WSb} \Theta_2 - \frac{1}{4} \, \partial_{\WSb}X^n\,\omega_{-n\Loa\Lob} \, \brGamma^{\Loa\Lob}\,\Theta_2\Bigr)
\\
 &\quad + \frac{\ii}{8} \,\Pg_{+}^{\WSa\WSb}\, \brTheta_{1}\, \Gamma_m\, \bisF \, \brGamma_n \Theta_{2} \, \partial_{\WSa}X^m\,\partial_{\WSb}X^n \,. 
\end{split}
\end{align}
where we have defined
\begin{align}
 \Pg_{\pm\WSa\WSb}\equiv \frac{\gga_{\WSa\WSb}\pm \varepsilon_{\WSa\WSb}}{2}\,,\qquad 
 \brTheta_2\equiv \Theta_2^\dagger\,\Gamma^0 \qquad \bigl(\brbrTheta_2\,\brGamma_m = \brTheta_2\,\Gamma_m\bigr)\,,
\end{align}
and used the explicit form of the spin connection \eqref{eq:spin-connections},
\begin{align}
 \Phi_{m\Loa\Lob} = \frac{1}{\sqrt{2}}\,\omega_{+m\Loa\Lob}\,,\qquad 
 \brPhi_{m\Lobra\Lobrb}= -\frac{1}{\sqrt{2}}\,\omega_{-m\Loa\Lob}\,, \qquad 
 \omega_{\pm m\Loa\Lob}\equiv \omega_{m\Loa\Lob} \pm \frac{1}{2}\, e_m{}^{\Loc}\,H_{\Loc\Loa\Lob}\,.
\label{eq:Phi-omega-pm}
\end{align}
Further using
\begin{align}
 \brGamma^{\Loa\Lob}=-\Gamma^{\Loa\Lob}\,,\qquad 
 \bisF \, \brGamma_m = \mp\,\bisF \, \Gamma_m \quad (\text{IIA/IIB})\,,
\end{align}
we obtain the type II superstring action
\begin{align}
\begin{split}
 \cL_{\text{IIA/IIB}} &= - \Pg_{-}^{\WSa\WSb}\, (\CG_{mn}+B_{mn})\, \partial_{\WSa}X^m\,\partial_{\WSb}X^n
\\ 
 &\quad - \ii\, \bigl(\Pg_{+}^{\WSa\WSb} \,\partial_{\WSa} X^m\, \brTheta_{1}\, \Gamma_m\, D_{+\WSb}\Theta_{1} 
  + \Pg_{-}^{\WSa\WSb} \,\partial_{\WSa} X^m\, \brTheta_{2}\, \Gamma_m\, D_{-\WSb}\Theta_{2} \bigr)
\\
 &\quad \mp \frac{\ii}{8}\, \Pg_{+}^{\WSa\WSb}\, \brTheta_{1}\, \Gamma_m\, \bisF \, \Gamma_n\, \Theta_{2} \, \partial_{\WSa} X^m\, \partial_{\WSb} X^n \,,
\end{split}
\label{eq:GS-action-rewriting}
\end{align}
where we defined
\begin{align}
 D_{\pm \WSa} \equiv \partial_{\WSa} + \frac{1}{4}\, \partial_{\WSa} X^m\, \omega_{\pm m}{}^{\Loa\Lob}\, \Gamma_{\Loa\Lob} \,.
\end{align}

\medskip

For type IIA superstring, defining $\Theta\equiv \Theta_1+\Theta_2$\,, we obtain a simple action
\begin{align}
\begin{split}
 \cL_{\text{IIA}} &=- \Pg_{-}^{\WSa\WSb}\,(\CG_{mn}+ B_{mn})\, \partial_{\WSa}X^m \, \partial_{\WSb}X^n 
\\
 &\quad - \frac{\ii}{2}\, \gga^{\WSa\WSb} \,\partial_{\WSa} X^m\,\brTheta\,\Gamma_m\, \bm{D}_{\WSb}\Theta 
  - \frac{\ii}{2}\, \varepsilon^{\WSa\WSb}\,\partial_{\WSa} X^m\,\brTheta\, \Gamma^{11} \, \Gamma_m\,\bm{D}_{\WSb}\Theta \,, 
\end{split}
\end{align}
where we defined
\begin{align}
 \bm{D}_{\WSa} \equiv \partial_{\WSa} + \frac{1}{4}\, \partial_{\WSa} X^m\, \omega_m{}^{\Loa\Lob}\, \Gamma_{\Loa\Lob} - \frac{1}{8}\,\partial_{\WSa} X^m\, H_m{}^{\Loa\Lob}\,\Gamma_{\Loa\Lob}\, \Gamma^{11}
 + \frac{1}{16} \, \partial_{\WSa}X^m\, \bisF \, \Gamma_m \,. 
\end{align}
On the other hand, for type IIB superstring, using the Pauli matrices $\sigma_i^{IJ}$ ($i=1,2,3$), we can rewrite the action in a familiar form
\begin{align}
\begin{split}
 \cL_{\text{IIB}} &= - \Pg_{-}^{\WSa\WSb}\, (\CG_{mn}+B_{mn})\, \partial_{\WSa}X^m\,\partial_{\WSb}X^n
\\
 &\quad - \frac{\ii}{2}\, \bigl(\gga^{\WSa\WSb}\,\delta^{IK}+\varepsilon^{\WSa\WSb}\,\sigma_3^{IK}\bigr) \,\brTheta_{I}\, \partial_{\WSa} X^m\, \Gamma_m\, D^{KJ}_{\WSb}\Theta_{J} \,, 
\end{split}
\label{eq:GS-action-conventional}
\end{align}
where we used \eqref{eq:bTGGGT} and defined
\begin{align}
\begin{split}
 D^{IJ}_{\WSa} &\equiv \delta^{IJ}\, \Bigl(\partial_{\WSa} + \frac{1}{4}\, \partial_{\WSa} X^m\, \omega_m{}^{\Loa\Lob}\, \Gamma_{\Loa\Lob} \Bigr) + \frac{1}{8}\,\sigma_3^{IJ}\, \partial_{\WSa} X^m\, H_{m\Loa\Lob}\,\Gamma^{\Loa\Lob}
\\
 &\quad - \frac{1}{8}\, \Big(\epsilon^{IJ}\, \bisF_1 + \sigma_1^{IJ}\, \bisF_3 +\frac{1}{2}\,\epsilon^{IJ}\, \bisF_5 \Big)\, \partial_{\WSa} X^n\, \Gamma_n \,,
\\
 \bisF_p&\equiv \frac{1}{p!}\,\hat{\cF}_{\Loa_1\cdots\Loa_p}\,\Gamma^{\Loa_1\cdots\Loa_p}\,,\qquad (\epsilon^{IJ}) \equiv \footnotesize{\begin{pmatrix} 0 & \,1\, \\ -1 & 0 \end{pmatrix}}\,. 
\end{split}
\end{align}

\medskip

As discussed around \eqref{eq:Psi-Tdual}, under a single $T$-duality along the $x^z$-direction, the fermionic variables transform as
\begin{align}
 \Theta_1 \ \to \ \Theta_1 \,,\qquad 
 \Theta_2 \ \to \ \frac{1}{\sqrt{\CG_{zz}}}\,\Gamma_z\, \Theta_2 \,. 
\end{align}
Since it flips the chirality of $\Theta_2$, it maps type IIA and IIB superstring to each other.

\section{YB deformations as $\beta$-deformations}
\label{sec:YB-beta-deform}

In this section, we show that the homogeneous YB deformation can be regarded as the $\beta$-transformation.

\subsection{YB deformations as $\beta$-deformations : formula}

As explained in the previous section \,\ref{sec:DFT},
the $\beta$-deformation (or $\beta$-transformation) belongs to a specific class of the $\OO(D,D)$ transformations.
The transformation is performed by shifting the $\beta$-field as
\begin{align}
 \beta_{0}^{mn}(x) \to \beta^{mn}(x) = \beta_{0}^{mn}(x) - \bmr^{mn}(x) \qquad (\bmr^{mn}=-\bmr^{nm})\,,
\end{align}
where $\beta_0$ is the $\beta$-field on the original background (see (\ref{eq:beta-rule-NS}), (\ref{eq:relation-open-closed}) for our convention).
The usual supergravity fields ($\CG_{mn}$, $B_{mn}$, $\Phi$, $\hat{F}$, $\hat{C}$) are transformed like
\begin{align}
\begin{split} 
 &\mathcal{H}'=\Exp{\bbeta^\rmT}\,\mathcal{H}\Exp{\bbeta}\,,\qquad\qquad\qquad
d'=d\,,
\\
 &\hat{F}'=\Exp{-B_2'\wedge}\Exp{-\beta\vee}\Exp{B_2\wedge}\check{F}\,,
\qquad
 \hat{C}'=\Exp{-B_2'\wedge}\Exp{-\beta\vee}\Exp{B_2\wedge}\check{C} \,,
\label{beta-formula-B}
\end{split}
\end{align}
where the matrix $\Exp{\bbeta}$ is
\begin{align}
\Exp{\bbeta}= (\Exp{\bbeta}{}_{MN})=
\begin{pmatrix}
~\delta^m_n ~& ~-\bmr^{mn}(x)~ \\
~ 0 ~& ~\delta_m^n ~
 \end{pmatrix} \,.
\end{align}
We should stress that unlike the $B$-field gauge transformations, the $\beta$-deformation is not a gauge transformation.
This fact implies that in general, the $\beta$-deformed background may not satisfy the (generalized) supergravity
equations (\ref{eq:GSEsecAdS}) even if the original background is a solution of the supergravity (or DFT).

\medskip

Now, let us explain a relation between the $\beta$-deformation and the YB deformation. For
this purpose, we concentrate on deformations of the AdS$_5\times$S$^5$ background.
Since the $B$-field vanishes on the AdS$_5\times$S$^5$ background,
the $\beta$-field in the original background also vanishes.
A homogeneous YB deformation is specified by taking a skew-symmetric classical $r$-matrix
\begin{align}
r=\frac{1}{2}r^{ij}T_i\wedge T_j\,.
\label{eq:beta-r-matrix}
\end{align}
Here we assume the generators $T_i$ are elements a bosonic subalgebra $\mathfrak{so}(2,4)\times \mathfrak{so}(6)$ of $\mathfrak{su}(2,2|4)$\,.
An important observation made in \cite{Sakamoto:2017cpu} is that a YB deformed background associated with the classical $r$-matrix (\ref{eq:beta-r-matrix}) can also be generated by
a $\beta$-deformation
\begin{align}
\begin{split}
 &\beta_{0}^{mn}(x)=0 \to \beta^{mn}(x) =-\bmr^{mn}(x) = 2\,\eta\,r^{ij}\,\hat{T}^{m}_i(x)\,\hat{T}_j^{n}(x) \,.
\label{YB-beta}
\end{split}
\end{align}
where $\hat{T}^{m}_i(x)$ are Killing vector fields associated with generators $T_i$ appearing in the $r$-matrix (\ref{eq:beta-r-matrix}).
The observation had been shown in \cite{Sakamoto:2018krs}.

\medskip

In terms of the usual supergravity fields ($\CG_{mn}$, $B_{mn}$, $\Phi$, $\hat{F}$, $\hat{C}$),
the YB-deformed background can be expressed as
\footnote{The deformed background can be reproduced from the requirement of the invariance of Non-zero Page forms and associated Page charges\cite{Araujo:2017enj}.}
\begin{align}
\begin{split}
 &\CG_{mn}'+B_{mn}'=\bigl[(G^{-1}-\beta)\bigr]^{-1}_{mn}\,,
\qquad d'=d\,,
\\
 &\hat{F}'=\Exp{-B_2'\wedge}\Exp{-\beta\vee}\check{F}\,,
\qquad
 \hat{C}'=\Exp{-B_2'\wedge}\Exp{-\beta\vee}\check{C} \,,
\label{beta-formula-noB}
\end{split}
\end{align}
where $G_{mn}$ is the metric of the AdS$_5\times$S$^5$ background.
The formula precisely describes the YB-deformed the AdS$_5\times$S$^5$ backgrounds which are read off from the deformed GS action (\ref{eq:YBsM}) described in the next section.
Since the action of the YB-deformed the AdS$_5\times$S$^5$ superstring has the $\kappa$-symmetry,
the deformed background (\ref{beta-formula-noB}) solves the (generalized) supergravity equations of motion \eqref{eq:GSEsecAdS}.
When the associated $r$-matrix is non-unimodular,
the Killing vector $I^m$ in the GSE is given by the divergent formula (\ref{div-formula}).
In this way, we can generate YB-deformed backgrounds by using the formulas (\ref{beta-formula-noB}), (\ref{div-formula}) with the $\beta$-filed (\ref{YB-beta}).

\medskip

If the original background has the $B$-field,
the formula (\ref{beta-formula-B}) should be employed in stead of (\ref{beta-formula-noB}).
The deformations of the AdS$_3\times$S$^3\times$T$^4$ supported by $H$-flux are discussed in the section \ref{sec:AdS3-YB}.

\subsubsection*{$R$-flux}

When there exists the $\beta$-field on a given background,
we can consider the associated tri-vector $R$ known as the non-geometric $R$-flux.
The flux is defined as 
\begin{align}
 R\equiv [\beta,\,\beta]_S\,,
\label{Jacobi}
\end{align}
where $[\,,\,]_S$ denotes the Schouten bracket.
The Schouten bracket is defined for a $p$-vector and a $q$-vector as
\begin{align}
\begin{split}
 &[a_1\wedge \cdots \wedge a_p,\,b_1\wedge \cdots \wedge b_q]_{\rmS} \\ 
\equiv &\sum_{i,j}(-1)^{i+j} [a_i,\,b_j] \wedge a_1\wedge \cdots \check{a_i}\cdots \wedge a_p 
\wedge b_1\wedge \cdots \check{b_j}\cdots \wedge b_q \,,
\label{eq:Sch-bra}
\end{split}
\end{align}
where the check $\check{a}_i$ denotes the omission of $a_i$\,.

\medskip

The $\beta$-field on the YB deformed backgrounds takes the form
\begin{align}
 \beta^{mn} = - \bmr^{mn} = 2\,\eta\,r^{ij}\,\hat{T}_i^m\, \hat{T}_j^n \,,\qquad 
 \beta = \frac{1}{2}\,\beta^{mn}\,\partial_m\wedge\partial_n 
    = 2\,\eta\,\biggl(\frac{1}{2}\,r^{ij}\,\hat{T}_i \wedge \hat{T}_j\biggr) \,.
\end{align}
By using the Lie bracket for the Killing vector fields $[\hat{T}_i,\,\hat{T}_j]=-f_{ij}{}^k\,\hat{T}_k$\,,
we obtain
\begin{align}
 R^{mnp} &=3\,\beta^{[m|q}\,\partial_q \beta^{|np]}\no\\
&= -8\,\eta^2\,\bigl(f_{l_1l_2}{}^i\,r^{jl_1}\,r^{kl_2} + f_{l_1l_2}{}^j\,r^{kl_1}\,r^{il_2} + f_{l_1l_2}{}^k\,r^{il_1}\,r^{jl_2}\bigr)\,\hat{T}_i^m\,\hat{T}_j^n\,\hat{T}_k^p = 0\,,
\end{align}
upon using the homogeneous CYBE \eqref{eq:CYBE-r}\cite{Sakamoto:2017cpu}.
This shows the absence of the $R$-flux in homogeneous YB-deformed backgrounds.

\subsection{YB deformed backgrounds from the GS action}
\label{subsec:YBfromGS}

In the following, we rewrite the YB-deformed action in the form of the conventional GS action,
and show that the target space is a $\beta$-deformed $\AdS{5} \times \rmS^5$ background. 
In order to determine the deformed background, it is sufficient to expand the action up to quadratic order in fermions,
\begin{align}
 S_{\YB}=S_{(0)}+S_{(2)}+\cO(\theta^4)\,.
\end{align}
In this subsection, we provide a general formula for the deformed background for an arbitrary $r$-matrix satisfying homogeneous CYBE, though our analysis is limited to the cases where the $r$-matrices are composed only of the bosonic generators of $\mathfrak{su}(2,2|4)$\,.\footnote{Rewriting of the YB sigma model action to the standard GS form based on the $\kappa$-symmetry was done in \cite{Borsato:2016ose} to full order in fermionic variables, and there, the deformed background associated with a general $r$-matrix was determined.}

\subsubsection{Preparations}

To expand the action (\ref{eq:YBsM}) of the YB sigma model,
we will introduce some notation.
Since the $r$-matrix is composed of bosonic generators only,
the dressed $R$-operator $R_{g_{\bos}}$ acts on each generators as
\begin{align}
\begin{split}
 R_{g_{\bos}}(\gP_{\Loa}) &=\lambda_{\Loa}{}^{\Lob}\,\gP_{\Lob}+\frac{1}{2}\,\lambda_{\Loa}{}^{\Lob\Loc}\,\gJ_{\Lob\Loc}\,, \\
 R_{g_{\bos}}(\gJ_{\Loa\Lob}) &=\lambda_{\Loa\Lob}{}^{\Loc}\,\gP_{\Loc}+\frac{1}{2}\,\lambda_{\Loa\Lob}{}^{\Loc\Lod}\,\gJ_{\Loc\Lod}\,, \\
 R_{g_{\bos}}(\gQ^I) &=0\,.
\label{eq:Rg-operation}
\end{split}
\end{align}
Since the (dressed) $R$-operator is skew-symmetric,
\begin{align}
 \str\bigl[R_{g_{\bos}}(X)\,Y\bigr] = - \str\bigl[X\,R_{g_{\bos}}(Y)\bigr] \,,
\end{align}
if we take $X$ and $Y$ as $\gP_{\Loa}$ or $\gJ_{\Loa\Lob}$\,, we obtain relations
\begin{align}
 \lambda_{\Loa\Lob} \equiv \lambda_{\Loa}{}^{\Loc}\,\eta_{\Loc\Lob} = -\lambda_{\Lob\Loa}\,,\quad 
 \lambda_{\Loa\Lob}{}^{\Loc}=-\frac{1}{2}\,\eta^{\Loc\Lod}\,R_{\Loa\Lob\Loe\Lof}\,\lambda_{\Lod}{}^{\Loe\Lof}\,,\quad 
 \lambda_{\Loa\Lob}{}^{\Loe\Lof}\,R_{\Loe\Lof\Loc\Lod} = - \lambda_{\Loc\Lod}{}^{\Loe\Lof}\,R_{\Loe\Lof\Loa\Lob}\,,
\label{eq:lambda-properties}
\end{align}
where $R_{\Loa\Lob\Loc\Lod}$ is the Riemann tensor in the tangent space of the $\AdS{5} \times \rmS^5$ background.

\medskip

For the later convenience, we will introduce deformed currents
\begin{align}
 J_{\pm} \equiv \cO_{\pm}^{-1}\,A_{\pm}\,.
\label{eq:J-O-inv-A}
\end{align}
By using the results in the Appendix \ref{app:expansion-O}, the currents can be expanded as
\begin{align}
 J_\pm&=\cO^{-1}_{\pm(0)}(A_{(0)})+\cO^{-1}_{\pm(0)}(A_{(1)})+\cO^{-1}_{\pm(1)}(A_{(0)})+\cO(\theta^2)
\nn\\
  &=e_{\pm}^{\Loa}\,\gP_{\Loa}-\frac{1}{2}\,W_{\pm}^{\Loa\Lob}\,\gJ_{\Loa\Lob} + \gQ^I\,D^{IJ}_{\pm}\theta_J +\cO(\theta^2)\,,
\label{eq:Jpm-expansion}
\end{align}
where we defined
\begin{align}
 &e_{\pm}^{\Loa}\equiv e^{\Lob}\,k_{\pm \Lob}{}^{\Loa}\,,\qquad 
 k_{\pm \Loa}{}^{\Lob} \equiv \bigl[(1\pm 2\,\eta\,\lambda)^{-1}\bigr]{}_{\Loa}{}^{\Lob}\,, \qquad
 W_{\pm}^{\Loa\Lob} \equiv \omega^{\Loa\Lob}\pm 2\,\eta\,e_{\pm}^{\Loc}\,\lambda_{\Loc}{}^{\Loa\Lob}\,, 
\label{eq:e-torsionful-spin-pm}
\\
 &D^{IJ}_{\pm}\equiv \delta^{IJ}\,D_{\pm} +\frac{\ii}{2}\,\epsilon^{IJ}\,e_{\pm}^{\Loa}\,\hat{\gamma}_{\Loa}\,, \qquad 
 D_{\pm}\equiv \rmd+\frac{1}{4}\,W_{\pm}^{\Loa\Lob}\, \gamma_{\Loa\Lob}\,. 
\end{align}
Here, $e_{\pm}^{\Loa}$ and $W_{\pm}^{\Loa\Lob}$ are two vielbeins on the deformed background and torsionful spin connections $\omega_{\pm}$ \eqref{eq:Phi-omega-pm}, respectively.
In fact, $e_{\pm}^{\Loa}$ satisfy
\begin{align}
g_{mn}'=\eta_{ab}e_{+}^{\Loa}e_{+}^{\Lob}=\eta_{ab}e_{-}^{\Loa}e_{-}^{\Lob}\,,
\end{align}
and describe the deformed metric $g_{mn}'$\,.
Also, $W_{\pm}^{\Loa\Lob}$ are given by
\begin{align}
W_{\pm \Loa\Lob}=\omega_{[\mp] \Loa\Lob} \pm\frac{1}{2}\,e_{\mp}^{\Loc}\,H'_{\Loc\Loa\Lob}\,,
\end{align}
where $\omega_{[\pm]}$ are spin connections associated with the vielbeins $e_{\pm}$ and
$H'_3$ is the $H$-flux on the deformed background.

\subsubsection{NS--NS sector}
\label{sec:YB-NS-NS}

\paragraph{metric and $B$-filed}

Let us first consider the metric and $B$-filed of the YB deformed action
\begin{align}
 S_{(0)}=-\frac{\dlT}{2}\int \rmd^2\sigma\,\Pg_{-}^{\WSa\WSb}\, \str\bigl[A_{\WSa (0)}\, d_-\circ\cO_{-(0)}^{-1}(A_{\WSb(0)})\bigr]\,. 
\end{align}
From \eqref{eq:Oinv-0}, the action can be rewritten as
\begin{align}
 S_{(0)}
 =-\dlT\int \rmd^2\sigma\,\Pg_{-}^{\WSa\WSb}\, \eta_{\Loa\Lob}\,e_{\WSa}{}^{\Loa}\,e_{\WSb}{}^{\Loc}\,k_{-\Loc}{}^{\Lob}\,.
\label{eq:action-order0}
\end{align}
By comparing it with the canonical form (\ref{eq:GS-action-rewriting}) of the GS action,
we can write down the expressions of the deformed metric and the $B$-field as
\begin{align}
 \CG'_{mn} = e_{(m}{}^{\Loa}\,e_{n)}{}^{\Lob}\, k_{+\Loa\Lob} \,,\qquad 
 B'_{mn} = e_{[m}{}^{\Loa}\,e_{n]}{}^{\Lob}\, k_{+\Loa\Lob} \,. 
\label{eq:G-B-prime}
\end{align}
Since the original $\AdS{5}\times\rmS^5$ background does not the $B$-field,
$E_{mn}=(g+B)_{mn}$ is simply
\begin{align}
 E_{mn}= \CG_{mn} = e_{m}{}^{\Loa}\,e_{n}{}^{\Lob}\,\eta_{\Loa\Lob}\,, \qquad 
 E^{mn}= \eta^{\Loa\Lob}\, e_{\Loa}{}^{m}\,e_{\Lob}{}^{n} \,. 
\end{align}
On the other hand, by using (\ref{eq:G-B-prime}), 
the inverse of $E_{mn}$ is deformed as
\begin{align}
 E'^{mn} \equiv \bigl[(\CG' + B')^{-1}\bigr]^{mn} = (k_+^{-1})^{\Loa\Lob}\, e_{\Loa}{}^{m}\,e_{\Lob}{}^{n} 
 = (\eta + 2\,\eta\,\lambda)^{\Loa\Lob}\, e_{\Loa}{}^{m}\,e_{\Lob}{}^{n} \,.
\end{align}
Therefore, the deformation can be summarized as
\begin{align}
 E^{mn} \ \to \ E'^{mn} = E^{mn} + 2\,\eta\,\lambda^{\Loa\Lob}\, e_{\Loa}{}^{m}\,e_{\Lob}{}^{n} \,.
\end{align}
By comparing it with the $\beta$-transformation rule \eqref{eq:beta-rule-NS},
the YB deformation can be regarded as the $\beta$-deformation with the parameter
\begin{align}
 {\bf r}^{mn} = 2\,\eta\,\lambda^{\Loa\Lob}\, e_{\Loa}{}^{m}\,e_{\Lob}{}^{n} \,. 
\label{eq:bmr-YB}
\end{align}
If we compute dual fields $G_{mn}\,, \beta_{mn}$ (\ref{eq:relation-open-closed}) in the deformed background, we obtain
\begin{align}
 \OG_{mn}&=\eta_{\Loa\Lob}\,e_m{}^{\Loa}\,e_n{}^{\Lob}\,, \qquad
 \beta^{mn}=-2\,\eta\,\lambda^{\Loa\Lob}\,e_{\Loa}{}^m\,e_{\Lob}{}^n\,.
\label{eq:beta}
\end{align}
The dual metric is invariant under the deformation $\OG_{0,mn}\to \OG_{0,mn} = \OG_{mn}$,
while the $\beta$-field, which is absent in the undeformed background, is shifted as $\beta_0^{mn}=0\ \to\ \beta'^{mn}= -\bmr^{mn}$\,.

\medskip

Moreover, let us rewrite $\bmr^{mn}$ of \eqref{eq:bmr-YB} by using the $r$-matrix instead of $\lambda^{\Loa\Lob}$\,.
From the definition, $\lambda^{\Loa\Lob}$ can be expressed as
\begin{align}
 \lambda^{\Loa\Lob} = \str\bigl[ R_{g_{\bos}}(\gP^{\Loa})\,\gP^{\Lob} \bigr] \,. 
\end{align}
By using the $r$-matrix $r=\frac{1}{2}\,r^{ij}\,T_i\wedge T_j$\,,
this can be expressed as
\begin{align}
 \str\bigl[ R_{g_{\bos}}(\gP^{\Loa})\,\gP^{\Lob} \bigr]
 &= r^{ij}\, \str\bigl(g_{\bos}^{-1}\,T_{i}\,g_{\bos}\,\gP_{\Lob})\, \str(g_{\bos}^{-1}\,T_{j}\,g_{\bos}\,\gP_{\Loa}) \no\\
 &= -r^{ij}\,\bigl[\Ad_{g_{\bos}^{-1}}\bigr]_i{}^{\Loa}\,\bigl[\Ad_{g_{\bos}^{-1}}\bigr]_i{}^{\Lob}\,,
\end{align}
and (\ref{eq:bmr-YB}) becomes
\begin{align}
 \bmr^{mn} = -2\,\eta\,r^{ij}\,\bigl[\Ad_{g_{\bos}^{-1}}\bigr]_i{}^{\Loa}\,\bigl[\Ad_{g_{\bos}^{-1}}\bigr]_j{}^{\Lob}\, e_{\Loa}{}^{m}\,e_{\Lob}{}^{n} \,. 
\end{align}
By using the Killing vectors \eqref{eq:Killing-Formula}, we obtain a very simple expression
\begin{align}
 \bmr^{mn} = -2\,\eta\,r^{ij}\,\hat{T}_i^m\, \hat{T}_j^n \,.
\label{eq:bmr-YB2}
\end{align}
This implies that $\beta$-field (\ref{eq:bmr-YB2}) is the bi-vector representation of the $r$-matrix characterizing a YB deformation.

\subsubsection*{dilaton}

Next let us see the YB deformed dilaton $\Phi'$\,.
The formula of the YB deformed dilaton $\Phi'$ had been proposed in \cite{Kyono:2016jqy,Borsato:2016ose} as
\begin{align}
 \Exp{\Phi'} = (\det k_{+})^{\frac{1}{2}}=(\det k_{-})^{\frac{1}{2}} \,. 
\label{eq:dilaton-YB}
\end{align}
Indeed, the formula is consistent to the equations of motion of SUGRA in the string frame
and reproduces the one on the well-known backgrounds ( e.g. Lunin-Maldacena-Frolov\cite{Lunin:2005jy}, Maldacena-Russo backgrounds\cite{Maldacena:1999mh,Hashimoto:1999ut}).

\medskip

In order to compare this with the $\beta$-transformation law of the dilation, we consider the two vielbeins $e_{\pm m}{}^{\Loa}=e_{m}{}^{\Lob}\,k_{\pm \Lob}{}^{\Loa}$ introduced in \eqref{eq:e-torsionful-spin-pm}. 
Here, we can rewrite $k_{\pm \Loa}{}^{\Lob}$ as
\begin{align}
 k_{\pm \Loa}{}^{\Lob} \equiv \bigl[(1\pm 2\,\eta\,\lambda)^{-1}\bigr]{}_{\Loa}{}^{\Lob}
 = e_{\Loa}{}^m\,\bigl[(1\pm \CG\,\bmr)^{-1}\bigr]_m{}^n\,e_n{}^{\Lob} 
 = e_{\Loa}{}^m\,\bigl[(1\pm E^{\rmT}\,\bmr)^{-1}\bigr]_m{}^n\,e_n{}^{\Lob} \,,
\end{align}
by using $\bmr^{mn}$ of \eqref{eq:bmr-YB} and $B_{mn}=0$ in the undeformed background. 
Then, $e_{\pm m}{}^{\Loa}$ become
\begin{align}
 e_{\pm m}{}^{\Loa} = \bigl[(E^{-\rmT} \pm \bmr)^{-1}\,E^{-\rmT}\bigr]{}_{m}{}^{n}\,e_n{}^{\Loa} \,. 
\end{align}
Comparing this with the $\beta$-transformation rule \eqref{eq:beta-rule-NS}, we can identify $e_{-m}{}^{\Loa}$ as the $\beta$-deformed vielbein $e'_m{}^{\Loa}$\,. 
Similarly, $e_{+m}{}^{\Loa}$ can be identified as the $\beta$-deformed barred vielbein $\bre'_m{}^{\Loa}$,
\begin{align}
 e_{-m}{}^{\Loa}\ \leftrightarrow\ e'_m{}^{\Loa}\,,\qquad 
 e_{+m}{}^{\Loa}\ \leftrightarrow\ \bre'_m{}^{\Loa}\,. 
\label{eq:e+-eebar}
\end{align}
Namely, we can express the deformed metric as
\begin{align}
 \CG'_{mn} = e_{+m}{}^{\Loa}\,e_{+n}{}^{\Lob}\,\eta_{\Loa\Lob}
    = e_{-m}{}^{\Loa}\,e_{-n}{}^{\Lob}\,\eta_{\Loa\Lob}\,,
\end{align}
and the invariance of $\Exp{-2d}=\Exp{-2\Phi}\sqrt{-\CG}$ under $\beta$-deformations shows
\begin{align}
 \Exp{-2\Phi'} = \frac{\sqrt{-\CG}}{\sqrt{-\CG'}} \Exp{-2\Phi} = \frac{\det (e_m{}^{\Loa})}{\det (e_{\pm m}{}^{\Loa})} \Exp{-2\Phi}
 = (\det k_{\pm})^{-1} \Exp{-2\Phi} \,. 
\end{align}
Recalling $\Phi=0$ in the undeformed background, the transformation rule \eqref{eq:dilaton-YB} can be understood as the $\beta$-transformation. 
Therefore, NS--NS fields are precisely $\beta$-deformed under the homogeneous YB deformation. 

\subsubsection{R--R sector}
\label{sec:YB-RR}

Next, we determine the R--R fields from the quadratic part $S_{(2)}$ of the YB deformed action,
and show that the R--R fields are also $\beta$-deformed with the $\bmr^{mn}$ given in (\ref{eq:bmr-YB2}).

\medskip

As noticed in \cite{Arutyunov:2015qva,Kyono:2016jqy}, the deformed action naively does not have the canonical form of the GS action \eqref{eq:GS-action-conventional}, and we need to choose the diagonal gauge and perform a suitable redefinition of the bosonic fields $X^m$\,. 
Since the analysis is considerably complicated, we relegate the details to Appendix \ref{app:deformed-Lagrangian}, and here we explain only the outline. 

\medskip

The quadratic part of the deformed action $S_{(2)}$ can be decomposed into two parts,
\begin{align}
 S_{(2)} = S_{(2)}^{\rmc} + \delta S_{(2)}\,.
\end{align}
For a while, we focus only on the first part $S_{(2)}^{\rmc}$
since the second part $\delta S_{(2)}$ is completely canceled after some field redefinitions.
The explicit expression of $S_{(2)}^{\rmc}$ is given by
\begin{align}
 S_{(2)}^{\rmc} &=-\ii\,\dlT\int \rmd^2\sigma\,
  \biggl[ \Pg_+^{\WSa\WSb}\,e_{-\WSa}{}^{\Loa}\,\brtheta_1\,\hat{\gamma}_{\Loa}\,D_{+\WSb}\theta_1
       +\Pg_-^{\WSa\WSb}\,e_{+\WSa}{}^{\Loa}\,\brtheta_2\,\hat{\gamma}_{\Loa}\,D_{-\WSb}\theta_2
\nn\\
&\qquad\qquad\qquad\qquad\quad
       +\ii\, \Pg_+^{\WSa\WSb}\,\epsilon^{IJ}\,\brtheta_1\,e_{-\WSa}{}^{\Loa}\,\hat{\gamma}_{\Loa}\,e_{+\WSb}{}^{\Lob}\,\hat{\gamma}_{\Lob}\,\theta_2\biggr]\,.
\label{eq:L-order2}
\end{align}
This action contains the two deformed vielbeins $e_{\pm m}{}^{\Loa}$ similar to the DSM action \eqref{eq:GS-DSM-ours} prior to taking the diagonal gauge. 
As we observed in \eqref{eq:e+-eebar}, these vielbeins $e_{-m}{}^{\Loa}$ and $e_{+m}{}^{\Loa}$ correspond to the two vielbeins $e'_m{}^{\Loa}$ and $\bre'_m{}^{\Lobra}$ introduced in \eqref{eq:dV-geometric}, respectively. 
In order to rewrite the action into the canonical form of the GS action, we need to choose the diagonal gauge $e_m{}^{\Loa}=\bre_m{}^{\Lobra}$\,. 

\medskip

For this purpose, we first rewrite the action \eqref{eq:L-order2} in terms of the $32\times 32$ gamma matrices. 
By using relations \eqref{eq:lift-32-AdS5-1}, \eqref{eq:lift-32-AdS5-2}, and \eqref{eq:lift-32-AdS5-3}, we obtain
\begin{align}
\begin{split}
 S^{\rmc}_{(2)}&=-\ii\,\dlT\int \rmd^2\sigma\,\biggl[
 \Pg_+^{\WSa\WSb}\,\brTheta_1\,e_{-\WSa}{}^{\Loa}\,\Gamma_{\Loa}\,D_{+\WSb}\Theta_1
 +\Pg_-^{\WSa\WSb}\,\brTheta_2\,e_{+\WSa}{}^{\Loa}\,\Gamma_{\Loa}D_{-\WSb}\Theta_2
\\
 &\qquad\qquad\qquad\qquad\quad -\frac{1}{8}\,\Pg_+^{\WSa\WSb}\,\brTheta_1\,e_{-\WSa}{}^{\Loa}\,\Gamma_{\Loa} \,\bisF_5\, e_{+\WSb}{}^{\Lob}\,\Gamma_{\Lob}\,\Theta_2 \biggr]\,,
\end{split}
\end{align}
where $D_{\pm\WSa}\Theta_I\equiv \bigl(\partial_{\WSa}+\frac{1}{4}\,W_{\pm\WSa}{}^{\Loa\Lob}\, \Gamma_{\Loa\Lob}\bigr)\,\Theta_I$ and $\bisF_5$ is the undeformed R--R $5$-form field strength
\begin{align}
 \bisF_5 =\frac{1}{5!}\,\bisF_{\Loa_1\cdots \Loa_5}\,\Gamma^{\Loa_1\cdots \Loa_5} =4\,\bigl(\Gamma^{01234}+\Gamma^{56789}\bigr)\,. 
\end{align}
Next, we eliminate the vielbein $e_{+m}{}^{\Loa}$ by using relations [recall the formula \eqref{eq:Hassan-formula}]
\begin{align}
 e_{+m}{}^{\Loa}=(\Lambda^{-1})^{\Loa}{}_{\Lob}\,e_{-m}{}^{\Lob}
 = \Lambda_{\Lob}{}^{\Loa}\,e_{-m}{}^{\Lob}\,,\qquad 
 \Lambda_{\Loa}{}^{\Lob} \equiv (k_-^{-1})_{\Loa}{}^{\Loc}\, k_{+\Loc}{}^{\Lob}\in SO(1,9)\,,
\label{eq:e--e+-relation}
\end{align}
which follows from \eqref{eq:e-torsionful-spin-pm}.
By further using the e identity
\begin{align}
 \Omega^{-1} \,\Gamma_{\Loa}\,\Omega = \Lambda_{\Loa}{}^{\Lob} \, \Gamma_{\Lob}\,, \qquad 
 \Omega =(\det k_-)^{\frac{1}{2}}\,\text{\AE}\bigl(-\eta\,\lambda^{\Loa\Lob}\,\Gamma_{\Loa\Lob}\bigr) \,,
\label{eq:omega}
\end{align}
the action becomes
\begin{align}
\begin{split}
 S^{\rmc}_{(2)}&=-\ii\,\dlT\int \rmd^2\sigma\,\biggl[
 \Pg_+^{\WSa\WSb}\,\brTheta_1\,e'_{\WSa}{}^{\Loa}\,\Gamma_{\Loa}\,D_{+\WSb}\Theta_1
 +\Pg_-^{\WSa\WSb}\,\brTheta_2\,\Omega^{-1}\,e'_{\WSa}{}^{\Loa}\,\Gamma_{\Loa}\,\Omega\,D_{-\WSb}\Theta_2
\\
 &\qquad\qquad\qquad\qquad\quad -\frac{1}{8}\,\Pg_+^{\WSa\WSb}\,\brTheta_1\,e'_{\WSa}{}^{\Loa}\,\Gamma_{\Loa} \,\bisF_5\,\Omega^{-1}\, e'_{\WSb}{}^{\Loc}\,\Gamma_{\Loc}\,\Omega\,\Theta_2 \biggr]\,.
\end{split}
\end{align}

\medskip

Then, we perform a redefinition of the fermionic vaiables $\Theta_I$\,,
\begin{align}
 \Theta'_1 \equiv \Theta_1\,,\qquad 
 \Theta'_2 \equiv \Omega \,\Theta_2 \,.
\label{eq:fermi-redef}
\end{align}
As the result of the redefinition, we obtain
\begin{align}
 S^{\rmc}_{(2)}&=-\dlT\int \rmd^2\sigma\,\biggl[
 \Pg_+^{\WSa\WSb}\,\ii\,\brTheta'_1 \,e'_{\WSa}{}^{\Loa}\,\Gamma_{\Loa}\,D'_{+\WSb} \Theta'_1 
 +\Pg_-^{\WSa\WSb}\,\ii\,\brTheta_2\,e'_{\WSa}{}^{\Loa}\,\Gamma_{\Loa}\,D'_{-\WSb} \Theta'_2 
\nn\\
 &\qquad\qquad\qquad\qquad\quad -\frac{1}{8}\,\Pg_+^{\WSa\WSb}\,\ii\,\brTheta_1\,e'_{\WSa}{}^{\Loa}\,\Gamma_{\Loa}\,\bisF_5\,\Omega^{-1}\, e'_{\WSb}{}^{\Lob}\,\Gamma_{\Lob}\,\Theta'_2 \biggr]\,,
\label{eq:ScYB2-2}
\end{align}
where the derivatives $D'_{\pm}$ are defined as
\begin{align}
\begin{split}
 D'_+ &\equiv D_+ = \rmd + \frac{1}{4}\,W_+^{\Loa\Lob}\,\Gamma_{\Loa\Lob}\,,
\\
 D'_- &\equiv \Omega\circ D_-\circ \Omega^{-1} = \rmd + \frac{1}{4}\,W_-^{\Loa\Lob}\,\Omega\,\Gamma_{\Loa\Lob}\,\Omega^{-1} + \Omega\,\rmd \Omega^{-1} 
\\
 &= \rmd + \frac{1}{4}\,\bigl[\Lambda^{\Loa}{}_{\Loc}\,\Lambda^{\Lob}{}_{\Lod}\,W_-^{\Loc\Lod} +(\Lambda\,\rmd\Lambda^{-1})^{\Loa\Lob}\bigr] \,\Gamma_{\Loa\Lob} \,. 
\end{split}
\end{align}
As we show in Appendix \ref{app:torsionful-spin-connections},
the spin connection $\omega'^{\Loa\Lob}$ associated with the deformed vielbein $e'^{\Loa}$ and the deformed $H$-flux $H'_{\Loa\Lob\Loc}$ satisfy
\begin{align}
\begin{split}
 \omega'^{\Loa\Lob}+\frac{1}{2}\,e'_{\Loc}\,H'^{\Loc\Loa\Lob} &=W_{+}^{\Loa\Lob}\,,
\\
 \omega'^{\Loa\Lob}-\frac{1}{2}\,e'_{\Loc}\,H'^{\Loc\Loa\Lob} &= \Lambda^{\Loa}{}_{\Loc}\,\Lambda^{\Lob}{}_{\Lod}\,W_-^{\Loc\Lod} +(\Lambda\,\rmd\Lambda^{-1})^{\Loa\Lob} \,,
\end{split}
\label{eq:torsionful-spin}
\end{align}
and $D'_{\pm}$ can be expressed as
\begin{align}
 D'_{\pm} = \rmd + \frac{1}{4}\,\Bigl(\omega'^{\Loa\Lob}\pm\frac{1}{2}\,e'_{\Loc}\,H'^{\Loc\Loa\Lob}\Bigr)\,\Gamma_{\Loa\Lob}\,.
\end{align}
Then, the deformed action \eqref{eq:ScYB2-2} becomes the conventional GS action at order $\cO(\theta^2)$
by identifying the deformed R--R field strengths as
\begin{align}
 \bisF' =\bisF_5\,\Omega^{-1} \,. 
\label{eq:YBRR0}
\end{align}
As shown in Appendix \ref{app:equivRR},
the transformation rule (\ref{eq:YBRR0}) is equivalent to the $\beta$-transformation (\ref{beta-formula-noB}) of the R--R field.
The transformation rule \eqref{eq:YBRR0} has originally been given in \cite{Borsato:2016ose}.
Another derivation based on the $\kappa$-symmetry variation is given in Appendix \ref{app:kappa}.

\medskip

Finally, let us consider the remaining part $\delta S_{(2)}$\,.
This is completely canceled by redefining the bosonic fields $X^m$ \cite{Arutyunov:2015qva,Kyono:2016jqy}\,,
\begin{align}
 X^m\ \to\ X^m + \frac{\eta}{4}\,\sigma_1^{IJ}\,e^{\Loc m}\,\lambda_{\Loc}{}^{\Loa\Lob}\,\brtheta_I\,\gamma_{\Loa\Lob}\,\theta_J + \cO(\theta^4)\,,
\label{eq:bosonic-shift}
\end{align}
as long as the $r$-matrix satisfies the homogeneous CYBE.
Indeed, this redefinition gives a shift $S_{(0)} \to S_{(0)} + \delta S_{(0)}$
and the summation of $\delta S_{(0)}$ and $\delta S_{(2)} $ has a quite simple expression
\begin{align}
 &\delta S_{(0)}+\delta S_{(2)} 
\nn\\
 &=\frac{\eta^2\,\dlT}{2}\int \rmd^2\sigma\,\, \Pg_-^{\WSa\WSb}\,\sigma_1^{IJ}\,
  \bigl[\CYBE^{(0)}_g\bigl(J_{+m}^{(2)},J_{-n}^{(2)}\bigr)\bigr]^{\Loa\Lob}\,\brtheta_I\, \gamma_{\Loa\Lob}\,\theta_J\,\partial_{\WSa}X^{m}\,\partial_{\WSb}X^n \,,
\label{eq:quadro-CYBE}
\end{align}
where $\CYBE^{(0)}_g (X,Y)$ represents the grade-$0$ component of $\CYBE_g (X,Y)$ defined in \eqref{eq:CYBE-g}.
This shows that $\delta S_{(2)}$ is completely canceled out by $\delta S_{(0)}$
when the $r$-matrix satisfies the homogeneous CYBE.
The derivation of (\ref{eq:quadro-CYBE}) is explained in Appendix \ref{app:bosonic-shift}.

\subsection{Killing vector $I$ and the divergent formula}
\label{sec:Killing-I-divformula}

Here, 
we derive the experimental formula (\ref{div-formula}) for YB deformed $\AdS5\times\rmS^5$ backgrounds.
In subsection \ref{sec:GSE-YBsec}, by using the formula (\ref{div-formula}) ,
the Killing vector $I^m$ were given by
\begin{align}
I^m= \sfD_n \bmr^{nm} = \eta\,r^{ij}\,f_{ij}{}^k\,\hat{T}_k^m\,.
\end{align}
A general formula for $I$ on the deformed $\AdS5\times\rmS^5$ backgrounds
was originally obtained in \cite{Borsato:2016ose}, and by neglecting contributions from fermionic generators,
we get a simple expression\cite{Wulff:2018aku}
\begin{align}
\begin{split}
 I&= -\frac{\eta}{2}\, \kappa^{ij}\,\str\bigl\{[T_i,\,R(T_j)]\,\Ad_g\,(J_+^{(2)}+J_-^{(2)})\bigr\} \,,
\\
 (\kappa^{ij})&\equiv \kappa^{-1}\,,\qquad 
 \kappa\equiv (\kappa_{ij})\,,\qquad 
 \kappa_{ij} \equiv \str(T_i\,T_j)\,.
\end{split}
\label{eq:BW-expression-I}
\end{align}
By substituting the definitions (\ref{eq:J-O-inv-A}) of the deformed currents $J_{\pm}$,
(\ref{eq:BW-expression-I}) becomes
\begin{align}
 I &= -\frac{\eta}{2}\, \kappa^{ij}\, r^{kl}\,\kappa_{jl}\,(e_+^{\Loa}+e_-^{\Loa})\,\str\bigl([T_i,\,T_k]\,g\,\gP_{\Loa}\,g^{-1}\bigr)
\nn\\
 &= \frac{\eta}{2}\, r^{ik}\, (e_+^{\Loa}+e_-^{\Loa})\,f_{ik}{}^j\,[\Ad_g]_{\Loa}{}^l\, \kappa_{jl} 
 = \eta \, r^{ij}\, f_{ij}{}^k\,\bigl[\Ad_{g^{-1}}\bigr]_k{}^{\Lob}\, \,k_{-\Lob}{}^{\Loa}\,e'_{\Loa} \equiv I^{\Loa}\,e'_{\Loa}\,,
\end{align}
where we have used
\begin{align}
\begin{split}
 &[\Ad_g]_{\Loa}{}^k\, \kappa_{ki} = \str(g\,\gP_{\Loa}\,g^{-1}\,T_i) = \str(\gP_{\Loa}\,g^{-1}\,T_i\,g) = \bigl[\Ad_{g^{-1}}\bigr]_i{}^{\Loc}\, \eta_{\Loc\Loa}\,, 
\\
 &e_+^{\Loa}+e_-^{\Loa} = (k_{-}^{-1}\,k_+)_{\Lob}{}^{\Loa}\,e'^{\Lob} + e'^{\Lob} = [(2-k_{+}^{-1})\,k_+]_{\Lob}{}^{\Loa}\,e'^{\Lob} + e'^{\Lob}
 = 2 \,k_{+\Lob}{}^{\Loa}\,e'^{\Lob} = 2 \,k_{-}^{\Loa\Lob}\,e'_{\Lob}\,. 
\end{split}
\end{align}
Then, the curved components become
\begin{align}
 I^m = I^{\Loa}\,e'_{\Loa}{}^m = I^{\Loa}\,(k_{-}^{-1})_{\Loa}{}^{\Lob}\,e_{\Lob}{}^m = \eta \, r^{ij}\, f_{ij}{}^k\,\,\bigl[\Ad_{g^{-1}}\bigr]_k{}^{\Lob}\,e_{\Lob}{}^m = \eta \, r^{ij}\, f_{ij}{}^k\,\hat{T}_k^m \,,
\label{eq:I-derived}
\end{align}
and the formula \eqref{eq:experimental} is reproduced. 
Here, it is noted that, although the right-hand side of \eqref{eq:I-derived} is expressed by using the Killing vectors $\hat{T}_i^m$ on the undeformed background, the Killing vector $I^m$ on the left-hand side should be understood as a vector field defined on the YB-deformed $\AdS5\times\rmS^5$ background.

\section{YB deformations as generalized diffeomorphisms}
\label{sec:YB-gdiff}

In the previous section, we have shown that the homogeneous YB deformations can be regarded as the $\beta$-transformations.
Since the deformed backgrounds are solutions of (generalized) supergravity,
the $\beta$-transformations are a gauge symmetry of DFT.
Therefore, one may expect that the transformations should be expressed as generalized diffeomorphisms.

\medskip

Our aim here is to find out generalized diffeomorphisms which produce various $\beta$-twists specified by various $r$-matrices satisfying the homogeneous CYBE. 
Namely, we construct a generalized diffeomorphisms parameter $V$ satisfying 
\begin{align}
 \cH_{MN}'=\Exp{\bbeta^\rmT}\,\mathcal{H}\Exp{\bbeta}=\Exp{\gLie_{V}}\cH_{MN}\,,\qquad 
d'=\Exp{\gLie_{V}}d=d\,. 
\label{eq:generalized-metric-diffe}
\end{align}
We will present strong evidence of the expectation for the following $r$-matrices:
\vspace*{-2mm}
\begin{enumerate}
\renewcommand{\labelenumi}{(\roman{enumi})}
\setlength{\itemsep}{0mm}
\item Abelian $r$-matrices
\item almost Abelian $r$-matrices and more general case
\item rank-$2$ non-unimodular $r$-matrices
\end{enumerate}
At this moment, the general expressions for the generalized diffeomorphisms corresponding any $r$-matrices have not been known.

\subsection{TsT-transformations as generalized diffeomorphisms}

Let us first to consider an Abelian $r$-matrix, 
\begin{align}
 r_1= \eta_1\,T_1\wedge T_2\,.
\label{eq:Abr-gdiff}
\end{align}
The associated YB deformation describes the TsT-transformation.
To obtain the description of TsT-transformations as generalized diffeomorphisms,
we will introduce the generalized Killing vectors $\hat{\mathsf{T}}_i$ associated with $T_i$\,. 
For the usual isometries, $\hat{\mathsf{T}}_i$ take the form $(\hat{\mathsf{T}}_i^M)=(\hat{T}_i^m,\,0)$ 
and satisfy
\begin{align}
 \eta_{MN}\,\hat{\mathsf{T}}_i^M \hat{\mathsf{T}}_j^N=0\,.
\end{align}
In addition, they are independent of the dual coordinates $\tilde{x}_i$\,. 

\medskip

Since we are considering the Abelian $r$-matrix (\ref{eq:Abr-gdiff}),
a coordinate system can always be found so that $\hat{T}_2 = \hat{T}_2^m\,\partial_m$ ($\hat{T}_2^m$: constant) is realized. 
In such coordinates, we consider 
\begin{align}
 V_{1}= \eta_1\, \hat{T}_2^m\,\tilde{x}_m\,\hat{\mathsf{T}}_1\,.
\end{align}
Thanks to $[\hat{T}_1, \hat{T}_2]=0$ and the Killing property of $\hat{T}_2$, 
$V_{1}$ satisfies the weak constraint $\partial_M\partial^M V_{1}=0$ and 
the strong constraint $\partial^M V_{1}^N\,\partial_M A=0$ 
where $A$ denotes $V_{1}$ or supergravity fields. 
Then, it is easy to show that 
\begin{align}
 \gLie_{V_{1}}\cH_{MN} = (\bfr^\rmT_1\,\cH + \cH\, \bfr_1)_{MN}\,, \ \
 \gLie_{V_{1}} d=0\,,
\label{eq:infinitesimal}
\end{align}
where $\bfr_1^{MN}\equiv 2\,\eta_1\,\hat{\mathsf{T}}^{[M}_1\, \hat{\mathsf{T}}^{N]}_2$\,. 
The R-R potentials $\check{C}$ and the field strengths $\check{F}$ are invariant. 
The finite transformation $\Exp{\gLie_{V_{1}}}$ gives \eqref{eq:generalized-metric-diffe} with $\bfr$ replaced with $\bfr_1$. 

\medskip

In order to demonstrate the relation to the usual TsT-transformation, 
let us consider an Abelian $r$-matrix and choose a coordinate system where $\hat{T}_i = \partial_i$ are realized. 
Then, our diffeomorphism parameter becomes 
\begin{align}
 V = \sum_{i=1}^N\eta_i\, \tilde{x}_{2i}\, \partial_{2i-1}\,,
\end{align}
and it generates a generalized diffeomorphism, 
\begin{align}
 x^M \to x'^M = \Exp{V}x^M\,,
\end{align}
or more explicitly, 
\begin{align}
 x'^{2i-1} = x^{2i-1} + \eta_i \, \tilde{x}_{2i}\,.
\end{align}
This is nothing but the TsT-transformation in the DFT language.

\subsection{Almost Abelian twist}

Next, we will consider almost Abelian twists which are generated by a $r$-matrix
\begin{align}
 r_N \equiv \sum_{i=1}^N \eta_i\,T_{2i-1} \wedge T_{2i}\,, \quad 
 [T_{2i-1},\, T_{2i}]=0\,,
\label{eq:r_k}
\end{align}
where $\eta_i$ $(i=1,\dotsc,N)$ are deformation parameters.
We can see the $r$-matrix satisfies the unimodularity condition (\ref{unimodular1}).
In particular, the \emph{almost Abelian} condition can be expressed as
\begin{align}
 [\hat{T}_{2k-1},\,\beta_{(r_{k-1})}]_{\rmS} = 0\,,\qquad 
[\hat{T}_{2k},\,\beta_{(r_{k-1})}]_{\rmS}=0\,, \qquad 
  (1\leq k\leq N)\,.
\label{eq:almost-Abelian}
\end{align}  
The bracket $[~,~]_{S}$ is the Schouten bracket (\ref{eq:Sch-bra}) and
$\beta_{(r_k)}$ represents the $\beta$-field associated with the $r$-matrix $r_k$\,.
This condition (\ref{eq:almost-Abelian}) ensures the homogeneous CYBE. 

\medskip

For the simplicity, we focus on the case of $N=2$\,,
\begin{align}
 r_2=r_1+\eta_2\, T_3\wedge T_4\,.
\end{align}
From the almost Abelian property, we can again find a coordinate system where $\hat{T}_4 = \hat{T}_4^m\,\partial_m$ ($\hat{T}_4^m$: constant) is realized, 
and perform a transformation $\Exp{\gLie_{V_{2}}}$ with 
\begin{align}
 V_{2} = \eta_2\,\hat{T}_4^m\,\tilde{x}_m\,\hat{\mathsf{T}}_3\,.
\end{align}
Repeating this procedure, we obtain the $\beta$-twisted background associated with the almost Abelian $r$-matrix, $r=r_N$\,. 

\medskip

As a non-trivial example, let us consider a deformation of $\AdS5\times \rmS^5$ background 
with the Poincar\'e metric,
\begin{align}
 \rmd s^2 =\frac{ - 2\,\rmd x^+ \rmd x^- +(\rmd x^2)^2+(\rmd x^3)^2+\rmd z^2}{z^2} + \rmd s^2_{\rmS^5} \,. 
\end{align}
We denote the translation, Lorentz, and dilatation generators by $P_\mu$, $M_{\mu\nu}$, and $D$ ($\mu,\nu=+,-,2,3$), respectively, 
and consider a rank-4 $r$-matrix, $r=r_2$, with
\begin{align}
 T_1=M_{+2}\,, \qquad T_2=P_3\,,\qquad T_3=D-M_{+-}\,, \qquad T_4=P_+ \,.
\end{align}
These satisfy $[T_3,\, T_1] = T_1$ and $[T_3,\, T_2] = -T_2$, and constitute an almost Abelian $r$-matrix. 
This case, we consider a sequence of finite transformations 
$\Exp{\gLie_{V_2}}\Exp{\gLie_{V_1}}$ with
\begin{align}
 V_1 \equiv \eta_1\,\tilde{x}_3\,\hat{M}_{+2} \,,\qquad 
 V_2 \equiv \eta_2\,\tilde{x}_+\, (\hat{D}-\hat{M}_{+-}) \,,
\end{align}
where hatted quantities like $\hat{M}_{+2}$ denote the generalized Killing vectors associated with the unhatted generators. 
These finite transformations produce the $\beta$-field,
\begin{align}
 \beta_{(r_2)}&=\eta_1\,\hat{M}_{+2}\wedge \hat{P}_3+\eta_2\,(\hat{D}-\hat{M}_{+-})\wedge \hat{P}_+ \no\\
&=\eta_1\,\bigl(x^2\,\partial_+ + x^-\,\partial_2\bigr)\wedge \partial_3
+ \eta_2\,\bigl(z\,\partial_z+2\,x^-\,\partial_-+x^2\,\partial_2+x^3\,\partial_3\bigr)\wedge\partial_+ \,,
\end{align}
and the deformed background is indeed a solution of type IIB supergravity. 
If one prefers to combine the transformations as a single one, 
$\Exp{\gLie_{V_2}}\Exp{\gLie_{V_1}}=\Exp{\gLie_{V_{12}}}$\,, 
the Baker-Campbell-Hausdorff formula \cite{Hohm:2012gk} would be useful.

\subsection{A more general class}

Let us consider a wider class of unimodular $r$-matrices, $r=r_N$ with \eqref{eq:r_k} satisfying
\begin{align}
 [\hat{T}_{2k-1},\,\beta_{(r_{k-1})}]_{\rmS} = 0\,,\qquad
 \hat{T}_{2k-1}\wedge [\hat{T}_{2k},\,\beta_{(r_{k-1})}]_{\rmS}=0\,,
\end{align}
for $1\leq k\leq N$, which covers all of the rank-4 unimodular $r$-matrices of $\AdS5$ \cite{Borsato:2016ose}, including the example where any TsT-like transformation has not been found.

\medskip

We explain a subtle issue in this class by considering the rank-4 example ($N=2$) 
where $[\hat{T}_3,\,\beta_{(r_1)}]_{\rmS} = 0$ but $[\hat{T}_4,\,\beta_{(r_1)}]_{\rmS} \neq 0$\,. 
Similar to the almost Abelian case, in coordinates where $\hat{T}_2=\partial_2$\,, 
we first consider a finite transformation
\begin{align}
 \cH^{(1)}_{MN}=\Exp{\gLie_{V_1}}\cH_{MN}\qquad {\rm with} \qquad V_1=\eta_1\,\tilde{x}_2\,\hat{\mathsf{T}}_1\,. 
\end{align}
Then, in coordinates where $e_4=\partial_4$, we perform the second transformation 
\begin{align}
 \cH^{(2)}_{MN}=\Exp{\gLie_{V_2}}\cH^{(1)}_{MN}\qquad {\rm with} \qquad V_2=\eta_2\,\tilde{x}_4\,\hat{\mathsf{T}}_3\,. 
\end{align}
According to $[\hat{T}_4,\,\beta_{(r_1)}]_{\rmS} \neq 0$, $\cH^{(1)}_{MN}$ depends on the $x^4$ coordinate 
and hence the second transformation breaks the strong constraint; 
\begin{align}
 \partial_K V_2\,\partial^K \cH^{(1)}_{MN} \neq 0\,.
\end{align}
In fact, the formula \eqref{beta-formula-B} itself does not require the strong constraint, 
and indeed $\Exp{\gLie_{V_2}}\Exp{\gLie_{V_1}}$ provides the desired background. 
The problem is that a generic strong-constraint-violating generalized diffeomorphism is not a gauge symmetry of DFT. 
Therefore, the deformed background may not be a solution of DFT. 
Interestingly, for all examples in the list presented in \cite{Borsato:2016ose} 
(which cover all inequivalent rank-4 deformations of $\AdS5$), 
one can check that the equations of motion transform covariantly under the diffeomorphisms. 
At the present stage, we are not aware of the clear reason why such diffeomorphisms are allowed. 
A more general formulation of DFT 
\cite{Geissbuhler:2013uka,Blumenhagen:2014gva,Blumenhagen:2015zma}, 
where the strong constraint is rather relaxed, may help us to answer the question.

\subsection{Non-unimodular cases}

The last type of homogeneous YB deformations is the non-unimodular one. 
For simplicity, we will here focus upon the rank-2 Jordanian $r$-matrix, 
\begin{align}
 r=\eta\,T_1\wedge T_2 \qquad \mbox{with}  \qquad [T_1,\,T_2] = T_1\,.
\end{align}
In this case, the formula \eqref{unimodular1} indicates that 
the unimodularity is broken: 
\begin{align}
 I^m = -\eta\,\hat{T}_1^m \neq 0\,.
\end{align}
For some non-unimodular cases, TsT-like transformations have been employed in \cite{Orlando:2016qqu} 
to reproduce the YB-deformed backgrounds on a case-by-case basis. 
Instead, we will here stick to our general strategy. 
In the present case, $[\hat{T}_1,\hat{T}_2]\neq 0$ introduces the $x^2$ dependence into $\hat{T}_1$ 
and the parameter $V=\eta\,\tilde{x}_2\,\hat{\mathsf{T}}_1$ breaks even the weak constraint: 
\begin{align}
 \partial_N\partial^N V^M\neq 0\,. 
\end{align}
However, the formula \eqref{eq:generalized-metric-diffe} still works 
due to the generalized Killing property of $\hat{\mathsf{T}}_1$ and the Jordanian property, $[\hat{T}_1, \hat{T}_2] = \hat{T}_1$\,. 
A subtle issue is again the covariance of the equations of motion, 
and, as we show below, they are not transformed covariantly in the non-unimodular case. 

\medskip

In DFT, the generalized connection $\Gamma_{MNK}$ is supposed to transform as 
\begin{align}
 \delta_V\Gamma_{MNK}=\gLie_V \Gamma_{MNK}-2\,\partial_M\partial_{[N}V_{K]}\,.
\end{align}
At the same time, it is defined to satisfy the condition 
\begin{align}
 \nabla_M d\equiv \partial_M d+\frac{1}{2}\,\Gamma_K{}^K{}_M=0\,.
\end{align}
By the consistency, 
\begin{align}
 \delta_V\nabla_M d=\gLie_V \nabla_M d + \partial^K V_M\,
 \partial_K d - \frac{1}{2}\,\partial_N\partial^N V_M\,,
\end{align}
must vanish. 
It indeed vanishes if the strong constraint is satisfied. 
In the present case, the first two terms on the right-hand side vanish 
but the last term does not vanish because $V^M$ breaks the weak constraint. 
A short calculation shows that 
\begin{align}
 \delta_V\nabla^M d= \eta\,[\hat{\mathsf{T}}_1,\,\hat{\mathsf{T}}_2]_\rmC^M\equiv -\bX^M\,. 
\end{align}
From the Jordanian property, the finite transformation gives rise to 
\begin{align}
 \Exp{\delta_V}\nabla^M d= -\bX^M\,. 
\end{align}
Namely, after performing the deformation, 
$\nabla^M d$ does not vanish but becomes (minus) the null generalized Killing vector, 
\begin{align}
 \bX^M=-\eta\,\hat{\mathsf{T}}_1^M=(I^m,\,0)\,.
\end{align}
This is the situation of the modified DFT (mDFT) \cite{Sakatani:2016fvh}, 
where the generalized connection is deformed by a null generalized Killing vector $\bX^M$. 
The details will be explained in section \ref{sec:(m)DFT}.

\medskip

In the R-R sector, we suppose that the potential $\ket{A}$ and the field strength $\ket{F}$ 
transform covariantly (see \cite{Sakamoto:2017wor} for our conventions):
\begin{align}
 \ket{A^{(r)}}=\Exp{\gLie_V} \ket{A} \qquad \mbox{and} \qquad \ket{F^{(r)}} 
 =\Exp{\gLie_V} \ket{F}\,.
\end{align}
However, the relation $\ket{F}=\sla{\partial}\ket{A}$ is deformed 
under the weak-constraint-violating generalized diffeomorphism as 
\begin{align}
 \ket{F^{(r)}} =\sla{\partial}\ket{A^{(r)}} -\bX_M\,\gamma^M\,\ket{A^{(r)}}\,,
\end{align}
which is again the same relation as the one known in mDFT \cite{Sakamoto:2017wor}. 
The Bianchi identities (or the equations of motion) for the R-R fields are also deformed in a similar manner. 
It is tough to evaluate the deviation of the generalized Ricci tensors, 
$(\delta_V -\gLie_V)\, \cS_{MN}$ and $(\delta_V -\gLie_V)\, \cS$\,, 
under the weak-constraint-violating generalized diffeomorphisms. 
In fact, they do not vanish. For all of the rank-2 examples listed in \cite{Orlando:2016qqu}, 
we have checked that the following relations are satisfied:
\begin{align}
\bigl(\Exp{\bfr^\rmT}\!\cS\Exp{\bfr}\bigr)_{MN}=\bS^{(r)}_{MN}\,,\qquad \cS=\bS^{(r)}\,. 
\label{eq:assumption}
\end{align}
Here, $\bS^{(r)}_{MN}$ and $\bS^{(r)}$ are modified generalized Ricci tensors 
\cite{Sakatani:2016fvh} in the deformed background. 
Then, since the stress-energy tensor obviously transforms covariantly, 
\begin{align}
 \bigl(\Exp{\bfr^\rmT}\!\cE\Exp{\bfr}\bigr)_{MN}=\cE^{(r)}_{MN}\,, 
\end{align}
the deformed background is a solution of mDFT. 
In fact, all solutions of mDFT can be mapped to solutions of DFT 
via a field redefinition \cite{Sakamoto:2017wor}, 
and the deformed background is still a solution of DFT. 

\medskip

To clearly see that the deformed background is indeed a solution of DFT, 
let us examine another route. For the example of $r=\eta\,P_-\wedge D$ \cite{Orlando:2016qqu} (see also (\ref{4.3})), 
instead of the weak-constraint-violating generalized diffeomorphism, 
we can find another generalized coordinate transformation 
which does not break the weak/strong constraint,
\begin{align}
\begin{split}
 z'&=(1 + \eta\,\tilde{x}_-)\,z\,,\qquad x'^+ =(1 + \eta\,\tilde{x}_-)\,x^+\,,
\\
 \rho'&=(1 + \eta\,\tilde{x}_-)\,\rho\,,\qquad \tilde{x}'_- = \eta^{-1}\,\log(1 + \eta\,\tilde{x}_-) \,.
\label{eq:gdiff-PmD}
\end{split}
\end{align}
Then, by employing Hohm and Zwiebach's finite transformation law \cite{Hohm:2012gk}, 
this transformation generates the same deformed background from the original $\AdS5\times \rmS^5$. 
In this route, instead of $\bX^M$, a linear $\tilde{x}_-$ dependence is introduced into the DFT-dilaton. 
This result is compatible with the (m)DFT picture discussed in \cite{Sakamoto:2017wor}.

\subsection{Twists by $\gamma$-fields}
{
So far, we have discussed only the homogeneous YB deformations, which always provide 
$\beta$-twists specified by the associated classical $r$-matrices. 
In the $U$-duality-covariant extension of DFT, called the \emph{exceptional field theory} (EFT) 
\cite{West:2001as,West:2003fc,West:2004st,Hillmann:2009ci,Berman:2010is,Berman:2011cg,Berman:2011jh,Berman:2012vc,Hohm:2013pua,Hohm:2013vpa,Hohm:2013uia,Aldazabal:2013via,Hohm:2014fxa}, 
we can consider a more general twisting via the $\gamma$-fields 
\cite{Aldazabal:2010ef,Chatzistavrakidis:2013jqa,Blair:2013gqa,Andriot:2014qla,Blair:2014zba,Sakatani:2014hba,Lee:2016qwn,Sakatani:2017nfr}. 
They are $p$-vectors dual to the R-R $p$-form potentials, and in particular, 
the bi-vector $\gamma^{mn}$ in type IIB theory is the $S$-dual of the $\beta^{mn}$ 
(see \cite{Sakatani:2017nfr} for the duality rules for these fields). 

\medskip

In \cite{Kyono:2016jqy}, a YB-deformed background associated with $r=P_+\wedge (D-M_{+-})$ 
has been determined including the R-R fields, and indeed it is a solution of GSE \cite{Orlando:2016qqu}. 
Interestingly, a solution of standard supergravity 
which has the same NS-NS fields but different R-R fields has been found in \cite{Sakatani:2017nfr}. 
In fact, the former is twisted by a $\beta$-field while the latter is twisted by a $\gamma$-field. 
At the supergravity level, the latter can be obtained by a combination of the TsT-transformations 
and the $S$-duality \cite{Matsumoto:2014ubv} (where the TsT-transformations generate 
a $\beta$-field and the last step converts the $\beta$-field to the $\gamma$-field). 
This deformation can also be realized as a generalized diffeomorphism in EFT. 
However, according to our current understanding, the YB deformation can produce only 
the former $\beta$-twisted background, and it is important to invent the extension, for example, 
by revealing YB deformations of the $S$-duality covariant $(p,q)$-string action \cite{Schwarz:1995dk}.

\section{``YB deformations'' of the $\AdS{3}\times $S$^3\times T^4$ superstring}
\label{sec:AdS3-YB}

In the previous section \ref{sec:YB-beta-deform},
we have shown that the YB sigma model on the $\AdS{5} \times \rmS^5$ background associated with an $r$-matrix $r=\frac{1}{2}\,r^{ij}\,T_i\wedge T_j$ can be regarded as the GS superstring theory defined on a $\beta$-deformed $\AdS{5} \times \rmS^5$ background with the $\beta$-deformation parameter $\bmr^{mn}=-2\,\eta\,r^{ij}\,\hat{T}_i^m\,\hat{T}_j^n$\,. 
The same conclusion will hold also for other backgrounds in string theory. 

\medskip

In this section, we study deformations of an AdS background with $H$-flux. 
In the presence of $H$-flux, it is not straightforward to define the YB sigma model
\footnote{There are several works \cite{Kawaguchi:2011mz,Kawaguchi:2013gma,Delduc:2014uaa,Delduc:2017fib,Demulder:2017zhz} where YB deformations of the WZ(N)W model based on the mCYBE have been studied.},
and we shall concentrate only on $\beta$-deformations. 
As an example, we here consider deformations of the $\AdS3\times \rmS^3\times \TT^4$ background supported by $H$-flux.

\subsection{The $\AdS{3}\times $S$^3\times T^4$ superstring with $H$-flux}

Let us start with a short review of the $\AdS3\times \rmS^3\times \TT^4$ background supported by $H$-flux.

\medskip

The background is given by
\begin{align}
\begin{split}
 &\rmd s^2 = \frac{-(\rmd x^0)^2+(\rmd x^1)^2+\rmd z^2}{z^2} + \rmd s^2_{\rmS^3} + \rmd s_{\TT^4}^2 \,,
\\
 &B_2= \frac{\rmd x^0\wedge \rmd x^1}{z^2} + \frac{1}{4}\,\cos\theta\,\rmd \phi \wedge \rmd \psi \,,\qquad \Phi=0\,,
\\
 &\rmd s^2_{\rmS^3} \equiv \frac{1}{4}\,\bigl[\rmd \theta^2 + \sin^2\theta\, \rmd \phi^2 + \bigl(\rmd \psi + \cos\theta\,\rmd \phi)^2\bigr] \,,
\end{split}
\label{eq:AdS3-S3-T4}
\end{align}
which contains the non-vanishing $H$-flux
\begin{align}
 H_3 =-2\,\frac{\rmd x^0\wedge \rmd x^1 \wedge \rmd z}{z^3}-\frac{1}{4}\,\sin\theta\rmd\phi \wedge \rmd \psi \wedge \rmd \theta\,. 
\end{align}
Using the Killing vectors $\hat{T}_i$ of the $\AdS3\times \rmS^3\times \TT^4$ background, we consider local $\beta$-deformations with deformation parameters of the form, $\bmr^{mn}=-2\,\eta\,r^{ij}\,\hat{T}_i^m\,\hat{T}_j^n$\,. 
We consider several $r$-matrices $r^{ij}$ satisfying the homogeneous CYBE, and show that all of the $\beta$-deformed backgrounds satisfy the equations of motion of (generalized) supergravity. 

\medskip

In order to find the Killing vectors explicitly, we introduce a group parameterization for $\AdS3\times \rmS^3$ (for simplicity, we do not consider the trivial $\TT^4$ directions)
\begin{align}
\begin{split}
 &g =g_{\AdS{3}}\cdot g_{\rmS^3}\cdot g_{\TT^4}\,, \qquad
 g_{\AdS{3}} =\exp(x^{\mu} P_{\mu})\cdot\exp(~\ln z\,D~)\qquad (\mu=0,1)\,,
\\
 &g_{\rmS^3} =\exp(\phi\,T^L_4)\cdot\exp(\theta\,T^L_3)\cdot\exp(\psi\,T^R_4) \,. 
\end{split}
\end{align}
Here, similar to the $\AdS{5}\times\rmS^5$ case (see Appendix \ref{app:psu-algebra}), we have introduced the $\alg{so}(2,2)\times \alg{so}(4)$ generators $(\gP_{\check{\Loa}}\,, \gP_{\hat{\Loa}}\,,\gJ_{\check{\Loa}\check{\Lob}}\,, \gJ_{\hat{\Loa}\hat{\Lob}})$ $(\check{\Loa},\,\check{\Lob}=0,1,2;\,\hat{\Loa},\,\hat{\Lob}=3,4,5)$ as the following $8\times 8$ supermatrices:
\begin{align}
\begin{alignedat}{3}
 \gP_{\check{\Loa}} &= 
 \begin{pmatrix}
  \frac{1}{2}\,\bm{\gamma}_{\check{\Loa}} & \bm{0_4} \\ 
  \bm{0_4} & \bm{0_4} 
 \end{pmatrix},& \qquad 
  \gJ_{\check{\Loa}\check{\Lob}} &= 
 \begin{pmatrix}
  -\frac{1}{2}\,\bm{\gamma}_{\check{\Loa}\check{\Lob}} & \bm{0_4} \\ 
  \bm{0_4} & \bm{0_4} 
 \end{pmatrix},& \qquad
 \bm{\gamma}_{\check{\Loa}} &\equiv \begin{pmatrix}
  +\gamma_{\check{\Loa}}& \bm{0_2} \\
  \bm{0_2} &-\gamma_{\check{\Loa}}
 \end{pmatrix},
\\
 \gP_{\hat{\Loa}} &= 
 \begin{pmatrix}
  \bm{0_4} & \bm{0_4} \\ 
  \bm{0_4} & -\frac{\ii}{2}\,\bm{\gamma}_{\hat{\Loa}} 
 \end{pmatrix}& , \qquad 
 \gJ_{\hat{\Loa}\hat{\Lob}} &= 
 \begin{pmatrix}
  \bm{0_4} & \bm{0_4} \\
  \bm{0_4} & -\frac{1}{2}\,\bm{\gamma}_{\hat{\Loa}\hat{\Lob}} 
 \end{pmatrix}& , \qquad 
 \bm{\gamma}_{\hat{\Loa}}&\equiv
 \begin{pmatrix}
  -\gamma_{\hat{\Loa}} & \bm{0_2} \\
  \bm{0_2} &+\gamma_{\hat{\Loa}}
 \end{pmatrix}, 
\end{alignedat}
\end{align}
where $2\times 2$ gamma matrices $\gamma_{\check{\Loa}}$ and $\gamma_{\hat{\Loa}}$ are defined as
\begin{align}
 \{\gamma_0,\,\gamma_1,\,\gamma_2\} = \{\ii\,\sigma_3\,, \sigma_1\,, \sigma_2\}\,,\qquad
 \{\gamma_3,\,\gamma_4,\,\gamma_5\} = \{\sigma_1\,, \sigma_2\,, \sigma_3\} \,.
\end{align}
We have also defined the conformal basis $\{P_{\mu}\,, M_{\mu\nu}\,, D\,, K_{\mu}\}$ as
\begin{align}
 P_{\mu}\equiv \gP_{\mu}+\gJ_{\mu2}\,,\qquad 
 K_{\mu}\equiv \gP_{\mu}-\gJ_{\mu2}\,,\qquad
 M_{\mu\nu}\equiv \gJ_{\mu\nu}\,,\qquad 
 D\equiv \gP_{2}\,.
\end{align}
The generators of $\alg{su}(2)_L\times \alg{su}(2)_R\simeq \alg{so}(4)$ are defined as
\begin{align}
\begin{split}
 T^L_3&=\frac{1}{2}\,(\gP_3-\gJ_{4,5})\,,\qquad
 T^R_3=\frac{1}{2}\,(\gP_3+\gJ_{4,5})\,,
\\
 T^L_4&=\frac{1}{2}\,(\gP_4-\gJ_{5,3})\,,\qquad
 T^R_4=\frac{1}{2}\,(\gP_4+\gJ_{5,3})\,,
\\
 T^L_5&=\frac{1}{2}\,(\gP_5-\gJ_{3,4})\,,\qquad
 T^R_5=\frac{1}{2}\,(\gP_5+\gJ_{3,4})\,,
\end{split}
\end{align}
which satisfy the commutation relations,
\begin{align}
 [T^{L}_{i},\,T^{L}_{j}]=-\epsilon_{ijk}\,T^{L}_{k}\,,\qquad [T^{R}_{i},\,T^{R}_{j}]=\epsilon_{ijk}\,T^{R}_{k}\,,\qquad [T^{L}_{i},\,T^{R}_{j}]=0\quad 
 (\epsilon_{345}=1)\,.
\end{align}
By computing the Maurer--Cartan 1-form $A=g^{-1}\,\rmd g$, and using the supertrace formula
\begin{align}
\begin{alignedat}{2}
 &\str(\gP_{\Loa}\,\gP_{\Lob}) =\eta_{\Loa\Lob}\,,&\qquad &\str(\gJ_{\Loa\Lob}\,\gJ_{\Loc\Lod})=R_{\Loa\Lob\Loc\Lod} \,,
\\
 &R_{\check{\Loa}\check{\Lob}}{}^{\check{\Loc}\check{\Lod}} \equiv -2\, \delta_{[\check{\Loa}}^{[\check{\Loc}}\,\delta_{\check{\Lob}]}^{\check{\Lod}]}\,,& \qquad 
 &R_{\hat{\Loa}\hat{\Lob}}{}^{\hat{\Loc}\hat{\Lod}} \equiv 2\, \delta_{[\hat{\Loa}}^{[\hat{\Loc}}\,\delta_{\hat{\Lob}]}^{\hat{\Lod}]} \,,
\end{alignedat}
\end{align}
we can reproduce the above metric \eqref{eq:AdS3-S3-T4}. 

\medskip

Then, we can find the Killing vectors $\hat{T}_i$ of this background associated with the generator $T_i$ by using the formula \eqref{eq:Killing-Formula}, or $\hat{T}_i^m = \str\bigl[ g^{-1}\,T^i\,g\,\gP_{\Loa} \bigr]\,e^{\Loa m}$\,. 
The result is summarized as
\begin{align}
\begin{split}
 \hat{P}_{\mu}&=\partial_{\mu}\,,\qquad
 \hat{M}_{\mu\nu}=x_{\mu}\,\partial_{\nu}-x_{\nu}\,\partial_{\mu}\,,\qquad
 \hat{D}=x^{\mu}\,\partial_{\mu}+z\,\partial_z\,,
\\
 \hat{K}_{\mu}& =(x^{\nu}\,x_{\nu}+z^2)\,\partial_{\mu}-2\,x_{\mu}\,(x^{\nu}\,\partial_{\nu}+z\,\partial_z)\,,
\\
 \hat{T}^L_3& =\cos\phi\,\partial_\theta+\sin\phi\,\Bigl(-\frac{1}{\tan\theta}\,\partial_\phi+\frac{1}{\sin\theta}\,\partial_{\psi}\Bigr) \,,
\\
 \hat{T}^R_3& =\cos\psi\,\partial_\theta+\sin\psi\,\Bigl(\frac{1}{\sin\theta}\,\partial_\phi-\frac{1}{\tan\theta}\,\partial_{\psi}\Bigr)\,,
\\
 \hat{T}^L_4&=\partial_{\phi}\,,\qquad
 \hat{T}^R_4=\partial_{\psi}\,,
\\
 \hat{T}^L_5& =\sin\phi\,\partial_\theta+\cos\phi\,\Bigl(\frac{1}{\tan\theta}\,\partial_\phi-\frac{1}{\sin\theta}\,\partial_{\psi}\Bigr) \,,
\\
 \hat{T}^R_5&=-\sin\psi\,\partial_\theta+\cos\psi\,\Bigl(\frac{1}{\sin\theta}\,\partial_\phi-\frac{1}{\tan\theta}\,\partial_{\psi}\Bigr) \,.
\end{split}
\end{align}

\medskip

We note that among the AdS isometries, $\hat{P}_{\mu}$, $\hat{M}_{01}$, and $\hat{D}$ are symmetry of the $B$-field,
\begin{align}
 \Lie_{\hat{P}_{\mu}}B_2 =\Lie_{\hat{M}_{01}}B_2 =\Lie_{\hat{D}}B_2=0\,,
\end{align}
while the special conformal generators $\hat{K}_{\mu}$ change the $B$-field by closed forms,
\begin{align}
 \Lie_{\hat{K}_{0}}B_2 = -\frac{2\,\rmd x^1\wedge \rmd z}{z}\,,\qquad
 \Lie_{\hat{K}_{1}}B_2 = \frac{2\,\rmd x^0\wedge \rmd z}{z}\,.
\label{eq:K-B2-closed}
\end{align}
In the following, we first study $\beta$-deformations by using Killing vectors $\hat{P}_{\mu}$, $\hat{M}_{01}$, $\hat{D}$, and $T^R_4$\,. 
Then, non-trivial cases using the Killing vectors $\hat{K}_{\mu}$ are studied in section \ref{sec:generalized-isometries}. 

\medskip

\subsection{Abelian deformations}

Let us begin by studying simple examples associated with Abelian $r$-matrices. 
As it has been known well \cite{Matsumoto:2014nra,Matsumoto:2014gwa,Matsumoto:2015uja,vanTongeren:2015soa,vanTongeren:2015uha,Kyono:2016jqy,Osten:2016dvf}, YB deformations associated with Abelian $r$-matrices can be also realized as TsT-transformations. 

\subsubsection{\mbox{\boldmath $1.~r=\frac{1}{2}\,P_0\wedge P_1$}}

Let us first consider an Abelian $r$-matrix
\begin{align}
 r=\frac{1}{2}\,P_0\wedge P_1 \,.
\end{align}
From $\hat{P}_0=\partial_0$ and $\hat{P}_1=\partial_1$, the $\beta$-transformation parameter is $\bmr^{mn}= -\eta\,(\delta_0^{m}\,\delta_1^{n}-\delta_1^{m}\,\delta_0^{n})$\,. 
After the $\beta$-transformation, we obtain the background
\begin{align}
\begin{split}
 \rmd s^2&=\frac{-(\rmd x^0)^2+(\rmd x^1)^2}{z^2+2\,\eta}+\frac{\rmd z^2}{z^2} +\rmd s_{\rmS^3}^2+\rmd s_{\TT^4}^2\,,
\\
 B_2&=\frac{\rmd x^0\wedge \rmd x^1}{z^2+2\,\eta}+\frac{1}{4}\,\cos \theta\,\rmd \phi\wedge \rmd \psi\,,\qquad 
 \Exp{-2\Phi} = \frac{z^2+2\,\eta}{z^2} \,.
\end{split}
\end{align}
As we have mentioned above, we can also obtain the background by a TsT transformation from the background \eqref{eq:AdS3-S3-T4}; (1) T-dualize along the $x^1$-direction, (2) (active) shift $x^0\to x^0+\eta\,x^1$\,, (3) T-dualize along the $x^1$-direction.
This background is of course a solution of supergravity. 

\medskip

As a side remark, noted that this background interpolates a linear dilaton background in the UV region ($z \sim 0$) and the undeformed $\AdS{3} \times \rmS^3\times \TT^4$ background in the IR region ($z \to \infty$). 
Indeed, by performing a coordinate transformation
\begin{align}
 x^{\pm}=\frac{x^0\pm x^1}{\sqrt{2}} \,,\qquad z=\Exp{\rho}\,,
\end{align}
the deformed background becomes
\begin{align}
\begin{split}
 \rmd s^2&=-\frac{2\Exp{-2\rho}}{1+2\,\eta\Exp{-2\rho}}\,\rmd x^+\,\rmd x^- + \rmd \rho^2 +\rmd s_{\rmS^3}^2+\rmd s_{\TT^4}^2\,,\qquad
 \Exp{-2\Phi} = 1+2\,\eta\,\Exp{-2\rho}\,,
\\
 B_2&=-\frac{\Exp{-2\rho}}{1+2\,\eta\Exp{-2\rho}}\,\rmd x^+\wedge\rmd x^- + \frac{1}{4}\,\cos \theta\,\rmd \phi\wedge \rmd \psi \,.
\label{eq:TT-AdS}
\end{split}
\end{align}
In the asymptotic region $\Exp{-2\rho} \gg \eta^{-1}$ (i.e.~$z\sim 0$), the background approaches to a solution that is independent of the deformation parameter $\eta$
\begin{align}
\begin{split}
 \rmd s^2&=-2\,\rmd x^+\rmd x^-+\rmd \rho^2 +\rmd s_{\rmS^3}^2+\rmd s_{\TT^4}^2\,,\qquad 
 \Phi =\rho\,,
\\
 B_2&=-\rmd x^+\wedge\rmd x^-+\frac{1}{4}\cos \theta\,\rmd \phi\wedge \rmd \psi \,,
\end{split}
\label{eq:TT-AdS-limit}
\end{align}
where we ignored the constant part of the dilaton and rescaled light-cone coordinates $x^{\pm}$ as
\begin{align}
x^{\pm}\to \sqrt{2\,\eta}\,x^{\pm}\,.
\end{align}
The $\AdS{3}$ part of the background \eqref{eq:TT-AdS} is precisely the geometry obtained via a null deformation of $SL(2,\mathbb{R})$ WZW model \cite{Forste:1994wp} (see also \cite{Israel:2003ry}), which is an exactly marginal deformation of the WZW model (see \cite{Giveon:2017nie,Giveon:2017myj,Asrat:2017tzd,Giribet:2017imm} for recent studies). 
Note also that, under a formal $T$-duality along the $\rho$-direction, the solution \eqref{eq:TT-AdS-limit} becomes the following solution in DFT:
\begin{align}
\begin{split}
 \rmd s^2&=-2\,\rmd x^+\rmd x^-+\rmd \rho^2 +\rmd s_{\rmS^3}^2+\rmd s_{\TT^4}^2\,,\qquad 
 \Phi =\tilde{\rho}\,,
\\
 B_2&=-\rmd x^+\wedge\rmd x^-+\frac{1}{4}\cos \theta\,\rmd \phi\wedge \rmd \psi \,,
\end{split}
\end{align}
where the dilaton depends linearly on the dual coordinate $\tilde{\rho}$\,. 
This background can be also interpreted as the following solution of GSE:
\begin{align}
\begin{split}
 \rmd s^2&=-2\,\rmd x^+\rmd x^-+\rmd \rho^2 +\rmd s_{\rmS^3}^2+\rmd s_{\TT^4}^2\,,\qquad \Phi=0\,,
\\
 B_2&=-\rmd x^+\wedge\rmd x^-+\frac{1}{4}\cos \theta\,\rmd \phi\wedge \rmd \psi \,,\qquad I =\partial_\rho \,. 
\end{split}
\end{align}

\subsubsection{\mbox{\boldmath $2.~r=\frac{1}{2}\,P_+\wedge T^R_4$}}

As the second example, let us consider an Abelian $r$-matrix
\begin{align}
 r = \frac{1}{2}\,P_+\wedge T^R_4\qquad \Bigl(P_+ \equiv \frac{P_0 + P_1}{\sqrt{2}}\Bigr)\,.
\end{align}
For convenience, let us change the coordinates such that the background \eqref{eq:AdS3-S3-T4} becomes
\begin{align}
\begin{split}
 \rmd s^2&=-2\Exp{2\rho} \rmd x^+\,\rmd x^- + \rmd \rho^2 +\rmd s_{\rmS^3}^2+\rmd s_{\TT^4}^2\,,
 \qquad \Phi =0\,,
\\
 B_2&=-\Exp{2\rho}\rmd x^+\wedge\rmd x^- + \frac{1}{4}\,\cos \theta\,\rmd \phi\wedge \rmd \psi \,.
\end{split}
\end{align}
In this coordinate system, the Killing vectors take the form, $\hat{P}_+=\partial_+$ and $\hat{T}^R_4=\partial_\psi$. 
Then, the associated $\beta$-deformed (or TsT-transformed) background is given by
\begin{align}
\begin{split}
 \rmd s^2&=-2\Exp{2\rho} \rmd x^+\,\rmd x^- + \rmd \rho^2 +\frac{\eta}{2}\Exp{2\rho} \rmd x^-\,(\rmd\psi+2\,\cos\theta\,\rmd\phi) +\rmd s_{\rmS^3}^2+\rmd s_{\TT^4}^2\,,
 \quad \Phi =0\,,
\\
 B_2&=-\Exp{2\rho}\rmd x^+\wedge\rmd x^- + \frac{1}{4}\,\cos \theta\,\rmd \phi\wedge \rmd \psi -\frac{\eta}{4}\,\Exp{2\rho}\rmd x^-\wedge (\rmd \psi+2\,\cos\theta\,\rmd\phi) \,.
\end{split}
\end{align}
This background has been studied in \cite{Azeyanagi:2012zd}, where the twist was interpreted as a spectral flow transformation of the original model in the context of the NS--R formalism. 

\subsubsection{\mbox{\boldmath $3.~r = \frac{1}{2}\,D\wedge M_{01}$}}
\label{sec:AdS3-DwM}

Let us also consider a slightly non-trivial example $r = \frac{1}{2}\,D\wedge M_{01}$, which is also an Abelian $r$-matrix. 
The associated $\beta$-deformed background is given by
\begin{align}
\begin{split}
 \rmd s^2&= \frac{\eta_{\mu\nu}\,\rmd x^\mu\,\rmd x^\nu - 2\,\eta\,z^{-1}\,x_\mu\,\rmd x^\mu\, \rmd z +(1+\frac{2\,\eta\,x_\mu\,x^\mu}{z^2}) \,\rmd z^2}{z^2-\eta\,(\eta-2)\,x_\mu\,x^\mu} 
 + \rmd s^2_{\rmS^3} + \rmd s_{\TT^4}^2 \,,
\\
 \Exp{-2\Phi} &= \frac{z^2-\eta\,(\eta-2)\,x_\mu\,x^\mu}{z^2}\,,
\\
 B_2 &= \frac{\rmd x^0\wedge\rmd x^1-\eta\,z^{-1}\, (x^1\,\rmd x^0-x^0\,\rmd x^1)\wedge \rmd z}{z^2-\eta\,(\eta-2)\,x_\mu\,x^\mu} + \frac{1}{4}\,\cos\theta\,\rmd \phi \wedge \rmd \psi \,. 
\end{split}
\end{align}
We can easily check that this is a solution of the supergravity. 
In order to obtain the same background by performing a TsT transformation, we should first change the coordinates such that the Killing vectors $\hat{D}$ and $\hat{M}_{01}$ become constant, and perform a TsT transformation, and then go back to the original coordinates. 
The $\beta$-transformation is much easier in this case. 

\medskip

In order to describe the same $\beta$-deformation in the global coordinates
\begin{align}
\begin{split}
 \rmd s^2&=-\cosh^2\rho\,\rmd \tau^2+\sinh^2\rho\,\rmd \chi^2+\rmd\rho^2 +\rmd s_{\rmS^3}^2+\rmd s_{\TT^4}^2\,,\qquad \Phi=0\,,
\\
 B_2&=\cosh^2\rho\,\rmd\tau\wedge \rmd \chi +\frac{1}{4}\,\cos \theta\,\rmd \phi\wedge \rmd \psi\,,
\end{split}
\end{align}
we change the group parameterization as
\begin{align}
 g_{\AdS{3}}= \exp(\ii\,\tau\, D+\ii\,\chi\, M_{01})\cdot\exp(\rho\, \gP_1)\,.
\end{align}
In this case, we can compute the Killing vectors as
\begin{align}
 \hat{D}=-\ii\,\partial_{\tau}\,,\qquad \hat{M}_{01}=-\ii\,\partial_{\chi}\,.
\end{align}
Then, the $\beta$-deformed background becomes
\begin{align}
\begin{split}
 \rmd s^2 
 &=\frac{-\cosh^2\rho\,\rmd \tau^2+\sinh^2\rho\,\rmd \chi^2}{1+\eta\,(\eta-2)\,\cosh^2\rho}+\rmd\rho^2
 +\rmd s_{\rmS^3}^2+\rmd s_{\TT^4}^2\,,
\\
 \Phi&=\frac{1}{2}\,\ln\Bigl[\frac{1}{1+\eta(\eta-2)\cosh^2\rho}\Bigr] \,,
\\
 B_2&=(1-\eta)\,\frac{\cosh^2\rho\,\rmd\tau\wedge \rmd \chi}{1+\eta\,(\eta-2)\,\cosh^2\rho}
 +\frac{1}{4}\,\cos \theta\,\rmd \phi\wedge \rmd \psi \,.
\end{split}
\end{align}
If the deformation parameter $\eta$ and the angular coordinate $\chi$ are replaced as
\begin{align}
 \eta\to 1-\sqrt{\alpha}\,,\qquad \chi\to \sqrt{\alpha}\, \chi\,,
\end{align}
the AdS part of this background reproduce the background obtained in \cite{Giveon:1993ph,Israel:2003ry} through a current-current deformation of the $SL(2,\mathbb{R})$ WZW model (see Eqs.~(5.1)--(5.3) in \cite{Israel:2003ry}). 

\medskip

\subsection{Non-unimodular deformations}
\label{sec:non-unimodular}

Let us next consider $\beta$-deformations associated with non-Abelian $r$-matrices. 
In particular, we consider non-unimodular $r$-matrices.
As was shown in \cite{Borsato:2016ose},
YB deformations associated with non-unimodular $r$-matrices give solutions of the GSE \cite{Arutyunov:2015mqj,Wulff:2016tju,Sakatani:2016fvh,Baguet:2016prz,Sakamoto:2017wor}, which include non-dynamical Killing vector $I^m$. 
As it was observed experimentally \cite{Araujo:2017jkb,Araujo:2017jap,Araujo:2017enj,Fernandez-Melgarejo:2017oyu}, the extra vector $I^m$ typically takes the form \cite{Araujo:2017jkb,Araujo:2017jap,Sakamoto:2017cpu,Araujo:2017enj,Fernandez-Melgarejo:2017oyu}
\begin{align}
 I^m = \sfD_n \bmr^{nm} \,,
\end{align}
where $\sfD_n$ is the usual covariant derivative associated with the undeformed $\AdS3\times \rmS^3\times \TT^4$ background and $\bmr^{mn} = -2\,\eta\,r^{ij}\,\hat{T}_i^m\, \hat{T}_j^n$\,. 

\medskip

Interestingly, as we explain in section \ref{sec:AdS3-S3-T4MwedgeP}, in some examples, even for non-unimodular $r$-matrices, the $\beta$-deformed backgrounds satisfy the usual supergravity equations of motion. 
Such example has not been observed in the case of the $\AdS5\times \rmS^5$ background,
\footnote{See a recent paper \cite{Wulff:2018aku} for a general analysis of such backgrounds, called the ``trivial solutions'' of GSE.} 
and this is due to a particular property of the $\AdS3\times \rmS^3\times \TT^4$ background as explained below.

\subsubsection{\mbox{\boldmath $1.~r = \frac{1}{2}\,\bar{c}^\mu\,D\wedge P_\mu$}}

Let us consider the simplest non-unimodular $r$-matrix $r = \frac{1}{2}\,\bar{c}^\mu\,D\wedge P_\mu$\,, satisfying
\begin{align}
 \cI = \eta\,\bar{c}^\mu\,[D,\,P_\mu] = c^\mu P_\mu \neq 0 \qquad \bigl(c^\mu\equiv\eta\,\bar{c}^\mu\bigr)\,.
\end{align}
The $\beta$-deformed background becomes
\begin{align}
\begin{split}
 \rmd s^2&= \frac{\eta_{\mu\nu}\,\rmd x^\mu\,\rmd x^\nu + 2\,z^{-1}\,(c^0\,\rmd x^1 -c^1\,\rmd x^0)\,\rmd z + \bigl[1+2\,z^{-2}\,(c^1\,x^0-c^0\,x^1)\bigr]\,\rmd z^2}{z^2 + c_\mu\,c^\mu + 2\,(c^1\,x^0 - c^0 \,x^1)} 
\\
 &\quad + \rmd s^2_{\rmS^3} + \rmd s_{\TT^4}^2 \,,\qquad 
 \Exp{-2\Phi} = \frac{z^2 + c_\mu\,c^\mu + 2\,(c^1\,x^0 - c^0 \,x^1)}{z^2}\,,
\\
 B_2 &= \frac{\rmd x^0\wedge\rmd x^1 - z^{-1}\, c_\mu\,\rmd x^\mu\wedge \rmd z}{z^2 + c_\mu\,c^\mu + 2\,(c^1\,x^0-c^0\,x^1)} + \frac{1}{4}\,\cos\theta\,\rmd \phi \wedge \rmd \psi \,,
\end{split}
\end{align}
where $c_\mu \equiv \eta_{\mu\nu}\,c^\nu$\,. 
Although this is not a solution of the usual supergravity, by introducing a Killing vector,
\begin{align}
 I =  \hat{\cI}=c^\mu \hat{P}_\mu = c^\mu\,\partial_\mu \,,
\end{align}
it becomes a solution of the GSE.

\subsubsection{\mbox{\boldmath $2.~r = \frac{1}{2}\,\bar{c}^\mu\,M_{01}\wedge P_\mu$}}
\label{sec:AdS3-S3-T4MwedgeP}

The next example is a non-unimodular $r$-matrix $r = \frac{1}{2}\,\bar{c}^\mu\,M_{01}\wedge P_\mu$\,, satisfying
\begin{align}
 \cI = - c^\mu P_\mu \neq 0 \qquad \bigl(c^\mu\equiv\eta\,\bar{c}^\mu\bigr)\,.
\end{align}
The $\beta$-deformed background becomes
\begin{align}
\begin{split}
 \rmd s^2&= \frac{\eta_{\mu\nu}\,\rmd x^\mu\,\rmd x^\nu}{z^2 - 2\, c_\mu\,x^\mu} + \frac{\rmd z^2}{z^2} + \rmd s^2_{\rmS^3} + \rmd s_{\TT^4}^2 \,,\qquad 
 \Exp{-2\Phi} = \frac{z^2 - 2\, c_\mu\,x^\mu}{z^2}\,,
\\
 B_2 &= \frac{\rmd x^0\wedge \rmd x^1}{z^2 - 2\, c_\mu\,x^\mu} + \frac{1}{4}\,\cos\theta\,\rmd \phi \wedge \rmd \psi \,. 
\label{eq:AdS3_M^P}
\end{split}
\end{align}
where $c_\mu \equiv \eta_{\mu\nu}\,c^\nu$\,. 
As usual, by introducing
\begin{align}
 I = - c^\mu \hat{P}_\mu = -c^\mu\,\partial_\mu \,,
\end{align}
this background satisfies the GSE. 

\medskip

Here, note that the defining properties of $I^m$,
\begin{align}
 \Lie_{I} \CG_{mn} = 0 \,,\qquad \Lie_{I} B_{mn} = 0 \,,
\end{align}
require that the parameters should satisfy $c^0=\pm c^1$\,. 
In terms of DFT, the above deformed background can be expressed as
\begin{align}
 (\cH_{MN})= \begin{pmatrix}
 (\CG-B\,\CG^{-1}\,B)_{mn} & (B\,\CG^{-1})_{m}{}^n \\
 -(\CG^{-1}\,B)^m{}_n & \CG^{mn} \end{pmatrix},\qquad 
 d = \Phi -\frac{1}{2}\,\ln \sqrt{-\CG} + I^\mu\,\tilde{x}_\mu \,. 
\end{align}
This solves the equations of motion of DFT for arbitrary parameters $c^\mu$\,, but they satisfy the strong constraint
\begin{align}
 \partial_P \cH_{MN}\,\partial^Pd=0\,,
\end{align}
only when $c^0=\pm c^1$\,. 

\medskip

In fact, this background has a distinctive feature that has not been observed before. 
According to the classification of \cite{Borsato:2016ose}, the condition for a YB-deformed background to be a standard supergravity background is the unimodularity condition. 
However, in this example, the background \eqref{eq:AdS3_M^P} satisfies the GSE even if we perform a rescaling $I^m \to \lambda\,I^m$ with arbitrary $\lambda\in\mathbb{R}$. 
In particular, by choosing $\lambda=0$, the background \eqref{eq:AdS3_M^P} without $I^m$ satisfies the usual supergravity equations of motion. 
As we explain below, the reason for the unusual behavior is closely related to the degeneracy of $(\CG \pm B)_{mn}$. 

\medskip

According to \cite{Sakatani:2016fvh}, the condition for a solution of the GSE to be a standard supergravity background is given by
\begin{align}
 \gLie_{\bm{Y}} \cH_{MN} = 0\,, \qquad 
 \cH_{MN}\,\bm{Y}^M\,\bm{Y}^N= \nabla_M \bm{Y}^M \,,
\label{eq:condition-GSE-to-SUGRA}
\end{align}
where $\nabla_M$ is the (semi-)covariant derivative in DFT and
\begin{align}
 \bm{X}^M \equiv \begin{pmatrix} I^m \\ 0 \end{pmatrix},\qquad 
 \bm{Y}^M \equiv \cH^M{}_N\,\bm{X}^N = \begin{pmatrix} -(\CG^{-1}\,B)^m{}_n\,I^n \\ (\CG-B\,\CG^{-1}\,B)_{mn}\,I^n \end{pmatrix} \,. 
\end{align}
In our example with $c^0=\pm c^1$, $(\CG\pm B)_{mn}\,I^n=0$ is satisfied, and this leads to $\bm{Y}^M = \pm \bm{X}^M$\,. 
Then, from the null and generalized Killing properties of $\bm{X}^M$
\begin{align}
 \cH_{MN}\,\bm{X}^M\,\bm{X}^N=0\,,\qquad 
 \gLie_{\bm{X}} \cH_{MN} = 0\,, \qquad 
 \nabla_M \bm{X}^M =0 \,,
\end{align}
the condition \eqref{eq:condition-GSE-to-SUGRA} is automatically satisfied, and our GSE solution is also a solution of the standard supergravity. 
If we regard the background \eqref{eq:AdS3_M^P} as a solution of supergravity, the strong constraint is satisfied for an arbitrary $c_\mu$ and it is not necessary to require $c^0=\pm c^1$. 

\subsubsection{\mbox{\boldmath $3.~r = \frac{1}{2}\,\bigl(\bar{a}^\mu D\wedge P_\mu + \bar{b}^\mu M_{01}\wedge P_\mu\bigr)$}}

As a more general class of $r$-matrices, let us consider
\begin{align}
 r = \frac{1}{2}\,\bigl(\bar{a}^\mu D\wedge P_\mu + \bar{b}^\mu M_{01}\wedge P_\mu\bigr) \,. 
\end{align}
The homogeneous CYBE requires
\begin{align}
 \bar{a}^0\,\bar{b}^1 - \bar{a}^1\,\bar{b}^0 = 0 \,,\qquad -\bar{a}^0\,\bar{b}^0 + a^1\,\bar{b}^1 = 0 \,,
\end{align}
and we consider a non-trivial solution
\begin{align}
 r = \frac{1}{2}\,\bigl(\bar{c}\,D + \bar{d}\,M_{01}\bigr)\wedge (P_0\pm P_1) \,. 
\end{align}
The non-unimodularity becomes
\begin{align}
 \cI = (c-d)\,(P_0\pm P_1) \qquad \bigl(c \equiv\eta\, \bar{c}\,,\quad d\equiv \eta\,\bar{d}\,\bigr)\,.
\end{align}
The corresponding $\beta$-deformed background is given by
\begin{align}
\begin{split}
 \rmd s^2&= \frac{\eta_{\mu\nu}\,\rmd x^\mu\,\rmd x^\nu \mp 2\,c \,z^{-1}\,(\rmd x^0 \mp \rmd x^1)\,\rmd z + \bigl[1\pm 2\,z^{-2}\,(c\pm d)(x^0\mp x^1)\bigr]\,\rmd z^2}{z^2 \pm 2\,(c\pm d)(x^0\mp x^1)} 
\\
 &\quad
 + \rmd s^2_{\rmS^3} + \rmd s_{\TT^4}^2 \,,\qquad 
 \Exp{-2\Phi} = \frac{z^2 \pm 2\,(c\pm d)(x^0\mp x^1)}{z^2}\,,
\\
 B_2 &= \frac{\rmd x^0\wedge\rmd x^1 + c \, z^{-1}\,(\rmd x^0\mp \rmd x^1) \wedge \rmd z}{z^2\pm 2\,(c\pm d)(x^0\mp x^1)} + \frac{1}{4}\,\cos\theta\,\rmd \phi \wedge \rmd \psi \,. 
\label{eq:D+M^P-pm}
\end{split}
\end{align}
By introducing a Killing vector,
\begin{align}
 I = (c-d)\,(\hat{P}_0\pm \hat{P}_1) = (c-d)\,(\partial_0\pm \partial_1) \,,
\end{align}
this background becomes a solution of the GSE. 
In particular, when $c=d$, this becomes a supergravity background. 

\medskip

Similar to the previous example, the Killing vector again satisfies $(\CG\pm B)_{mn}\,I^n=0$, and even if we rescale the Killing vector as $I^m\to \lambda\,I^m$, this is still a solution of the GSE. 
As a particular case $\lambda=0$, the background \eqref{eq:D+M^P-pm} becomes a solution of the usual supergravity.

\subsection{$\beta$-deformations with generalized isometries}
\label{sec:generalized-isometries}

In the previous subsections, we have not considered the special conformal generators $\hat{K}_{\mu}$\,. 
As in the case of $\AdS5\times\rmS^5$ background, if there is no $B$-field, we can obtain various solutions from $\beta$-deformations using $\hat{K}_{\mu}$. 
However, in the $\AdS3\times \rmS^3\times \TT^4$ background, we cannot naively use $\hat{K}_\mu$ according to $\Lie_{\hat{K}_{\mu}}B_2\neq 0$\,. 
Indeed, even for a simple Abelian $r$-matrix, such as $r=\frac{1}{2}\,K_0\wedge K_1$ or $r=\frac{1}{2}\,K_+\wedge P_+$, the $\beta$-deformed background does not satisfy the supergravity equations of motion. 
In this subsection, we explain how to utilize the special conformal generators, and obtain several solutions from (generalization of) $\beta$-deformations. 

\medskip

In the canonical section $\tilde{\partial}^m=0$, if there exists a pair $(v^m,\,\tilde{v}_m)$ satisfying
\begin{align}
 \Lie_v \CG_{mn} = 0 \,,\qquad \Lie_v B_2 + \rmd \tilde{v}_1 = 0\,,\qquad \Lie_v \Phi =0 \,,
\end{align}
it means that the background admits a generalized Killing vector $(V^M)=(v^m,\,\tilde{v}_m)$ satisfying
\begin{align}
 \Lie_V \cH_{MN} = 0\,,\qquad \Lie_V d = 0\,. 
\end{align}
Then, the equation \eqref{eq:K-B2-closed} shows that there exist generalized Killing vectors $\hat{\mathsf{K}}_\mu^M$ associated with the Killing vectors $\hat{K}_\mu^m$\,. 
Since a generalized vector of the form $V^M=\partial^M f(x)$, which we call a trivial Killing vector, is always a generalized Killing vector, there is ambiguity in the definition of the generalized Killing vector. 
Using the ambiguity, we can find a set of generalized Killing vectors $\hat{\mathsf{T}}_i=(\hat{\mathsf{T}}_i^M)$ that satisfy
\begin{align}
 \gLie_{\hat{\mathsf{T}}_i}\cH_{MN} = \gLie_{\hat{\mathsf{T}}_i}d =0\,,\qquad 
 \gLie_{\hat{\mathsf{T}}_i}\hat{\mathsf{T}}_j^M + \gLie_{\hat{\mathsf{T}}_j}\hat{\mathsf{T}}_i^M =0 \,,
\end{align}
as well as the conformal algebra $\alg{so}(2,2)$ by means of the C-bracket
\begin{align}
 [V,\,W]_{\rmC}^M \equiv \frac{1}{2}\, (\gLie_V W - \gLie_W V)^M \,. 
\end{align}
Note that, according the requirement $\gLie_{\hat{\mathsf{T}}_i}\hat{\mathsf{T}}_j^M + \gLie_{\hat{\mathsf{T}}_j}\hat{\mathsf{T}}_i^M =0$, the C-bracket coincides with the D-bracket, $[V,\,W]_{\rmD}^M\equiv \gLie_V W^M$. 
We can find the following set of generalized Killing vectors:
\begin{align}
\begin{split}
 &\hat{\mathsf{D}} \equiv x^+\,\partial_+ + x^-\,\partial_- + z\,\partial_z \,, \qquad
 \hat{\mathsf{P}}_+ \equiv \partial_+\,, \qquad
 \hat{\mathsf{P}}_- \equiv \partial_-\,, 
\\
 &\hat{\mathsf{M}}_{+-} = x^+\,\partial_+ - x^-\,\partial_- + z^{-1}\,\tilde{\partial}^z \,,
\\
 &\hat{\mathsf{K}}_+ = z^2\,\partial_+ + 2\,(x^-)^2\,\partial_- + 2\,x^-\,z\,\partial_z 
           + 2\,\tilde{\partial}^- -\frac{2\, x^-}{z}\,\tilde{\partial}^z \,,
\\
 &\hat{\mathsf{K}}_- = 2\,(x^+)^2\,\partial_+ + z^2\,\partial_- + 2\, x^+\, z\,\partial_z 
           -2\,\tilde{\partial}^+ + \frac{2\,x^+}{z}\,\tilde{\partial}^z \,,
\end{split}
\end{align}
which satisfy
\begin{align}
 \eta_{MN}\,\hat{\mathsf{K}}_\pm^M\, \hat{\mathsf{P}}_\mp^N = \pm 2 \,, \qquad 
 \eta_{MN}\,\hat{\mathsf{D}}^M\, \hat{\mathsf{M}}_{+-}^N = 1\,, \qquad 
 \eta_{MN}\,\hat{\mathsf{T}}_i^M\,\hat{\mathsf{T}}_j^N = 0 \quad (\text{others}) \,. 
\end{align}
If we could find generators $\hat{\mathsf{T'}}_i$ which satisfy
\begin{align}
 \eta_{MN}\,\hat{\mathsf{T'}}_i^M\,\hat{\mathsf{T'}}_j^N = 0\,, 
\end{align}
they are on a common $D$-dimensional section, and we can find a duality frame where the generalized Killing vectors take the form $(\mathsf{T'}_i^M) =(\mathsf{T'}_i^m,\,0)$. 
If it is possible, the generalized Killing vectors reduces to the usual Killing vector and we can consider the usual $\beta$-deformations in such duality frame. 
However, it seems unlikely to be the case in the $\AdS3\times \rmS^3\times \TT^4$ background, and in the following, we employ the above set of generalized Killing vectors. 

\subsubsection{\mbox{\boldmath $1.~r=\frac{1}{8}\,\mathsf{K}_+\wedge \mathsf{P}_+$}}

Let us first consider an Abelian $r$-matrix $r=\frac{1}{8}\,\mathsf{K}_+\wedge \mathsf{P}_+$ associated with the Abelian generalized isometries; $[\hat{\mathsf{K}}_+,\, \hat{\mathsf{P}}_+]_{\rmC}=0$. 
Since $\hat{\mathsf{K}}_+$ has the dual components, it is not clear how to perform a ``$\beta$-deformation.'' 
We thus change the generalized coordinates such that the dual components disappear. 

\medskip

We here employ the simple coordinate transformation law by Hohm and Zwiebach \cite{Hohm:2012gk}. 
Namely, under a generalized coordinate transformation $x^M\to x'^M$, the generalized tensors are transformed as
\begin{align}
\begin{split}
 &\cH'_{MN}(x') = \cF_M{}^K(x',x)\,\cF_N{}^L(x',x)\,\cH_{KL}(x) \,,\qquad 
 \Exp{-2d'(x')} = \Bigl\lvert\det\frac{\partial x^M}{\partial x'^N}\Bigr\rvert\,\Exp{-2d'(x')}\,,
\\
 &\cF_M{}^N(x',x)\equiv \frac{1}{2}\,\biggl(\frac{\partial x'_M}{\partial x_P}\,\frac{\partial x^N}{\partial x'^P}+\frac{\partial x^P}{\partial x'^M}\,\frac{\partial x'_P}{\partial x_N}\biggr)\,. 
\end{split}
\end{align}
We can easily check that a generalized coordinate transformation
\begin{align}
 \tilde{z}' = \tilde{z} + \frac{\ln x^-}{z}\,, \qquad x'^M = x^M \quad (\text{others})\,,
\label{eq:gen-coord-trsf}
\end{align}
indeed removes the dual components; $(\mathsf{K'}_+^M)=(\hat{K'}_+^m,\,0)$ and $(\mathsf{P'}_+^M)=(\hat{P'}_+^m,\,0)$. 
In fact, this transformation $\cH'_{MN} = \cF_M{}^K\,\cF_N{}^L\,\cH_{KL}$ with
\begin{align}
 (\cF_M{}^N) = \begin{pmatrix} \bm{1_{10}} & \bmq_{mn} \\ 0 & \bm{1_{10}} \end{pmatrix}\,,\qquad \bmq_{-z}=-\bmq_{z-} = -\frac{1}{x^-\,z}\,,
\end{align}
is precisely a $B$-field gauge transformation,
\begin{align}
 B_2 \ \to \ B_2 - \frac{\rmd x^- \wedge \rmd z}{x^-\,z} \,. 
\end{align}
In the transformed background, the $B$-field is shifted
\begin{align}
\begin{split}
 \rmd s^2&= \frac{-2\,\rmd x^+\,\rmd x^- + \rmd z^2}{z^2} + \rmd s^2_{\rmS^3} + \rmd s_{\TT^4}^2 \,,
\\
 B_2&= \frac{\rmd x^- \wedge (x^-\,\rmd x^+ - z\,\rmd z)}{x^-\,z^2} + \frac{1}{4}\,\cos\theta\,\rmd \phi \wedge \rmd \psi \,, 
 \qquad \Exp{-2\Phi}=1\,,
\end{split}
\end{align}
we can check the isometries
\begin{align}
 \Lie_{\hat{K}_+}\CG_{mn}=\Lie_{\hat{K}_+}B_{mn}=\Lie_{\hat{K}_+}\Phi=0\,,\qquad 
 \Lie_{\hat{P}_+}\CG_{mn}=\Lie_{\hat{P}_+}B_{mn}=\Lie_{\hat{P}_+}\Phi=0\,.
\end{align}
Then, we can perform the usual $\beta$-deformation associated with $r=\frac{1}{8}\,K_+\wedge P_+$,
\begin{align}
\begin{split}
 \rmd s^2&= \frac{-2\,\rmd x^+\,\rmd x^- + \eta\,\rmd x^-\,(\rmd x^- - 2\,x^-\,z^{-1}\,\rmd z)}{z^2 + \eta\,(x^-)^2} + \frac{\rmd z^2}{z^2} + \rmd s^2_{\rmS^3} + \rmd s_{\TT^4}^2 \,,
\\
 B_2&= \frac{\rmd x^- \wedge (x^-\,\rmd x^+ - z\,\rmd z)}{x^-\,[z^2 + \eta\,(x^-)^2]} + \frac{1}{4}\,\cos\theta\,\rmd \phi \wedge \rmd \psi \,, 
 \qquad \Exp{-2\Phi}=\frac{z^2 + \eta\,(x^-)^2}{z^2}\,. 
\end{split}
\end{align}
Finally, we go back to the original coordinates, $\cH_{MN} = (\cF^{-1})_M{}^K\,(\cF^{-1})_N{}^L\,\cH_{KL}$, and obtain
\begin{align}
\begin{split}
 \rmd s^2&= \frac{-2\,\rmd x^+\,\rmd x^- + \eta\,\rmd x^-\,(\rmd x^- - 2\,x^-\,z^{-1}\,\rmd z)}{z^2 + \eta\,(x^-)^2} + \frac{\rmd z^2}{z^2} + \rmd s^2_{\rmS^3} + \rmd s_{\TT^4}^2 \,,
\\
 B_2&= \frac{\rmd x^- \wedge \rmd x^+ + \eta\,x^-\,z^{-1}\,\rmd x^- \wedge \rmd z}{z^2 + \eta\,(x^-)^2} + \frac{1}{4}\,\cos\theta\,\rmd \phi \wedge \rmd \psi \,, 
 \qquad \Exp{-2\Phi}=\frac{z^2 + \eta\,(x^-)^2}{z^2}\,. 
\end{split}
\label{eq:K_+wP_+}
\end{align}
This is a new solution of the usual supergravity. 

\subsubsection{General procedure}

In general, it is not easy to find a generalized coordinate transformation like \eqref{eq:gen-coord-trsf}, which removes the dual components of the generalized Killing vectors. 
However, in fact, it is not necessary to find such a coordinate transformation. 
As it is clear from the above procedure, for an $r$-matrix, $r=\frac{1}{2}\,r^{ij}\,\mathsf{T}_i\wedge \mathsf{T}_j$, associated with the generalized Killing vectors, the previous deformation is simply a transformation\cite{Sakamoto:2018krs,Araujo:2018rho}
\begin{align}
 \cH_{MN} \ \to \ \cH'_{MN} = h_M{}^K\,h_N{}^L\,\cH_{KL}\,,\qquad h_{M}{}^N =(e^{-\eta\,r^{ij}T_{ij}})_M{}^N\,,
\label{eq:general-Odd}
\end{align}
where we defined 
\begin{align}
(T_{ij}) _{M}{}^N \equiv\hat{\mathsf{T}}_{iM}\,\hat{\mathsf{T}}_j^N-\hat{\mathsf{T}}_{jM}\,\hat{\mathsf{T}}_i^N
=-(T_{ji})_{M}{}^{N}\,.
\end{align}
We can easily see that the transformation matrix $h_{M}{}^N$ is an $O(D,D)$ matrix. 
In general, this $O(D,D)$ transformation is a combination of a $\beta$-transformation and diffeomorphisms, but in particular, when all of $\hat{\mathsf{T}}_{i}$ do not have the dual components, this $h_{M}{}^N$ reduces to the usual $\beta$-transformation matrix. 
We can easily check that the above solution \eqref{eq:K_+wP_+} can be obtained from the original background in the single step \eqref{eq:general-Odd}. 

\medskip

When we consider a non-unimodular $r$-matrix, we suppose that the formula \eqref{eq:cI-formula} will be correct in a duality frame where $\hat{\mathsf{T}}_{i}$ take the form $(\hat{\mathsf{T}}_{i}^M)=(\hat{T}_{i}^m,\,0)$. 
Then, the deformed background will be a solution of modified DFT (mDFT) \cite{Sakatani:2016fvh} with
\begin{align}
 \bm{X}^M=\hat{\cI}^M \equiv \begin{pmatrix} \hat{\cI}^m\\ \hat{\cI}_m \end{pmatrix} 
 \equiv \eta\,r^{ij}\,[\hat{\mathsf{T}}_i,\,\hat{\mathsf{T}}_j]_{\rmC}^M \,. 
\end{align}
In terms of the GSE, it is a solution with $I^m=\hat{\cI}^m$ and $Z_m=\partial_m\Phi +I^n\,B_{nm} + \hat{\cI}_m$\,. 

\subsubsection{\mbox{\boldmath $2.~r=\frac{1}{2}\,\mathsf{K}_+\wedge \mathsf{K}_-$}}

For an Abelian $r$-matrix $r=\frac{1}{2}\,\mathsf{K}_{+}\wedge \mathsf{K}_-$, we do not find a generalized coordinate system where dual components of both $\hat{\mathsf{K}}_+^M$ and $\hat{\mathsf{K}}_-^M$ vanish. 
However, from the general procedure \eqref{eq:general-Odd}, we can easily obtain the deformed background
\begin{align}
\begin{split}
 \rmd s^2 &= \frac{-2\,\rmd x^+\,\rmd x^- + \rmd z^2 + 2\,\eta\,[2\,(x^-\,\rmd x^+ + x^+ \,\rmd x^-) - (2\,x^+\,x^- + z^2)\,\frac{\rmd z}{z}]^2}{z^2+2\,\eta\,(z^2- 2\,x^- \, x^+)^2} 
\\
 &\quad+ \rmd s^2_{\rmS^3} + \rmd s_{\TT^4}^2\,, \qquad
 \Exp{-2\Phi} =\frac{z^2+2\,\eta\,(2\,x^+\,x^- -z^2)^2}{z^2}\,,
\\
 B_2&= \frac{\rmd x^-\wedge \rmd x^+ -4\,\eta\,(2\,x^+\,x^- - z^2) (\rmd x^- \wedge\rmd x^+ + (x^-\,\rmd x^+ - x^+\,\rmd x^-)\wedge\frac{\rmd z}{z}}{z^2+2\,\eta\,(z^2- 2\,x^- \, x^+)^2} 
\\
 &\quad + \frac{1}{4}\,\cos\theta\,\rmd \phi \wedge \rmd \psi \,. 
\end{split}
\end{align}
We can easily see that this is a solution of the usual supergravity. 

\subsubsection{\mbox{\boldmath $3.~r=\frac{1}{2}\,\mathsf{M}_{+-}\wedge \mathsf{K}_+$ or $r=\frac{1}{2}\,\mathsf{D}\wedge \mathsf{K}_+$}}

Finally, we consider a non-unimodular $r$-matrix $r=\frac{1}{2}\,\mathsf{M}_{+-}\wedge \mathsf{K}_+$, satisfying
\begin{align}
 \cI = \eta\, [\mathsf{M}_{+-},\,\mathsf{K}_+] = \eta\,\mathsf{K}_+ \,.
\end{align}
In this case, the deformed background
\begin{align}
 \rmd s^2 &= \frac{-2\,\rmd x^-\,\rmd x^+ + \rmd z^2 - 2\,\eta\, (\rmd x^- - x^-\,\frac{\rmd z}{z})\,[2\,x^+\,\rmd x^- + 2\,x^-\, \rmd x^+ - (2\,x^+\, x^- + z^2)\,\frac{\rmd z}{z}]}{z^2 - 2\,\eta\,x^-\, (2 x^- x^+ - z^2)}
\nn\\
 &\quad+ \rmd s^2_{\rmS^3} + \rmd s_{\TT^4}^2\,, \qquad
 \Exp{-2\Phi} = \frac{z^2-2\,\eta\,x^-\,(2\,x^+\,x^- -z^2)}{z^2} \,,
\nn\\
 B_2&= \frac{1}{4}\,\cos\theta\,\rmd \phi \wedge \rmd \psi
\nn\\
 &\quad + \frac{(1+2\,\eta\,x^-)\,\rmd x^-\wedge \rmd x^+ + \eta\, [\,2\,(x^-)^2\,\rmd x^+ - (4\,x^+\,x^- - z^2)\,\rmd x^-\,]\wedge \frac{\rmd z}{z}}{z^2 - 2\,\eta\,x^-\, (2\,x^+\,x^- -z^2)} \,,
\label{eq:MwedgeK+}
\end{align}
satisfies the equations of motion of mDFT with $\bm{X}^M =\eta\,\hat{\mathsf{K}}_+^M$\,. 

\medskip

Similar to the example studied in section \ref{sec:AdS3-S3-T4MwedgeP}, we can freely rescale $\bm{X}^M$ as $\bm{X}^M\to \lambda\,\bm{X}^M$ $(\lambda\in\mathbb{R})$, and in a particular case $\lambda=0$, \eqref{eq:MwedgeK+} can be regarded as a solution of the usual supergravity. 

\medskip

Interestingly, we can obtain the same background also by considering an $r$-matrix $r=\frac{1}{2}\,\mathsf{D}\wedge \mathsf{K}_+$, satisfying $\cI = \eta\, [\mathsf{D},\,\mathsf{K}_+] = -\eta\,\mathsf{K}_+$\,.
This also may be related to the degeneracy of $(\CG \pm B)_{mn}$ in the $\AdS3\times \rmS^3\times \TT^4$ background.

\subsubsection{Short summary}

Let us summarize this subsection. 
Usually, we prepare a bi-vector $\hat{r}=\frac{1}{2}\,r^{ij}\,\hat{T}_i\wedge \hat{T}_j$ satisfying the homogeneous CYBE (or the Poisson condition)
\begin{align}
 [\hat{r},\,\hat{r}]_{\rmS} \equiv r^{ij}\,r^{kl}\, [\hat{T}_i,\,\hat{T}_k]\wedge T_j\wedge T_l
 = - r^{ij}\,r^{kl}\,f_{ik}{}^m\,\hat{T}_m\wedge \hat{T}_j\wedge \hat{T}_l = 0 \,,
\end{align}
where $[\cdot,\cdot]_{\rmS}$ is the Schouten bracket, and perform a local $\beta$-transformation
\begin{align}
 h_M{}^N = \begin{pmatrix} \bm{{1_{10}}} & \bm{{0_{10}}} \\ \bmr^{mn}(x) & \bm{{1_{10}}} \end{pmatrix},\qquad
 \bmr^{mn}(x) = -2\,\eta\,r^{ij}\,\hat{T}_i^m\, \hat{T}_j^n \,. 
\end{align}
In this subsection, in order to allow for the non-standard Killing vectors $\hat{K}_{\mu}$ satisfying \eqref{eq:K-B2-closed}, we have generalized the Killing vectors $\hat{T}_i$ into the generalized Killing vectors $\hat{\mathsf{T}}_i$\,. 
The generalized Killing vectors $\hat{\mathsf{T}}_i$ are defined such that their C-bracket satisfy the same commutation relations as those of $\hat{T}_i$\,. 
The homogeneous CYBE is generalized by replacing the usual Lie bracket with the C-bracket, and performing $O(D,D)$ transformations
\begin{align}
\begin{split}
 h_{M}{}^N =(e^{-\eta\,r^{ij}T_{ij}})_M{}^N\,,\qquad
(T_{ij}) _{M}{}^N =\hat{\mathsf{T}}_{iM}\,\hat{\mathsf{T}}_j^N-\hat{\mathsf{T}}_{jM}\,\hat{\mathsf{T}}_i^N\,,
\end{split}
\end{align}
we have obtained several new solutions of DFT. 
In particular, when all of the generalized Killing vectors $\hat{\mathsf{T}}_{i}^M$ do not have the dual components, this generalized transformation reduces to the usual local $\beta$-transformations. 


\chapter{$T$-folds from YB deformations}
\label{Ch:YB-T-fold}

In this chapter, we will concentrate on YB deformations of Minkowski and 
$\AdS5\times \rmS^5$ backgrounds,
and show that the deformed backgrounds we consider here 
belong to a specific class of non-geometric backgrounds, called $T$-folds \cite{Hull:2004in}. 
Moreover, it is worth noting that our examples have an intriguing feature that 
the R-R fields are also twisted by the $T$-duality monodromy, in comparison to 
the well-known $T$-folds which include no R-R fields.

\medskip 

This chapter is organized as follows. 
Section \ref{sec:T-fold} provides a brief review of $T$-folds, including two well-known examples in the literature. 
In Section \ref{sec:Non-geometric-YB},
we consider YB deformed Minkowski and $\AdS5\times \rmS^5$ backgrounds, 
and argue that these deformed backgrounds are regarded as $T$-folds. 
The deformed Minkowski backgrounds discussed in Section \ref{sec:non-geometry-Minkowski} can be reproduced by applying the modified Penrose limit to the deformed $\AdS5\times \rmS^5$ backgrounds.
The modified Penrose limit is explained in Appendix \ref{sec:Penrose-limit}.
In addition, we study a solution of GSE that is obtained by a non-Abelian $T$-duality 
but not as a YB deformation, and show that this can also be regarded as a $T$-fold.

\section{A brief review of $T$-folds}
\label{sec:T-fold}

In this subsection, we briefly explain what is $T$-fold.
A $T$-fold is supposed to be a generalization of the usual manifold. 
It locally looks like a Riemannian manifold, but which is glued together not just 
by diffeomorphisms but also by $T$-duality. 
It plays a significant role in studying non-geometric fluxes 
beyond the effective supergravity description.
As illustrative examples, 
we revisit two well-known cases in the literature, corresponding to 
a chain of duality transformations \cite{Kachru:2002sk,Shelton:2005cf} 
and to the codimension-1 $5_2^2$-brane solution \cite{Hassler:2013wsa}. 

\medskip 

It is conjectured that string theories are related by some discrete dualities. 
One thing that can occur is that, by duality transformations, a flux configuration transforms 
into a non-geometric flux configuration, which means that it cannot be realized in terms of 
the usual fields in 10/11-dimensional supergravities. Therefore, dualities suggest that 
we need to go beyond the usual geometric isometries to fully understand 
the arena of flux compactifications.

\medskip 

For the case of $T$-duality, one proposal to address this problem is the so-called doubled formalism. 
This construction consists of a manifold in which all the local patches are geometric. 
However, the transition functions that are needed to glue these patches not only include usual diffeomorphisms and gauge transformations, but also $T$-duality transformations.

\medskip 

$T$-fold backgrounds are formulated in an enlarged space with a $T^n\times \tilde T^n$ fibration. The tangent space is the doubled torus $T^n\times \tilde T^n$ and is described by a set of coordinates $Y^M=(y^m, y_m)$ which transforms in the fundamental representation of $\OO(n,n)$. 
The physical internal space arises as a particular choice of a subspace of the double torus, $T^n_{\text{phys}}\subset T^n\times \tilde T^n$. 
Then $T$-duality transformations $\OO(n,n;\mathbb{Z})$ act by changing the physical subspace $T^n_{\text{phys}}$ to a different subspace of the enlarged $T^n \times \tilde T^n$. 
For a geometric background, we have a spacetime which is a geometric bundle, $T^n_{\text{phys}}=T^n$.\footnote{We can also have $T^n_{\text{phys}}=\tilde T^n$, which corresponds to a dual geometric description. 
} 
Nevertheless non-geometric backgrounds do not fit together to form a conventional manifold. 
That is to say, despite of they are locally well-defined, their global description is not valid. 
Instead, they are globally well-defined as $T$-folds.

\medskip 

This formulation is manifestly invariant under the $T$-duality group $O(n,n;\mathbb{Z})$. 
However, to make contact with the conventional formulation, one needs to choose a polarization, 
\emph{i.e.}, a particular choice of $T^n_{\text{phys}}\subset T^n\times \tilde T^n$. 
This means that we have to break the $\OO(n,n;\mathbb{Z})$ and pick $n$ coordinates out of 
the $2n$ coordinates $(y^m, \tilde y_m)$. Then, $T$-duality transformations allow to identify 
the backgrounds that belong to the same physical configuration or duality orbit and just differ 
on a choice of polarization\footnote{These orbits have been determined in terms of a classification of gauged supergravities in \cite{Dibitetto:2012rk}.}.

%

\medskip 

Let us now review some examples of $T$-folds 
that have been studied in the literature.

\subsection{A toy example}

We start by reviewing a toy model example that involves several duality transformations 
of a given background. 
This example has been discussed in \cite{Kachru:2002sk,Shelton:2005cf}. 
Before introducing a $T$-fold example, we will present geometric cases 
like a twisted torus and a torus with $H$-flux as simple exercises. 

\subsubsection{Twisted torus}

Let us consider the metric of a twisted torus,
\begin{align}
 \rmd s^2 
 = 
 \rmd x^2 + \rmd y^2 + (\rmd z-m\,x\,\rmd y)^2 
 \, ,
 \qquad (m\in\mathbb{Z})
 \,.
\label{eq:twisted-torus-metric}
\end{align}
Note that this is not a supergravity solution for $m\neq 0$, 
but still is a useful example to reveal a non-geometric global property. 
As this background has isometries along $y$ and $z$ directions, 
these directions can be compactified with certain boundary conditions. For example, let us take
\begin{align}
 (x,\,y,\,z) \sim (x,\,y+1,\,z)\,,\qquad 
 (x,\,y,\,z) \sim (x,\,y,\,z+1)\,. 
\end{align}
Apparently, there is no isometry along the $x$ direction, 
but there actually exists a deformed Killing vector,
\begin{align}
 k = \partial_x + m\,y\,\partial_z \,. 
\end{align}
Thus, this isometry direction can be compactified as
\begin{align}
 (x,\,y,\,z) \sim \Exp{k}(x,\,y,\,z)=(x+1,\,y,\,z+m\,y)\,.
\label{eq:twisted-torus-identification}
\end{align}
According to this identification, a 1-form $e_z\equiv \rmd z-m\,x\,\rmd y$ 
is globally well-defined \cite{Kachru:2002sk}, 
and the metric \eqref{eq:twisted-torus-metric} is also globally well-defined. 

\medskip 

When this background is regarded as a 2-torus $T^2_{y,z}$ fibered over a base $\rmS^1_x$\,, 
the metric of the 2-torus takes the form
\begin{align}
 (\CG_{mn}) = 
 \begin{pmatrix}
 1 & -m\,x \\
 0 & 1
 \end{pmatrix}
 \begin{pmatrix}
 1 & 0 \\
 0 & 1 
 \end{pmatrix}
 \begin{pmatrix}
 1 & 0 \\
 -m\,x & 1 
 \end{pmatrix} \,. 
\end{align}
Then, as one moves around the base $\rmS_x^1$ , the metric is transformed 
by a $\GL(2)$ rotation. That is to say, for $x\to x+1$, the metric is given by
\begin{align}
 \CG_{mn}(x+1) = \bigl[\Omega^\rmT\,\CG(x)\,\Omega\bigr]_{mn} \,,\qquad
 \Omega^m{}_n \equiv
 \begin{pmatrix}
 1 & 0 \\
 -m & 1 
 \end{pmatrix} \,. 
\end{align}
This monodromy twist can be compensated by a coordinate transformation
\begin{align}
 y = y'\,, \qquad z = z'+m\,y'\,.
\end{align}
Thus the metric is single-valued up to the above coordinate transformation. 
Then this background can be understood to be geometric 
because general coordinate transformations belong to the gauge group of supergravity.

\subsubsection{Torus with $H$-flux}

When a $T$-duality is formally performed on the twisted torus \eqref{eq:twisted-torus-metric} 
along the $x$ direction, we obtain the following background 
\begin{align}
 \rmd s^2 = \rmd x^2+\rmd y^2+\rmd z^2\,,\qquad B_2=-m\,x\,\rmd y\wedge\rmd z\,,
\end{align}
equipped with the $H$-flux,
\begin{align}
 H_3=\rmd B_2 = -m\,\rmd x\wedge\rmd y\wedge\rmd z\,.
\end{align}
If we consider the generalized metric \eqref{eq:H-geometric} on the doubled torus 
$(y,z,\tilde{y},\tilde{z})$ associated to this background, 
then we can easily identify the induced monodromy when $x \to x+1$. 
In this case, the monodromy matrix is given by
\begin{align}
 \cH_{MN}(x+1) = \bigl[\Omega^\rmT\,\cH(x)\,\Omega\bigr]_{MN}\,,\qquad 
 \Omega^M{}_N = \begin{pmatrix}
 \delta^m_n & 0 \\
 2\,m\,\delta_{[m}^y\,\delta_{n]}^z & \delta_m^n
 \end{pmatrix} \in \OO(2,2;\mathbb{Z})\,.
\end{align}
Then, the induced monodromy can be compensated by a constant shift in the $B$-field,
\begin{align}
 B_{yz} ~~\to~~ B_{yz}-m. 
\end{align}
This shift transformation, which makes the background single-valued, 
belongs to the gauge transformations of supergravity. 
Hence we conclude that the background is geometric.

\subsubsection{$T$-fold}

Finally, let us perform another $T$-duality transformation along the $y$-direction 
on the twisted torus \eqref{eq:twisted-torus-metric}. 
Then we obtain the following background \cite{Kachru:2002sk}:
\begin{align}
 \rmd s^2 = \rmd x^2 + \frac{\rmd y^2+\rmd z^2}{1+m^2\,x^2} \,, 
\qquad B_2= \frac{m\,x}{1+m^2\,x^2}\,\rmd y\wedge\rmd z\,. 
 \label{eq:H-Q-flux}
\end{align}
In this case, neither general coordinate transformations nor $B$-field gauge transformations 
are enough to remove the multi-valuedness of the background. 
This can also be seen by calculating the monodromy matrix. 
The associated generalized metric is given by
\begin{align}
 \cH(x)=
\begin{pmatrix}
 \delta_m^p & 0 \\
 -2\,m\,x\,\delta_y^{[m}\,\delta_z^{p]} & \delta^m_p
\end{pmatrix}
\begin{pmatrix}
 \delta_{pq} & 0 \\
 0 & \delta^{pq} 
\end{pmatrix}
\begin{pmatrix}
 \delta^q_n & 2\,m\,x\,\delta_y^{[q}\,\delta_z^{n]} \\
 0 & \delta_q^n
\end{pmatrix} \,. 
\label{eq:H-Q-flux-simple}
\end{align}
Then, we find that, upon the transformation $x\to x+1$, the induced monodromy is 
\begin{align}
 \cH_{MN}(x+1) = \bigl[\Omega^\rmT\,\cH(x)\,\Omega\bigr]_{MN} \,,
\qquad \Omega^M{}_N \equiv \begin{pmatrix}
 \delta^m_n & 2\,m\,\delta_y^{[m}\,\delta_z^{n]} \\
 0 & \delta_m^n
\end{pmatrix}\in \OO(2,2;\mathbb{Z})\,. 
\label{eq:monodromy-simple}
\end{align}
The present $\OO(2,2;\mathbb{Z})$ monodromy matrix $\Omega$ takes an upper-triangular form 
(called a $\beta$-transformation) which is not part of the gauge group of supergravity. 
Hence, to keep the background globally well defined, the transition functions that glue the local patches 
should be extended to the full set of $\OO(2,2;\mathbb{Z})$ transformations beyond 
general coordinate transformations and B-field gauge transformations. 
This is what happens to the $T$-fold case. 

\medskip

In summary, we conclude that a non-geometric background 
with a non-trivial $\OO(n,n;\mathbb{Z})$ monodromy transformation, 
such as a $\beta$-transformation, is a $T$-fold. 
The background \eqref{eq:H-Q-flux} is a simple example.

\medskip 

From a viewpoint of DFT, by choosing a suitable solution of the section condition, 
the $\beta$-transformations can be realized as the gauge symmetries. 
Indeed, the above $\OO(2,2;\mathbb{Z})$ monodromy matrix $\Omega$ can be canceled 
by a generalized coordinate transformation on the double torus coordinates $(y,z,\tilde y, \tilde z)$,
\begin{align}
 y=y'+m\,\tilde{z}'\,,\qquad z=z'\,,\qquad \tilde{y}=\tilde{y}'\,,\qquad \tilde{z}=\tilde{z}' \,.
\label{eq:gen-coord-transf-simple}
\end{align}
In this sense, the twisted doubled torus is globally well-defined in DFT.

\medskip 

In addition, it is also possible to make the single-valuedness manifest by introducing the dual fields ($\OG_{mn}$\,, $\beta^{mn}$\,, $\tilde{\phi}$) \cite{Shapere:1988zv,Giveon:1988tt,Duff:1989tf,Tseytlin:1990nb,Giveon:1994fu} 
defined by (\ref{eq:H-non-geometric}) or (\ref{eq:relation-open-closed}), (\ref{eq:DFT-dilaton}).
The dual metric $\OG_{mn}$ is precisely the same as the open-string metric \cite{Seiberg:1999vs}, 
and the original metric $\CG_{mn}$ may be called the closed-string metric. 
In the non-geometric parameterization (\ref{eq:H-non-geometric}),
the background \eqref{eq:H-Q-flux-simple} becomes
\begin{align}
 \rmd s_{\text{dual}}^2 \equiv \OG_{mn}\,\rmd x^m\,\rmd x^n 
= \rmd x^2+\rmd y^2+\rmd z^2\,,\qquad \beta^{yz} = m\,x \,,
\end{align}
and the $\OO(2,2;\mathbb{Z})$ monodromy matrix \eqref{eq:monodromy-simple} corresponds to 
a constant shift in the $\beta$ field; $\beta^{yz}\to \beta^{yz} + m$\,. 
Namely, up to a constant $\beta$-shift, which is a gauge symmetry 
\eqref{eq:gen-coord-transf-simple} of DFT, the background becomes single-valued. 

\medskip 

We shall define a non-geometric $Q$-flux as \cite{Grana:2008yw}
\begin{align}
 Q_p{}^{mn} \equiv \partial_p \beta^{mn} \,.
\end{align}
Then, upon a transformation $x \to x+1$, the induced monodromy on the $\beta$-field is measured 
by an integral of the $Q$-flux,
\begin{align}
 \beta^{mn}(x+1)-\beta^{mn}(x) = \int_x^{x+1} \rmd x'^p\,\partial_p \beta^{mn}(x') 
= \int_x^{x+1} \rmd x'^p\,Q_p{}^{mn}(x') \,. 
\end{align}
This expression plays the central role in our argument. 

\medskip 

After this illustrative example we conclude that $Q$-flux backgrounds are globally well-defined 
as $T$-folds. In the next subsection, let us explain a codimension-1 example 
of the exotic $5_2^2$-brane by using the above $Q$-flux. 

\subsection{Codimension-1 $5^2_2$-brane background}

The second example is a supergravity solution studied in \cite{Hassler:2013wsa}. 
It is obtained by smearing the codimension-2 exotic $5^2_2$-brane solution 
\cite{LozanoTellechea:2000mc,deBoer:2010ud}, which is related to the NS5-brane solution 
by two $T$-duality transformations. 
It is also referred to as a $Q$-brane, as it is a source of $Q$-flux, as we are going to check. 
The codimension-1 version of this solution is given by
\begin{align}
\begin{split}
 \rmd s^2 &= m\, x\,(\rmd x^2+\rmd y^2) + \frac{x\,(\rmd z^2+\rmd w^2)}{m\,(x^2+z^2)} 
+ \rmd s_{\mathbb{R}^6}^2 \,,
\\
 B_2&= \frac{x}{m\,(x^2+z^2)}\rmd z\wedge\rmd w \,,\qquad \Phi 
= \frac{1}{2}\,\ln\biggl[\frac{x}{m\,(x^2+z^2)}\biggr] \,. 
\end{split}
\end{align}
With the non-geometric parameterization \eqref{eq:non-geometric-parameterization}, 
this solution is simplified as 
\begin{align}
\begin{split}
 \rmd s_{\text{dual}}^2 &= m\, x\,(\rmd x^2+\rmd y^2) 
+ \frac{\rmd z^2+\rmd w^2}{m\,x} + \rmd s_{\mathbb{R}^6}^2 \,, 
\\
 \beta^{zw} &=m\,y \,,\qquad \tilde{\phi} = \frac{1}{2}\,\ln \biggl[\frac{1}{m\,x}\biggr] \,. 
\end{split}
\end{align}
Assuming that the $y$ direction is compactified with $y\sim y+1$, 
the monodromy under $y\to y+1$ is given by a constant $\beta$-shift; 
\begin{align}
 \beta^{zw}\to \beta^{zw}+m\,.
\end{align}
As the background is twisted by a $\beta$-shift, this example can be considered as a $T$-fold. 
In terms of the $Q$-flux, this solution has a constant $Q$-flux,
\begin{align}
 Q_y{}^{zw} = m \,. 
\end{align}
Finally, the monodromy matrix is given by
\begin{align}
 \cH_{MN}(y+1) = \bigl[\Omega^\rmT\,\cH(y)\,\Omega\bigr]_{MN} \,,\qquad 
 \Omega^M{}_N \equiv \begin{pmatrix}
 \delta^m_n & 2\,m\,\delta_z^{[m}\,\delta_w^{n]} \\
 0 & \delta_m^n
\end{pmatrix}\in \OO(10,10;\mathbb{Z})\,. 
\end{align}

\medskip

By employing the knowledge on $T$-folds introduced in this section, 
we will elaborate on a non-geometric aspect of YB-deformed backgrounds as $T$-folds.

\section{Non-geometric aspects of YB deformations}
\label{sec:Non-geometric-YB}

Let us show that various YB-deformed backgrounds can be regarded as T-folds.

\medskip

In Subsec. \ref{sec:monodromy-YB},
the general structure of T-duality monodromy is revealed for the YB-deformed backgrounds studied in this section.
In Subsec. \ref{sec:non-geometry-Minkowski},
various T-folds are obtained as YB-deformations of Minkowski spacetime.
Then, in Subsec. \ref{sec:non-geometry-NATD},
we study a certain background which is obtained by a non-Abelian T-duality but is not described as a Yang-Baxter deformation.
It is shown that this background is a solution of GSE and can also be regarded as a T-fold.
Finally, in Subsec. \ref{sec:non-geometry-AdS5xS5},
in order to study a more non-trivial class of T-folds with
R-R fields, we consider some backgrounds obtained as YB-deformations of the $\AdS5\times \rmS^5$ background.

\subsection{$T$-duality monodromy of YB-deformed background}
\label{sec:monodromy-YB}

As we explained in section (\ref{sec:YB-beta-deform}),
the YB-deformed background described by $(\cH,\,d,F)$ always has the following structure:
\begin{align}
\begin{split}
 \cH &= \Exp{\bbeta^\rmT}\check{\cH}^{(0)}\Exp{\bbeta}\,,\qquad 
 d = d^{(0)} \,, \qquad 
 F = \Exp{-\beta\vee} \check{F}^{(0)} \,,
\\
\Exp{\bbeta}  &=  \begin{pmatrix}
 \delta^m_n & \beta^{mn} \\
 0 & \delta_m^n
 \end{pmatrix} \,,\qquad 
 \beta^{mn} = 2\,\eta\,r^{ij}\, \hat{T}_i^{m}\,\hat{T}_j^{n} \,,
\end{split}
\end{align}
where $(\check{\cH}^{(0)},\,d^{(0)},\,\check{F}^{(0)})$ represent the undeformed background. 
In the following examples, $B$-field vanishes in the undeformed background.
At this stage, we know only the local property of the YB-deformed background. 

\medskip

In the examples considered in this section, the bi-vector $\beta^{mn}$ always has a linear-coordinate dependence. 
Suppose that $\beta^{mn}$ depends on a coordinate $y$ linearly like,
\begin{align}
 \beta^{mn}=\mathsf{r}^{mn}\,y + \mathsf{\bar{r}}^{mn} \qquad (\mathsf{r}^{mn}:\text{ constant},\quad \mathsf{\bar{r}}^{mn}:\text{ independent of $y$}) \,,
\end{align}
and the $\beta$-untwisted fields are independent of $y$. 
Then, from the Abelian property,
\begin{align}
 \Exp{\bbeta_1+\bbeta_2}=\Exp{\bbeta1}\Exp{\bbeta_2}=\Exp{\bbeta_2}\Exp{\bbeta_1}\,,\qquad
 \Exp{-(\beta_1+\beta_2)\vee}=\Exp{-\beta_1\vee}\Exp{-\beta_2\vee}=\Exp{-\beta_2\vee}\Exp{-\beta_1\vee}\,,
\end{align}
we obtain
\begin{align}
\begin{split}
 \cH_{MN}(y+a) &= \bigl[\Omega_a^\rmT\cH(y)\,\Omega_a\bigr]_{MN} \,,\quad 
 d(y+a) = d(y) \,, \quad 
 F(y+a) = \Exp{-\omega_a\vee} F(y) \,,
\\
 (\Omega_a)^M{}_N &\equiv\begin{pmatrix}
 \delta^m_n & a\,\mathsf{r}^{mn} \\
 0 & \delta_m^n
 \end{pmatrix}\,,\qquad 
 (\omega_a)^{mn} \equiv a\,\mathsf{r}^{mn} \,.
\end{split}
\end{align}
If we can find out an $a_0$ (where the $\OO(10,10;\,\mathbb{R})$ matrix $\Omega_{a_0}$ 
is an element of $\OO(10,10;\mathbb{Z})$), 
the background allows us to compactify the $y$ direction as $y\sim y + a_0$\,. 
This is because $\OO(10,10;\mathbb{Z})$ is a gauge symmetry of String Theory and the background can be identified up to the gauge transformation. 
In this example of $T$-fold, the monodromy matrices for the generalized metric and R-R fields 
are $\Omega_{a_0}$ and $\Exp{-\omega_{a_0}\vee}$, respectively, 
while the dilaton $d$ is single-valued. 
Note that the R-R potential $A$ has the same monodromy as $F$\,.

\subsection{YB-deformed Minkowski backgrounds}
\label{sec:non-geometry-Minkowski}

In this subsection, we study YB-deformations of Minkowski spacetime 
\cite{Matsumoto:2015ypa,Borowiec:2015wua}. 
We begin by a simple example of the Abelian YB deformation. 
Then two purely NS-NS solutions of GSE are presented and are shown to be $T$-folds. 
These backgrounds have vanishing R-R fields and are the first examples of 
purely NS-NS solutions of GSE. 

\subsubsection{Abelian example}

Let us consider a simple Abelian $r$-matrix \cite{Matsumoto:2015ypa} 
\begin{align}
 r=-\frac{1}{2}\,P_1\wedge M_{23}\,. \label{Melvin-r}
\end{align}
The corresponding YB-deformed background becomes 
\begin{align}
 \rmd s^2&= -(\rmd x^0)^2+\frac{(\rmd x^1)^2+\bigl[1+(\eta\, x^2)^2\bigr]\,(\rmd x^2)^2 + \bigl[1+(\eta\, x^3)^2\bigr]\,(\rmd x^3)^2 + 2\,\eta^2\,x^2\,x^3\,\rmd x^2\,\rmd x^3}{1+\eta^2 \, \bigl[(x^2)^2+(x^3)^2\bigr]} 
\nn\\
 & \qquad +\sum_{i=4}^9(\rmd x^i)^2\,,
\nn\\
 B_2 &= \frac{\eta\,\rmd x^1\wedge \bigl(x^2\,\rmd x^3 - x^3\, \rmd x^2\bigr)}{1+\eta^2\, \bigl[(x^2)^2+(x^3)^2\bigr]}\,,\qquad
 \Phi = \frac{1}{2} \ln\biggl[\frac{1}{1+\eta^2\, \bigl[(x^2)^2+(x^3)^2\bigr]}\biggr]\,. 
 \label{Melvin}
\end{align}
It seems very messy, but after moving to an appropriate polar coordinate system 
(see Sec.\,3.1 of \cite{Matsumoto:2015ypa}), this background \eqref{Melvin} 
is found to be the well-known Melvin background \cite{Tseytlin:1994ei,Gibbons:1987ps,Hashimoto:2004pb}. 
In \cite{Matsumoto:2015ypa}, it was reproduced as a Yang-Baxter deformation 
with the classical $r$-matrix \eqref{Melvin-r}. 
For later convenience, we will keep the expression in \eqref{Melvin}. 

\medskip

The dual parameterization of this background is given by 
\begin{align}
 \rmd s^2_{\text{dual}} = -(\rmd x^0)^2 + \sum_{i=1}^9(\rmd x^i)^2 \,, \quad
 \beta = \eta\,\bigl(x^2 \,\partial_1\wedge \partial_3 -x^3\,\partial_1\wedge \partial_2 \bigr)\,, \quad 
 \tilde{\phi}=0 \,. 
\end{align}
Hence, under a shift $x^2\to x^2+\eta^{-1}$\,, the background receives the $\beta$-transformation,
\begin{align}
 \beta ~~\to~~ \beta + \partial_1\wedge \partial_3 \,. 
\end{align}
Therefore, if the $x^2$ direction is compactified with the period $\eta^{-1}$, 
then the monodromy matrix becomes
\begin{align}
 \cH_{MN}(x^2+\eta^{-1}) = \bigl[\Omega^\rmT\cH(x^2)\,\Omega\bigr]_{MN} \,,\qquad 
 \Omega^M{}_N \equiv \begin{pmatrix}
 \delta^m_n & 2\,\delta_1^{[m}\,\delta_3^{n]} \\
 0 & \delta_m^n \end{pmatrix} \in \OO(10,10;\mathbb{Z})\,. 
\end{align}
Thus this background has been shown to be a $T$-fold. 

\medskip

When the $x^3$ direction is also identified with the period $\eta^{-1}$, 
the corresponding monodromy matrix becomes
\begin{align}
 \cH_{MN}(x^3+\eta^{-1}) = \bigl[\Omega^\rmT\cH(x^3)\,\Omega\bigr]_{MN} \,,\qquad 
 \Omega^M{}_N \equiv \begin{pmatrix}
 \delta^m_n & - 2\,\delta_1^{[m}\,\delta_2^{n]} \\
 0 & \delta_m^n \end{pmatrix} \in \OO(10,10;\mathbb{Z})\,. 
\end{align}

\medskip 

In terms of non-geometric fluxes, this background has a constant $Q$-flux. 
In the examples of $T$-folds presented in Sec.\,\ref{sec:T-fold}, 
a background with a constant $Q$-flux, $Q_p{}^{mn}$, 
is mapped to another background with a constant $H$-flux, $H_{pmn}$, 
under a double $T$-duality along $x^m$ and $x^n$ directions. 
On the other hand, in the present example, the background has two types of constant $Q$-fluxes, $Q_2{}^{13}$ and $Q_3{}^{12}$\,, 
but we cannot perform $T$-dualities to make the background a constant-$H$-flux background 
because $x^2$ and $x^3$ directions are not isometry directions. 

\subsubsection{Non-unimodular example 1: \ $r = \frac{1}{2} \, (P_0-P_1) \wedge M_{01}$}
\label{sec:Minkowski-example1}

Let us consider a non-unimodular classical $r$-matrix\footnote{As far as we know, 
this example has not been discussed anywhere so far. } 
\begin{align}
 r = \frac{1}{2} \, (P_0-P_1) \wedge M_{01}\,.
\end{align}
The corresponding YB-deformed background becomes
\begin{align}
\begin{split}
 \rmd s^2 &= \frac{-(\rmd x^0)^2+(\rmd x^1)^2}{1- \eta^2\,(x^0+x^1)^2} + \sum_{i=2}^9(\rmd x^i)^2\,, 
\\ 
 B_2 &= -\frac{\eta\, (x^0+x^1)}{1-\eta^2\,(x^0+x^1)^2}\, \rmd x^0 \wedge \rmd x^1\,, \quad 
 \Phi = \frac{1}{2}\ln \biggl[\frac{1}{1-\eta^2\,(x^0+x^1)^2}\biggr]\,. 
\end{split}
\label{flatpen}
\end{align}
Apparently, this background has a coordinate singularity at $x^0+x^1=\pm 1/\eta$. 
But when the dual parameterization \eqref{eq:non-geometric-parameterization} is employed, 
the dual fields are given by 
\begin{align}
 \rmd s^2_{\text{dual}} = -(\rmd x^0)^2+\sum_{i=1}^9(\rmd x^i)^2 \,, \quad 
 \beta = \eta\,(x^0+x^1)\,\partial_0\wedge \partial_1 \,,\quad \tilde{\phi} = 0 \,,
\end{align}
and they are regular everywhere.%
\footnote{A similar resolution of singularities in the dual parameterization has been argued in \cite{Malek:2012pw,Malek:2013sp} in the context of the exceptional field theory.}

\medskip

By introducing a Killing vector $I$ with the help of the divergence formula \eqref{div-formula} as 
\begin{align}
 I = \tilde{\sfD}_{n} \beta^{mn}\,\partial_m = \partial_{n} \beta^{mn}\,\partial_m 
 = \eta\,(\partial_0 -\partial_1) \,,
\end{align}
the background \eqref{flatpen} with this $I$ solves GSE. 
Here $\tilde{\sfD}_{n} $ is the covariant derivative associated with the original Minkowski spacetime.

\medskip

Since the $\beta$-field depends on $x^1$ linearly, as one moves along the $x^1$ direction, 
the background is twisted by the $\beta$-transformation. 
In particular, when the $x^1$ direction is identified with period $1/\eta$, 
this background becomes a $T$-fold with an $\OO(10,10;\mathbb{Z})$ monodromy,
\begin{align}
 \cH_{MN}(x^1+\eta^{-1}) = \bigl[\Omega^\rmT\cH(x^1)\,\Omega\bigr]_{MN} \,,\qquad 
 \Omega^M{}_N\equiv \begin{pmatrix} \delta^m_n & 2\,\delta_0^{[m}\,\delta_1^{n]} \\ 
 0 & \delta_m^n \end{pmatrix} \,. 
\end{align}

\medskip

Note that an arbitrary solution of GSE can be regarded as a solution of DFT \cite{Sakamoto:2017wor}. 
Indeed, by introducing the light-cone coordinates and a rescaled deformation parameter as
\begin{align}
 x^\pm \equiv \frac{x^0 \pm x^1}{\sqrt{2}} \,,\qquad \bar{\eta} = \sqrt{2}\,\eta \,,
\end{align}
the present YB-deformed background can be regarded as the following solution of DFT:
\begin{align}
 \cH = \begin{pmatrix}
 0 & -1 & -\bar{\eta}\,x^+ & 0 \\
 -1 & 0 & 0 & \bar{\eta}\, x^+ \\
 -\bar{\eta}\, x^+ & 0 & 0 & (\bar{\eta}\,x^+)^2 -1 \\
 0 & \bar{\eta}\,x^+ & (\bar{\eta}\,x^+)^2 - 1 & 0 
\end{pmatrix}\,,\qquad d=\bar{\eta}\,\tilde{x}_-\,, 
\end{align}
where only $(x^+,x^-,\tilde{x}_+,\tilde{x}_-)$-components of $\cH_{MN}$ are displayed. 
Note here that the dilaton has an explicit dual-coordinate dependence 
because we are now considering a non-standard solution of the section condition 
which makes this background a solution of GSE rather than the usual supergravity. 

\medskip

Before perfoming this YB deformation (\emph{i.e.}~$\bar{\eta}=0$), 
there is a Killing vector $\chi \equiv\partial_+$\,, 
but the associated isometry is broken for non-zero $\bar{\eta}$\,. 
However, even after deforming the geometry, there exists a generalized Killing vector
\begin{align}
 \chi \equiv \Exp{\bar{\eta}\,\tilde{x}_-}\partial_+ \qquad 
 \bigl(\gLie_{\chi}\cH_{MN}=0\,,\quad \gLie_{\chi}d=0\bigr) \,,
\end{align}
which goes back to the original Killing vector in the undeformed limit, $\bar{\eta}\to 0$\,. 
In order to make the generalized isometry manifest, 
let us consider a generalized coordinate transformation,
\begin{align}
 x'^+ = \Exp{-\bar{\eta} \,\tilde{x}_-} x^+ \,,\qquad \tilde{x}'_- 
 = -\bar{\eta}^{-1}\Exp{-\bar{\eta}\,\tilde{x}_-}\,,\qquad 
 x'^M = x^M\quad (\text{others})\,. 
\end{align}
By employing Hohm and Zwiebach's finite transformation matrix \cite{Hohm:2012gk},
\begin{align}
 \cF_M{}^N = \frac{1}{2}\,\Bigl(\frac{\partial x^K}{\partial x'^M}\frac{\partial x'_K}{\partial x_N}
        +\frac{\partial x'_M}{\partial x_{K}}\frac{\partial x^N}{\partial x'^K}\Bigr) \,,
\end{align}
the generalized Killing vector in the primed coordinates becomes constant, $\chi = \partial'_+$\,. 
We can also check that the generalized metric in the primed coordinate system 
is precisely the undeformed background. Namely, at least locally, 
the YB deformation can be undone by the generalized coordinate 
transformation\footnote{In the study of YB deformations of AdS$_5$\,, 
the similar phenomenon has already been observed in \cite{Orlando:2016qqu}.}. 
This fact is consistent with the fact that YB deformations can be realized 
as the generalized diffeomorphism \cite{Sakamoto:2017cpu}. 

\paragraph*{Non-Riemannian background:} 
Since the above background has a linear coordinate dependence on $\tilde{x}_-$\,, 
let us rotate the solution to the canonical section 
(\emph{i.e.}~a section in which all of the fields are independent of the dual coordinates). 
By performing a $T$-duality along the $x^-$ direction, we obtain
\begin{align}
 \cH = \begin{pmatrix}
 0 & 0 & -\bar{\eta}\, x^+ & -1 \\
 0 & 0 & (\bar{\eta}\,x^+)^2-1 & \bar{\eta}\,x^+ \\
 -\bar{\eta}\,x^+ & (\bar{\eta}\,x^+)^2 -1 & 0 & 0 \\
 -1 & \bar{\eta}\,x^+ & 0 & 0 
\end{pmatrix}\,,\qquad 
 d=\bar{\eta}\,x^- \,.
\label{eq:non-Riemannian}
\end{align}
The resulting background is indeed a solution of DFT defined on the canonical section. 
However, this solution cannot be parameterized in terms of $(\CG_{mn},\,B_{mn})$ and 
is called a non-Riemannian background in the terminology of \cite{Lee:2013hma}. 
This background does not even allow the dual parameterization \eqref{eq:non-geometric-parameterization} in terms of $(\OG_{mn},\,\beta^{mn})$\footnote{For another example of non-Riemannian backgrounds, see \cite{Lee:2013hma}. A classification of non-Riemannian backgrounds in DFT has been made in \cite{Morand:2017fnv}. In the context of the exceptional field theory, non-Riemannian backgrounds have been found in \cite{Malek:2013sp} even before \cite{Lee:2013hma}. There, the type IV generalized metrics do not allow both the conventional and dual parameterizations similar to our solution \eqref{eq:non-Riemannian}.}.

\subsubsection{Non-unimodular example 2: \ $r=\frac{1}{2\sqrt{2}}\, \sum_{\mu=0}^4 \bigl(M_{0\mu}-M_{1\mu}\bigr) \wedge P^\mu$}
\label{sec:Minkowski-example2}

The next example is the classical $r$-matrix\cite{Borowiec:2015wua}, 
\begin{align}
 r=\frac{1}{2\sqrt{2}}\, \sum_{\mu=0}^4 \bigl(M_{0\mu}-M_{1\mu}\bigr) \wedge P^\mu\,. 
\end{align}
This classical $r$-matrix is a higher dimensional generalization of the light-cone 
$\kappa$-Poincar\'e $r$-matrix in the four dimensional one. 

\medskip

By using the light-cone coordinates,
\begin{align}
 x^\pm \equiv \frac{x^0 \pm x^1}{\sqrt{2}} \,,
\end{align}
the corresponding YB-deformed background becomes
\begin{align}
\begin{split}
 \rmd s^2&= \frac{-2\,\rmd x^+\, \rmd x^- - \eta^2\,\rmd x^+ \bigl[\sum_{i=2}^4(x^i)^2 \, \rmd x^+ - 2\,x^+ \sum_{i=2}^4x^i\,\rmd x^i \bigr]}{1-(\eta \, x^+)^2} + \sum_{i=2}^9(\rmd x^i)^2\,,
\\
 B_2&= \frac{\eta\,\rmd x^+\wedge \bigl(x^+\,\rmd x^- -\sum_{i=2}^4x^i\,\rmd x^i \bigr)}{1-(\eta \, x^+)^2} \,,\quad 
 \Phi = \frac{1}{2}\ln\biggl[\frac{1}{1-(\eta \, x^+)^2}\biggr] \,.
\end{split}
\label{eq:penHvT}
\end{align}
In terms of the dual parameterization, this background becomes
\begin{align}
\begin{split}
 &\rmd s^2_{\text{dual}} = -2\,\rmd x^+\,\rmd x^- + \sum_{i=2}^9(\rmd x^i)^2 \,,\quad \tilde{\phi}=0 \,, 
\\
 &\beta = \eta\,\sum_{\mu=0}^4 \hat{M}_{-\mu} \wedge \hat{P}^\mu = \eta\, \partial_-\wedge \bigl(x^+\,\partial_+ + {\textstyle\sum}_{i=2}^4\, x^i\, \partial_i \bigr) \,.
\end{split}
\end{align}
Again, by introducing a Killing vector from the divergence formula \eqref{div-formula} as 
\begin{align}
 I = 4\,\eta\,\partial_- \,, 
\end{align}
the background \eqref{eq:penHvT} with this $I$ solves GSE. 

\medskip

This background can also be regarded as the following solution of DFT:
\begin{align}
\begin{split}
 & \cH 
 = {\tiny\begin{pmatrix}
 0 & -1 & 0 & 0 & 0 & -\eta\,x^+ & 0 & -\eta\,x^2 & -\eta\,x^3 & -\eta\,x^4 \\
 -1 & 0 & 0 & 0 & 0 & 0 & \eta\,x^+ & 0 & 0 & 0 \\
 0 & 0 & 1 & 0 & 0 & 0 & -\eta\,x^2 & 0 & 0 & 0 \\
 0 & 0 & 0 & 1 & 0 & 0 & -\eta\,x^3 & 0 & 0 & 0 \\
 0 & 0 & 0 & 0 & 1 & 0 & -\eta\,x^4 & 0 & 0 & 0 \\
 -\eta\,x^+ & 0 & 0 & 0 & 0 & 0 & (\eta\,x^+)^2 -1 & 0 & 0 & 0 \\
 0 & \eta\,x^+ & -\eta\,x^2 & -\eta\,x^3 & -\eta\,x^4 & (\eta\,x^+)^2 -1 
 & \eta^2\,\sum_{i=2}^4(x^i)^2 & \eta^2\, x^+\,x^2 & \eta^2 \, x^+\,x^3 & \eta^2\,x^+\,x^4 \\
 -\eta\,x^2 & 0 & 0 & 0 & 0 & 0 & \eta^2\, x^+\,x^2 & 1 & 0 & 0 \\
 -\eta\,x^3 & 0 & 0 & 0 & 0 & 0 & \eta^2\, x^+\,x^3 & 0 & 1 & 0 \\
 -\eta\,x^4 & 0 & 0 & 0 & 0 & 0 & \eta^2\, x^+\,x^4 & 0 & 0 & 1 
\end{pmatrix}} \,,
\\
 &d= 4\,\eta\,\tilde{x}_- \,, 
\end{split}
\end{align}
where only $(x^+,\,x^-,\,x^2,\,x^3,\,x^4,\,\tilde{x}_+,\,\tilde{x}_-,\,\tilde{x}_2,\,
\tilde{x}_3,\,\tilde{x}_4)$-components of $\cH_{MN}$ are displayed. 

\medskip

When one of the $(x^2,\,x^3,\,x^4)$-coordinates, say $x^2$, is compactified 
with the period $x^2 \sim x^2 + \eta^{-1}$, 
the monodromy matrix is given by 
\begin{align}
 \cH_{MN}(x^2 +\eta^{-1}) = \bigl[\Omega^\rmT\cH(x^2)\,\Omega\bigr]_{MN} \,,\qquad 
 \Omega^M{}_N\equiv \begin{pmatrix} \delta^m_n & 2\,\delta_-^{[m}\,\delta_2^{n]} \\ 0 & \delta_m^n \end{pmatrix} \in \OO(10,10;\mathbb{Z})\,,
\end{align}
and in this sense the compactified background is a $T$-fold. 
In terms of the non-geometric $Q$-flux, 
this background has the following components of it: 
\begin{align}
 Q_+{}^{-+} = Q_2{}^{-2} = Q_3{}^{-3} = Q_4{}^{-4} = \eta \,.
\end{align}

\subsection{A non-geometric background from non-Abelian $T$-duality}
\label{sec:non-geometry-NATD}

Before considering YB-deformations of $\AdS5\times\rmS^5$\,, let us consider another example 
of purely NS-NS background, 
which was found in \cite{Gasperini:1993nz} via a non-Abelian $T$-duality. 

\medskip 

The background takes the form,
\begin{align}
\begin{split}
 \rmd s^2&= -\rmd t^2 + \frac{(t^4+y^2)\,\rmd x^2-2\,x\,y\,\rmd x\,\rmd y+(t^4+x^2)\,\rmd y^2+t^4\,\rmd z^2}{t^2\,(t^4+x^2+y^2)} + \rmd s_{T^6}^2\,,
\\
 B_2 &= \frac{(x\,\rmd x+y\,\rmd y)\wedge \rmd z}{t^4+x^2+y^2}\,,\qquad 
 \Phi= \frac{1}{2} \ln\biggl[\frac{1}{t^2\,(t^4+x^2+y^2)}\biggr] \,, 
\end{split}
\label{eq:bgNATD}
\end{align}
where $\rmd s_{T^6}^2$ is the flat metric on a 6-torus. 
In terms of the dual parameterization, this background takes 
a Friedmann-Robertson-Walker-type form,
\begin{align}
\begin{split}
 \rmd s^2_{\text{dual}}&= -\rmd t^2 + t^{-2}\,\bigl(\rmd x^2+ \rmd y^2 
 + \rmd z^2\bigr) + \rmd s_{T^6}^2\,,
\\
 \beta &= (x\,\partial_x+y\,\partial_y)\wedge \partial_z\,,\qquad 
 \tilde{\phi} = -\ln t^3 \,. \label{3.69}
\end{split}
\end{align}
Note here that this background cannot be represented by a coset or a Lie group itself. 
This is because the background \eqref{eq:bgNATD} contains a curvature singularity 
and is not homogeneous. 
Hence the background \eqref{eq:bgNATD} cannot be realized as a Yang-Baxter deformation 
and is not included in the discussion of \cite{Hoare:2016wsk,Borsato:2016pas,Hoare:2016wca}. 

\medskip

It is easy to see that the associated $Q$-flux is constant on this background \eqref{3.69},
\begin{align}
 Q_y{}^{xy} = Q_z{}^{xz} = -1\,. 
\end{align}
Therefore, if the $x$-direction is compactified as $x\sim x+1$, 
the background fields are twisted by an $\OO(10,10;\mathbb{Z})$ transformation as
\begin{align}
 \cH_{MN}(x+1) = \bigl[\Omega^\rmT\cH(x)\,\Omega\bigr]_{MN} \,,\qquad 
 \Omega^M{}_N \equiv \begin{pmatrix} \delta^m_n & 2\,\delta_x^{[m} \,\delta_z^{n]} \\ 0 
 & \delta_m^n \end{pmatrix}\,,\quad 
 d(x+1)=d(x) \,. 
\end{align}
Thus the background can be interpreted as a $T$-fold. 
If the $z$-direction is also compactified as $z\sim z+1$, another twist is realized as 
\begin{align}
 \cH_{MN}(y+1) = \bigl[\Omega^\rmT\cH(y)\,\Omega\bigr]_{MN} \,,\qquad 
 \Omega^M{}_N \equiv \begin{pmatrix} \delta^m_n & 2\,\delta_y^{[m} \,\delta_z^{n]} \\ 
 0 & \delta_m^n \end{pmatrix}\,,\quad 
 d(y+1)=d(y) \,. 
\end{align}

\medskip

As stated in \cite{Gasperini:1993nz}, this background is not a solution of the usual supergravity. 
However, by using the divergence formula $I^m = \tilde{\sfD}_n\beta^{mn}$ again and introducing 
a vector field as 
\begin{align}
 I=-2\,\partial_z \,,
\end{align}
we can see that the background \eqref{eq:bgNATD} together with this vector field $I$ 
satisfies GSE. 
Thus, this background can also be regarded as a $T$-fold solution of DFT. 

\medskip

In this subsection, we have considered just one example of non-Abelian $T$-duality, 
but it would be interesting to study a lot of examples as a new technique 
to generate GSE solutions. In fact, it is well-known that non-Abelian $T$-duality 
is a systematic method to construct $T$-fold solutions in DFT.

\subsection{YB-deformed $\AdS5\times\rmS^5$ backgrounds}
\label{sec:non-geometry-AdS5xS5}

We show that various YB deformations of the $\AdS5\times\rmS^5$ background are $T$-folds. 
We consider here examples associated with the following five classical $r$-matrices:
\begin{enumerate}
\setlength{\leftskip}{0.7cm} 
\item \quad $r = \frac{1}{2\,\eta}\,\bigl[\eta_1\,(D+M_{+-})\wedge P_+ 
+ \eta_2\,M_{+2}\wedge P_3 \bigr]$\,, 

\item \quad $r = \frac{1}{2}\,P_0\wedge D$\,, 

\item \quad $r = \frac{1}{2}\,\bigl[P_0\wedge D + P^i \wedge (M_{0i}+ M_{1i})\bigr]$\,, 

\item \quad $r = \frac{1}{2\eta}\, P_- \wedge (\eta_1\,D-\eta_2\,M_{+-})$\,, 

\item \quad $r = \frac{1}{2}\, M_{-\mu}\wedge P^\mu$\,. 
\end{enumerate}
The classical $r$-matrices other than the first one are non-unimodular. 
Note here that the $\rmS^5$ part remains undeformed and only the $\AdS5$ part is deformed. 
As shown in App.\,\ref{sec:Penrose-limit}, through the (modified) Penrose limit, 
the second and third examples are reduced to the two examples discussed in the previous subsection. 

\subsubsection{Non-Abelian unimodular $r$-matrix}

Let us consider a non-Abelian unimodular $r$-matrix 
(see $R_5$ in Tab.\,$1$ of \cite{Borsato:2016ose}),
\begin{align}
 r = \frac{1}{2\,\eta}\,\bigl[\eta_1\,(D+M_{+-})\wedge P_+ + \eta_2\,M_{+2}\wedge P_3 \bigr] \,,
\end{align}
where, for simplicity, it is written in terms of the light-cone coordinates,%
\footnote{In the following, our light-cone convention is taken as $\varepsilon_{z+-23r\xi\phi_1\phi_2\phi_3} = + \sqrt{\abs{\CG}}$\,.}
\begin{align}
 x^\pm \equiv \frac{x^0 \pm x^1}{\sqrt{2}} \,. 
\end{align}
The corresponding YB-deformed background is given by 
\begin{align}
 \rmd s^2 &= 
  \frac{-2\,z^2\,\rmd x^+\,\rmd x^- + 4\,\eta_1^2\,z^{-1}\,x^-\,\rmd z\,\rmd x^-}{z^4- (2\,\eta_1\,x^-)^2}+\frac{z^2\,[(\rmd x^2)^2+(\rmd x^3)^2]}{z^4+(\eta_2\,x^-)^2} +\frac{\rmd z^2}{z^2} 
\nn\\
 &\quad + \frac{2\,\{[x^2 (2\,\eta_1^2+\eta_2^2)-\eta_1\,\eta_2\,x^3]\,z^2\,x^-\,\rmd x^2+\eta_1\,(2\,\eta_1\,x^3-\eta_2\,x^2)\,\rmd x^3\}\,\rmd x^-}{[z^4-(2\,\eta_1\,x^-)^2]\,[z^4+(\eta_2\,x^-)^2]}
\nn\\
 &\quad - \frac{(\eta_1^2+\eta_2^2)\,(z\,x^2)^2 -2\,\eta_1\,\eta_2\,z^2\,x^2\,x^3 + \eta_1^2\,[z^4 + (z\,x^3)^2+ (\eta_2\,x^-)^2]}{[z^4-(2\,\eta_1\,x^-)^2]\,[z^4+(\eta_2\,x^-)^2]}\,(\rmd x^-)^2
 + \rmd s_{\rmS^5}^2\,,
\nn\\
 B_2 &= -\biggl[\frac{\eta_1\,\{x^2\,[z^4+2\,(\eta_2\,x^-)^2]-2\,\eta_1\,\eta_2\,(x^-)^2\,x^3\}\,\rmd x^2+\{\eta_1\,z^4\,x^3-\eta_2\,x^2\,[z^4-2\,(\eta_1\,x^-)^2]\}\,\rmd x^3}{[z^4- (2\,\eta_1\,x^-)^2]\,[z^4+(\eta_2\,x^-)^2]}
\nn\\
 &\qquad +\frac{\eta_1\,(z\,\rmd z - 2\, x^-\,\rmd x^+)}{z^4-(2\,\eta_1\,x^-)^2}\biggr]\wedge \rmd x^-
 +\frac{\eta_2\,x^-\,\rmd x^2 \wedge \rmd x^3}{z^4+(\eta_2\,x^-)^2} \,,
\nn\\
 \Phi &=\frac{1}{2} \ln \biggl[\frac{z^8}{[z^4- (2\,\eta_1\,x^-)^2]\,[z^4+ (\eta_2\,x^-)^2]}\biggr]\,,
\nn\\
 \hat{F}_1 &=\frac{4\,\eta_1\,\eta_2\,x^-\,(2\,x^-\,\rmd z -z\,\rmd x^-)}{z^5} \,,
\nn\\
 \hat{F}_3 &= -B_2\wedge F_1 +\frac{4\,\eta_1}{z^5}\, \bigl(2\,x^-\,\rmd z - z\,\rmd x^-\bigr)\wedge \rmd x^2 \wedge \rmd x^3 
\nn\\
 &\quad +\frac{4}{z^5}\, \rmd z \wedge \rmd x^- \wedge \bigl[\eta_1\,(x^3\,\rmd x^2- x^2\,\rmd x^3) + \eta_2\,(x^-\,\rmd x^+ - x^2\,\rmd x^2) \bigr] \,,
\nn\\
 \hat{F}_5&=4\,\biggl[\frac{z^8}{[z^4-(2\,\eta_1\,x^-)^2]\,[z^4+ (\eta_2\,x^-)^2]}\,\omega_{\AdS5} + \omega_{\rmS^5}\biggr] \,,
\nn\\
 \hat{F}_7&= -B_2\wedge F_5 \,,\qquad 
 \hat{F}_9= -\frac{1}{2}\,B_2\wedge F_7 \,. 
\end{align}
In terms of the dual fields, we obtain the following expression:
\begin{align}
\begin{split}
 \rmd s_{\text{dual}}^2 &= \frac{ -2\,\rmd x^+\,\rmd x^-+ (\rmd x^2)^2
 +(\rmd x^3)^2+\rmd z^2}{z^2} + \rmd s_{\rmS^5}^2\,, \qquad 
 \tilde{\phi}=0\,, \\
 \beta &=\eta_1\,\bigl(2\,x^-\,\partial_-+x^2\,\partial_2+x^3\,\partial_3+z\,\partial_z\bigr)
 \wedge\partial_+ + \eta_2\,\bigl(x^2\,\partial_+ + x^-\,\partial_2\bigr)\wedge \partial_3 \,.
\end{split}
\end{align}
It is straightforward to check that the R-R field strengths are given by 
\begin{align}
 \hat{F} = \Exp{-B_2\wedge}F\,,\qquad F \equiv \Exp{-\beta\vee}\check{F}\,,\qquad 
 \check{F} = 4\,\bigl(\omega_{\AdS5} + \omega_{\rmS^5}\bigr)\,.
\end{align}
Namely, as advocated in Sec.\,\ref{sec:YB-beta-deform}, 
the $\beta$-untwisted R-R fields $\check{F}$ are invariant 
under the YB deformation. 

\medskip

This background has the following components of $Q$-flux:
\begin{align}
 Q_z{}^{z+} = \eta_1\,,\quad 
 Q_-{}^{-+} = 2\,\eta_1\,,\quad 
 Q_2{}^{2+} = \eta_1\,,\quad 
 Q_3{}^{3+} = \eta_1\,,\quad 
 Q_2{}^{+3} = \eta_2\,,\quad 
 Q_-{}^{23} = \eta_2\,. 
\end{align}
Accordingly, for example, when the $x^3$ direction is compactified with a period 
$x^3\sim x^3+\eta_1^{-1}$\,, this background becomes a $T$-fold with the monodromy,
\begin{align}
 \cH_{MN}(x^3 +\eta_1^{-1}) = \bigl[\Omega^\rmT\cH(x)\,\Omega\bigr]_{MN} \,,\qquad 
 \Omega^M{}_N\equiv \begin{pmatrix} \delta^m_n & 2\,\delta_3^{[m}\,\delta_+^{n]} \\ 0 & \delta_m^n \end{pmatrix} \in \OO(10,10;\mathbb{Z})\,. 
\end{align}
The R-R fields $F$ are also twisted by the same monodromy,
\begin{align}
 F(x^3+\eta_1^{-1}) = \Exp{- \omega\vee}F(x^3)\,,\qquad 
 \omega^{mn} = 2\,\delta_3^{[m} \,\delta_+^{n]} \,. 
\end{align}
Note that the R-R potentials are twisted by the same monodromy as well, 
though their explicit forms are not written down here. 

\subsubsection{$r=\frac{1}{2}\,P_0\wedge D$}
\label{sec:AdS-P0-D}

Let us next consider a classical $r$-matrix\cite{vanTongeren:2015uha,Orlando:2016qqu},
\begin{align}
 r=\frac{1}{2}\,P_0\wedge D\,.
\end{align}
Because $[P_0, D] \neq 0$\,, this classical $r$-matrix does not satisfy the unimodularity condition. 

\medskip

Although we have considered in the subsection \ref{subsec:Non-unimodular-sol},
we once again present the full background.
By introducing the polar coordinates,
\begin{align}
 x^1 =\rho\sin\theta \cos\phi\,, \qquad x^2 =\rho\sin\theta\sin\phi\,, 
 \qquad x^3 =\rho\cos\theta\,,
\end{align}
the deformed background can be rewritten as \cite{Orlando:2016qqu}\footnote{Only the metric 
and NS-NS two-form were computed in \cite{vanTongeren:2015uha}.}
\begin{align}
\label{eq:AdS-P0-D}
\begin{split}
 \rmd s^2 &= \frac{z^2\,\bigl[ -(\rmd x^0)^2+\rmd\rho^2+\rmd z^2\bigr]-\eta^2\,(\rmd \rho -\rho\,z^{-1}\,\rmd z)^2}{z^4-\eta^2\,(z^2+\rho^2)}
    +\frac{\rho^2\,(\rmd \theta^2+\sin^2\theta\,\rmd \phi^2)}{z^2}\\
  &\quad  +\rmd s_{\rmS^5}^2 \,, 
\\
    B_2 &= -\eta\,\frac{\rmd x^0 \wedge ( \rho\,\rmd \rho+z\,\rmd z )}{z^4-\eta^2\,(z^2+\rho^2)}\,, \quad 
 \Phi = \frac{1}{2}\ln \biggl[\frac{z^4}{z^4-\eta^2\,(z^2+\rho^2)}\biggr]\,,\quad 
 I = -\eta\,\partial_0 \,,
\\
 \hat{F}_1 &=0\,,\quad 
 \hat{F}_3 = \frac{4\,\eta\,\rho^2\sin\theta}{z^5}\, (z\,\rmd \rho - \rho \, \rmd z)\wedge \rmd \theta\wedge \rmd \phi \,,
\\
 \hat{F}_5 &= 4\,\biggl[\frac{z^4}{z^4-\eta^2\, (z^2+\rho^2)}\,\omega_{\AdS5} + \omega_{\rmS^5}\biggr]\,,
\\
 \hat{F}_7&=\frac{4\,\eta\,\rmd x^0 \wedge (\rho\,\rmd \rho+z\,\rmd z)}{z^4-\eta^2\, (z^2+\rho^2)}\wedge \omega_{\rmS^5}\,,\quad \hat{F}_9=0\,. 
\end{split}
\end{align}
This background is not a solution of the usual type IIB supergravity, 
but that of GSE \cite{Arutyunov:2015mqj}. 
By setting $\eta=0$, this background reduces to the original $\AdS5\times \rmS^5$. 

\medskip

In the dual parameterization, the dual metric, the $\beta$ field and the dual dilaton 
are given by 
\begin{align}
\begin{split}
 \rmd s_{\text{dual}}^2 &= \frac{\rmd z^2 -(\rmd x^0)^2+(\rmd x^1)^2+(\rmd x^2)^2+(\rmd x^3)^2}{z^2} +\rmd s_{\rmS^5}^2 \,,\qquad \tilde{\phi}=0\,,
\\
 \beta &= \eta\,\hat{P}_0\wedge \hat{D}
=\eta\,\partial_0\wedge (  x^1\,\partial_1 + x^2\,\partial_2 + x^3\,\partial_3+z\,\partial_z) \no\\
 &= \eta\,\partial_0\wedge (\rho\,\partial_\rho+z\,\partial_z)\,. 
\end{split}
\end{align}
The Killing vector $I^m$ satisfies the divergence formula,
\begin{align}
 I^0 = -\eta = \sfD_{m}\beta^{0m} \,. 
\end{align}
The $Q$-flux has the following non-vanishing components:
\begin{align}
 Q_z{}^{0z} = Q_1{}^{01} = Q_2{}^{02} = Q_3{}^{03} = \eta\,. 
\end{align}
Thus, when at least one of the $(x^1,x^2,x^3)$ directions is compactified, 
the background can be interpreted as a $T$-fold. 
For example, when the $x^1$ direction is compactified, the monodromy is given by 
\begin{align}
 \cH_{MN}(x^1+\eta^{-1}) = \bigl[\Omega^\rmT\cH(x^1)\,\Omega\bigr]_{MN} \,,\qquad 
 \Omega^M{}_N \equiv \begin{pmatrix} \delta^m_n & 2\,\delta_0^{[m} \,\delta_1^{n]} \\ 0 & \delta_m^n \end{pmatrix}\,. 
\label{eq:monodromy-AdS-P0-D}
\end{align}

\medskip

From \eqref{eq:AdS-P0-D}, the R-R potentials can be found as follows:
\begin{align}
\begin{split}
 \hat{C}_0&=0\,,\quad 
 \hat{C}_2= \frac{\eta\,\rho^3\sin\theta}{z^4}\,\rmd \theta\wedge \rmd \phi\,, 
\\
 \hat{C}_4&= \frac{\rho^2\sin\theta}{z^4}\,\rmd x^0\wedge \rmd \rho\wedge \rmd \theta \wedge \rmd\phi + \omega_4 -B_2 \wedge \hat{C}_2 \,,
\\
 \hat{C}_6&= -B_2\wedge \omega_4 \,, \quad
 \hat{C}_8 = 0 \,,
\end{split}
\end{align}
where the $4$-form $\omega_{4}$ satisfies $\omega_{\rmS^5}=\rmd\omega_{4}$\,,
Providing the $B$-twist, we obtain
\begin{align}
\begin{split}
 F_1&=0\,,\quad 
 F_3 = \frac{4\,\eta \,\rho^2 \sin \theta}{z^5}\, (\rho\,\rmd z- z\,\rmd \rho) \wedge \rmd \theta \wedge \rmd \phi\,,
\\
 F_5&=4\,\bigl(\omega_{\AdS5}+\omega_{\rmS^5} \bigr)\,,\quad 
F_7 =0\,,\quad 
F_9 =0\,,
\\
 A_0&=0\,,\quad 
A_2 = \frac{\eta\,\rho^3 \sin \theta}{z^4}\,\rmd \theta \wedge\rmd \phi\,,
\\
 A_4&= \frac{\rho^2 \sin\theta}{z^4}\,\rmd x^0 \wedge \rmd \rho \wedge \rmd \theta \wedge \rmd \phi+\omega_4 \,,\quad 
A_6=0\,,\quad 
A_8=0\,. 
\end{split}
\end{align}
We can further compute the $\beta$-untwisted fields,
\begin{align}
\begin{split}
 \check{F}_1 &=0\,,\quad 
 \check{F}_3 = 0\,, \quad
 \check{F}_5 =4\,\bigl(\omega_{\AdS5}+\omega_{\rmS^5}\bigr)\,,\quad 
\check{F}_7 =0\,,\quad \check{F}_9 =0\,,
\\
 \check{C}_0 &=0\,,\quad \check{C}_2 = 0\,, \quad
 \check{C}_4 = \frac{\rho^2 \sin\theta}{z^4}\,\rmd x^0 \wedge \rmd \rho \wedge \rmd \theta \wedge \rmd \phi +\omega_4 \,,\quad \check{C}_6=0\,,\quad \check{C}_8=0\,.
\end{split}
\end{align}
As expected, the $\beta$-untwisted R-R fields are precisely the R-R fields in the undeformed background, and they are single-valued. 
In terms of the twisted R-R fields, $(F,\,A)$, the R-R fields have the same monodromy as \eqref{eq:monodromy-AdS-P0-D},
\begin{align}
 A(x^1+\eta^{-1})= \Exp{- \omega \vee} A(x^1)\,,\quad 
F(x^1+\eta^{-1}) = \Exp{- \omega \vee}F(x^1)\,,\quad 
 \omega^{mn} = 2\,\delta_0^{[m} \,\delta_1^{n]} \,. 
\end{align}

\subsubsection{A scaling limit of the Drinfeld--Jimbo $r$-matrix}
\label{sec:AdS-Drinfeld--Jimbo}

Let us consider a classical $r$-matrix \cite{Hoare:2016hwh,Orlando:2016qqu},
\begin{align}
 r = \frac{1}{2}\,\bigl[P_0\wedge D + P^i \wedge (M_{0i}+ M_{1i})\bigr] \,,
\end{align}
which can be obtained as a scaling limit of the classical $r$-matrix of 
Drinfeld-Jimbo type \cite{Drinfeld:1985rx,Jimbo:1985zk}. 
By using the polar coordinates $(\rho,\theta)$,
\begin{align}
 (\rmd x^2)^2 + (\rmd x^3)^2 = \rmd \rho^2+\rho^2\,\rmd \theta^2 \,,
\end{align}
the YB-deformed background, which satisfies GSE, is given by \cite{Hoare:2016hwh,Orlando:2016qqu}
\begin{align}
\begin{split}
 \rmd s^2 &= \frac{-(\rmd x^0)^2+\rmd z^2}{z^2- \eta^2}
    +\frac{z^2\bigl[(\rmd x^1)^2+\rmd \rho^2\bigr]}{z^4+ \eta^2\,\rho^2}
    +\frac{\rho^2\,\rmd \theta^2}{z^2}+ \rmd s^2_{\rmS^5}\,, \\
 B_2 &= \eta\,\biggl[\frac{-\rmd x^0\wedge \rmd z }{z\,(z^2-\eta^2)} - \frac{\rho\,\rmd x^1 \wedge \rmd \rho}{z^4+ \eta^2\,\rho^2}\biggr]\,, \\
 \Phi &= \frac{1}{2}\ln\biggl[\frac{z^6}{(z^2- \eta^2)(z^4+ \eta^2\,\rho^2)}\biggr]\,,\qquad 
 I = -\eta\,(4\,\partial_0 + 2\,\partial_1) \,, \\
 \hat{F}_1&= -\frac{4\,\eta^2\,\rho^2}{z^4}\, \rmd\theta\,, \\
 \hat{F}_3 &= 4\,\eta\,\rho\,\biggl(\frac{-\rho\, \rmd x^0\wedge \rmd z }{z\,(z^4-\eta^2\,z^2)} + \frac{\rmd x^1 \wedge \rmd \rho}{z^4+ \eta^2\,\rho^2}\biggr)\wedge \rmd \theta \,,\\
 \hat{F}_5&= 4\,\biggl[\frac{z^6}{(z^2- \eta^2) (z^4 + \eta^2\, \rho^2)}\,\omega_{\AdS5} +\omega_{\rmS^5}\biggr]\,,\\
 \hat{F}_7&=4\,\eta\,\biggl(\frac{\rmd x^0\wedge \rmd z }{z\,(z^2- \eta^2)}+\frac{\rho\,\rmd x^1 \wedge \rmd \rho}{z^4+ \eta^2\,\rho^2}\biggr) \wedge \omega_{\rmS^5} \,,\\
 \hat{F}_9&=\frac{4\,\eta^2\,\rho}{z\,(z^2- \eta^2) (z^4+ \eta^2\,\rho^2)}\,
\rmd x^0 \wedge \rmd x^1\wedge \rmd \rho \wedge\rmd z \wedge \omega_{\rmS^5}\,. 
\label{eq:HvT}
\end{split}
\end{align}
The R-R potentials can be found as follows:
\begin{align}
\begin{split}
 \hat{C}_0&= 0\,, \quad
 \hat{C}_2 = -\frac{\eta\,\rho^2}{z^4}\,\rmd x^0\wedge \rmd \theta\,,
\quad
 \hat{C}_4 = \frac{\rho}{z^4+ \eta^2\,\rho^2}\, \rmd x^0\wedge \rmd x^1\wedge \rmd\rho \wedge \rmd\theta + \omega_4 \,,
\\
 \hat{C}_6&= -B_2\wedge \omega_4 \,, \quad
 \hat{C}_8 = \frac{\eta^2\,\rho}{z\,(z^2-\eta^2)(z^4+\eta^2\,\rho^2)}\,\rmd x^0\wedge \rmd x^1\wedge \rmd\rho\wedge \rmd z\wedge \omega_4 \,.
\end{split}
\end{align}

\medskip

Then the corresponding dual fields in the NS-NS sector are given by 
\begin{align}
\begin{split}
 \rmd s_{\text{dual}}^2 &= \frac{ - (\rmd x^0)^2+(\rmd x^1)^2+\rmd \rho^2+\rho^2\,\rmd \theta^2+\rmd{z}^2}{z^2} + \rmd s^2_{\rmS^5}\,,\qquad \tilde{\phi}=0\,, 
\\
 \beta &= \eta\,\bigl[\hat{P}_0\wedge \hat{D}+\hat{P}^i\wedge (M_{0i}+ M_{1i})\bigr]
 = \eta\,( - x^2\,\partial_1\wedge\partial_2 - x^3\,\partial_1\wedge\partial_3+z\,\partial_0\wedge \partial_z)
\nn\\
 &=\eta\,( -\rho\,\partial_1\wedge\partial_\rho+z\,\partial_0\wedge \partial_z) \,, 
\end{split}
\end{align}
and the Killing vector $I^m$ again satisfies the divergence formula,
\begin{align}
 I^0 = -4\,\eta = \sfD_m\beta^{0m} \,,\qquad I^1 = -2\,\eta = \sfD_m\beta^{1m} \,.
\end{align}

Providing the $B$-twist to the R-R field strengths, we obtain
\begin{align}
\begin{split}
 F_1&=-\frac{4\,\eta^2 \rho^2}{z^4}\,\rmd \theta \,,\quad 
 F_3 = \frac{4\,\eta\,\rho}{z^5}\, \bigl(\rho\,\rmd z\wedge \rmd x^0 + z\,\rmd x^1\wedge \rmd\rho\bigr) \wedge \rmd \theta \,,
\\
 F_5&=4\,\bigl(\omega_{\AdS5}+\omega_{\rmS^5} \bigr)\,,
\quad F_7 =0\,,\quad F_9 =0\,,
\\
 A_0 &=0\,,\quad 
 A_2 = -\frac{\eta\,\rho^2}{z^4}\, \rmd x^0\wedge \rmd \theta \,,
\\
 A_4 &= \frac{\rho}{z^4} \, \rmd x^0 \wedge \rmd x^1 \wedge \rmd \rho \wedge \rmd \theta +\omega_4 \,, \quad
 A_6 =0\,,\quad 
 A_8 =0\,. 
\end{split}
\end{align}
Furthermore, the $\beta$-untwist leads to the following expressions: 
\begin{align}
\begin{split}
 \check{F}_1 &=0\,,\quad 
 \check{F}_3 = 0\,, \quad
 \check{F}_5 =4\,\bigl(\omega_{\AdS5}+\omega_{\rmS^5}\bigr)\,,\quad 
\check{F}_7 =0\,,\quad \check{F}_9 =0\,,
\\
 \check{C}_0 &=0\,,\quad \check{C}_2 = 0\,, \quad
 \check{C}_4 = \frac{\rho}{z^4} \, \rmd x^0 \wedge \rmd x^1 \wedge \rmd \rho \wedge \rmd \theta +\omega_4 \,,\quad \check{C}_6=0\,,\quad \check{C}_8=0\,. 
\end{split}
\end{align}
These are the same as the undeformed R-R potentials. 

\medskip

Then the non-zero component of $Q$-flux are given by 
\begin{align}
 Q_z{}^{0z} = \eta\,,\qquad Q_2{}^{12} = -\eta\,,\qquad Q_3{}^{13} = -\eta\,. 
\end{align}
When the $x^2$-direction is compactified as $x^2 \sim x^2 +\eta^{-1}$, 
this background becomes a $T$-fold with the monodromy,
\begin{align}
\begin{split}
 \cH_{MN}(x^2+\eta^{-1}) &= \bigl[\Omega^\rmT\cH(x^2)\,\Omega\bigr]_{MN} \,,\qquad 
 \Omega^M{}_N \equiv \begin{pmatrix} \delta^m_n & -2\,\delta_1^{[m} \,\delta_2^{n]} \\ 0 & \delta_m^n \end{pmatrix}\,,
\\
 F(x^2+\eta^{-1}) &= \Exp{- \omega \vee}F(x^2)\,,\qquad 
 \omega^{mn} = -2\,\delta_1^{[m} \,\delta_2^{n]} \,. 
\end{split}
\end{align}

\subsubsection{$r=\frac{1}{2\,\eta}\,P_-\wedge (\eta_1\,D-\eta_2\,M_{+-})$}

Let us consider a non-unimodular $r$-matrix\footnote{This $r$-matrix includes the known examples studied in Sec.\,4.3 ($\eta_1=-\eta_2=-\eta$) 
and 4.4 ($\eta_1=-\eta$, $\eta_2=0$) of \cite{Orlando:2016qqu} as special cases. },
\begin{align}
 r=\frac{1}{2\,\eta}\,P_-\wedge (\eta_1\,D-\eta_2\,M_{+-})\,. 
\end{align}
Here we have introduced the light-cone coordinates and polar coordinates as
\begin{align}
 x^\pm \equiv \frac{x^0 \pm x^1}{\sqrt{2}} \,, \qquad 
 (\rmd x^2)^2 + (\rmd x^3)^2 = \rmd \rho^2+\rho^2\,\rmd \theta^2 \,. 
\end{align}

\medskip 

The YB-deformed background is given by
\begin{align}
\begin{split}
 \rmd s^2&= \frac{-2\, z^2\,\rmd x^+\,\rmd x^-}{z^4 - (\eta_1 + \eta_2)^2\,(x^+)^2} 
+\frac{ \rmd \rho^2 + \rho^2\,\rmd \theta^2+\rmd z^2 }{z^2}\\
 &\quad + \eta_1\,\rmd x^+\,\frac{2\,x^+\,(\eta_1 + \eta_2)\,(z\,\rmd z + \rho\,\rmd \rho) - \eta_1\,(z^2 + \rho^2)\,\rmd x^+}{z^2\,[z^4 - (\eta_1 + \eta_2)^2\,(x^+)^2]} + \rmd s^2_{\rmS^5}\,,\\
 B_2 &=-\frac{\bigl[\eta_1\,(  \rmd x^+\wedge (\rho\,\rmd \rho+z\,\rmd z)-x^+\,\rmd x^+\wedge \rmd x^- ) - \eta_2\,x^+\,\rmd x^+ \wedge \rmd x^-\bigr]}{z^4 - (\eta_1 + \eta_2)^2\,(x^+)^2}
   \,,\\
 \Phi &= \frac{1}{2} \ln\biggl[\frac{z^4}{z^4 - (\eta_1 + \eta_2)^2\, (x^+)^2}\biggr]\,,\qquad 
 I = -(\eta_1 - \eta_2)\,\partial_-\,,\\
 \hat{F}_1&= 0\,,\\
 \hat{F}_3 &= -\frac{4\,\rho\,\bigl[\eta_1\,(\rmd x^+ \wedge (z\,\rmd \rho-\rho\, \rmd z)  - x^+\,\rmd z\wedge \rmd \rho)
                                    - \eta_2\,x^+\,\rmd z\wedge \rmd \rho\bigr]\wedge \rmd \theta}{z^5}\,,\\
 \hat{F}_5&= 4 \biggl[\frac{z^4}{z^4 - (\eta_1 + \eta_2)^2\,(x^+)^2}\,\omega_{\AdS5} + \omega_{\rmS^5}\biggr] \,,\\
 \hat{F}_7&= \frac{4\,\bigl[\eta_1\,(  \rmd x^+\wedge (\rho\,\rmd \rho+z\,\rmd z)-x^+\,\rmd x^+\wedge \rmd x^- ) - \eta_2\,x^+\,\rmd x^+ \wedge \rmd x^-\bigr]\wedge \omega_{\rmS^5}}{z^4 - (\eta_1+\eta_2)^2\,(x^+)^2} \,,\\
 \hat{F}_9&= 0\,. 
\end{split}
\end{align}
The R-R potentials are also given by
\begin{align}
\begin{split}
 \hat{C}_0&= 0\,,\qquad
 \hat{C}_2 = \frac{\rho\,[\eta_1\,\rho\,\rmd x^+ - (\eta_1 + \eta_2)\,x^+\,\rmd \rho]\wedge \rmd \theta}{z^4}\,,
\\
 \hat{C}_4&= \frac{\rho\,\rmd x^+\wedge [z^3\,\rmd x^- - \eta_1\,(\eta_1 + \eta_2)\,x^+\,\rmd z]\wedge \rmd\rho \wedge \rmd\theta}{z^3\,[z^4 - (\eta_1 + \eta_2)^2\,(x^+)^2]} + \omega_4\,,
\\
 \hat{C}_6&= -B_2\wedge \omega_4\,,\qquad 
 \hat{C}_8 = 0\,.
\end{split}
\end{align}

\medskip

The dual fields are given by
\begin{align}
\begin{split}
 \rmd s_{\text{dual}}^2&= \frac{ -2\,\rmd x^+\,\rmd x^- + \rmd \rho^2 + \rho^2\,\rmd \theta^2+\rmd z^2}{z^2} 
+ \rmd s^2_{\rmS^5} \,,\qquad 
 \tilde{\phi}=0 \,,
\\
 \beta &= \hat{P}_-\wedge (\eta_1\,\hat{D}+\eta_2\,\hat{M}_{+-}) 
= \eta_1 \, \partial_- \wedge ( x^+\,\partial_+ + \rho \,\partial_\rho+z\,\partial_z)
         +\eta_2 \,x^+\, \partial_- \wedge \partial_+ 
\\
 &= \eta_1 \, \partial_- \wedge ( x^+\,\partial_+ + x^2\,\partial_2 + x^3\,\partial_3+z\,\partial_z)
         +\eta_2 \,x^+\, \partial_- \wedge \partial_+ \,,
\end{split}
\end{align}
and the $Q$-flux has the following non-vanishing components:
\begin{align}
 Q_z{}^{-z} = Q_+{}^{-+} = Q_2{}^{-2} = Q_3{}^{-3} = \eta_1\,,\qquad 
 Q_+{}^{-+} = \eta_2 \,. 
\end{align}
In a similar manner as the previous examples, by compactifying one of the $x^1$, $x^2$, and $x^3$ directions with a certain period, this background can also be regarded as a $T$-fold. 
For example, if we make the identification, $x^3\sim x^3 + \eta_1^{-1}$, 
the associated monodromy becomes
\begin{align}
\begin{split}
 \cH_{MN}(x^3+\eta_1^{-1}) &= \bigl[\Omega^\rmT\cH(x^3)\,\Omega\bigr]_{MN} \,,\qquad 
 \Omega^M{}_N \equiv \begin{pmatrix} \delta^m_n & 2\,\delta_-^{[m} \,\delta_3^{n]} \\ 0 & \delta_m^n \end{pmatrix}\,,
\\
 F(x^3+\eta_1^{-1}) &= \Exp{- \omega \vee}F(x^3)\,,\qquad 
 \omega^{mn} = 2\,\delta_-^{[m} \,\delta_3^{n]} \,. 
\end{split}
\end{align}

\paragraph{A solution of Generalized Type IIA Supergravity Equations:}

In the background \eqref{eq:HvT}, by performing a $T$-duality along the $x^1$-direction (see \cite{Sakamoto:2017wor} for the duality transformation rule), 
we obtain the following solution of the generalized type IIA equations of motion:
\begin{align}
\begin{split}
 \rmd s^2 &= \frac{ - (\rmd x^0)^2+\rmd z^2}{z^2 - \eta^2} + z^2\,(\rmd x^1)^2 
 + \frac{(\rmd \rho + \eta\, \rho\,\rmd x^1)^2 + \rho^2\,\rmd \theta^2}{z^2} + \rmd s_{\rmS^5} \,,\\
 B_2 &= -\frac{\eta\, \rmd x^0 \wedge \rmd z}{z\,(z^2 - \eta^2)}\,,\qquad 
 \Phi = -2\,\eta\,x^1 - \frac{1}{2} \ln\Bigl(\frac{z^2 - \eta^2}{z^4}\Bigr)\,,\qquad I=-4\,\eta\,\partial_0\,,\\
 \hat{F}_2&= \frac{4\,\eta\Exp{2\,\eta\,x^1}\rho\,(\rmd\rho + \eta\,\rho\,\rmd x^1)\wedge \rmd \theta}{z^4} \,,\\
 \hat{F}_4&= \frac{4\Exp{2\,\eta\,x^1}\rho\,\rmd x^0 \wedge (\rmd \rho+\eta\,\rho\,\rmd x^1) \wedge \rmd \theta\wedge \rmd z}{z^3\,(z^2- \eta^2)} \,,\\
 \hat{F}_6&=-4 \Exp{2\,\eta\,x^1} \rmd x^1 \wedge \omega_{\rmS^5}\,,\qquad
 \hat{F}_8 = \frac{4\,\eta\Exp{2\,\eta\,x^1}\rmd x^0 \wedge \rmd x^1 \wedge \rmd z \wedge \omega_{\rmS^5}}{z\,(z^2 - \eta^2)} \,. 
\end{split}
\end{align}
Here the R-R potentials are given by 
\begin{align}
\begin{split}
 \hat{C}_1 &= 0 \,, \qquad
 \hat{C}_3 = \Exp{2\,\eta\,x^1} \frac{\rho\,\rmd x^0 \wedge (\rmd \rho + \eta\,\rho\, \rmd x^1)\wedge \rmd \theta}{z^4} \,,
\\
 \hat{C}_5 &= \Exp{2\,\eta\,x^1} \rmd x^1 \wedge \omega_4 \,, \qquad
 \hat{C}_7 = -\Exp{2\,\eta\,x^1} \frac{\eta\,\rmd z\wedge \rmd x^0 \wedge \rmd x^1 \wedge \omega_4}{z\,(z^2- \eta^2)} \,.
\end{split}
\end{align}
This background cannot be regarded as a $T$-fold, 
but it is the first example of the solution for the generalized type IIA supergravity equations.

\subsubsection{$r = \frac{1}{2}\, M_{-\mu}\wedge P^\mu$}

The final example is associated with the $r$-matrix\cite{Orlando:2016qqu}
\begin{align}
 r = \frac{1}{2}\, M_{-\mu}\wedge P^\mu\,.
\end{align}
This $r$-matrix is called the light-like $\kappa$-Poincar\'e.
Again, by introducing the coordinates,
\begin{align}
 x^\pm \equiv \frac{x^0 \pm x^1}{\sqrt{2}} \,, \qquad 
 (\rmd x^2)^2 + (\rmd x^3)^2 = \rmd \rho^2+\rho^2\,\rmd \theta^2 \,,
\end{align}
the YB-deformed background is given by (see Sec.\,4.5 of \cite{Orlando:2016qqu})
\footnote{The metric and NS-NS two form were computed in~\cite{vanTongeren:2015uha}. }
\begin{align}
\begin{split}
 \rmd s^2&=\frac{z^2( -2\,\rmd x^+\,\rmd x^-+\rmd z^2) }{z^4 - (\eta\,x^+)^2}
- \eta^2 \,\frac{\rho^2 (\,\rmd x^+)^2 -2\,x^+ \rho\,\rmd x^+\,\rmd \rho+(x^+)^2\,\rmd z^2}{z^2(z^4 - (\eta\,x^+)^2)} \\
&\quad+ \frac{\rmd \rho^2 + \rho^2\,\rmd \theta^2}{z^2} 
 + \rmd s_{\rmS^5}^2 \,,\\
 B_2 &= \frac{\eta\,\rmd x^+\wedge (x^+\,\rmd x^- -\rho\,\rmd\rho)}{z^4-(\eta \,x^+)^2}\,,\qquad 
 \Phi =\frac{1}{2} \ln\biggl[\frac{z^4}{z^4-(\eta\,x^+)^2}\biggr] \,,\qquad 
 I^- = 3\,\eta\,,\\
 \hat{F}_1&=0 \,,\quad
 \hat{F}_3= -\frac{4\,\eta\,\rho}{z^5}\,  \bigl(\rho\,\rmd x^+ - x^+\,\rmd \rho\bigr)\wedge \rmd\theta \wedge \rmd z \,,\\
 \hat{F}_5&= 4\,\biggl[\frac{z^4}{z^4-(\eta\,x^+)^2}\,\omega_{\AdS5}+\omega_{\rmS^5} \biggr] \,,\\
 \hat{F}_7&= - \frac{4\,\eta}{z^4-(\eta\,x^+)^2}\,\rmd x^+ \wedge \bigl(x^+\,\rmd x^- - \rho\,\rmd \rho\bigr)\wedge \omega_{\rmS^5} \,,\qquad
 \hat{F}_9= 0\,. 
\end{split}
\end{align}
The R-R potentials can be found as follows:
\begin{align}
\begin{split}
 \hat{C}_0&=0\,,\quad 
 \hat{C}_2= \frac{\eta\,\rho}{z^4}\,\bigl(\rho\,\rmd x^+-x^+\,\rmd \rho\bigr)\wedge\rmd \theta \,,
\\
 \hat{C}_4& =\frac{\rho}{z^4-(\eta\,x^+)^2}\, \rmd x^+\wedge \rmd x^-\wedge \rmd\rho \wedge \rmd\theta + \omega_4\,, \quad 
 \hat{C}_6 = -B_2\wedge \omega_4 \,, \quad 
 \hat{C}_8=0 \,.
\end{split}
\end{align}

\medskip

The corresponding dual fields are given by
\begin{align}
\begin{split}
 \rmd s_{\text{dual}}^2
 &= \frac{-2\rmd x^+\,\rmd x^- + \rmd \rho^2 + \rho^2\,\rmd \theta^2+\rmd z^2 }{z^2} + \rmd s^2_{\rmS^5} \,, \qquad \tilde{\phi}=0 \,,
\\
 \beta &= \eta\, \hat{M}_{-\mu}\wedge \hat{P}^\mu
= \eta\,\partial_-\wedge (x^+\,\partial_+ + \rho\,\partial_\rho) \no\\ 
 &= \eta\,\partial_-\wedge (x^+\,\partial_+ + x^2\,\partial_2 + x^3\,\partial_3) \,,
\end{split}
\end{align}
and it is easy to check that the divergence formula is satisfied: 
\begin{align}
 I^- = 3\,\eta = \sfD_m \beta^{-m} \,.
\end{align}

\medskip 

We can calculate other types of the R-R field fields as
\begin{align}
\begin{split}
 F_1&=0\,,\quad 
 F_3 =-\frac{4\,\eta\,\rho}{z^5}\, (\rho\,\rmd x^+ -x^+\,\rmd \rho)\wedge \rmd \theta\wedge \rmd z  \,,
\\
 F_5&=4\,\bigl(\omega_{\AdS5}+\omega_{\rmS^5} \bigr)\,,\quad 
F_7 =0\,,\quad F_9 =0\,,
\\
 A_0 &=0\,,\quad 
 A_2 = \frac{\eta\,\rho}{z^4}\,\bigl(\rho\,\rmd x^+ -x^+\,\rmd \rho\bigr)\wedge \rmd \theta \,,
\\
 A_4 &= \frac{\rho}{z^4} \, \rmd x^+ \wedge \rmd x^- \wedge \rmd \rho \wedge \rmd \theta +\omega_4 \,, \quad
 A_6 =0\,,\quad 
 A_8 =0\,,
\end{split}
\end{align}
and
\begin{align}
\begin{split}
 \check{F}_1 &=0\,,\quad 
 \check{F}_3 = 0\,, \quad
 \check{F}_5 =4\,\bigl(\omega_{\AdS5}+\omega_{\rmS^5}\bigr)\,,\quad 
\check{F}_7 =0\,,\quad \check{F}_9 =0\,,
\\
 \check{C}_0 &=0\,,\quad \check{C}_2 = 0\,, \quad
 \check{C}_4 =\frac{\rho}{z^4} \, \rmd x^+ \wedge \rmd x^-\wedge \rmd \rho \wedge \rmd \theta +\omega_4 \,,\quad 
\check{C}_6=0\,,\quad \check{C}_8=0\,,
\end{split}
\end{align}
and the $\beta$-twisted fields are again invariant under the YB deformation. 

\medskip 

The non-geometric $Q$-flux has the non-vanishing components,
\begin{align}
 Q_+{}^{-+}=Q_2{}^{-2}=Q_3{}^{-3}= \eta \,,
\end{align}
and again by compactifying one of the $x^1$, $x^2$, and $x^3$ directions, 
this background becomes a $T$-fold. 
Namely, if we compactify the $x^3$-direction as, $x^3\sim x^3 + \eta^{-1}$, 
the associated monodromy becomes
\begin{align}
\begin{split}
 \cH_{MN}(x^3+\eta^{-1}) &= \bigl[\Omega^\rmT\cH(x^3)\,\Omega\bigr]_{MN} \,,\qquad 
 \Omega^M{}_N \equiv \begin{pmatrix} \delta^m_n & 2\,\delta_-^{[m} \,\delta_3^{n]} \\ 0 & \delta_m^n \end{pmatrix}\,,
\\
 \sfF(x^3+\eta^{-1}) &= \Exp{- \omega \vee}\sfF(x^3)\,,\qquad 
 \omega^{mn} = 2\,\delta_-^{[m} \,\delta_3^{n]} \,. 
\end{split}
\end{align}


\chapter{Weyl invariance for generalized supergravity}
\label{Ch:Weyl-GSE}

In this chapter,
we discuss Weyl invariance of bosonic string theories on generalized supergravity backgrounds.
For this purpose,
we begin to show that generalized supergravity can be reproduced by the usual DFT.
After that, we construct a possible counterterm to cancel the Weyl anomaly and
show that the counterterm is definitely local.

\medskip

The chapter is organized as follows.
In section \ref{sec:(m)DFT},
after giving a short review of the mDFT \cite{Sakatani:2016fvh}, 
we reinterpret the mDFT as the usual DFT with a modified section.
In section \ref{sec:R-R(m)DFT},
we review the R-R sector of the DFT and then 
reproduce the generalized type IIA and IIB supergravity equations by choosing a non-standard section. 
We also present the $T$-duality transformation rule for solutions of the generalized type II supergravities. 
In section \ref{sec:Weyl},
we discuss the Weyl invariance of the string sigma model defined on a solution of the GSE. 

\section{NS--NS sector of (m)DFT}
\label{sec:(m)DFT}

In this section, we introduce the mDFT proposed in \cite{Sakatani:2016fvh}, 
in which only the NS--NS sector was studied and the R--R fields have not been included yet. 
We first give a short introduction to the mDFT. 
Then, we show that the mDFT can be regarded as the conventional DFT 
with a non-standard solution of the strong constraint. 

\subsection{A brief review of mDFT}

Let us give a short introduction to the mDFT.
In the absence of the R--R fields, the set of GSE in $D$ dimensions takes the following form: 
\begin{align}
\begin{split}
 &R_{mn} - \frac{1}{4}\,H_{mpq}\,H_n{}^{pq} + \sfD_m X_n + \sfD_n X_m = 0 \,,
\\
 &\frac{1}{2}\,\sfD^k H_{kmn} - \bigl(X^k H_{kmn} + \sfD_m X_n - \sfD_n X_m \bigr) = 0 \,,
\\
 &R - \frac{1}{2}\, \abs{H_3}^2 + 4\,\sfD_m X^m - 4\,X^m X_m = 0 \,,\qquad 
 X_m\equiv I_m +Z_m\,. 
\end{split}
\label{eq:GSE-NSNS}
\end{align}
Here we have defined $\abs{\alpha_p}^2\equiv \frac{1}{p!}\,
\alpha_{m_1\cdots m_p}\,\alpha^{m_1\cdots m_p}$\,. 
A vector field $I^m$ and a 1-form $Z_m$ are defined so as to satisfy
\begin{align}
 \sfD_m I_n + \sfD_n I_m =0\,, \qquad 
 I^k\, H_{kmn} + \sfD_m Z_n - \sfD_n Z_m =0\,,\qquad I^m\,Z_m = 0\,. 
\label{eq:GSE-conditions}
\end{align}
The conventional dilaton is included in $Z_m$ as follows:
\begin{align}
 Z_m = \partial_m \Phi + U_m \,. 
\end{align}
Note that the equations of motion in \eqref{eq:GSE-NSNS} reduce to the usual supergravity ones if $I^m=0$ and $U_m=0$ are satisfied. 
Since the GSE depend on $\Phi$ and $U_m$ only through the combination $Z_m$,
there is an ambiguity in the decomposition of $Z=\rmd\Phi+U$ into $\rmd\Phi$ and $U$. 
Namely, at the level of the equations of motion, there is a local symmetry,
\begin{align}
 \Phi(x)\to \Phi(x) + \omega(x)\,,\qquad U(x)\to U(x)-\rmd \omega(x) \,.
\end{align}
Therefore, for a given solution of the GSE, we can always choose the dilaton to satisfy
\begin{align}
 \Lie_I \Phi = I^m\,\partial_m \Phi = 0 \,. 
\label{eq:dilaton-isometry}
\end{align}

\medskip 

In \cite{Sakatani:2016fvh}, by using techniques developed in the DFT, 
the above equations of motion have been reformulated 
in a manifestly $\OO(D,D)$ $T$-duality-covariant form (see \cite{Sakatani:2016fvh} for more details),
\begin{align}
 \bS_{MN} = 0 \,,\qquad \bS = 0 \,,\qquad 
 \gLie_\bX \cH_{MN}=0\,,\qquad \gLie_\bX d=0\,,\qquad \bX^M \bX_M =0\,,
\label{eq:mDFT-eom}
\end{align}
where $\cH_{MN}$ and $d(x)$ are the generalized metric (\ref{eq:H-geometric}) and the DFT dilaton (\ref{eq:DFT-dilaton}), respectively.
As in the usual case, we suppose that all of the fields and gauge parameters satisfy the so-called strong and weak constraints (\ref{eq:strong-const}), (\ref{eq:weak-const}).

\medskip

Let us explain new ingredients ($\bX^M$\,, $\bS_{MN}$\, $\bS$) in the equations (\ref{eq:dilaton-isometry}).
A generalized vector field $\bX^M$, which is absent in the conventional DFT, is parameterized as
\begin{align}
 \bigl(\bX^M\bigr) = \begin{pmatrix} I^m \\ U_m + B_{mn}\, I^n
 \end{pmatrix} \,,
\label{eq:X-parameterization}
\end{align}
where $I^m$ and $U_n$ here are identified with the ones appearing in the GSE. 
The last three equations in \eqref{eq:mDFT-eom}, which reproduce \eqref{eq:GSE-conditions} and \eqref{eq:dilaton-isometry}, indicate that the generalized vector $\bX^M$ is a null generalized Killing vector. 
On the other hand, the first two equations in \eqref{eq:mDFT-eom} describe 
the dynamics of $\cH_{MN}(x)$ and $d(x)$\,. 
In particular, the first equation reproduces the first two equations in \eqref{eq:GSE-NSNS} 
and the second equation leads to the last equation in \eqref{eq:GSE-NSNS}. 
In fact, $\bS_{MN}$ and $\bS$ are the generalized Ricci tensor/scalar associated 
with the covariant derivative satisfying (see \cite{Sakatani:2016fvh} for more details)
\begin{align}
 \bnabla_K \eta_{MN}=0\,,\qquad 
 \gLie_V=\gLie^{\nabla}_V\,,\qquad 
 \bnabla_K \cH_{MN}=0\,,\qquad 
 \bnabla_M d + \bX_M = 0 \,. 
\label{eq:generalized-connection}
\end{align}
The explicit expressions of the modified quantities, the generalized connection 
$\bGamma_{MNK}$, the generalized Ricci tensor $\bS_{MN}$, and the generalized Ricci scalar $\bS$, 
in terms of $(\cH_{MN},\, d,\, \bX^M)$ or $(\CG_{mn},\,B_{mn},\,\Phi,\,I^m,\,U_m)$, 
can be found in \cite{Sakatani:2016fvh} (see Sects.~3.2, 4.1, and 4.2 therein). 

\medskip

The corresponding quantities in the conventional DFT, $\Gamma_{MNK}$, $\cS_{MN}$, 
and $\cS$, can be reproduced from these modified quantities by setting $\bX^M=0$. 
Conversely, the modified quantities $(\bGamma_{MNK},\,\bS_{MN},\,\bS)$ can be obtained 
from $(\Gamma_{MNK},\,\cS_{MN},\,\cS)$ with the replacement
\begin{align}
 \partial_M d\ \to\ \partial_M d +\bX_M \,. 
\label{eq:modification}
\end{align}
The meaning of this shift will be clarified in the next subsection. 

\subsection{(m)DFT for ``DFT on a modified section''}
\label{sec:mDFT-is-DFT}

In this subsection, we show that the mDFT, which was reviewed in the previous subsection, 
is equivalent to the conventional DFT with a non-standard solution of the strong constraint. 
In this sense, the mDFT should rather be called the (m)DFT. 

\medskip 

Let us first prove that by performing a certain generalized coordinate transformation, 
the null generalized Killing vector $\bX^M$ can always be brought into the following form:%
\footnote{Our proof partially follows the discussion given in Sect.~3.1 of \cite{Hull:2006qs}.}
\begin{align}
 \bX^M \equiv \begin{pmatrix} I^m \\ U_m + B_{mn}\, I^n
 \end{pmatrix} = \begin{pmatrix} I^m \\ 0
 \end{pmatrix}\qquad (I^m:\text{constant})\,. 
\label{eq:X-gauge}
\end{align}
This statement can be regarded as a generalization of the well-known fact 
in the Riemannian geometry that we can always find a certain coordinate system 
where the components of a Killing vector are constant (see, for example, \cite{Wald}). 

\medskip 

From the strong constraint, we can always find a section where all of the fields 
$(\cH_{MN},\,d,\,\bX^M)$ are independent of the dual coordinates. 
With this choice of section, the null and the generalized Killing properties 
lead to the conditions \eqref{eq:GSE-conditions} and \eqref{eq:dilaton-isometry}. 
Since $I^m$ is a Killing vector field, we can always find a certain coordinate system $(x^m)=(x^\mu,\,y)$ 
in which the Killing vector is a coordinate basis: $I^m= c\,\delta^m_y$\,, where $c$ is a constant. 
In such a coordinate system, both $\CG_{mn}$ and $\Phi$ are independent of $y$\,. 
The 3-form $H_3$ is also independent of $y$, as we can easily show $\Lie_I H_3=0$ from \eqref{eq:GSE-conditions}. 
Thus, we can generally expand $H_3$ as
\begin{align}
 H_3 = h_3 + c^{-1}\,\iota_I H_3\wedge \rmd y = h_3 - c^{-1}\,\rmd Z \wedge \rmd y \qquad \bigl(\iota_I h_3=0\bigr)\,,
\end{align}
where we used \eqref{eq:GSE-conditions}, and $h_3$ should satisfy $\Lie_I h_3=0$ 
that follows from $\Lie_I H_3=0$ and $\Lie_I Z=0$\,. 
From this expansion, we find an expansion of the $B$-field satisfying $H_3=\rmd B_2$\,,
\begin{align}
 B_2 = b_2 - c^{-1}\,U \wedge \rmd y \qquad \bigl(\iota_I b_2=0\,, \quad h_3=\rmd b_2 \bigr)\,,
\label{eq:B_2-solution}
\end{align}
where we used $\rmd Z=\rmd U$, and $b_2$ can always be chosen 
such that $\Lie_I b_2=0$ is satisfied. 
This shows that we can always take a gauge (for generalized diffeomorphisms) so that $B_{mn}$ is also independent of $y$ (i.e., $\Lie_I B_2=0$), 
and all of the NS--NS fields are now independent of $y$\,. 
From \eqref{eq:B_2-solution} and $\iota_I U=0$, which comes from \eqref{eq:GSE-conditions} and \eqref{eq:dilaton-isometry}, we also find the relation
\begin{align}
 \iota_I B_2 - U=0 \,.
\label{eq:i_IB=U}
\end{align}
This completes the proof that a null generalized Killing vector $\bX^M$ can always be brought into the form \eqref{eq:X-gauge}. 

\medskip 

Then, since all of the fields are independent of $y$\,, 
the $\tilde{y}$ dependence can sneak in without violating the strong constraint. 
Indeed, in a coordinate system where \eqref{eq:X-gauge} is realized, 
the shift \eqref{eq:modification} from the DFT to the mDFT can be interpreted 
as an implicit introduction of the linear $\tilde{y}$ dependence 
into the dilaton $d_*(x)$ or $\Phi_*(x)$:
\begin{align}
 (\partial_M d)=\begin{pmatrix}
 \partial_m d\\ 0
 \end{pmatrix}\quad \to\quad (\partial_M d+\bX_M)=\begin{pmatrix}
 \partial_m d \\ I^m
 \end{pmatrix} = 
 (\partial_M d_*) \,,\qquad 
 d_* \equiv d+ c\,\tilde{y} \,. 
\end{align}
When the dilaton does not depend on $\tilde{y}$ (i.e., $c=0$), 
$I^m$ vanishes and the modification disappears. 
From the above argument, solutions of the mDFT can always be described 
as solutions of the DFT in which the dilaton has a linear dual-coordinate dependence. 

\medskip 

Conversely, let us consider a solution of the DFT where the DFT dilaton has a linear dual-coordinate dependence, 
$d_*=d(x)+c^m \,\tilde{y}_m$ with constant $c^m$\,. 
From the identification
\begin{align}
 \partial_M d + \bX_M = \partial_M d_*\,,
\end{align}
we obtain that 
\begin{align}
 I^m=c^m\,, \qquad U_m + B_{mn}\, I^n=0\,.
\end{align}
Note here that $d_*$, say the (m)DFT dilaton, can have only linear dependence 
on the dual coordinates with constant coefficients if we prefer to avoid 
the explicit appearance of the dual coordinates in $I^m$ and $\partial_m d$\,. 
Suppose that all of the fields, collectively denoted by $\varphi$, are independent of $\tilde{y}$\,. 
Then the strong constraint requires that
\begin{align}
 0 = \partial^M d\,\partial_M \varphi = I^m \,\partial_m \varphi = \Lie_I \varphi \,. 
\end{align}
Here, in the last equality we have used the fact that $I^m$ is constant 
in order to express the condition in a covariant form. 
Then, the following conditions, namely \eqref{eq:GSE-conditions} and \eqref{eq:dilaton-isometry}, are automatically satisfied:
\begin{align}
 \Lie_I \CG_{mn} = 0\,, \quad I^m\,\partial_m\Phi = 0\,, \quad \iota_I H_3 + \rmd U =0 \,, 
\quad I^m\, U_m =0 \,. 
\label{eq:condition-section}
\end{align}
Here, the dilaton $\Phi$ is defined through the relation
\begin{align}
 \Exp{-2d}=\Exp{-2\Phi}\sqrt{\abs{\CG}}\,,
\end{align}
and it is independent of $\tilde{y}$\,. 
When the R--R fields are also introduced, they should also satisfy
\begin{align}
 \Lie_I \cF_p = 0\,. 
\end{align}

\medskip 

From the above viewpoint, it is not necessary to look for the action for the GSE. 
The DFT action supplies the $2D$-dimensional equations of motion of the DFT. 
If the DFT dilaton has the dual-coordinate dependence, the equations of motion take the form of the GSE. 

\section{Ramond--Ramond sector of (m)DFT}
\label{sec:R-R(m)DFT}

In this section, we introduce the R--R fields by following the well-established formulation of the DFT 
\cite{Hohm:2011zr,Hohm:2011dv,Geissbuhler:2011mx,Jeon:2012kd,Geissbuhler:2013uka}.
\footnote{The approaches in \cite{Hohm:2011zr,Hohm:2011dv}, \cite{Jeon:2012kd}, 
and \cite{Geissbuhler:2011mx,Geissbuhler:2013uka}, respectively, are slightly different from each other. 
In this section, we basically follow the approach of \cite{Hohm:2011zr,Hohm:2011dv}. 
It is also useful to follow \cite{Jeon:2012kd} when we consider the type II supersymmetric DFT 
\cite{Jeon:2012hp}.}
The whole bosonic part of the GSE is reproduced from the equations of motion of the DFT 
by choosing a modified section. 

\subsection{Ramond--Ramond sector of DFT}

\paragraph*{Gamma matrices and O$(D,D)$ spinors\\}

It is convenient to introduce the gamma matrices $\{\gamma^M\} = \{\gamma^m,\,\gamma_m\}$ 
satisfying the $\OO(D,D)$ Clifford algebra,%
\footnote{In this thesis, we call the Pin$(D,D)$ group simply $\OO(D,D)$\,. }
\begin{align}
 \bigl\{\gamma^M,\, \gamma^N \bigr\} = \eta^{MN}\,. 
\end{align}
Here, the gamma matrices are real and satisfy $\gamma^m=(\gamma_m)^\rmT$\,. 
The gamma matrices with multi-indices are defined as 
\begin{align}
 \gamma^{M_1\cdots M_p}\equiv \gamma^{[M_1}\cdots \gamma^{M_p]}\,, \qquad 
 \gamma^{m_1\cdots m_p}\equiv \gamma^{[m_1}\cdots \gamma^{m_p]} = \gamma^{m_1}\cdots \gamma^{m_p}\,.
\end{align}
From the anti-commutation relations,
\begin{align}
 \{\gamma^m,\,\gamma_n\}=\delta^m_n\,,\qquad \{\gamma^m,\,\gamma^n\}=0
=\{\gamma_m,\,\gamma_n\} \,,
\end{align}
the $\gamma_m$ can be regarded as fermionic annihilation operators. 
The Clifford vacuum $\ket{0}$ is defined so as to satisfy
\begin{align}
 \gamma_m\ket{0}=0\,,\qquad \langle 0\vert 0\rangle = 1 \,,
\end{align}
where $\bra{0}\equiv \ket{0}^\rmT$. 
By acting $\gamma^m$ matrices on $\ket{0}$, an $\OO(D,D)$ spinor can be constructed as 
\begin{align}
 \ket{\alpha_p} \equiv \frac{1}{p!}\,\alpha_{m_1\dots m_p}\,\gamma^{m_1\cdots m_p}\ket{0}\,,
\end{align}
and it is in one-to-one correspondence with a $p$-form,
$\alpha_p \equiv\frac{1}{p!}\,\alpha_{m_1\dots m_p}\,\rmd x^{m_1}\wedge\cdots\wedge \rmd x^{m_p}$\,. 
A formal sum of $\OO(D,D)$ spinors,
\begin{align}
 \ket{\alpha}\equiv \sum_p \frac{1}{p!}\,\alpha_{m_1\dots m_p}\,\gamma^{m_1\cdots m_p}\ket{0}\,,
\end{align}
corresponds to a poly-form $\alpha=\sum_p \alpha_p$\,. 

\medskip

The $\OO(D,D)$ transformations are generated by $\gamma^{MN}$ satisfying 
\begin{align}
 [\gamma^{MN},\,\gamma_L] = \gamma_K\,(T^{MN})^K{}_L\,,\qquad 
 (T^{MN})^K{}_L \equiv 2\,\eta^{K[M}\,\delta^{N]}_L \,. 
\end{align}
By utilizing the generators, one can define the following quantities: 
\begin{align}
 S_{e^\rmT} \equiv \Exp{\frac{1}{2}\,h^m{}_n\,[\gamma_m,\,\gamma^n]} \quad 
\bigl[e_m{}^n\equiv (\Exp{h^\rmT})_m{}^n\bigr]\,,\qquad
 \Exp{\bB} \equiv \Exp{\frac{1}{2}\,B_{mn}\,\gamma^{mn}} \,,\qquad
 \Exp{\bbeta} \equiv \Exp{\frac{1}{2}\,\beta^{mn}\,\gamma_{mn}} \,. 
\end{align}
We can easily show that they satisfy
\begin{align}
\begin{split}
 S_{e^\rmT} \,\gamma_N S_e^{-1} &= \gamma_M\,(\Lambda_{e^\rmT})^M{}_N\,,\quad
 \Exp{\bB}\, \gamma_N \Exp{-\bB} = \gamma_M\,(\Lambda_B)^M{}_N \,,\quad
 \Exp{\bbeta}\, \gamma_N \Exp{-\bbeta} = \gamma_M\,(\Lambda_\beta)^M{}_N \,,
\\
 \Lambda_{e^\rmT}&\equiv\begin{pmatrix}
 (e^\rmT)^m{}_n & 0 \\ 0 & (e^{-1})_m{}^n
 \end{pmatrix} \,,\quad
 \Lambda_B\equiv\begin{pmatrix}
 \delta^m_n & 0 \\ B_{mn} & \delta_m^n
 \end{pmatrix} \,,\quad
 \Lambda_\beta\equiv\begin{pmatrix}
 \delta^m_n & \beta^{mn} \\ 0 & \delta_m^n
 \end{pmatrix} \,.
\end{split}
\end{align}
It is helpful to define the correspondent of the flat metric as 
\begin{align}
\begin{split}
 &S_k\equiv \gamma_0\,\gamma^0 - \gamma^0\,\gamma_0 = S_k^{-1}= S_k^\rmT \,,
\qquad 
 S_k\,\gamma_N\,S_k^{-1} = \gamma_M\,k^M{}_N \,,
\\
 &(k^M{}_N)\equiv \begin{pmatrix}
 k^m{}_n & 0 \\ 0 & k_m{}^n
 \end{pmatrix} \,, \quad 
 (k^m{}_n) \equiv \diag(-1,+1,\dotsc,+1) \equiv (k_m{}^n) \equiv (k_{mn}) \,. 
\end{split}
\end{align}
The correspondent of the $B$-untwisted metric,%
\footnote{See \cite{Hull:2014mxa} for discussions of the untwisted form of generalized tensors.}
\begin{align}
 (\hat{\cH}_{MN})\equiv \begin{pmatrix} \CG_{mn} & 0 \\ 0 & \CG^{mn} \end{pmatrix}\,,\qquad 
 \CG_{mn}\equiv e_m{}^k\,e_n{}^l\,k_{kl}\,,\qquad (\CG^{mn})\equiv (\CG_{mn})^{-1}\,,
\end{align}
can also be defined as
\begin{align}
 S_{\hat{\cH}} \equiv S_e\,S_k\,S_{e^\rmT} =S_{\hat{\cH}}^\rmT\,,\quad 
S_e\equiv (S_{e^\rmT})^\rmT\,, \quad 
 S_{\hat{\cH}} \,\gamma_N\,S_{\hat{\cH}}^{-1} = (\gamma^M)^\rmT\,\hat{\cH}_{MN}\,. 
\end{align}
This gives a natural metric,
\begin{align}
 \bra{\alpha}S_{\hat{\cH}} \ket{\beta} &= \sum_{p,q}\frac{1}{p!\,q!}\,
\alpha_{m_1\cdots m_p}\,\beta_{n_1\cdots n_q}\bra{0}\,\gamma_{m_p\cdots m_1} \,S_{\hat{\cH}} \, 
\gamma^{n_1\cdots n_q} \ket{0}
\nn\\
 &= \sum_{p,q}\frac{1}{p!\,q!}\,\alpha_{m_1\cdots m_p}\,\beta_{n_1\cdots n_q}\bra{0}\,
\gamma_{m_p\cdots m_1}\,(\gamma_{l_1\cdots l_q})^\rmT \,S_{\hat{\cH}}\, \ket{0}\,\CG^{l_1n_1}
\cdots \CG^{l_qn_q}
\nn\\
 &= \sqrt{\abs{\CG}}\sum_p \frac{1}{p!}\, \CG^{m_1n_1}\cdots 
\CG^{m_pn_p}\,\alpha_{m_1\cdots m_p}\,\beta_{n_1\cdots n_p} 
 = \bra{\beta}S_{\hat{\cH}}\ket{\alpha} \,.
\label{eq:inner-product}
\end{align}

\medskip

The charge conjugation matrix is defined as 
\begin{align}
\begin{split}
 &C \equiv (\gamma^0\pm \gamma_0) \cdots (\gamma^{D-1}\pm \gamma_{D-1}) 
\qquad (D:\text{even/odd})\,, 
\\
 &C\,\gamma^M\,C^{-1} = -(\gamma^M)^\rmT \,,\qquad C^{-1}= (-1)^{\frac{D(D+1)}{2}}C = C^\rmT \,.
\end{split}
\end{align}
By using $C$ and $S_{\hat{\cH}}$, the Hodge dual can be constructed as 
\begin{align}
 C\,S_{\hat{\cH}} \, \gamma^{m_1\cdots m_p}\ket{0} 
 &= \sqrt{\abs{\CG}}\, (-1)^p \CG^{m_1n_1}\cdots \CG^{m_pn_p}\, \gamma_{n_1\cdots n_p} \, C \ket{0} 
\nn\\
 &= \frac{1}{(D-p)!}\, (-1)^{\frac{p(p+1)}{2}}\, 
\varepsilon^{m_1\cdots m_p}{}_{n_1\cdots n_{D-p}} \, \gamma^{n_1\cdots n_{D-p}} \ket{0} \,, 
\label{eq:SH-Gamma}
\end{align}
where $\varepsilon_{01\cdots (D-1)}=+\sqrt{\abs{\CG}}$ and indices are raised or lowered 
with $\CG_{mn}$.

\medskip

The correspondent of the generalized metric is defined as 
\begin{align}
\begin{split}
 &S_\cH \equiv \Exp{-\bB^\rmT} S_{\hat{\cH}} \Exp{-\bB} = S_\cH^\rmT\,,\qquad 
 S_\cH \,\gamma_N\,S_\cH^{-1} = (\gamma^M)^\rmT\,\cH_{MN}\,,
\\
 &(\cH_{MN}) = \begin{pmatrix} \delta_m^k & B_{mk} \\
  0 & \delta^m_k 
 \end{pmatrix}\begin{pmatrix} \CG_{kl} & 0 \\
  0 & \CG^{kl} 
 \end{pmatrix}\begin{pmatrix} \delta^l_n & 0 \\
  -B_{ln} & \delta_l^n 
 \end{pmatrix}\,. 
\end{split}
\end{align}
As stressed in \cite{Hohm:2011dv}, $S_\cH$ is a particular parameterization of the fundamental field 
$\bbS$ that corresponds to the generalized metric $\cH_{MN}$ before providing a parameterization. 
If we take another parameterization of the generalized metric,
\begin{align}
 (\cH_{MN}) = \begin{pmatrix} \delta_m^k & 0 \\
  -\beta^{mk} & \delta^m_k 
 \end{pmatrix} \begin{pmatrix} \OG_{kl} & 0 \\
  0 & \OG^{kl} 
 \end{pmatrix} \begin{pmatrix} \delta^l_n & \beta^{kn} \\
  0 & \delta_l^n 
 \end{pmatrix} \,,
\label{eq:non-geometric-parameterization}
\end{align}
then the field $\bbS$ is parameterized as
\begin{align}
 S_{\tilde{\cH}} \equiv \Exp{\bbeta^\rmT} S_{\check{\cH}} \Exp{\bbeta}\,,\qquad 
 S_{\check{\cH}} \equiv S_{\tilde{e}}\,S_k\,S_{\tilde{e}^\rmT} \,.
\end{align}

\medskip

For later discussion, it is also convenient to define the following quantity,
\begin{align}
 \cK \equiv -C\,\bbS = C\,\cK^\rmT\,C \,, \qquad \cK\,\gamma_M\,\cK^{-1} 
= -\cH_M{}^N \, \gamma_N\,,
\end{align}
and the chirality operator
\begin{align}
 \gamma^{D+1} \equiv (-1)^{N_F}\,,\qquad N_F\equiv \gamma^m\,\gamma_m \,. 
\end{align}
The chirality operator acts on an $\OO(D,D)$ spinor as
\begin{align}
 \gamma^{D+1} \,\ket{\alpha} = \sum_p (-1)^p \ket{\alpha_p} \,.
\end{align}
A chiral/anti-chiral spinor corresponds to a poly-form with even/odd degree.

\paragraph*{The classical action and equations of motion\\}

The dynamical fields that correspond to the R--R fields are introduced as $\OO(D,D)$ spinors,
\begin{align}
 \ket{A} \equiv \sum_p \frac{1}{p!}\, A_{m_1\cdots m_p}\,\gamma^{m_1\cdots m_p}\ket{0} \,,
\end{align}
which transform under generalized diffeomorphisms \cite{Hohm:2011zr,Hohm:2011dv} as follows:
\begin{align}
 \delta_V \ket{A} = \gLie_V \ket{A}
 \equiv V^M\,\partial_M \ket{A} + \partial_M V_N\, \gamma^M\,\gamma^N \,\ket{A} \,.
\label{eq:generalized-diffeo-RR}
\end{align}
Depending on the type IIA or IIB theory, $\ket{A}$ takes a definite chirality,
\begin{align}
 \gamma^{11}\ket{A} = \mp \ket{A} \quad (\text{IIA/IIB})\,.
\label{eq:chirality}
\end{align}
It is easy to show that under the strong constraint, the R--R field strength,
\begin{align}
 \ket{F} \equiv \sla{\partial} \ket{A} \equiv \gamma^M\, \partial_M \ket{A} \,,
\label{eq:F-DA}
\end{align}
transforms covariantly under generalized diffeomorphisms. 

\medskip

From the strong constraint, the operator $\sla{\partial}$ is nilpotent, and one can readily see 
that the field strength is invariant under the gauge transformations for the R--R fields,
\begin{align}
 \delta_\lambda \ket{A} = \sla{\partial} \ket{\lambda}\,,
\end{align}
where $\ket{\lambda}$ is an arbitrary $\OO(D,D)$ spinor which respects the chirality \eqref{eq:chirality}. 
The Bianchi identity also follows from the nilpotency $\sla{\partial}^2=0$,
\begin{align}
 \sla{\partial} \ket{F} = \sla{\partial}^2 \ket{A} = 0 \,. 
\label{eq:Bianchi-id}
\end{align}

\medskip

In the democratic formulation \cite{Fukuma:1999jt,Bergshoeff:2001pv}, the self-duality relation should be imposed 
at the level of the equations of motion. 
In our convention, it takes the form
\begin{align}
 \ket{F} = \cK\, \ket{F} \,. 
\label{eq:self-duality}
\end{align}
From this expression and the Bianchi identity, we obtain the following relation:
\begin{align}
 \sla{\partial} \cK\, \ket{F} = 0 \,. 
\label{eq:eom-RR}
\end{align}
This is nothing but the equation of motion for the R--R field, as we will see below. 

\medskip

Now, let us write down the bosonic part of the (pseudo-)action for the type II supergravity,
\begin{align}
 \cL = \Exp{-2d} \cS -\frac{1}{4}\,\bra{F}\, \bbS\, \ket{F} 
 = \Exp{-2d} \cS + \frac{1}{4}\,\overline{\bra{F}}\, \cK\, \ket{F} \,,
\label{eq:Lagrangian-full}
\end{align}
where we have defined $\overline{\bra{F}}\equiv \bra{F}\,C^\rmT$. 
Taking a variation with respect to $A$, we obtain
\begin{align}
 \delta_A\cL = \frac{1}{2}\,\overline{\bra{\delta A}}\,\sla{\partial} \cK\, \ket{F}
 - \partial_M \Bigl[ \frac{1}{2}\,\overline{\bra{\delta A}}\, \gamma^M\,\cK\, \ket{F}\Bigr] \,,
\end{align}
and as expected, the equations of motion for $A$ reproduce \eqref{eq:eom-RR}. 
The variation with respect to the DFT dilaton becomes
\begin{align}
 \delta_d \cL = - 2\Exp{-2d} \cS\,\delta d 
 + \partial_M \bigl(4\Exp{-2d} \cH^{MN}\, \partial_N \delta d \bigr) \,, 
\end{align}
and there is no contribution from the R--R fields. 
Finally, the variation with respect to the generalized metric gives rise to 
\begin{align}
 \delta_\cH \cL &= - \frac{1}{2} \Exp{-2d} \cS^{MN}\, \delta \cH_{MN} 
 - \partial_M \bigl[\Exp{-2d} \nabla_N\,\delta \cH^{MN} \bigr]
 + \frac{1}{4}\, \overline{\bra{F}}\, \delta \cK\, \ket{F} 
\nn\\
 &= - \frac{1}{2} \Bigl(\Exp{-2d} \cS^{MN}- \frac{1}{4}\, \cH^{(M}{}_K\,\overline{\bra{F}}\,
\gamma^{N)K}\,\cK\, \ket{F}\Bigr)\, \delta \cH_{MN} 
 - \partial_M \bigl[\Exp{-2d} \nabla_N\,\delta \cH^{MN} \bigr] 
\nn\\
 &= - \frac{1}{2}\,\Exp{-2d} \bigl(\cS^{MN} + \cE^{MN}\bigr)\, \delta \cH_{MN} 
 - \partial_M \bigl[\Exp{-2d} \nabla_N\,\delta \cH^{MN} \bigr] \,,
\end{align}
where we have employed the identity \cite{Hohm:2011dv}
\begin{align}
 \delta \cK = \frac{1}{2}\,\cH^{(M}{}_K\,\gamma^{N)K}\,\cK\,\delta\cH_{MN} \,, 
\end{align} 
in the second equality, and defined the ``energy--momentum tensor'' \cite{Hohm:2011zr,Hohm:2011dv}
\begin{align}
 \cE^{MN} \equiv \frac{1}{4} \Exp{2d}\,\Bigl[\bra{F} \,(\gamma^{(M})^\rmT\,\bbS\,\gamma^{N)}\, 
\ket{F} - \frac{1}{2}\,\cH^{MN}\,\bra{F}\,\bbS\,\ket{F}\Bigr] \,.
\end{align}

\medskip

In summary, the equations of motion of the DFT are given by 
\begin{align}
 \cS_{MN} + \cE_{MN} = 0 \,,\qquad \cS = 0 \,, \qquad 
 \sla{\partial} \, \cK\, \ket{F} = 0\,.
\label{eq:EOM-DFT}
\end{align}

\paragraph*{The classical action and equations of motion in the conventional formulation\\}

Next, let us show that the expressions we obtained above indeed reproduce 
the well-known expressions in conventional supergravity 
by choosing \eqref{eq:H-geometric}, \eqref{eq:DFT-dilaton}, and $\tilde{\partial}^m=0$. 

\medskip

From \eqref{eq:generalized-diffeo-RR} and $\tilde{\partial}^m=0$, 
the transformation of the R--R field under a generalized diffeomorphism becomes
\begin{align}
 \delta_V\ket{A}
 &= \bigl[v^m\,\partial_m + \partial_m v^n\,\gamma^m\,\gamma_n 
+ \tfrac{1}{2}\,(\partial_m\tilde{v}_n-\partial_n\tilde{v}_m)\,\gamma^{mn}\bigr]\,\ket{A}
\nn\\
 &= \ket{\Lie_v A + \rmd \tilde{v}\wedge A} \,,
\end{align}
and it is equivalent to a conventional diffeomorphism and a $B$-field gauge transformation. 
The field strength \eqref{eq:F-DA} and the Bianchi identity \eqref{eq:Bianchi-id} take the following form: 
\begin{align}
 \ket{\cF} = \sla{\partial} \ket{A} = \gamma^m\, \partial_m \ket{A} = \ket{\rmd A} \,, \qquad 
 \sla{\partial}\ket{\cF} = \ket{\rmd\cF} =0 \,. 
\end{align}
The self-duality relation for the R--R field strength becomes
\begin{align}
 \ket{F} = -C\,\Exp{-\bB^\rmT} S_{\hat{\cH}} \Exp{-\bB}\, 
\ket{F} = - \Exp{\bB} C\, S_{\hat{\cH}} \Exp{-\bB}\, \ket{F} \,. 
\label{eq:self-dual0}
\end{align}
If the $B$-untwisted field strength is defined as 
\begin{align}
 \ket{\hat{F}} \equiv \Exp{-\bB}\, \ket{F} \,,
\end{align}
which is invariant under the $B$-field gauge transformations, \eqref{eq:self-dual0} can be rewritten as
\begin{align}
 \ket{\hat{F}} = - C\, S_{\hat{\cH}} \, \ket{\hat{F}} \,. 
\end{align}
From \eqref{eq:SH-Gamma}, we obtain the following relations:
\begin{align}
 - C\, S_{\hat{\cH}} \, \ket{\hat{F}} 
 = - \sum_p \frac{1}{p!}\, \hat{F}_{m_1\cdots m_p}\, C\, S_{\hat{\cH}}\gamma^{m_1\cdots m_p}\ket{0} 
 = \sum_p (-1)^{\frac{p(p+1)}{2}+1}\, \ket{*\hat{F}_p} \,, 
\end{align}
and the self-duality relation \eqref{eq:self-duality} becomes
\begin{align}
 * \hat{F}_p = (-1)^{\frac{p(p+1)}{2}+1} \hat{F}_{10-p} \,,\qquad 
 \hat{F}_p = (-1)^{\frac{p(p-1)}{2}} * \hat{F}_{10-p} \,. 
\label{eq:self-duality-form}
\end{align}
From this relation and the Bianchi identity for $\hat{F}$,
\begin{align}
 \rmd \hat{F} + H_3\wedge \hat{F} = 0 \,,
\end{align}
the equation of motion becomes
\begin{align}
 \rmd * \hat{F} - H_3\wedge * \hat{F} =0 \,,
\end{align}
where, compared to the Bianchi identity, the sign in front of $H_3$ is flipped, 
according to the sign in \eqref{eq:self-duality-form}. 

\medskip

From \eqref{eq:inner-product}, the action \eqref{eq:Lagrangian-full} becomes
\begin{align}
 \cL =\sqrt{\abs{\CG}} \biggl[\Exp{-2\Phi} \Bigl(R + 4\,\sfD^m \partial_m \Phi - 4\,
\abs{\partial \Phi}^2 - \frac{1}{2}\,\abs{H_3}^2\Bigr) - \frac{1}{4} \sum_p \abs{\hat{F}_p}^2\biggr] \,. 
\end{align}
In order to evaluate the equations of motion \eqref{eq:EOM-DFT}, 
let us recall that $\cS_{MN}$ takes the form \cite{Sakatani:2016fvh}
\begin{align}
\begin{split}
 (\cS_{MN})
 &=\begin{pmatrix}
 2\,\CG_{(m|k}\, s^{[kl]}\, B_{l|n)} - s_{(mn)} 
 - B_{mk}\, s^{(kl)}\, B_{ln} & B_{mk}\,s^{(kn)} - \CG_{mk}\, s^{[kn]} \\
 s^{[mk]}\, \CG_{kn} -s^{(mk)}\, B_{km} & s^{(mn)}
 \end{pmatrix} \,,
\\
 s_{(mn)} &\equiv R_{mn}-\frac{1}{4}\,H_{mpq}\,H_n{}^{pq} + 2 \sfD_m \partial_n \Phi \,,
\qquad
 s_{[mn]} \equiv - \frac{1}{2}\,\sfD^k H_{kmn} + \partial_k\Phi\,H^k{}_{mn} \,. 
\end{split}
\end{align}
In fact, $\cE_{MN}$ also takes a similar form. 
From the rewriting
\begin{align}
 \cE_{MN} = \frac{1}{4}\Exp{2d}\,\bra{\hat{F}}\,\bigl[(\gamma_K)^\rmT\,S_{\hat{\cH}}\,\gamma_L 
-\tfrac{1}{2}\,\hat{\cH}_{KL}\,S_{\hat{\cH}}\bigr]\,\ket{\hat{F}} \,(\Lambda_{-B^\rmT})^K{}_M
\,(\Lambda_{-B})^L{}_N\,,
\end{align}
it is straightforward to derive 
\begin{align}
\begin{split}
 (\cE_{MN})&= \begin{pmatrix}
 -2\,\CG_{(m|k}\, \cK^{kl}\, B_{l|n)} + T_{mn} 
 + B_{mk}\, T^{kl}\, B_{ln} & -B_{mk}\,T^{kn} + \CG_{mk}\, \cK^{kn} \\
 -\cK^{mk}\, \CG_{kn} + T^{mk}\, B_{km} & -T^{mn}
 \end{pmatrix} \,,
\\
 T_{mn} &\equiv \frac{1}{4} \Exp{2\Phi}\, \sum_p \Bigl[ \frac{1}{(p-1)!}\, 
\hat{F}_{(m}{}^{k_1\cdots k_{p-1}} \hat{F}_{n) k_1\cdots k_{p-1}} 
- \frac{1}{2}\, \CG_{mn}\,\abs{\hat{F}_p}^2 \Bigr] \,,
\\
 \cK_{mn}&\equiv \frac{1}{4}\Exp{2\Phi}\, \sum_p \frac{1}{(p-2)!}\, 
\hat{F}_{k_1\cdots k_{p-2}}\, \hat{F}_{mn}{}^{k_1\cdots k_{p-2}} \,. 
\end{split}
\end{align}
Thus, the equations of motion \eqref{eq:EOM-DFT} are summarized as
\begin{align}
\begin{alignedat}{2}
 &R_{mn}-\frac{1}{4}\,H_{mpq}\,H_n{}^{pq} + 2 \sfD_m \partial_n \Phi = T_{mn}\,, \quad&
 &R + 4\,\sfD^m \partial_m \Phi - 4\,\abs{\partial \Phi}^2 - \frac{1}{2}\,\abs{H_3}^2 = 0 \,,
\\
 &- \frac{1}{2}\,\sfD^k H_{kmn} + \partial_k\Phi\,H^k{}_{mn} = \cK_{mn} \,,\quad& 
 &\rmd * \hat{F}_p - H_3\wedge * \hat{F}_{p+2} =0 \,,
\end{alignedat}
\end{align}
where $p$ is even/odd for type IIA/IIB supergravity.

\subsection{Generalized type IIA/IIB equations}

As discussed in Sect.~\ref{sec:mDFT-is-DFT}, 
the equations of motion for the (m)DFT can be obtained 
by introducing a linear $\tilde{x}_m$ dependence 
into the DFT dilaton or the conventional dilaton,
\begin{align}
 d ~~\to~~ d_*\equiv d + I^m\,\tilde{x}_m \,,\qquad \Phi ~~\to~~ \Phi_* \equiv \Phi + I^m\,\tilde{x}_m \,. 
\end{align}
According to this replacement, the equations of motion for the NS--NS sector 
are modified \cite{Sakatani:2016fvh}. 
On the other hand, because there is no dilaton dependence in the equations of motion for the R--R sector, 
one may deduce that the R--R sector should not be modified. 
However, as we see in various examples \cite{Fernandez-Melgarejo:2017oyu}, the R--R fields $\ket{A}$ or $\ket{F}$ 
already include a non-linear dual-coordinate dependence through the dilaton $\Phi_*$,
\begin{align}
 \ket{A} = \Exp{-\Phi_*} \ket{\cA} \,,\qquad \ket{F} = \Exp{-\Phi_*} \ket{\cF} \,, 
\end{align}
where $\cA$ and $\cF$, to be called the $\Phi_*$-untwisted fields, are supposed 
to be independent of the dual coordinates. 
In fact, these rescaled fields, $\cA$ and $\cF$, appear more naturally 
in the vielbein formulations of the DFT discussed in 
\cite{Jeon:2012kd} and \cite{Geissbuhler:2011mx,Geissbuhler:2013uka}.

\medskip

In terms of the $\Phi_*$-untwisted fields, we obtain the relation
\begin{align}
 \ket{\cF} &= \Exp{\Phi_*}\sla{\partial}(\Exp{-\Phi_*} \ket{\cA}) 
 = \sla{\partial} \ket{\cA} - (\sla{\partial}\Phi_*)\,\ket{\cA} 
 = \gamma^m\,(\partial_m - \partial_m\Phi)\,\ket{\cA} - I^m \,\gamma_m\,\ket{\cA} 
\nn\\
 &= \ket{\rmd\cA - \rmd\Phi\wedge \cA - \iota_I \cA} \,. 
\end{align}
The Bianchi identity is also modified in a similar manner,
\begin{align}
 0 = \Exp{\Phi_*}\sla{\partial}(\Exp{-\Phi_*} \ket{\cF}) 
   = \ket{\rmd\cF - \rmd\Phi\wedge \cF - \iota_I \cF} \,.
\end{align}
More explicitly, we obtain
\begin{align}
 \cF_{p+1} =\rmd\cA_p - \rmd\Phi\wedge \cA_p - \iota_I \cA_{p+2}\,,\qquad 
 \rmd\cF_p - \rmd\Phi\wedge \cF_p - \iota_I \cF_{p+2} =0 \,.
\end{align}

\medskip

If we further introduce the $(\Phi_*,B)$-untwisted fields%
\footnote{The flat components of $\hat{\cC}$ correspond to the fundamental fields in 
\cite{Geissbuhler:2011mx,Geissbuhler:2013uka} 
by taking the conventional parameterization of the generalized vielbein 
in terms of the vielbein for $\CG_{mn}$ and $B_{mn}$.} defined as 
\begin{align}
 \hat{\cF} \equiv \Exp{-B_2\wedge} \cF \,,\qquad \hat{\cC} \equiv \Exp{-B_2\wedge} \cA \,, 
\end{align}
the relation between the potential and the field strength is represented by 
\begin{align}
 \hat{\cF} &= \Exp{-B_2\wedge} \bigl[\rmd (\Exp{B_2\wedge}\hat{\cC}) 
- \rmd\Phi\wedge (\Exp{B_2\wedge}\hat{\cC}) - \iota_I (\Exp{B_2\wedge}\hat{\cC})\bigr] 
\nn\\
 &= \rmd \hat{\cC} - Z\wedge \hat{\cC} - \iota_I \hat{\cC} + H_3\wedge \hat{\cC} \qquad (Z\equiv \rmd\Phi +\iota_I B_2)\,. 
\end{align}
Namely, the $(p+1)$-form field strength is given by
\begin{align}
 \hat{\cF}_{p+1} = \rmd \hat{\cC}_p - Z \wedge \hat{\cC}_p - \iota_I \hat{\cC}_{p+2} + H_3 \wedge \hat{\cC}_{p-2} \,. 
\end{align}
The Bianchi identity becomes
\begin{align}
 \rmd \hat{\cF}_p - Z \wedge \hat{\cF}_p - \iota_I \hat{\cF}_{p+2} + H_3 \wedge \hat{\cF}_{p-2} = 0\,.
\end{align}
The gauge transformation of the R--R potential is expressed as 
\begin{align}
 \delta_\lambda \hat{\cC} = \rmd \hat{\lambda} - Z \wedge \hat{\lambda} - \iota_I \hat{\lambda} 
+ H_3 \wedge \hat{\lambda} \,,
\end{align}
and the invariance of the field strength requires the nilpotency
\begin{align}
 0=(\rmd - Z \wedge - \iota_I + H_3 \wedge)^2 \hat{\lambda} 
= - \Lie_I \hat{\lambda} - \bigl(\rmd Z + \iota_I H_3 - \iota_I Z\bigr) \wedge \hat{\lambda} \,. 
\label{eq:nilpotency}
\end{align}
But this is indeed satisfied from the strong constraint 
\eqref{eq:condition-section} and $\Lie_I \lambda=0$\,. 

\medskip

The equations of motion become (see \cite{Sakatani:2016fvh} for the modification of the NS--NS sector)
\begin{align}
 &R_{mn}-\frac{1}{4}\,H_{mpq}\,H_n{}^{pq} + 2 \sfD_m \partial_n \Phi + \sfD_m U_n +\sfD_n U_m = T_{mn} \,,
\nn\\
 &R + 4\,\sfD^m \partial_m \Phi - 4\,\abs{\partial \Phi}^2 - \frac{1}{2}\,\abs{H_3}^2 
 - 4\,\bigl(I^m I_m+U^m U_m + 2\,U^m\,\partial_m \Phi - \sfD_m U^m\bigr) =0 \,,
\nn\\
 &-\frac{1}{2}\,\sfD^k H_{kmn} + \partial_k\Phi\,H^k{}_{mn} + U^k\,H_{kmn} + \sfD_m I_n - \sfD_n I_m = \cK_{mn} \,,
\label{eq:GSE}
\\
 &\rmd *\hat{\cF}_p - Z \wedge * \hat{\cF}_p -\iota_I * \hat{\cF}_{p-2} 
- H_3\wedge * \hat{\cF}_{p+2} =0 \,,
\nn
\end{align}
where we have introduced the following quantities: 
\begin{align}
\begin{split}
 T_{mn} &= \frac{1}{4} \sum_p \Bigl[ \frac{1}{(p-1)!}\, 
\hat{\cF}_{(m}{}^{k_1\cdots k_{p-1}} \hat{\cF}_{n) k_1\cdots k_{p-1}} - \frac{1}{2}\, 
\CG_{mn}\,\abs{\hat{\cF}_p}^2 \Bigr] \,,
\\
 \cK_{mn}&= \frac{1}{4} \sum_p \frac{1}{(p-2)!}\, \hat{\cF}_{k_1\cdots k_{p-2}}\, 
\hat{\cF}_{mn}{}^{k_1\cdots k_{p-2}} \,. 
\end{split}
\end{align}
For the type IIB case, these equations are nothing but 
the generalized type IIB equations (in the democratic form).

\subsection{$T$-duality transformation rules}

In this subsection, we present the $T$-duality transformation rule 
in a coordinate system in which \eqref{eq:X-gauge} is realized. 
If the fields $(\cH_{MN},\,d,\,A_p)$ are independent of a coordinate $z$, 
the following $T$-duality transformation along the $z$-direction maps 
a solution of the (m)DFT to another one of the (m)DFT:
\begin{align}
\begin{split}
 &\cH_{MN}\to \cH'_{MN}=(\Lambda^\rmT\,\cH\,\Lambda)_{MN} \,,\qquad 
 \Lambda\equiv (\Lambda^M{}_N) \equiv \begin{pmatrix} \delta^m_n - \delta^m_z\,\delta_n^z 
& \delta^m_z\,\delta^n_z \\ \delta_m^z\,\delta_n^z 
& \delta_m^n - \delta_m^z\,\delta^n_z \end{pmatrix}\,, 
\\
 &\ket{A}_{\text{IIA}} \to \ket{\cA'}_{\text{IIB}} = (\gamma_z -\gamma^z)\,\ket{\cA}_{\text{IIA}} \,,\qquad 
 \ket{A}_{\text{IIB}} \to \ket{\cA'}_{\text{IIA}} = (\gamma^z -\gamma_z)\,\ket{\cA}_{\text{IIB}} \,, 
\\
 &d\to d'= d + I^z\,z\,, \qquad I^z \to I'^z=0\,,\qquad I^i \to I'^i=I^i \,.
\end{split}
\end{align}
Here, $x^i$ is an arbitrary coordinate other than $z$\,. 
In the component expression, the above rule is represented by the following map:
\begin{align}
\begin{split}
 &\CG'_{ij} = \CG_{ij} - \frac{\CG_{iz}\, \CG_{jz}- B_{iz}\, B_{jz}}{\CG_{zz}}\,,\qquad 
 \CG'_{iz} = \frac{B_{iz}}{\CG_{zz}}\,,\qquad 
 \CG'_{zz}=\frac{1}{\CG_{zz}}\,,
\\
 &B'_{ij} = B_{ij} - \frac{B_{iz}\, \CG_{jz}- \CG_{iz}\, B_{jz}}{\CG_{zz}}\,,\qquad 
 B'_{iz} = \frac{\CG_{iz}}{\CG_{zz}} \,, 
\\
 &\Phi' =\Phi + \frac{1}{4}\,\ln\Bigl\lvert \frac{\det \CG'_{mn}}{\det \CG_{mn}} \Bigr\rvert 
+ I^z z \,,\qquad 
 I'^i = I^i\,,\qquad I'^z = 0 \,,
\\
 &\cA'_{i_1\cdots i_{p-1}z} = \cA_{i_1\cdots i_{p-1}} \,,
\qquad
 \cA'_{i_1\cdots i_p} = \cA_{i_1\cdots i_pz} \,. 
\end{split}
\end{align}
For the other R--R fields, the transformation rules are given by 
\begin{align}
\begin{split}
&\hat{\cC}'_{i_1\cdots i_{p-1}z} = \hat{\cC}_{i_1\cdots i_{p-1}} 
- (p-1)\,\frac{\hat{\cC}_{[i_1\cdots i_{p-2}|z|}\, \CG_{i_{p-1}]z}}{\CG_{zz}} \,,
\\
 &\hat{\cC}'_{i_1\cdots i_p} = \hat{\cC}_{i_1\cdots i_pz} + p\, 
\hat{\cC}_{[i_1\cdots i_{p-1}}\, B_{i_p]z} + p\,(p-1)\,\frac{\hat{\cC}_{[i_1\cdots i_{p-2}|z|}\, 
B_{i_{p-1}|z|}\, \CG_{i_p]z}}{\CG_{zz}} \,. 
\end{split}
\end{align}

\section{Weyl invariance of string theories in generalized supergravity backgrounds}
\label{sec:Weyl}

In this section, we consider Weyl invariance of bosonic strings in generalized supergravity backgrounds.

\subsection{The basics on Weyl invariance of bosonic string}
\label{sec:review}

Let us first recall the basics on Weyl invariance of bosonic string theory in $D=26$ dimensions. 

\medskip

We shall begin with the conventional string sigma model on general backgrounds, 
\begin{align}
 S =-\frac{1}{4\pi\alpha'} \int \rmd^2\sigma \sqrt{-\gamma}\,\bigl(g_{mn}\,\gamma^{\alpha\beta} - B_{mn}\,\varepsilon^{\alpha\beta}\bigr)\, \partial_\alpha X^m\, \partial_\beta X^n \,,
\label{eq:string-action}
\end{align}
where $\varepsilon^{01}=1/\sqrt{-\gamma}$\,. 
Then the Weyl anomaly of this system takes the form,
\begin{align}
 2\alpha'\,\langle T^\alpha{}_\alpha\rangle = \bigl(\beta^{g}_{mn}\,\gamma^{\alpha\beta} - \beta^{B}_{mn}\, \varepsilon^{\alpha\beta}\bigr)\, \partial_\alpha X^m\, \partial_\beta X^n \,.
\label{eq:Weyl-anomaly}
\end{align}
Here, the $\beta$-functions at the one-loop level have been computed (for example in \cite{Hull:1985rc}) as
\begin{align}
 \beta^{\CG}_{mn} = \alpha'\,\Bigl(R_{mn}-\frac{1}{4}\,H_{mpq}\,H_n{}^{pq}\Bigr) \,,\qquad
 \beta^{B}_{mn} = \alpha'\,\Bigl(- \frac{1}{2}\,\sfD^k H_{kmn}\Bigr) \,.
\end{align}
If the trace of the energy-momentum tensor (\ref{eq:Weyl-anomaly}) vanishes,
the system is Weyl invariant.

\medskip

To obtain a Weyl invariant worldsheet theory, it is not necessary to require $\beta^{g}_{mn}=\beta^{B}_{mn}=0$
as explained below. 
As long as they take the form
\begin{align}
 \beta^{g}_{mn} = - 2\,\alpha'\,\sfD_{m} \partial_{n} \Phi\,,\qquad 
 \beta^{B}_{mn} = - \alpha'\, \partial_k\Phi\,H^k{}_{mn} \,,
\label{eq:SUGRA}
\end{align}
the Weyl anomaly has a simple form under the equations of motion
\begin{align}
 \langle T^\alpha{}_\alpha\rangle = - \cD^\alpha \partial_\alpha\Phi + \sfD^m \Phi\,\frac{2\pi\alpha'}{\sqrt{-\gamma}}\frac{\delta S}{\delta X^m} \ \overset{\text{e.o.m.}}{\sim}\ - \cD^\alpha \partial_\alpha\Phi \,,
\end{align}
where $\cD_\alpha$ is the covariant derivative associated with $\gamma_{\alpha\beta}$ and $\overset{\text{e.o.m.}}{\sim}$ represents the equality up to the equations of motion. 
This can be canceled out by adding a counterterm, the so-called Fradkin--Tseytlin term \cite{Fradkin:1984pq},
\begin{align}
 S_{\text{FT}} = \frac{1}{4\pi}\int \rmd^2\sigma \, \sqrt{-\gamma}\,R^{(2)}\,\Phi\,,
\label{eq:F-T}
\end{align}
to the original action \eqref{eq:string-action}. 
Compared to the sigma model action,
the counterterm (1.1) is higher order in $\alpha'$ and it should be regarded as a quantum correction. 

\medskip 

Therefore, as long as the target space satisfies the equations \eqref{eq:SUGRA}, namely the supergravity equations of motion,
\begin{align}
 R_{mn}-\frac{1}{4}\,H_{mpq}\,H_n{}^{pq} + 2\, \sfD_{m} \partial_{n} \Phi =0 \,,\qquad
 - \frac{1}{2}\,\sfD^k H_{kmn} + \partial_k\Phi\,H^k{}_{mn} =0 \,.
\label{eq:eom-sugra}
\end{align}
the Weyl invariance is ensured. 
As discussed in \cite{Callan:1985ia}, equations \eqref{eq:eom-sugra} imply that
\begin{align}
 \beta^{\Phi} \equiv R + 4\,\sfD^m \partial_m \Phi - 4\,\abs{\partial \Phi}^2 - \frac{1}{12}\,H_{mnp}\,H^{mnp} \,,
\end{align}
is constant, and by choosing $\beta^{\Phi}=0$\,, we obtain the usual dilaton equation of motion. 

\medskip

The main observation of \cite{Fernandez-Melgarejo:2018wpg,Sakamoto:2017wor} is that the requirement \eqref{eq:SUGRA} is a sufficient condition for the Weyl invariance but is not necessary. 

\subsection{Weyl invariance}
\label{sec:Doubled-Weyl}

Let us consider a milder requirement,
\begin{align}
 \beta^{g}_{mn} = - 2\,\alpha'\,D_{(m} Z_{n)}\,,\qquad 
 \beta^{B}_{mn} = - \alpha'\,\bigl(Z^k\,H_{kmn} + 2\,D_{[m} I_{n]}\bigr) \,,
\label{eq:scale-GSE}
\end{align}
where $I_m$ and $Z_m$ are certain vector fields in the target space, which are functions of $X^m(\sigma)$\,. 
The condition \eqref{eq:scale-GSE} reduces to \eqref{eq:SUGRA} when $Z_m=\partial_m\Phi$ and $I^m=0$\,. 

\medskip

When the $\beta$-functions have the form \eqref{eq:scale-GSE}, the Weyl anomaly \eqref{eq:Weyl-anomaly} becomes
\begin{eqnarray}
 \langle T^\alpha{}_\alpha \rangle \! &=& \! -\cD_\alpha\bigl[(Z_m\,\gamma^{\alpha\beta} - I_m\,\varepsilon^{\alpha\beta})\,\partial_\beta X^m\bigr] 
 +Z^m\,\frac{2\pi\alpha'}{\sqrt{-\gamma}}\frac{\delta S}{\delta X^m} 
\nn\\
 \! &\overset{\text{e.o.m.}}{\sim}& \! -\cD_\alpha\bigl[(Z_m\,\gamma^{\alpha\beta} - I_m\,\varepsilon^{\alpha\beta})\,\partial_\beta X^m\bigr] \,. 
\label{eq:Weyl-general}
\end{eqnarray}
Then, there is a rigid scale invariance \cite{Hull:1985rc}, but it has been believed that 
the Weyl invariance could be broken because the counterterm \eqref{eq:F-T} cannot cancel out the anomaly \eqref{eq:Weyl-general}. 
However, we will construct a {\it modified} local counterterm such that \eqref{eq:Weyl-general} vanishes on-shell. 

\medskip

First of all, by considering the doubled spacetime, 
we show that the Fradkin--Tseytlin term can completely cancel the Weyl anomaly 
if $I^m$ and $Z_m$ satisfy the conditions \eqref{eq:GSE-conditions} and \eqref{eq:dilaton-isometry}. 
When these conditions are satisfied, $I^m$ and $Z_m$ can be replaced 
by the (m)DFT dilaton $d_*$ or $\Phi_*$, which has dual-coordinate dependence. 

\medskip

In order to treat the dual coordinates, we consider the DSM in which 
the number of the embedding functions is doubled: $(\bbX^M)=(X^m,\,\tilde{X}_m)$\,. 
For convenience, we choose the coordinates
\begin{align}
 (X^m) = (X^\mu,\,Y^i) \quad (\mu=0,\dotsc,D-N-1;\ i=1,\dotsc,N) 
\end{align}
on the target space such that the background fields $(\CG_{mn},\,B_{mn},\,\Phi)$ 
depend only on $X^\mu$, and the modified dilaton $\Phi_*$ has the form $\Phi_*=\Phi +I^i\,\tilde{Y}_i$\,. 

\medskip

Then, the essence of our argument is that the contribution of the Fradkin--Tseytlin term (with $\Phi$ replaced by $\Phi_*$) to the trace $\langle T\rangle$ can be written as
\begin{align}
 &\frac{4\pi}{\sqrt{-\gamma}}\,\gamma^{\alpha\beta}\,\frac{\delta S_{\text{FT}}}{\delta \gamma^{\alpha\beta}}
 = \cD^\alpha \bigl[\partial_M\Phi_*(\bbX)\,\partial_\alpha \bbX^M \bigr] 
 = \cD^\alpha \bigl[\partial_\mu\Phi(X)\, \partial_\alpha X^\mu + I^i\,\partial_\alpha \tilde{Y}_i \bigr] 
\nn\\
 &= \cD^\alpha \bigl(Z_m\,\partial_\alpha X^m \bigr) - \partial_{[m} I_{n]} \,\varepsilon^{\alpha\beta}\,\partial_\alpha X^m\,\partial_\beta X^n 
  + \cD^\alpha \bigl[I^i\,(\partial_\alpha \tilde{Y}_i - \CG_{in}\, \varepsilon^\beta{}_\alpha\,\partial_\beta X^n - B_{in}\,\partial_\alpha X^n) \bigr] 
\nn\\
 &= \frac{1}{2}\, \bigl[2\,D_{(m} Z_{n)}\,\gamma^{\alpha\beta} - (Z^k\,H_{kmn}+2\,\partial_{[m} I_{n]})\,
\varepsilon^{\alpha\beta} \bigr]\,\partial_\alpha X^m \, \partial_\beta X^n 
\nn\\
 &\quad +Z^m \,\frac{2\pi\alpha'}{\sqrt{-\gamma}}\frac{\delta S}{\delta X^m}
 + I^i\,\cD^\alpha (\partial_\alpha \tilde{Y}_i - \CG_{in}\, \varepsilon^b{}_\alpha\,\partial_\beta X^n - B_{in}\,\partial_\alpha X^n) \,. 
\label{eq:Weyl-inv}
\end{align}
By using the equations of motion for $X^m$ and the self-duality relations \cite{Duff:1989tf,Hull:2004in}
\begin{align}
 \partial_\alpha \tilde{Y}_i = \CG_{in}\, \varepsilon^\beta{}_\alpha\,\partial_\beta X^n + B_{in}\,\partial_\alpha X^n \,,
\label{eq:self-duality-components}
\end{align}
which are also obtained as the equations of motion of the DSM, as we will explain below, the contribution from the Fradkin--Tseytlin term can completely cancel the Weyl anomaly \eqref{eq:Weyl-general}. 

\medskip 

In order to explain the self-duality relation, let us consider Hull's double sigma model,
\begin{align}
 S = \frac{1}{4\pi\alpha'}\int \Bigl[ \frac{1}{2}\,\cH_{MN}(X)\,\cP^M\wedge *_\gamma 
\cP^N - \bigl(\rmd \tilde{X}_m+C_m\bigr)\wedge \rmd X^m \Bigr] \,,
\end{align}
where we have introduced the quantities 
\begin{align}
 \cP^M(\sigma) \equiv \rmd \bbX^M(\sigma) + C^M(\sigma) \,,\qquad 
(C^M)=\begin{pmatrix} 0 \\ C_m \end{pmatrix}\,,
\end{align}
and the generalized metric $\cH_{MN}(X)$ are supposed to be independent of 
$(Y^I)\equiv (Y^i,\,\tilde{Y}_i)$\,. 
The equations of motion for $C_m$ give rise to 
\begin{align}
 \rmd \tilde{X}_m + C_m = \CG_{mn}\,*_\gamma \rmd X^n + B_{mn}\,\rmd X^n \,,
\end{align}
which is equivalent to the well-known self-duality relation
\begin{align}
 \cP_M = *_\gamma \cH_M{}^N\,\cP_N \,.
\end{align}
Using the above equations of motion for $C_m$ and eliminating the combination 
$\rmd \tilde{X}_m + C_m$ from the above action, 
we obtain the conventional string sigma model action for $X^m$\,,
\begin{align}
 S = \frac{1}{4\pi\alpha'}\int \bigl(\CG_{mn}\,\rmd X^m\wedge *_\gamma \rmd X^n + B_{mn}\,\rmd X^m\wedge\rmd X^n \bigr) \,. 
\end{align}
Thus, the DSM is classically equivalent to the conventional sigma model. 
By combining the equations of motion for $Y^i$ and $C_m$, we can show that 
\begin{align}
 \rmd C_i = \rmd \bigl(\CG_{in}\,*_\gamma \rmd X^n + B_{in}\,\rmd X^n\bigr) = 0\,,
\end{align}
and $C_i$ is a closed form. 
Thus, using the local symmetry
\begin{align}
 \tilde{Y}_i(\sigma) \to \tilde{Y}_i(\sigma) + v_i(\sigma)\,,\qquad 
C_i(\sigma) \to C_i -\rmd v_i(\sigma)\,,
\end{align}
we can (at least locally) set $C_i=0$, and the equations of motion for $C_i$ become
\begin{align}
 \rmd \tilde{Y}_i = \CG_{in}\,*_\gamma \rmd X^n + B_{in}\,\rmd X^n \,. 
\end{align}
This is nothing but the key relation \eqref{eq:self-duality-components}. 

\medskip 

The one-loop $\beta$-function of the DSM is computed in 
\cite{Berman:2007xn,Berman:2007yf,Copland:2011yh} by using the background field method, 
and the result is consistent with the conventional string sigma model. 
There, in order to cancel the Weyl anomaly, the $T$-duality-invariant Fradkin--Tseytlin term \cite{Hull:2006va},%
\footnote{This reduces to the conventional one after integrating over the auxiliary fields $C_m$ \cite{Hull:2006va}.}
\begin{align}
 S_{\text{FT}} = \frac{1}{8\pi}\int \rmd^2\sigma \, \sqrt{-\gamma}\,R^{(2)}\,d(X)\,,
\end{align}
has been introduced, though the dilaton is supposed to be independent of the dual coordinates 
and the equations of motion of the (m)DFT have not been reproduced. 
Now let us replace $d(X)$ with $d_*(X,\,\tilde{Y})$ in the $T$-duality-invariant Fradkin--Tseytlin term,
\begin{align}
 S_{\text{FT}} = \frac{1}{8\pi}\int \rmd^2\sigma \, \sqrt{-\gamma}\,R^{(2)}\,d_*(X)\,.
\label{eq:mFT-term}
\end{align}
By following from the same calculation as \eqref{eq:Weyl-inv},
we can show the Weyl invariance.
Therefore, the bosonic string sigma model is Weyl invariant, as long as the background fields 
satisfy the GSE. 

%

%

\medskip 

Before closing this section, let us comment on the central charge identity \cite{Callan:1985ia,Curci:1986hi}, 
namely the constancy of $\cS$. 
As discussed in \cite{Sakatani:2016fvh}, one can show the identity only 
from the differential Bianchi identity and the equations of motion for the generalized metric, $\cS_{MN}=0$,
\begin{align}
 \partial_M \cS = 2\,\cH_{MN}\,\nabla_K \cS^{KN} = 0\,. 
\end{align}
In \cite{Sakatani:2016fvh}, the differential Bianchi identity has not been proven 
in the presence of $\bX^M$, and it has not been clear 
whether $\bS_{MN}=0$ can generally show the central charge identity. 
However, as we have found that the mDFT is just the conventional DFT, 
the differential Bianchi identity and $\bS_{MN}=0$ indicate that $\bS$ is constant.

\subsection{Local counterterm for generalized supergravity}
\label{sec:counterterm}

Although we showed the Weyl invariance of bosonic string in generalized supergravity backgrounds,
the equations of motion of
the DSM  (\ref{eq:self-duality-components}) imply that $Y^i$ is a non-local function of $X^m$\,.
Therefore, one may suspect
that the proposed counterterm (\ref{eq:mFT-term}) is non-local as well.
However, as we show in this subsection, 
we can construct a similar local counterterm by considering that the two-dimensional Ricci scalar
$R^{(2)}$ is a total derivative and $I$ is a Killing vector. 

\medskip

Recalling that the two-dimensional Einstein--Hilbert action is a total derivative,
\begin{align}
 \sqrt{-\gamma}\,R^{(2)} = \partial_\alpha w^\alpha \,,
\end{align}
we define a vector density $w^\alpha$ that transforms as\footnote{Although the explicit form of $w^\alpha$ is not important here, it is given in Section \ref{sec:alpha}.}
\begin{align}
 \delta_\xi w^\alpha = \Lie_\xi w^\alpha 
= \xi^\beta\,\partial_\beta w^\alpha - w^\beta\,\partial_\beta \xi^\alpha + \partial_\beta \xi^\beta\,w^\alpha\,,
\end{align}
under diffeomorphisms on the world-sheet. 
We then introduce the counterterm as
\begin{align}
 S^{(I,Z)}_{\text{FT}}= -\frac{1}{4\pi}\int \rmd^2\sigma\, w^\alpha\, \bigl(Z_m\,\partial_\alpha X^m - I_m\,\varepsilon_\alpha{}^{\beta}\,\partial_\beta X^m\bigr) \,. 
\label{eq:FT-I}
\end{align}
Note that this reduces to the Fradkin--Tseytlin term \eqref{eq:F-T} when $I^m=0$ and $Z_m=\partial_m\Phi$\,,
\begin{align}
 S^{(0,\rmd\Phi)}_{\text{FT}} = -\frac{1}{4\pi}\int \rmd^2\sigma\, w^\alpha\, \partial_\alpha \Phi 
 = \frac{1}{4\pi}\int \rmd^2\sigma\,\sqrt{-\gamma}\, R^{(2)}\, \Phi \,,
\end{align}
where we supposed the world-sheet has no boundary. 
Assuming that $Z_m$ and $I_m$ are independent of $\gamma_{\alpha\beta}$\,, if we vary the counterterm with respect to $\gamma^{\alpha\beta}$, we obtain 
\begin{align}
 \frac{4\pi}{\sqrt{-\gamma}}\,\frac{\delta S^{(I,Z)}_{\text{FT}}}{\delta\gamma^{\alpha\beta}} 
 &= -\cD_{(\alpha|} \bigl(Z_m\,\partial_{|\beta)} X^m - I_m\,\varepsilon_{|\beta)}{}^\gamma\,\partial_\gamma X^m\bigr) 
  + \gamma_{\alpha\beta} \,\cD^\gamma \bigl(Z_m\,\partial_\gamma X^m - I_m\,\varepsilon_\gamma{}^{\delta}\,\partial_\delta X^m\bigr) 
\nn\\
 &\quad + \frac{\alpha^\gamma}{\sqrt{-\gamma}}\, I_m\,\bigl(\varepsilon_{\gamma(\alpha}\,\partial_{\beta)} X^m-\tfrac{1}{2}\, \gamma_{\alpha\beta}\,\varepsilon_{\gamma}{}^\delta\,\partial_\delta X^m\bigr) 
\\
 &\quad - \varphi_{\alpha\beta}\,\cD_\gamma \bigl[(I_m\, \gamma^{\gamma\delta} - Z_m\,\varepsilon^{\gamma\delta})\, \partial_\delta X^m \bigr] \,.
\nn
\end{align}
Here, suggested by the identity in two dimensions,
\begin{align}
 \delta\bigl(\sqrt{-\gamma}\,R^{(2)}\bigr) = \partial_\gamma\bigl[\sqrt{-\gamma}\,\bigl(\gamma^{\gamma\alpha}\,\cD^\beta\delta\gamma_{\alpha\beta}-\gamma^{\alpha\beta}\cD^\gamma\delta\gamma_{\alpha\beta}\bigr)\bigr]\,,
\end{align}
we have used the variation
\begin{align}
 \delta w^\gamma = \sqrt{-\gamma}\,\bigl(\gamma^{\gamma\alpha}\,\cD^\beta\delta\gamma_{\alpha\beta}-\gamma^{\alpha\beta}\cD^\gamma\delta\gamma_{\alpha\beta}\bigr) + \epsilon^{\gamma\delta}\,\partial_\delta (\varphi^{\alpha\beta}\,\delta \gamma_{\alpha\beta}) \,,
\label{eq:delta-alpha}
\end{align}
where $\epsilon^{01}=+1$ and $\varphi^{\alpha\beta}$ is a symmetric tensor made of the fundamental fields and their derivatives. 
Then, the contribution of the counterterm \eqref{eq:FT-I} to the Weyl anomaly becomes
\begin{align}
 \langle T^{\alpha}{}_{\alpha} \rangle_{\text{FT}} 
 &= \frac{4\pi}{\sqrt{-\gamma}}\,\gamma^{\alpha\beta}\,\frac{\delta S^{(I,Z)}_{\text{FT}}}{\delta \gamma^{\alpha\beta}}
\nn\\
 &= \cD_\alpha \bigl[\bigl(Z_m \,\gamma^{\alpha\beta} - I_m\,\varepsilon^{\alpha\beta}\bigr)\,\partial_\beta X^m \bigr]
  - \varphi^\alpha{}_\alpha \,\cD_\gamma \bigl[(I_m\, \gamma^{\gamma\delta} - Z_m\,\varepsilon^{\gamma\delta})\, \partial_\delta X^m \bigr] \,. 
\end{align}
In fact, the divergence in the last term vanishes
\begin{align}
 \cD_\gamma \bigl[(I_m\, \gamma^{\gamma\delta} - Z_m\,\varepsilon^{\gamma\delta})\, \partial_\delta X^m \bigr]=\cD_\gamma J^\gamma \overset{\text{e.o.m.}}{\sim}0\,,
\end{align}
by using the on-shell conservation law of a Noether current (see Section \ref{sec:linear-dilaton}), and we obtain
\begin{align}
 \langle T^\alpha{}_{\alpha} \rangle_{\text{FT}} \ \overset{\text{e.o.m.}}{\sim}\ \cD_\alpha \bigl[\bigl(Z_m \,\gamma^{\alpha\beta} - I_m\,\varepsilon^{\alpha\beta}\bigr)\,\partial_\beta X^m \bigr] \,. 
\end{align}
Therefore, the anomaly \eqref{eq:Weyl-general} completely cancels. 

\medskip 

Actually, the requirement \eqref{eq:scale-GSE} was derived as the condition for the one-loop finiteness of string sigma model \cite{Hull:1985rc}. 
Now, we have proven that the Weyl symmetry can also be preserved upon introducing the above counterterm, so it is reasonable to expect that string theory can be consistently defined with the relaxed condition \eqref{eq:GSE}. 
From \eqref{eq:beta} and \eqref{eq:scale-GSE},
we can express the condition for the Weyl invariance as modified supergravity equations of motion.

\medskip

In earlier works, many solutions of GSE have been obtained from the $q$-deformation \cite{Arutyunov:2015qva}, homogeneous Yang--Baxter deformations \cite{Kyono:2016jqy,Orlando:2016qqu,Fernandez-Melgarejo:2017oyu,Sakamoto:2018krs}, and non-Abelian $T$-duality \cite{Gasperini:1993nz,Fernandez-Melgarejo:2017oyu,Hong:2018tlp} (see also \cite{Hoare:2016wsk}), despite there was not guarantee that these are string backgrounds. However, the cancellation of the Weyl anomaly that we provide here is an important step towards that direction\footnote{The solution obtained from $q$-deformation includes an imaginary Ramond--Ramond field, and may not be a string background.}.

\medskip 

As discussed in the previous section \cite{Sakatani:2016fvh,Sakamoto:2017wor},
we can regard solutions of GSE as solutions of DFT. 
In the solutions of DFT, by using adapted coordinates where the Killing vector $I^m$ is constant, we find that the dilaton has a linear dependence on the dual coordinate $\tilde{x}_m$ \cite{Sakamoto:2017wor}. 
Moreover, if we perform a formal $T$-duality\footnote{A formal $T$-duality means the factorized $T$-duality along a non-isometry direction $x^z$\,, which maps the coordinate $x^z$ into the dual coordinate $\tilde{x}_z$\,. Such transformation is a symmetry of the equations of motion of DFT.} along the $I^m$-direction, an arbitrary solution of GSE is mapped to a solution of the conventional supergravity that has a linear coordinate dependence in the dilaton \cite{Hoare:2015wia,Arutyunov:2015mqj,Sakamoto:2017wor} (see the subsection \ref{subsec:Non-unimodular-sol} for examples). 
In the next section, we sketch the origin of the linear dilaton by introducing the Noether current associated with the Killing vector $I^m$\,. 

\subsection{Linear dilaton in generalized supergravity backgrounds}
\label{sec:linear-dilaton}

In this section, we shall discuss a relation between the generalized supergravities and linear dilatons. 
These are closely related to each other intrinsically as we show below. 

\medskip 

An arbitrary solution to the GSE admits a Killing vector $I^m$ by the definition of GSE,
\begin{align}
 \Lie_I g_{mn} = 0 \,,\qquad 
 \Lie_I B_{mn} + \partial_m \tilde{I}_n - \partial_n \tilde{I}_m = 0 \,,\qquad 
 \Lie_I \Phi = 0\,.
\end{align}
In this case, $Z_m$ is parameterized as
\begin{align}
 Z_m = \partial_m \Phi + I^nB_{nm} + \tilde{I}_m \, .
\label{eq:Z-Phi-IB}
\end{align}
Due to the existence of the Killing vector, the string sigma model has a conserved current associated with the global symmetry 
\begin{equation}
X^m\to X^m + \epsilon\,I^m\,, 
\end{equation} 
where $\epsilon$ is an infinitesimal constant. 
Under an infinitesimal variation, $\delta X^m=\epsilon\,I^m$\,, we obtain the (on-shell conserved) Noether current $J^\alpha$\,,
\begin{align}
 J^\alpha \equiv \bigl[ I^m\,\bigl(g_{mn}\,\gamma^{\alpha\beta}-B_{mn}\,\varepsilon^{\alpha\beta}\bigr) - \tilde{I}_n\,\varepsilon^{\alpha\beta} \bigr]\, \partial_\beta X^n \,,\qquad 
 \cD_\alpha J^\alpha \ \overset{\text{e.o.m.}}{\sim}\ 0 \,. 
\label{eq:Noether-current}
\end{align}
Then, by recalling the parameterization \eqref{eq:Z-Phi-IB}, our counterterm \eqref{eq:FT-I} can be written as
\begin{align}
 S^{(I,Z)}_{\text{FT}}
 = \frac{1}{4\pi}\int \rmd^2\sigma\,\bigl(\sqrt{-\gamma}\, R^{(2)}\,\Phi + \varepsilon_{\alpha\beta}\,w^\alpha\, J^b\bigr)\,.
\label{eq:FT-I1}
\end{align}
From the conservation law, $J^\alpha$ can be represented by using a certain function $\tilde{Z}(\sigma)$ as
\begin{align}
 J^\alpha\ \overset{\text{e.o.m.}}{\sim}\ \varepsilon^{\beta\alpha}\,\partial_\beta \tilde{Z} \,.
\label{eq:Ja-dZ}
\end{align}
Then, the counterterm \eqref{eq:FT-I1} can be further rewritten as 
\begin{align}
 S^{(I,Z)}_{\text{FT}} \overset{\text{e.o.m.}}{\sim} \frac{1}{4\pi}\int \rmd^2\sigma\,\sqrt{-\gamma}\, R^{(2)}\,(\Phi + \tilde{Z}) \,.
\label{eq:gFT-Z}
\end{align}

\medskip

Now, let us choose a particular gauge $\tilde{I}_m=0$\,. 
In this case, by comparing the relation \eqref{eq:Ja-dZ} with the equations of motion of the double sigma model \eqref{eq:self-duality-components}, we can identify $\tilde{Z}$ with a combination of the dual coordinates $I^m\,\tilde{X}_m$\,. 
Then, \eqref{eq:gFT-Z} is precisely the counterterm (\ref{eq:mFT-term})\cite{Sakamoto:2017wor}, 
namely the Fradkin--Tseytlin term with the modified dilaton
\begin{align}
 \Phi_* = \Phi + I^m\,\tilde{X}_m \,. 
\end{align}
In this manner, thanks to the Killing property of $I^m$\,, the difference between the standard Fradkin--Tseytlin term and our counterterm \eqref{eq:FT-I1} can always be expressed as a linear dual-coordinate dependence in the dilaton. 
As we concretely showed in the section \ref{subsec:Non-unimodular-sol},
by performing a formal $T$-duality in DFT along the Killing direction, this linear dual-coordinate dependence becomes a linear dependence on the physical coordinate $X^m$\,.

\subsection{Constructions of local $w^\alpha$}
\label{sec:alpha}

In this section, we explain two ways to construct the vector density $w^\alpha$\,. 
Naively, from the defining relation,
\begin{align}
 \sqrt{-\gamma}\,R^{(2)} = \partial_\alpha w^\alpha \,,
\end{align}
one might expect that $w^\alpha$ can be expressed in terms of the metric $\gamma_{\alpha\beta}$\,. 
However, it is not the case as it was clearly discussed in \cite{Deser:1995ne,Yale:2010jy}. 
In order to construct $w^\alpha$ in terms of the metric $\gamma_{\alpha\beta}$\,, we need to break the general covariance on the worldsheet. 
Indeed, the general solution obtained in \cite{Yale:2010jy} takes the form
\begin{align}
\begin{split}
 w^0 &= \frac{1}{\sqrt{-\gamma}}\,\Bigl[-\lambda\,\frac{\gamma_{01}}{\gamma_{00}}\,\partial_1\gamma_{00} + (1-\lambda)\,\frac{\gamma_{01}}{\gamma_{11}}\,\partial_1\gamma_{11}-2\,(1-\lambda)\,\partial_1\gamma_{01}+\partial_0\gamma_{11}\Bigr]\,,
\\
 w^1 &= \frac{1}{\sqrt{-\gamma}}\,\Bigl[-(1-\lambda)\,\frac{\gamma_{01}}{\gamma_{11}}\,\partial_0\gamma_{11} + \lambda \,\frac{\gamma_{01}}{\gamma_{00}}\,\partial_0\gamma_{00}-2\,\lambda \,\partial_0\gamma_{01}+\partial_1\gamma_{00}\Bigr]\,,
\end{split}
\end{align}
where $\lambda$ is an arbitrary parameter that is coming from the ambiguity of $w^\alpha$
\begin{align}
 w^\alpha \ \to \ w^\alpha + \epsilon^{\alpha\beta}\,\partial_\beta f \,. 
\label{eq:alpha-ambiguity}
\end{align}
If we consider its variation under an infinitesimal diffeomorphism $\delta_v \gamma_{\alpha\beta}=\Lie_v \gamma_{\alpha\beta}$ (with $\delta_v\lambda=0$), we find it is not covariant,
\begin{align}
 \delta_v w^\alpha \neq \Lie_v w^\alpha\,. 
\end{align}
Therefore, if $w^\alpha$ is only written in terms of the metric and its derivatives, it will not be covariant. 
On the other hand, similar to the approach of \cite{Deser:1995ne}, if we introduce a zweibein $e_{a}{}^\alpha$ on the worldsheet, we find another expression up to the ambiguity \eqref{eq:alpha-ambiguity}
\begin{align}
 w^\alpha = - 2 \sqrt{-\gamma}\, e_{a}{}^\alpha\,\omega_{b}{}^{ba}\,,
\end{align}
where $\omega_{a}{}^{bc}$ is the spin connection. 
In this case, despite $w^\alpha$ is manifestly covariant under diffeomorphisms, it is not covariant under the local Lorentz symmetry. 
In the following, we explain two ways to provide covariant definitions of $w^\alpha$\,. 

\subsubsection{A construction using the Noether current}

The first approach is based on the approach explained in Section II.B. of \cite{Yale:2010jy}. 
In two dimensions, if there exists a normalized vector field $n^\alpha$ ($\gamma_{\alpha\beta}\,n^\alpha\,n^\beta=\pm1\equiv\sigma$), we can show that
\begin{align}
 \sqrt{-\gamma}\,R^{(2)} = 2\,\sigma\,\partial_\alpha\bigl[\sqrt{-\gamma}\,\bigl(n^\beta\,\cD_\beta n^\alpha - n^\alpha\,\cD_\beta n^\beta\bigr)\bigr] \,. 
\end{align}
In string theory on generalized supergravity backgrounds, we have a natural vector field on the worldsheet, which is the Noether current $J^\alpha$ in \eqref{eq:Noether-current}. 
Supposing $J^\alpha$ is not a null vector on the worldsheet, we define the vector field $n^\alpha$ as $n^\alpha\equiv \frac{J^\alpha}{\sqrt{\sigma\,\gamma_{\gamma\delta}\,J^\gamma\,J^\delta}}$\,. 
Then $w^\alpha$ is defined as
\begin{align}
 w^\alpha \equiv 2\,\sigma\, \sqrt{-\gamma}\,\bigl(n^\beta\,\cD_\beta n^\alpha - n^\alpha\,\cD_\beta n^\beta\bigr)\,,
\label{eq:alpha-unit-vector}
\end{align}
which is manifestly covariant and a local function of the fundamental fields. 
Moreover, by taking a variation of this $w^\alpha$ in terms of $\gamma_{\alpha\beta}$\,, where the Noether current transforms as
\begin{align}
 \delta (\sqrt{-\gamma}\,J^\alpha) = \delta (\sqrt{-\gamma}\,\gamma^{\alpha\beta})\,\partial_\beta X^m\,I_m \,,
\end{align}
after a tedious computation,\footnote{We repeatedly use the identity $2\,A^{\cdots [\alpha}\,B^{\beta]\cdots}=-\varepsilon^{\alpha\beta}\,\varepsilon_{\gamma\delta}\,A^{\cdots [\gamma}\,B^{\delta]\cdots}$ satisfied in two dimensions.} 
we find the desired variation formula \eqref{eq:delta-alpha} with $\varphi^{\alpha\beta}$ given by
\begin{align}
 \varphi^{\alpha\beta} = \sigma\,\biggl[n^\gamma\,\varepsilon_\gamma{}^{(\alpha}\,n^{\beta)} + \frac{2}{\sqrt{\sigma\,\gamma_{\rho\sigma}\,J^\rho\,J^\sigma}}\,\varepsilon^{(\alpha}{}_{\!\!(\gamma}\,\delta^{\beta)}_{\delta)}\,\cD^\gamma X^m\,I_m\,n^\delta \biggr] \,. 
\end{align}
Therefore, this fully determines the variation of $w^\alpha$, for which the Weyl anomaly cancels out in generalized supergravity backgrounds.

\subsubsection{A construction in the gauged sigma model}

In the second approach, we introduce auxiliary fields to construct $w^\alpha$\,. 
For simplicity, here we choose a gauge $\tilde{I}_m=0$\,. 

\medskip 

Let us consider the action of a gauged sigma model
\begin{align}
 S' =-\frac{1}{4\pi\alpha'} \int \rmd^2\sigma \sqrt{-\gamma}\,\bigl[\bigl(g_{mn}\,\gamma^{\alpha\beta} - B_{mn}\,\varepsilon^{\alpha\beta}\bigr)\, D_\alpha X^m\, D_\beta X^n - \tilde{Z}\,\varepsilon^{\alpha\beta}\, F_{\alpha\beta}\bigr] \,,
\end{align}
where $D_\alpha X^m \equiv \partial_\alpha X^m - I^m\,A_\alpha$\,, $F_{ab}\equiv\partial_\alpha A_\beta-\partial_\beta A_\alpha$\,, and $I\equiv I^m\,\partial_m$ satisfies the Killing equations. 
This theory has a local symmetry,
\begin{align}
 X^m(\sigma) \to X^m(\sigma) + I^m\,v(\sigma)\,,\qquad 
 A_\alpha(\sigma) \to A_\alpha(\sigma) + \partial_\alpha v(\sigma) \,. 
\label{eq:local-symmetry}
\end{align}
The action reproduces the standard one \eqref{eq:string-action} after integrating out the auxiliary field $\tilde{Z}$\,. 
In order to cancel out the one-loop Weyl anomaly, we add the following local term to $S'$:
\begin{align}
 S_c \equiv \frac{1}{4\pi}\int \rmd^2\sigma \, \sqrt{-\gamma}\,R^{(2)}\,(\Phi + \tilde{Z}) \,,
\end{align}
which is higher order in $\alpha'$\,. 
The contribution to the trace of the energy-momentum tensor coming from $S_c$ is
\begin{align}
 \langle T \rangle_c 
 = \frac{4\pi}{\sqrt{-\gamma}}\,\gamma^{\alpha\beta}\,\frac{\delta S_c}{\delta \gamma^{\alpha\beta}}
 \overset{\text{e.o.m.}}{\sim} \cD^\alpha (\partial_\alpha \Phi + \partial_\alpha \tilde{Z})\,. 
\label{eq:Tc}
\end{align}

\medskip

The equations of motion for $A_\alpha$ and $\tilde{Z}$ give
\begin{align}
 \partial_\alpha \tilde{Z} = \varepsilon^\beta{}_\alpha\,J_\beta - \abs{I}^2\,\varepsilon^\beta{}_\alpha\, A_\beta\,, \qquad
 \epsilon^{\alpha\beta}\, F_{\alpha\beta} = \alpha'\,\sqrt{-\gamma}\,R^{(2)} \,,
\end{align}
where $J_\alpha$ is the Noether current defined in \eqref{eq:Noether-current}. 
Since the field strength $F_{\alpha\beta}$ vanishes to the leading order in $\alpha'$\,, by using the local symmetry \eqref{eq:local-symmetry}, we can find a gauge where the order $\cO(\alpha'^0)$ term vanishes
\begin{align}
 A_\alpha = 0 + \alpha'\,\cA_\alpha \,,\qquad \epsilon^{\alpha\beta}\,(\partial_\alpha \cA_\beta - \partial_\beta \cA_\alpha) = - \sqrt{-\gamma} \,R^{(2)}\,.
\label{eq:A-solution}
\end{align}
Here, $\cA_\alpha$ is a quantity of order $\cO(\alpha'^0)$\,. 
Then the trace \eqref{eq:Tc} becomes
\begin{align}
 \langle T \rangle_c 
 \overset{\text{e.o.m.}}{\sim} \cD^\alpha \bigl(\partial_\alpha \Phi +\varepsilon^\beta{}_\alpha\,J_\beta\bigr) + \cO(\alpha') 
 = \cD_\alpha \bigl[\bigl(Z_m \,\gamma^{\alpha\beta} - I_m\,\varepsilon^{\alpha\beta}\bigr)\,\partial_\beta X^m \bigr] + \cO(\alpha') \,. 
\end{align}
This completely cancels the one-loop Weyl anomaly \eqref{eq:Weyl-general}, which will be coming from $S'$\,. 

\medskip 

After eliminating the auxiliary field $\tilde{Z}$\,, the action $S'+S_c$ becomes
\begin{align}
 S'+S_c &=-\frac{1}{4\pi\alpha'} \int \rmd^2\sigma \sqrt{-\gamma}\, \bigl(g_{mn}\,\gamma^{\alpha\beta} - B_{mn}\,\varepsilon^{\alpha\beta}\bigr)\, \partial_\alpha X^m\, \partial_\beta X^n 
\nn\\
 &\quad + \frac{1}{4\pi}\int \rmd^2\sigma \,\sqrt{-\gamma}\,\bigl[ R^{(2)}\, \Phi + \epsilon_{\alpha\beta}\, (-2\,\epsilon^{\alpha\gamma}\,\cA_\gamma)\,J^\beta - \alpha'\,\abs{I}^2\,\gamma^{\alpha\beta}\,\cA_\alpha\,\cA_\beta\bigr] \,.
\end{align}
As it is clear from \eqref{eq:A-solution}, the gauge field $\cA_\alpha$ plays the role of the desired $w^\alpha$ via $w^\alpha=-2\,\epsilon^{\alpha\beta}\,\cA_\beta$\,. 
Then, we obtain
\begin{align}
 S'+S_c &=-\frac{1}{4\pi\alpha'} \int \rmd^2\sigma \sqrt{-\gamma}\, \bigl(g_{mn}\,\gamma^{\alpha\beta} - B_{mn}\,\varepsilon^{\alpha\beta}\bigr)\, \partial_\alpha X^m\, \partial_\beta X^n 
\nn\\
 &\quad + \frac{1}{4\pi}\int \rmd^2\sigma \,\sqrt{-\gamma}\,\biggl[ R^{(2)}\, \Phi + \epsilon_{\alpha\beta}\, w^\alpha\,J^\beta + \alpha'\,\frac{\abs{I}^2}{4\sqrt{-\gamma}^2}\,\gamma^{\alpha\beta}\,w_\alpha\,w_\beta\biggr] \,,
\label{eq:Sprime-Sc}
\end{align}
and by neglecting the higher order term in $\alpha'$\,, this is precisely the same as the standard sigma model action including our local counterterm $S^{(I,Z)}_{\text{FT}}$ \eqref{eq:FT-I}. 

\medskip 

It is noted that the second line in the action \eqref{eq:Sprime-Sc} is the same as Eq.~(5.13) of \cite{Wulff:2018aku}. 
There, it was obtained by rewriting the non-local piece of the effective action $S_{\text{non-local}}$ of \cite{Elitzur:1994ri} through the identifications of $I_m$ and $Z_m$ with some quantities in Yang--Baxter sigma model. 
In \cite{Elitzur:1994ri}, the non-local action $S_{\text{non-local}}$ appeared in the process of non-Abelian $T$-duality, and it played an important role to show the tracelessness of $T_{\alpha\beta}$\,. 
However, according to the non-local nature of the effective action, by truncating the non-linear term by hand, it was concluded in \cite{Elitzur:1994ri} that the string model (called the B'-model) is scale invariant but not Weyl invariant. 
On the other hand, the action \eqref{eq:Sprime-Sc} or our local counterterm \eqref{eq:FT-I} with $w^\alpha$ defined as \eqref{eq:alpha-unit-vector} is local and free from the Weyl anomaly.


\chapter{Conclusion and discussion}
\label{Ch:concl}

Let us summarize the results described in chapter $2$-$5$ and give open problems for future directions.

\section{Summery of the thesis}

\paragraph{Chapter \ref{Ch:YB-AdS5}}
In this chapter, we started to give a brief review of the $\AdS{5}\times \rmS^5$ superstring.
Next,
we have introduced the homogeneous YB deformations of the $\AdS{5}\times \rmS^5$ superstring
with the $\kappa$-symmetry
and showed the classical integrability of the deformed model.
We further provided a simple formula \eqref{pre-beta-formula-noB} for general deformed backgrounds.
Then, we gave the condition of the $r$-matrix, the unimodulartiy condition,
in order that the YB deformed $\AdS{5}\times \rmS^5$ backgrounds solve the usual supergravity equations.
If a given $r$-matrix doesn't satisfy the condition,
the deformed background is a solution of the GSE presented in subsection \ref{sec:GSE-YBsec}.

\medskip

In section \ref{sec:ExampleYBAdS5},
we presented full expressions of the deformed backgrounds associated with various classical $r$-matrices.
We reproduced the well-known examples by using the formula presented in the previous section.
For non-unimodular $r$-matrices,
the associated YB deformed backgrounds are solutions of the GSE.
Moreover, we observed that some of them are reduced to the original background after performing the generalized TsT transformations.

\paragraph{Chapter \ref{Ch:YB-duality}}
Main purpose of this chapter was to explain that the homogeneous YB deformation can be interpreted as a string duality transformation. 
We first reviewed the double-vielbein formalism of the DFT and
gave the $\beta$-transformation rules for the supergravity fields.
Moreover, the action of the double sigma model for type II superstring that reproduces the
conventional GS superstring action was presented.

\medskip

After that, in section \ref{sec:YB-beta-deform},
we have shown that the YB deformation with 
a classical $r$-matrix $r=\frac{1}{2}\,r^{ij}\,T_i\wedge T_j$ satisfying the homogeneous CYBE, 
is equivalent to the $\beta$-deformation with the deformation parameter 
\begin{align}
 \bmr^{mn} = 2\,\eta\,r^{ij}\, \hat{T}^{m}_i\,\hat{T}_j^{n}\,.
\end{align}
Therefore, the YB deformations can be regarded as a string duality transformation.
Furthermore, in section \ref{sec:YB-gdiff},
we have found the generalized diffeomorphism parameters which produce various $\beta$-twisted backgrounds, 
including all rank-4 deformations classified in \cite{Borsato:2016ose}. 

\medskip

Finally, in section \ref{sec:AdS3-YB},
we have considered $\beta$-deformations of $\AdS3\times \rmS^3\times \TT^4$ backgrounds supported by $H$-flux
and obtained various solutions as (generalized) supergravity.
In several examples of non-unimodular $r$-matrices,
we unexpectedly obtained solutions of the usual supergravity. 
In this way,
the $\beta$-deformations are certain duality transformations in string theory.

\paragraph{Chapter \ref{Ch:YB-T-fold}}
We have first reviewed the notion of $T$-folds by showing two well-known examples:
(1) a toy model which shows how to obtain a $T$-fold background upon a chain of dualizations of a geometric torus and (2) the co-dimension 1 exotic $5_2^2$-brane background.

\medskip

In section.\ref{sec:Non-geometric-YB},
We have computed monodromy matrices for various YB-deformed backgrounds 
and a non-Abelian $T$-dual background. 
Using the formulas \eqref{beta-formula-noB} and \eqref{div-formula},
we have showed that YB deformations can be regarded as non-geometric backgrounds involving $Q$-fluxes. 
Importantly, as long as the $r$-matrix solves the homogeneous CYBE,
the deformed background is a solution of DFT.
Therefore, the YB deformation is a systematic way to obtain 
solutions with $Q$-fluxes in DFT.

\paragraph{Chapter \ref{Ch:Weyl-GSE}}
In section \ref{sec:(m)DFT}, 
we have shown that the bosonic part of the GSE are reproduced from the DFT
by choosing a non-standard section. 
When all of the fields are independent of the dual coordinates, 
the equations of motion of the DFT lead to the conventional supergravity equations, 
while when the dilaton has a linear dual-coordinate dependence, the GSE are reproduced. 

\medskip

Next, in section \ref{sec:Weyl}, 
we have considered the Weyl invariance of bosonic string theories in generalized supergravity backgrounds.
We stated to review the basics on Weyl invariance of bosonic string theory in general backgrounds.
After that, by introducing the dual-coordinate dependence into the Fradkin-Tseytlin term,
we have constructed a possible counterterm \eqref{eq:FT-I}
that cancels out the Weyl anomaly of bosonic string theory defined on generalized supergravity backgrounds.
Furthermore, by rewriting the counterterm,
we have shown that the resulting counterterm is definitely local and 
it is not necessary for introducing the doubled formalism. 
In this sense, string theory is consistently defined on generalized supergravity backgrounds.

\section{Open problems}

Since until now YB deformations of the ${\rm AdS}_5\times {\rm S}^5$ superstring have been investigated in detail,
a natural next step is to consider the dual gauge theories side and to performa a direct verification of the correspondences by using integrability methods.
In several works \cite{vanTongeren:2015soa,vanTongeren:2015uha,Araujo:2017jkb,Araujo:2017jap},
it was discussed that when we only consider a YB deformation of the ${\rm AdS}_5$ part,
the dual gauge theory is defined on a non-commutative space specified by an $r$-matrix used in the deformation.
Since the deformed string sigma models are classical integrable,
one might infer the corresponding gauge theories are also planar integrable.
As the undeformed case, it would be desirable to explore the structure of integrable spin chains 
underlying the deformed systems.

\medskip

Another possible future direction is to further study quantum aspects of string theory on generalized supergravity backgrounds.
We have shown that, in this thesis, Weyl invariance of string theory on generalized supergravity backgrounds at the one-loop level.
In this scene, string theory may be consistently defined on such background.
However, to establish the claim,
it would be necessary to study some quantum aspects of the associated world-sheet theory in more detail,
such as the Weyl invariance at the higher order, the absence of negative norm states, and the 
modular invariance of partition functions. 

\medskip

For this purpose, it would be useful to construct tractable and solvable string sigma models on generalized supergravity backgrounds without R--R fluxes.
The $\beta$-deformations of Minkowski and pp-wave backgrounds might lead to such solvable string theories on generalized supergravity backgrounds (for example, see  (\ref{flatpen})). 

\medskip

As a more nontrivial example,
it may be useful to consider $\beta$-deformations of $\AdS3\times \rmS^3\times \TT^4$ background with pure $H$-flux.
In this case, we could use integrability methods to study the quantum spectrum of the deformed systems
\footnote{By using integrability techniques, the spectral problem in the undeformed case was recently solved  (for example, see \cite{Dei:2018mfl}).
The obtained results are consistent the previous results \cite{Maldacena:2000hw}.
}.
Indeed, in the case of TsT transformations (i.e. Abelian $r$-matrices),
the classical integrability of the deformed system was shown in \cite{Delduc:2017fib} by constructing the Lax pair explicitly.
This would be extended to deformations with non-Abelian $r$-matrices,
and we may study the quantum aspects of the deformed systems in details.

\medskip

Finally, it is interesting to consider whether it is possible to uplift the generalized type IIA supergravity equations to eleven-dimensional (generalized?) supergravity.
In addition, it would also be an important work to clarify relations between string sigma model action on generalized supergravity backgrounds and the membrane action.

\newpage

\chapter*{Acknowledgements}

I would like to express my gratitude to many people who helped me during the Ph.D. course.

\medskip

First of all, I would like to express the most profound appreciation of Kentaroh Yoshida
for the fruitful collaborations, and his guidance and generous support.
I am benefited from the stimulating discussions with him and impressed with his approach to physics research.
Thanks to him, I could have an exciting and fruitful time during my Ph.D. course.

\medskip

I also want to thank all my collaborators on various projects:
Thiago~Araujo, Ilya~Bakhmatov, Andrzej~Borowiec, Eoin~\'O~Colg\'ain, Jose~J.~Fern\'andez-Melgarejo, Takashi~Kameyama, Hideki~Kyono, Jerzy~Lukierski, Takuya~Matsumoto, Domenico~Orlando, Susanne~Reffert, Yuho~Sakatani, and Mohammad~M.~Sheikh-Jabbari.
I have learned a lot of things from them, and this thesis would not exist without them.
I am especially grateful to Domenico~Orlando, Susanne~Reffert, and Yuho~Sakatani
for many fruitful discussions and their support we had.

\medskip

I would also like to express my gratitude to all people at Kyoto University.
I would like to especially thank Masafumi~Fukuma and Hikaru~Kawai
who have given insightful comments and taught me a great deal of essence about QFT and string theory.
I am also grateful to our secretary Kiyoe~Yokota.

\medskip

I would like to thank all former and current members at Osaka City University.
I particularly appreciate the support and encouragement from Hiroshi~Itoyama, Nobuhito~Maru, Sanefumi~Moriyama and Yukinori~Yasui.

%

\medskip

Last but not least, I am grateful to my parents and my family for their constant support.

\newpage


\appendix

\chapter{Conventions and Formulas}
\label{app:conventions}

In this Appendix, we explain our conventions and formulas for gamma matrices and spinors.

\section{Differential form and curvature}
\label{app-section:conventions-form}

The antisymmeterization is defined as
\begin{align}
 A_{[m_1\cdots m_n]} \equiv \frac{1}{n!}\,\bigl(A_{m_1\cdots m_n} \pm \text{permutations}\bigr) \,.
\end{align}
For conventions of differential forms, we use
\begin{align}
\begin{split}
 &\varepsilon^{01}= \frac{1}{\sqrt{-\gga}}\,,\qquad 
 \varepsilon_{01}= - \sqrt{-\gga} \,, \qquad 
 \rmd^{2}\sigma = \rmd \tau \wedge\rmd \sigma \,,
\\
 &(*_{\gga} \alpha_q)_{\WSa_1\cdots\WSa_{p+1-q}} =\frac{1}{q!}\,\varepsilon^{\WSb_1\cdots\WSb_q}{}_{\WSa_1\cdots\WSa_{p+1-q}}\,\alpha_{\WSb_1\cdots\WSb_q} \,,
\\
 & *_{\gga} (\rmd \sigma^{\WSa_1}\wedge \cdots \wedge \rmd \sigma^{\WSa_q}) 
 = \frac{1}{(p+1-q)!}\,\varepsilon^{\WSa_1\cdots\WSa_q}{}_{\WSb_1\cdots\WSb_{p+1-q}}\,\rmd \sigma^{\WSb_1}\wedge \cdots \wedge \rmd \sigma^{\WSb_{p+1-q}} \,,
\end{split}
\end{align}
on string worldsheet while in the spacetime, we define
\begin{align}
\begin{split}
 &\varepsilon^{1\cdots D}=-\frac{1}{\sqrt{-\CG}}\,,\qquad 
 \varepsilon_{1\cdots D}= \sqrt{-\CG} \,, \qquad 
 \epsilon^{1\cdots D} = - 1\,,\qquad 
 \epsilon_{1\cdots D} = 1 \,, 
\\
 &(* \alpha_q)_{m_1\cdots m_{D-q}} =\frac{1}{q!}\,\varepsilon^{n_1\cdots n_q}{}_{m_1\cdots m_{D-q}}\,\alpha_{n_1\cdots n_q} \,,\qquad 
 \rmd^{D}x = \rmd x^1\wedge\cdots\wedge\rmd x^D \,,
\\
 &* (\rmd x^{m_1}\wedge \cdots \wedge \rmd x^{m_q}) = \frac{1}{(p+1-q)!}\,\varepsilon^{m_1\cdots m_q}{}_{n_1\cdots n_{p+1-q}}\,\rmd x^{n_1}\wedge \cdots \wedge \rmd x^{n_{p+1-q}} \,,
\\
 &(\iota_v \alpha_n) = \frac{1}{(n-1)!}\,v^n\,\alpha_{nm_1\cdots m_{n-1}}\,\rmd x^{m_1}\wedge\cdots\wedge \rmd x^{m_{n-1}}\,. 
\end{split}
\end{align}

\medskip

The spin connection is defined as
\begin{align}
 \omega_m{}^{\Loa\Lob} \equiv 2\,e^{n[\Loa}\,\partial_{[m} e_{n]}{}^{\Lob]} - e^{p[\Loa}\,e^{\Lob]q}\,\partial_{[p} e_{q]}{}^{\Loc}\,e_{m\Loc} \,,
\end{align}
which satisfies
\begin{align}
 \rmd e^{\Loa} + \omega^{\Loa}{}_{\Lob}\wedge e^{\Lob} = 0\,, 
\end{align}
where $e^{\Loa}\equiv e_m{}^{\Loa}\,\rmd x^m$ and $\omega^{\Loa}{}_{\Lob}\equiv \omega_m{}^{\Loa}{}_{\Lob}\,\rmd x^m$\,. 
The Riemann curvature tensor is defined as
\begin{align}
 R^{\Loa}{}_{\Lob} \equiv \frac{1}{2}\,R^{\Loa}{}_{\Lob\Loc\Lod}\,e^{\Loc}\wedge e^{\Lod} \equiv \rmd \omega^{\Loa}{}_{\Lob} + \omega^{\Loa}{}_{\Loc}\wedge \omega^{\Loc}{}_{\Lob} \,,\qquad 
 R^{\Loa}{}_{\Lob\Loc\Lod} = e_m{}^{\Loa}\,e_{\Lob}{}^n\,e_{\Loc}{}^p\,e_{\Lod}{}^q\,R^m{}_{npq} \,. 
\end{align}

\section{Formulas for gamma matrices and spinors}

Products of antisymmetrized $32\times 32$ gamma matrices satisfy
\begin{align}
\begin{split}
 &\Gamma^{\Loa_1\cdots \Loa_p}\,\Gamma_{\Lob_1\cdots \Lob_q} 
 =\sum^{p+q}_{r=\abs{p-q}}\frac{(-1)^{\frac{u(u-1)}{2}}p!\,q!}{u!\,v!\,w!}\,\eta^{[\underline{\Loa_1}\Loc_1}\cdots\eta^{\underline{\Loa_v}\Loc_v}\,\delta_{[\Lob_1}^{\underline{\Loa_{v+1}}}\cdots \delta_{\Lob_u}^{\underline{\Loa_{p}}]}\,\Gamma_{|\Loc_1\cdots\Loc_v|\Lob_{u+1}\cdots \Lob_q]} 
\\
 &\Bigl[u\equiv \frac{1}{2}\,(p+q-r)\,,\quad
 v\equiv \frac{1}{2}\,(p-q+r)\,,\quad
 w\equiv \frac{1}{2}\,(-p+q+r)\Bigr]\,,
\label{eq:Gamma-p-q}
\end{split}
\end{align}
where the under-barred indices are totally antisymmetrized and the integer $r$ takes values
\begin{align}
 r=\abs{p-q}\,,\ \abs{p-q}+2\,,\ \dotsc \,,\ p+q-2\,,\ p+q\,,
\end{align}
and $u$, $v$ and $w$ are non-negative integers.
As particular cases, we obtain
\begin{align}
 &\Gamma^{\Loa_1\cdots \Loa_n}\,\Gamma^{\Lob}
 = \Gamma^{\Loa_1\cdots \Loa_n\Lob} + n\, \Gamma^{[\Loa_1\cdots \Loa_{n-1}}\,\eta^{\Loa_n]\Lob} \,. 
\label{eq:Gamma-n-1}
\\
 &\Gamma^{\Loa}\,\Gamma^{\Lob_1\cdots \Lob_n} 
 = \Gamma^{\Loa\Lob_1\cdots \Lob_n} + n\, \eta^{\Loa[\Lob_1} \,\Gamma^{\Lob_2\cdots \Lob_n]} \,. 
\end{align}

\medskip

Arbitrary 32-component Majorana spinors $\Theta$ and $\Psi$ satisfy
\begin{align}
 &\brTheta\,\Gamma_{\Loa_1\Loa_2\cdots \Loa_n}\,\Psi = (-1)^{\frac{n(n+1)}{2}} \bar{\Psi}\,\Gamma_{\Loa_1\Loa_2\cdots \Loa_n}\,\Theta\,, 
\\
 &\brTheta\,\Gamma^{\Loa_1 \cdots \Loa_n}\,\Psi = 0 \qquad (n=1,\,2,\,5,\,6,\,9,\,10)\,. 
\end{align}
For spinors with a definite chirality, $\Gamma^{11}\,\Psi_{\pm}=\pm \Psi_{\pm}$ and $\Gamma^{11}\,\Theta_{\pm}=\pm \Theta_{\pm}$, we have
\begin{align}
\begin{split}
 \brTheta_{+}\,\Gamma_{\Loa_1\Loa_2\cdots \Loa_n}\,\Psi_{\pm} 
 &= \begin{cases}
 (-1)^{\frac{n(n+1)}{2}} \bar{\Psi}_{\pm}\,\Gamma_{\Loa_1\Loa_2\cdots \Loa_n}\,\Theta_{+} & (n\text{:odd/even})
\\
 0 & (n\text{:even/odd})
\end{cases}\,,
\\
 \brTheta_{-}\,\Gamma_{\Loa_1\Loa_2\cdots \Loa_n}\,\Psi_{\mp} 
 &= \begin{cases}
 (-1)^{\frac{n(n+1)}{2}} \bar{\Psi}_{\mp}\,\Gamma_{\Loa_1\Loa_2\cdots \Loa_n}\,\Theta_{-} & (n\text{:odd/even})
\\
 0 & (n\text{:even/odd})
\end{cases}\,,
\end{split}
\\
\begin{split}
 \brTheta_{+}\,\Gamma_{\Lob_1}\,\Gamma_{\Loa_1\Loa_2\cdots \Loa_n}\,\Gamma_{\Lob_2}\,\Psi_{\pm} 
 &= \begin{cases}
 (-1)^{\frac{n(n+1)}{2}} \bar{\Psi}_{\pm}\,\Gamma_{\Lob_2}\,\Gamma_{\Loa_1\Loa_2\cdots \Loa_n}\,\Gamma_{\Lob_1}\,\Theta_{+} & (n\text{:odd/even})
\\
 0 & (n\text{:even/odd})
\end{cases}\,,
\\
 \brTheta_{-}\,\Gamma_{\Lob_1}\,\Gamma_{\Loa_1\Loa_2\cdots \Loa_n}\,\Gamma_{\Lob_2}\,\Psi_{\mp} 
 &= \begin{cases}
 (-1)^{\frac{n(n+1)}{2}} \bar{\Psi}_{\mp}\,\Gamma_{\Lob_2}\,\Gamma_{\Loa_1\Loa_2\cdots \Loa_n}\,\Gamma_{\Lob_1}\,\Theta_{-} & (n\text{:odd/even})
\\
 0 & (n\text{:even/odd})
\end{cases}\,.
\end{split}
\label{eq:bTGGGT}
\end{align}

\chapter{$\alg{psu}(2,2|4)$ algebra}
\label{app:psu-algebra}

In this appendix, we collect our conventions and useful formulas on the $\alg{psu}(2,2|4)$ algebra (see for example \cite{Arutyunov:2009ga} for more details). 

\section{Matrix realization}

\subsection*{$8 \times 8$ supermatrix representation}

The super Lie algebra $\alg{su}(2,2|4)$ can be realized by using $8 \times 8$ supermatrices $\cM$ satisfying $\str\cM =0$ and the reality condition
\begin{align}
 \cM^\dagger\, H+H\,\cM =0\,,\qquad 
 \cM = \begin{pmatrix} A & B \\ C & D \end{pmatrix} \,,
\label{eq:reality}
\end{align}
where $\str\cM\equiv \Tr A -\Tr D$ and the hermitian matrix $H$ is defined as
\begin{align}
 H\equiv \begin{pmatrix} \Sigma & \bm{0_4} \\ \bm{0_4} & \bm{1_4} \end{pmatrix} \,,\qquad
 \Sigma \equiv \begin{pmatrix} \bm{0_2} & -\ii\,\sigma_3 \\ \ii\,\sigma_3 & \bm{0_2} \end{pmatrix}=\sigma_2\otimes\sigma_3 \,. 
\end{align}
A trivial element satisfying the above requirement is the $\alg{u}(1)$ generator
\begin{align}
 Z = \ii \begin{pmatrix} \bm{1_4} & \bm{0_4} \\ \bm{0_4} & \bm{1_4} \end{pmatrix} \,,
\label{eq:gZ-def}
\end{align}
and the $\mathfrak{psu}(2,2|4)$ is defined as the quotient $\mathfrak{su}(2,2|4)/\mathfrak{u}(1)$\,. 

\medskip

The $\mathfrak{psu}(2,2|4)$ has an automorphism $\Omega$ defined as
\begin{align}
 \Omega(\cM)=-\cK \,\cM^{\ST}\,\cK^{-1}\,,\qquad
 \cK = \begin{pmatrix} K & \bm{0_4} \\ \bm{0_4} & K \end{pmatrix} \,,
\end{align}
where $K$ is a $4\times 4$ matrix
\begin{align}
 K\equiv {\footnotesize\begin{pmatrix} 0 & -1 & 0 & 0 \\ 1 & 0 & 0 & 0 \\ 0 & 0 & 0 & -1 \\ 0 & 0 & 1 & 0 \end{pmatrix}}\,,\qquad 
 K^{-1} = - K \,,
\end{align}
and $\cM^{\ST}$ represents the supertranspose of $\cM$ defined as
\begin{align}
 \cM^{\ST} = \begin{pmatrix} A^\rmT & -C^{\rmT} \\ B^{\rmT} & D^{\rmT} \end{pmatrix} \,. 
\end{align}
By using the automorphism $\Omega$ (of order four), we decompose $\mathfrak{g}=\mathfrak{psu}(2,2|4)$ as
\begin{align}
 \mathfrak{g}=\mathfrak{g}^{(0)}\oplus\mathfrak{g}^{(1)}\oplus\mathfrak{g}^{(2)}\oplus\mathfrak{g}^{(3)}\,,
\end{align}
where $\Omega(\mathfrak{g}^{(k)})=\ii^k\,\alg{g}^{(k)}$ ($k=0,1,2,3$) and the projector to each vector space $\mathfrak{g}^{(k)}$ can be expressed as
\begin{align}
 P^{(k)}(\cM) \equiv \frac{1}{4}\,\bigl[\, \cM + \ii^{3k}\,\Omega(\cM)+\ii^{2k}\, \Omega^2(\cM) +\ii^k\,\Omega^3(\cM) \,\bigr]\,. 
\label{eq:P-i-projector}
\end{align}

\subsection*{Bosonic generators}

The bosonic generators of $\alg{psu}(2,2|4)$ algebra, $\gP_{\Loa}$ and $\gJ_{\Loa\Lob}$, can be represented by the following $8\times 8$ supermatrices:
\begin{align}
\begin{split}
 &\{\gP_{\Loa}\}\equiv \{\gP_{\check{\Loa}}\,, \gP_{\hat{\Loa}}\}\,,\qquad 
 \{\gJ_{\Loa\Lob}\}\equiv \{\gJ_{\check{\Loa}\check{\Lob}}\,, \gJ_{\hat{\Loa}\hat{\Lob}}\}\,,
\\
 &\gP_{\check{\Loa}} = 
 \begin{pmatrix}
 \frac{1}{2}\,\bm{\gamma}_{\check{\Loa}} & \bm{0_4} \\ 
 \bm{0_4} & \bm{0_4} 
 \end{pmatrix} \,, \qquad 
 \gJ_{\check{\Loa}\check{\Lob}} = 
 \begin{pmatrix}
 -\frac{1}{2}\,\bm{\gamma}_{\check{\Loa}\check{\Lob}} & \bm{0_4} \\ 
 \bm{0_4} & \bm{0_4} \end{pmatrix}
 \qquad (\check{\Loa},\,\check{\Lob}=0,\dotsc,4)\,, 
\\
 &\gP_{\hat{\Loa}} = 
 \begin{pmatrix}
  \bm{0_4} & \bm{0_4} \\ 
  \bm{0_4} & -\frac{\ii}{2}\,\bm{\gamma}_{\hat{\Loa}} 
 \end{pmatrix} \,, \qquad 
 \gJ_{\hat{\Loa}\hat{\Lob}} = 
 \begin{pmatrix}
  \bm{0_4} & \bm{0_4} \\ 
  \bm{0_4} & -\frac{1}{2}\, \bm{\gamma}_{\hat{\Loa}\hat{\Lob}} 
 \end{pmatrix}
 \qquad (\hat{\Loa},\,\hat{\Lob}=5,\dotsc,9)\,, 
\end{split}
\label{eq:P-J-super}
\end{align}
where we defined $4\times 4$ matrices $\bm{\gamma}_{\check{\Loa}} \equiv (\bm{\gamma}_{\check{\Loa}\check{i}}{}^{\check{j}})$ $(\check{i},\check{j}=1,\dotsc,4)$ and $\bm{\gamma}_{\check{\Loa}} \equiv (\bm{\gamma}_{\check{\Loa}\check{i}}{}^{\check{j}})$ $(\hat{i},\hat{j}=1,\dotsc,4)$
\begin{align}
 &\{\bm{\gamma}_{\check{\Loa}}\} \equiv \bigl\{\brgamma_0\,,\brgamma_1\,,\brgamma_2\,,\brgamma_3\,,\brgamma_5\,\bigr\} \,, \qquad 
 \{\bm{\gamma}_{\hat{\Loa}}\} \equiv \bigl\{-\brgamma_4\,,-\brgamma_1\,,-\brgamma_2\,,-\brgamma_3\,,-\brgamma_5\,\bigr\} \,,
\\[1mm]
 \begin{split}
 &\brgamma_1=
 {\footnotesize\begin{pmatrix}
  0 & 0 & 0 & -1\\
  0 & 0 & 1 & 0\\
  0 & 1 & 0 & 0\\
  -1& 0 & 0 & 0 
 \end{pmatrix}} , \quad 
 \brgamma_2=
 {\footnotesize\begin{pmatrix}
 0 & 0 & 0 & \ii \\
 0 & 0 & \ii & 0 \\
 0 & -\ii& 0 & 0 \\
 -\ii& 0 & 0 & 0
 \end{pmatrix}}, \quad 
 \brgamma_3=
 {\footnotesize\begin{pmatrix}
 0 & 0 & 1 & 0 \\
 0 & 0 & 0 & 1 \\
 1 & 0 & 0 & 0 \\
 0 & 1 & 0 & 0 
 \end{pmatrix}},
\\
 &\brgamma_0= -\ii\,\brgamma_4=
 {\footnotesize\begin{pmatrix}
  0 & 0 & 1 & 0 \\
  0 & 0 & 0 &-1 \\
  -1& 0 & 0 & 0 \\
  0 & 1 & 0 & 0 
 \end{pmatrix}}, \quad 
 \brgamma_5=\ii\,\brgamma_1\brgamma_2\brgamma_3\brgamma_0=
 {\footnotesize\begin{pmatrix}
  1 & 0 & 0 & 0 \\
  0 & 1 & 0 & 0 \\
  0 & 0 &-1 & 0 \\
  0 & 0 & 0 &-1
\end{pmatrix}},
\end{split}
\end{align}
and their antisymmeterizations $\bm{\gamma}_{\check{\Loa}\check{\Lob}} \equiv \bm{\gamma}_{[\check{\Loa}}\,\bm{\gamma}_{\check{\Lob}]}$ and $\bm{\gamma}_{\hat{\Loa}\hat{\Lob}} \equiv \bm{\gamma}_{[\hat{\Loa}}\,\bm{\gamma}_{\hat{\Lob}]}$. 
Here, $\brgamma_{\mu}$ ($\mu=0,\dotsc,3$) and $(\bm{\gamma}_{\Loa})\equiv (\bm{\gamma}_{\check{\Loa}},\,\bm{\gamma}_{\hat{\Loa}})$ satisfy
\begin{align}
 \{\brgamma_{\mu}\,, \brgamma_{\nu}\} = 2\,\eta_{\mu\nu}\,,\qquad (\eta_{\mu\nu}) \equiv \diag(-1,1,1,1)\,,\qquad
 (\bm{\gamma}_{\Loa})^{\rmT}= K\,\bm{\gamma}_{\Loa}\,K^{-1}\,. 
\end{align}

\medskip

The conformal basis, $\{P_{\mu},\, M_{\mu\nu},\,D,\,K_{\mu}\}$\,, of a bosonic subalgebra $\alg{su}(2,2)\cong \alg{so}(2,4)$ that corresponds to the AdS isometries, can be constructed from $\gP_{\check{\Loa}}$ and $\gJ_{\check{\Loa}\check{\Lob}}$ as
\begin{align}
 P_\mu \equiv \gP_\mu + \gJ_{\mu 4}\,,\qquad K_\mu \equiv \gP_\mu - \gJ_{\mu 4}\,,\qquad M_{\mu\nu}\equiv \gJ_{\mu\nu}\,,\qquad D\equiv \gP_4\,,
\end{align}
where $P_{\mu}$, $M_{\mu\nu}$, $D$, and $K_{\mu}$ represent the translation generators, the Lorentz generators, the dilatation generator, and the special conformal generators, respectively. 
On the other hand, a bosonic subalgebra $\alg{su}(4)\cong\alg{so}(6)$ that corresponds to the isometries of $\rmS^5$ are generated by $\gP_{\hat{\Loa}}$ and $\gJ_{\hat{\Loa}\hat{\Lob}}$\,. 
We choose the Cartan generators of $\alg{su}(4)$ as follows
\begin{align}
 h_1 \equiv \gJ_{57}\,,\qquad h_2 \equiv \gJ_{68}\,,\qquad h_3 \equiv \gP_9\,.
\end{align}

\medskip

For later convenience, let us also define $16\times 16$ matrices $\gamma_{\Loa}$, $\hat{\gamma}_{\Loa}$, and $\gamma_{\Loa\Lob}$ as
\begin{align}
\begin{split}
 (\gamma_{\Loa})&\equiv(\gamma_{\check{\Loa}},\, \gamma_{\hat{\Loa}}) 
 =(\bm{\gamma}_{\check{\Loa}}\otimes \bm{1_4},\, \bm{1_4}\otimes \bm{\gamma}_{\hat{\Loa}})\,,
\\
 (\hat{\gamma}_{\Loa})&\equiv(\hat{\gamma}_{\check{\Loa}},\, \hat{\gamma}_{\hat{\Loa}}) 
 =(\bm{\gamma}_{\check{\Loa}}\otimes \bm{1_4},\, \bm{1_4}\otimes \ii\,\bm{\gamma}_{\hat{\Loa}})\,,
\\
 (\gamma_{\Loa\Lob})&\equiv (\gamma_{\check{\Loa}\check{\Lob}},\,\gamma_{\hat{\Loa}\hat{\Lob}})
 =(\bm{\gamma}_{\check{\Loa}\check{\Lob}}\otimes \bm{1_4}\,, \bm{1_4}\otimes\bm{\gamma}_{\hat{\Loa}\hat{\Lob}}) \,,
\end{split}
\label{eq:gamma-16}
\end{align}
which satisfy
\begin{align}
\begin{split}
 &(\gamma_{\check{\Loa}})^\dagger = \gamma_{\check{0}}\,\gamma_{\check{\Loa}}\,\gamma_{\check{0}} \,,\qquad 
 (\gamma_{\hat{\Loa}})^\dagger = -\gamma_{\check{0}}\,\gamma_{\hat{\Loa}}\,\gamma_{\check{0}} \,,\qquad 
 (\gamma_{\Loa})^\rmT = (K\otimes K)^{-1}\, \gamma_{\Loa}\,(K\otimes K) \,, 
\\
 &(\hat{\gamma}_{\check{\Loa}})^\dagger = \gamma_{\check{0}}\,\hat{\gamma}_{\check{\Loa}}\,\gamma_{\check{0}} \,,\qquad 
 (\hat{\gamma}_{\Loa})^\rmT = (K\otimes K)^{-1}\, \hat{\gamma}_{\Loa}\,(K\otimes K)\,,
\\
 &\{\gamma_{\Loa},\,\gamma_{\Lob}\}=2\,\eta_{\Loa\Lob}\,,\qquad
 \{\hat{\gamma}_{\check{\Loa}},\,\hat{\gamma}_{\check{\Lob}}\}=2\,\eta_{\check{\Loa}\check{\Lob}}\,,\qquad
 \{\hat{\gamma}_{\hat{\Loa}},\,\hat{\gamma}_{\hat{\Lob}}\}=-2\,\delta_{\hat{\Loa}\hat{\Lob}}\,.
\end{split}
\end{align}
We can easily see $\gamma_{\check{\Loa}\check{\Lob}}= \gamma_{[\check{\Loa}}\,\gamma_{\check{\Lob}]}$ and $\gamma_{\hat{\Loa}\hat{\Lob}}=\gamma_{[\hat{\Loa}}\, \gamma_{\hat{\Lob}]}$\,. 
If we also define $\hat{\gamma}_{\check{\Loa}\check{\Lob}}\equiv \hat{\gamma}_{[\check{\Loa}}\,\hat{\gamma}_{\check{\Lob}]}$ and $\hat{\gamma}_{\hat{\Loa}\hat{\Lob}}\equiv \hat{\gamma}_{[\hat{\Loa}}\,\hat{\gamma}_{\hat{\Lob}]}$\,, they satisfy
\begin{align}
 \hat{\gamma}_{\Loa\Lob} = -\frac{1}{2}\,R_{\Loa\Lob}{}^{\Loc\Lod}\, \gamma_{\Loc\Lod}\,,
\end{align}
where $R_{\Loa\Lob}{}^{\Loc\Lod}$ are the tangent components of Riemann tensor in $\AdS{5}\times\rmS^5$, whose non-vanishing components are
\begin{align}
 R_{\check{\Loa}\check{\Lob}}{}^{\check{\Loc}\check{\Lod}} = -2\, \delta_{[\check{\Loa}}^{[\check{\Loc}}\,\delta_{\check{\Lob}]}^{\check{\Lod}]} \,,\qquad 
 R_{\hat{\Loa}\hat{\Lob}}{}^{\hat{\Loc}\hat{\Lod}} = 2\, \delta_{[\hat{\Loa}}^{[\hat{\Loc}}\,\delta_{\hat{\Lob}]}^{\hat{\Lod}]} \,.
\end{align}

\subsection*{Fermionic generators}

The fermionic generators $(\gQ^I)^{\check{\SPa}\hat{\SPa}}$ $(\check{\SPa}, \hat{\SPa}=1,\dotsc, 4)$ are given by
\begin{align}
 (\gQ^1)^{\check{\SPa}\hat{\SPa}} &=
 \begin{pmatrix}
  \bm{0_4} & \ii\,\delta^{\check{\SPa}}_{\check{i}}\,K^{\hat{j}\hat{\SPa}} \\
  -\delta_{\hat{i}}^{\hat{\SPa}}\, K^{\check{\SPa}\check{j}} & \bm{0_4}
 \end{pmatrix} ,
\qquad 
 (\gQ^2)^{\check{\SPa}\hat{\SPa}} =
 \begin{pmatrix}
  \bm{0_4} & - \delta^{\check{\SPa}}_{\check{i}} \,K^{\hat{j}\hat{\SPa}} \\
  \ii\,\delta_{\hat{i}}^{\hat{\SPa}}\, K^{\check{\SPa}\check{j}}& \bm{0_4}
 \end{pmatrix}\,. 
\label{eq:Q-matrix}
\end{align}
As discussed in \cite{Arutyunov:2015qva}, these matrices do not satisfy the reality condition \eqref{eq:reality} but rather their redefinitions $\mathcal{Q}^I$ do. 
The choice, $\gQ^I$ or $\mathcal{Q}^I$, is a matter of convention, and we here employ $\gQ^I$ by following \cite{Arutyunov:2015qva}. 
We also introduce Grassmann-odd coordinates $\theta_I\equiv (\theta_{\check{\SPa}\hat{\SPa}})_I$ which are 16-component Majorana--Weyl spinors satisfying
\begin{align}
 (\gQ^I\,\theta_I)^{\dagger}\, H+H\,(\gQ^I\,\theta_I) = 0 \,. 
\label{eq:Q-theta-reality}
\end{align}
Since the matrices $\gQ^I$ satisfy
\begin{align}
\begin{split}
 &(\gQ^I)^{\dagger}_{\check{\SPa}\hat{\SPa}}
 =-\ii \,K^{-1}_{\check{\SPa}\check{\SPb}}\,(\gQ^I)^{\check{\SPb}\hat{\SPb}}\,K^{-1}_{\hat{\SPb}\hat{\SPa}} \,,
\\
 &H\,(\gQ^I)^{\check{\SPa}\hat{\SPa}}\,H^{-1}=\ii\,(\bm{\gamma}^0)_{\check{\SPb}}{}^{\check{\SPa}}\,(\gQ^I)^{\check{\SPb}\hat{\SPa}}\,,
\end{split}
\end{align}
the condition \eqref{eq:Q-theta-reality} is equivalent to the Majorana condition
\begin{align}
 \brtheta_I \equiv\theta^\dagger_I\, \gamma^0 = \theta_I^{\rmT}\,(K\otimes K)\,,
\end{align}
or more explicitly,
\begin{align}
 \brtheta_I^{\check{\SPa}\hat{\SPa}} = \theta_{I\check{\SPb}\hat{\SPb}}\, K^{\check{\SPb}\check{\SPa}}\,K^{\hat{\SPb}\hat{\SPa}}\,. 
\end{align}

\subsection*{Commutation relations}

The generators of $\alg{su}(2,2|4)$ algebra, $\gP_{\Loa}$, $\gJ_{\Loa\Lob}$, $\gQ^I$, and $Z$ satisfy the following commutation relations:
\begin{align}
\begin{split}
 [\gP_{\Loa},\,\gP_{\Lob}] &= \frac{1}{2}\,R_{\Loa\Lob}{}^{\Loc\Lod}\,\gJ_{\Loc\Lod}\,, 
\qquad 
 [\gJ_{\Loa\Lob},\,\gP_{\Loc}] = \eta_{\Loc\Loa}\,\gP_{\Lob} - \eta_{\Loc\Lob}\,\gP_{\Loa} \,, 
\\
 [\gJ_{\Loa\Lob},\,\gJ_{\Loc\Lod}] 
 &= \eta_{\Loa\Loc}\,\gJ_{\Lob\Lod}-\eta_{\Loa\Lod}\,\gJ_{\Lob\Loc}-\eta_{\Lob\Loc}\,\gJ_{\Loa\Lod}+\eta_{\Lob\Lod}\,\gJ_{\Loa\Loc} \,, 
\\
 [\gQ^I\,\theta_I,\,\gP_{\Loa}] &= \frac{\ii}{2}\,\epsilon^{IJ}\,\gQ^J\,\hat{\gamma}_{\Loa}\,\theta_I\,, \qquad 
 [\gQ^I\,\theta_I,\,\gJ_{\Loa\Lob}] = \frac{1}{2}\,\delta^{IJ}\,\gQ^I\,\gamma_{\Loa\Lob}\,\theta_J\,,
\\
 [\gQ^I\,\theta_I,\,\gQ^J\,\psi_J] 
 &= -\ii\, \delta^{IJ}\, \brtheta_I\,\hat{\gamma}^{\Loa}\,\psi_J\, \gP_{\Loa} 
   - \frac{1}{4}\,\epsilon^{IJ}\, \brtheta_I\,\gamma^{\Loa\Lob}\,\psi_J\,R_{\Loa\Lob}{}^{\Loc\Lod}\, \gJ_{\Loc\Lod} 
   - \frac{1}{2}\, \delta^{IJ}\, \brtheta_I\,\psi_J\,Z \,,
\end{split}
\label{eq:su(2,2|4)}
\end{align}
and the $\alg{psu}(2,2|4)$ algebra is obtained by dropping the last term proportional to $Z$\,. 

On the other hand, the bosonic generators $\{P_{\mu},\, M_{\mu\nu},\,D,\,K_{\mu}\}$ satisfy the $\alg{so}(2,4)$ algebra,
\begin{align}
\begin{split}
 [P_\mu,\, K_\nu]&= 2\,\bigl(\eta_{\mu\nu}\, D - M_{\mu\nu}\bigr)\,,\quad 
 [D,\, P_{\mu}]= P_\mu\,,\quad [D,\,K_\mu]= -K_\mu\,,
\\
 [M_{\mu\nu},\, P_\rho] &= \eta_{\mu\rho}\, P_\nu-\eta_{\nu\rho}\, P_\mu \,,\quad 
 [M_{\mu\nu},\, K_\rho] = \eta_{\mu\rho}\,K_\nu-\eta_{\nu\rho}\,K_\mu\,, 
\\
 [M_{\mu\nu},\,M_{\rho\sigma}]&= \eta_{\mu\rho}\,M_{\nu\sigma}-\eta_{\mu\sigma}\,M_{\nu\rho} - \eta_{\nu\rho}\,M_{\mu\sigma}+\eta_{\nu\sigma}\,M_{\mu\rho}\,.
\end{split}
\label{eq:so(2-4)-algebra}
\end{align}

\subsection*{Supertrace and Projections}

For generators of the $\alg{psu}(2,2|4)$ algebra, the supertrace become
\begin{align}
\begin{split}
 &\str(\gP_{\Loa}\,\gP_{\Lob})=\eta_{\Loa\Lob}\,,\qquad
 \str(\gJ_{\Loa\Lob}\,\gJ_{\Loc\Lod}) =R_{\Loa\Lob\Loc\Lod}\,,
\\
 &\str(\gQ^I\theta_I\,\gQ^J\lambda_J) =-2\,\epsilon^{IJ}\,\brtheta_I\,\lambda_J\,, 
\end{split}
\end{align}
where $R_{\Loa\Lob\Loc\Lod}\equiv R_{\Loa\Lob}{}^{\Loe\Lof}\,\eta_{\Loe\Loc}\,\eta_{\Lod\Lof}$ and
\begin{align}
 \eta_{\Loa\Lob} \equiv \begin{pmatrix} \eta_{\check{\Loa}\check{\Lob}} & 0 \\ 0 & \eta_{\hat{\Loa}\hat{\Lob}} \end{pmatrix}\,, \quad 
 \eta_{\check{\Loa}\check{\Lob}} \equiv \diag (-1,1,1,1,1)\,,\quad 
 \eta_{\hat{\Loa}\hat{\Lob}} \equiv \diag (1,1,1,1,1)\,. 
\end{align}

\medskip

Each $\mathbb{Z}_4$-component $\alg{g}^{(i)}$ is spanned by the following generators:
\begin{align}
 \alg{g}^{(0)}\! = \Span_{\mathbb{R}}\{\gJ_{\Loa\Lob}\}\,,\quad 
 \alg{g}^{(1)}\! = \Span_{\mathbb{R}}\{\gQ^1\}\,,\quad 
 \alg{g}^{(2)}\! = \Span_{\mathbb{R}}\{\gP_{\Loa}\}\,,\quad 
 \alg{g}^{(3)}\! = \Span_{\mathbb{R}}\{\gQ^2\}\,. 
\end{align}
Then, from the definition of $d_{\pm}$ \eqref{eq:dpm},
\begin{align}
 d_{\pm} \equiv \mp P^{(1)}+2\,P^{(2)}\pm P^{(3)}\,.
\end{align}
we obtain
\begin{align}
 d_\pm(\gP_{\Loa}) = 2\, \gP_{\Loa}\,,\qquad d_\pm(\gJ_{\Loa\Lob}) =0\,,\qquad 
 d_\pm(\gQ^I) = \mp \sigma_3^{IJ}\,\gQ^J \,. 
\end{align}

\section{Connection to ten-dimensional quantities}

By using the $16\times 16$ matrices $\gamma_{\Loa}$ defined in \eqref{eq:gamma-16}, the $32\times 32$ gamma matrices $(\Gamma_{\Loa})^{\SPa}{}_{\SPb}$ are realized as
\begin{align}
 (\Gamma_{\Loa}) \equiv \bigl(\Gamma_{\check{\Loa}},\, \Gamma_{\hat{\Loa}}\bigr) 
 \equiv \bigl(\sigma_1\otimes \gamma_{\check{\Loa}},\, \sigma_2\otimes \gamma_{\hat{\Loa}}\bigr)\,. 
\end{align}
We can also realize the charge conjugation matrix as
\begin{align}
 C= \ii\,\sigma_2\otimes K \otimes K \,. 
\end{align}

\medskip

The $32$-component Majorana--Weyl fermions $\Theta_I$ expressed as
\begin{align}
 \Theta_I = \begin{pmatrix} 1 \\ 0\end{pmatrix}\otimes \theta_I \,,
\label{eq:Theta-theta}
\end{align}
which satisfies the chiral conditions
\begin{align}
 \Gamma^{11}\,\Theta_I = \Theta_I\,. 
\end{align}
The Majorana condition is given by
\begin{align}
 \brTheta_I = \Theta_{I}^\rmT\,C = \begin{pmatrix} 0 & 1 \end{pmatrix}\otimes \brtheta_I \,.
\label{eq:brTheta-brtheta}
\end{align}
This decomposition leads to the following relations between $32$- and $8$-component fermions:
\begin{align}
 &\brtheta_I \hat{\gamma}_{\Loa} \theta_J = \brTheta_I \Gamma_{\Loa} \Theta_J \,,
\label{eq:lift-32-AdS5-1}
\\
 &\brtheta_I\,\hat{\gamma}_{\Loa}\,\hat{\gamma}_{\Lob}\,\theta_J 
 = -\ii\,\brTheta_I\,\Gamma_{\Loa}\, \Gamma_{01234}\,\Gamma_{\Lob}\,\Theta_J 
 = \ii\,\brTheta_I\,\Gamma_{\Loa}\, \Gamma_{56789}\,\Gamma_{\Lob}\,\Theta_J \,,
\\
&\ii\,\sigma_1\otimes \bm{1_4}\otimes \bm{1_4} = \Gamma_{01234}\,,\qquad 
 \sigma_2\otimes \bm{1_4}\otimes \bm{1_4} = \Gamma_{56789} \,,
\end{align}
The second relation plays an important role for a supercoset construction of the $\AdS{5} \times \rmS^5$ background since the R--R bispinor in the $\AdS{5}\times \rmS^5$ background takes the form
\begin{align}
 \bisF_5 
 =\frac{1}{5!}\,\bisF_{\Loa_1\cdots \Loa_5}\,\Gamma^{\Loa_1\cdots \Loa_5}
 =4\,(\Gamma^{01234}+\Gamma^{56789})\,.
\end{align}
Indeed, we obtain
\begin{align}
 \brtheta_I\,\hat{\gamma}_{\Loa}\,\hat{\gamma}_{\Lob}\theta_J
 =\frac{\ii}{8}\,\brTheta_I\,\Gamma_{\Loa}\,\bisF_5\,\Gamma_{\Lob}\,\Theta_J\,. 
\label{eq:lift-32-AdS5-2}
\end{align}
We can also show the following relations:\footnote{Recall that $\gamma_{\Loa\Lob}$ has only the components $(\gamma_{\Loa\Lob})=(\gamma_{\check{\Loa}\check{\Lob}},\,\gamma_{\hat{\Loa}\hat{\Lob}})$.}
\begin{align}
 &\brtheta_I\,\hat{\gamma}_{\Loa}\, \gamma_{\Lob\Loc}\,\theta_J = \brTheta_I\,\Gamma_{\Loa}\, \Gamma_{\Lob\Loc} \,\Theta_J \,, 
\label{eq:lift-32-AdS5-3}
\\
 &\brtheta_I\, \gamma_{\Loa\Lob} \,\theta_J
 = -\ii\,\brTheta_I\,\Gamma_{01234}\, \Gamma_{\Loa\Lob} \,\Theta
 = -\ii\,\brTheta_I\,\Gamma_{56789}\, \Gamma_{\Loa\Lob} \,\Theta_J \,, 
\\
 & \brtheta_I\, \gamma_{\Loa\Lob}\,\gamma_{\Loc\Lod}\,\theta_J
 = -\ii\,\brTheta_I\,\Gamma_{01234}\, \Gamma_{\Loa\Lob}\,\Gamma_{\Loc\Lod} \,\Theta_J 
 = -\ii\,\brTheta_I\,\Gamma_{56789}\, \Gamma_{\Loa\Lob}\,\Gamma_{\Loc\Lod} \,\Theta_J \,. 
\end{align}

\chapter{Geometry of reductive homogeneous space}
\label{app:homogeneous-space}

In this appendix, we review geometry of reductive homogeneous spaces (see for example \cite{Castellani:1991et,Ortin:2004ms} for more details). 

\medskip

\section{Generalities}

Let us consider a homogeneous space $G/H$ and decompose the Lie algebra as a direct sum of vector spaces, $\alg{g}=\alg{h}\oplus\alg{k}$. 
If $[\alg{k},\,\alg{h}]\subset \alg{k}$ is satisfied, $G/H$ is called reductive, and if $[\alg{k},\,\alg{k}]\subset \alg{h}$ is further satisfied, $G/H$ is called symmetric. 
We denote the basis of $\alg{h}$ as $\{\sfJ_{\sfi}\}$ $(\sfi=1,\dotsc,\dim G-\dim H)$ and those of $\alg{k}$ as $\{\sfP_{\sfa}\}$ $(\sfa=1,\dotsc,\dim G-\dim H)$\,. 

\medskip

We choose a gauge where the coset representative $g(x)$ is expanded only in terms of $\sfK_{m}$, like $g(x)=\exp(x^{m}\,\sfK_{m})$ ($m=1,\dotsc,\dim G-\dim H$)\,. 
Here, $\{\sfK_{m}\}$ is arbitrary as long as $\{\sfK_{m}\}$ and $\{\sfJ_{\sfi}\}$ span the vector spaces $\alg{g}$\,. 
An obvious choice is $\{\sfK_{m}\}=\{\sfP_{\sfa}\}$, but it is not necessary to choose in that way. 
Once we fix the set $\{\sfK_{m}\}$, in order to maintain the gauge choice under a left multiplication $g(x)\to g_L\,g(x)$, we need to simultaneously perform a local right multiplication,
\begin{align}
 g(x) \ \to \ g(x') = g_L\,g(x)\,h^{-1}(x)\qquad (h\in H) \,.
\label{eq:left-G}
\end{align}
Then, if we expand the left-invariant Maurer--Cartan 1-form as
\begin{align}
 A\equiv g^{-1}\,\rmd g = e^{\sfa}\,\sfP_{\sfa} - \Omega^{\sfi}\,\sfJ_{\sfi} \,, 
\label{eq:MC-1-form-decomp}
\end{align}
we obtain the following transformation laws under the left multiplication \eqref{eq:left-G}:
\begin{align}
 e^{\sfa}(x)\ \to \ e'^{\sfa}(x') = \Lambda^{\sfa}{}_{\sfb}\,e^{\sfb}(x) \,,\qquad 
 \Omega^{\sfi}(x)\ \to \ \Omega'^{\sfi}(x') = \bigl[\Lambda^{\sfi}{}_{\sfj} \, \Omega^{\sfj} - (h^{-1}\,\rmd h)^{\sfi}\bigr](x) \,,
\end{align}
where we have defined $h\,\sfP_{\sfa}\,h^{-1}=\Lambda^{\sfb}{}_{\sfa}\,\sfP_{\sfb}$ and $h\,\sfJ_{\sfi}\,h^{-1}\equiv \Lambda^{\sfj}{}_{\sfi}\,\sfP_{\sfj}$\,. 
This shows that $\Omega^{\sfi}$ behaves as a connection of $H$\,. 
From the decomposition \eqref{eq:MC-1-form-decomp}, the Maurer--Cartan equations become
\begin{align}
\begin{split}
 0 =\rmd A + A\wedge A 
 &= \bigl(\rmd e^{\sfa} - \Omega^{\sfi} \wedge e^{\sfb}\, f_{\sfi\sfb}{}^{\sfa} 
 + \frac{1}{2}\, e^{\sfb}\wedge e^{\sfc}\,f_{\sfb\sfc}{}^{\sfa}\bigr)\,\sfP_{\sfa}
\\
 &\quad - \Bigl(\rmd \Omega^{\sfi} - \frac{1}{2}\,\Omega^{\sfj}\wedge \Omega^{\sfk}\,f_{\sfj\sfk}{}^{\sfi} - \frac{1}{2}\, e^{\sfb}\wedge e^{\sfc}\,f_{\sfb\sfc}{}^{\sfi} \Bigr) \,\sfJ_{\sfi}\,. 
\end{split}
\end{align}
If we regard $e^{\sfa}$ as the vielbein on $G/H$ and suppose the absence of torsion
\begin{align}
 T^{\sfa} \equiv \rmd e^{\sfa} + \omega^{\sfa}{}_{\sfb}\wedge e^{\sfb} = 0\,,
\end{align}
the Maurer--Cartan equations show that the spin connection can be expressed as
\begin{align}
 \omega^{\sfa}{}_{\sfb} = - \Omega^{\sfi} \, f_{\sfi\sfb}{}^{\sfa} + \frac{1}{2}\, e^{\sfc} \,f_{\sfc\sfb}{}^{\sfa} \,.
\label{eq:spin-connection-Omega}
\end{align}
Moreover, the associated Riemann curvature tensor is expressed as
\begin{align}
\begin{split}
 R^{\sfa}{}_{\sfb} &\equiv \frac{1}{2}\,e^{\sfc}\wedge e^{\sfd}\,R_{\sfc\sfd}{}^{\sfa}{}_{\sfb} 
 \equiv \rmd \omega^{\sfa}{}_{\sfb} + \omega^{\sfa}{}_{\sfc}\wedge \omega^{\sfc}{}_{\sfb} 
\\
 &= - \frac{1}{2}\, e^{\sfe}\wedge e^{\sff}\,\Bigl(f_{\sfe\sff}{}^{\sfi}\, f_{\sfi\sfb}{}^{\sfa} +\frac{1}{2}\,f_{\sfe\sff}{}^{\sfc}\,f_{\sfc\sfb}{}^{\sfa}-\frac{1}{2}\,f_{\sfe\sfd}{}^{\sfa}\,f_{\sff\sfb}{}^{\sfd} \Bigr) \,,
\\
 R^{\sfa}{}_{\sfb\sfc\sfd} &= - \Bigl(f_{\sfc\sfd}{}^{\sfi}\, f_{\sfi\sfb}{}^{\sfa} +\frac{1}{2}\,f_{\sfc\sfd}{}^{\sfe}\,f_{\sfe\sfb}{}^{\sfa}-\frac{1}{2}\,f_{\sfc\sfe}{}^{\sfa}\,f_{\sfd\sfb}{}^{\sfe} \Bigr)\,.
\end{split}
\label{eq:Riemann-reductive}
\end{align}

\medskip

In order to obtain the Killing vectors on $G/H$, let us consider an infinitesimal left multiplication
\begin{align}
 g_L=1+\epsilon^i\,T_i\,,\qquad h=1-\epsilon^i\,W_i{}^{\sfi}\,\sfJ_{\sfi}\,, 
\end{align}
under which the coordinates are supposed to transform as
\begin{align}
 x'^{m} = x^{m} + \epsilon^i\,\hat{T}_i^m \,. 
\end{align}
We obtain
\begin{align}
\begin{split}
 \epsilon^i\,\bigl( T_i\,g + g\,W_i{}^{\sfi}\,\sfJ_{\sfi}\bigr) 
 &= \delta_{\epsilon} g = g(x+ \epsilon^i\,\hat{T}_i)-g(x)
 = \epsilon^i\, \hat{T}_i^m\, \partial_m g 
\\
 &= \epsilon^i\, \hat{T}_i^m\, g\,\bigl(e_m{}^{\sfa}\,\sfP_{\sfa} - \Omega_m{}^{\sfi}\,\sfJ_{\sfi}\bigr)\,,
\end{split}
\end{align}
and this leads to
\begin{align}
 \bigl[\Ad_{g^{-1}}\bigr]_i{}^j\,T_j \equiv g^{-1}\,T_i\,g = \hat{T}_i^{\sfa}\,\sfP_{\sfa} - \bigl(\hat{T}_i^m\,\Omega_{m}{}^{\sfi}\,\sfJ_{\sfi} + W_i{}^{\sfi}\bigr)\,\sfJ_{\sfi} \,,
\end{align}
where $\hat{T}_i^{\sfa} \equiv \hat{T}_i^m\, e_m{}^{\sfa}$\,. 
We thus obtain the following expression:
\begin{align}
 \hat{T}_i^{\sfa} = \bigl[\Ad_{g^{-1}}\bigr]_i{}^{\sfa} \,,\qquad 
 W_i{}^{\sfi} = - \hat{T}_i^m\,\Omega_{m}{}^{\sfi} - \bigl[\Ad_{g^{-1}}\bigr]_i{}^{\sfi} \,.
\label{eq:Killing-formula}
\end{align}
Under the same variation, we obtain
\begin{align}
 \delta_\epsilon A&= \epsilon^i\,\bigl[e^{\sfa}\,W_i{}^{\sfi}\,f_{\sfa\sfi}{}^{\sfb}\,\sfP_{\sfb} + (\rmd W_i{}^{\sfi}-\Omega^{\sfj}\,W_i{}^{\sfk}\,f_{\sfj\sfk}{}^{\sfi})\,\sfJ_{\sfi}\bigr]\,,
\\
 \delta_\epsilon e^{\sfa}&=\epsilon^i\, e^{\sfb}\,W_i{}^{\sfi}\,f_{\sfb\sfi}{}^{\sfa}\,,\qquad 
 \delta_\epsilon \Omega^{\sfi} = \epsilon^i\,\bigl(\Omega^{\sfj}\,W_i{}^{\sfk}\,f_{\sfj\sfk}{}^{\sfi} - \rmd W_i{}^{\sfi}\bigr) \,.
\end{align}
If we define the metric on $G/H$ as
\begin{align}
 g_{mn} \equiv e_m{}^{\sfa}\,e_n{}^{\sfb}\,\kappa_{\sfa\sfb} \,,
\end{align}
by using a constant matrix $\kappa_{\sfa\sfb}$ satisfying
\begin{align}
 f_{\sfi(\sfa}{}^{\sfc}\,\kappa_{\sfb)\sfc} = 0 \,,
\end{align}
the metric is invariant under the variation,
\begin{align}
 \delta_\epsilon g_{mn} = -2\,\epsilon^i\, e_{(m}{}^{\sfa}\,e_{n)}{}^{\sfb}\,W_i{}^{\sfi}\,f_{\sfi(\sfa}{}^{\sfc}\,\kappa_{\sfb)\sfc} = 0 \,. 
\end{align}
We can check that the variation is the same as the Lie derivative,
\begin{align}
 \delta_\epsilon e_m{}^{\sfa} = \epsilon^i\,\Lie_{\hat{T}_i} e_m{}^{\sfa} = \epsilon^i\,\bigl(\hat{T}_i^n\,\partial_n e_m{}^{\sfa} + \partial_m \hat{T}_i^n\,e_n{}^{\sfa}\bigr) \,,
\end{align}
and the invariance of the metric indicates that $\hat{T}_i^m$ are Killing vectors associated with the generator $T_i$\,. 

\medskip

From $T_i\,g = \hat{T}_i^m\, \partial_m g -g\,W_i{}^{\sfi}\,\sfJ_{\sfi}$\,, we can calculate commutators of two variations as
\begin{align}
 [T_i,\,T_j]\,g = -(\Lie_{\hat{T}_i} \hat{T}_j)^m\, \partial_m g 
 + \bigl(\hat{T}_i^n\,\partial_n W_j{}^{\sfi} - \hat{T}_j^n\,\partial_n W_i{}^{\sfi}+ W_i{}^{\sfj}\, \,W_j{}^{\sfk}\,f_{\sfj\sfk}{}^{\sfi} \bigr)\, g\,\sfJ_{\sfi} \,. 
\end{align}
On the other hand, from $[T_i,\,T_j]=f_{ij}{}^k\,T_k$\,, we can also express the left-hand side as
\begin{align}
 [T_i,\,T_j]\,g = f_{ij}{}^k\,T_k\,g = f_{ij}{}^k\,\hat{T}_k^m\, \partial_m g -f_{ij}{}^k\,g\,W_k{}^{\sfi}\,\sfJ_{\sfi}\,,
\end{align}
and by comparing these, we obtain
\begin{align}
\begin{split}
 &[\hat{T}_i,\, \hat{T}_j]^m = (\Lie_{\hat{T}_i} \hat{T}_j)^m = - f_{ij}{}^k\,\hat{T}_k^m\,,
\\
 &\hat{T}_i^n\,\partial_n W_j{}^{\sfi} - \hat{T}_j^n\,\partial_n W_i{}^{\sfi}+ W_i{}^{\sfj}\, \,W_j{}^{\sfk}\,f_{\sfj\sfk}{}^{\sfi} = -f_{ij}{}^k\,g\,W_k{}^{\sfi}\,. 
\end{split}
\end{align}

\section{$\AdS{5}\times \rmS^5$}

In the case of the (bosonic) coset
\begin{align}
 \AdS{5}\times \rmS^5 = \frac{\SO(2,4)}{\SO(1,4)}\times \frac{\SO(6)}{\SO(5)}\,,
\end{align}
the two sets of generators are given by
\begin{align}
 \{\sfP_{\sfa}\} = \bigl\{\gP_{\Loa}\bigr\}\,,\qquad 
 \{\sfJ_{\sfi}\} = \bigl\{\gJ_{\Loa\Lob}/\sqrt{2!} \bigr\}\,,
\end{align}
and it is a symmetric coset space ($f_{\Loa\Lob}{}^{\Loc}=0$). 
The normalization $\tfrac{1}{\sqrt{2!}}$ is introduced to prevent overcounting coming from the summation of antisymmetrized indices. 
Quantities with the index $\sfi$ always contains the factor $\tfrac{1}{\sqrt{2!}}$ and, for example, the Maurer--Cartan 1-form \eqref{eq:MC-1-form-decomp} is expressed as
\begin{align}
 A = e^{\sfa}\,\sfP_{\sfa} - \Omega^{\sfi}\,\sfJ_{\sfi}\qquad\leftrightarrow\qquad
 A = e^{\Loa}\,\gP_{\Loa} - \frac{1}{2}\,\Omega^{\Loa\Lob}\,\gJ_{\Loa\Lob} \,. 
\end{align}
From \eqref{eq:spin-connection-Omega} and $f_{\Loa\Lob}{}^{\Loc}=0$, the spin connection becomes
\begin{align}
 \omega^{\Loa}{}_{\Lob} = - \frac{1}{2}\, \Omega^{[\Loc\Lod]} \, f_{[\Loc\Lod]\Lob}{}^{\Loa} 
 = \Omega^{\Loa\Loc} \, \eta_{\Loc\Lob} \,,
\label{eq:index-convention}
\end{align}
where we used $f_{[\Loc\Lod]\Lob}{}^{\Loa}=2\,\eta_{\Lob[\Loc}\,\delta_{\Lod]}^{\Loa}$ (see $[\gJ,\,\gP]$-commutator of \eqref{eq:su(2,2|4)}) and we obtain
\begin{align}
 A = e^{\Loa}\,\gP_{\Loa} - \frac{1}{2}\,\omega^{\Loa\Lob}\,\gJ_{\Loa\Lob} \,,
\end{align}
independent of the explicit parameterization of $g$ like \eqref{eq:group-parameterization}. 

\medskip

From \eqref{eq:Riemann-reductive} and $f_{\Loa\Lob}{}^{\Loc}=0$, the Riemann curvature tensor becomes
\begin{align}
 R^{\Loa}{}_{\Lob\Loc\Lod} = - \frac{1}{2}\, f_{\Loc\Lod}{}^{[\Loe\Lof]}\, f_{[\Loe\Lof]\Lob}{}^{\Loa} 
 = f_{\Loc\Lod}{}^{[\Loa\Loe]}\, \eta_{\Loe\Lob} \,.
\end{align}
This explains why the $[\gP,\,\gP]$-commutator in \eqref{eq:su(2,2|4)} is expressed in terms of the Riemann tensor; $f_{\Loc\Lod}{}^{[\Loa\Lob]}=R^{\Loa\Lob}{}_{\Loc\Lod}=R_{\Loc\Lod}{}^{\Loa\Lob}$\,.

\chapter{The $\beta$-transformation of R--R field strengths}

In this Appendix, we give proofs of some formulas related to the $\beta$-transformation of R--R field strengths.
We stat to prove that the formula of YB deformed R--R field strengths (\ref{eq:R-R-relation1}) is equivalent to
the $\beta$-transformation rule \eqref{eq:R-R-relation1}.
In the Appendix \ref{app:omega},
we show the formula \eqref{eq:Hassan-formula} in an arbitrary even dimension $D$
which gives the spinor representation of a local Lorentz transformation. 

\section{Equivalence of \eqref{eq:R-R-relation1} and \eqref{eq:R-R-relation2}}
\label{app:equivRR}

In the subsection \ref{subsec:YBfromGS},
YB deformed R--R field strengths are given by a formula \eqref{eq:R-R-relation2}
\begin{align}
 \bisF = \check{\bm{\cF}}\,\Omega_0^{-1}\,,\qquad 
 \Omega^{-1}_0 =(\det \Einv_{\Loa}{}^{\Lob})^{-\frac{1}{2}} 
\text{\AE}\Bigl(-\frac{1}{2}\,\beta^{\Loa\Lob}\,\Gamma_{\Loa\Lob}\Bigr) \,,
\end{align}
where $\text{\AE}\Bigl(-\frac{1}{2}\,\beta^{\Loa\Lob}\,\Gamma_{\Loa\Lob}\Bigr)$ is defined by
\begin{align}
 \text{\AE}\bigl(\tfrac{1}{2}\,\beta^{\Loa\Lob}\,\Gamma_{\Loa\Lob}\bigr) \equiv \sum^5_{p=0}\frac{1}{2^{p}\,p!}\, \beta_{\Loa_1\Loa_2}\cdots\beta_{\Loa_{2p-1}\Loa_{2p}}\,\Gamma^{\Loa_1\cdots \Loa_{2p}}\,. 
\end{align}
We here prove that the relation \eqref{eq:R-R-relation1},
\begin{align}
 \hat{\cF} = \Exp{\Phi-\tilde{\phi}}\Exp{-B_2\wedge} \Exp{-\beta\vee} \check{\cF} \,,
\end{align}
is equivalent to the formula \eqref{eq:R-R-relation2}.
Since $\Exp{\Phi-\tilde{\phi}} = (\det \Einv_{\Loa}{}^{\Lob})^{-\frac{1}{2}}$ from \eqref{eq:two-dilatons},
we here show the equivalence of two formulas,
\begin{align}
 \hat{F} = \Exp{-B_2\wedge} \Exp{-\beta\vee} \check{F} \qquad \Leftrightarrow\qquad 
 \hat{\bmF} = \check{\bmF}\,\brOmega_0^{-1} \,,\quad \brOmega_0^{-1} \equiv \text{\AE}\bigl(-\tfrac{1}{2}\,\beta^{\Loa\Lob}\,\Gamma_{\Loa\Lob}\bigr) \,,
\end{align}
where $\hat{\bmF}\equiv \Exp{-\Phi}\bisF$ and $\check{\bmF}\equiv \Exp{-\tilde{\phi}}\check{\bm{\cF}}$\,. 

\medskip

Let us first evaluate $\Exp{-B_2\wedge}\Exp{-\beta\vee}\check{F}$\,. 
This can be expanded as
\begin{align}
 \Exp{-B_2\wedge}\Exp{-\beta\vee}\check{F}
 &=\sum_{k:\Atop{\text{even}}{\text{odd}}} \sum_{t=0}^{[\frac{k}{2}]} \sum_{s=0}^{t+[\frac{D-k}{2}]} \frac{(-1)^{s}}{2^{s+t}\,s!\,t!\,(k-2\,t)!}\,B_{m_1m_2}\cdots B_{m_{2s-1}m_{2s}}\,\beta^{n_1n_2}\cdots\beta^{n_{2t-1}n_{2t}}
\nn\\
 &\quad\times
 \check{F}_{n_1\cdots n_{2t}m_{2s+1}\cdots m_{2s+k-2t}}\, \rmd x^{m_1}\wedge\cdots\wedge\rmd x^{m_{2s+k-2t}}
\nn\\
 &=\sum_{r:\Atop{\text{even}}{\text{odd}}} \sum_{s=0}^{[\frac{r}{2}]} \sum_{t=0}^{s+[\frac{D-r}{2}]} \frac{(-1)^s}{2^{s+t}\,s!\,t!\,(r-2s)!}\,\beta^{\Loc_1\Loc_2}\cdots\beta^{\Loc_{2t-1}\Loc_{2t}}\,
 \beta_{\Lob_1\Lob_2}\cdots \beta_{\Lob_{2s-1}\Lob_{2s}}
\nn\\
 &\quad\times e_{m_1}{}^{\Lob_1}\cdots e_{m_{2s}}{}^{\Lob_{2s}}\,
 \check{F}_{\Loc_1\cdots \Loc_{2t}m_{2s+1}\cdots m_r}\,\rmd x^{m_1}\wedge\cdots\wedge\rmd x^{m_r} \,,
\end{align}
where the square bracket $[n]$ denotes the integral part of $n$\,, and in the second equality, we have used relations \eqref{eq:dual-RR-flat-components} and \eqref{eq:g-G-B-beta}. 
Then, $\hat{F}$ with flat indices becomes
\begin{align}
 \hat{F}_{\Loa_1\cdots \Loa_k} &\equiv e_{\Loa_1}{}^{m_1}\cdots e_{\Loa_k}{}^{m_k}\,\hat{F}_{m_1\cdots m_k} 
\nn\\
 &=\sum_{s=0}^{[\frac{k}{2}]} \sum_{t=0}^{s+[\frac{D-k}{2}]} \frac{(-1)^{s}\,k!}{2^{s+t}\,s!\,t!\,(k-2s)!}\,
 \beta^{\Loc_{1}\Loc_{2}}\cdots\beta^{\Loc_{2t-1}\Loc_{2t}} 
\nn\\
&\quad\times
 \beta_{[\Loa_1\Loa_2}\cdots \beta_{\Loa_{2s-1}\Loa_{2s}}\, (\Einv^\rmT)_{\Loa_{2s+1}}{}^{\Lob_{2s+1}}\cdots (\Einv^\rmT)_{\Loa_{k}]}{}^{\Lob_{k}}\,
 \check{F}_{\Lob_{2s+1}\cdots \Lob_{k} \Loc_1\cdots \Loc_{2t}}
\nn\\
 &=\sum_{s=0}^{[\frac{k}{2}]} \sum_{t=0}^{s+[\frac{D-k}{2}]} \sum_{u=0}^{k-2s} \frac{(-1)^{s}\,k!}{2^{s+t}\,s!\,t!\,u!\,(k-2s-u)!}\,
 \beta^{\Loc_{1}\Loc_{2}}\cdots\beta^{\Loc_{2t-1}\Loc_{2t}} 
\nn\\
 &\quad\times \beta_{[\Loa_1\Loa_2}\cdots \beta_{\Loa_{2s-1}\Loa_{2s}}\,
 \beta_{\Loa_{2s+1}}{}^{\Lob_{1}}\cdots \beta_{\Loa_{2s+u}}{}^{\Lob_{u}}\,
 \check{F}_{|\Lob_{1}\cdots \Lob_{u}|\Loa_{2s+u+1}\cdots \Loa_k] \Loc_1\cdots \Loc_{2t}}\,,
\label{eq:Fhat-Fcheck-flat}
\end{align}
where we used $e_{\Loa}{}^m = (\Einv^{\rmT})_{\Loa}{}^{\Lob}\, \tilde{e}_{\Lob}{}^m$ and $(\Einv^\rmT)_{\Loa}{}^{\Lob}=\delta_{\Loa}^{\Lob}+\beta_{\Loa}{}^{\Lob}$\,. 

\medskip

Next, by using the definitions,
\begin{align}
 \check{\bmF} = \sum_{k:\Atop{\text{even}}{\text{odd}}}\frac{1}{k!}\, \check{F}_{\Loa_1\cdots \Loa_k}\,\Gamma^{\Loa_1\cdots \Loa_k} \,,\qquad
 \brOmega_0^{-1} = \sum_{\ell=0}^{[\frac{D}{2}]} \frac{(-1)^\ell}{2^\ell\,\ell!}\,
 \beta^{\Lob_1 \Lob_2}\cdots\beta^{\Lob_{2\ell-1} \Lob_{2\ell}}\,\Gamma_{\Lob_1\cdots \Lob_{2\ell}} \,,
\end{align}
let us expand the right-hand side of $\hat{\bmF}=\check{\bmF} \,\brOmega_0^{-1}$ as
\begin{align}
 \hat{\bmF} 
 &= \sum^{[\frac{D}{2}]}_{\ell=0}\sum_{k:\Atop{\text{even}}{\text{odd}}}\frac{(-1)^\ell}{2^\ell\,\ell!\,k!}\,
 \check{F}_{\Loa_1\cdots \Loa_k}\,
 \beta^{\Lob_1\Lob_2}\cdots\beta^{\Lob_{2\ell-1}\Lob_{2\ell}}\,
 \Gamma^{\Loa_1\cdots \Loa_k}\,\Gamma_{\Lob_1\cdots \Lob_{2\ell}}
\nn\\
 &= \sum^{[\frac{D}{2}]}_{\ell=0}\sum_{k:\Atop{\text{even}}{\text{odd}}}\sum^{2\ell+k}_{s=\abs{2\ell-k}} \frac{(-1)^{\frac{r(r-1)}{2}}\,{}_{s}\mathrm{C}_{r}\,(2\ell)!}{2^\ell\,\ell!\,s!\,(2\ell-r)!}\,
 \beta^{[\Lob_1\Lob_2}\cdots \beta^{\Lob_{2\ell-1} \Lob_{2\ell}]}\,
 \eta_{\Lob_{1}\Loc_{1}}\cdots \eta_{\Lob_{r}\Loc_{r}}\,
 \check{F}_{\Lob_{r+1}\cdots \Lob_{2\ell} \Loc_{r+1}\cdots \Loc_{s}}\,
 \Gamma^{\Loc_1\cdots \Loc_s}\,,
\end{align}
where we used the formula \eqref{eq:Gamma-p-q} and defined $r\equiv \frac{2\ell-k+s}{2}$\,. 
Then, the R--R field strength $\hat{F}_{\Loa_1\cdots \Loa_k}$ with flat indices becomes
\begin{align}
 \hat{F}_{\Loa_1\cdots \Loa_k}
 =\sum_{\ell,\,r} \frac{(-1)^{\frac{r(r-1)}{2}}\,{}_k\mathrm{C}_r\,(2\ell)!}{2^\ell\,\ell!\,(2\ell-r)!}\, \beta^{[\Lob_1 \Lob_2}\cdots\beta^{\Lob_{2\ell-1} \Lob_{2\ell}]}
 \,\eta_{\Lob_1[\underline{\Loa_{1}}}\cdots \eta_{\Lob_{r}\underline{\Loa_{r}}}\, \check{F}_{\Lob_{r+1}\cdots\Lob_{2\ell}\underline{\Loa_{r+1}\cdots\Loa_{k}}]} \,,
\label{eq:hatF-flat}
\end{align}
where the under-barred indices are totally antisymmetrized and non-negative integers $\ell$ and $r$ run over the region where the following relations are satisfied:
\begin{align}
 0\leq 2\,\ell-r \,,\qquad 0\leq k-r \,, \qquad 0\leq k+2\,\ell-2\,r\leq D\,.
\end{align}
We can further expand the right-hand side of \eqref{eq:hatF-flat} as
\footnote{We used the identity for arbitrary totally antisymmetric tensors $C^{\Loa\Lob}$\,, $A_{\Loa_1\cdots \Loa_r}$\,, and $B_{\Loa_{r+1}\cdots \Loa_{2\ell}}$,
\begin{align*}
 &C^{[\Loa_1\Loa_2}\cdots C^{\Loa_{2\ell-1}\Loa_{2\ell}]}\, A_{\Loa_1\cdots \Loa_r}\,B_{\Loa_{r+1}\cdots \Loa_{2\ell}}
\\
 &=\sum_{u}\frac{(-1)^{\frac{u(u-1)}{2}}\,2^u\,\ell!}{s!\,t!\,u!}\,\frac{r!\,(2\ell-r)!}{(2\ell)!}\, C^{\Lob_1\Lob_2}\cdots C^{\Lob_{2t-1}\Lob_{2t}} 
\\
 &\quad\times \bigl(C^{\Loa_1\Loa_2}\cdots C^{\Loa_{2s-1}\Loa_{2s}}\bigr)
 \bigl(C^{\Loc_1\Lod_1}\cdots C^{\Loc_u\Lod_u}\bigr)
 \,A_{\Loa_1\cdots \Loa_{2s}\Loc_1\cdots \Loc_u}\,B_{\Lod_1\cdots \Lod_u\Lob_1\cdots \Lob_{2t}}\,,
\end{align*}
where $0\leq r \leq 2 \ell$,\ $2 s\equiv r-u$, and $2 t\equiv 2 \ell-r-u$.
The range of the summation over $u$ is as follows:
\begin{align*}
 u= \begin{cases}
    0\,, 2\,, \dotsc\,, \min (r,2\ell-r) & \text{for $r$ even}\,, \\
    1\,, 3\,, \dotsc\,, \min (r,2\ell-r) & \text{for $r$ odd}\,.
  \end{cases}
\end{align*}}
\begin{align}
 \hat{F}_{\Loa_1\cdots \Loa_k} 
 &=\sum_{\ell,\,r,\,u} \frac{(-1)^{s}\,2^u\,k!}{2^\ell\,s!\,t!\,u!\,(k-r)!} \,
 \beta^{\Lob_{1} \Lob_{2}}\cdots\beta^{\Lob_{2t-1}\Lob_{2t}}
\nn\\
 &\quad\times 
 \beta_{[\underline{\Loa_{1}\Loa_{2}}}\cdots\beta_{\underline{\Loa_{2s-1}\Loa_{2s}}} \,
 \beta_{\underline{\Loa_{2s+1}}}{}^{\Loc_{1}}\cdots \beta_{\underline{\Loa_{r}}}{}^{\Loc_{u}}\,
 \check{F}_{|\Loc_{1}\cdots\Loc_{u}| \underline{\Loa_{r+1}\cdots\Loa_k}]\Lob_{1}\cdots\Lob_{2t}} \,,
\end{align}
where $s$ and $t$ are defined as
\begin{align}
 2\,s\equiv r-u \,,\qquad 2\,t\equiv 2\,\ell-r-u \,,
\end{align}
and non-negative integers $\ell$, $r$, and $u$ run over the region where
\begin{align}
 0\leq s\,,\quad 0\leq t\,,\quad r\leq k \,,\quad 0\leq k+2\,t-2\,s \leq D \,,
\end{align}
are satisfied. 
If we change the variables, we obtain a more explicit expression,
\begin{align}
 \hat{F}_{\Loa_1\cdots \Loa_k} 
 &=\sum_{s=0}^{[\frac{k}{2}]} \sum_{t=0}^{s+[\frac{D-k}{2}]} \sum_{u=0}^{k-2s} \frac{(-1)^{s}\,k!}{2^{s+t}\,s!\,t!\,u!\,(k-2s-u)!} \,
 \beta^{\Loc_{1}\Loc_{2}}\cdots\beta^{\Loc_{2t-1}\Loc_{2t}}
\nn\\
 &\times 
 \beta_{[\underline{\Loa_{1}\Loa_{2}}}\cdots\beta_{\underline{\Loa_{2s-1}\Loa_{2s}}} \,
 \beta_{\underline{\Loa_{2s+1}}}{}^{\Lod_{1}}\cdots \beta_{\underline{\Loa_{r}}}{}^{\Lod_{u}}\,
 \check{F}_{|\Lod_{1}\cdots\Lod_{u}| \underline{\Loa_{r+1}\cdots\Loa_k}]\Loc_{1}\cdots\Loc_{2t}} \,. 
\end{align}
This precisely matches with \eqref{eq:Fhat-Fcheck-flat} and the equivalence has been proven.

\section{The spinor rotation $\Omega$}
\label{app:omega}

In this Appendix, we prove the formula \eqref{eq:Hassan-formula} in an arbitrary even dimension $D$. 
Namely, we prove that the spinor representation of a local Lorentz transformation $\Lambda^{\Loa}{}_{\Lob} \equiv (O_+^{-1}\,O_-)^{\Loa}{}_{\Lob}$ ($O_\pm \equiv \delta^{\Loa}_{\Lob} \pm a^{\Loa}{}_{\Lob}$, $a^{\Loa\Lob}=-a^{\Lob\Loa}$) is given by
\begin{align}
 \Omega_{(a)} = (\det O_{\pm})^{-\frac{1}{2}} \text{\AE}\bigl(-\tfrac{1}{2}\,a_{\Loa\Lob}\,\Gamma^{\Loa\Lob}\bigr)\,,
 \qquad 
 \Omega_{(a)}^{-1} =(\det O_{\pm})^{-\frac{1}{2}} \text{\AE}\bigl(\tfrac{1}{2}\,a_{\Loa\Lob}\,\Gamma^{\Loa\Lob}\bigr) \,.
\end{align}

\medskip

If we define matrices
\begin{align}
 \Omega_\pm \equiv \text{\AE}\bigl(\pm\tfrac{1}{2}\,a_{\Loa\Lob}\,\Gamma^{\Loa\Lob}\bigr)
 = \sum^{5}_{p=0}\frac{(\pm 1)^p}{2^p\,p!}\, a_{\Lob_1 \Lob_2}\cdots a_{\Lob_{2p-1}\Lob_{2p}}\,\Gamma^{\Lob_1\Lob_2\cdots \Lob_{2p-1}\Lob_{2p}}\,,
\end{align}
we can easily show the identity,
\begin{align}
 (O_\mp)^{\Loa}{}_{\Lob}\,\Gamma^{\Lob}\,\Omega_\pm - (O_{\pm})^{\Loa}{}_{\Lob}\,\Omega_\pm\,\Gamma^{\Lob}
 = [\Gamma^{\Loa},\,\Omega_\pm] \mp a^{\Loa}{}_{\Lob}\,\{\Gamma^{\Lob},\,\Omega_\pm\} =0\,,
\end{align}
and this leads to
\begin{align}
 \Omega^{-1}_\pm\,\Gamma^{\Loa}\,\Omega_\pm = (O_{\mp}^{-1}\,O_{\pm})^{\Loa}{}_{\Lob}\,\Gamma^{\Lob} \,. 
\label{eq:Omega-pm}
\end{align}
Choosing the lower sign, we obtain the desired relation,
\begin{align}
 \Omega^{-1}_-\,\Gamma^{\Lob}\,\Omega_- = \Lambda^{\Loa}{}_{\Lob}\,\Gamma^{\Lob} \,.
\end{align}
In the following, we rescale $\Omega_-$ and define $\Omega_{(a)}$ such that $\Omega_{(a)}^{-1}=\Omega_{(-a)}$\,. 
The relation \eqref{eq:Omega-pm} implies that $\Omega^{-1}_\pm$ is proportional to $\Omega_{\mp}$\,, and we denote their relation as
\begin{align}
 \Omega^{-1}_- = \frac{1}{\abs{\Omega}^2}\,\Omega_+ \,. 
\end{align}
We shall show $\abs{\Omega}^2 =\det (\delta_{\Loa}^{\Lob} - a_{\Loa}{}^{\Lob})= \det O_- \, (= \det O_+)$\,, and then we find that $\Omega_{(a)} \equiv \abs{\Omega}^{-1}\,\Omega_-$ satisfies the relation $\Omega_{(a)}^{-1}=\Omega_{(-a)}$\,. 

\medskip

We can compute $\abs{\Omega}^2 = \Omega_-\,\Omega_+$ as
\begin{align}
 \abs{\Omega}^2
 &= \sum_{p=0}^{D/2}\, \frac{(2p)!}{2^{2p}\,(p!)^2}\, a^{[\Lob_1 \Lob_2}\cdots a^{\Lob_{2p-1} \Lob_{2p}]}\, a_{[\Lob_{1} \Lob_{2}}\cdots a_{\Lob_{2p-1} \Lob_{2p}]}
\nn\\
 &=\sum_{p=0}^{D/2}\ \sum_{0\leq \Lob_1<\cdots<\Lob_{2p}\leq D-1}\ \sum_{\sigma\in S_{2p}}
 \Bigl(\frac{\sgn(\sigma)}{2^p\,p!}\,
 a_{\Lob_{\sigma(1)}\Lob_{\sigma(2)}}\cdots a_{\Lob_{\sigma(2p-1)} \Lob_{\sigma(2p)}}\Bigr)
\nn\\
 &\qquad\qquad\qquad\times \sum_{\sigma'\in S_{2p}}\Bigl(\frac{\sgn(\sigma')}{2^p\,p!}\,
 a^{\Lob_{\sigma'(1)}\Lob_{\sigma'(2)}}\cdots a^{\Lob_{\sigma'(2p-1)} \Lob_{\sigma'(2p)}}\Bigr)
\nn\\
 &=\sum_{p=0}^{D/2}\ \sum_{0\leq \Lob_1<\cdots<\Lob_{2p}\leq D-1} \varepsilon_{\Lob_1}\,\Pf\bigl[a(\Lob_1,\dotsc,\Lob_{2p})\bigr]^2\,,
\end{align}
where $S_{2p}$ is the symmetric group on a set of $2p$ indices, $\sgn(\sigma)$ is the sign of a permutation $\sigma\in S_{2p}$ and $\varepsilon_{\Lob_1} \equiv \eta_{\Lob_1\Lob_1}$ is $-1$ for $\Lob_1=0$ and $+1$ for $\Lob_1\geq 1$. 
The Pfaffian
\begin{align}
 \Pf[A(\Lob_1,\dotsc,\Lob_{2p})]
 \equiv \sum_{\sigma\in S_{2p}} \biggl(\frac{\sgn(\sigma)}{2^p\,p!}
 A_{\Lob_{\sigma(1)}\Lob_{\sigma(2)}}\cdots A_{\Lob_{\sigma(2p-1)} \Lob_{\sigma(2p)}}\biggr)\,,
\end{align}
is the polynomial in matrix elements of the antisymmetric matrix $A(\Lob_1,\dotsc,\Lob_{2p})$ which is defined by
\begin{align}
 A(\Lob_1,\dotsc,\Lob_{2p})=
 \begin{pmatrix} 0~& A_{{\Lob}_1{\Lob}_2} & \ldots & A_{{\Lob}_1{\Lob}_{2p-1}}& A_{{\Lob}_1{\Lob}_{2p}} \\
      -A_{{\Lob}_1{\Lob}_2} & 0 & \ldots & A_{{\Lob}_2{\Lob}_{2p-1}}& A_{{\Lob}_2{\Lob}_{2p}} \\
      \vdots & \vdots & \ddots & \vdots& \vdots \\
      -A_{{\Lob}_1{\Lob}_{2p-1}} & -A_{{\Lob}_2{\Lob}_{2p-1}} & \ldots & 0& A_{{\Lob}_{2p-1}{\Lob}_{2p}} \\
      -A_{{\Lob}_{1}{\Lob}_{2p}} &-A_{{\Lob}_{2}{\Lob}_{2p}} & \ldots & -A_{{\Lob}_{2p-1}{\Lob}_{2p}} & 0
 \end{pmatrix} \,.
\end{align}
As it is well known, the square of the Pfaffian $\Pf[A(\Lob_1,\dotsc,\Lob_{2p})]^2$ coincides with $\det [A(\Lob_1,\dotsc,\Lob_{2p})]$\,.

\medskip

If we define a matrix function $p_A(x)$ $(x\in \mathbb{R})$ of a $D\times D$ antisymmetric matrix $A$ as
\begin{align}
 p_A(x) =-\det (x\,\eta_{\Loa\Lob}-A_{\Loa\Lob})\,,
\end{align}
its Taylor series around $x=0$ is
\begin{align}
 p_A(x) = x^{D} + c_2\, x^{D-2} + \cdots + c_{D-2}\,x^{2}+c_{D}\,,
\end{align}
where the coefficients $c_{2p}$ $(p=0\,,1\,,\dotsc,\,D/2)$ are given by
\begin{align}
 c_{2p}=\frac{1}{(D-2p)!}\,p^{(D-2p)}_A(0) =\sum_{0\leq \Lob_1<\cdots<\Lob_{2p}\leq D-1} \varepsilon_{\Lob_1}\,\det \bigl[A(\Lob_1,\dotsc,\Lob_{2p})\bigr]\,.
\end{align}
From this, we finally obtain
\begin{align}
 \abs{\Omega}^2 
 &=\sum_{p=0}^{D/2}\ \sum_{0\leq \Lob_1<\cdots<\Lob_{2p}\leq D-1} \varepsilon_{\Lob_1}\,\det \bigl[a(\Lob_1,\dotsc,\Lob_{2p})\bigr]
\nn\\
 &=p_a(1)= -\det \bigl(\eta_{\Loa\Lob}-a_{\Loa\Lob}\bigr) = \det \bigl(\delta_{\Loa}^{\Lob}-a_{\Loa}{}^{\Lob}\bigr) = \det O_- \,.
\end{align}

\chapter{Details of the supercoset construction}

We here corrected detailed calculations of the supercoset construction.

\section{Expansion of $\cO^{-1}_{\pm}$}
\label{app:expansion-O}

In this appendix, we expand the operators $\cO^{-1}_{\pm}\equiv (1\pm\eta\,R_g\circ d_\pm)^{-1}$ in terms of $\theta$\,. 
To this end, we first use the parameterization $g=g_{\bos}\cdot g_{\fer}$, and expand $R_g(X)$ as
\begin{align}
 R_g(X) &= g_{\fer}^{-1}\,g_{\bos}^{-1}\,R(g_{\bos}\,g_{\fer}\,X\,g_{\fer}^{-1}\,g_{\bos}^{-1})\,g_{\bos}\,g_{\fer}
\nn\\
 &= R_{g_{\bos}}(X) - [\chi,\,R_{g_{\bos}}(X)] + R_{g_{\bos}}([\chi,\,X])
\nn\\
 &\quad + \frac{1}{2}\,R_{g_{\bos}}([\chi,\,[\chi,\,X]]) + \frac{1}{2}\,[\chi,\,[\chi,\,R_{g_{\bos}}(X)]] - [\chi,\,R_{g_{\bos}}([\chi,\,X])] + \cO(\theta^3)\,, 
\end{align}
where $R_{g_{\bos}}(X) \equiv g_{\bos}^{-1}\,R(g_{\bos}\,X\,g_{\bos}^{-1})\,g_{\bos}$ and $\chi\equiv \gQ^I\, \theta_I$\,. 
We can then expand $\cO_{\pm}$ as
\begin{align}
\begin{split}
 \cO_{\pm}&=1\pm \eta\, R_g\circ d_\pm =\cO_{\pm(0)} + \cO_{\pm(1)} + \cO_{\pm(2)} + \cO(\theta^3)\,,
\\
 \cO_{(0)}(X)&= 1 \pm \eta\, R_{g_{\bos}}(d_\pm(X)) \,,
\\
 \cO_{\pm(1)}(X)&=\pm \eta\, R_{g_{\bos}}([\chi,\, d_{\pm}(X)])\mp \eta\,[\chi,\,R_{g_{\bos}}\circ d_{\pm}(X)]\,,
\\
 \cO_{\pm(2)}(X)&=\mp\frac{\eta}{2}\,\bigl([\chi,\,[\chi,\,R_{g_{\bos}}\circ d_{\pm}(X)]]-R_{g_{\bos}}([\chi,\,[\chi,\,d_{\pm}(X)]])\bigr) -[\chi,\,\cO_{\pm(1)}(X)]\,.
\end{split}
\end{align}
The inverses can be also expanded as
\begin{align}
\begin{split}
 \cO_{\pm}^{-1}&=\frac{1}{1\pm\eta\,R_g\circ d_\pm}
 = \cO_{\pm(0)}^{-1}+\cO_{\pm(1)}^{-1}+\cO_{\pm(2)}^{-1}+\cO(\theta^3)\,,
\\
 \cO_{\pm(0)}^{-1}&= \frac{1}{1\pm \eta\, R_{g_{\bos}}\circ d_{\pm}}\,,
\\
 \cO_{\pm(1)}^{-1}&= -\cO_{\pm(0)}^{-1}\circ\cO_{\pm(1)}\circ\cO_{\pm(0)}^{-1}\,,
\\
 \cO_{\pm(2)}^{-1}&= -\cO_{\pm(0)}^{-1}\circ\cO_{\pm(2)}\circ\cO_{\pm(0)}^{-1} -\cO_{\pm(1)}^{-1}\circ\cO_{\pm(1)}\circ\cO_{\pm(0)}^{-1}\,.
\end{split}
\end{align}

\paragraph*{\underline{Order $\cO(\theta^0)$:}}

The leading order part $\cO^{-1}_{\pm(0)}$ of the inverse operators act as
\begin{align}
\begin{split}
 \cO^{-1}_{\pm(0)}(\gP_{\Loa})
 &=k_{\pm\Loa}{}^{\Lob}\,\gP_{\Lob} \mp \eta\,k_{\pm\Loa}{}^{\Lob}\,\lambda_{\Lob}{}^{\Loc\Lod}\,\gJ_{\Loc\Lod}\,,
\\
 \cO^{-1}_{\pm(0)}(\gJ_{\Loa\Lob})&=\gJ_{\Loa\Lob}\,,\qquad
 \cO^{-1}_{\pm(0)}(\gQ^I)=\gQ^I\,,
\end{split}
\label{eq:Oinv-0}
\end{align}
where we have used \eqref{eq:Rg-operation} and defined $k_{\pm\Loa}{}^{\Lob}$ as
\begin{align}
 k_{\pm\Loa}{}^{\Lob} \equiv \bigl[(1\pm 2\,\eta\,\lambda)^{-1}\bigr]{}_{\Loa}{}^{\Lob} \,. 
\end{align}
Note that $k_{\pm\Loa}{}^{\Lob}$ satisfies $k_{\pm\Loa\Lob}\equiv k_{\pm\Loa}{}^{\Loc}\,\eta_{\Loc\Lob}=k_{\mp\Lob\Loa}$ due to the antisymmetry of $\lambda_{\Loa\Lob}$ given in \eqref{eq:lambda-properties}. 

\paragraph*{\underline{Order $\cO(\theta^1)$:}}

At the next order, we obtain
\begin{align}
\begin{split}
 \cO_{\pm(1)}(\gP_{\Loa})
 &=\pm \eta\,\gQ^I\,\Bigl(\ii\,\epsilon^{IJ}\,\lambda_{\Loa}{}^{\Lob}\,\hat{\gamma}_{\Lob}-\frac{1}{2}\,\delta^{IJ}\,\lambda_{\Loa}{}^{\Lob\Loc}\,\gamma_{\Lob\Loc}\Bigr)\,\theta_J\,, \qquad
 \cO_{\pm(1)}(\gJ_{\Loa\Lob}) =0\,,
\\
 \cO_{\pm(1)}(\gQ^I\psi_I) 
 &=-\frac{\ii}{2}\,\eta\,\brtheta_I\,
 \bigl(2\,\sigma_3^{IJ}\,\lambda_{\Lob}{}^{\Loc}\,\hat{\gamma}_{\Loc}+\ii\,\sigma_1^{IJ}\,\lambda_{\Lob}{}^{\Loc\Lod}\,\gamma_{\Loc\Lod}\bigr)\,\psi_J\,\eta^{\Lob\Loa}\,\gP_{\Loa} + (\gJ\text{-term})\,,
\end{split}
\end{align}
where ``$(\gJ\text{-term})$'' represents terms proportional to $\gJ_{\Loa\Lob}$ that are not relevant to our computation. 
Then, the operations of the inverse operators are
\begin{align}
\begin{split}
 \cO^{-1}_{\pm(1)}(\gP_{\Loa})
 &=\mp\eta\,\gQ^I\, k_{\pm\Loa}{}^{\Lob}
 \Bigl(\ii\,\epsilon^{IJ}\,\lambda_{\Lob}{}^{\Loc}\,\hat{\gamma}_{\Loc}
 -\frac{1}{2}\,\delta^{IJ}\,\lambda_{\Lob}{}^{\Loc\Lod}\,\gamma_{\Loc\Lod}\Bigr)\,\theta_J\,, \qquad
 \cO^{-1}_{\pm(1)}(\gJ_{\Loa\Lob}) =0\,,
\\
 \cO^{-1}_{\pm(1)}(\gQ^I\psi_I)
 &=\frac{\ii}{2}\eta\,\brtheta_I\, k_\pm^{\Lob\Loa}
 \bigl(2\,\sigma_3^{IJ}\,\lambda_{\Lob}{}^{\Loc}\,\hat{\gamma}_{\Loc}+\ii\,\sigma_1^{IJ}\,\lambda_{\Lob}{}^{\Loc\Lod}\,\gamma_{\Loc\Lod}\bigr)\,\psi_J\gP_{\Loa}
 +(\gJ\text{-term})\,.
\end{split}
\end{align}

\paragraph*{\underline{Order $\cO(\theta^2)$:}}

Finally, the operators at the quadratic order are given by
\begin{align}
\begin{split}
 \cO_{\pm(2)}(\gP_{\Loa})&=\pm\frac{\ii}{2}\,\eta\,\brtheta_I\,\hat{\gamma}^{\Lob}
 \Bigl(\ii\,\epsilon^{IJ}\,\lambda_{\Loa}{}^{\Loc}\,\hat{\gamma}_{\Loc}-\frac{1}{2}\,\delta^{IJ}\,\lambda_{\Loa}{}^{\Loc\Lod}\,\gamma_{\Loc\Lod}\Bigr)\,\theta_J\,\gP_{\Lob}
\nn\\
 &\quad\mp\frac{\ii}{2}\,\eta\,\brtheta_I\,
 \Bigl(\ii\,\epsilon^{IJ}\,\lambda^{\Lob}{}_{\Loc}\,\hat{\gamma}^{\Loc}-\frac{1}{2}\,\delta^{IJ}\,\eta^{\Lob\Loe}\,\lambda_{\Loe}{}^{\Loc\Lod}\,\gamma_{\Loc\Lod}\Bigr)\,\hat{\gamma}_{\Loa}\,\theta_J\,\gP_{\Lob} +(\gJ\text{-term}) \,,
\\
 \cO_{\pm(2)}(\gJ_{\Loa\Lob})&=0\,. 
\end{split}
\end{align}
The inverses are
\begin{align}
 \cO^{-1}_{\pm(2)}(\gP_{\Loa})
 &= \mp\frac{\ii}{2}\,\brtheta_I\,k_{\pm}^{\Lob\Loh}\,k_{\pm\Loa}{}^{\Lod}\,
 \biggl[ (\delta^{IJ}\,\delta_{\Lob}^{\Loc}+2\,\eta\,\sigma_3^{IJ}\,\lambda_{\Lob}{}^{\Loc})\,\hat{\gamma}_{\Loc}\,\Bigl(-\frac{\eta}{2}\,\lambda_{\Lod}{}^{\Loe\Lof}\, \gamma_{\Loe\Lof}\Bigr)
\nn\\
 &\qquad+ \Bigl(\frac{\eta}{2}\,\lambda_{\Lob}{}^{\Loe\Lof}\, \gamma_{\Loe\Lof}\Bigr)\,(\delta^{IJ}\,\delta_{\Lod}^{\Loc}+2\,\eta\,\sigma_3^{IJ}\,\lambda_{\Lod}{}^{\Loc})\,\hat{\gamma}_{\Loc}
\nn\\
 &\qquad+\frac{\ii}{2}\,\epsilon^{IJ}\,\bigl[\hat{\gamma}_{\Lob}\,(2\,\eta\, \lambda_{\Lod}{}^{\Loc}\,\hat{\gamma}_{\Loc})
 -(2\,\eta\,\lambda_{\Lob}{}^{\Loc}\,\hat{\gamma}_{\Loc})\,\hat{\gamma}_{\Lod}\bigr]
\nn\\
 &\qquad
 +\ii\,\sigma_1^{IJ}\,(\eta\,\lambda_{\Lob}{}^{\Loe\Lof}\, \gamma_{\Loe\Lof})\,\Bigl(-\frac{\eta}{2}\,\lambda_{\Lod}{}^{\Lof\Log}\, \gamma_{\Lof\Log}\Bigr)
\nn\\
 &\qquad+\frac{\ii}{2}\,\sigma_1^{IJ}\,(2\,\eta\,\lambda_{\Lob}{}^{\Lod}\,\hat{\gamma}_{\Lod})\,(2\,\eta\,\lambda_{\Lod}{}^{\Loc}\,\hat{\gamma}_{\Loc})
 \biggr]\,\theta_J\,\gP_{\Loh} +(\gJ\text{-term})\,.
\end{align}
Operators of $\cO^{-1}_{\pm(2)}$ on other generators are not necessary for the computation of the action.

\section{Deformed torsionful spin connections}
\label{app:torsionful-spin-connections}

In this appendix, we show that two torsionful spin connections $W_{\pm}^{\Loa\Lob}$ introduced in \eqref{eq:e-torsionful-spin-pm} satisfy the following relations (we basically follow the discussion of \cite{Borsato:2016ose}):
\begin{align}
 \omega'^{\Loa\Lob}_+ &\equiv \omega'^{\Loa\Lob} + \frac{1}{2}\,e'_{\Loc}\,H'^{\Loc\Loa\Lob} =W_{+}^{\Loa\Lob}\,,
\label{eq:Lorntz-connection1}
\\
 \omega'^{\Loa\Lob}_- &\equiv \omega'^{\Loa\Lob}-\frac{1}{2}\,e'_{\Loc}\,H'^{\Loc\Loa\Lob}
  =\Lambda^{\Loa}{}_{\Loc}\,\Lambda^{\Lob}{}_{\Lod}\,W_{-}^{\Loc\Lod} + (\Lambda\,\rmd\Lambda^{-1})^{\Loa\Lob} \,,
\label{eq:Lorntz-connection2}
\end{align}
which are assumed in \eqref{eq:torsionful-spin}.

\subsection{Two expressions of the deformed $H$-flux}
\label{app:deformed-H-flux}

In order to show \eqref{eq:Lorntz-connection1} and \eqref{eq:Lorntz-connection2}, we here obtain two expressions for the deformed $H$-flux.
Let us begin by considering two expressions of the deformed $B$-field [recall \eqref{eq:G-B-prime} and \eqref{eq:e-torsionful-spin-pm}]
\begin{align}
\begin{split}
 B_2'&= -\eta\,\lambda_{\Loa\Lob}\,e_{+}^{\Loa}\wedge e_{+}^{\Lob} =\eta\,\str \bigl[J_+^{(2)}\wedge R_g(J_+^{(2)})\bigr]\bigr\rvert_{\theta=0} 
\\
 &= - \eta\,\lambda_{\Loa\Lob}\,e_{-}^{\Loa}\wedge e_{-}^{\Lob} =\eta\,\str \bigl[J_-^{(2)}\wedge R_g(J_-^{(2)})\bigr]\bigr\rvert_{\theta=0} \,,
\end{split}
\label{eq:B-YB-2}
\end{align}
where $J_{\pm}$ are defined in \eqref{eq:J-O-inv-A} and $J_{\pm}^{(n)}\equiv P^{(n)}\,J_{\pm}$\,. 
Since we are only interested in the $B$-field at order $\cO(\theta^0)$\,, in the following computation, we ignore terms involving the grade-$1$ and $3$ components of $A$ and $J_{\pm}$ (where we have $d_{\pm}\sim 2\,P^{(2)}$). 

\medskip

The exterior derivatives of the two expressions in Eq.~\eqref{eq:B-YB-2} become
\begin{align}
 H'_3&\equiv \frac{1}{3!}\,H'_{\Loa\Lob\Loc}\,e'^{\Loa}\wedge e'^{\Lob}\wedge e'^{\Loc}\equiv \rmd B'_2
\nn\\
 &=\eta\,\rmd\,\str \bigl[J_{\pm}^{(2)}\wedge R_g(J_{\pm}^{(2)})\bigr]\bigr\rvert_{\theta=0}
\nn\\
 &=2\,\eta\,\str \bigl[\rmd J_{\pm}^{(2)}\wedge R_g(J_{\pm}^{(2)})+J_{\pm}^{(2)}\wedge \{A,\,R_g(J_{\pm}^{(2)})\}\bigr]\bigr\rvert_{\theta=0}
\nn\\
 &=2\,\eta\,\str \bigl[\rmd J_{\pm}^{(2)}\wedge R_g(J_{\pm}^{(2)})+J_{\pm}^{(2)}\wedge \{J_{\pm}^{(0)}+J_{\pm}^{(2)},\,R_g(J_{\pm}^{(2)})\}\bigr]\bigr\rvert_{\theta=0}
\nn\\
 &\quad \pm 4\,\eta^2\,\str \bigl[J_{\pm}^{(2)}\wedge \{R_g(J_{\pm}^{(2)}),\,R_g(J_{\pm}^{(2)})\}\bigr]\bigr\rvert_{\theta=0} \,.
\label{eq:H3-eq1}
\end{align}
Here, in the third line, we have used a relation
\begin{align}
 \rmd \bigl[ R_g(B) \bigr] = R_g(\rmd B) -\{A,\,R_g(B)\}+R_g(\{A,\,B\}) \qquad 
 \bigl[\,\{B,\,C\} \equiv B\wedge C + C\wedge B \,\bigr]\,. 
\end{align}
for $\alg{g}$-valued 1-forms $B$ and $C$, and in the last equality, we have used the relation
\begin{align}
 A\rvert_{\theta=0} = \cO_{\pm} (J_{\pm})\bigr\rvert_{\theta=0} = J_{\pm}^{(0)}+J^{(2)}_{\pm} \pm 2\,\eta\,R_g(J_{\pm}^{(2)})\bigr\rvert_{\theta=0}\,. 
\end{align}
It is easy to see that the last term in \eqref{eq:H3-eq1} vanishes by using the cyclic property of the supertrace and the homogeneous CYBE
\begin{align}
 \{R_g(J_{\pm}^{(2)}),\, R_g(J_{\pm}^{(2)})\}-2\,R_g\{R_g(J_{\pm}^{(2)}),\, J_{\pm}^{(2)}\}=0\,.
\end{align}
Now, we utilize the deformed structure equation \cite{Borsato:2016ose}
\begin{align}
 \rmd J_\pm &= \rmd\bigl(\cO_{\pm}^{-1}\,A\bigr) = -\cO_{\pm}^{-1}\,\rmd\cO_{\pm} \,\cO_{\pm}^{-1}\,A + \cO_{\pm}^{-1}\,(\rmd A)
\nn\\
 &= \mp \eta\,\cO_{\pm}^{-1}\,(\rmd R_g)\,d_{\pm}\,J_{\pm} - \cO_{\pm}^{-1}\,(A\wedge A)
\nn\\
 &= \mp \eta\,\cO_{\pm}^{-1}\,\bigl[-\{A,\,R_g(d_{\pm}\,J_{\pm})\} + R_g\,\{A,\,d_{\pm}\,J_{\pm}\} \bigr]\,d_{\pm}\,J_{\pm} - \frac{1}{2}\,\cO_{\pm}^{-1}\,\{A,\, A\}
\nn\\
 &= - \frac{1}{2}\,\cO_{\pm}^{-1}\,\{J_{\pm},\, J_{\pm}\} \mp \eta\,\cO_{\pm}^{-1}\,R_g\,\{J_{\pm},\,d_{\pm}\,J_{\pm}\}
\nn\\
 &\quad + \frac{\eta^2}{2}\,\cO_{\pm}^{-1}\,\bigl[\{R_g(d_{\pm}\,J_{\pm}),\,R_g(d_{\pm}\,J_{\pm})\}-2\,R_g\bigl(\{R_g(d_{\pm}\,J_{\pm}),\,d_{\pm}\,J_{\pm}\}\bigr)\bigr]
\nn\\
 &= - \frac{1}{2}\,\cO_{\pm}^{-1}\,\{J_{\pm},\, J_{\pm}\} \mp \eta\,\cO_{\pm}^{-1}\,R_g\,\{J_{\pm},\,d_{\pm}\,J_{\pm}\} \,,
\end{align}
where we have repeatedly used $A=\cO_{\pm}(J_{\pm})$\,, and in the last equality, we have used the homogeneous CYBE. 
In the following computation, since terms involving $J_{\pm}^{(1)}$ or $J_{\pm}^{(3)}$ are irrelevant, we have
\begin{align}
 \rmd J_{\pm} =-\frac{1}{2}\,\{J_{\pm},\,J_{\pm}\} \mp 2\,\eta\,\cO_{\pm}^{-1}\,R_g\{J_{\pm}^{(2)},\,J_{\pm}^{(2)}\} \,,
\label{eq:dJ+-}
\end{align}
and then \eqref{eq:H3-eq1} is simplified as
\begin{align}
 H'_3&=2\,\eta\,\str\bigl[\bigl(\{J_{\pm},\,J_{\pm}^{(2)}\}-\frac{1}{2}\,P^{(2)}\,\{J_{\pm},\,J_{\pm}\}
  \mp 2\,\eta\, P^{(2)}\,\cO_{\pm}^{-1}\, R_g\{J_{\pm}^{(2)},\,J_{\pm}^{(2)}\} \bigr)\wedge R_g(J_{\pm}^{(2)})\bigr]\bigr\rvert_{\theta=0}
\nn\\
 &=2\,\eta\,\str\bigl[ \{J_{\pm}^{(2)},\,J_{\pm}^{(2)}\}\wedge \cO_{\mp}^{-1}\,R_g(J_{\pm}^{(2)})\bigr]\bigr\rvert_{\theta=0} \,,
\label{eq:H3-eq2}
\end{align}
where, in the last equality, we have used relations\footnote{The transpose of an operator $\mathsf{O}$ is defined as $\str[\,A\,\mathsf{O}(B)\,]=\str[\,\mathsf{O}^{\rmT}(A)\,B\,]$\,. Since $R_g$ is defined to be antisymmetric, $R_g^\rmT=-R_g$, and $d_{\pm}$ satisfies $d_{\pm}^\rmT=d_{\mp}$ [see \eqref{eq:dpm-transpose}], we can for example show $\cO_{\pm}^\rmT=1\mp\eta\,d_{\mp}\,R_g$, $d_{\pm}\,\cO_{\pm}=\cO_{\mp}^\rmT\,d_{\pm}$, and $d_{\pm}\,\cO_{\pm}^{-1}=\cO_{\mp}^{-\rmT}\,d_{\pm}$\,.}
\begin{align}
\begin{split}
 &\pm\eta\,d_{\pm}\,\cO_{\pm}^{-1}\,R_g = \pm\eta\,\cO_{\mp}^{-\rmT}\,d_{\pm}\,R_g = 1-\cO_{\mp}^{-\rmT} \,,
\\
 &\{J_+,\,J_+^{(2)}\} - \frac{1}{2}\,P^{(2)}\{J_+,\,J_+\} = \{J^{(2)}_+,\,J^{(2)}_+\} \,.
\end{split}
\end{align}
We can further rewrite the expression \eqref{eq:H3-eq2} by using the operator $M=\cO_-^{-1}\,\cO_+$ and its inverse $M^{-1}=\cO_+^{-1}\,\cO_-$\,. 
From
\begin{align}
 P^{(0)}\,M^{\pm1}\,P^{(2)} = P^{(0)}\,\cO_{\mp}^{-1}\,(\cO_{\mp}\pm 4\,\eta \, R_{g})\,P^{(2)}=\pm 4\,\eta\, P^{(0)}\,\cO_{\mp}^{-1}\,R_g\, P^{(2)}\,,
\end{align}
we can rewrite \eqref{eq:H3-eq2} as
\begin{align}
 H'_3 = \pm \frac{1}{2}\,\str\bigl[\{J_{\pm}^{(2)},\,J_{\pm}^{(2)}\}\wedge M^{\pm1}(J_{\pm}^{(2)})\bigr]\bigr\rvert_{\theta=0} \,.
\end{align}
Finally, by introducing a notation
\begin{align}
\begin{split}
 &M(\gP_{\Loc})\rvert_{\theta=0} = (\Lambda^{-1})_{\Loc}{}^{\Loa} \, \gP_{\Loa} + \frac{1}{2}\,M_{\Loc}{}^{\Loa\Lob}\,\gJ_{\Loa\Lob} \,,\quad
 M^{-1}(\gP_{\Loc})\rvert_{\theta=0} = \Lambda_{\Loc}{}^{\Loa} \, \gP_{\Loa} + \frac{1}{2}\,(M^{-1})_{\Loc}{}^{\Loa\Lob}\,\gJ_{\Loa\Lob} \,,
\\
 &\bigl[\,\Lambda_{\Loa}{}^{\Lob}= (k_-^{-1})_{\Loa}{}^{\Loc}\, k_{+\Loc}{}^{\Lob}\,,\quad 
 M_{\Loc}{}^{\Loa\Lob} \equiv 4\,\eta\,k_{-\Loc}{}^{\Lod}\,\lambda_{\Lod}{}^{\Loa\Lob} \,,\quad
 (M^{-1})_{\Loc}{}^{\Loa\Lob} \equiv -4\,\eta\,k_{+\Loc}{}^{\Lod}\,\lambda_{\Lod}{}^{\Loa\Lob}\,\bigr] \,,
\end{split}
\end{align}
and using $e_+^{\Loa}=\Lambda_{\Lob}{}^{\Loa}\,e'^{\Lob}$\,, we obtain two expressions for the deformed $H$-flux
\begin{align}
 H'_3 &= \frac{1}{2}\,\Lambda_{[\Loc}{}^{\Log}\,\Lambda_{\Loa}{}^{\Loe}\,\Lambda_{\Lob]}{}^{\Lof}\,M_{\Log,\Loe\Lof}\,e'^{\Loa}\wedge e'^{\Lob}\wedge e'^{\Loc}\qquad 
 \bigl[\,M_{\Loc,\Loa\Lob} \equiv M_{\Loc}{}^{\Loe\Lof}\,\eta_{\Loe\Loa}\,\eta_{\Lof\Lob}\,\bigr]
\label{eq:Hprime1}
\\
  &=-\frac{1}{2}\,M_{[\Loc,\Loa\Lob]}^{-1} \,e'^{\Loa}\wedge e'^{\Lob}\wedge e'^{\Loc}\qquad 
 \bigl[\,M^{-1}_{\Loc,\Loa\Lob} \equiv (M^{-1})_{\Loc}{}^{\Loe\Lof}\,\eta_{\Loe\Loa}\,\eta_{\Lof\Lob}\,\bigr]\,.
\label{eq:Hprime2}
\end{align}

\subsection{Deformed torsionful spin connections}

By considering the leading order part $\cO(\theta^{0})$ of \eqref{eq:dJ+-}, we obtain
\begin{align}
 \rmd e_{\pm}^{\Loa}+(\omega_{[\pm]})^{\Loa}{}_{\Lob}\wedge e_{\pm}^{\Lob}=0\,,
\label{eq:de+-}
\end{align}
where the spin connections $(\omega_{[\pm]})_{\Loa\Lob}$ associated with the deformed vielbeins $e_{\pm}^{\Loa}$ have the form
\begin{align}
 \omega_{[\pm]\Loa\Lob} = W_{\pm\Loa\Lob}+\frac{1}{2}\,e_{\pm}^{\Loc}\,\bigl(M^{\pm1}_{\Loa,\Lob\Loc}+M^{\pm1}_{\Lob,\Loc\Loa} - M^{\pm1}_{\Loc,\Loa\Lob}\bigr)\,.
\label{eq:deformed-spin-2}
\end{align}
In particular, for the spin connection $\omega'^{\Loa\Lob}\equiv\omega_{[-]}^{\Loa\Lob}$ associated with the deformed vielbeins $e'^{\Loa}=e_{-}^{\Loa}$, using the formula $H'_{\Loa\Lob\Loc} =-3\,M_{[\Loc,\Loa\Lob]}^{-1}$ in \eqref{eq:Hprime2}, we obtain the first relation \eqref{eq:Lorntz-connection1} as
\begin{align}
 &\omega'^{\Loa\Lob}+\frac{1}{2}\,e'_{\Loc}\,H'^{\Loc\Loa\Lob}
 =\omega'^{\Loa\Lob}-\frac{1}{2}\,e'_{\Loc}\,\bigl[(M^{-1})_{\Loa}{}^{\Lob\Loc}+(M^{-1})_{\Lob}{}^{\Loc\Loa}+(M^{-1})_{\Loc}{}^{\Loa\Lob}\bigr]
\nn\\
 &=W_{-}^{\Loa\Lob}-e'^{\Loc}\,(M^{-1})_{\Loc}{}^{\Loa\Lob} = W_{-}^{\Loa\Lob} + 4\,\eta\,e'^{\Loc}\,k_{+\Loc}{}^{\Lod}\,\lambda_{\Lod}{}^{\Loa\Lob}
  =W_{+}^{\Loa\Lob}\,,
\end{align}
where in the last equality we have used\footnote{In order to show the relation $W_{+}^{\Loa\Lob}=W_{-}^{\Loa\Lob}-e'^{\Loc}\,(M^{-1})_{\Loc}{}^{\Loa\Lob}$\,, it will be easier to observe the bosonic part of the relation $J_{+}^{(0)}=P^{(0)}\,M^{-1}(J_{-})$\,.}
\begin{align}
 W_{+}^{\Loa\Lob} &= W_{-}^{\Loa\Lob} + 2\,\eta\,(e_+^{\Loc}+e_-^{\Loc})\,\lambda_{\Loc}{}^{\Loa\Lob}
 = W_{-}^{\Loa\Lob} + 2\,\eta\,e'^{\Loc}\,(\Lambda_{\Loc}{}^{\Lod}+\delta_{\Loc}^{\Lod})\,\lambda_{\Lod}{}^{\Loa\Lob}
\nn\\
 &= W_{-}^{\Loa\Lob} + 2\,\eta\,e'^{\Loc}\,[(k_-^{-1})_{\Loc}{}^{\Loe}+(k_+^{-1})_{\Loc}{}^{\Loe}]\,k_{+\Loe}{}^{\Lod}\,\lambda_{\Lod}{}^{\Loa\Lob}
  = W_{-}^{\Loa\Lob} + 4\,\eta\,e'^{\Loc}\,k_{+\Loc}{}^{\Lod}\, \lambda_{\Lod}{}^{\Loa\Lob} \,. 
\end{align}

\medskip

On the other hand, if we take the upper sign in \eqref{eq:de+-}, from $e_+^{\Loa}=\Lambda_{\Lob}{}^{\Loa}\,e'^{\Lob}$, we obtain
\begin{align}
 \rmd e'^{\Loa} + \bigl[(\Lambda^{-1})_{\Loc}{}^{\Loa} \,\rmd \Lambda_{\Lob}{}^{\Loc} + \Lambda^{\Loa\Lod}\, \omega_{+\Lod\Loe}\, \Lambda_{\Loc}{}^{\Loe}\bigr]\wedge e'^{\Loc}=0\,. 
\label{eq:deprime-2}
\end{align}
From the upper sign of \eqref{eq:deformed-spin-2}, $H'_{\Loa\Lob\Loc} = 3\,\Lambda_{\Loa}{}^{\Lod}\,\Lambda_{\Lob}{}^{\Loe}\,\Lambda_{\Loc}{}^{\Lof}\,M_{[\Lod,\Loe\Lof]}$ in \eqref{eq:Hprime1}, and the identity $\Lambda_{\Loa}{}^{\Lod}\,M_{\Lod,\Lob\Loc}=-(M^{-1})_{\Loa,\Lob\Loc}$\,, we can show
\begin{align}
 \Lambda_{\Loa}{}^{\Lod} \,\omega_{+\Lod\Loe}\,\Lambda_{\Loc}{}^{\Loe} 
 &= \Lambda_{\Loa}{}^{\Lod} \,W_{+\Lod\Loe}\,\Lambda_{\Loc}{}^{\Loe} + \frac{1}{2}\,\Lambda_{\Loa}{}^{\Lod} \, \Lambda_{\Loc}{}^{\Loe}\,\Lambda_{\Lob}{}^{\Lof}\,e'^{\Lob}\,\bigl(M_{\Lod,\Loe\Lof}+M_{\Loe,\Lof\Lod} - M_{\Lof,\Lod\Loe}\bigr)
\nn\\
 &= \Lambda_{\Loa}{}^{\Lod} \,\Lambda_{\Loc}{}^{\Loe} \, \bigl(W_{+\Lod\Loe} - \Lambda_{\Lob}{}^{\Lof}\,e'^{\Lob}\, M_{\Lof,\Lod\Loe} \bigr)
 + \frac{1}{2}\,e'^{\Lob}\,H'_{\Lob\Loa\Loc} 
\nn\\
 &= \Lambda_{\Loa}{}^{\Lod} \,\Lambda_{\Loc}{}^{\Loe} \, W_{-\Lod\Loe} + \frac{1}{2}\,e'^{\Lob}\,H'_{\Lob\Loa\Loc} \,. 
\end{align}
This together with \eqref{eq:deprime-2} shows the second relation \eqref{eq:Lorntz-connection2},
\begin{align}
 \omega'^{\Loa\Lob} = \Lambda^{\Loa}{}_{\Lod} \,\Lambda^{\Lob}{}_{\Loe} \, W_{-}^{\Lod\Loe} + (\Lambda\,\rmd \Lambda^{-1})^{\Loa\Lob} + \frac{1}{2}\,e'_{\Loc}\,H'^{\Loc\Loa\Lob} \,. 
\end{align}

\section{Deformed Lagrangian at order $\cO(\theta^2)$}
\label{app:deformed-Lagrangian}

In this appendix, we show that the YB-deformed sigma model action can be rewritten as the conventional GS superstring action (up to quadratic order in fermions) by performing suitable field redefinitions. 

\subsection{A derivation of the deformed Lagrangian at $\cO(\theta^2)$}

Let us start with a straightforward computation of the deformed Lagrangian $\cL_{(2)}$ by using the results obtained in Appendix \ref{app:expansion-O}. 
For convenience, we decompose $\cL_{(2)}$ as
\begin{align}
 \cL_{(2)}=\cL_{(2,0,0)}+\cL_{(0,0,2)}+\cL_{(1,1,0)}+\cL_{(0,1,1)}+\cL_{(0,2,0)}+\cL_{(1,0,1)}+\cO(\theta^4)\,,
\end{align}
where we have defined
\begin{align}
 \cL_{(a,b,c)} \equiv -\frac{\dlT}{2}\,\Pg_-^{\WSa\WSb}\,\str[A_{\WSa(a)}\,d_-\circ \cO^{-1}_{-(b)}(A_{\WSb(c)})] \,, 
\end{align}
and $A_{\WSa(a)}$ ($a=0,1,2$) have the following form as we can see from \eqref{eq:A-AdS5xS5}:
\begin{align}
\begin{split}
 A_{\WSa(0)} &\equiv e_{\WSa}^{\Loa}\,\gP_{\Loa} -\frac{1}{2}\,\omega_{\WSa}{}^{\Loa\Lob} \,\gJ_{\Loa\Lob}\,,\qquad 
 A_{\WSa(1)} \equiv \gQ^I\,D_{\WSa}^{IJ}\theta_J \,,
\\
 A_{\WSa(2)} &\equiv \frac{\ii}{2}\,\brtheta_I\,\hat{\gamma}^{\Loa}\,D_{\WSa}^{IJ}\theta_J \,\gP_{\Loa} + \frac{1}{8}\,\epsilon^{IK}\,\brtheta_I\,\gamma^{\Loc\Lod}\,R_{\Loc\Lod}{}^{\Loa\Lob}\,D_{\WSa}^{KJ}\theta_J \,\gJ_{\Loa\Lob} \,.
\end{split}
\end{align}
Each part is given by
\begin{align}
 &\cL_{(2,0,0)}+\cL_{(0,0,2)}
 = -\ii\,\frac{\dlT}{2}\,\Pg_-^{\WSa\WSb}\,\brtheta_I\,\bigl(e_{+\WSa}{}^{\Loa}\,\hat{\gamma}_{\Loa}\,D^{IJ}_{\WSb}+e_{-\WSb}{}^{\Loa}\,\hat{\gamma}_{\Loa}\,D^{IJ}_{\WSa}\bigr)\theta_J\,.
\label{eq:part-1}\\
 &\cL_{(1,1,0)}= -\ii\,\frac{\dlT}{2}\,\Pg_-^{\WSa\WSb}\, \brtheta_I\,
 \biggl[\Bigl(2\,\eta\,\sigma_3^{IJ}\,e_{-\WSb}{}^{\Loa}\,\lambda_{\Loa}{}^{\Lob}\,\hat{\gamma}_{\Lob}
 -\frac{\ii}{2}\,\sigma_1^{IJ}\,\delta W_{-\WSb}{}^{\Lob\Loc}\,\gamma_{\Lob\Loc}\Bigr)\,D_{\WSa}\theta_J
\nn\\
 &\qquad\qquad\qquad +\frac{\ii}{2}\, \sigma_1^{IJ}\,\bigl(2\,\eta\,e_{-\WSb}{}^{\Loa}\,\lambda_{\Loa}{}^{\Loc}\,\hat{\gamma}_{\Loc}\bigr)\bigl(e_{\WSa}{}^{\Lod}\,\hat{\gamma}_{\Lod}\bigr)\, \theta_J
 -\Bigl(\frac{1}{4}\,\delta W_{-\WSb}{}^{\Lob\Loc}\,\gamma_{\Lob\Loc}\Bigr) \bigl(\sigma_3^{IJ}\,e_{\WSa}{}^{\Lod}\,\hat{\gamma}_{\Lod}\bigr)\, \theta_J \biggr]\,.
\label{eq:part-2}
\\
 &\cL_{(0,1,1)} =-\ii\,\frac{\dlT}{2}\,\Pg_-^{\WSa\WSb}\, \brtheta_I\,
 \biggl[\Bigl(2\,\eta\,\sigma_3^{IJ}\,e_{+\WSa}{}^{\Loa}\,\lambda_{\Loa}{}^{\Lob}\,\hat{\gamma}_{\Lob}
 +\frac{\ii}{2}\,\sigma_1^{IJ}\,\delta W_{+\WSa}{}^{\Lob\Loc}\,\gamma_{\Lob\Loc}\Bigr)\,D_{\WSb}\theta_J
\nn\\
 &\qquad\qquad\qquad +\frac{\ii}{2}\,\sigma_1^{IJ}\,\bigl(2\,\eta\,e_{+\WSa}{}^{\Loa}\,\lambda_{\Loa}{}^{\Loc}\,\hat{\gamma}_{\Loc}\bigr)\,\bigl(e_{\WSb}{}^{\Lod}\,\hat{\gamma}_{\Lod}\bigr)\,\theta_J
 +\Bigl(\frac{1}{4}\,\delta W_{+\WSa}{}^{\Lob\Loc}\,\gamma_{\Lob\Loc}\Bigr)\bigl(\sigma_3^{IJ}\,e_{\WSb}{}^{\Lod}\,\hat{\gamma}_{\Lod}\bigr)\,\theta_J \biggr]\,,
\label{eq:part-3}
\\
 &\cL_{(0,2,0)}=-\ii\,\frac{\dlT}{2}\,\Pg_-^{\WSa\WSb}\,\brtheta_I\,\biggl[
 e_{+\WSa}{}^{\Loa}\,\bigl(\delta^{IJ}\,\delta_{\Loa}^{\Lob}+2\,\eta\,\sigma_3^{IJ}\,\lambda_{\Loa}{}^{\Lob}\bigr)\,\hat{\gamma}_{\Lob}\,
 \Bigl(\frac{1}{4}\,\delta W_{-\WSb}{}^{\Loc\Lod}\,\gamma_{\Loc\Lod}\Bigr)
\nn\\
 &\qquad\qquad\qquad\qquad\qquad +\Bigl(\frac{1}{4}\,\delta W_{+\WSa}{}^{\Loc\Lod}\,\gamma_{\Loc\Lod}\Bigr)
 e_{-\WSb}{}^{\Loa}\,\bigl(\delta^{IJ}\,\delta_{\Loa}^{\Lob}+2\,\eta\,\sigma_3^{IJ}\,\lambda_{\Loa}{}^{\Lob}\bigr)\,\hat{\gamma}_{\Lob}
\nn\\
 &\qquad\qquad\qquad\qquad\qquad +\frac{\ii}{2}\,\epsilon^{IJ}\,\bigl[\bigl(e_{+\WSa}{}^{\Loa}\,\hat{\gamma}_{\Loa}\bigr)\bigl(2\,\eta\,e_{-\WSb}{}^{\Loc}\,\lambda_{\Loc}{}^{\Lod}\,\hat{\gamma}_{\Lod}\bigr)
 -\bigl(2\,\eta\,e_{+\WSa}{}^{\Loa}\,\lambda_{\Loa}{}^{\Lob}\,\hat{\gamma}_{\Lob}\bigr)\bigl(e_{-\WSb}{}^{\Loc}\,\hat{\gamma}_{\Loc}\bigr)\bigr]
\nn\\
 &\qquad\qquad\qquad\qquad\qquad +\frac{\ii}{8}\,\sigma_1^{IJ}\,\bigl(\delta W_{+\WSa}{}^{\Loa\Lob}\,\gamma_{\Loa\Lob}\bigr)\bigl(\delta W_{-\WSb}{}^{\Loc\Lod}\,\gamma_{\Loc\Lod}\bigr)
\nn\\
 &\qquad\qquad\qquad\qquad\qquad +\frac{\ii}{2}\,\sigma_1^{IJ}\,\bigl(2\,\eta\,e_{+\WSa}{}^{\Loa}\,\lambda_{\Loa}{}^{\Lob}\,\hat{\gamma}_{\Lob}\bigr)\bigl(2\,\eta\,e_{-\WSb}{}^{\Loc}\,\lambda_{\Loc}{}^{\Lod}\,\hat{\gamma}_{\Lod}\bigr) \biggr]\,\theta_J\,.
\label{eq:part-4}
\\
 &\cL_{(1,0,1)}=-\ii\,\frac{\dlT}{2}\,\Pg_-^{\WSa\WSb}\,\bigl(-2\,\sigma_3^{IK}\,\brtheta_I\,e_{[\WSa}{}^{\Loa}\,\hat{\gamma}_{\Loa}\,D_{\WSb]}^{KJ}\theta_J\bigr)\,.
\label{eq:part-5}
\end{align}
Here, we have defined $\delta W_{\pm}^{\Loa\Lob}$ as
\begin{align}
 \delta W_{\pm}^{\Loa\Lob}=\pm 2\,\eta\,e_{\pm}^{\Loc}\, \lambda_{\Loc}{}^{\Loa\Lob}\,,
\end{align}
which are parts of the torsionful spin connections
\begin{align}
 W_{\pm}^{\Loa\Lob}=\omega^{\Loa\Lob}+\delta W_{\pm}^{\Loa\Lob}\,.
\end{align}

\medskip

Gathering the results \eqref{eq:part-1}--\eqref{eq:part-5}, we can calculate $\cL_{(2)}$\,. 
In the following computation, it may be useful to use the following identities:
\begin{align}
\begin{split}
 \brtheta_I\,\hat{\gamma}_{\Loa}\,\gamma_{\Loc\Lod}\,\theta_J
 &=\brtheta_J\,\gamma_{\Loc\Lod}\,\hat{\gamma}_{\Loa}\,\theta_I\,,\qquad
 \brtheta_I\hat{\gamma}_{\Loa}\,\hat{\gamma}_{\Lob}\theta_J
 =-\brtheta_J\,\hat{\gamma}_{\Lob}\,\hat{\gamma}_{\Loa}\,\theta_I\,,
\\
 \brtheta_I\,\gamma_{\Loa\Lob}\,\theta_J
 &=\brtheta_J\,\gamma_{\Loa\Lob}\,\theta_I\,,\qquad
 \brtheta_I\,\gamma_{\Loa\Lob}\,\gamma_{\Loc\Lod}\,\theta_J
 = - \brtheta_J \gamma_{\Loc\Lod}\,\gamma_{\Loa\Lob}\,\theta_I\,.
\label{eq:thetaflip}
\end{split}
\end{align}
The result is
\begin{align}
 \cL_{(2)}&=-\ii\,\frac{\dlT}{2}\,\Pg_-^{\WSa\WSb}\,\brtheta_I\,
 \biggl\{ \bigl[\sigma^{IJ}_3\,e_{\WSb}{}^{\Loa}+e_{-\WSb}{}^{\Lob}\,(\delta^{IJ}\,\delta_{\Lob}^{\Loa}+2\,\eta\,\sigma_3^{IJ}\,\lambda_{\Lob}{}^{\Loa})\bigr]\,\hat{\gamma}_{\Loa}\,D_{+\WSa}
\nn\\
 &\qquad +\bigl[-\sigma^{IJ}_3\,e_{\WSa}{}^{\Loa} + e_{+\WSa}{}^{\Lob}\,(\delta^{IJ}\,\delta_{\Lob}^{\Loa}+2\,\eta\,\sigma_3^{IJ}\,\lambda_{\Lob}{}^{\Loa})\bigr]\,\hat{\gamma}_{\Loa}\,D_{-\WSb}
\nn\\
 &\qquad +\frac{\ii}{2}\,\epsilon^{IJ}\,\Bigl[\bigl(e_{+\WSa}{}^{\Loa}\,\hat{\gamma}_{\Loa}\bigr)\bigl(e_{\WSb}{}^{\Lob}\,\hat{\gamma}_{\Lob}+2\,\eta\,e_{-\WSb}{}^{\Loc}\,\lambda_{\Loc}{}^{\Lod}\,\hat{\gamma}_{\Lod}\bigr)
 +\bigl(e_{-\WSb}{}^{\Loa}\,\hat{\gamma}_{\Loa}\bigr)\bigl(e_{\WSa}{}^{\Lob}\,\hat{\gamma}_{\Lob}-2\,\eta\,e_{+\WSa}{}^{\Loa}\,\lambda_{\Loa}{}^{\Lob}\,\hat{\gamma}_{\Lob}\bigr) \Bigr]
\nn\\
 &\qquad +\frac{\ii}{2}\,\sigma_1^{IJ}\,\Bigl[\bigl(2\,\eta\,e_{-\WSb}{}^{\Loa}\,\lambda_{\Loa}{}^{\Loc}\,\hat{\gamma}_{\Loc}\bigr)\bigl(e_{\WSa}{}^{\Lod}\,\hat{\gamma}_{\Lod}\bigr)
 +\bigl(2\,\eta\,e_{+\WSa}{}^{\Loa}\,\lambda_{\Loa}{}^{\Loc}\,\hat{\gamma}_{\Loc}\bigr)\bigl(e_{\WSb}{}^{\Lod}\,\hat{\gamma}_{\Lod}\bigr)
\nn\\
 &\qquad\qquad\qquad +\bigl(2\,\eta\,e_{+\WSa}{}^{\Loa}\,\lambda_{\Loa}{}^{\Lob}\,\hat{\gamma}_{\Lob}\bigr)\bigl(2\,\eta\,e_{-\WSb}{}^{\Loc}\,\lambda_{\Loc}{}^{\Lod}\,\hat{\gamma}_{\Lod}\bigr)\Bigr]
\nn\\
 &\qquad +\frac{\ii}{2}\,\sigma_1^{IJ}\,
  \Bigl[\delta W_{+\WSa}{}^{\Loa\Lob}\,\gamma_{\Loa\Lob}\,D_{\WSb}
       -\delta W_{-\WSb}{}^{\Loa\Lob}\,\gamma_{\Loa\Lob}\,D_{\WSa}
       +\frac{1}{4}\,\bigl(\delta W_{+\WSa}{}^{\Loa\Lob}\,\gamma_{\Loa\Lob}\bigr)\bigl(\delta W_{-\WSb}{}^{\Loc\Lod}\,\gamma_{\Loc\Lod}\bigr)\Bigr]\biggr\}\,\theta_J
\nn\\
 &=-\ii\,\frac{\dlT}{2}\,\Pg_-^{\WSa\WSb}\,\brtheta_I\,
 \biggl\{2\,\pi^{IJ}_+\,e_{-\WSb}{}^{\Loa}\,\hat{\gamma}_{\Loa}\,D_{+\WSa}
      +2\,\pi^{IJ}_-\,e_{+\WSa}{}^{\Loa}\,\hat{\gamma}_{\Loa}\,D_{-\WSb}
      +\ii\,\epsilon^{IJ}\,e_{-\WSb}{}^{\Loa}\,\hat{\gamma}_{\Loa}\,e_{+\WSa}{}^{\Lob}\,\hat{\gamma}_{\Lob}
\nn\\
 &\qquad\qquad +\frac{\ii}{2} \,\sigma_1^{IJ}\,\Bigl[\bigl(e_{+\WSa}{}^{\Loa}\,\hat{\gamma}_{\Loa}\bigr)\bigl(e_{-\WSb}{}^{\Lob}\,\hat{\gamma}_{\Lob}\bigr) -\bigl(e_{\WSa}{}^{\Loa}\,\hat{\gamma}_{\Loa}\bigr)\bigl(e_{\WSb}{}^{\Lob}\,\hat{\gamma}_{\Lob}\bigr)
\nn\\
 &\qquad\qquad +\delta W_{+\WSa}{}^{\Loa\Lob}\, \gamma_{\Loa\Lob}\,D_{\WSb}
  -\delta W_{-\WSb}{}^{\Loa\Lob}\,\gamma_{\Loa\Lob}\,D_{\WSa}
  +\frac{1}{4}\,\bigl(\delta W_{+\WSa}{}^{\Loa\Lob}\,\gamma_{\Loa\Lob}\bigr)\bigl(\delta W_{-\WSb}{}^{\Loc\Lod}\,\gamma_{\Loc\Lod}\bigr)\Bigr]\biggr\}\,\theta_J\,.
\end{align}
In the second equality, we have used
\begin{align}
\begin{split}
 2\,\pi^{IJ}_+\, e_{-\WSa}{}^{\Loa}=& \sigma^{IJ}_3\,e_{\WSb}{}^{\Loa}+e_{-\WSb}{}^{\Lob}\,\bigl(\delta^{IJ}\,\delta_{\Lob}^{\Loa}+2\,\eta\,\sigma_3^{IJ}\,\lambda_{\Lob}{}^{\Loa}\bigr)\,,
\\
 2\,\pi^{IJ}_-\, e_{+\WSa}{}^{\Loa}=&-\sigma^{IJ}_3\,e_{\WSa}{}^{\Loa}+e_{+\WSa}{}^{\Lob}\,\bigl(\delta^{IJ}\,\delta_{\Lob}^{\Loa}+2\,\eta\,\sigma_3^{IJ}\,\lambda_{\Lob}{}^{\Loa}\bigr)\,,
\end{split}
\end{align}
where the projection operators $\pi_{\pm}^{IJ}$ are defined by
\begin{align}
 \pi_{\pm}^{IJ}=\frac{\delta^{IJ} \pm \sigma_3^{IJ}}{2} \,.
\end{align}
Now, we decompose the deformed Lagrangian into two parts
\begin{align}
 \cL_{(2)} =\cL_{(2)}^{\rmc}+\delta\cL_{(2)} \,,
\end{align}
where $\cL_{(2)}^{\rmc}$ takes the form of the canonical GS Lagrangian after taking the diagonal gauge (see section \ref{sec:YB-RR}) while $\delta\cL_{(2)}$ is the remaining part. 
The explicit form of $\cL_{(2)}^{\rmc}$ is given by
\begin{align}
 \cL_{(2)}^{\rmc} &=-\ii\,\dlT\,\brtheta_I\,\biggl[
   \Pg_-^{\WSa\WSb}\,\pi_+^{IJ}\,e_{-\WSb}{}^{\Loa}\,\hat{\gamma}_{\Loa}\,D_{+\WSa}
   +\Pg_-^{\WSa\WSb}\,\pi_-^{IJ}\,e_{+\WSa}{}^{\Loa}\,\hat{\gamma}_{\Loa}\,D_{-\WSb}
   +\frac{\ii}{2}\,\epsilon^{IJ}\,e_{-\WSb}{}^{\Loa}\,\hat{\gamma}_{\Loa}\,e_{+\WSa}{}^{\Lob}\,\hat{\gamma}_{\Lob}\biggr]\,\theta_J
\nn\\
  &=-\ii\,\dlT\,\biggl[
   \Pg_+^{\WSa\WSb}\,e_{-\WSa}{}^{\Loa}\,\brtheta_1\,\hat{\gamma}_{\Loa}\,D_{+\WSb}\theta_1
   +\Pg_-^{\WSa\WSb}\,e_{+\WSa}{}^{\Loa}\,\brtheta_2\,\hat{\gamma}_{\Loa}\,D_{-\WSb}\theta_2 
   +\ii\, \Pg_+^{\WSa\WSb}\,\epsilon^{IJ}\,\brtheta_1\,e_{-\WSa}{}^{\Loa}\,\hat{\gamma}_{\Loa}\,e_{+\WSb}{}^{\Lob}\,\hat{\gamma}_{\Lob}\,\theta_2\biggr] \,.
\end{align}
On the other hand, the remaining term $\delta\cL_{(2)}$ has the form
\begin{align}
 \delta\cL_{(2)}
 &= \frac{\dlT}{4}\,\Pg_-^{\WSa\WSb}\,\sigma_1^{IJ}\,\brtheta_I\,
  \biggl[\delta W_{+\WSa}{}^{\Loa\Lob}\,\hat{\gamma}_{\Loa\Lob}\,D_{\WSb}
  -\delta W_{-\WSb}{}^{\Loa\Lob}\,\hat{\gamma}_{\Loa\Lob}\,D_{\WSa}
\nn\\
 &\qquad +\frac{1}{4}\,\bigl(\delta W_{+\WSa}{}^{\Loa\Lob}\,\gamma_{\Loa\Lob}\bigr)\bigl(\delta W_{-\WSb}{}^{\Loc\Lod}\,\gamma_{\Loc\Lod}\bigr)
 +\bigl(e_{+\WSa}{}^{\Loa}\,\hat{\gamma}_{\Loa}\bigr)\bigl(e_{-\WSb}{}^{\Lob}\,\hat{\gamma}_{\Lob}\bigr)
 -\bigl(e_{\WSa}{}^{\Loa}\,\hat{\gamma}_{\Loa}\bigr)\bigl(e_{\WSb}{}^{\Lob}\,\hat{\gamma}_{\Lob}\bigr)\biggr]\,\theta_J\,.
\end{align}
By using \eqref{eq:thetaflip}, this can be rewritten as
\begin{align}
 \delta\cL_{(2)} &= \frac{\dlT}{4}\,\Pg_-^{\WSa\WSb}\,\sigma_1^{IJ}\,\brtheta_I\,
 \biggl[\bigl(\delta W_{+\WSa}{}^{\Loa\Lob}\,\partial_{\WSb}-\delta W_{-\WSb}{}^{\Loa\Lob}\,\partial_{\WSa}\bigr)\,\gamma_{\Loa\Lob}
\nn\\
 &\quad\qquad\qquad\qquad +\frac{1}{8}\,\bigl(\delta W_{+\WSa}{}^{\Loa\Lob}\,\omega_{\WSb}{}^{\Loc\Lod}
 -\delta W_{-\WSb}{}^{\Loa\Lob}\,\omega_{\WSa}{}^{\Loc\Lod}+\delta W_{+\WSa}{}^{\Loa\Lob}\,\delta W_{-\WSb}{}^{\Loc\Lod}\bigr)\, \bigl[\gamma_{\Loa\Lob},\,\gamma_{\Loc\Lod}\bigr]
\nn\\
 &\quad\qquad\qquad\qquad +\frac{1}{2}\,(e_{+\WSa}{}^{\Loa}\,e_{-\WSb}{}^{\Lob}
 -e_{\WSa}{}^{\Loa}\,e_{\WSb}{}^{\Lob})\,[\hat{\gamma}_{\Loa},\hat{\gamma}_{\Lob}] \biggr]\,\theta_J
\nn\\
 &=\frac{\dlT}{4}\,\Pg_-^{\WSa\WSb}\,\sigma_1^{IJ}\,
 \brtheta_I\,\biggl[
 \delta W_{+\WSa}{}^{\Loa\Lob}\,\partial_{\WSb}-\delta W_{-\WSb}{}^{\Loa\Lob}\,\partial_{\WSa}
 +(\delta W_{+\WSa})^{\Loa}{}_{\Loc}\,\omega_{\WSb}{}^{\Loc\Lob}
 -(\delta W_{-\WSb})^{\Loa}{}_{\Loc}\,\omega_{\WSa}{}^{\Loc\Lob}
\nn\\
 &\quad\qquad\qquad\qquad +(\delta W_{+\WSa})^{\Loa}{}_{\Loc}\,\delta W_{-\WSb}{}^{\Loc\Lob}
     -\frac{1}{2}\,(e_{+\WSa}{}^{\Loc}\,e_{-\WSb}{}^{\Lod}
     -e_{\WSa}{}^{\Loc}\, e_{\WSb}{}^{\Lod})\,R_{\Loc\Lod}{}^{\Loa\Lob}\biggr]\,\gamma_{\Loa\Lob}\theta_J\,,
\label{eq:delta-cL2}
\end{align}
where we used
\begin{align}
 \hat{\gamma}_{\Loa\Lob} =-\frac{1}{2}\,R_{\Loa\Lob}{}^{\Loc\Lod}\, \gamma_{\Loc\Lod}\,,\qquad 
 [\gamma_{\Loa\Lob},\,\gamma_{\Loc\Lod}]
 =-2\,\bigl(\eta_{\Loa\Loc}\,\gamma_{\Lob\Lod}-\eta_{\Lob\Loc}\,\gamma_{\Loa\Lod}-\eta_{\Loa\Lod}\,\gamma_{\Lob\Loc}+\eta_{\Lob\Lod}\,\gamma_{\Loa\Loc}\bigr)\,.
\end{align}
In the following, we show that $\delta\cL_{(2)}$ are completely canceled by performing an appropriate redefinition of the bosonic fields $X^m$\,. 

\medskip

\subsection{Bosonic shift}
\label{app:bosonic-shift}

We consider the redefinition of $X^m$\,,
\begin{align}
 X^m\ \to\ X^m + \xi^m\,,\qquad \xi^m\equiv \frac{\eta}{4}\,\sigma_1^{IJ}\, e^{\Loc m}\,\lambda_{\Loc}{}^{\Loa\Lob}\,\brtheta_I\, \gamma_{\Loa\Lob}\,\theta_J +\cO(\theta^4)\,. 
\label{eq:shift}
\end{align}
This was originally considered in \cite{Arutyunov:2015qva,Kyono:2016jqy} such that the unwanted terms involving $\sigma_1^{IJ}\, \brtheta_I\, \gamma_{\Loa\Lob}\,\partial_{\WSa}\theta_J$ in \eqref{eq:delta-cL2} are canceled out by the deviation of the Lagrangian under the shift \eqref{eq:shift}, $\delta\cL_{\YB}=\delta\cL_{(0)}+\cO(\theta^2)$ where
\begin{align}
\begin{split}
 &\cL_{(0)}\equiv -\dlT\,\Pg_{-}^{\WSa\WSb}\,E'_{mn}\,\partial_{\WSa} X^m\,\partial_{\WSb} X^n 
\\
 &\bigl(E'_{mn}=\CG'_{mn}+B'_{mn} = e_m{}^{\Loa}\,e_n{}^{\Lob}\,k_{+\Loa\Lob} = \str [A_{m}\,\cO_{-}^{-1}(A_{n})]\rvert_{\theta=0}\bigr)\,. 
\end{split}
\end{align}
As we show below, in fact, $\delta\cL_{(2)}$ is completely canceled out under the redefinition \eqref{eq:shift} when the $r$-matrix satisfies the homogeneous CYBE, which has been checked for specific examples in the previous works. 

\medskip

For simplicity, we introduce a shorthand notation
\begin{align}
 \frac{1}{\sqrt{2!}}\,\gJ_{\Loa\Lob} \ \rightarrow \ \gJ_{\sfi} \,,
\end{align}
with combinatoric factors discussed around \eqref{eq:index-convention}. 
In this notation, the commutation relations of bosonic generators $\{\gP_{\Loa},\, \gJ_{\sfi}\}$ and matrices $\{\hat{\gamma}_{\Loa},\,\gamma_{\sfi}\}$ become
\begin{align}
\begin{split}
 [\gP_{\Loa},\,\gP_{\Lob}] &=f_{\Loa\Lob}{}^{\sfi}\,\gJ_{\sfi}\,,\qquad
 [\gJ_{\sfi},\,\gJ_{\sfj}]= f_{\sfi\sfj}{}^{\sfk}\, \gJ_{\sfk}\,,\qquad
 [\gP_{\Loa},\,\gP_{\sfi}]=f_{\Loa\sfi}{}^{\Lob}\, \gP_{\Lob}\,,
\\
 [\hat{\gamma}_{\Loa},\,\hat{\gamma}_{\Lob}] &=-2\,f_{\Loa\Lob}{}^{\sfi}\, \gamma_{\sfi}\,,\qquad
 [\gamma_{i},\, \gamma_{j}]=-2\,f_{\sfi\sfj}{}^{\sfk}\, \gamma_{\sfk} \,.
\end{split}
\end{align}
Then, $\delta\cL_{(2)}$ in \eqref{eq:delta-cL2} can be expressed as
\begin{align}
 \delta\cL_{(2)}
 &=\frac{\dlT}{2}\,\Pg_-^{\WSa\WSb}\sigma_1^{IJ}
 \brtheta_I\, \gamma_{\sfk}\,\bigl(\delta W_{+\WSa}{}^{\sfk}\,\partial_{\WSb}\theta_J-\delta W_{-\WSb}{}^{\sfk}\,\partial_{\WSa}\theta_J\bigr) 
\nn\\
 &\quad -\frac{\dlT}{4}\,\Pg_-^{\WSa\WSb}\,\biggl[\bigl(\delta W_{+\WSa}{}^{\sfi}\,\omega_{\WSb}{}^{\sfj}
 +\omega_{\WSa}{}^{\sfi}\,\delta W_{-\WSb}{}^{\sfj}+\delta W_{+\WSa}{}^{\sfi}\,\delta W_{-\WSb}{}^{\sfj}\bigr)\, f_{\sfi\sfj}{}^{\sfk}
\nn\\
 &\quad\qquad\qquad +\bigl(e_{+\WSa}{}^{\Loa}\,e_{-\WSb}{}^{\Lob}-e_{\WSa}{}^{\Loa}\,e_{\WSb}{}^{\Lob}\bigr)\,f_{\Loa\Lob}{}^{\sfk} \biggr]\,\sigma_1^{IJ}\,\brtheta_I\, \gamma_{\sfk}\,\theta_J\,.
\label{eq:L2-shift}
\end{align}

\subsubsection*{A computation of $\delta \cL_{(0)}$}

A straightforward computation shows
\begin{align}
 \delta \cL_{(0)}&= -\dlT\,\Pg_{-}^{\WSa\WSb}\,\Lie_{\xi}E'_{mn}\,\partial_{\WSa} X^m\,\partial_{\WSb} X^n
\nn\\
 &=-\dlT\,\Pg_-^{\WSa\WSb}\,\bigl( \xi^p\,\partial_p E'_{mn} + \partial_m \xi^p\, E'_{pn}
 +\partial_{n}\xi^p \,E'_{mp}\bigr)\,\partial_{\WSa}X^{m}\,\partial_{\WSb}X^n
\nn\\
 &=-\frac{\dlT}{2}\,\Pg_-^{\WSa\WSb}\,\sigma_1^{IJ}\, \brtheta_I\,\gamma_{\sfk}\,\bigl(\delta W_{+m}{}^{\sfk}\,\partial_n\theta_J -\delta W_{-n}{}^{\sfk}\,\partial_m\theta_J\bigr)\,\partial_{\WSa} X^m\,\partial_{\WSb} X^n 
\nn\\
 &\quad -\frac{\eta\,\dlT}{2}\,\Pg_-^{\WSa\WSb}\,\sigma_1^{IJ}\,\brtheta_I\,\gamma_{\sfk}\,\theta_J\,\Bigl[\lambda_{\Loc}{}^{\sfk}\,e_m{}^{\Loa}\,e_n{}^{\Lob}\,e^{\Loc p}\,\partial_p k_{+\Loa\Lob} + \partial_m\lambda_{\Loa}{}^{\sfk}\,e_{-n}{}^{\Loa} + \partial_n\lambda_{\Loa}{}^{\sfk}\,e_{+m}{}^{\Loa} 
\nn\\
 &\quad\qquad\qquad\qquad\qquad\qquad 
  + 2\,e^{\Loc p}\,\lambda_{\Loc}{}^{\sfk}\,\bigl(\partial_{[p}e_{m]\Loa}\,e_{-n}{}^{\Loa}+ \partial_{[p}e_{n]\Loa}\,e_{+m}{}^{\Loa}\bigr)\Bigr]\,\partial_{\WSa} X^m\,\partial_{\WSb} X^n \,,
\end{align}
where we have used $e_{\pm m}{}^{\Loa}=k_{\pm \Lob}{}^{\Loa}\,e_m{}^{\Lob}$ and $\delta W_{\pm m}{}^{\sfk}=\pm 2\,\eta\,e_{\pm m}{}^{\Loc}\,\lambda_{\Loc}{}^{\sfk}$\,. 
From this, we can easily see that the terms involving $\sigma_1^{IJ}\, \brtheta_I\, \gamma_{\sfk}\,\partial_{\WSa}\theta_J$ in $\delta \cL_{(0)}$ and $\delta\cL_{(2)}$ cancel each other out. 

\medskip

We can further compute $\partial_p k_{+\Loa\Lob}$ and $\partial_m\lambda_{\Loa}{}^{\sfk}$ as follows by recalling their original definitions. 
The $\partial_m\lambda_{\Loa}{}^{\sfk}$ can be obtained from
\begin{align}
 \partial_{m}\lambda_{\Loc}{}^{\sfi}\,\gJ_{\sfi} &=\partial_{m}\bigl[P^{(0)}R_{g_{\bos}}(\gP_{\Loc})\bigr]\bigr\rvert_{\theta=0}
 =-P^{(0)} [A_{m},\,R_{g_{\bos}}(\gP_{\Loc})]\bigr\rvert_{\theta=0}+P^{(0)}R_{g_{\bos}}\bigl([A_{m},\,\gP_{\Loc}]\bigr)\bigr\rvert_{\theta=0}
\nn\\
 &=\bigl[e_m{}^{\Loa}\,(\lambda_{\sfk}{}^{\sfi}\,f_{\Loa\Loc}{}^{\sfk} - \lambda_{\Loc}{}^{\Lob}\,f_{\Loa\Lob}{}^{\sfi})
 +\omega_m{}^{\sfj}\,(\lambda_{\Loc}{}^{\sfk}\,f_{\sfj\sfk}{}^{\sfi} - \lambda_{\Lod}{}^{\sfi}\,f_{\sfj\Loc}{}^{\Lod})\bigr]\,\gJ_{\sfi}\,.
\end{align}
where we have used
\begin{align}
\begin{split}
 &\partial_{m}R_g(\,\cdot\,)=-[A_{m},\,R_g(\cdot)\,]+R_g([A_{m},\,\cdot\,]) \,, \qquad 
 A_m\rvert_{\theta=0}=e_m{}^{\Loa}\,\gP_{\Loa}-\omega_m{}^{\sfk}\,\gJ_{\sfk} \,,
\\
 &R_g(\gP_{\Loa})\rvert_{\theta=0}=\lambda_{\Loa}{}^{\Lob}\,\gP_{\Lob}-\lambda_{\Loa}{}^{\sfk}\,\gJ_{\sfk} \,,\qquad 
 R_g(\gJ_{\sfi})\rvert_{\theta=0}=\lambda_{\sfi}{}^{\Lob}\,\gP_{\Lob}-\lambda_{\sfi}{}^{\sfk}\,\gJ_{\sfk} \,. 
\end{split}
\end{align}
Similarly, we obtain
\begin{align}
 \partial_{m}k_{+\Loa\Lob}&=\partial_{m}\str\bigl[\gP_{\Lob}\,\cO_+^{-1}(\gP_{\Loa})\bigr]\bigr\rvert_{\theta=0}
 = -\frac{1}{2}\,\str\bigl[d_-(\gP_{\Lob})\,\cO_+^{-1}\circ\partial_{m}\cO_+\circ\cO_+^{-1}(\gP_{\Loa})\bigr]\bigr\rvert_{\theta=0}
\nn\\
 &= -\frac{1}{2}\,\str\bigl[\cO_+^{-\rmT}\circ d_-(\gP_{\Lob})\, \partial_{m}\cO_+\circ\cO_+^{-1}(\gP_{\Loa})\bigr]\bigr\rvert_{\theta=0}
\nn\\
 &= -\frac{\eta}{2}\,\str\bigl[d_-\,\cO_-^{-1}(\gP_{\Lob})\, \partial_{m} R_g\circ d_+\circ\cO_+^{-1}(\gP_{\Loa})\bigr]\bigr\rvert_{\theta=0}
\nn\\
 &= -2\,\eta\,k_{+\Loa}{}^{\Loc}\,k_{-\Lob}{}^{\Lod}\,\str\bigl[\gP_{\Lod}\,\bigl\{-[A_m,\,R_g(\gP_{\Loc})] + R_g([A_m,\,\gP_{\Loc}]) \bigr\}\bigr]\bigr\rvert_{\theta=0}
\nn\\
 &=-2\,\eta\,k_{+\Loa}{}^{\Loc}\,k_{-\Lob\Lod}\,\bigl[e_m{}^{\Loe}\,(\lambda_{\sfk}{}^{\Lod}\,f_{\Loe\Loc}{}^{\sfk}-\lambda_{\Loc}{}^{\sfk}\,f_{\Loe\sfk}{}^{\Lod}) +\omega_m{}^{\sfk}\,(\lambda_{\Loc}{}^{\Loe}\,f_{\sfk\Loe}{}^{\Lod}-\lambda_{\Loe}{}^{\Lod}\,f_{\sfk\Loc}{}^{\Loe})\bigr] \,.
\end{align}
By using several identities, such as
\begin{align}
 e_m{}^{\Loa}=e_{\pm m}{}^{\Lob}\,\bigl(\delta_{\Lob}^{\Loa}\pm 2\,\eta\,\lambda_{\Lob}{}^{\Loa}\bigr)\,,\qquad
 \partial_{[m}e_{n]}{}^{\Loa}=-\omega_{[m}{}^{\Loa\Lob}\,e_{n]\Lob}\,,
\end{align}
and the explicit forms of the structure constants, we can straightforwardly obtain
\begin{align}
 \delta \cL_{(0)}&=-\frac{\dlT}{2}\,\Pg_-^{\WSa\WSb}\,\sigma_1^{IJ}\,\brtheta_I\,\gamma_{\sfk}\,\bigl(\delta W_{+m}{}^{\sfk}\,\partial_{n}\theta_J -\delta W_{-n}{}^{\sfk}\,\partial_{m}\theta_J\bigr) \,\partial_{\WSa}X^{m}\,\partial_{\WSb}X^n
\nn\\
 &\quad +\frac{\dlT}{4}\,\Pg_-^{\WSa\WSb}\,\Bigl[\bigl(\delta W_{+m}{}^{\sfi}\,\omega_{n}{}^{\sfj} +\omega_m{}^{\sfi}\,\delta W_{-n}{}^{\sfj}\bigr)\,f_{\sfi\sfj}{}^{\sfk}
  +2\,\eta\,\bigl(e_{m}{}^{\Loa}\,e_{-n}{}^{\Loc}\,\lambda_{\Loc}{}^{\Lob}- e_{+m}{}^{\Loc}\,\lambda_{\Loc}{}^{\Loa}\,e_{n}{}^{\Lob}\bigr)\,f_{\Loa\Lob}{}^{\sfk}
\nn\\
 &\qquad +4\,\eta^2\,\bigl(e_{+m}{}^{\Loc}\,\lambda_{\Loc}{}^{\sfj}\,e_{-n}{}^{\Lob}- e_{+m}{}^{\Lob}\,e_{-n}{}^{\Loc}\,\lambda_{\Loc}{}^{\sfj}\bigr)\,
  f_{\Lob\sfj}{}^{\Loa}\,\lambda_{\Loa}{}^{\sfk}
\nn\\
 &\qquad -4\,\eta^2\,\bigl(e_{+m}{}^{\Loc}\,\lambda_{\Loc}{}^{\Loa}\,e_{-n}{}^{\Lob}+e_{+m}{}^{\Loa}\,e_{-n}{}^{\Loc}\,\lambda_{\Loc}{}^{\Lob}\bigr)\,f_{\Loa\Lob}{}^{\sfj}\,\lambda_{\sfj}{}^{\sfk} \Bigr]\,\sigma_1^{IJ}\,\brtheta_I\,\gamma_{\sfk}\,\theta_J\,\partial_{\WSa}X^{m}\,\partial_{\WSb}X^n\,.
\label{eq:L0-shift}
\end{align}

\subsubsection*{A computation of $\delta \cL_{(0)}+\delta\cL_{(2)}$}

Now, the sum of \eqref{eq:L0-shift} and \eqref{eq:L2-shift} becomes
\begin{align}
 &\delta \cL_{(0)}+\delta\cL_{(2)}
\nn\\
 &= \frac{\dlT}{4}\,\Pg_-^{\WSa\WSb}\,\biggl[ 4\,\eta^2\,(e_{+m}{}^{\Loa}\,\lambda_{\Loa}{}^{\sfi})(e_{-n}{}^{\Lob}\,\lambda_{\Lob}{}^{\sfj})\,f_{\sfi\sfj}{}^{\sfk}
\nn\\
 &\quad +2\,\eta\,\bigl(e_{m}{}^{\Loa}\,e_{-n}{}^{\Loc}\,\lambda_{\Loc}{}^{\Lob}- e_{+m}{}^{\Loc}\,\lambda_{\Loc}{}^{\Loa}\,e_{n}{}^{\Lob}\bigr)\,f_{\Loa\Lob}{}^{\sfk}
 -\bigl(e_{+m}{}^{\Loa}\,e_{-n}{}^{\Lob}-e_{m}{}^{\Loa}\,e_{n}{}^{\Lob}\bigr)\,f_{\Loa\Lob}{}^{\sfk}
\nn\\
 &\quad +4\,\eta^2\,\bigl(e_{+m}{}^{\Loc}\,\lambda_{\Loc}{}^{\sfj}\,e_{-n}{}^{\Lob}- e_{+m}{}^{\Lob}\,e_{-n}{}^{\Loc}\,\lambda_{\Loc}{}^{\sfj}\bigr)\,f_{\Lob\sfj}{}^{\Loa}\,\lambda_{\Loa}{}^{\sfk}
\nn\\
 &\quad -4\,\eta^2\,\bigl(e_{+m}{}^{\Loc}\,\lambda_{\Loc}{}^{\Loa}\,e_{-n}{}^{\Lob}+e_{+m}{}^{\Loa}\,e_{-n}{}^{\Loc}\,\lambda_{\Loc}{}^{\Lob}\bigr)\,f_{\Loa\Lob}{}^{\sfj}\,\lambda_{\sfj}{}^{\sfk}\biggr]\,\sigma_1^{IJ}\,\brtheta_I\,\gamma_{\sfk}\,\theta_J\,\partial_{\WSa}X^{m}\,\partial_{\WSb}X^n\,,
\end{align}
where we used $\delta W_{\pm}^{\sfi}=\pm 2\,\eta\,e_{\pm}^{\Loc}\, \lambda_{\Loc}{}^{\sfi}$ in the first line.
Moreover, the second line is simplified as
\begin{align}
 2\,\eta\,\bigl(e_{m}{}^{\Loa}\, e_{-n}{}^{\Loc}\,\lambda_{\Loc}{}^{\Lob}- e_{+m}{}^{\Loc}\,\lambda_{\Loc}{}^{\Loa}\,e_{n}{}^{\Lob}\bigr)
 -\bigl(e_{+m}{}^{\Loa}\,e_{-n}{}^{\Lob}-e_{m}{}^{\Loa}\,e_{n}{}^{\Lob}\bigr)
 =4\,\eta^2\,e_{+m}{}^{\Loc}\,\lambda_{\Loc}{}^{\Loa}\,e_{-n}{}^{\Lod}\,\lambda_{\Lod}{}^{\Lob}\,,
\end{align}
and we obtain
\begin{align}
\begin{split}
 \delta \cL_{(0)}+\delta\cL_{(2)}&=\eta^2\,\dlT\,\Pg_-^{\WSa\WSb}\,
 \Bigl[(e_{+m}{}^{\Loc}\,\lambda_{\Loc}{}^{\Loa})(e_{-n}{}^{\Lod}\,\lambda_{\Lod}{}^{\Lob})\,f_{\Loa\Lob}{}^{\sfk}
 +(e_{+m}{}^{\Loa}\,\lambda_{\Loa}{}^{\sfi})(e_{-n}{}^{\Lob}\,\lambda_{\Lob}{}^{\sfj})\,f_{\sfi\sfj}{}^{\sfk}
\\
 &\quad +\bigl(e_{+m}{}^{\Loc}\,\lambda_{\Loc}{}^{\sfj}\,e_{-n}{}^{\Lob} - e_{+m}{}^{\Lob}\, e_{-n}{}^{\Loc}\,\lambda_{\Loc}{}^{\sfj}\bigr)\,f_{\Lob\sfj}{}^{\Loa}\,\lambda_{\Loa}{}^{\sfk}
\\
 &\quad -\bigl(e_{+m}{}^{\Loc}\,\lambda_{\Loc}{}^{\Loa}\,e_{-n}{}^{\Lob}+e_{+m}{}^{\Loa}\,e_{-n}{}^{\Loc}\,\lambda_{\Loc}{}^{\Lob}\bigr)\,f_{\Loa\Lob}{}^{\sfj}\,\lambda_{\sfj}{}^{\sfk}\Bigr]\,\sigma_1^{IJ}\,\brtheta_I\,\gamma_{\sfk}\,\theta_J\,\partial_{\WSa}X^{m}\,\partial_{\WSb}X^n\,.
\end{split}
\label{eq:sum-shift2}
\end{align}
Remarkably, the quantities in the square bracket of \eqref{eq:sum-shift2} are precisely the grade-$0$ component of $\CYBE_g(J_{+m}^{(2)},\,J_{-n}^{(2)})$\,,
\begin{align}
\begin{split}
 \bigl[\CYBE^{(0)}_g(J_{+m}^{(2)},\,J_{-n}^{(2)})\bigr]^{\sfk}
 &= (e_{+m}{}^{\Loc}\,\lambda_{\Loc}{}^{\Loa})(e_{-n}{}^{\Lod}\,\lambda_{\Lod}{}^{\Lob})\,f_{\Loa\Lob}{}^{\sfk}
  +(e_{+m}{}^{\Loa}\,\lambda_{\Loa}{}^{\sfi})(e_{-n}{}^{\Lob}\,\lambda_{\Lob}{}^{\sfj})\,f_{\sfi\sfj}{}^{\sfk}
\\
 &\quad +\bigl(e_{+m}{}^{\Loc}\,\lambda_{\Loc}{}^{\sfj}\,e_{-n}{}^{\Lob}- e_{+m}{}^{\Lob}\,e_{-n}{}^{\Loc}\,\lambda_{\Loc}{}^{\sfj}\bigr)\, f_{\Lob\sfj}{}^{\Loa}\,\lambda_{\Loa}{}^{\sfk}
\\
 &\quad -(e_{+m}{}^{\Loc}\,\lambda_{\Loc}{}^{\Loa}\,e_{-n}{}^{\Lob}+e_{+m}{}^{\Loa}\,e_{-n}{}^{\Loc}\,\lambda_{\Loc}{}^{\Lob})\,f_{\Loa\Lob}{}^{\sfj}\,\lambda_{\sfj}{}^{\sfk}\,,
\end{split}
\end{align}
where we have used $J_{\pm m}^{(2)}=e_{\pm m}{}^{\Loa}\,\gP_{\Loa}$ [see \eqref{eq:Jpm-expansion}]. 
Therefore, we obtain 
\begin{align}
 \delta \cL_{(0)}+\delta\cL_{(2)}
 =\eta^2\,\dlT\,\Pg_-^{\WSa\WSb}\,\sigma_1^{IJ}\,
 \bigl[\CYBE^{(0)}_g(J_{+m}^{(2)},\,J_{-n}^{(2)})\bigr]^{\sfi}\, \brtheta_I\,\hat{\gamma}_{\sfi}\,\theta_J\,\partial_{\WSa}X^{m}\,\partial_{\WSb}X^n\,,
\end{align}
which shows that $\delta \cL_{(0)}+\delta\cL_{(2)}$ vanishes when the $r$-matrix satisfies the homogeneous CYBE.

\chapter{The $\kappa$-symmetry transformation}
\label{app:kappa}

As shown in the subsection \ref{sec:kappa-YB},
the YB sigma model action \eqref{eq:YBsM} is invariant under the following $\kappa$-symmetry variations:
\begin{align}
\mathcal{O}_-^{-1}g^{-1}\delta_{\kappa}g&=
P^{\alpha\beta}_-\{\gQ^1\kappa_{1\alpha},J_{-\beta}^{(2)}\}
+P^{\alpha\beta}_+\{\gQ^2\kappa_{2\alpha},J_{+\beta}^{(2)}\}
\,, \label{eq:kappa1} \\
\delta_\kappa(\sqrt{-\ga} \ga^{\alpha\beta})&=
\frac{1}{4}\sqrt{-\ga}\,{\rm Str}\biggl[\Upsilon\Bigl([\gQ^1\kappa^\alpha_{1(+)},J_{+(+)}^{(1)\beta}]
+[\gQ^2\kappa^\alpha_{2(-)},J_{-(-)}^{(3)\beta}]\Bigr)+(\alpha\leftrightarrow \beta)\biggr]\,.
\label{eq:kappa2}
\end{align}
In this appendix, following the procedure of \cite{Arutyunov:2015qva}, we rewrite the $\kappa$-variations \eqref{eq:kappa1} and \eqref{eq:kappa2} as the standard $\kappa$-variations in the GS type IIB superstring \cite{Grisaru:1985fv},
\begin{align}
 \delta_\kappa X^m &=-\frac{\ii}{2}\,e'^{\Loa m}\,\brTheta_I\,\Gamma_{\Loa}\,\delta_\kappa\Theta'_I + \cO(\Theta'^3)\,, 
\label{eq:kappaX1}
\\
 \delta_\kappa\Theta'_I
 &= \frac{1}{4}\,(\delta^{IJ}\,\gga^{\WSa\WSb}-\sigma_3^{IJ}\,\epsilon^{\WSa\WSb})\,e'_{\WSb}{}^{\Loa}\,\Gamma_{\Loa}\,K'_{J\WSa} + \cO(\Theta'^2)\,,
\label{eq:kappath1}
\\
 \frac{1}{\sqrt{-\gga}}\,\delta_{\kappa}\bigl(\sqrt{-\gga}\,\gga^{\WSa\WSb}\bigr)
 &= -2\,\ii\,\bar{K}'^{(\WSa}_{1(+)}\,D'^{\WSb)}_{+(+)}\Theta'_1 - 2\,\ii\,\bar{K}'^{(\WSa}_{2(-)}\,D'^{\WSb)}_{-(-)}\Theta'_2
\nn\\
 &\quad + \frac{\ii}{8}\,\biggl[\bar{K}'^{(\WSa}_{1(+)}\,\bisF'\,e'^{\WSb)\Loa}_{(+)}\,\Gamma_{\Loa}\,\Theta'_2 -\brTheta'_1\,e'^{(\WSb|\Loa}_{(-)}\,\Gamma_{\Loa}\,\bisF'\,K'^{|\WSa)}_{2(-)}\biggr] + \cO(\Theta'^3)\,,
\label{eq:kappaga1}
\end{align}
where the detailed notations are explained below. 
In the course of the rewriting, we need to identify the supergravity background $(e'_m{}^{\Loa},\,B'_{mn},\,\bisF')$ as the $\beta$-deformed $\AdS{5}\times\rmS^5$ background. 
In this sense, the following computation serves as a non-trivial check of the equivalence between YB deformations and local $\beta$-deformations. 

\section{Bosonic fields}

We first consider the $\kappa$-symmetry transformation of the bosonic fields $X^m$\,, which can be extracted from the grade-$2$ component of \eqref{eq:kappa1}.
From \eqref{eq:kappa1}, we can easily see
\begin{align}
 P^{(2)}\circ\cO_-^{-1}\,g^{-1}\,\delta_{\kappa}g=0\,.
\end{align}
The left-hand side can be expanded as 
\begin{align}
 P^{(2)}\circ\cO_-^{-1}\,g^{-1}\,\delta_{\kappa}g
 &=P^{(2)}\,\biggl[\Bigl(e_{m}{}^{\Loa}\,\delta_{\kappa}X^m+\frac{\ii}{2}\,\brtheta_I\,\hat{\gamma}^{\Loa}\,\delta_{\kappa}\theta_I \Bigr)\,\cO_{-(0)}^{-1}(\gP_{\Loa})
 +\cO_{-(1)}^{-1}\bigl(\gQ^{I}\,\delta_{\kappa}\theta_{I}\bigr) +\cO(\theta^3) \biggr]
\nn\\
 &=\biggl[e_{m\Lob}\,\delta_{\kappa}X^m +\frac{\ii}{2}\,\bigl(\delta^{IJ}\,\delta_{\Lob}^{\Loc} + 2\,\eta\,\sigma_3^{IJ}\,\lambda_{\Lob}{}^{\Loc}\bigr)\,\brtheta_I\,\hat{\gamma}_{\Loc}\,\delta_\kappa\,\theta_J\biggr]\,k_-^{\Lob\Loa}\,\gP_{\Loa}
\nn\\
 &\quad -\frac{\eta}{2}\,\sigma_1^{IJ}\,\brtheta_I\,\lambda_{\Lob}{}^{\Loc\Lod}\,\gamma_{\Loc\Lod}\,\delta_\kappa\theta_J\,k_{-}^{\Lob\Loa}\,\gP_{\Loa} + \cO(\theta^3) \,.
\end{align}
Now, by performing the redefinition \eqref{eq:bosonic-shift} of $X^m$\,, the term proportional to $\sigma^{IJ}_1$ disappears and we obtain
\begin{align}
 0=\Bigl[ e_{m \Lob}\,\delta_{\kappa}X^m + \frac{\ii}{2}\,\bigl(\delta^{IJ}\,\delta_{\Lob}^{\Loc}+2\,\eta\,\sigma_3^{IJ}\,\lambda_{\Lob}{}^{\Loc}\bigr)\,\brtheta_I\,\hat{\gamma}_{\Loc}\, \delta_\kappa\theta_J\Bigr]\,k_-^{\Lob\Loa}\,\gP_{\Loa}+\cO(\theta^3)\,.
\label{eq:X-kappa-expansion}
\end{align}
Then, solving the equation \eqref{eq:X-kappa-expansion} for $\delta_{\kappa}X^{m}$\,, we obtain
\begin{align}
 \delta_{\kappa}X^{m}
 &= -\frac{\ii}{2}\,(k_-^{-1})^{\Loa}{}_{\Lob}\,e^{\Lob m}\,\brtheta_1\,\hat{\gamma}_{\Loa}\,\delta_{\kappa}\theta_1
 -\frac{\ii}{2}\,(k_{+}^{-1})^{\Loa}{}_{\Lob}\,e^{\Lob m}\,\brtheta_2\,\hat{\gamma}_{\Loa}\,\delta_{\kappa}\theta_2 + \cO(\theta^3)
\nn\\
 &=-\frac{\ii}{2}\,e_-^{\Loa m}\,\brTheta_1\,\Gamma_{\Loa}\,\delta_{\kappa}\Theta_1 - \frac{\ii}{2}\,e_+^{\Loa m}\,\brTheta_2\,\Gamma_{\Loa}\,\delta_{\kappa}\Theta_2+\cO(\Theta^3)\,,
\label{eq:kappa-X}
\end{align}
where it is note that the inverse of $e_{\pm m}{}^{\Loa}=e_{m}{}^{\Lob}\,k_{\pm \Lob}{}^{\Loa}$ is $e_{\pm \Loa}{}^{m}=(k_\pm^{-1})_{\Loa}{}^{\Lob}\,e_{\Lob}{}^{m}$\,. 
Finally, by using $\Lambda_{\Loa}{}^{\Lob} \, \Gamma_{\Lob}=\Omega^{-1} \,\Gamma_{\Loa}\,\Omega$ and the redefined fermions $\Theta'_I$ given in \eqref{eq:fermi-redef}, \eqref{eq:kappa-X} becomes
\begin{align}
 \delta_{\kappa}X^{m} &=-\frac{\ii}{2}\,e'^{\Loa m}\,\brTheta_1\,\Gamma_{\Loa}\,\delta_{\kappa}\Theta_1 -\frac{\ii}{2}\,e'^{\Loa m}\,\brTheta_2\,\Omega^{-1}\,\Gamma_{\Loa}\,\Omega\,\delta_{\kappa}\Theta_2+\cO(\Theta^3)
\nn\\
 &=-\frac{\ii}{2}\,e'^{\Loa m}\,\brTheta'_I\,\Gamma_{\Loa}\,\delta_{\kappa}\Theta'_I + \cO(\Theta^3) \,,
\end{align}
which is the usual $\kappa$-variation \eqref{eq:kappaX1} of $X^m$\,.

\section{Fermionic fields}

Next, let us consider the $\kappa$-variations of fermionic variables.
These can be found from
\begin{align}
\begin{split}
 P^{(1)}\,\cO^{-1}_-\,g^{-1}\,\delta_{\kappa}g&=\Pg^{\WSa\WSb}_-\,\{\gQ^1\,\kappa_{1\WSa},\,J_{-\WSb}^{(2)}\}\,,
\\
 P^{(3)}\,\cO^{-1}_-\,g^{-1}\,\delta_{\kappa}g&=\Pg^{\WSa\WSb}_+\,\{\gQ^2\,\kappa_{2\WSa},\,J_{+\WSb}^{(2)}\}\,.
\end{split}
\label{eq:P1P3Ogdg}
\end{align}
Indeed, the left-hand side gives
\begin{align}
 P^{(1)}\,\cO^{-1}_-\,g^{-1}\,\delta_{\kappa}g = \gQ^1\,\delta_{\kappa}\theta_1 + \cO(\theta^2)\,,\qquad
 P^{(3)}\,\cO^{-1}_-\,g^{-1}\,\delta_{\kappa}g = \gQ^2\,\delta_{\kappa}\theta_2 + \cO(\theta^2)\,. 
\end{align}
In order to evaluate the right-hand side, we use the following relations \cite{Arutyunov:2015qva}:
\begin{align}
 \gQ^I\,\gP_{\check{\Loa}}+\gP_{\check{\Loa}}\,\gQ^I=\frac{1}{2}\,\gQ^I\,\hat{\gamma}_{\check{\Loa}}\,,\qquad
 \gQ^I\,\gP_{\hat{\Loa}}+\gP_{\hat{\Loa}}\,\gQ^I=-\frac{1}{2}\,\gQ^I\,\hat{\gamma}_{\hat{\Loa}} \,,
\end{align}
which can be verified by using the matrix representations of $\gP_{\Loa}$ and $\gQ_I$ given in \eqref{eq:P-J-super} and \eqref{eq:Q-matrix}. 
Then, the transformations \eqref{eq:P1P3Ogdg} become
\begin{align}
\begin{split}
 \gQ^1\delta_{\kappa}\theta_1&=\frac{1}{2}\,\Pg_-^{\WSa\WSb}\,\gQ^1\,\bigl(e_{-\WSb}{}^{\check{\Loa}}\,\hat{\gamma}_{\check{\Loa}}- e_{-\WSb}{}^{\hat{\Loa}}\,\hat{\gamma}_{\hat{\Loa}}\bigr)\,\kappa_{1\WSa} + \cO(\theta^2)\,,
\\
 \gQ^2\delta_{\kappa}\theta_2&=\frac{1}{2}\,\Pg_+^{\WSa\WSb}\,\gQ^2\,\bigl(e_{+\WSb}{}^{\check{\Loa}}\,\hat{\gamma}_{\check{\Loa}}- e_{+\WSb}{}^{\hat{\Loa}}\,\hat{\gamma}_{\hat{\Loa}}\bigr)\,\kappa_{2\WSa} + \cO(\theta^2)\,.
\end{split}
\end{align}
By using relations \eqref{eq:Theta-theta} and \eqref{eq:brTheta-brtheta}, these can be rewritten as
\begin{align}
\begin{split}
 \delta_{\kappa}\Theta_1&=\frac{1}{2}\,\Pg^{\WSa\WSb}_-\,e_{-\WSb}{}^{\Loa}\,\Gamma_{\Loa}\,K_{1\WSa} + \cO(\Theta^2)\,,
\\
 \delta_{\kappa}\Theta_2&=\frac{1}{2}\,\Pg^{\WSa\WSb}_+\,e_{+\WSb}{}^{\Loa}\,\Gamma_{\Loa}\,K_{2\WSa} + \cO(\Theta^2)\,,
\end{split}
\end{align}
where we have introduced 
\begin{align}
 K_I\equiv \begin{pmatrix} 0\\ 1 \end{pmatrix} \otimes \kappa_I\,,\qquad
 \bar{K}_I\equiv \begin{pmatrix} 1 & 0 \end{pmatrix} \otimes \bar{\kappa}_I\,,
\end{align}
and used
\begin{align}
 \Gamma_{\check{\Loa}}\,K_{I} = \begin{pmatrix} 1\\ 0 \end{pmatrix} \otimes \hat{\gamma}_{\check{\Loa}}\,\kappa_I \,,\qquad 
 \Gamma_{\hat{\Loa}}\,K_{I} = - \begin{pmatrix} 1\\ 0 \end{pmatrix} \otimes \hat{\gamma}_{\hat{\Loa}}\,\kappa_I \,.
\end{align}
Finally, using the redefined fermions $\Theta'_I$ and considering redefinitions of $K_I$\,,
\begin{align}
 K'_1=K_1\,,\qquad K'_2=\Omega\,K_2 \,,
\label{eq:fermi-redef-K}
\end{align}
we obtain the standard $\kappa$-variations of the fermions \eqref{eq:kappath1}.

\section{Worldsheet metric}

Finally, we rewrite the $\kappa$-variation \eqref{kappa2} of $\gga_{\WSa\WSb}$ into the standard form \eqref{eq:kappaga1}.

\medskip

By using the expansion \eqref{eq:Jpm-expansion} of the deformed currents $J_{\pm}$\,, the variation \eqref{kappa2} can be expanded as
\begin{align}
 \frac{1}{\sqrt{-\gga}}\,\delta_{\kappa}\bigl(\sqrt{-\gga}\,\gga^{\WSa\WSb}\bigr) 
 = -\ii\,\bar{\kappa}_{1(+)}^{\WSa}\,D^{\WSb\,1J}_{+(+)}\theta_J
  -\ii\,\bar{\kappa}_{2(-)}^{\WSa}\,D^{\WSb\,2J}_{-(-)}\theta_J
  +(\WSa\leftrightarrow \WSb)+\cO(\theta^3)\,,
\end{align}
where we have used commutation relations of $\alg{su}(2,2|4)$ algebra
\begin{align}
\begin{split}
 \bigl[\gQ^1\,\kappa^{\WSa}_{1(+)},\,J_{+(+)}^{(1)\WSb}\bigr]
 &=\Bigl(-\frac{1}{2}\,\bar{\kappa}^{\WSa}_{1(+)}\,D^{\WSb\,1J}_{+(+)}\theta_J + \cO(\theta^3)\Bigr)\,Z + (\gP\text{-term}) \,,
\\
 \bigl[\gQ^2\kappa^{\WSa}_{2(-)},\,J_{-(-)}^{(3)\WSb}\bigr]
 &=\Bigl(-\frac{1}{2}\,\bar{\kappa}^{\WSa}_{2(-)}\,D^{\WSb\,2J}_{-(-)}\theta_J + \cO(\theta^3)\Bigr)\,Z + (\gP\text{-term})\,,
\end{split}
\end{align}
and supertrace formulas
\begin{align}
 \str\bigl[\Upsilon\,Z\bigr]=8\,\ii\,,\qquad \str\bigl[\Upsilon\,(\text{other generators})\bigr]=0\,.
\end{align}
It is noted that the redefinition \eqref{eq:bosonic-shift} of $X^m$ does not affect the variation of the worldsheet metric at the leading order in $\theta$.

\medskip

Then, by using the $32\times 32$ gamma matrices, the variation can be expressed as
\begin{align}
 \frac{1}{\sqrt{-\gga}}\,\delta_{\kappa}\bigl(\sqrt{-\gga}\,\gga^{\WSa\WSb}\bigr)
 &= -\ii\,\bar{K}_{1(+)}^{\WSa}\,D^{\WSb}_{+(+)}\Theta_1 -\ii\,\bar{K}_{2(-)}^{\WSa}\,D^{\WSb}_{-(-)}\Theta_2
\nn\\
 &\quad +\frac{\ii}{16}\,\biggl[\bar{K}_{1(+)}^{\WSa}\,\bisF_5\,e^{\WSb\Loa}_{+(+)}\,\Gamma_{\Loa}\,\Theta_2 - \bar{K}_{2(-)}^{\WSa}\,\bisF_5\,e^{\WSb\Loa}_{-(-)}\,\Gamma_{\Loa}\,\Theta_1 \biggr]
\nn\\
 &\quad +(\WSa \leftrightarrow \WSb) + \cO(\Theta^3)\,,
\end{align}
where we have used relations
\begin{align}
 \bar{\kappa}_{I(\pm)}^{\WSa}\,\gamma_{\Loa\Lob}\,\theta_J =\bar{K}_{I(\pm)}^{\WSa}\,\Gamma_{\Loa\Lob}\Theta_J \,,\qquad
 \bar{\kappa}_{I(\pm)}^{\WSa}\,\hat{\gamma}_{\Loa}\,\theta_J =\frac{\ii}{8}\,\bar{K}_{I(\pm)}^{\WSa}\,\bisF_5\,\Gamma_{\Loa}\,\Theta_J\,.
\end{align}
Finally, we perform the redefinitions \eqref{eq:fermi-redef} and \eqref{eq:fermi-redef-K}. 
By using relations \eqref{eq:torsionful-spin} and \eqref{eq:YBRR0}, the variation of the worldsheet metric becomes
\begin{align}
 \frac{1}{\sqrt{-\gga}}\,\delta_\kappa\bigl(\sqrt{-\gga}\,\gga^{\WSa\WSb}\bigr)
 &= -\ii\,\bar{K}'^{\WSa}_{1(+)}\,D'^{\WSb}_{+(+)}\Theta'_1 - \ii\,\bar{K}'^{\WSa}_{2(-)}\,D'^{\WSb}_{-(-)}\Theta'_2
\nn\\
 &\quad +\frac{\ii}{16}\,\biggl[\bar{K}'^{\WSa}_{1(+)}\,\bigl(\bisF_5\,\Omega^{-1}\bigr)\,e'^{\WSb \Loa}_{(+)}\,\Gamma_{\Loa}\,\Theta'_2
 -\brTheta'_1\,e'^{\WSb\Loa}_{(-)}\,\Gamma_{\Loa}\,\bigl(\bisF_5\,\Omega^{-1}\bigr)\,K'^{\WSa}_{2(-)}\biggr]
\nn\\
 &\quad +(\WSa \leftrightarrow \WSb)+\cO(\Theta'^3)
\nn\\
 &=-\ii\,\bar{K}'^{\WSa}_{1(+)}\,D'^{\WSb}_{+(+)}\Theta'_1 - \ii\,\bar{K}'^{\WSa}_{2(-)}\,D'^{\WSb}_{-(-)}\Theta'_2
\nn\\
 &\quad +\frac{\ii}{16}\,\biggl[\bar{K}'^{\WSa}_{1(+)}\,\bisF'\,e'^{\WSb\Loa}_{(+)}\,\Gamma_{\Loa}\,\Theta'_2 - \brTheta'_1\,e'^{\WSb\Loa}_{(-)}\,\Gamma_{\Loa}\,\bisF'\,K'^{\WSa}_{2(-)}\biggr]
\nn\\
 &\quad +(\WSa \leftrightarrow \WSb)+\cO(\Theta'^3) \,.
\end{align}
In this way, we have obtained the standard $\kappa$-variation of the worldsheet metric \eqref{eq:kappaga1}.

\chapter{Generating GSE solutions with Penrose limits}
\label{sec:Penrose-limit}

In this appendix, we consider Penrose limit \cite{Penrose,Gueven:2000ru} of 
YB-deformed $\AdS5\times\rmS^5$ backgrounds and reproduce solutions of GSE 
studied in Section \ref{sec:non-geometry-Minkowski}. 
The R-R fluxes in the YB-deformed $\AdS5\times\rmS^5$ backgrounds may disappear under the Penrose limit. In that case, the resulting backgrounds become purely NS-NS solutions of GSE. 

\medskip

Penrose limit \cite{Penrose,Gueven:2000ru} is formulated for the standard supergravity. 
But, at least so far, there is no general argument on Penrose limit for the GSE case. 
Hence, it is quite non-trivial whether it can be extended to GSE or not. 
Here, we will not discuss a general theory of Penrose limit for GSE, 
but explain how to take a scaling of the extra vector $I$. 
The point here is that a YB-deformed background contains a deformations parameter 
and $I$ is proportional to it. 
Hence, there is a freedom to scale the deformation parameter 
in taking a Penrose limit. Without scaling the deformation parameter, 5D Minkowski spacetime 
is obtained as in the undeformed case. On the other hand, by taking an appropriate scaling of 
the deformation parameter, one can obtain a non-trivial solution of GSE with non-vanishing extra vector fields. 
We refer to the latter manner as {\it the modified Penrose limit}.
As a result, this modified Penrose limit may be regarded as a technique to generate solutions 
of GSE\footnote{Without any general argument, it is not ensured that the resulting background 
should satisfy the GSE. However, this point can be overcome by directly checking the GSE 
for the resulting background. As far as we have checked, it seems likely that 
this procedure works well. }. 

\section{Penrose limit of Poincar\'e $\AdS5$} 
\label{sec:Penrose-Ads5}

Let us first recall how to take a Penrose limit of the Poincar\'e metric of $\AdS5$\,. 

\medskip

The metric is given by 
\begin{align}
 \dd s^2 = \frac{r^2}{R^2}\,\bigl(-\dd t^2 + \dd \vec{x}^2\bigr) + R^2\,\frac{\dd r^2}{r^2}\,,
\end{align}
where $\vec{x}=(x^1,x^2,x^3)$\,. 

\medskip 

The first task is to determine a null geodesic. 
Here we are interested in a radial null geodesic described by 
\begin{align}
 \Bigl(\frac{\dd s}{\dd \tau}\Bigr)^2 
 = R^2\,\frac{\dot{r}^2}{r^2} + \frac{r^2}{R^2}\,(-\dot{t}^2) =0\,.
\label{null}
\end{align}
Here $\tau$ is an affine parameter and the symbol ``$\cdot$'' denotes a derivative 
in terms of $\tau$\,. 
From the energy conservation, we obtain that 
\begin{align}
 \frac{r^2}{R^2}\,\dot{t} \equiv E \quad (\mbox{constant})\,.
\end{align}
Hereafter, we will set $E=1$ by rescaling $\tau$\,. 
Then the equation \eqref{null} can be rewritten as 
\begin{align}
\dot{r}^2 =1\,. 
\end{align}
Hence, we will take a solution as 
\begin{align}
r = -\tau\,,
\end{align}
by adjusting an integration constant to be zero. 
Then $t$ can also be determined as follows: 
\begin{align}
 t = - \frac{R^2}{\tau}\,. 
\end{align}
As a result, the radial null geodesic is described as 
\begin{align}
t = \frac{R^2}{r}\,. \label{ng}
\end{align}

\medskip

Let us take a Penrose limit by employing the radial null geodesic \eqref{ng}. 
The first step is to introduce a new variable $\tilde{t}$ as a fluctuation around the null geodesic as 
\begin{align}
t = \frac{R^2}{r} - \tilde{t}\,.
\end{align}
Then, the metric of Poincar\'e $\AdS5$ is rewritten into the pp-wave form: 
\begin{align}
 \dd s^2 = -2\,\dd r\,\dd \tilde{t} - \frac{r^2}{R^2}\,\dd \tilde{t}^2 + \frac{r^2}{R^2}\,\dd \vec{x}^2\,.
\end{align}

\medskip

Next, by further transforming the coordinates as
\begin{align}
 \vec{x} = \frac{R}{r}\,\vec{y}\,, \qquad \tilde{t} = v - \frac{1}{2r}\,\vec{y}^2\,, 
\end{align}
the metric can be rewritten as 
\begin{align}
\dd s^2 = -2\,\dd r\, \dd v + \dd \vec{y}^2 + \cO\bigl(1/R^2\bigr)\,. 
\end{align}
Finally, by taking the $R \to \infty$ limit, the metric of 5D Minkowski spacetime is obtained. 

\section{Penrose limits of YB-deformed $\AdS5\times\rmS^5$}
\label{sec:Penrose-YB-deform}

Our aim here is to consider the modified Penrose limit of YB-deformed $\AdS5\times\rmS^5$ 
with classical $r$-matrices satisfying the homogeneous CYBE. 
In the following, we will focus upon two examples of non-unimodular classical $r$-matrices. 

\subsubsection*{\underline{Example 1) \ [solution of section \ref{sec:AdS-P0-D}]$\quad\overset{\text{Penrose limit}}{\longrightarrow}$\quad [solution of section \ref{sec:Minkowski-example1}]}}

The first example is a YB-deformed background associated with $r=\frac{1}{2}\,P_0\wedge D$, 
which was studied in section \ref{sec:AdS-P0-D}. 
To take a Penrose limit of the background \eqref{eq:AdS-P0-D}, let us rescale the fields as follows:
\begin{align}
\begin{alignedat}{2}
 \rmd s^2&~~\to~~ \rmd\tilde{s}^2 = R^2\,\rmd s^2\,,\qquad &
 B_2& ~~\to~~ \tilde{B}_2= R^2\,B_2\,,
\\
 F_3&~~\to~~ \tilde{F}_3= R^2\,F_3\,,&
 F_5&~~\to~~ \tilde{F}_5= R^4\,F_5\,.
\end{alignedat}
\end{align}
After performing a coordinate transformation for the radial direction,
\begin{align}
 z=\frac{R^2}{r}\,,
\end{align}
the radial null geodesic is given by 
\begin{align}
 x^0=\frac{R^2}{r}\label{pdng1}\,.
\end{align}
This expression coincides with the one \eqref{ng} even after performing the deformation. 

\medskip

As in the case of Poincar\'e $\AdS5$\,, a new variable $\tilde{t}$ is introduced 
as a fluctuation around the null geodesic \eqref{pdng1}:
\begin{align}
 x^0=\frac{R^2}{r}-\tilde{t}\,.
\end{align}
Let us perform a further coordinate transformation,
\begin{align}
 \rho=\frac{R}{r}\, p\,,\qquad
 \tilde{t}=v-\frac{p^2}{2r}\,.
\end{align}
If the $R\to \infty$ limit is taken naively, one can perform the usual Penrose limit, 
but it again leads to 5D Minkowski spacetime as in the case of the Poincar\'e $\AdS5$\,. 

\medskip 

It is interesting to add a modification to the usual process. 
That is to rescale the deformation parameter $\eta$ as well,
\begin{align}
 \eta=R^2\,\xi\,.
\end{align}
We refer to this modification as the modified Penrose limit. 

\medskip 

By taking the $R\to\infty$ limit and also the flat limit of the $\rmS^5$ part, 
we obtain the YB-deformed Minkowski background \eqref{flatpen} with the following identifications:
\begin{align}
 \{x^+,\,x^-,\,x,\,y,\,z,\,\eta\} \ ~~\longleftrightarrow~~ \ \{r,\, v,\, \rho\sin\theta \cos\phi,\, 
 \rho\sin\theta\sin\phi,\, z,\,\xi\} \,.
\end{align}
Remarkably, all of the R-R fluxes have vanished under this modified Penrose limit.

\subsubsection*{\underline{Example 2) \ [solution of section \ref{sec:AdS-Drinfeld--Jimbo}]$\quad\overset{\text{Penrose limit}}{\longrightarrow}$\quad [solution of section \ref{sec:Minkowski-example2}]}}

Let us next consider another YB-deformed background studied 
in Sec.\,\ref{sec:AdS-Drinfeld--Jimbo}.
To consider a Penrose limit of the background \eqref{eq:HvT}, 
let us rescale the fields as follows: 
\begin{align}
\begin{alignedat}{2}
 \rmd s^2& ~~\to~~ \rmd\tilde{s}= R^2\,\rmd s^2\,,\qquad&
 B_2& ~~\to~~ \tilde{B}_2= R^2\,B_2\,, 
\\
 F_3& ~~\to~~ \tilde{F}_3= R^2\,F_3\,,\qquad&
 F_5& ~~\to~~ \tilde{F}_5= R^4\,F_5\,.
\end{alignedat}
\end{align}
After performing a coordinate transformation,
\begin{align}
 z=\frac{R^2}{r}\,,
\end{align}
we obtain a radial null geodesic, which again takes the form,
\begin{align}
 x^0=\frac{R^2}{r}
\label{pdng}\,.
\end{align}

\medskip

Let us next introduce a new variable $\tilde{t}$ as a fluctuation around the null geodesic \eqref{pdng}:
\begin{align}
x^0=\frac{R^2}{r}-\tilde{t}\,.
\end{align}
Then, we perform a further coordinate transformation 
\begin{align}
 x^1=\frac{R}{r}\,z\,,\qquad 
 \rho=\frac{R}{r}\,p\,,\qquad
 \tilde{t}=v-\frac{p^2+z^2}{2r} \,.
\end{align}
As in the previous case, the deformation parameter is rescaled as
\begin{align}
 \eta= R^2\,\xi\,.
\end{align}
After taking the $R\to \infty$ limit, the resulting background is given by \eqref{eq:penHvT} 
with the following replacements:
\begin{align}
 \{r,\,v,\,p,\,\theta,\,\xi\}\to \{x^+,\,x^-,\,\sqrt{x^2+y^2}\,,\arctan(y/x)\,,\eta\}\,.
\end{align}
Note that all of the R-R fluxes have vanished again as in the previous example \eqref{flatpen}.


\begin{thebibliography}{99}




\bibitem{Maldacena:1997re}
  J.~M.~Maldacena,
  ``The Large N limit of superconformal field theories and supergravity,''
  Int.\ J.\ Theor.\ Phys.\  {\bf 38} (1999) 1113
   [Adv.\ Theor.\ Math.\ Phys.\  {\bf 2} (1998) 231]
  [hep-th/9711200].

\bibitem{Gubser:1998bc}
  S.~S.~Gubser, I.~R.~Klebanov and A.~M.~Polyakov,
  ``Gauge theory correlators from noncritical string theory,''
  Phys.\ Lett.\ B {\bf 428} (1998) 105
  [hep-th/9802109].

\bibitem{Witten:1998qj}
  E.~Witten,
  ``Anti-de Sitter space and holography,''
  Adv.\ Theor.\ Math.\ Phys.\  {\bf 2} (1998) 253
  [hep-th/9802150].

\bibitem{tHooft:1973alw}
  G.~'t Hooft,
  ``A Planar Diagram Theory for Strong Interactions,''
  Nucl.\ Phys.\ B {\bf 72} (1974) 461.


\bibitem{Minahan:2002ve}
  J.~A.~Minahan and K.~Zarembo,
  ``The Bethe ansatz for $\mathcal{N}=4$ superYang-Mills,''
  JHEP {\bf 0303} (2003) 013
  [hep-th/0212208].


\bibitem{Metsaev:1998it} 
  R.~R.~Metsaev and A.~A.~Tseytlin,
  ``Type IIB superstring action in AdS$_{5} \times$ S$^{5}$ background,''
  Nucl.\ Phys.\ B {\bf 533}, 109 (1998)
  [hep-th/9805028].


\bibitem{Bena:2003wd}
  I.~Bena, J.~Polchinski and R.~Roiban,
  ``Hidden symmetries of the AdS$_{5} \times$ S$^{5}$ superstring,''
  Phys.\ Rev.\ D {\bf 69} (2004) 046002
  [hep-th/0305116].

\bibitem{Gromov:2009bc}
  N.~Gromov, V.~Kazakov, A.~Kozak and P.~Vieira,
  ``Exact Spectrum of Anomalous Dimensions of Planar N = 4 Supersymmetric Yang-Mills Theory: TBA and excited states,''
  Lett.\ Math.\ Phys.\  {\bf 91} (2010) 265
  [arXiv:0902.4458 [hep-th]].

\bibitem{Bombardelli:2009ns}
  D.~Bombardelli, D.~Fioravanti and R.~Tateo,
  ``Thermodynamic Bethe Ansatz for planar AdS/CFT: A Proposal,''
  J.\ Phys.\ A {\bf 42} (2009) 375401
  [arXiv:0902.3930 [hep-th]].

\bibitem{Arutyunov:2009ur}
  G.~Arutyunov and S.~Frolov,
  ``Thermodynamic Bethe Ansatz for the AdS(5) x S(5) Mirror Model,''
  JHEP {\bf 0905} (2009) 068
  [arXiv:0903.0141 [hep-th]].


\bibitem{Gromov:2009tv}
  N.~Gromov, V.~Kazakov and P.~Vieira,
  ``Exact Spectrum of Anomalous Dimensions of Planar N=4 Supersymmetric Yang-Mills Theory,''
  Phys.\ Rev.\ Lett.\  {\bf 103} (2009) 131601
  [arXiv:0901.3753 [hep-th]].


\bibitem{Cavaglia:2010nm}
  A.~Cavaglia, D.~Fioravanti and R.~Tateo,
  ``Extended Y-system for the $AdS_5/CFT_4$ correspondence,''
  Nucl.\ Phys.\ B {\bf 843} (2011) 302
  [arXiv:1005.3016 [hep-th]].


\bibitem{Gromov:2013pga}
  N.~Gromov, V.~Kazakov, S.~Leurent and D.~Volin,
  ``Quantum Spectral Curve for Planar $\mathcal{N} = 4$ Super-Yang-Mills Theory,''
  Phys.\ Rev.\ Lett.\  {\bf 112} (2014) no.1,  011602
  [arXiv:1305.1939 [hep-th]].


\bibitem{Gromov:2014caa}
  N.~Gromov, V.~Kazakov, S.~Leurent and D.~Volin,
  ``Quantum spectral curve for arbitrary state/operator in AdS$_{5}$/CFT$_{4}$,''
  JHEP {\bf 1509} (2015) 187
  [arXiv:1405.4857 [hep-th]].

%

\bibitem{Basso:2015zoa}
  B.~Basso, S.~Komatsu and P.~Vieira,
  ``Structure Constants and Integrable Bootstrap in Planar N=4 SYM Theory,''
  arXiv:1505.06745 [hep-th].

\bibitem{Eden:2016xvg}
  B.~Eden and A.~Sfondrini,
  ``Tessellating cushions: four-point functions in $\mathcal{N} $ = 4 SYM,''
  JHEP {\bf 1710} (2017) 098
  [arXiv:1611.05436 [hep-th]].

\bibitem{Fleury:2016ykk}
  T.~Fleury and S.~Komatsu,
  ``Hexagonalization of Correlation Functions,''
  JHEP {\bf 1701} (2017) 130
  [arXiv:1611.05577 [hep-th]].


\bibitem{Beisert:2010jr}
   N.~Beisert  et al., ``Review of AdS/CFT Integrability: An Overview,''
   Lett.\ Math.\ Phys.{\bf 99} (2012)  3--32,
   arXiv:1012.3982 [hep-th].







\bibitem{Sfetsos:2013wia}
  K.~Sfetsos,
  ``Integrable interpolations: From exact CFTs to non-Abelian T-duals,''
  Nucl.\ Phys.\ B {\bf 880} (2014) 225
  [arXiv:1312.4560 [hep-th]].

\bibitem{Demulder:2015lva}
  S.~Demulder, K.~Sfetsos and D.~C.~Thompson,
  ``Integrable $\lambda$-deformations: Squashing Coset CFTs and $AdS_5\times S^5$,''
  JHEP {\bf 1507} (2015) 019
  [arXiv:1504.02781 [hep-th]].

\bibitem{Hollowood:2014qma}
  T.~J.~Hollowood, J.~L.~Miramontes and D.~M.~Schmidtt,
  ``An Integrable Deformation of the $AdS_5 \times S^5$ Superstring,''
  J.\ Phys.\ A {\bf 47} (2014) no.49,  495402
  [arXiv:1409.1538 [hep-th]].

\bibitem{Hollowood:2014rla}
  T.~J.~Hollowood, J.~L.~Miramontes and D.~M.~Schmidtt,
  ``Integrable Deformations of Strings on Symmetric Spaces,''
  JHEP {\bf 1411} (2014) 009
  [arXiv:1407.2840 [hep-th]].

\bibitem{Borsato:2016ose} 
  R.~Borsato and L.~Wulff,
  ``Target space supergeometry of $\eta$ and $\lambda$-deformed strings,''
  JHEP {\bf 1610}, 045 (2016)
  [arXiv:1608.03570 [hep-th]].




\bibitem{Vicedo:2015pna}
  B.~Vicedo,
  ``Deformed integrable $\sigma$-models, classical R-matrices and classical exchange algebra on Drinfel'd doubles,''
  J.\ Phys.\ A {\bf 48} (2015) no.35,  355203
  [arXiv:1504.06303 [hep-th]].


\bibitem{Hoare:2015gda}
  B.~Hoare and A.~A.~Tseytlin,
  ``On integrable deformations of superstring sigma models related to $AdS_n \times S^n$ supercosets,''
  Nucl.\ Phys.\ B {\bf 897} (2015) 448
  [arXiv:1504.07213 [hep-th]].


\bibitem{Sfetsos:2015nya}
  K.~Sfetsos, K.~Siampos and D.~C.~Thompson,
  ``Generalised integrable $\lambda$- and $\eta$-deformations and their relation,''
  Nucl.\ Phys.\ B {\bf 899} (2015) 489
  [arXiv:1506.05784 [hep-th]].

\bibitem{Klimcik:2015gba}
  C.~Klimcik,
  ``$\eta$ and $\lambda$ deformations as $\mathcal{E}$-models,''
  Nucl.\ Phys.\ B {\bf 900} (2015) 259
  [arXiv:1508.05832 [hep-th]].


\bibitem{Klimcik:1995ux}
  C.~Klimcik and P.~Severa,
  ``Dual nonAbelian duality and the Drinfeld double,''
  Phys.\ Lett.\ B {\bf 351} (1995) 455
  [hep-th/9502122].

\bibitem{Klimcik:1995jn}
  C.~Klimcik,
  ``Poisson-Lie T duality,''
  Nucl.\ Phys.\ Proc.\ Suppl.\  {\bf 46} (1996) 116
  [hep-th/9509095].

\bibitem{Klimcik:1995dy}
  C.~Klimcik and P.~Severa,
  ``Poisson-Lie T duality and loop groups of Drinfeld doubles,''
  Phys.\ Lett.\ B {\bf 372} (1996) 65
  [hep-th/9512040].




\bibitem{Hassler:2017yza}
  F.~Hassler,
  ``Poisson-Lie T-Duality in Double Field Theory,''
  arXiv:1707.08624 [hep-th].


\bibitem{Jurco:2017gii}
  B.~Jurco and J.~Vysoky,
  ``Poisson-Lie T-duality of string effective actions: A new approach to the dilaton puzzle,''
  J.\ Geom.\ Phys.\  {\bf 130} (2018) 1
  [arXiv:1708.04079 [hep-th]].

\bibitem{Hoare:2017ukq}
  B.~Hoare and F.~K.~Seibold,
  ``Poisson-Lie duals of the $\eta$ deformed symmetric space sigma model,''
  JHEP {\bf 1711} (2017) 014
  [arXiv:1709.01448 [hep-th]].

\bibitem{Hoare:2018ebg}
  B.~Hoare and F.~K.~Seibold,
  ``Poisson-Lie duals of the $\eta$-deformed $\mathrm{AdS}_2 \times \mathrm{S}^2 \times \mathrm{T}^6$ superstring,''
  JHEP {\bf 1808} (2018) 107
  [arXiv:1807.04608 [hep-th]].

\bibitem{Severa:2018pag}
  P.~Ševera and F.~Valach,
  ``Courant algebroids, Poisson-Lie T-duality, and type II supergravities,''
  arXiv:1810.07763 [math.DG].


\bibitem{Demulder:2018lmj}
  S.~Demulder, F.~Hassler and D.~C.~Thompson,
  ``Doubled aspects of generalised dualities and integrable deformations,''
  JHEP {\bf 1902} (2019) 189
  [arXiv:1810.11446 [hep-th]].


\bibitem{Sakatani:2019jgu}
  Y.~Sakatani,
  ``Type II DFT solutions from Poisson-Lie T-duality/plurality,''
  arXiv:1903.12175 [hep-th].






\bibitem{Klimcik:2002zj} 
  C.~Klimcik,
  ``Yang-Baxter sigma models and dS/AdS T duality,''
  JHEP {\bf 0212}, 051 (2002)
  [hep-th/0210095].


\bibitem{Klimcik:2008eq} 
  C.~Klimcik,
  ``On integrability of the Yang-Baxter sigma-model,''
  J.\ Math.\ Phys.\  {\bf 50}, 043508 (2009)
  [arXiv:0802.3518 [hep-th]].


\bibitem{Drinfeld:1985rx}
   V.~G. Drinfeld, ``Hopf algebras and the quantum Yang-Baxter equation,''
   Sov.\ Math.\ Dokl. {\bf 32} (1985)  254--258.
   [Dokl. Akad. Nauk Ser. Fiz.283,1060(1985)].


\bibitem{Jimbo:1985zk}
   M.~Jimbo, ``A $q$ difference analog of $U(g)$ and the Yang-Baxter equation,''
   Lett.\ Math.\ Phys. {\bf 10} (1985)  63--69.


\bibitem{Delduc:2013fga} 
  F.~Delduc, M.~Magro and B.~Vicedo,
  ``On classical $q$-deformations of integrable sigma-models,''
  JHEP {\bf 1311}, 192 (2013)
  [arXiv:1308.3581 [hep-th]].


\bibitem{Delduc:2013qra} 
  F.~Delduc, M.~Magro and B.~Vicedo,
  ``An integrable deformation of the AdS$_{5} \times$ S$^{5}$ superstring action,''
  Phys.\ Rev.\ Lett.\  {\bf 112}, no. 5, 051601 (2014)
  [arXiv:1309.5850 [hep-th]].


\bibitem{Delduc:2014kha} 
  F.~Delduc, M.~Magro and B.~Vicedo,
  ``Derivation of the action and symmetries of the $q$-deformed $AdS_{5} \times S^{5}$ superstring,''
  JHEP {\bf 1410}, 132 (2014)
  [arXiv:1406.6286 [hep-th]].


\bibitem{Arutyunov:2013ega}
  G.~Arutyunov, R.~Borsato and S.~Frolov,
  ``S-matrix for strings on $\eta$-deformed AdS5 x S5,''
  JHEP {\bf 1404} (2014) 002
  [arXiv:1312.3542 [hep-th]].


\bibitem{Arutyunov:2015qva} 
  G.~Arutyunov, R.~Borsato and S.~Frolov,
  ``Puzzles of $\eta$-deformed AdS$_5 \times$ S$^5$,''
  JHEP {\bf 1512}, 049 (2015)
  [arXiv:1507.04239 [hep-th]].

\bibitem{Hoare:2018ngg}
  B.~Hoare and F.~K.~Seibold,
  ``Supergravity backgrounds of the $\eta$-deformed AdS$_2 \times S^2 \times T^6 $ and AdS$_5 \times S^5$ superstrings,''
  JHEP {\bf 1901} (2019) 125
  [arXiv:1811.07841 [hep-th]].

\bibitem{Arutyunov:2015mqj} 
  G.~Arutyunov, S.~Frolov, B.~Hoare, R.~Roiban and A.~A.~Tseytlin,
  ``Scale invariance of the $\eta$-deformed $AdS_5\times S^5$ superstring, T-duality and modified type II equations,''
  Nucl.\ Phys.\ B {\bf 903}, 262 (2016)
  [arXiv:1511.05795 [hep-th]].


\bibitem{Wulff:2016tju} 
  A.~A.~Tseytlin and L.~Wulff,
  ``Kappa-symmetry of superstring sigma model and generalized $10d$ supergravity equations,''
  JHEP {\bf 1606}, 174 (2016)
  [arXiv:1605.04884 [hep-th]].



\bibitem{Matsumoto:2015jja} 
  T.~Matsumoto and K.~Yoshida,
  ``Yang-Baxter sigma models based on the CYBE,''
  Nucl.\ Phys.\ B {\bf 893}, 287 (2015)
  [arXiv:1501.03665 [hep-th]].


\bibitem{Kawaguchi:2014qwa} 
  I.~Kawaguchi, T.~Matsumoto and K.~Yoshida,
  ``Jordanian deformations of the AdS$_{5} \times$ S$^{5}$ superstring,''
  JHEP {\bf 1404}, 153 (2014)
  [arXiv:1401.4855 [hep-th]].


\bibitem{Matsumoto:2014nra} 
  T.~Matsumoto and K.~Yoshida,
  ``Lunin-Maldacena backgrounds from the classical Yang-Baxter equation - towards the gravity/CYBE correspondence,''
  JHEP {\bf 1406}, 135 (2014)
  [arXiv:1404.1838 [hep-th]].


\bibitem{Matsumoto:2014gwa} 
  T.~Matsumoto and K.~Yoshida,
  ``Integrability of classical strings dual for noncommutative gauge theories,''
  JHEP {\bf 1406}, 163 (2014)
  [arXiv:1404.3657 [hep-th]].


\bibitem{Matsumoto:2015uja} 
  T.~Matsumoto and K.~Yoshida,
  ``Schr\"odinger geometries arising from Yang-Baxter deformations,''
  JHEP {\bf 1504}, 180 (2015)
  [arXiv:1502.00740 [hep-th]].



\bibitem{Kyono:2016jqy} 
  H.~Kyono and K.~Yoshida,
  ``Supercoset construction of Yang-Baxter deformed AdS$_5\times$S$^5$ backgrounds,''
  PTEP {\bf 2016}, no. 8, 083B03 (2016)
  [arXiv:1605.02519 [hep-th]].


\bibitem{Lunin:2005jy} 
  O.~Lunin and J.~M.~Maldacena,
  ``Deforming field theories with $U(1) \times U(1)$ global symmetry and their gravity duals,''
  JHEP {\bf 0505}, 033 (2005)
  [hep-th/0502086].


\bibitem{Frolov:2005dj} 
  S.~Frolov,
  ``Lax pair for strings in Lunin-Maldacena background,''
  JHEP {\bf 0505}, 069 (2005)
  [hep-th/0503201].


\bibitem{Hashimoto:1999ut} 
  A.~Hashimoto and N.~Itzhaki,
  ``Noncommutative Yang-Mills and the AdS / CFT correspondence,''
  Phys.\ Lett.\ B {\bf 465}, 142 (1999)
  [hep-th/9907166].


\bibitem{Maldacena:1999mh} 
  J.~M.~Maldacena and J.~G.~Russo,
  ``Large N limit of noncommutative gauge theories,''
  JHEP {\bf 9909}, 025 (1999)
  [hep-th/9908134].


\bibitem{Herzog:2008wg} 
  C.~P.~Herzog, M.~Rangamani and S.~F.~Ross,
  ``Heating up Galilean holography,''
  JHEP {\bf 0811}, 080 (2008)
  [arXiv:0807.1099 [hep-th]].



\bibitem{Maldacena:2008wh} 
  J.~Maldacena, D.~Martelli and Y.~Tachikawa,
  ``Comments on string theory backgrounds with non-relativistic conformal symmetry,''
  JHEP {\bf 0810}, 072 (2008)
  [arXiv:0807.1100 [hep-th]].



\bibitem{Adams:2008wt} 
  A.~Adams, K.~Balasubramanian and J.~McGreevy,
  ``Hot Spacetimes for Cold Atoms,''
  JHEP {\bf 0811}, 059 (2008)
  [arXiv:0807.1111 [hep-th]].


\bibitem{Osten:2016dvf} 
  D.~Osten and S.~J.~van Tongeren,
  ``Abelian Yang-Baxter deformations and TsT transformations,''
  Nucl.\ Phys.\ B {\bf 915}, 184 (2017)
  [arXiv:1608.08504 [hep-th]].


\bibitem{Orlando:2016qqu} 
  D.~Orlando, S.~Reffert, J.~Sakamoto and K.~Yoshida,
  ``Generalized type IIB supergravity equations and non-Abelian classical $r$-matrices,''
  J.\ Phys.\ A {\bf 49}, no. 44, 445403 (2016)
  [arXiv:1607.00795 [hep-th]].


\bibitem{Hoare:2016wsk} 
  B.~Hoare and A.~A.~Tseytlin,
  ``Homogeneous Yang-Baxter deformations as non-abelian duals of the AdS$_5$ sigma-model,''
  J.\ Phys.\ A {\bf 49}, no. 49, 494001 (2016)
  [arXiv:1609.02550 [hep-th]].


\bibitem{Borsato:2016pas} 
  R.~Borsato and L.~Wulff,
  ``Integrable Deformations of T-Dual $\sigma$ Models,''
  Phys.\ Rev.\ Lett.\  {\bf 117}, no. 25, 251602 (2016)
  [arXiv:1609.09834 [hep-th]].


\bibitem{Borsato:2017qsx} 
  R.~Borsato and L.~Wulff,
  ``On non-abelian T-duality and deformations of supercoset string sigma-models,''
  JHEP {\bf 1710}, 024 (2017)
  [arXiv:1706.10169 [hep-th]].


\bibitem{Sakamoto:2017cpu} 
  J.~Sakamoto, Y.~Sakatani and K.~Yoshida,
  ``Homogeneous Yang-Baxter deformations as generalized diffeomorphisms,''
  J.\ Phys.\ A {\bf 50}, no. 41, 415401 (2017)
  [arXiv:1705.07116 [hep-th]].


\bibitem{Sakamoto:2018krs}
  J.~Sakamoto and Y.~Sakatani,
  ``Local $\beta$-deformations and Yang-Baxter sigma model,''
  JHEP {\bf 1806} (2018) 147
  [arXiv:1803.05903 [hep-th]].

\bibitem{Araujo:2018rho}
  T.~Araujo, E.~\'O Colg\'ain, Y.~Sakatani, M.~M.~Sheikh-Jabbari and H.~Yavartanoo,
  ``Holographic integration of $T \bar{T}$ \& $J \bar{T}$ via $O(d,d)$,''
  arXiv:1811.03050 [hep-th].


\bibitem{Borsato:2018spz}
  R.~Borsato and L.~Wulff,
  ``Marginal deformations of WZW models and the classical Yang-Baxter equation,''
  arXiv:1812.07287 [hep-th].


\bibitem{Hoare:2016wca} 
  B.~Hoare and D.~C.~Thompson,
  ``Marginal and non-commutative deformations via non-abelian T-duality,''
  JHEP {\bf 1702}, 059 (2017)
  [arXiv:1611.08020 [hep-th]].

\bibitem{Sakamoto:2016ppx}
  J.~Sakamoto and K.~Yoshida,
  ``Yang-Baxter deformations of $W_{2,4}\times T^{1,1}$ and the associated T-dual models,''
  Nucl.\ Phys.\ B {\bf 921} (2017) 805
  [arXiv:1612.08615 [hep-th]].

\bibitem{Bakhmatov:2017joy}
  I.~Bakhmatov, \"O.~Kelekci, E.~\'O.~Colg\'ain and M.~M.~Sheikh-Jabbari,
  ``Classical Yang-Baxter Equation from Supergravity,''
  Phys.\ Rev.\ D {\bf 98} (2018) no.2,  021901
  [arXiv:1710.06784 [hep-th]].


\bibitem{Bakhmatov:2018apn} 
  I.~Bakhmatov, E.~\'O.~Colg\'ain, M.~M.~Sheikh-Jabbari and H.~Yavartanoo,
  ``Yang-Baxter Deformations Beyond Coset Spaces (a slick way to do TsT),''
  JHEP {\bf 1806} (2018) 161
  [arXiv:1803.07498 [hep-th]].XIV:1803.07498;

%

\bibitem{Lust:2018jsx}
  D.~L\"ust and D.~Osten,
  ``Generalised fluxes, Yang-Baxter deformations and the O(d,d) structure of non-abelian T-duality,''
  JHEP {\bf 1805} (2018) 165
  [arXiv:1803.03971 [hep-th]].


\bibitem{Araujo:2018rbc}
  T.~Araujo, E.~Ó.~Colgáin and H.~Yavartanoo,
  ``Embedding the modified CYBE in Supergravity,''
  Eur.\ Phys.\ J.\ C {\bf 78} (2018) no.10,  854
  [arXiv:1806.02602 [hep-th]].


\bibitem{Borsato:2018idb}
  R.~Borsato and L.~Wulff,
  ``Non-abelian T-duality and Yang-Baxter deformations of Green-Schwarz strings,''
  JHEP {\bf 1808} (2018) 027
  [arXiv:1806.04083 [hep-th]].


\bibitem{Bakhmatov:2018bvp}
  I.~Bakhmatov and E.~T.~Musaev,
  ``Classical Yang-Baxter equation from $\beta$-supergravity,''
  JHEP {\bf 1901} (2019) 140
  [arXiv:1811.09056 [hep-th]].


\bibitem{Orlando:2018kms}
  D.~Orlando, S.~Reffert, Y.~Sekiguchi and K.~Yoshida,
  ``Killing spinors from classical $r$-matrices,''
  J.\ Phys.\ A {\bf 51} (2018) no.39,  395401
  [arXiv:1805.00948 [hep-th]].


\bibitem{Orlando:2018qaq}
  D.~Orlando, S.~Reffert, Y.~Sekiguchi and K.~Yoshida,
  ``SUSY and the bi-vector,''
  arXiv:1811.11764 [hep-th].


\bibitem{delaOssa:1992vci} 
  X.~C.~de la Ossa and F.~Quevedo,
  ``Duality symmetries from nonAbelian isometries in string theory,''
  Nucl.\ Phys.\ B {\bf 403}, 377 (1993)
  [hep-th/9210021].


\bibitem{Giveon:1993ai} 
  A.~Giveon and M.~Rocek,
  ``On nonAbelian duality,''
  Nucl.\ Phys.\ B {\bf 421}, 173 (1994)
  [hep-th/9308154].



\bibitem{Alvarez:1994np} 
  E.~Alvarez, L.~Alvarez-Gaume and Y.~Lozano,
  ``On nonAbelian duality,''
  Nucl.\ Phys.\ B {\bf 424}, 155 (1994)
  [hep-th/9403155].



\bibitem{Sfetsos:2010uq} 
  K.~Sfetsos and D.~C.~Thompson,
  ``On non-abelian T-dual geometries with Ramond fluxes,''
  Nucl.\ Phys.\ B {\bf 846}, 21 (2011)
  [arXiv:1012.1320 [hep-th]].



\bibitem{Lozano:2011kb} 
  Y.~Lozano, E.~O Colgain, K.~Sfetsos and D.~C.~Thompson,
  ``Non-abelian T-duality, Ramond Fields and Coset Geometries,''
  JHEP {\bf 1106}, 106 (2011)
  [arXiv:1104.5196 [hep-th]].


\bibitem{Matsumoto:2015ypa}
  T.~Matsumoto, D.~Orlando, S.~Reffert, J.~Sakamoto and K.~Yoshida,
  ``Yang-Baxter deformations of Minkowski spacetime,''
  JHEP {\bf 1510} (2015) 185
  [arXiv:1505.04553 [hep-th]].


\bibitem{Borowiec:2015wua}
  A.~Borowiec, H.~Kyono, J.~Lukierski, J.~Sakamoto and K.~Yoshida,
  ``Yang-Baxter sigma models and Lax pairs arising from $\kappa$-Poincaré $r$-matrices,''
  JHEP {\bf 1604} (2016) 079
  [arXiv:1510.03083 [hep-th]].


\bibitem{Pachol:2015mfa}
  A.~Pachoł and S.~J.~van 
,
  ``Quantum deformations of the flat space superstring,''
  Phys.\ Rev.\ D {\bf 93} (2016) 026008
  [arXiv:1510.02389 [hep-th]].


\bibitem{Kyono:2015zeu}
  H.~Kyono, J.~Sakamoto and K.~Yoshida,
  ``Lax pairs for deformed Minkowski spacetimes,''
  JHEP {\bf 1601} (2016) 143
  [arXiv:1512.00208 [hep-th]].


\bibitem{Fernandez-Melgarejo:2017oyu} 
  J.~J.~Fernandez-Melgarejo, J.~Sakamoto, Y.~Sakatani and K.~Yoshida,
  ``T-folds from Yang-Baxter deformations,''
  JHEP {\bf 1712}, 108 (2017)
  [arXiv:1710.06849 [hep-th]].


\bibitem{Crichigno:2014ipa}
  P.~M.~Crichigno, T.~Matsumoto and K.~Yoshida,
  ``Deformations of $T^{1,1}$ as Yang-Baxter sigma models,''
  JHEP {\bf 1412} (2014) 085
  [arXiv:1406.2249 [hep-th]].


\bibitem{Sakamoto:2016ppx}
  J.~Sakamoto and K.~Yoshida,
  ``Yang-Baxter deformations of $W_{2,4}\times T^{1,1}$ and the associated T-dual models,''
  Nucl.\ Phys.\ B {\bf 921} (2017) 805
  [arXiv:1612.08615 [hep-th]].


\bibitem{Negron:2018btz}
  R.~Negrón and V.~O.~Rivelles,
  ``Yang-Baxter deformations of the $AdS_4\times\mathbb{CP}^3$ superstring sigma model,''
  JHEP {\bf 1811} (2018) 043
  [arXiv:1809.01174 [hep-th]].


\bibitem{Jackiw:1984je}
  R.~Jackiw,
  ``Lower Dimensional Gravity,''
  Nucl.\ Phys.\ B {\bf 252} (1985) 343.

\bibitem{Teitelboim:1983ux}
  C.~Teitelboim,
  ``Gravitation and Hamiltonian Structure in Two Space-Time Dimensions,''
  Phys.\ Lett.\  {\bf 126B} (1983) 41.


\bibitem{Kyono:2017jtc}
  H.~Kyono, S.~Okumura and K.~Yoshida,
  ``Deformations of the Almheiri-Polchinski model,''
  JHEP {\bf 1703} (2017) 173
  [arXiv:1701.06340 [hep-th]].


\bibitem{Kyono:2017pxs}
  H.~Kyono, S.~Okumura and K.~Yoshida,
  ``Comments on 2D dilaton gravity system with a hyperbolic dilaton potential,''
  Nucl.\ Phys.\ B {\bf 923} (2017) 126
  [arXiv:1704.07410 [hep-th]].


\bibitem{Okumura:2018xbh}
  S.~Okumura and K.~Yoshida,
  ``Weyl transformation and regular solutions in a deformed Jackiw–Teitelboim model,''
  Nucl.\ Phys.\ B {\bf 933} (2018) 234
  [arXiv:1801.10537 [hep-th]].


\bibitem{Roychowdhury:2018clp}
  D.~Roychowdhury,
  ``$q$ SYK models from deformed Almheiri-Polchinski gravity,''
  arXiv:1810.09404 [hep-th].


\bibitem{Lala:2018yib}
  A.~Lala and D.~Roychowdhury,
  ``SYK/AdS duality with Yang-Baxter deformations,''
  JHEP {\bf 1812} (2018) 073
  [arXiv:1808.08380 [hep-th]].



\bibitem{Howe:1983sra}
  P.~S.~Howe and P.~C.~West,
  ``The Complete N=2, D=10 Supergravity,''
  Nucl.\ Phys.\ B {\bf 238} (1984) 181.


\bibitem{Grisaru:1985fv} 
  M.~T.~Grisaru, P.~S.~Howe, L.~Mezincescu, B.~Nilsson and P.~K.~Townsend,
  ``N=2 Superstrings in a Supergravity Background,''
  Phys.\ Lett.\  {\bf 162B}, 116 (1985).


\bibitem{Sakamoto:2017wor} 
  J.~Sakamoto, Y.~Sakatani and K.~Yoshida,
  ``Weyl invariance for generalized supergravity backgrounds from the doubled formalism,''
  PTEP {\bf 2017}, no. 5, 053B07 (2017)
  [arXiv:1703.09213 [hep-th]].


\bibitem{Fradkin:1984pq} 
  E.~S.~Fradkin and A.~A.~Tseytlin,
  ``Effective Field Theory from Quantized Strings,''
  Phys.\ Lett.\  {\bf 158B}, 316 (1985).

\bibitem{Fernandez-Melgarejo:2018wpg}
  J.~J.~Fern\'andez-Melgarejo, J.~Sakamoto, Y.~Sakatani and K.~Yoshida,
  ``Weyl Invariance of String Theories in Generalized Supergravity Backgrounds,''
  Phys.\ Rev.\ Lett.\  {\bf 122} (2019) no.11,  111602
  [arXiv:1811.10600 [hep-th]].

\bibitem{Muck:2019pwj}
  W.~M\"ack,
  ``Generalized Supergravity Equations and Generalized Fradkin-Tseytlin Counterterm,''
  arXiv:1904.06126 [hep-th].

%
%








\bibitem{Benitez:2018xnh}
  H.~A.~Benítez and V.~O.~Rivelles,
  ``Yang-Baxter deformations of the $AdS_{5}\times S^{5}$ pure spinor superstring,''
  JHEP {\bf 1902} (2019) 056
  [arXiv:1807.10432 [hep-th]].

\bibitem{Cvetic:1999zs} 
  M.~Cvetic, H.~Lu, C.~N.~Pope and K.~S.~Stelle,
  ``$T$ duality in the Green-Schwarz formalism, and the massless / massive IIA duality map,''
  Nucl.\ Phys.\ B {\bf 573}, 149 (2000)
  [hep-th/9907202].


\bibitem{vanTongeren:2019dlq}
  S.~J.~van Tongeren,
  ``Unimodular jordanian deformations of integrable superstrings,''
  arXiv:1904.08892 [hep-th].


\bibitem{Stolin1}
  A.~Stolin,
  ``On rational solutions of Yang-Baxter equation for $\mathfrak{sl}(n)$,''
  Math.\ Scand.\ {\bf 69}, 57 (1991).

\bibitem{Stolin2}
  A. Stolin,
  ``Rational solutions of the classical Yang-Baxter equation and quasi Frobenius Lie algebras,''
  Journal of Pure and Applied Algebra\ {\bf 137}, 285 (1999),


\bibitem{Abelian-compact}  
A.~Lichnerowicz and A.~Medina, 
 ``On Lie groups with left-invariant symplectic or K\''ahlerian structures'',
 Letters in Mathematical Physics {\bf 16}, 225 (1988),


\bibitem{vanTongeren:2016eeb}
  S.~J.~van Tongeren,
  ``Almost abelian twists and AdS/CFT,''
  Phys.\ Lett.\ B {\bf 765} (2017) 344
  [arXiv:1610.05677 [hep-th]].


\bibitem{Tolstoy2004}
  V.~N.~Tolstoy, ``Chains of extended Jordanian twists for Lie superalgebras,''
  [math/0402433].


\bibitem{Sakatani:2016fvh} 
  Y.~Sakatani, S.~Uehara and K.~Yoshida,
  ``Generalized gravity from modified DFT,''
  JHEP {\bf 1704}, 123 (2017)
  [arXiv:1611.05856 [hep-th]].


\bibitem{Baguet:2016prz} 
  A.~Baguet, M.~Magro and H.~Samtleben,
  ``Generalized IIB supergravity from exceptional field theory,''
  JHEP {\bf 1703}, 100 (2017)
  [arXiv:1612.07210 [hep-th]].


\bibitem{Leigh:1995ep}
  R.~G.~Leigh and M.~J.~Strassler,
  ``Exactly marginal operators and duality in four-dimensional N=1 supersymmetric gauge theory,''
  Nucl.\ Phys.\ B {\bf 447} (1995) 95
  [hep-th/9503121].

\bibitem{Guica:2017mtd}
  M.~Guica, F.~Levkovich-Maslyuk and K.~Zarembo,
  ``Integrability in dipole-deformed $\boldsymbol{\mathcal{N}=4}$ super Yang–Mills,''
  J.\ Phys.\ A {\bf 50} (2017) no.39,  394001
  [arXiv:1706.07957 [hep-th]].


\bibitem{Hoare:2015wia}
  B.~Hoare and A.~A.~Tseytlin,
  ``Type IIB supergravity solution for the T-dual of the $\eta$-deformed AdS$_{5} \times$ S$^{5}$ superstring,''
  JHEP {\bf 1510} (2015) 060
  [arXiv:1508.01150 [hep-th]].


\bibitem{vanTongeren:2015uha} 
  S.~J.~van Tongeren,
  ``Yang-Baxter deformations, AdS/CFT, and twist-noncommutative gauge theory,''
  Nucl.\ Phys.\ B {\bf 904}, 148 (2016)
  [arXiv:1506.01023 [hep-th]].

\bibitem{Hoare:2016hwh}
  B.~Hoare and S.~J.~van Tongeren,
  ``On jordanian deformations of AdS$_5$ and supergravity,''
  J.\ Phys.\ A {\bf 49} (2016) no.43,  434006
  [arXiv:1605.03554 [hep-th]].


\bibitem{Alday:2005ww}
  L.~F.~Alday, G.~Arutyunov and S.~Frolov,
  ``Green-Schwarz strings in TsT-transformed backgrounds,''
  JHEP {\bf 0606} (2006) 018
  [hep-th/0512253].


\bibitem{Kameyama:2015ufa}
  T.~Kameyama, H.~Kyono, J.~Sakamoto and K.~Yoshida,
  ``Lax pairs on Yang-Baxter deformed backgrounds,''
  JHEP {\bf 1511} (2015) 043
  [arXiv:1509.00173 [hep-th]].




\bibitem{Hubeny:2005qu}
  V.~E.~Hubeny, M.~Rangamani and S.~F.~Ross,
  ``Causal structures and holography,''
  JHEP {\bf 0507} (2005) 037
  [hep-th/0504034].


\bibitem{Kawaguchi:2011wt}
  I.~Kawaguchi and K.~Yoshida,
  ``Classical integrability of Schrodinger sigma models and $q$-deformed Poincare symmetry,''
  JHEP {\bf 1111} (2011) 094
  [arXiv:1109.0872 [hep-th]].


\bibitem{Kawaguchi:2012ug}
  I.~Kawaguchi and K.~Yoshida,
  ``Exotic symmetry and monodromy equivalence in Schrodinger sigma models,''
  JHEP {\bf 1302} (2013) 024
  [arXiv:1209.4147 [hep-th]].


\bibitem{Kawaguchi:2013lba}
  I.~Kawaguchi, T.~Matsumoto and K.~Yoshida,
  ``Schroedinger sigma models and Jordanian twists,''
  JHEP {\bf 1308} (2013) 013
  [arXiv:1305.6556 [hep-th]].



\bibitem{vanTongeren:2015soa} 
  S.~J.~van Tongeren,
  ``On classical Yang-Baxter based deformations of the AdS$_{5} \times$ S$^{5}$ superstring,''
  JHEP {\bf 1506}, 048 (2015)
  [arXiv:1504.05516 [hep-th]].



\bibitem{Siegel:1993xq} 
  W.~Siegel,
  ``Two vierbein formalism for string inspired axionic gravity,''
  Phys.\ Rev.\ D {\bf 47}, 5453 (1993)
  [hep-th/9302036].

\bibitem{Jeon:2011cn} 
  I.~Jeon, K.~Lee and J.~H.~Park,
  ``Stringy differential geometry, beyond Riemann,''
  Phys.\ Rev.\ D {\bf 84}, 044022 (2011)
  [arXiv:1105.6294 [hep-th]].



\bibitem{Jeon:2011vx} 
  I.~Jeon, K.~Lee and J.~H.~Park,
  ``Incorporation of fermions into double field theory,''
  JHEP {\bf 1111}, 025 (2011)
  [arXiv:1109.2035 [hep-th]].



\bibitem{Jeon:2011sq} 
  I.~Jeon, K.~Lee and J.~H.~Park,
  ``Supersymmetric Double Field Theory: Stringy Reformulation of Supergravity,''
  Phys.\ Rev.\ D {\bf 85}, 081501 (2012)
  Erratum: [Phys.\ Rev.\ D {\bf 86}, 089903 (2012)]
  [arXiv:1112.0069 [hep-th]].


\bibitem{Jeon:2012hp} 
  I.~Jeon, K.~Lee, J.~H.~Park and Y.~Suh,
  ``Stringy Unification of Type IIA and IIB Supergravities under $N=2 D=10$ Supersymmetric Double Field Theory,''
  Phys.\ Lett.\ B {\bf 723}, 245 (2013)
  [arXiv:1210.5078 [hep-th]].


\bibitem{Jeon:2012kd}
  I.~Jeon, K.~Lee and J.~H.~Park,
  ``Ramond-Ramond Cohomology and $O(D,D)$ $T$-duality,''
  JHEP {\bf 1209} (2012) 079
  [arXiv:1206.3478 [hep-th]].

\bibitem{Hohm:2010xe} 
  O.~Hohm and S.~K.~Kwak,
  ``Frame-like Geometry of Double Field Theory,''
  J.\ Phys.\ A {\bf 44}, 085404 (2011)
  [arXiv:1011.4101 [hep-th]].



\bibitem{Jeon:2010rw} 
  I.~Jeon, K.~Lee and J.~H.~Park,
  ``Differential geometry with a projection: Application to double field theory,''
  JHEP {\bf 1104}, 014 (2011)
  [arXiv:1011.1324 [hep-th]].



\bibitem{Hohm:2011si} 
  O.~Hohm and B.~Zwiebach,
  ``On the Riemann Tensor in Double Field Theory,''
  JHEP {\bf 1205}, 126 (2012)
  [arXiv:1112.5296 [hep-th]].


\bibitem{Fukuma:1999jt} 
  M.~Fukuma, T.~Oota and H.~Tanaka,
  ``Comments on T dualities of Ramond-Ramond potentials on tori,''
  Prog.\ Theor.\ Phys.\  {\bf 103}, 425 (2000)
  [hep-th/9907132].


\bibitem{Bergshoeff:2001pv} 
  E.~Bergshoeff, R.~Kallosh, T.~Ortin, D.~Roest and A.~Van Proeyen,
  ``New formulations of $D = 10$ supersymmetry and D8 - O8 domain walls,''
  Class.\ Quant.\ Grav.\  {\bf 18}, 3359 (2001)
  [hep-th/0103233].

\bibitem{Siegel:1993th} 
  W.~Siegel,
  ``Superspace duality in low-energy superstrings,''
  Phys.\ Rev.\ D {\bf 48}, 2826 (1993)
  [hep-th/9305073].


\bibitem{Hull:2009mi}
  C.~Hull and B.~Zwiebach,
  ``Double Field Theory,''
  JHEP {\bf 0909} (2009) 099
  [arXiv:0904.4664 [hep-th]].


\bibitem{Hull:2009zb} 
  C.~Hull and B.~Zwiebach,
  ``The Gauge algebra of double field theory and Courant brackets,''
  JHEP {\bf 0909}, 090 (2009)
  [arXiv:0908.1792 [hep-th]].


\bibitem{Hohm:2012gk} 
  O.~Hohm and B.~Zwiebach,
  ``Large Gauge Transformations in Double Field Theory,''
  JHEP {\bf 1302}, 075 (2013)
  [arXiv:1207.4198 [hep-th]].


\bibitem{Hohm:2013bwa} 
  O.~Hohm, D.~L\"ust and B.~Zwiebach,
  ``The Spacetime of Double Field Theory: Review, Remarks, and Outlook,''
  Fortsch.\ Phys.\  {\bf 61}, 926 (2013)
  [arXiv:1309.2977 [hep-th]].



\bibitem{Hassan:1999mm} 
  S.~F.~Hassan,
  ``$SO(d,d)$ transformations of Ramond-Ramond fields and space-time spinors,''
  Nucl.\ Phys.\ B {\bf 583}, 431 (2000)
  [hep-th/9912236].

\bibitem{Hohm:2011dv}
  O.~Hohm, S.~K.~Kwak and B.~Zwiebach,
  ``Double Field Theory of Type II Strings,''
  JHEP {\bf 1109} (2011) 013
  [arXiv:1107.0008 [hep-th]].


\bibitem{Duff:1989tf} 
  M.~J.~Duff,
  ``Duality Rotations in String Theory,''
  Nucl.\ Phys.\ B {\bf 335}, 610 (1990).



\bibitem{Tseytlin:1990nb} 
  A.~A.~Tseytlin,
  ``Duality Symmetric Formulation of String World Sheet Dynamics,''
  Phys.\ Lett.\ B {\bf 242}, 163 (1990).



\bibitem{Tseytlin:1990va} 
  A.~A.~Tseytlin,
  ``Duality symmetric closed string theory and interacting chiral scalars,''
  Nucl.\ Phys.\ B {\bf 350}, 395 (1991).



\bibitem{Hull:2004in} 
  C.~M.~Hull,
  ``A Geometry for non-geometric string backgrounds,''
  JHEP {\bf 0510}, 065 (2005)
  [hep-th/0406102].



\bibitem{Hull:2006va} 
  C.~M.~Hull,
  ``Doubled Geometry and $T$-Folds,''
  JHEP {\bf 0707}, 080 (2007)
  [hep-th/0605149].



\bibitem{Copland:2011wx} 
  N.~B.~Copland,
  ``A Double Sigma Model for Double Field Theory,''
  JHEP {\bf 1204}, 044 (2012)
  [arXiv:1111.1828 [hep-th]].



\bibitem{Lee:2013hma} 
  K.~Lee and J.~H.~Park,
  ``Covariant action for a string in "doubled yet gauged" spacetime,''
  Nucl.\ Phys.\ B {\bf 880}, 134 (2014)
  [arXiv:1307.8377 [hep-th]].



\bibitem{Park:2016sbw} 
  J.~H.~Park,
  ``Green-Schwarz superstring on doubled-yet-gauged spacetime,''
  JHEP {\bf 1611}, 005 (2016)
  [arXiv:1609.04265 [hep-th]].



\bibitem{Blair:2013noa} 
  C.~D.~A.~Blair, E.~Malek and A.~J.~Routh,
  ``An $O(D, D)$ invariant Hamiltonian action for the superstring,''
  Class.\ Quant.\ Grav.\  {\bf 31}, no. 20, 205011 (2014)
  [arXiv:1308.4829 [hep-th]].



\bibitem{Bandos:2015cha} 
  I.~Bandos,
  ``Superstring in doubled superspace,''
  Phys.\ Lett.\ B {\bf 751}, 408 (2015)
  [arXiv:1507.07779 [hep-th]].


\bibitem{Driezen:2016tnz} 
  S.~Driezen, A.~Sevrin and D.~C.~Thompson,
  ``Aspects of the Doubled Worldsheet,''
  JHEP {\bf 1612}, 082 (2016)
  [arXiv:1609.03315 [hep-th]].



\bibitem{Bandos:2016jez} 
  I.~Bandos,
  ``Type II superstring in doubled superspace,''
  Fortsch.\ Phys.\  {\bf 64}, 361 (2016).


\bibitem{Geissbuhler:2013uka}
  D.~Geissbuhler, D.~Marques, C.~Nunez and V.~Penas,
  ``Exploring Double Field Theory,''
  JHEP {\bf 1306} (2013) 101
  [arXiv:1304.1472 [hep-th]].

\bibitem{Blumenhagen:2014gva}
  R.~Blumenhagen, F.~Hassler and D.~L\"ust,
  ``Double Field Theory on Group Manifolds,''
  JHEP {\bf 1502} (2015) 001
  [arXiv:1410.6374 [hep-th]].

\bibitem{Blumenhagen:2015zma}
  R.~Blumenhagen, P.~du Bosque, F.~Hassler and D.~Lust,
  ``Generalized Metric Formulation of Double Field Theory on Group Manifolds,''
  JHEP {\bf 1508} (2015) 056
  [arXiv:1502.02428 [hep-th]].



\bibitem{West:2001as}
  P.~C.~West,
  ``$E(11)$ and M theory,''
  Class.\ Quant.\ Grav.\  {\bf 18} (2001) 4443
  [hep-th/0104081].

\bibitem{West:2003fc}
  P.~C.~West,
  ``$E(11), SL(32)$ and central charges,''
  Phys.\ Lett.\ B {\bf 575} (2003) 333
  [hep-th/0307098].

\bibitem{West:2004st}
  P.~C.~West,
  ``The IIA, IIB and eleven-dimensional theories and their common $E(11)$ origin,''
  Nucl.\ Phys.\ B {\bf 693} (2004) 76
  [hep-th/0402140].

\bibitem{Hillmann:2009ci}
  C.~Hillmann,
  ``Generalized $E_{7(7)}$ coset dynamics and $D=11$ supergravity,''
  JHEP {\bf 0903} (2009) 135
  [arXiv:0901.1581 [hep-th]].

\bibitem{Berman:2010is}
  D.~S.~Berman and M.~J.~Perry,
  ``Generalized Geometry and M theory,''
  JHEP {\bf 1106} (2011) 074
  [arXiv:1008.1763 [hep-th]].

\bibitem{Berman:2011cg}
  D.~S.~Berman, H.~Godazgar, M.~Godazgar and M.~J.~Perry,
  ``The Local symmetries of M-theory and their formulation in generalised geometry,''
  JHEP {\bf 1201} (2012) 012
  [arXiv:1110.3930 [hep-th]].

\bibitem{Berman:2011jh}
  D.~S.~Berman, H.~Godazgar, M.~J.~Perry and P.~West,
  ``Duality Invariant Actions and Generalised Geometry,''
  JHEP {\bf 1202} (2012) 108
  [arXiv:1111.0459 [hep-th]].
  
\bibitem{Berman:2012vc}
  D.~S.~Berman, M.~Cederwall, A.~Kleinschmidt and D.~C.~Thompson,
  ``The gauge structure of generalised diffeomorphisms,''
  JHEP {\bf 1301} (2013) 064
  [arXiv:1208.5884 [hep-th]].

\bibitem{Hohm:2013pua}
  O.~Hohm and H.~Samtleben,
  ``Exceptional Form of $D=11$ Supergravity,''
  Phys.\ Rev.\ Lett.\  {\bf 111} (2013) 231601
  [arXiv:1308.1673 [hep-th]].

\bibitem{Hohm:2013vpa}
  O.~Hohm and H.~Samtleben,
  ``Exceptional Field Theory I: $E_{6(6)}$ covariant Form of M-Theory and Type IIB,''
  Phys.\ Rev.\ D {\bf 89} (2014) no.6,  066016
  [arXiv:1312.0614 [hep-th]].

\bibitem{Hohm:2013uia}
  O.~Hohm and H.~Samtleben,
  ``Exceptional field theory. II. E$_{7(7)}$,''
  Phys.\ Rev.\ D {\bf 89} (2014) 066017
  [arXiv:1312.4542 [hep-th]].

\bibitem{Aldazabal:2013via}
  G.~Aldazabal, M.~Gra\~na, D.~Marqu\'es and J.~A.~Rosabal,
  ``The gauge structure of Exceptional Field Theories and the tensor hierarchy,''
  JHEP {\bf 1404} (2014) 049
  [arXiv:1312.4549 [hep-th]].

\bibitem{Hohm:2014fxa}
  O.~Hohm and H.~Samtleben,
  ``Exceptional field theory. III. E$_{8(8)}$,''
  Phys.\ Rev.\ D {\bf 90} (2014) 066002
  [arXiv:1406.3348 [hep-th]].

\bibitem{Aldazabal:2010ef}
  G.~Aldazabal, E.~Andres, P.~G.~Camara and M.~Grana,
  ``U-dual fluxes and Generalized Geometry,''
  JHEP {\bf 1011} (2010) 083
  [arXiv:1007.5509 [hep-th]].
  
\bibitem{Chatzistavrakidis:2013jqa}
  A.~Chatzistavrakidis, F.~F.~Gautason, G.~Moutsopoulos and M.~Zagermann,
  ``Effective actions of nongeometric five-branes,''
  Phys.\ Rev.\ D {\bf 89} (2014) no.6,  066004
  [arXiv:1309.2653 [hep-th]].

\bibitem{Blair:2013gqa}
  C.~D.~A.~Blair, E.~Malek and J.~H.~Park,
  ``M-theory and Type IIB from a Duality Manifest Action,''
  JHEP {\bf 1401} (2014) 172
  [arXiv:1311.5109 [hep-th]].

\bibitem{Andriot:2014qla}
  D.~Andriot and A.~Betz,
  ``Supersymmetry with non-geometric fluxes, or a $\beta$-twist in Generalized Geometry and Dirac operator,''
  JHEP {\bf 1504} (2015) 006
  [arXiv:1411.6640 [hep-th]].

\bibitem{Blair:2014zba}
  C.~D.~A.~Blair and E.~Malek,
  ``Geometry and fluxes of $SL(5)$ exceptional field theory,''
  JHEP {\bf 1503} (2015) 144
  [arXiv:1412.0635 [hep-th]].

\bibitem{Sakatani:2014hba}
  Y.~Sakatani,
  ``Exotic branes and non-geometric fluxes,''
  JHEP {\bf 1503} (2015) 135
  [arXiv:1412.8769 [hep-th]].

\bibitem{Lee:2016qwn}
  K.~Lee, S.~J.~Rey and Y.~Sakatani,
  ``Effective Action for Non-Geometric Fluxes from Duality Covariant Actions,''
  arXiv:1612.08738 [hep-th].


\bibitem{Sakatani:2017nfr}
  Y.~Sakatani and S.~Uehara,
  ``Connecting M-theory and type IIB parameterizations in Exceptional Field Theory,''
  PTEP {\bf 2017} (2017) no.4,  043B05
  [arXiv:1701.07819 [hep-th]].


\bibitem{Matsumoto:2014ubv}
  T.~Matsumoto and K.~Yoshida,
  ``Yang-Baxter deformations and string dualities,''
  JHEP {\bf 1503} (2015) 137
  [arXiv:1412.3658 [hep-th]].


\bibitem{Schwarz:1995dk}
  J.~H.~Schwarz,
  ``An $SL(2,Z)$ multiplet of type IIB superstrings,''
  Phys.\ Lett.\ B {\bf 360} (1995) 13
   Erratum: [Phys.\ Lett.\ B {\bf 364} (1995) 252]
  [hep-th/9508143].




\bibitem{Kawaguchi:2011mz} 
  I.~Kawaguchi, D.~Orlando and K.~Yoshida,
  ``Yangian symmetry in deformed WZNW models on squashed spheres,''
  Phys.\ Lett.\ B {\bf 701}, 475 (2011)
  [arXiv:1104.0738 [hep-th]].


\bibitem{Kawaguchi:2013gma}
  I.~Kawaguchi and K.~Yoshida,
  ``A deformation of quantum affine algebra in squashed Wess-Zumino-Novikov-Witten models,''
  J.\ Math.\ Phys.\  {\bf 55} (2014) 062302
  [arXiv:1311.4696 [hep-th]].


\bibitem{Delduc:2014uaa} 
  F.~Delduc, M.~Magro and B.~Vicedo,
  ``Integrable double deformation of the principal chiral model,''
  Nucl.\ Phys.\ B {\bf 891}, 312 (2015)
  [arXiv:1410.8066 [hep-th]].


\bibitem{Delduc:2017fib} 
  F.~Delduc, B.~Hoare, T.~Kameyama and M.~Magro,
  ``Combining the bi-Yang-Baxter deformation, the Wess-Zumino term and $TsT$ transformations in one integrable $\sigma$-model,''
  JHEP {\bf 1710}, 212 (2017)
  [arXiv:1707.08371 [hep-th]].


\bibitem{Demulder:2017zhz} 
  S.~Demulder, S.~Driezen, A.~Sevrin and D.~C.~Thompson,
  ``Classical and Quantum Aspects of Yang-Baxter Wess-Zumino Models,''
  JHEP {\bf 1803}, 041 (2018)
  [arXiv:1711.00084 [hep-th]].



\bibitem{Forste:1994wp} 
  S.~Forste,
  ``A Truly marginal deformation of $SL(2, R)$ in a null direction,''
  Phys.\ Lett.\ B {\bf 338}, 36 (1994)
  [hep-th/9407198].



\bibitem{Israel:2003ry} 
  D.~Israel, C.~Kounnas and M.~P.~Petropoulos,
  ``Superstrings on NS5 backgrounds, deformed AdS$_3$ and holography,''
  JHEP {\bf 0310}, 028 (2003)
  [hep-th/0306053].



\bibitem{Giveon:2017nie} 
  A.~Giveon, N.~Itzhaki and D.~Kutasov,
  ``$ \mathrm{T}\overline{\mathrm{T}} $ and LST,''
  JHEP {\bf 1707}, 122 (2017)
  [arXiv:1701.05576 [hep-th]].



\bibitem{Giveon:2017myj} 
  A.~Giveon, N.~Itzhaki and D.~Kutasov,
  ``A solvable irrelevant deformation of AdS$_{3}$/CFT$_{2}$,''
  JHEP {\bf 1712}, 155 (2017)
  [arXiv:1707.05800 [hep-th]].



\bibitem{Asrat:2017tzd} 
  M.~Asrat, A.~Giveon, N.~Itzhaki and D.~Kutasov,
  ``Holography Beyond AdS,''
  arXiv:1711.02690 [hep-th].



\bibitem{Giribet:2017imm} 
  G.~Giribet,
  ``$T\bar{T}$-deformations, AdS/CFT and correlation functions,''
  JHEP {\bf 1802}, 114 (2018)
  [arXiv:1711.02716 [hep-th]].



\bibitem{Azeyanagi:2012zd} 
  T.~Azeyanagi, D.~M.~Hofman, W.~Song and A.~Strominger,
  ``The Spectrum of Strings on Warped AdS$_3$ x S$^3$,''
  JHEP {\bf 1304}, 078 (2013)
  [arXiv:1207.5050 [hep-th]].



\bibitem{Giveon:1993ph} 
  A.~Giveon and E.~Kiritsis,
  ``Axial vector duality as a gauge symmetry and topology change in string theory,''
  Nucl.\ Phys.\ B {\bf 411}, 487 (1994)
  [hep-th/9303016].


\bibitem{Araujo:2017jkb} 
  T.~Araujo, I.~Bakhmatov, E.~\'O.~Colg\'ain, J.~Sakamoto, M.~M.~Sheikh-Jabbari and K.~Yoshida,
  ``Yang-Baxter $\sigma$-models, conformal twists, and noncommutative Yang-Mills theory,''
  Phys.\ Rev.\ D {\bf 95}, no. 10, 105006 (2017)
  [arXiv:1702.02861 [hep-th]].


\bibitem{Araujo:2017jap} 
  T.~Araujo, I.~Bakhmatov, E.~\'O.~Colg\'ain, J.~Sakamoto, M.~M.~Sheikh-Jabbari and K.~Yoshida,
  ``Conformal Twists, Yang-Baxter $\sigma$-models \& Holographic Noncommutativity,''
  arXiv:1705.02063 [hep-th].



\bibitem{Araujo:2017enj} 
  T.~Araujo, E.~\'O Colg\'ain, J.~Sakamoto, M.~M.~Sheikh-Jabbari and K.~Yoshida,
  ``$I$ in generalized supergravity,''
  Eur.\ Phys.\ J.\ C {\bf 77}, no. 11, 739 (2017)
  [arXiv:1708.03163 [hep-th]].


\bibitem{Wulff:2018aku} 
  L.~Wulff,
  ``Trivial solutions of generalized supergravity vs non-abelian T-duality anomaly,''
  arXiv:1803.07391 [hep-th].


\bibitem{Hull:1989jk} 
  C.~M.~Hull and B.~J.~Spence,
  ``The Gauged Nonlinear $\sigma$ Model With {Wess-Zumino} Term,''
  Phys.\ Lett.\ B {\bf 232}, 204 (1989).



\bibitem{Hull:1990ms} 
  C.~M.~Hull and B.~J.~Spence,
  ``The Geometry of the gauged sigma model with Wess-Zumino term,''
  Nucl.\ Phys.\ B {\bf 353}, 379 (1991).




\bibitem{Shelton:2005cf}
  J.~Shelton, W.~Taylor and B.~Wecht,
  ``Nongeometric flux compactifications,''
  JHEP {\bf 0510} (2005) 085
  [hep-th/0508133].


\bibitem{Kachru:2002sk}
  S.~Kachru, M.~B.~Schulz, P.~K.~Tripathy and S.~P.~Trivedi,
  ``New supersymmetric string compactifications,''
  JHEP {\bf 0303} (2003) 061
  [hep-th/0211182].


\bibitem{Hassler:2013wsa}
  F.~Hassler and D.~Lust,
  ``Non-commutative/non-associative IIA (IIB) $Q$- and $R$-branes and their intersections,''
  JHEP {\bf 1307} (2013) 048
  [arXiv:1303.1413 [hep-th]].


\bibitem{Dibitetto:2012rk}
  G.~Dibitetto, J.~J.~Fernandez-Melgarejo, D.~Marques and D.~Roest,
  ``Duality orbits of non-geometric fluxes,''
  Fortsch.\ Phys.\  {\bf 60} (2012) 1123
  [arXiv:1203.6562 [hep-th]].



\bibitem{Shapere:1988zv}
  A.~D.~Shapere and F.~Wilczek,
  ``Selfdual Models with Theta Terms,''
  Nucl.\ Phys.\ B {\bf 320} (1989) 669.


\bibitem{Giveon:1988tt}
  A.~Giveon, E.~Rabinovici and G.~Veneziano,
  ``Duality in String Background Space,''
  Nucl.\ Phys.\ B {\bf 322} (1989) 167.

\bibitem{Giveon:1994fu}
  A.~Giveon, M.~Porrati and E.~Rabinovici,
  ``Target space duality in string theory,''
  Phys.\ Rept.\  {\bf 244} (1994) 77
  [hep-th/9401139].


\bibitem{Seiberg:1999vs}
  N.~Seiberg and E.~Witten,
  ``String theory and noncommutative geometry,''
  JHEP {\bf 9909} (1999) 032
  [hep-th/9908142].



\bibitem{Grana:2008yw}
  M.~Grana, R.~Minasian, M.~Petrini and D.~Waldram,
  ``T-duality, Generalized Geometry and Non-Geometric Backgrounds,''
  JHEP {\bf 0904} (2009) 075
  [arXiv:0807.4527 [hep-th]].


\bibitem{LozanoTellechea:2000mc}
  E.~Lozano-Tellechea and T.~Ortin,
  ``$7$-branes and higher Kaluza-Klein branes,''
  Nucl.\ Phys.\ B {\bf 607} (2001) 213
  [hep-th/0012051].


\bibitem{deBoer:2010ud}
  J.~de Boer and M.~Shigemori,
  ``Exotic branes and non-geometric backgrounds,''
  Phys.\ Rev.\ Lett.\  {\bf 104} (2010) 251603
  [arXiv:1004.2521 [hep-th]].


\bibitem{Tseytlin:1994ei}
  A.~A.~Tseytlin,
  ``Melvin solution in string theory,''
  Phys.\ Lett.\ B {\bf 346} (1995) 55
  [hep-th/9411198].


\bibitem{Gibbons:1987ps}
  G.~W.~Gibbons and K.~Maeda,
  ``Black Holes and Membranes in Higher Dimensional Theories with Dilaton Fields,''
  Nucl.\ Phys.\ B {\bf 298} (1988) 741.


\bibitem{Hashimoto:2004pb}
  A.~Hashimoto and K.~Thomas,
  ``Dualities, twists, and gauge theories with non-constant non-commutativity,''
  JHEP {\bf 0501} (2005) 033
  [hep-th/0410123].


\bibitem{Malek:2013sp}
  E.~Malek,
  ``Timelike U-dualities in Generalised Geometry,''
  JHEP {\bf 1311} (2013) 185
  [arXiv:1301.0543 [hep-th]].


\bibitem{Malek:2012pw}
  E.~Malek,
  ``U-duality in three and four dimensions,''
  Int.\ J.\ Mod.\ Phys.\ A {\bf 32} (2017) no.27,  1750169
  [arXiv:1205.6403 [hep-th]].


\bibitem{Morand:2017fnv}
  K.~Morand and J.~H.~Park,
  ``Classification of non-Riemannian doubled-yet-gauged spacetime,''
  Eur.\ Phys.\ J.\ C {\bf 77} (2017) no.10,  685
  [arXiv:1707.03713 [hep-th]].


\bibitem{Gasperini:1993nz} 
  M.~Gasperini, R.~Ricci and G.~Veneziano,
  ``A Problem with nonAbelian duality?,''
  Phys.\ Lett.\ B {\bf 319}, 438 (1993)
  [hep-th/9308112].





\bibitem{Hull:2006qs}
  C.~M.~Hull,
  ``Global aspects of T-duality, gauged sigma models and T-folds,''
  JHEP {\bf 0710} (2007) 057
  [hep-th/0604178].


\bibitem{Wald}
  R.~M.~Wald,
  “General Relativity,”
  University of Chicago Press (Chicago, 1984).


\bibitem{Hohm:2011zr}
  O.~Hohm, S.~K.~Kwak and B.~Zwiebach,
  ``Unification of Type II Strings and T-duality,''
  Phys.\ Rev.\ Lett.\  {\bf 107} (2011) 171603
  [arXiv:1106.5452 [hep-th]].


\bibitem{Geissbuhler:2011mx}
  D.~Geissbuhler,
  ``Double Field Theory and N=4 Gauged Supergravity,''
  JHEP {\bf 1111} (2011) 116
  [arXiv:1109.4280 [hep-th]].


\bibitem{Hull:2014mxa}
  C.~M.~Hull,
  ``Finite Gauge Transformations and Geometry in Double Field Theory,''
  JHEP {\bf 1504} (2015) 109
  [arXiv:1406.7794 [hep-th]].






\bibitem{Callan:1985ia} 
  C.~G.~Callan, Jr., E.~J.~Martinec, M.~J.~Perry and D.~Friedan,
  ``Strings in Background Fields,''
  Nucl.\ Phys.\ B {\bf 262}, 593 (1985).


\bibitem{Hull:1985rc}
  C.~M.~Hull and P.~K.~Townsend,
  ``Finiteness and Conformal Invariance in Nonlinear $\sigma$ Models,''
  Nucl.\ Phys.\ B {\bf 274} (1986) 349.


\bibitem{Berman:2007xn}
  D.~S.~Berman, N.~B.~Copland and D.~C.~Thompson,
  ``Background Field Equations for the Duality Symmetric String,''
  Nucl.\ Phys.\ B {\bf 791} (2008) 175
  [arXiv:0708.2267 [hep-th]].



\bibitem{Berman:2007yf}
  D.~S.~Berman and D.~C.~Thompson,
  ``Duality Symmetric Strings, Dilatons and $O(d,d)$ Effective Actions,''
  Phys.\ Lett.\ B {\bf 662} (2008) 279
  [arXiv:0712.1121 [hep-th]].



\bibitem{Copland:2011yh}
  N.~B.~Copland,
  ``Connecting T-duality invariant theories,''
  Nucl.\ Phys.\ B {\bf 854} (2012) 575
  [arXiv:1106.1888 [hep-th]].


\bibitem{Curci:1986hi}
  G.~Curci and G.~Paffuti,
  ``Consistency Between the String Background Field Equation of Motion and the Vanishing of the Conformal Anomaly,''
  Nucl.\ Phys.\ B {\bf 286} (1987) 399.


\bibitem{Hong:2018tlp} 
  M.~Hong, Y.~Kim and E.~\'O Colg\'ain,
  ``On non-Abelian T-duality for non-semisimple groups,''
  arXiv:1801.09567 [hep-th].


\bibitem{Hohm:2010pp} 
  O.~Hohm, C.~Hull and B.~Zwiebach,
  ``Generalized metric formulation of double field theory,''
  JHEP {\bf 1008}, 008 (2010)
  [arXiv:1006.4823 [hep-th]].


\bibitem{Deser:1995ne}
  S.~Deser and R.~Jackiw,
  ``Energy momentum tensor improvements in two-dimensions,''
  Int.\ J.\ Mod.\ Phys.\ B {\bf 10} (1996) 1499
  [hep-th/9510145].


\bibitem{Yale:2010jy}
  A.~Yale and T.~Padmanabhan,
  ``Structure of Lanczos-Lovelock Lagrangians in Critical Dimensions,''
  Gen.\ Rel.\ Grav.\  {\bf 43} (2011) 1549
  [arXiv:1008.5154 [gr-qc]].


\bibitem{Elitzur:1994ri} 
  S.~Elitzur, A.~Giveon, E.~Rabinovici, A.~Schwimmer and G.~Veneziano,
  ``Remarks on nonAbelian duality,''
  Nucl.\ Phys.\ B {\bf 435}, 147 (1995)
  [hep-th/9409011].




\bibitem{Dei:2018mfl}
  A.~Dei and A.~Sfondrini,
  ``Integrable spin chain for stringy Wess-Zumino-Witten models,''
  JHEP {\bf 1807} (2018) 109
  [arXiv:1806.00422 [hep-th]].

\bibitem{Maldacena:2000hw}
  J.~M.~Maldacena and H.~Ooguri,
  ``Strings in AdS(3) and SL(2,R) WZW model 1.: The Spectrum,''
  J.\ Math.\ Phys.\  {\bf 42} (2001) 2929
  [hep-th/0001053].


\bibitem{Bergshoeff:2002nv}
  E.~Bergshoeff, T.~de Wit, U.~Gran, R.~Linares and D.~Roest,
  ``(Non)Abelian gauged supergravities in nine-dimensions,''
  JHEP {\bf 0210} (2002) 061
  [hep-th/0209205].


\bibitem{FernandezMelgarejo:2011wx}
  J.~J.~Fernandez-Melgarejo, T.~Ortin and E.~Torrente-Lujan,
  ``The general gaugings of maximal d=9 supergravity,''
  JHEP {\bf 1110} (2011) 068
  [arXiv:1106.1760 [hep-th]].


\bibitem{Arutyunov:2009ga} 
  G.~Arutyunov and S.~Frolov,
  ``Foundations of the AdS$_5$ x S$^5$ Superstring. Part I,''
  J.\ Phys.\ A {\bf 42}, 254003 (2009)
  [arXiv:0901.4937 [hep-th]].


\bibitem{Castellani:1991et} 
  L.~Castellani, R.~D'Auria and P.~Fre,
  ``Supergravity and superstrings: A Geometric perspective. Vol. 1: Mathematical foundations,''
  Singapore, Singapore: World Scientific (1991) 1-603


\bibitem{Ortin:2004ms} 
  T.~Ortin,
  ``Gravity and strings,''
  Cambridge University Press (2004).

\bibitem{Penrose}
R.~Penrose, ``Any space-time has a plane wave as a limit,''
  Differential Geometry and Relativity: A Volume in Honour of Andr{\'e} Lichnerowicz on His
  60th Birthday (1976)  271--275..

\bibitem{Gueven:2000ru}
   R.~Gueven, 
   ``Plane wave limits and T duality,''
   Phys. Lett. {\bf B482} (2000)  255--263,
   arXiv:hep-th/0005061[hep-th].




%
%
%

%
%
%
%
%
%
%
%
%
%
%
%
%

\end{thebibliography}
\end{document}